\documentclass[longauth]{aa}  
\usepackage{natbib}
\usepackage{graphicx, color}		% wanted by A&A
\usepackage{rotating}		% sideways of figures & tables
\usepackage{lscape}		% landscape pages
\usepackage{amssymb,amsmath}	% AMS Math
\usepackage{morefloats}		% enhance no. of floats
\usepackage{hyperref}		% To add links in your PDF file, use the package "hyperref"
\usepackage{enumitem}
\usepackage{multicol}
\usepackage{txfonts}		% wanted by A&A
\usepackage{graphbox}		% to align figures vertically (option align=c in includegraphics)
\usepackage{dcolumn}        % for special columns alignment in tables

\bibpunct{(}{)}{;}{a}{}{,} 	% to follow the A&A style 

\mathchardef\mhyphen="2D

\extrafloats{200}

\newlength{\imagewidth}
\newlength{\imageheight}

\begin{document}
%%%%%%%%%%%%%%%%%%%%%%%%%%%%%%%%%%%%%%%%%
%	Title & authors			%
%%%%%%%%%%%%%%%%%%%%%%%%%%%%%%%%%%%%%%%%%

\title{A panchromatic view of N2CLS GOODS-N: The evolution of the dust cosmic density since $z\sim7$}

\author{S.~Berta \inst{\ref{IRAMF}}\thanks{\email{berta@iram.fr}}
        \and  G.~Lagache \inst{\ref{LAM}}
        \and  A.~Beelen \inst{\ref{LAM}}
        \and  R.~Adam \inst{\ref{OCA}}
        \and  P.~Ade \inst{\ref{Cardiff}}
        \and  H.~Ajeddig \inst{\ref{CEA}}
        \and  S.~Amarantidis \inst{\ref{IRAME}}
        \and  P.~Andr\'e \inst{\ref{CEA}}
        \and  H.~Aussel \inst{\ref{CEA}}
        \and  A.~Beno\^it \inst{\ref{Neel}}
        \and  M.~Bethermin \inst{\ref{UniStra}}
        \and  L.-J.~Bing \inst{\ref{Sussex}}
        \and  A.~Bongiovanni \inst{\ref{IRAME}}
        \and  J.~Bounmy \inst{\ref{LPSC}}
        \and  O.~Bourrion \inst{\ref{LPSC}}
        \and  M.~Calvo \inst{\ref{Neel}}
        \and  A.~Catalano \inst{\ref{LPSC}}
        \and  D.~Ch\'erouvrier \inst{\ref{LPSC}}
        \and  L.~Ciesla \inst{\ref{LAM}}
        \and  M.~De~Petris \inst{\ref{Roma}}
        \and  F.-X.~D\'esert \inst{\ref{IPAG}}
        \and  S.~Doyle \inst{\ref{Cardiff}}
        \and  E.~F.~C.~Driessen \inst{\ref{IRAMF}}
        \and  G.~Ejlali \inst{\ref{Teheran}}
        \and  D.~Elbaz \inst{\ref{CEA}}
        \and  A.~Ferragamo \inst{\ref{Roma}, \ref{Napoli}}
        \and  A.~Gomez \inst{\ref{CAB}} 
        \and  J.~Goupy \inst{\ref{Neel}}
        \and  C.~Hanser \inst{\ref{LPSC}}
        \and  S.~Katsioli \inst{\ref{AthenObs}, \ref{AthenUniv}}
        \and  F.~K\'eruzor\'e \inst{\ref{Argonne}}
        \and  C.~Kramer \inst{\ref{IRAMF}}
        \and  B.~Ladjelate \inst{\ref{IRAME}} 
        \and  S.~Leclercq \inst{\ref{IRAMF}}
        \and  J.-F.~Lestrade \inst{\ref{LERMA}}
        \and  J.~F.~Mac\'ias-P\'erez \inst{\ref{LPSC}}
        \and  S.~C.~Madden \inst{\ref{CEA}}
        \and  A.~Maury \inst{\ref{ICE}}, \inst{\ref{ICREA}}, \inst{\ref{CEA}}
        \and  F.~Mayet \inst{\ref{LPSC}}
        \and  H.~Messias \inst{\ref{ALMACL}, \ref{ESOCL}}
        \and  A.~Monfardini \inst{\ref{Neel}}
        \and  A.~Moyer-Anin \inst{\ref{LPSC}}
        \and  M.~Mu\~noz-Echeverr\'ia \inst{\ref{IRAP}}
        \and  I.~Myserlis \inst{\ref{IRAME}}
        \and  R.~Neri \inst{\ref{IRAMF}} 
        \and  A.~Paliwal \inst{\ref{Roma}}
        \and  L.~Perotto \inst{\ref{LPSC}}
        \and  G.~Pisano \inst{\ref{Roma}}
        \and  N.~Ponthieu \inst{\ref{IPAG}}
        \and  V.~Rev\'eret \inst{\ref{CEA}}
        \and  A.~J.~Rigby \inst{\ref{Leeds}}
        \and  A.~Ritacco \inst{\ref{TorVergata}, \ref{ENS}}  %%%\ref{INAF_CA}
        \and  H.~Roussel \inst{\ref{IAP}}
        \and  F.~Ruppin \inst{\ref{IP2I}}
        \and  M.~S\'anchez-Portal \inst{\ref{IRAME}}
        \and  S.~Savorgnano \inst{\ref{LPSC}}
        \and  K.~Schuster \inst{\ref{IRAMF}}
        \and  A.~Sievers \inst{\ref{IRAME}}
        \and  C.~Tucker \inst{\ref{Cardiff}}
        \and  M.-Y.~Xiao \inst{\ref{UniGe}}
        \and  R.~Zylka \inst{\ref{IRAMF}}
}         

% P. Oesch,  M.-Y. Xiao, A. Weibel: they provided the JWST data and did the source extraction.
% Oesch & Weibel: do not want to be co-author.
% L. Ciesla: She did the CIGALE SED fitting of all sources.
% R. Neri: He reduced the NOEMA follow-up data
% H. Messias: He helped developing the SED fitting code SED3FIT-hz
% D. Elbaz: for his contribution in the PNCG & telescope proposals.
% Longji
% Matthieu

%
%
% 
\institute{
     Institut de Radioastronomie Millim\'etrique (IRAM), 300 rue de la Piscine, 38400 Saint-Martin-d'H{\`e}res, France
     \label{IRAMF}
     \and 
     Aix Marseille Univ., CNRS, CNES, LAM (Laboratoire d'Astrophysique de Marseille), Marseille, France
     \label{LAM}
     \and
     Universit\'e C\^ote d'Azur, Observatoire de la C\^ote d'Azur, CNRS, Laboratoire Lagrange, France 
     \label{OCA}
     \and
     School of Physics and Astronomy, Cardiff University, Queen’s Buildings, The Parade, Cardiff, CF24 3AA, UK 
     \label{Cardiff}
     \and
     Universit\'e Paris-Saclay, Universit\'e Paris Cit\'e, CEA, CNRS, AIM, 91191, Gif-sur-Yvette, France
     \label{CEA}
     \and
     Institut de Radioastronomie Millim\'etrique (IRAM), Avenida Divina Pastora 7, Local 20, E-18012, Granada, Spain
     \label{IRAME}
     \and
     Institut N\'eel, CNRS, Universit\'e Grenoble Alpes, France
     \label{Neel}
     \and
     Universit\'e de Strasbourg, CNRS, Observatoire astronomique de Strasbourg, UMR 7550, 67000 Strasbourg, France
     \label{UniStra}
     \and
     Astronomy Centre, Department of Physics and Astronomy, University of Sussex, Brighton BN1 9QH
     \label{Sussex}
     \and
     Univ. Grenoble Alpes, CNRS, Grenoble INP, LPSC-IN2P3, 53, avenue des Martyrs, 38000 Grenoble, France
     \label{LPSC}
     \and	
     Dipartimento di Fisica, Sapienza Universit\`a di Roma, Piazzale Aldo Moro 5, I-00185 Roma, Italy
     \label{Roma}
     \and
     Univ. Grenoble Alpes, CNRS, IPAG, 38000 Grenoble, France 
     \label{IPAG}
     \and
     Institute for Research in Fundamental Sciences (IPM), School of Astronomy, Tehran, Iran
     \label{Teheran}
     \and
     Physics Department ``Ettore Pancini'', Universit\`a degli Studi di Napoli ``Federico II'', Via Cintia 21, I-80126 Napoli, Italy
     \label{Napoli}
     \and
     Centro de Astrobiolog\'ia (CSIC-INTA), Torrej\'on de Ardoz, 28850 Madrid, Spain
     \label{CAB}
     \and
     National Observatory of Athens, Institute for Astronomy, Astrophysics, Space Applications and Remote Sensing, Ioannou Metaxa
     and Vasileos Pavlou GR-15236, Athens, Greece
     \label{AthenObs}
     \and
     Department of Astrophysics, Astronomy \& Mechanics, Faculty of Physics, University of Athens, Panepistimiopolis, GR-15784
     Zografos, Athens, Greece
     \label{AthenUniv}
     \and
     High Energy Physics Division, Argonne National Laboratory, 9700 South Cass Avenue, Lemont, IL 60439, USA
     \label{Argonne}
     \and  
     LERMA, Observatoire de Paris, PSL Research University, CNRS, Sorbonne Universit\'e, UPMC, 75014 Paris, France  
     \label{LERMA}
     \and
     Institute of Space Sciences (ICE), CSIC, Campus UAB, Carrer de Can Magrans s/n, E-08193 Barcelona, Spain
     \label{ICE}
     \and
     ICREA, Pg. Llu\'\i{}s Companys 23, Barcelona, Spain
     \label{ICREA}
     \and
     Joint ALMA Observatory, Alonso de C\'ordova 3107, Vitacura 763-0355, Santiago, Chile
     \label{ALMACL}
     \and
     European Southern Observatory, Alonso de C\'ordova 3107, Vitacura, Casilla 19001, Santiago de Chile, Chile
     \label{ESOCL}
     \and
     IRAP, CNRS, Universit\'e de Toulouse, CNES, UT3-UPS, (Toulouse), France 
     \label{IRAP}
     \and
     School of Physics and Astronomy, University of Leeds, Leeds LS2 9JT, UK
     \label{Leeds}
     \and
     Dipartimento di Fisica, Universit\`a di Roma ``Tor Vergata'', Via della Ricerca Scientifica 1, I-00133 Roma, Italy
     \label{TorVergata}
     \and
     Laboratoire de Physique de l’\'Ecole Normale Sup\'erieure, ENS, PSL Research University, CNRS, Sorbonne Universit\'e, Universit\'e de Paris, 75005 Paris, France 
     \label{ENS}
     \and
     Institut d'Astrophysique de Paris, CNRS (UMR7095), 98 bis boulevard Arago, 75014 Paris, France
     \label{IAP}
     \and
     University of Lyon, UCB Lyon 1, CNRS/IN2P3, IP2I, 69622 Villeurbanne, France
     \label{IP2I}
     \and 
     Department of Astronomy, University of Geneva, Chemin Pegasi 51, 1290 Versoix, Switzerland
     \label{UniGe}
}

     %%% OLD Ritacco
     %INAF-Osservatorio Astronomico di Cagliari, Via della Scienza 5, 09047 Selargius, IT
     %\label{INAF_CA}
     %\and    

\date{Received 5 November 2024 / Accepted 27 February 2025}

% -----------------------------------------------------------------------

\abstract{

%%%%% max 300 words
%%%% SB 2025/02/01: I have made the abstract to be 300 words again
%%%% SB 2025/03/08 after language editing, the number of words is no more <=300

To understand early star formation, it is essential to determine the dust mass budget of high-redshift galaxies. Sub-millimeter rest-frame emission, dominated by cold dust, is an unbiased tracer of dust mass.
The New IRAM KID Arrays 2 (NIKA2) conducted a deep blank field survey at 1.2 and 2.0\,mm in the GOODS-N field as part of the NIKA2 Cosmological Legacy Survey (N2CLS), detecting 65 sources with $S/N\ge4.2$. Thanks to a dedicated interferometric program with NOEMA and other high-angular resolution data, we identified the multi-wavelength counterparts of these sources and resolved them into 71 individual galaxies. 
We built detailed spectral energy distributions (SEDs) and assigned a redshift to 68 of them over the range $0.6<z<7.2$. %, with median $z=2.819$. 
We fit these SEDs using modified blackbody and \citet{DL07} models and the panchromatic approaches MAGPHYS, CIGALE, and SED3FIT, thus deriving their dust mass ($M_\textrm{dust}$), infrared luminosity ($L_\textrm{IR}$), and stellar mass ($M_\star$). 
Eight galaxies require an active galactic nucleus torus component, and another six require an unextinguished young stellar population. A significant fraction of our galaxies are classified as starbursts based on their position on the $M_\star$ versus star formation rate plane or their depletion timescales.
We computed the dust mass function in three redshift bins ($1.6<z\le2.4$, $2.4<z\le4.2$ and $4.2<z\le7.2$) and determined the Schechter function that best describes it. The dust cosmic density, $\rho_\textrm{dust}$, increases by at least an order of magnitude from $z\sim7$ to $z\sim1.5$, as predicted by theoretical works. At lower redshifts, the evolution flattens. Nonetheless, significant differences exist between results obtained with different selections and methods. 
The superb GOODS-N data set enabled a systematic investigation into the dust properties of distant galaxies. N2CLS holds promise for combining these deep field findings with the wide COSMOS field into a self-consistent analysis of dust in galaxies both near and far.
}


\authorrunning{S.~Berta et al.}

\titlerunning{Panchromatic N2GN dust}

\maketitle

% -----------------------------------------------------------------------

\section{Introduction}\label{sect:intro}

Dust plays a central role in galaxy evolution. It functions as a catalyst in transforming atomic hydrogen into the molecular hydrogen from which stars are formed, while chemical reactions on the surface of dust grains define the structure of the interstellar medium \citep[ISM; e.g.,][]{hollenbach1971, mathis1990, wolfire1995, wd01, draine2003}. Dust is mainly produced in the envelopes of asymptotic giant branch stars \citep[AGB;][]{gehrz1989, ferrarotti2006, sargent2010, nanni2013, schneider2014} and at the end of the life of massive stars during supernovae (SNe) explosions \citep{todini2001, rho2008, dunne2009}. On the other hand, SNe shocks destroy dust grains \citep{draine1979, mckee1987, jones1994, bianchi2005, nozawa2007}, which can form again in the ISM through an accretion process \citep{tielens1998, zhukovska2008}. 
Dust can also be ejected from galaxies by powerful winds, providing an additional cooling channel and contributing to the metal enrichment of the intergalactic medium \citep[IGM; e.g.,][]{ostriker1973, bouche2007}.

Dust absorbs the ultraviolet (UV) emission of young OB-type stars, further acting as a coolant to condense gas and form new stars. This radiation is reemitted at long wavelengths in the mid-infrared (MIR) and far-infrared (FIR) spectral domains up to the millimeter-wavelength regime
\citep[e.g.,][]{puget1996,fixsen1998,dole2006,driver2008}.
Thus, it has become clear that studying the cosmic dust budget and how it has evolved from the primordial Universe to the current epoch plays a key role in understanding the  evolution of galaxies through cosmic history.
So far, not many studies have been dedicated to the statistics of the dust content of galaxies. Direct measurements of the dust mass function (DMF) of galaxies are hindered by the difficulty of performing an unbiased selection on the basis of their dust mass.

The first attempts to measure the DMF in the local Universe were those by \citet{dunne2000} and \citet{dunne2001}, who used a sample of IRAS bright galaxies observed with the Sub-millimeter Common User Bolometer Array (SCUBA) at 450 and 850 $\mu$m, and by \citet{vlahakis2005}, who added a sample of optically selected galaxies. These early studies were extended up to redshift $z\sim0.5$ by \citet{dunne2011}, exploiting a sample of 250 $\mu$m-selected {\it Herschel}-sources with a reliable counterpart in the Sloan Digital Sky Survey catalog.
It was not until the works by \citet{driver2018} and \citet{pozzi2020} that the study of the DMF reached redshift $z\sim2$. The former computed the DMF of more than $5\times10^5$ optical and near-infrared (NIR) selected galaxies at $z<1.75$ in {\it Herschel} extragalactic surveys. The latter studied the DMF of 160 $\mu$m selected {\it Herschel} galaxies in the redshift range from $z=0.2$ to 2.5. 

All the works mentioned so far based the determination of the dust mass in galaxies, the DMF, and the dust cosmic density on fitting the FIR-millimeter spectral energy distributions (SEDs) of galaxies with a model describing their dust emission.  
Several alternative approaches have also been employed. \citet{fukugita2011} computed the local dust cosmic density by integrating the star formation rate density (SFRD). \citet{menard2012} measured the amount of dust residing in Mg{\sc ii} absorbers along the lines of sight to distant quasars. \citet{peroux2020} derived the comoving dust density of galaxies from the gas cosmic density, adopting a dust-to-gas mass ratio.

An unbiased and direct dust-driven selection implies using an observable tracer univocally linked to the amount of dust in galaxies. Such an indicator must trace the bulk of the dust mass, minimizing losses or biases. A natural choice is the rest-frame sub-millimeter or millimeter emission of galaxies tracing the cold dust component that dominates their dust mass budget. So far, few studies have attempted to determine the DMF and comoving dust density in this way, especially at high redshift.
\citet{dudzeviciute2021} based their study on an SED fitting of 450 and 850 $\mu$m selected galaxies detected by SCUBA2 and the Atacama Large Millimeter/submillimeter Array (ALMA). \citet{magnelli2020} studied the contribution of NIR $H$ band selected galaxies to the cosmic dust density up to $z=3.2$ by performing stacking on deep 1.2 mm ALMA maps. \citet{pozzi2021} followed a similar approach for UV-selected galaxies at $z=4.4-5.9$ in the ALMA/ALPINE survey 
(\citealt{lefevre2020}). Recently, \citet{eales2024} exploited a stacking of optically selected galaxies on 850 $\mu$m SCUBA-2 maps \citep{millard2020} to study the evolution of dust up to $z=5.5$.

In order to access the rest-frame sub-millimeter emission of dust in high-$z$ galaxies, it is necessary to carry out observations in the millimeter wavelength domain. \citet{traina2024} derived the DMF of 189 galaxies from the ALMA A$^3$COSMOS collection \citep{liu2019b,adscheid2024}, finding a smooth evolution of the dust mass density over the whole redshift range $0.5<z\le6.0$, which is at odds with the sudden drop at $z>3$ predicted by models. 

The New IRAM KID Arrays 2 \citep[NIKA2;][]{perotto2020, adam2018, calvo2016, bourrion2016, monfardini2014} is a dual-band millimeter continuum camera operating at 1.2 and 2.0~mm (260 and 150 GHz) installed at the IRAM 30~m telescope in Spain. It consists of three arrays of kinetic inductance detectors, two at 1.2~mm (with 1140 detectors each) and one at 2.0~mm (616 detectors), covering an effective field of view of $\sim6.5$\,arcmin in diameter.

As part of the guaranteed time program, the NIKA2 Cosmological Legacy Survey (N2CLS) dedicated 300 hours to the observation of the GOODS-N and COSMOS fields \citep{dickinson2001,scoville2007} to obtain a systematic census of dusty star forming galaxies (DSFGs) at 1.2 and 2.0~mm \citep{bing2023}. Because of the long wavelength sampled by this blind survey, the detection of high redshift galaxies is favored \citep{blain2002,lagache2005,lutz2014,bethermin2015b}.
The GOODS-N field benefits from a very rich multi-wavelength database spanning from the X-rays to radio frequencies and including observations with space telescopes such as {\it Hubble}, {\it James Webb}, {\it Spitzer}, {\it Herschel}, {\it Chandra}, and ground-based facilities such as the Gran Telescopio Canarias (GTC), Keck, the Large Binocular Telescope (LBT), Subaru, the Canada-France Hawaii Telescope (CFHT), the Kitt Peak National Observatory (KPNO), the Very Large Array (VLA), and others.
We matched the NIKA2 catalog to the available multi-wavelength data, starting from high-resolution sub-millimeter and radio observations of the Sub-millimeter Array (SMA) and VLA down to NIR and optical through FIR and MIR.  
In case of missing high-resolution long-wavelength data or redshift estimates, in order to pinpoint the positions of the NIKA2 sources and measure their redshifts, we performed further millimeter interferometric observations with the IRAM NOrthern Extended Millimeter Array (NOEMA).

We exploited this exceptional data set, driven by the NIKA2 1.2 and 2.0~mm observations, and performed a panchromatic SED fitting, with the ultimate goal being to derive the dust mass of the N2CLS galaxies. In this way, we studied the evolution of the DMF and of the dust cosmic density from $z\sim 7$ to $z\sim1.5$ in an unbiased way, as close as possible to a direct dust mass selection.
This paper is organized as follows. Section \ref{sect:data} describes the NIKA2 and NOEMA data as well as all the ancillary data sets available in the GOODS-N region. The details about how the NIKA2 catalogs have been matched to all other multi-wavelength photometry are given in Sect. \ref{sect:ID}. In Sect. \ref{sect:sed_fit}, the eight methods adopted to fit the SEDs of our sources are presented. A comparison between the results of the different methods is in Appendix \ref{app:comparison_codes}. The overall properties of the N2CLS GOODS-N galaxies, as found with the SED fitting, are summarized in Sect. \ref{sect:properties}. The comoving number density of these galaxies is computed in Sect. \ref{sect:comoving_N}, and the evolution of the DMF and of the dust cosmic density are discussed in Sect. \ref{sect:discussion}. Finally, Sect. \ref{sect:summary} summarizes our findings. Throughout this manuscript, we adopt a $\Lambda$CDM cosmology with $H_0=67.4$ km\;s$^{-1}$\;Mpc$^{-1}$, $\Omega_\textrm{m}=0.315$, and $\Omega_\Lambda=0.685$ \citep{planck2020_cosmo}, a \citet{chabrier2003} stellar initial mass function (IMF), and the \citet{draine2003} frequency-dependent dust absorption coefficient, $\kappa_\nu$, renormalized as indicated by \citet[][i.e., $\kappa_{850\mu\textrm{m}}=0.047$\,m$^2$\,kg$^{-1}$ as reference]{draine2014}.\\
All the products from this paper have been released online on the survey's home page.\footnote{Data release: \url{https://data.lam.fr/n2cls/home}.} These comprise the N2CLS final maps and catalogs, the NOEMA follow-up data, and the matched catalog, which includes the identification, multi-wavelength counterparts, and redshifts.

% --------------------------------------------------------------------

\section{N2CLS and multi-wavelength data}\label{sect:data}

The Great Observatories Origins Deep Survey Northern field \citep[GOODS-N, R.A. 12:36:55, Dec. +62:14:19;][]{dickinson2001} is centered on the now-historic {\it Hubble} Deep Field North \citep[HDF;][]{williams1996} and benefits from a very rich multi-wavelength coverage, ranging from the X-rays to radio frequencies, including observations carried out with all major facilities in the northern hemisphere and in space.
This section presents an overview of all data sets that have been considered when building the SEDs of the N2CLS sources in GOODS-N, starting from the driving sample detected by NIKA2 at the IRAM 30~m telescope and by NOEMA.

\subsection{N2CLS observations and source extraction}

The N2CLS observations of the GOODS-N field ($\sim$\,160\,arcmin$^2$) were carried out with NIKA2 between October 2017 and January 2023, under project ID 192-16, for a total of 86.15 hours of telescope time \citep{bing2023}. NIKA2 observes simultaneously in two photometric bands at wavelengths of 1.2 and 2.0\,mm, with half power beam widths (HPBW) of $\sim$\,12 and 18\,arcsec, respectively. The data have been reduced with the IRAM PIIC software\footnote{Pointing and Imaging In Continuum, found at \url{https://publicwiki.iram.es/PIIC} and \url{https://www.iram.fr/IRAMFR/GILDAS/}.} \citep{zylka2013,piic2024} following the standard iterative procedure for deep fields. We defer to \citet{bing2023} for a detailed description of the data set and its reduction. 

The final NIKA2 maps of GOODS-N reach noise levels of 0.11\,mJy/beam, and 0.031\,mJy/beam in the deepest, central area, at 1.2 and 2.0\,mm, respectively. At these depths, NIKA2 hits the photometric confusion limit at 2.0 mm and reaches within a factor of 2 from the confusion limit at 1.2\,mm (Ponthieu et al., in prep.). The noise in the map is not uniform and increases toward the outer regions of the field. 

Source extraction was performed with a dedicated software on the matched-filter PIIC maps, using circular 2D Gaussian kernels matching the HPBW. Sources were identified as peaks on the signal to noise ratio (S/N) maps and measured using PSF fitting on the signal maps \citep{bing2023}. Possible systematic effects introduced by the data reduction were evaluated with simulations, by injecting an artificial sky model into the NIKA2 data timelines and performing the data reduction in the same way as for the original data. In this way the purity and photometric completeness of the extracted catalog were determined. The GOODS-N catalog reaches a purity of 80\% at S/N=3.0 and 2.9, at 1.2 and 2.0\,mm, respectively, and $>95$\% at S/N$>$4.2 and $>$4.1. This procedure allowed us also to correct the effect of flux boosting/filtering, that is the impact of noise and data reduction on the measured source fluxes, as well as to determine the ``effective area'' of the survey ``seen'' by each detected object, given the non-uniform noise level across the maps.

\subsection{N2CLS sample selection}

The sample analyzed in this work consists of the N2CLS sources extracted from the NIKA2 1.2 and 2.0\,mm GOODS-N maps, with S/N$\ge$4.2 in at least one NIKA2 band. This translates in a flux cut of $\sim$0.7\,mJy at 1.2\,mm, with a few sources that are slightly fainter because lying in the deepest central region of the maps. In the statistical analysis that follows, the non-uniform coverage and noise level are taken into account in a natural way by using the effective area mentioned above.

The NIKA2 sample analyzed here includes a total of 65 sources selected with S/N$\ge$4.2 at either 1.2 or 2.0\,mm \citep[corresponding to $>$95\% purity;][]{bing2023}. Out of these, 63 are detected at 1.2\,mm and 26 at 2.0\,mm, with S/N$\ge$4.2. Two sources are detected only at 2.0\,mm. We complement the flux at either 1.2 or 2\,mm with our blind catalog of S/N$>$3 detected sources.
Table \ref{tab:catalogs} summarizes the number of sources and the number of matches to all multi-wavelength catalogs considered here. In what follows, we name this sample and catalog ``N2GN''.

\begin{table*}[!t]
\footnotesize
\centering
\caption{\label{tab:catalogs} Statistics of the multi-wavelength counterparts adopted in the N2GN catalog.}
\begin{tabular}{llcl}
\hline
\hline
Survey & Instrument & Nr. & Ref.\\
& and band & sources & \\
\hline
N2CLS GOODS-N & Total & 71 & This work \\
& With $z$ & 68 & This work \\
& NIKA2 & 65 & \citet{bing2023} \\
& 1.2 mm & 63  & \citet{bing2023} \\ % 4sigma detections in the original NIKA2 catalog (no deboost)
& 2.0 mm & 26 & \citet{bing2023} \\ % 4sigma detections in the original NIKA2 catalog (no deboost)
& NOEMA 1.2 or 2.0\,mm & 31$^\ast$ &  W21CV, this work \\
& NOEMA 1.2 or 2.0\,mm & 2 & W16EE, E16AI, \citet{bingthesis} \\ 
\hline
Others sub-mm to radio & SMA 860 $\mu$m & 22 & \citet{cowie2017} \\ 
& SCUBA2 850 $\mu$m & 46 & \citet{cowie2017} \\ 
& SCUBA2 450 $\mu$m & 31 & \citet{barger2022} \\ 
& NOEMA  & 5 & Aa.Vv. (Sect. \ref{sect:others_mm}) \\ 
& VLA 1-2 GHz & 53 & \citet{owen2018} \\ 
\hline
CANDELS/GOODS & Optical U to K & 59 & \citet{barro2019} \\ 
FRESCO, JADES, PANORAMIC, Congress & JWST F090W to F770W & 24 & New extraction \\ CANDELS/GOODS & IRAC 3.6 to 8.0 $\mu$m & 59 & \citet{barro2019, liu2018} \\ 
PeakUp IRS & IRS 16\,$\mu$m & 10 & \citet{liu2018}\\
GOODS & MIPS 24 $\mu$m & 43 & \citet{barro2019, liu2018} \\  
PEP + GOODS-H & PACS 100, 160 $\mu$m & 28& \citet{magnelli2013, liu2018}\\ 
GOODS-H/HerMES & SPIRE 250-500 $\mu$m & 22 & \citet{elbaz2011,oliver2012,liu2018}\\ 
CDFN & {\it Chandra} 0.5 to 7.0 keV & 5 & \citet{evans2024} \\
\hline
\multicolumn{4}{l}{$^\ast$ 26/27 NIKA2 detected/targeted; five are split in two components by NOEMA}\\
\end{tabular}
\end{table*}

\subsection{NOEMA follow-up observations \label{sect:data_NOEMA}}

With the current sensitivity, N2CLS is constructing the most complete sample
of high-z IR-luminous massive galaxies in GOODS-N. However, the angular resolution of NIKA2 at 1.2 and 2\,mm makes it difficult to unambiguously identify the multi-wavelength counterparts of the N2CLS sources. As already demonstrated by the follow-ups of SCUBA-2 sources with the SMA \citep[e.g.,][]{cowie2017} or ALMA \citep[e.g.,][]{stach2019}, the combination of (sub-)mm single-dish and interferometer surveys is by far the most efficient way of constraining the dusty star formation at $z>2$. The high resolution and sensitivity of NOEMA were thus used to provide accurate position measurements on N2CLS sources that had ambiguous proxy for multi-wavelength identification or no counterpart (Sect.\ref{sect:ID}).\\

\begin{figure}[t]
\centering
\includegraphics[width=0.24\textwidth]{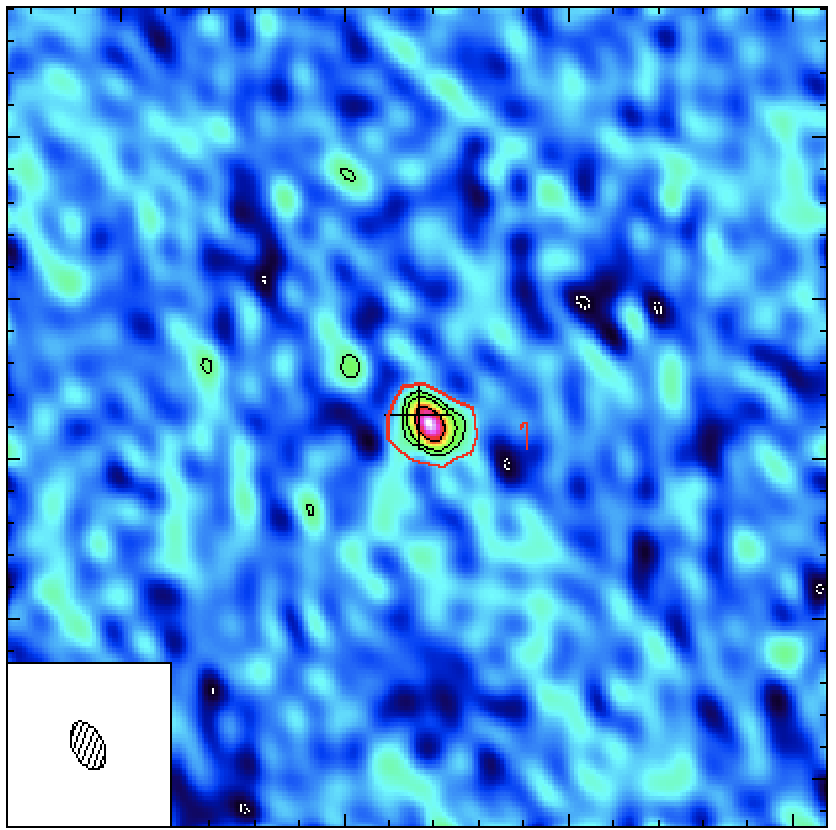}
\includegraphics[width=0.24\textwidth]{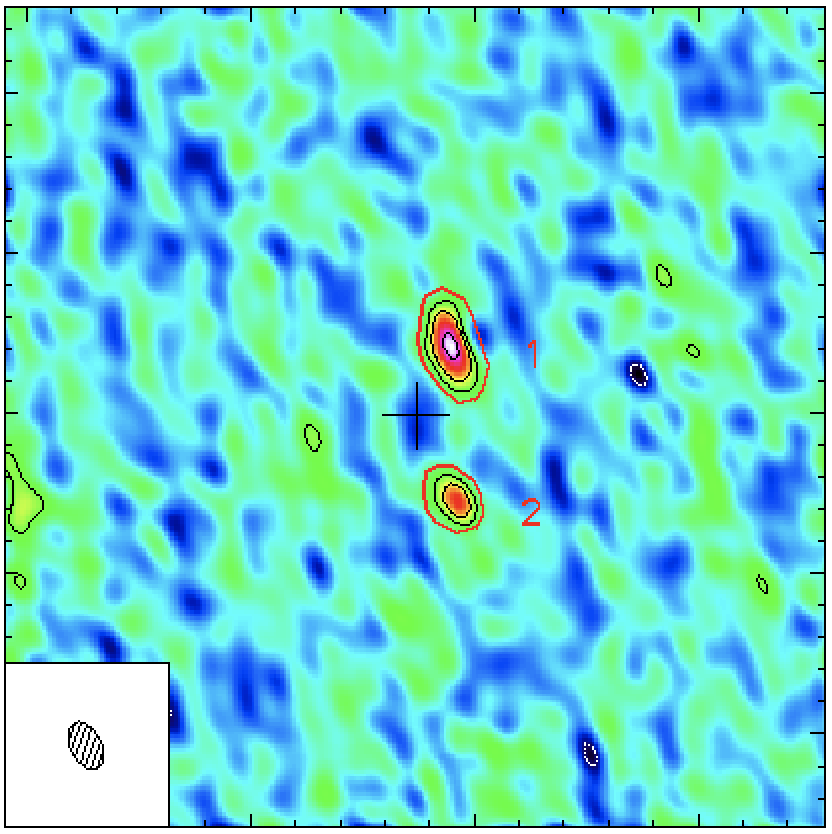}
\caption{NOEMA 150\,GHz maps for N2GN\_1\_13 (left) and  N2GN\_1\_17 (right) that reveal one and two counterparts, respectively. The beam shown at the bottom left has a size of 1.61\arcsec$\times$0.93\arcsec. Multi-wavelength postage stamps, including NIKA2, are shown in Fig. \ref{fig:indiv_glxs}.}
\label{fig_NOEMA_maps}
\end{figure}

\begin{figure}[t]	% h
\centering
\rotatebox{-90}{\includegraphics[height=0.47\textwidth]{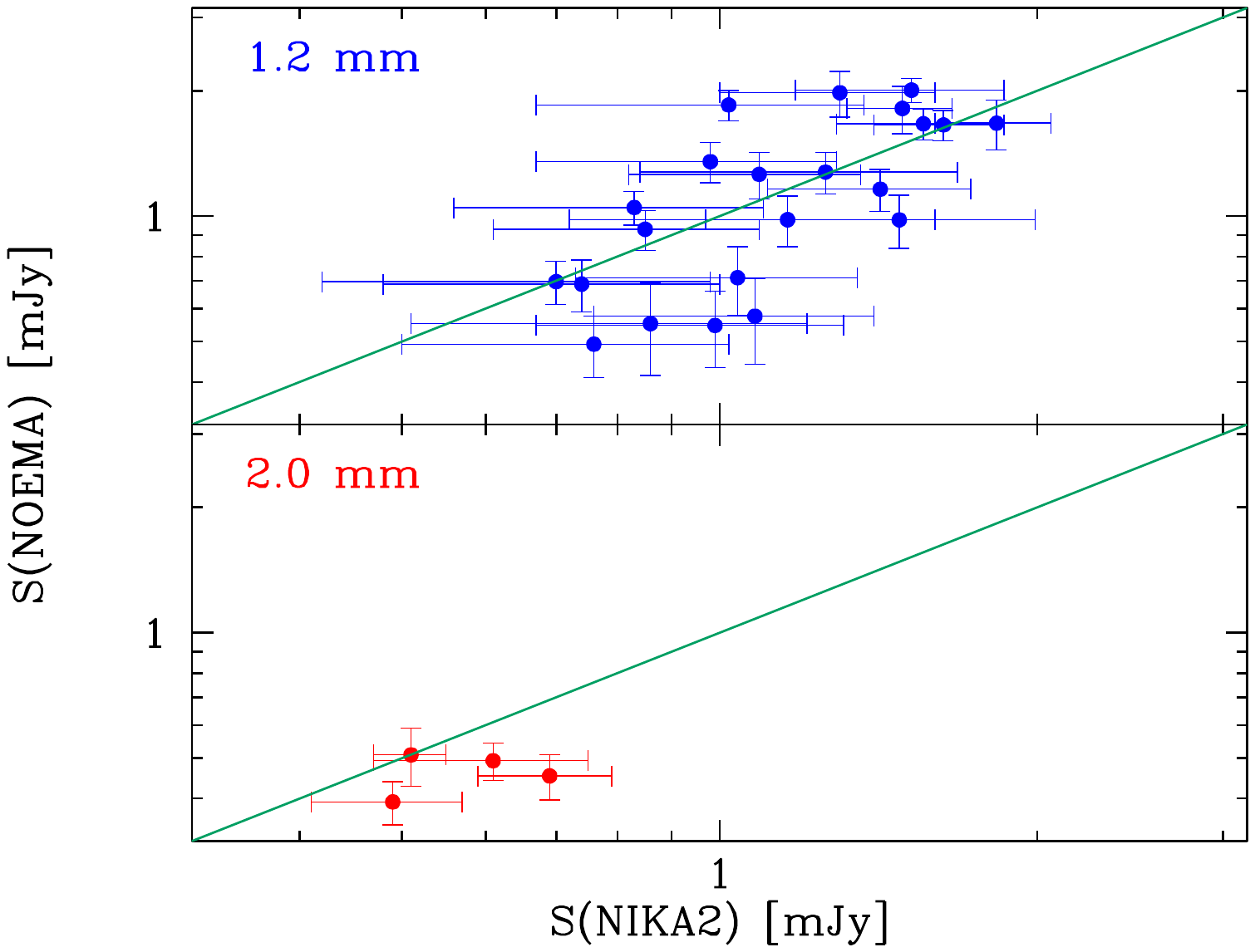}}
\caption{Comparison between NIKA2 and NOEMA flux densities for the sources observed in program W21CV. The green solid diagonal lines marks the 1:1 correspondence between the two instruments.}
\label{fig:cfr_NIKA2_NOEMA}
\end{figure}

We observed 27 N2GN sources (program W21CV, P.I. L.~Bing, for 39.5\,hours in C-configuration) either at 255\,GHz (23 sources) or 150\,GHz (4 sources) in the continuum,  matching the frequencies of the two NIKA2 bands. We reached beam sizes of 0.84\arcsec$\times$0.61\arcsec and 1.61\arcsec$\times$0.93\arcsec at 255 and 150\,GHz, respectively.
The observations reveal at least one counterpart at S/N$>$4 for all but one source (N2GN\_1\_59 which has a NOEMA source at the very edge of the beam, at 9.7\arcsec from the NIKA2 position). Five N2CLS sources break into two millimeter sources (N2GN\_1\_17, N2GN\_1\_24, N2GN\_1\_27, N2GN\_1\_34 and N2GN\_1\_56). Our NOEMA data for N2GN\_1\_05 also reveal two broad CO lines, giving a redshift similar to that obtained by 3D-HST ($z=1.996$). 
Figure\,\ref{fig_NOEMA_maps} shows the resulting continuum maps of two N2GN example sources. 

The NOEMA data were reduced and calibrated in the standard way with the GILDAS\footnote{Grenoble Image and Line Data Analysis Software, \url{https://www.iram.fr/IRAMFR/GILDAS/}} software, producing uv-tables of the lower and upper sidebands (LSB, USB). The main calibrators adopted were MWC349 and LkH$\alpha$101. The absolute flux uncertainty is 10\% and the positional error is 0.2\,arcsec.  

All channels of LSB and USB were combined together, taking care to exclude possible spectral lines, into double side band (DSB) continuum maps. These maps were cleaned using natural weights and support masks including all detected sources. Aperture fluxes were measured using ad hoc extraction polygons, taking into account primary beam losses. Flux statistical uncertainties were measured by rescaling the noise to the same apertures and finally source positions were computed as barycenters of the signal within the polygonal apertures.  Figure\,\ref{fig:cfr_NIKA2_NOEMA} compares the 1.2 and 2.0\,mm flux densities measured on the NIKA2 and NOEMA maps. No bandpass correction has been applied. In case of multiple sources, we have summed the contribution of the different NOEMA components.

Furthermore, as a N2CLS pilot project, we previously observed with the NIKA pathfinder \citep{Adam2014, Catalano2014} a small 3.5’$\times$2.5’ field (proposal 230-14) centered at (RA, Dec)=(12:36:27, 62:12:18) on AzGN10 \citep{Penner2011}, a z$\sim$6.5 candidate galaxy \citep{liu2018}. We followed up with the Plateau de Bure Interferometer (PdBI) two sources that are now included in N2CLS: N2GN\_1\_09 and N2GN\_1\_12. These two sources were observed using the WideX correlator in band3 (255\,GHz) with the D configuration (project W16EE). We further followed-up N2GN\_1\_09 at 3~mm with the D configuration (project E16AI) which provided a continuum flux at 109\,GHz \citep{bingthesis}. N2GN\_1\_12 has two interferometric counterparts.
Summarizing, our N2CLS selected sample comprises 71 individual millimeter galaxies.

\subsection{Other (sub-)millimeter data}\label{sect:others_mm}

Further NOEMA observations by other teams allow us to get an accurate position and/or complementary data for N2GN\_1\_01 \citep[GN10;][]{riechers2020}, N2GN\_1\_04 \citep[GN20;][]{daddi2009}, N2GN\_1\_06 \citep[HDF850.1;][]{neri2014}, N2GN\_1\_09 \citep[ID12646;][]{jin2022}, and N2GN\_1\_44 \citep[z=7.2;][]{fujimoto2022}.
We also use the SCUBA-2 850\,$\rm{\mu}$m source catalog by \citet{cowie2017}. This catalog comprises 186 sources with S/N$\ge$4, over an area of 450 arcmin$^2$. In the central region, the 850\,$\rm{\mu}$m observations cover the field to near the confusion limit of $\sim$1.65\,mJy, while over the wider region, they have a median 4$\sigma$ limit of 3.5\,mJy. The James Clerk Maxwell Telescope (JCMT) at 850\,$\rm{\mu}$m has an angular resolution of $\sim$14\arcsec FWHM.  

To find accurate positions and identify the true optical, NIR, and MIR counterparts of the SCUBA-2 sources, \citet{cowie2017} conducted interferometric follow-up observations with the SMA at 860\,$\rm{\mu}$m. They observed nearly all of the bright SCUBA-2 sources with the SMA. Including archival data from other SMA programs,
they have identifications for 33 SCUBA-2 sources with the SMA at 4$\sigma$.
All sources but GN20 were observed in the compact mode of the SMA with a spatial resolution of 2\arcsec at 860\,$\rm{\mu}$m. 

\citet{barger2022} presented the deepest SCUBA-2 450\,$\rm{\mu}$m data, achieving a central r.m.s. of 1.14\,mJy. They detect 79 sources with an S/N above four, on 175 arcmin$^2$. The JCMT at 450\,$\rm{\mu}$m has an angular resolution of 7.5\arcsec FWHM.

\subsection{Radio data}

\citet{owen2018} obtained new wide-band continuum observations in the 1-2\,GHz band using the Karl G. Jansky Very Large Array. The best image with an effective frequency of 1525\,MHz reaches an r.m.s. noise in the field center of 2.2\,$\rm{\mu}$Jy, with 1.6\arcsec angular resolution. We used their catalog, that contains 795 sources with an S/N=5 or greater, covering a radius of 9\,arcmin centered near the nominal center for the GOODS-N field. This area covers all our N2CLS sources. %The absolute radio positions are known to 0."1-0."2 r.m.s. 

\subsection{Mid- and far-infrared data}

\paragraph{{\it Herschel:}} The {\it Herschel} space telescope \citep{pilbratt2010} observed the GOODS-N field with the Photodetector Array Camera and Spectrometer \citep[PACS;][]{poglitsch2010} and the Spectral and Photometric Imaging REceiver \citep[SPIRE;][]{griffin2010}. The field was included in the Guaranteed Time surveys PACS Evolutionary Probe \citep[PEP, in the 100 and 160 $\mu$m bands;][]{lutz2011} and Herschel Multi-tiered Extragalactic Survey \citep[HerMES, at 250, 350 and 500\,$\rm{\mu}$m;][]{oliver2012}. The open time GOODS-{\it Herschel} survey \citep[][]{elbaz2011} reached the confusion limit in GOODS-N at 160\,$\rm{\mu}$m. 
In this work, we make use of the combined PEP plus GOODS-{\it Herschel} PACS maps and catalogs by \citet{magnelli2013} and the GOODS-{\it Herschel} SPIRE data from \citet{elbaz2011}.

\paragraph{{\it Spitzer:}}
The GOODS-N field has been observed at MIR wavelengths with {\it Spitzer}'s Multiband Imaging Photometer (MIPS) at 24 and 70\,$\rm{\mu}$m as part of the Guaranteed Time Observers (GTO) and GOODS surveys \citep{dickinson2003b, frayer2006}.
We retrieved the MIPS data from the \citet{barro2019} multi-wavelength catalog. They use the photometric catalogs in both bands described
in \citet{pg2005,pg2008}, which are based on the reduced and mosaicked data. 

\paragraph{{\it Deblended catalog:}} 
Source confusion and clustering can introduce substantial biases in photometric works \citep[e.g.,][]{bethermin2017}. {\it Herschel} SPIRE images have point spread functions (PSFs) that are several times larger (17.6–35.2\arcsec) than those of PACS images (7–12\arcsec), which are also several times higher than optical and NIR images. The fluxes of individual galaxies are often difficult to measure because of blending from close neighbors. When appropriate (i.e., when the N2CLS counterpart is clearly blended and we visually saw from the cutouts that an accurate flux could be measured using deblending techniques), we used the flux from the super-deblended catalog by \citet{liu2018}. Finally, the super-deblended catalog provides the additional 16\,$\rm{\mu}$m flux of ten sources in our N2CLS selected sample. The photometry at 16\,$\rm{\mu}$m was measured from the {\it Spitzer} IRS peak-up image by \citet{teplitz2011}.

\subsection{Optical and near-infrared data}

We used the multi-wavelength catalog from \citet{barro2019}, which is selected in the WFC3 F160W (H-band) in the GOODS-N CANDELS (Cosmic Assembly Near-IR Deep Extragalactic Legacy Survey) field. The multi-wavelength photometry includes broad-band data from UV (U-band from KPNO and LBT), optical ({\it Hubble Space Telescope}, Advanced Camera for Surveys, HST/ACS F435W, F606W, F775W, F814W, and F850LP), NIR-to-MIR (HST/WFC3 F105W, F125W, F140W and F160W, Subaru/MOIRCS Ks, CFHT/Megacam K, and {\it Spitzer} InfraRed Array Camera, IRAC 3.6, 4.5, 5.8, 8.0 \,$\rm{\mu}$m). Similarly to the case of MIR to FIR data, for the K and IRAC bands we used instead fluxes from the super-deblended catalog by \citet{liu2018}, when appropriate. 

\subsection{JWST photometric data}

We used the available public imaging of the {\it James Webb} Space Telescope (JWST) in the GOODS-N field (i.e., the mosaics v7.3 obtained from the DAWN JWST Archive, DJA\footnote{\url{https://dawn-cph.github.io/dja/index.html}}), including the surveys of FRESCO \citep[First Reionization Epoch Spectroscopically Complete Observations;][]{oesch2023}, JADES \citep[JWST Advanced Deep Extragalactic Survey;][]{eisenstein2023}, PANORAMIC (Williams et al., in prep), and Congress (Egami et al., in prep.). From shortest to longest wavelengths, we used the NIRCam (Near Infrared Camera) and MIRI (Mid-InfraRed Instrument) filters F090W, F115W, F150W, F182M, F200W, F210M, F277W, F335M, F356W, F410M, F444W, and F770W. The photometry measurements are similar to those described by \citet{weibel2024} and \citet{xiao2024}. Specifically, we used SExtractor \citep{bertin1996} in dual-image mode and the longest wavelength NIRCam filter, F444W, as the detection image. The fluxes were measured in a circular aperture of radius 0\farcs35.  Prior to flux measurement, all images were co-aligned and drizzled to a common grid of 40 mas/pixel.  All NIRCam filters were PSF-matched to the F444W band. For MIRI/F770W, whose PSF is broader than the NIRCam/F444W PSF, we further computed the matching kernel from F444W to F770W and produced PSF-matched flux and r.m.s. images. We measured the flux of F770W based on the original F444W map and in accordance with the PSF-matched color between F770W and PSF-matched F444W. Finally, we performed an aperture correction based on the flux measured through the Kron aperture in F444W, and scaled the fluxes to total by computing the encircled energy of the Kron ellipse on the F444W PSF.\\

The circular aperture of radius 0\farcs35 is appropriate for the typical population of massive dusty sources at high redshift ($z>3$). We observe that some of our $z\sim2$ dusty galaxies are larger, with very complex and highly wavelength-dependent morphologies. For these kinds of sources, we do not consider our JWST photometry and rather rely on previous HST+IRAC published catalogs. For HDF850.1 we adopted the fluxes published in \citet{sun2024}.
In summary, we complemented our multi-wavelength SEDs with JWST fluxes for 24 sources, including 10 optically dark galaxies \citep[notice that for N2GN\_1\_61 we only have one photometric data point, F444W, from][]{sun2024}.

\subsection{X-ray data}\label{sect:xray}

%%% Summary of X-rays known:

%N2GN\_1\_03      in Barro+2019 (matched to Alexander+2003) + 2020 CDC, AGN fit
%N2GN\_1\_05      in Barro+2019 (matched to Alexander+2003), no need for AGN 
%N2GN\_1\_10      in Barro+2019 (matched to Alexander+2003) + 2020 CDC, AGN fit
%N2GN\_1\_11      in the 2020 CDC catalog, AGN fit
%N2GN\_1\_14      in Barro+2019 (matched to Alexander+2003), no need for AGN
%N2GN\_1\_19	    in Wang+2016, AGN fit
%N2GN\_1\_21      in Barro+2019 (matched to Alexander+2003), no need for AGN
%N2GN\_1\_25      in Barro+2019 (matched to Alexander+2003), no need for AGN
%N2GN\_1\_32      in Barro+2019 (matched to Alexander+2003) + 2020 CDC, AGN fit
%N2GN\_1\_37      in Barro+2019 (matched to Alexander+2003), no need for AGN
%N2GN\_1\_39      in Barro+2019 (matched to Alexander+2003), no need for AGN
%N2GN\_1\_45      in Barro+2019 (matched to Alexander+2003), no need for AGN
%N2GN\_1\_46	    Brandt+2001 + WISE + IRS 16um, AGN fit
%N2GN\_1\_47      in Barro+2019 (matched to Alexander+2003), no need for AGN
%N2GN\_1\_51      in Barro+2019 (matched to Alexander+2003), no need for AGN
%N2GN\_1\_54      in Barro+2019 (matched to Alexander+2003), no need for AGN
%N2GN\_1\_56_a    in Barro+2019 (matched to Alexander+2003), no need for AGN

GOODS-N is included in the 2 Ms {\it Chandra} observations of the {\it Chandra} Deep Field North (CDFN), in the soft and hard X-ray bands ($0.5-2.0$ and $2.0-7.0$ keV). \citet{alexander2003} published the catalog of CDFN, including a total of 503 X-ray sources, 348 of which lie in the IRAC GOODS-N area \citep{rovilos2010}.

The GOODS-N matched catalog by \citet{barro2019} includes the X-ray ID in the \citet{alexander2003} catalog. In addition, we also used the {\it Chandra} Source Catalog \citep[CSC 2.0;][]{evans2024} and the NASA Extragalactic Database (NED\footnote{\url{http://ned.ipac.caltech.edu/}. NED is funded by the National Aeronautics and Space Administration and operated by the California Institute of Technology.}) to find additional X-ray counterparts of the N2GN sources and identify potential active galactic nuclei (AGN). 
\begin{figure*}[!t]
\centering
\settowidth{\imagewidth}{\includegraphics{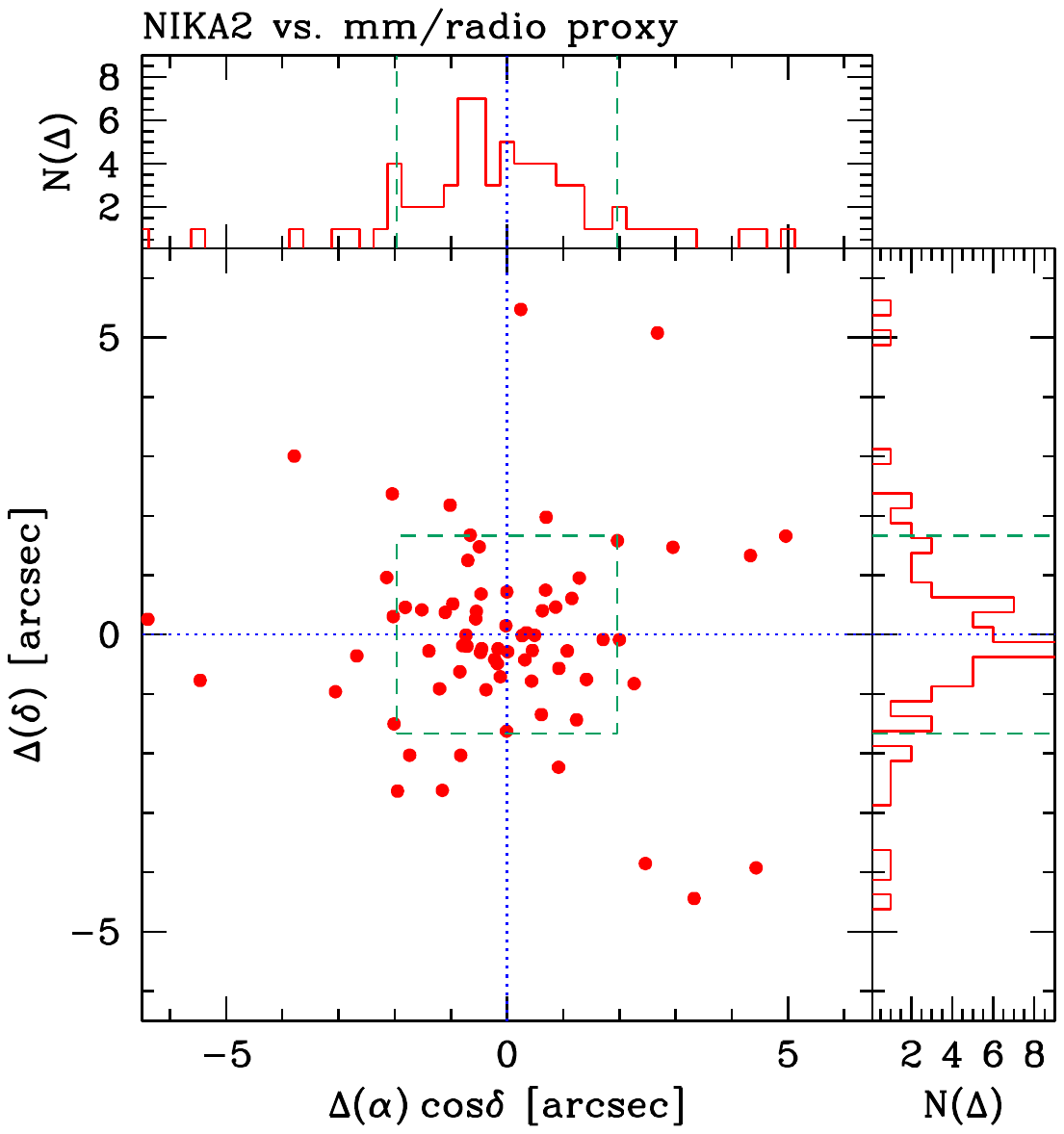}}
\includegraphics[trim=0 0.22\imagewidth{} 0 0.10\imagewidth{}, clip, width=0.47\textwidth]{figs1/plot_coords_diff.pdf}
\includegraphics[trim=0 0.22\imagewidth{} 0 0.10\imagewidth{}, clip, width=0.47\textwidth]{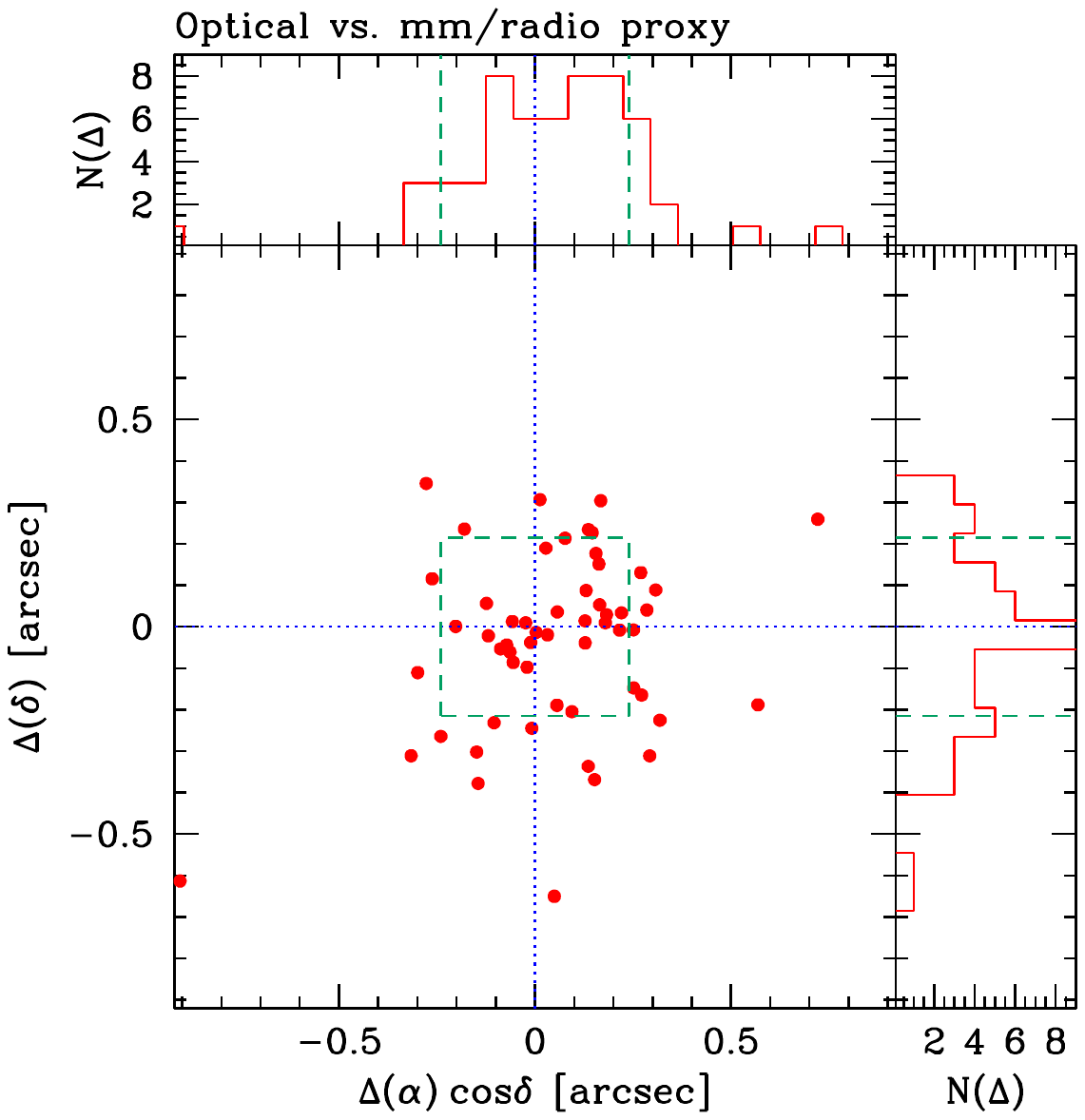}
\caption{Differences between the coordinates of the N2GN sources and their counterparts. {\em Left panel}: Distance between the NIKA2 coordinates (all from the 1.2\,mm position but two) and the matched proxy (Sect.\,\ref{sect:ID_process}). {\em Right panel}: Distance between the matched proxy and the N2GN optical counterparts. The green dashed lines and boxes mark the r.m.s. of the distribution.}
\label{fig:ID_distances}
\end{figure*}
As a result of this search, 17 N2GN sources have a X-ray counterpart \citep{evans2024, barro2019, wang2016, alexander2003, brandt2001}. Out of these, only six require an AGN-torus component to reproduce their observed  optical-to-radio SEDs (Sect. \ref{sect:sed_fit}), namely  N2GN\_1\_03, 10, 11, 19, 32, and 46. Appendix \ref{app:indiv_srcs} gives more details about the sources.

% --------------------------------------------------------------------

\section{Identification of N2CLS sources}\label{sect:ID}

Due to the relatively low angular resolution of NIKA2, it is challenging to confidently match our N2CLS galaxies with any counterparts in the rich ancillary GOODS-N data set. This is a notoriously difficult problem for (sub-)millimeter galaxies found with blind single-dish surveys. In this Section, we describe the process of identification and the procedure followed to build the UV-to-radio SEDs of the N2GN sources.

\subsection{Identification process \label{sect:ID_process}}

Based on the empirical relation observed between the FIR and radio emission in star-forming galaxies, the radio emission has often been used as a high-angular resolution proxy to get the position of the rest-frame FIR emission observed in the (sub-)mm \citep[][]{smail2000, ivison2002, borys2004}. The radio proxy has also been combined with MIR imaging from {\it Spitzer} \citep[e.g.,][]{ivison2007}. However, when observed at these radio or 24\,$\rm{\mu}m$ wavelengths, galaxies are subject to a positive $k$-correction and therefore become fainter with increasing redshift. 
This technique thus biases the counterpart identification toward lower redshift \citep[e.g.,][]{chapman2005}.

As demonstrated by the follow-up observations of SCUBA-2 \citep[e.g.,][]{cowie2017, simpson2017, stach2019, simpson2020} or South Pole Telescope \citep[SPT; e.g.,][]{hezaveh2013, reuter2020} sources with the SMA or ALMA, the synergy between single-dish and interferometer surveys at similar wavelengths is the most effective approach to obtain large unbiased samples of dusty star-forming galaxies, pinpoint their position and obtain their redshift. 
Following these previous studies, we located the precise positions of our galaxies using, by order of priority:
\begin{itemize}
\item Known millimeter counterparts from previous PdBI (e.g., HDF850.1) or NOEMA observations: six sources.
\item Our own NOEMA follow-up observations (see Sect.\,\ref{sect:data_NOEMA}): 31 sources.
\item The SMA follow-up observations of SCUBA-2 sources from \citet{cowie2017}: 15 sources. %%%\GL{check why 22 SMA in Table 1} SB: the 22 are not the identification, but are how many have a 860 S/N>=3 flux in the catalog
\item Two forced identifications: one with a high-redshift optical galaxy revealed by JWST and one with the closest VLA counterpart to the N2CLS position\footnote{These are N2GN\_1\_33 and N2GN\_1\_55, as reported in Appendix\,\ref{app:indiv_srcs}.}.
\item The very deep VLA radio catalog by \citet{owen2018}: 17 sources. 
\end{itemize}

The list of sources with their proxy for identification is given in Table\,\ref{tab:ID_z_Md_LIR}. In the left panel of Fig.\,\ref{fig:ID_distances}, we show the distribution of the distances between the N2CLS coordinates and the precise position of the matched proxy, as obtained through our identification process. Given the small angles involved, the total distance is computed with the Pythagorean theorem. The median total distance is 1.56\arcsec. Of the ten sources with a distance higher than 4\arcsec, six are double, three are located with VLA, and one is located with NOEMA.  

When a NIKA2 source has multiple associations in high-resolution NOEMA data, all such associated objects are used. Their low spatial resolution, blended photometry (e.g. NIKA2 and {\it Herschel}) is excluded and different SEDs are built for each identified source. None of the SMA identifications is multiple. Finally, in the case of radio identifications, 16 out of the 17 NIKA2 sources identified through VLA data have only one radio counterpart. The remaining one, N2GN\_1\_55, turns out to be an optically dark galaxy with no redshift information yet (Appendix \ref{app:indiv_srcs}).

\subsection{Multi-wavelength spectral energy distribution}

We used the multi-wavelength data presented in Sect.\,\ref{sect:data} to build the SED of each galaxy. We automatically searched for multi-wavelength counterparts at a given distance $d$ given by
\begin{equation}\label{eq:distance_to_proxy}
d = \sqrt{\rm{HWHM_{proxy}}^2 + \rm{HWHM_{\lambda}}^2},
\end{equation}
where $\rm{HWHM_{proxy}}$ is the half width at half maximum (HWHM) of the beam of the proxy used to get the precise position (e.g., NOEMA) and $\rm{HWHM_{\lambda}}$ is the HWHM of the beam of the multi-wavelength observations. The right-hand panel of Fig. \ref{fig:ID_distances} presents the distances between the matched proxy (Sect. \ref{sect:ID_process}) and the optical identification of the N2GN sources. The median distance is 0.25\arcsec. Five sources out of 71 have a double entry in the optical catalogs when applying Eq.~(\ref{eq:distance_to_proxy}): N2GN\_1\_05, 15, 24a, 26, and 32. However, the two optical counterparts correspond to: {\em i)} the same galaxy in the first two cases; {\em ii)} a very-close merger in the second two cases. For the last source (N2GN\_1\_32), we chose the closest optical counterpart (0.22\,arcsec against 0.93\,arsec), which is also the brightest one.

Using the multi-wavelength cutouts, we checked one by one the multi-wavelength counterparts and removed the photometric point for blended sources (e.g., SPIRE ``blobs'') or replace their fluxes with the deblended photometry by \citet{liu2018}, when appropriate.
The SEDs of all individual galaxies are shown in Appendix\,\ref{app:indiv_srcs}. 
The numbers of the multi-wavelength counterparts used for the SED fitting are given in Table\,\ref{tab:catalogs}.

\subsection{Redshifts}\label{sect:redshift}

\begin{figure}[t]
\centering
\settowidth{\imagewidth}{\includegraphics{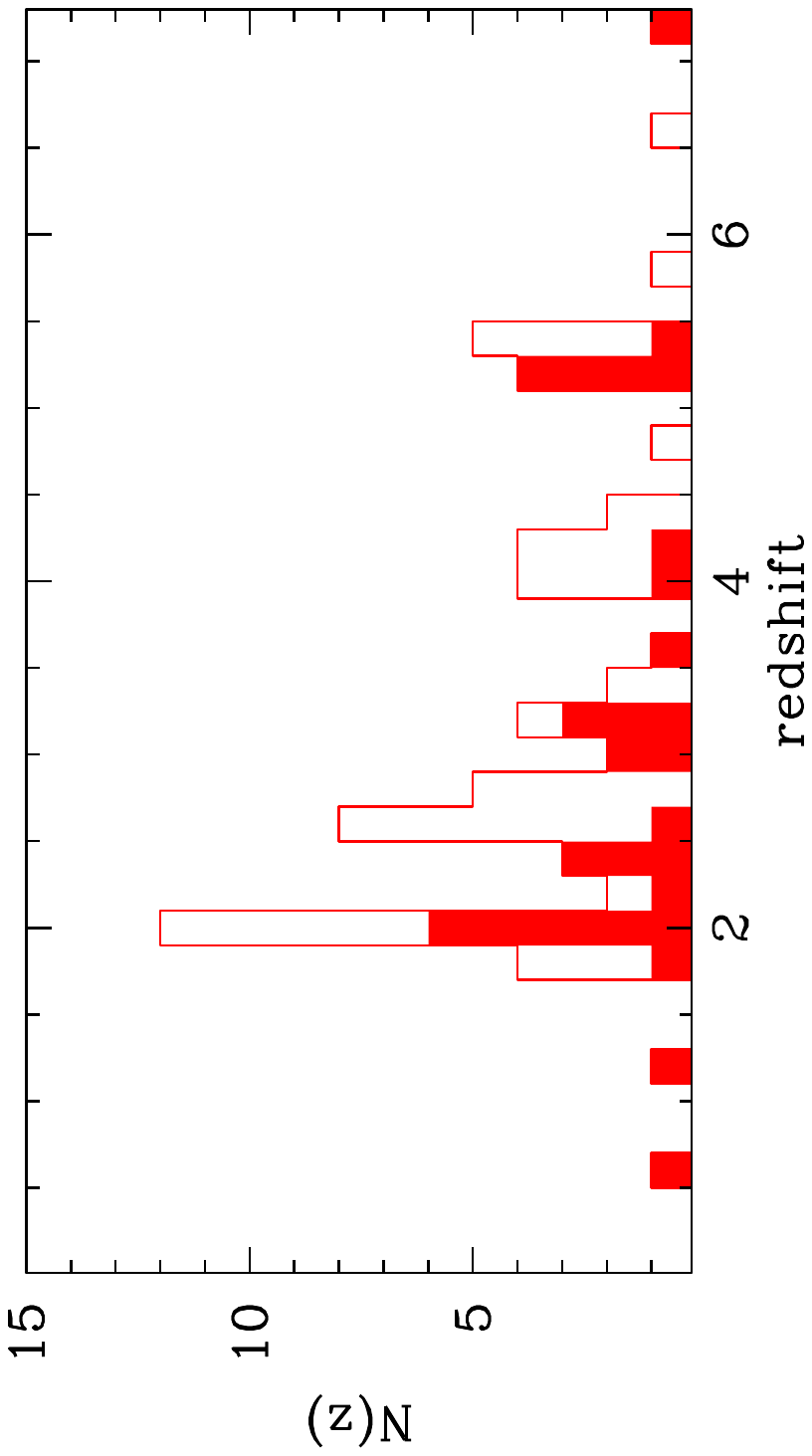}}
\rotatebox{-90}{\includegraphics[trim=0 0 0.27\imagewidth{} 0, clip, height=0.48\textwidth]{figs1/zdistr_20250128.pdf}}
\caption{Redshift distribution of the N2GN sources. The solid histogram includes spectroscopic redshifts, while the open histogram contains all sources in the sample.}
\label{fig:zdistr}
\end{figure}

We gathered all the redshift information found in the dedicated surveys and catalogs from the literature. These are from \citet{steidel2003, skelton2014, bouwens2015, cowie2017, arrabal_haro2018, owen2018, barro2019, kodra2023}. 
We complemented this survey approach with a dedicated search for individual sources, to obtain the redshifts measured from e.g., PdBI or NOEMA, VLA, or JWST observations. We also used the NED to check some individual objects. 

For four sources (N2GN\_1\_08, N2GN\_1\_12\_a, N2GN\_1\_16, N2GN\_1\_36), we found some discrepancies between different photometric redshifts from the literature. We also have three sources (N2GN\_1\_18, N2GN\_1\_34\_b, and N2GN\_1\_43) that previously lacked known redshift values but now have sufficient photometric data points to determine a redshift. For these seven sources, we used The Code Investigating GALaxy Emission (CIGALE) software \citep{burgarella2005,noll2009,boquien2019} to determine a photometric redshift. 

The redshift of each source is given in Table\,\ref{tab:ID_z_Md_LIR}. We have spectroscopic redshifts for 29 sources ($\sim43$\% of the sample) and photometric redshift for 39 sources. Three galaxies (N2GN\_1\_17\_b, N2GN\_1\_34\_a and N2GN\_1\_55),  identified with NOEMA (the first two) and VLA (the third), have no redshift.

Figure\,\ref{fig:zdistr} presents the redshift distribution of the N2GN sources, distinguishing between spectroscopic (solid histogram) and photometric redshifts. The median redshift of the N2GN sample is $z=2.819$, with a median absolute deviation (M.A.D.) of 0.831.
Compared to the redshift distribution expected for the N2CLS galaxies from the Simulated Infrared Dusty Extragalactic Sky (SIDES) simulations \citep[][B\'ethermin et al., in prep.]{bethermin2017}, we clearly see an excess of galaxies at z$\sim$5.1-5.3, linked to the complex overdense environment hosting the famous HDF850.1 (N2GN\_1\_06) and GN10 (N2GN\_1\_01) sub-millimeter galaxies \citep{sun2024, xiao2024}. A dedicated analysis of the N2GN galaxies in this overdensity will be presented in Lagache et al. (in prep.).

% --------------------------------------------------------------------

\section{Fitting the spectral energy distributions}\label{sect:sed_fit}

The SEDs of the N2GN sources consist of up to 34 broad-band photometric measurements, spanning over the wavelength range from 3600 \AA\ to 21 cm. We reproduced the SEDs with different modeling approaches, with the main goal to derive as robust dust properties as possible.

The observed FIR to millimeter SEDs were studied with a modified blackbody (MBB; Sect. \ref{sect:MBB}) in the optically thin approximation and in its general form, as well as with the \citet[][DL07, Sect. \ref{sect:dl07}]{DL07} model.
The panchromatic optical to radio SEDs were modeled with the MAGPHYS and SED3FIT codes, in their original and high-$z$ versions \citep[][Sects. \ref{sect:magphys} and \ref{sect:sed3fit}]{dacunha2008,dacunha2015,battisti2020,berta2013}, as well as with CIGALE \citep[][Sect. \ref{sect:CIGALE}]{burgarella2005,noll2009,boquien2019}.

Appendix \ref{app:comparison_codes} compares the results of the eight SED fitting methods presented here, in terms of the derived $M_\textrm{dust}$, $L_\textrm{IR}$ and $M_\star$ of the N2GN sources. The different approaches lead to consistent results, within their limitations. In the main analysis of this work we adopt the results obtained with MAGPHYS high-$z$ for purely star-forming galaxies and with  SED3FIT-h$z$ in case an additional component (AGN or a young stellar population) is needed to reproduce the observed SED.

\subsection{Modified blackbody}\label{sect:MBB}

The emitted FIR-to-millimeter dust SED of a galaxy can be approximated with a single-temperature MBB, described as
\begin{equation}
L=\Omega\epsilon_\nu B_\nu\left(T_\textrm{dust}\right)\textrm{,}
\end{equation}
where $\Omega$ is the solid angle of emission, $\epsilon_\nu$ is the dust emissivity coefficient and $B_\nu\left(T_\textrm{dust}\right)$ is the Planck function of temperature $T_\textrm{dust}$. 
For a uniform medium of optical depth $\tau_\nu$, radiative transfer theory leads to $\epsilon_\nu=1-\exp\left(-\tau_\nu\right)$. Relating $\tau_\nu$ to the mass absorption coefficient $\kappa_\nu$ and assuming that the dusty medium is optically thin, the broadly used expression of the MBB is obtained \citep[see][for a thorough derivation]{berta2016}:
\begin{equation}\label{eq:mbb_mass}
L_\nu=4\pi M_\textrm{dust}\kappa_\nu B_\nu\left(T_\textrm{dust}\right)\textrm{,}
\end{equation}
which is used to model the observed SEDs. 
The absorption coefficient depends on frequency as a power law: $\kappa_\nu=\kappa_0\left(\nu/\nu_0\right)^\beta$. We adopt the values of $\kappa_\nu$ tabulated by \citet{draine2003}, with the correction indicated by \citet{draine2014}. In this assumption, the reference value is $\kappa_{850\mu\textrm{m}}=0.047$ m$^2$ kg$^{-1}$.

The MBB-thin fit includes three free parameters, namely normalization, $T_\textrm{dust}$ and $\beta$. The dust mass of the given galaxy, $M_\textrm{dust}$, was computed by inverting Eq. (\ref{eq:mbb_mass}) and evaluating the model at the longest wavelength covered by the observed SED, such to minimize the effect of possible optically thick dust at shorter wavelengths and to avoid extrapolations beyond the available data.
In computing $M_\textrm{dust}$, we took care to account for the correction described by \citet{bianchi2013} and \citet{berta2016}, related to the fact that the best fit value of $\beta$ is not necessarily the same as the one in the tabulated $\kappa_\nu$. We defer to these works for more details. 

The general form of the MBB differs from the optically thin approximation by the factor $1-\exp\left(-\tau_\nu\right)$, which can be expressed in terms of the wavelength $\lambda_\textrm{thick}$ at which the medium becomes optically thick or in terms of the size of the absorbing medium \citep{ismail2023}. Here we adopted the former approach: a fourth additional free parameter is hence $\lambda_\textrm{thick}$.

\citet{ismail2023} modeled the FIR-to-millimeter SEDs of a sample of 125 high-$z$ dusty galaxies with an exquisite wavelength coverage with up to 12 bands between 250\,$\mu$m and 3.5\,mm in the observed frame \citep{cox2023}. They studied the general MBB in two different formulations: one adopting $\lambda_\textrm{thick}$ as free parameter; the other based on the size $A$ of the dust emitting region. The two are linked by the expression $\lambda_\textrm{thick}=\lambda_0\left(\kappa_0 M_\textrm{dust}/A\right)^{1/\beta}$. 
With the use of mock simulated SEDs, \citet{ismail2023} demonstrated that, adopting the first formulation, $\lambda_\textrm{thick}$ is hardly constrained. Furthermore, these authors showed that a reliable estimate of $T_\textrm{dust}$, $\beta$ and $M_\textrm{dust}$ with the general MBB requires an independent knowledge of $A$.  The degeneracies that affect $\lambda_\textrm{thick}$ can induce a significant underestimation of $M_\textrm{dust}$ of 20\% to 50\%, depending on dust mass itself.

The MBB fit was limited to data at rest-frame wavelengths $\ge 50$\,$\mu$m, because below this limit the contribution of warm dust to the SED becomes non-negligible, and is not included in the MBB models. The effect of the cosmic microwave background (CMB) was taken into account as described by \citet{dacunha2013}. Radio data were not included in the fit, but they were compared to a synchrotron model obtained with a simple power law $S(\nu)\propto\nu^\alpha$ (spectral index $\alpha=-0.8$) normalized such to obey the radio-FIR correlation for star formation \citep{magnelli2015,delhaize2017}. Sources that show a significant radio excess with respect to the radio-FIR correlation are very likely to host a radio-AGN.

\subsection{The \citet{DL07} dust model}\label{sect:dl07}

The \citet{DL07} models are an upgrade of those originally developed by \citet{DL01}, \citet{LD01} and \citet{wd01}. Interstellar dust is described as a mixture of carbonaceous and amorphous silicate grains, whose size distributions are chosen to mimic the observed extinction law in the Milky Way (MW), Large Magellanic Cloud (LMC) or Small Magellanic Cloud (SMC) bar region. 

The dust distribution is divided in two components: the diffuse ISM, responsible for the bulk of the dust mass budget; and dust enclosed in photo-dissociation regions (PDRs). The former is heated by a radiation field of constant intensity $U_\textrm{min}$. The latter, representing a fraction $\gamma$ of the total amount of dust, is exposed to starlight with intensities within the range $U_\textrm{min}$ to $U_\textrm{max}$. Although PDRs usually provide a small fraction of the total dust mass, they can contribute to the majority of the MIR dust emission.
The properties of grains are parameterized by the index $q_\textrm{PAH}$, defined as the fraction of dust mass in the form of PAH (polycyclic aromatic hydrocarbons) molecules. Finally the fraction of dust in PDRs heated by starlight with an intensity $U$ is a power law of $U$ with index $-\alpha$.

The DL07 model thus has six free parameters: $U_\textrm{min}$, $U_\textrm{max}$, $\gamma$. $q_\textrm{PAH}$, $\alpha$, and $M_\textrm{dust}$. Studying local {\it Spitzer} galaxies, \citet{draine2007} demonstrated that some of the parameters can be limited to a restricted range of values. We adopted these as commonly done in the literature \citep[e.g.,][]{magdis2012, magnelli2012, santini2014, berta2016}.
Also in the case of DL07 modeling, we adopted the renormalization of $\kappa_\nu$ prescribed by \citet{draine2014}.

\subsection{MAGPHYS}\label{sect:magphys}

The Multi-wavelength Analysis of Galaxy Physical Properties (MAGPHYS) software reproduces the observed SEDs of galaxies linking together their optical stellar emission to their dust component through energy balance (no radiation transfer involved). The energy absorbed by dust in stellar birth clouds and in the ISM is re-distributed to the dust emission at infrared wavelengths. Here we used two different versions: the original by \citet{dacunha2008}; and one dedicated to $z>1$ galaxies \citep[named ``high-$z$ v2'';][]{dacunha2015,battisti2020}.

The code combines the \citet[][BC03]{bc03} optical/NIR stellar library, with the MIR/FIR dust emission computed by \citet{dacunha2008}. The adopted star formation history (SFH) is a continuous delayed exponential function of the form $\psi(t)\propto \frac{t}{\tau^2}\exp\left(-t/\tau\right)$, where $t$ is the model age and $\tau$ the star formation timescale in units of Gyr. Superimposed to this continuous SFH are  bursts of star formation of random duration and age. Dust attenuation is described with the  recipe by \citet{charlot2000}. The main assumptions are that the energy re-radiated by dust is equal to that absorbed, and that starlight is the only significant source of dust heating. 

The power re-radiated by dust in stellar birth clouds is computed as the sum of three components: PAHs; a MIR continuum describing the emission of hot grains with temperatures $T = 130$ to 250 K; and grains in thermal equilibrium with $T = 30$ to 60 K. The ``ambient'' ISM is modeled by fixing the relative proportions of these three components to reproduce the cirrus emission of the Milky Way, and adding a component of cold grains in thermal equilibrium, with adjustable temperature in the range $T = 15$ to 25 K.

Different combinations of star formation histories, metallicities and dust content can lead to similar amounts of energy absorbed by dust in the stellar birth clouds, and these energies can be distributed in wavelength using different combinations of dust parameters. Consequently, in the process of fitting, a wide range of optical models is associated with a wide range of infrared spectra. We defer to \citet{dacunha2008} for a formal description of how the SED models are built.

\begin{figure*}[t]  % !ht
\centering
\rotatebox{-90}{\includegraphics[height=0.47\textwidth]{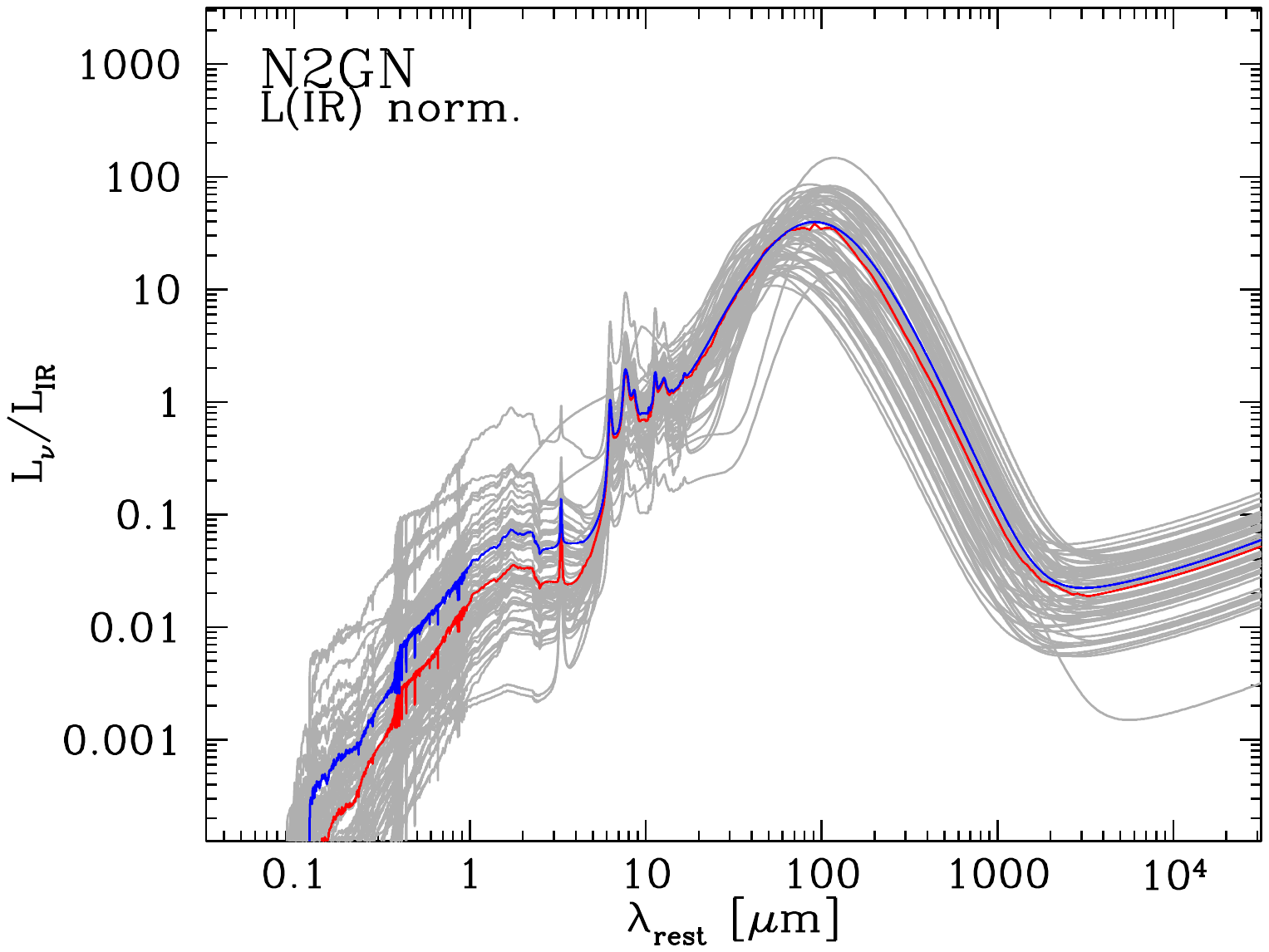}}
\rotatebox{-90}{\includegraphics[height=0.47\textwidth]{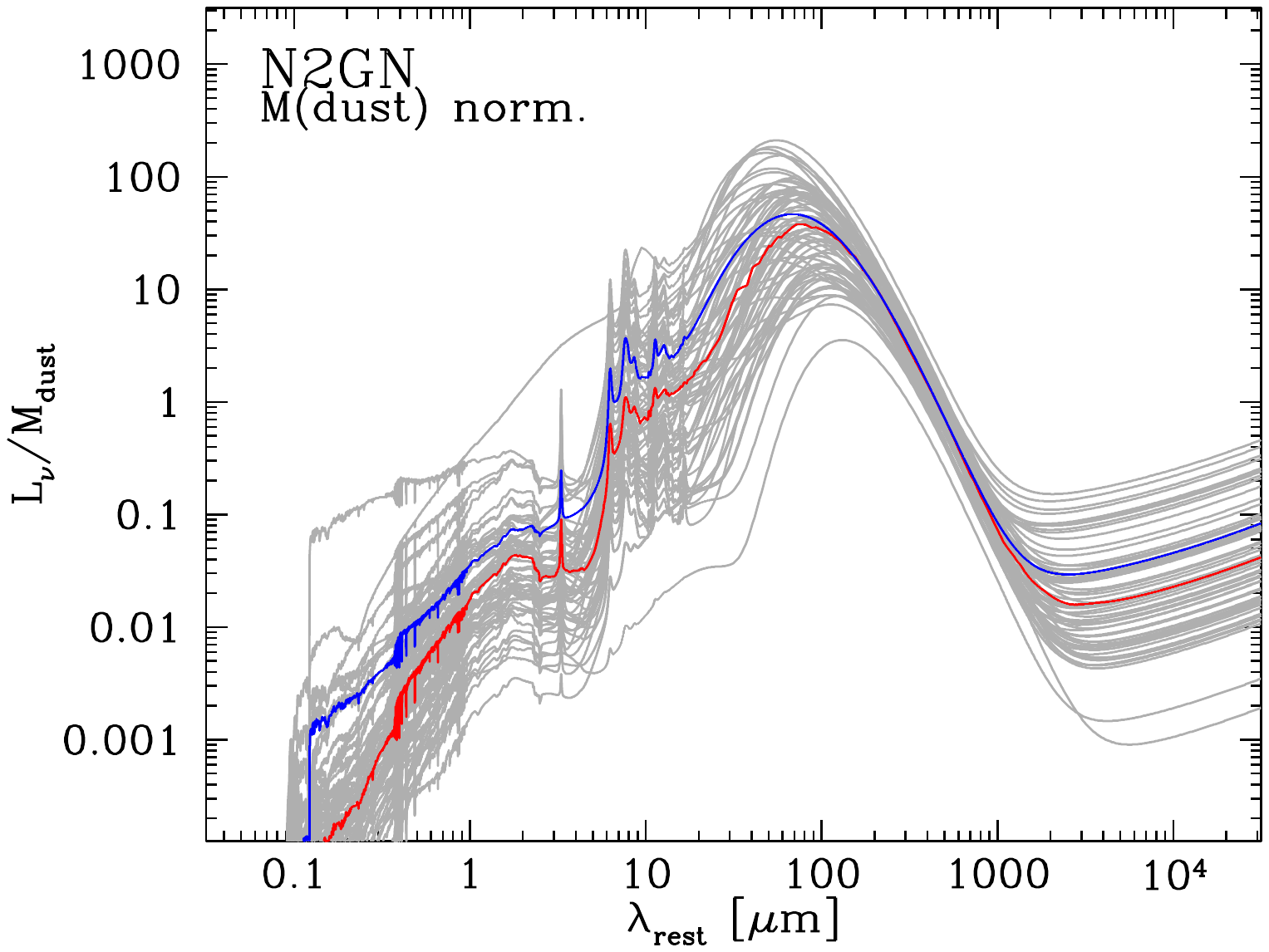}}
\caption{Spectral energy distributions of all N2GN sources obtained with the UV-to-radio SED fitting (gray lines). The red and blue lines represent the median and average SEDs, respectively, as obtained by combining all models.}
\label{fig:med_avg_SED}
\end{figure*}

In the high-$z$ version of the code, \citet{dacunha2015} extended the parameter space of the models, such that they include higher dust optical depths, higher SFRs, and younger ages. Also the dust properties have been modified to allow for higher values of dust attenuation both in the stellar birth clouds and in the diffuse ISM. Moreover, the high-$z$ MAGPHYS takes into account the UV absorption by the IGM including Lyman series line blanketing and Lyman-continuum absorption.

At radio frequencies, a thermal and a non-thermal components have been added to the model. The former ({\em Bremsstrahlung}, or free-free) is a power-law $L\propto\nu^{-0.1}$, normalized such to have a fixed contribution of 10\% to the rest-frame 20 cm emission. The latter (synchrotron) is a power-law with spectral index -0.8, linked to the FIR emission by the local radio-FIR correlation ($q_\textrm{FIR}=2.34$ with a 1$\sigma$ dispersion of 0.25).

Finally, \citet{battisti2020} introduced a flexible 2175 \AA\ absorption feature to the diffuse ISM attenuation curve, with strength depending on the actual $A_V/E(B-V)$ of the model. These authors also upgraded the IGM absorption from the \citet{madau1995} to the \citet{inoue2014} prescription.

The code produces the marginalized probability distribution of the derived quantities. To our aims, among the many products available, we focus the analysis on dust mass, $M_\textrm{dust}$, infrared luminosity, $L_\textrm{IR}$, and stellar mass, $M_\star$.

\subsection{CIGALE \label{sect:CIGALE}}

The CIGALE software \citep{burgarella2005,noll2009,boquien2019} is a public and versatile code that models galaxy SEDs from X-rays to radio taking into account the balance between the energy absorbed by dust in the UV-optical and reemitted in the IR. The code can be used to fit photometry and spectroscopy, or to create mock SEDs thanks to its large library of models. 
CIGALE allows for different kinds of star formation histories (SFH, parametric and non-parametric), stellar population models \citep[e.g.,][]{bc03,maraston2005}, dust attenuation laws \citep[e.g.,][]{calzetti2000,charlot2000}, IR emission models \citep[e.g.,][]{DL07,dale2014}, AGN components \citep[e.g.,][]{fritz2006,ciesla2015}, X-rays and radio emission. 

SED fitting was performed on the N2GN sources adopting the \citet{bc03} stellar models, with a \citet{chabrier2003} IMF and solar metallicity, the \citet{charlot2000} attenuation law, the \citet{DL07} dust emission models, renormalized as in \citet{draine2014}, and \citet{stalevski2016,stalevski2012} AGN models when hinted by the X-ray, UV, MIR or radio emission. A delayed and truncated SFH was used, with e-folding time of the main stellar population of 0.1, 0.5, 1, 5, 10\,Gyr, and truncation ages of 50 or 100\,Myr.

\subsection{SED3FIT}\label{sect:sed3fit}

Originally inspired by MAGPHYS, SED3FIT \citep{berta2013} combines three emission components simultaneously to reproduce the observed SEDs of galaxies. It adds a third component to the MAGPHYS stellar and dust models \citep{bc03,dacunha2008,charlot2000}. 
The third component can be an AGN-torus model or any other kind of emission that might need to be considered in addition to the stellar+dust model.
Also SED3FIT exists in the original version by \citet{berta2013} and in a new high-$z$ implementation (named SED3FIT-h$z$).

Due to the huge number of possible model combinations that arise from adding a third library to the already big stellar and dust collections, as well as from the free normalization of the stars+dust component, the code samples the parameters space and the libraries of models randomly instead of systematically test all possibilities. A total of $10^{11}$ models are compared to the observed SED of each N2GN galaxy. 
SED3FIT is here used in two cases:
\begin{enumerate}[label=\alph*)]
\item sources with evidence of AGN activity, such as radio excess with respect to the radio-FIR correlation, X-ray emission, or MIR excess or power-law SED. In this case, we adopted the AGN-torus library by \citet{fritz2006}, updated by \citet{feltre2012}, that includes both the emission of the dusty torus, heated by the central AGN engine, and the emission of the accretion disc. 
\item sources with the short-wavelength photometry not reproduced by the standard MAGPHYS or CIGALE fit, thus having a rest-frame UV excess. In this case, we used SED3FIT with a library of non-extinguished young simple stellar populations (SSPs), of solar metallicity, \citet{chabrier2003} IMF, and with ages between 10 and 100\,Myr, drawn from the stellar library by \citet{bc03}. This choice is aimed at schematically simulating a young stellar component producing the observed excess and does not affect the estimate of dust masses.
\end{enumerate}

Out of the whole N2GN sample, eight sources require an AGN-torus component to fit their MIR and UV emission. Six of them are also detected in the X-rays \citep{evans2024,barro2019,alexander2003}. More objects are matched to X-rays sources, for a total number of 17, and in six cases the observed radio emission is in excess to the radio-FIR correlation of star forming galaxies. A detailed description of each individual source is given in Appendix \ref{app:indiv_srcs}.

\begin{figure}[!t] % !ht
\centering
\settowidth{\imagewidth}{\includegraphics{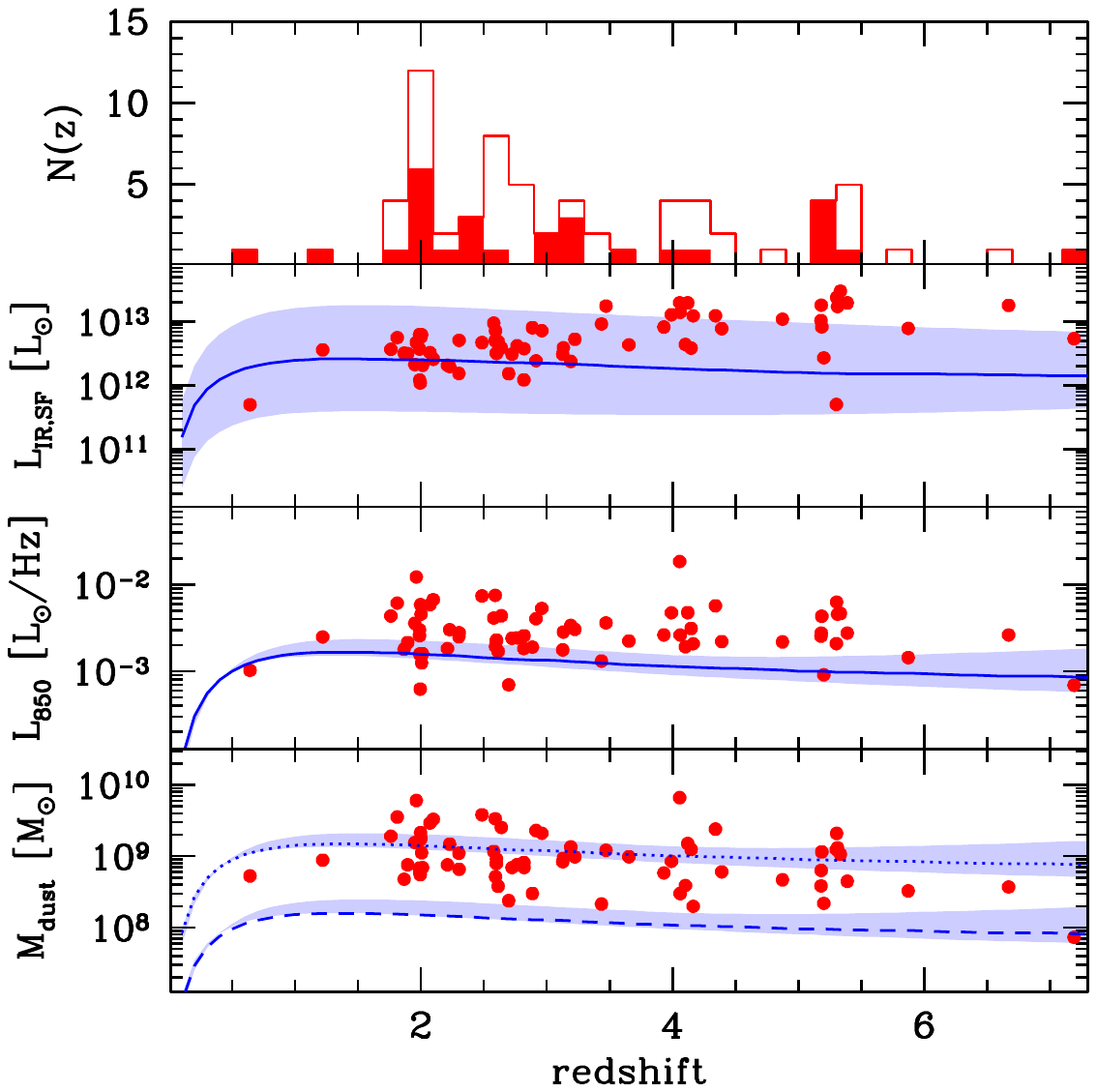}}
\includegraphics[trim=0 0.22\imagewidth{} 0 0.10\imagewidth{}, clip, width=0.47\textwidth]{figs1/LM_z3_20250128.pdf}
\caption{Distribution of the N2GN galaxies, their luminosity, and dust mass as a function of redshift. 
{\em Top panel}: Redshift distribution. The filled histogram includes the sources with spectroscopic redshift, while the open histogram includes all sources. 
{\em Second} panel: Infrared luminosity of the star forming component, integrated between 8 and 1000 $\mu$m.  The blue solid line represents the expected luminosity for a source of 0.7\,mJy in the NIKA2 1.2\,mm band, assuming its emission is described by the median SED shown in the left panel of Fig. \ref{fig:med_avg_SED}. The shaded area is given by the best fit models of all sources. {\em Third}: 850 $\mu$m monochromatic luminosity. {\em Bottom} panel: Dust mass distribution as a function of redshift. The dashed and dotted lines represent the minimum dust mass and maximal-completeness 
dust mass obtained rescaling the $L_{850}$ versus $z$ line adopting the minimum and maximum $M_\textrm{dust}/L_{850}$ ratio in the sample (Sect. \ref{sect:mass_compl}).}
\label{fig:L_z}
\end{figure}

\section{Derived properties of the N2CLS sources}\label{sect:properties}

The quantities produced by the SED fitting, of interest for this work, are the dust mass, $M_\textrm{dust}$, the infrared luminosity, $L_\textrm{IR}$, and the stellar mass, $M_\star$, of the N2GN galaxies.   
The best fit of each individual galaxy is shown in Fig. \ref{fig:indiv_glxs}.

The best models of all N2GN sources (obtained with the UV-to-radio SED fitting) are also collected in Fig. \ref{fig:med_avg_SED}, normalized by their $L_\textrm{IR}$ ({\em left}) and by their $M_\textrm{dust}$ ({\em right}).  
The normalization by $M_\textrm{dust}$ is very similar to a sub-millimeter normalization along the Raileigh-Jeans (RJ) tail of the dust emission (e.g., at 850 $\mu$m in the rest frame). This is easily explained by the bulk of the dust mass budget being locked into the cold dust that dominates the RJ tail. The scatter is driven by the fact that the mass-to-light ratio is not unique, since each galaxy comes with its own value of $M_\textrm{dust}/L_\textrm{sub-mm}$. 
Consequently, because of this relation between sub-millimeter emission and $M_\textrm{dust}$,  dust masses derived by SED fitting without the knowledge of the actual rest-frame sub-millimeter emitted luminosity are subject to large uncertainties \citep[e.g.,][]{berta2016} and/or need important assumptions in the SED modeling (e.g. fixing $\beta$ in a MBB analysis). 

The two middle panels of Figure \ref{fig:L_z} present the integrated infrared luminosity ($L_\textrm{IR,SF}$, 8-1000 $\mu$m, excluding the possible AGN-torus contribution), and the monochromatic 850 $\mu$m luminosity of the N2GN galaxies ($L_{850}$), as a function of redshift. The effects of the strong negative $k$-correction induced by the steep RJ tail of the dust SED are evident in both panels, and are more accentuated for $L_{850}$: on average the sub-millimeter luminosity of the sources keeps almost constant (and even slightly decreases) as redshift increases.
The blue line in the $L_\textrm{IR,SF}$ panel shows the expected luminosity limit, given the median SED template based on the N2GN sources and the 0.7\,mJy flux cut at 1.2\,mm.
The infrared luminosity is affected by a much larger scatter than $L_{850}$, related to the variety of observed SED colors (Fig. \ref{fig:med_avg_SED}).

The bottom panel of Fig.\,\ref{fig:L_z} reports the distribution of $M_\textrm{dust}$ as a function of redshift. Thanks to the negative sub-millimeter $k$-correction, there is no evident dependence of $M_\textrm{dust}$ on redshift. The dust masses of the N2GN galaxies span over the range between $\sim10^8$ and $\sim6\times10^9$ $M_\odot$.

\begin{figure}[!t]
\centering
\rotatebox{-90}{\includegraphics[height=0.47\textwidth]{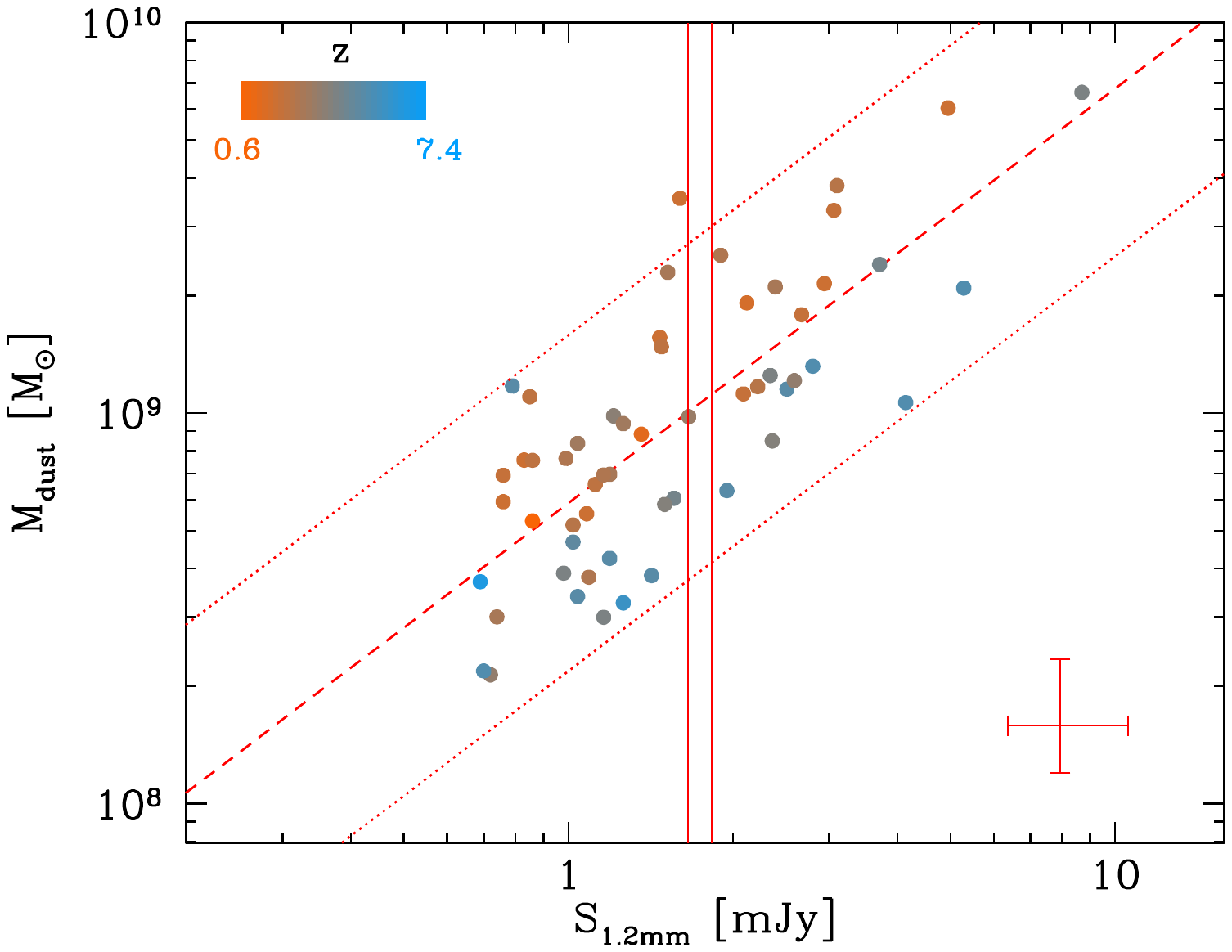}}
\caption{Distribution of the N2GN galaxies as a function of dust mass and 1.2 mm flux density, color coded on the basis of their redshift. Median error bars are shown in the bottom-right corner. The red dashed line is a least squares fit to the data. The red dotted lines represent the same line scaled to the 2.5th and 97.5th percentiles of the residuals distribution. The red, solid vertical lines mark the flux of the sources with no redshift measurement and multi-wavelength counterparts available.}
\label{fig:no_z}
\end{figure}

\subsection{Sources without redshift}

Three N2GN sources have no match to multi-wavelength catalogs, except to radio data, namely N2GN\_1\_17b, 34a and 55. Therefore, these sources have no $M_\textrm{dust}$, nor a redshift determination. N2GN\_1\_17b and 34a are sub-components of NIKA2 sources and are detected by NOEMA at 2.0\,mm and 1.2\,mm, respectively. N2GN\_1\_55 is a single source, detected by NIKA2 at 1.2 mm. 
We derived the dust mass of these objects by analyzing the dependence of $M_\textrm{dust}$ on millimeter flux of all other N2GN galaxies.

Figure \ref{fig:no_z} shows the derived $M_\textrm{dust}$ versus the observed 1.2~mm flux of the galaxies. The red dashed line represents a simple least square fit to the sample and the dotted lines are the same fit, rescaled $\pm1.96\times$ the residuals r.m.s. (corresponding to the 2.5-97.5 percentile range) around the fit. The red, solid vertical lines mark the 1.2 mm flux of the two sources without a redshift determination detected at 1.2 mm. A similar analysis was carried out at 2.0 mm for N2GN\_1\_17b.
As a result, the range of dust masses associated to these three sources is roughly as broad as one order of magnitude. Table \ref{tab:ID_z_Md_LIR} includes these values.

\subsection{Dust mass to light ratio}

\begin{figure}[!t]	%[!ht]
\centering
\settowidth{\imagewidth}{\includegraphics{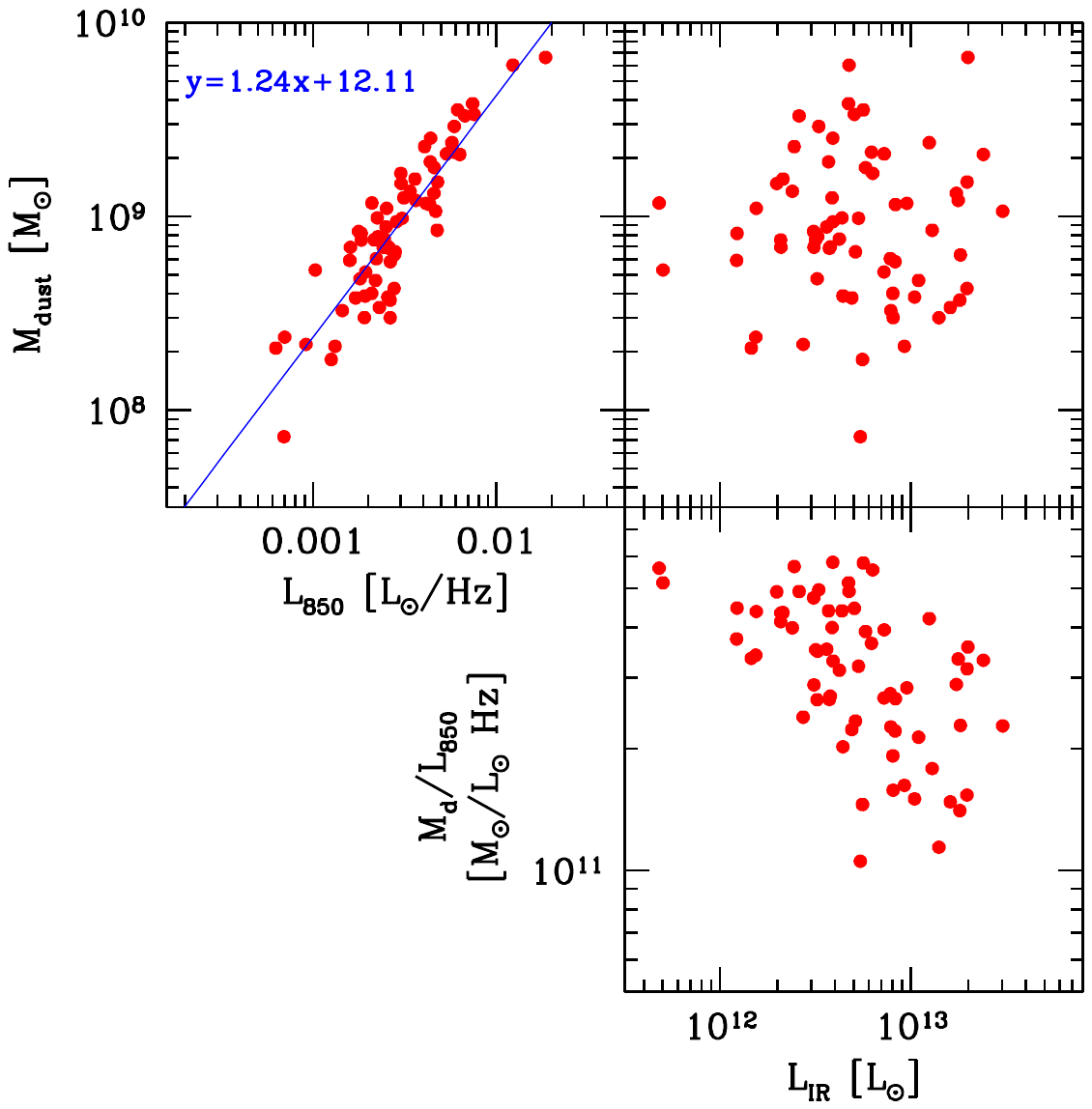}}
\includegraphics[trim=0 0.22\imagewidth{} 0 0.10\imagewidth{}, clip, width=0.47\textwidth]{figs1/pl_ML850abc.pdf}
\caption{Study of the $M_\textrm{dust}/L_{850}$ ratio. {\em Top-left panel}: least squares fit to the correlation between $M_\textrm{dust}$ and $L_{850}$. {\em Right-hand panels}: $M_\textrm{dust}$ and $M_\textrm{dust}/L_{850}$ as a function of IR luminosity.}
\label{fig:ML850}
\end{figure}

For many dusty infrared galaxies, the absence of a fully sampled SED hinders the possibility to perform a detailed SED fitting and to derive a reliable estimate of -- among other quantities -- its dust mass.
Thanks to the exquisite SED sampling of the N2GN sources, here we study the dust mass to rest-frame sub-millimeter luminosity ratio, with the goal to define a relation between these two quantities, to be used in cases that do not benefit from such a good wavelength coverage as the GOODS-N field. To this aim, Fig. \ref{fig:ML850} shows the $M_\textrm{dust}/L_{850\mu\textrm{m}}$ ratio, based on the $\kappa_\nu$ normalization by \citet{draine2014} as usual.

We quantified the correlation  between $M_\textrm{dust}$ and $L_{850}$ with a simple least squares fit to the data, leading to the relation $\log(M_\textrm{dust}) = (1.24 \pm 0.08) \log(L_{850}) + (12.11 \pm 0.20)$. The Spearman rank correlation coefficient is $r_s=0.83$, with 66 degrees of freedom and a Student's $t$ distribution of $t=12.1$, translating in a probability of correlation larger than $99.9$\%.

No correlation is seen between $M_\textrm{dust}$ and $L_\textrm{IR}$ (top-right panel), thus testifying that no artificial dependence between the two quantities has been introduced by the SED fitting underlying assumptions.
On the other hand, there seem to be an apparent anti-correlation between $M_\textrm{dust}/L_{850}$ and $L_\textrm{IR}$ (bottom-right panel), but it is driven by the avoidance zone dictated by the N2GN selection and by the more obvious correlation between $L_{850}$ and $L_\textrm{IR}$.

\subsection{Starbursty nature of the N2GN galaxies}\label{sect:starbursty}

\begin{figure}[t]
\centering
\rotatebox{-90}{\includegraphics[height=0.47\textwidth]{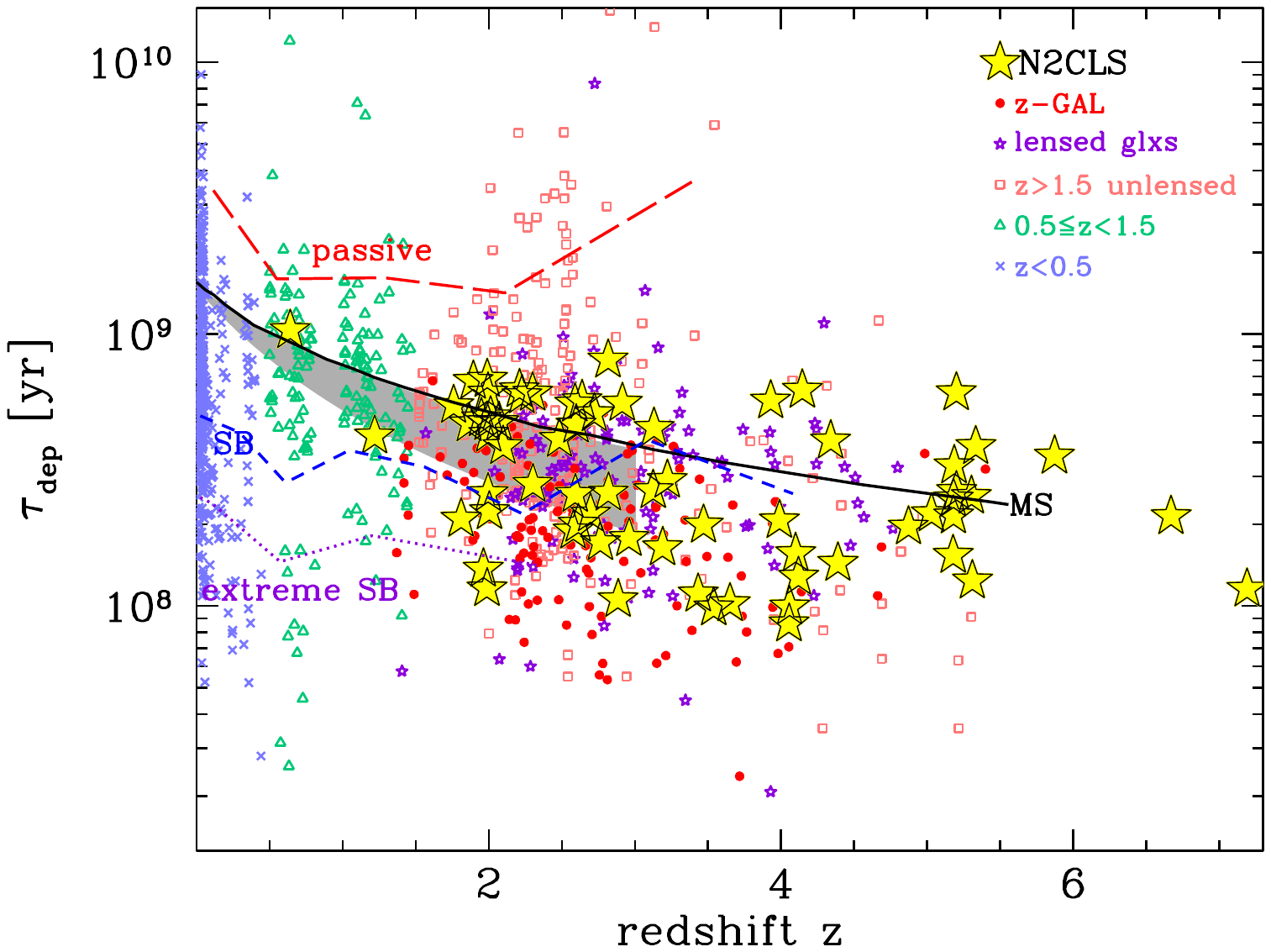}}
\caption{Depletion timescale of the N2GN galaxies obtained applying the \citet{tacconi2020} scaling relation. 
The gray shaded area is the trend found by \citet[][]{saintonge2013}. The different lines represent the trends found by \citet{tacconi2020} for: main sequence galaxies (selected within $\Delta_\textrm{MS}=\pm0.6$ dex, black solid line); starburst galaxies ($\Delta_\textrm{MS}>0.6$ dex, blue dashed line); extreme starbursts  ($\Delta_\textrm{MS}>1.2$ dex, purple dotted line); below-MS galaxies ($\Delta_\textrm{MS}<0.4$ dex, red long-dashed line).
Literature data are from \citet{alaghbandzadeh2013}, \citet{aravena2016, aravena2014, aravena2013}, \citet{bakx2020}, \citet{berta2023}, \citet{bothwell2017, bothwell2013}, \citet{carilli2010}, \citet{chung2009}, \citet{combes2011, combes2013}, \citet{dannerbauer2019}, \citet{decarli2016, decarli2019}, \citet{dunne2021, dunne2020b}, \citet{fujimoto2017}, \citet{freundlich2019}, \citet{geach2011}, \citet{genzel2015, genzel2003}, \citet{george2013}, \citet{hagimoto2023}, \citet{harris2012, harris2010}, \citet{ivison2013, ivison2011, ivison2010}, \citet{penney2020}, \citet{riechers2020, riechers2011}, \citet{rudnick2017}, \citet{sharon2016}, \citet{solomon1997}, \citet{tacconi2018, tacconi2013}, \citet{thomson2012}, \citet{valentino2018}, \citet{villanueva2017}, \citet{wang2018}, and \citet{yang2017}.
}
\label{fig:taudep}
\end{figure}

Adopting the scaling relation by \citet{tacconi2020}, that links $M_\star$, redshift and distance from the main sequence (MS) of star forming galaxies to molecular gas mass ($M_\textrm{mol}$) and molecular gas depletion timescale ($\tau_\textrm{dep}$), we computed $\tau_\textrm{dep}$ of the N2GN galaxies. Figure \ref{fig:taudep} compares the result to a collection of galaxies from the literature with $M_\textrm{mol}$ derived from CO data \citep[][and references therein]{tacconi2020,berta2023}.

The N2GN galaxies occupy mainly the locus of starbursts, with depletion timescales of the order of 0.1 to 1.0\,Gyr. It is worth recalling that the \citet{tacconi2020} relation uses the MS parametrization by \citet{speagle2014} as reference, that does not include the bending of the MS at high stellar masses.

%%%%%%% taudep fig with many references %%%%%%

%%%%%%% end of taudep fig %%%%%%%

\begin{figure}[!ht]
\centering
\rotatebox{-90}{\includegraphics[height=0.47\textwidth]{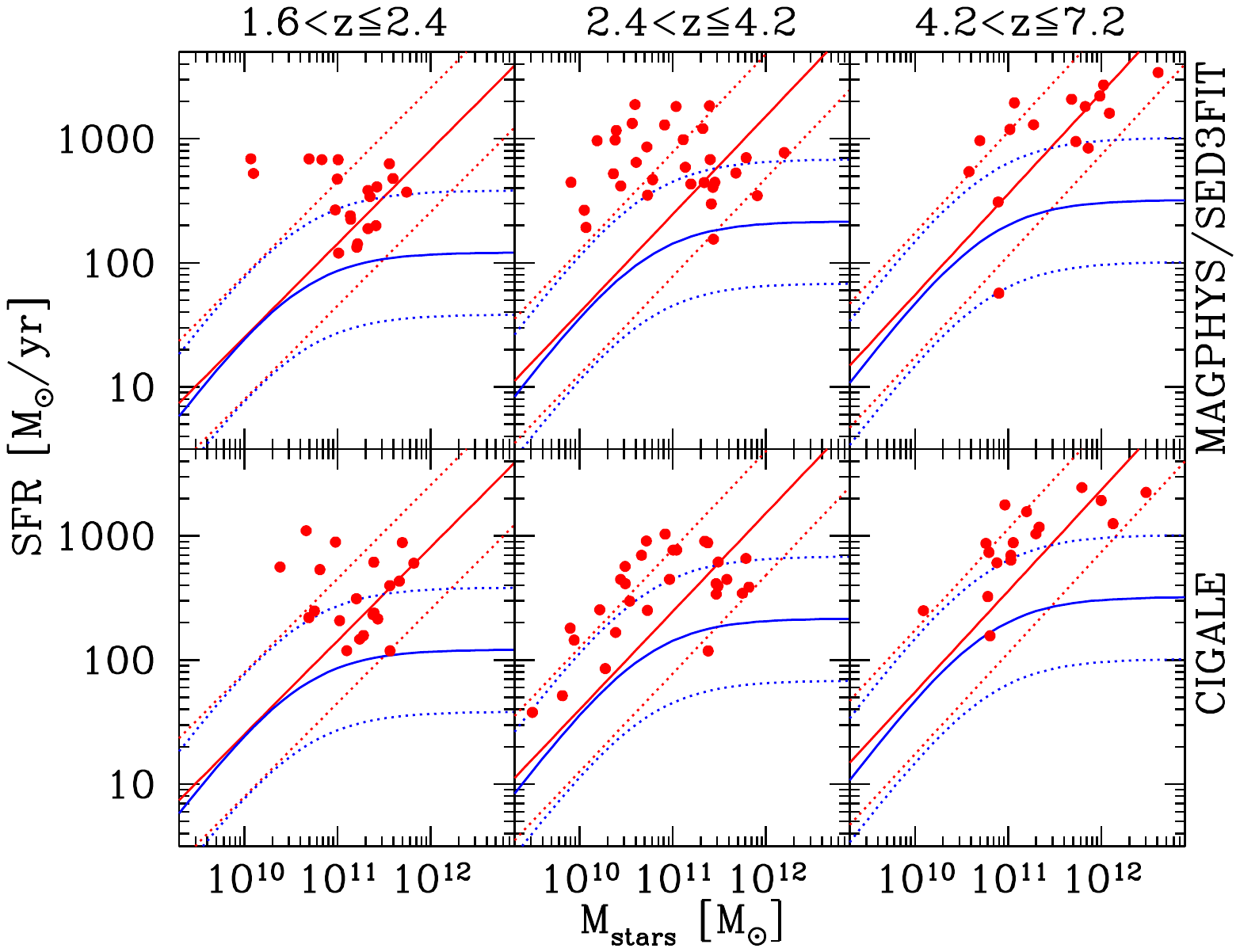}}
\caption{Position of the N2CLS GOODS-N galaxies in the $M_\star$ versus SFR plane, as obtained with the SED fitting results, in different redshift bins. Each galaxy is positioned on the basis of its $\Delta\left(\log\left(\textrm{SFR}\right)\right)_\textrm{MS}$ computed at its actual redshift with respect to the \citet{popesso2023} MS, translated to the average redshift of each bin. The solid blue lines represent the parametrization of the MS of star forming galaxies by \citet{popesso2023}. For comparison the solid red lines refer to the one by  \citet{speagle2014}. The dotted lines are placed at $\pm0.5$ dex from the solid ones. {\em Top row}: results obtained with the fiducial fit (MAGPHYS or SED3FIT high-$z$). {\em Bottom row}: results obtained with CIGALE for comparison.}
\label{fig:MS}
\end{figure}

Figure \ref{fig:MS} shows the position of the N2GN sources in the $M_\star$ versus SFR space, as obtained with our fiducial UV-to-radio SED fitting (MAGPHYS or SED3FIT high-$z$) and for comparison with CIGALE. Two different parametrizations of the MS are overlaid to the data: those by \citet[][red lines]{speagle2014} and \citet[][blue lines]{popesso2023}. The latter includes the well-known flattening at large stellar masses, while the former does not.
Depending on the adopted reference MS, most of the N2GN sources are outliers of the MS \citep[i.e., starbursts, case of][]{popesso2023} or include also a non-negligible number of MS galaxies \citep[case of][]{speagle2014}. 

Noteworthy, the stellar mass of several N2GN high-redshift galaxies is of the order of a few $10^{11}\, M_\odot$. Similar very large values of $M_\star$ at high redshift have been found in the recent literature, when exploiting deep JWST data \citep[e.g.,][]{xiao2024}. They represent a challenge to galaxy formation models, as such high stellar masses would require very efficient star formation in the early stages of galaxy evolution, and would account for about 17\% of the cosmic SFRD at $z=5-6$ \citep{xiao2024}. 

For some of them, significant differences between MAGPHYS/SED3FIT and CIGALE are evident in Fig. \ref{fig:MS}. Appendix \ref{app:comparison_codes} discusses the direct comparison of the derived stellar masses, revealing a good overall agreement but also the presence of a few important outliers, with differences in the best fit $M_\star$ estimate as large as one order of magnitude. The most critical cases are very red galaxies at $z>4$. Appendix \ref{app:taudep_change} shows the consequence on $\tau_\textrm{dep}$ of adopting the $M_\star$ value derived with CIGALE instead of the one by MAGPHYS/SED3FIT. The main result of short depletion timescales is not affected.
The most likely cause of these $M_\star$  differences is a large difference of $A_V$ for very extinguished galaxies. 

\begin{figure}[!t]
\centering
\settoheight{\imagewidth}{\includegraphics{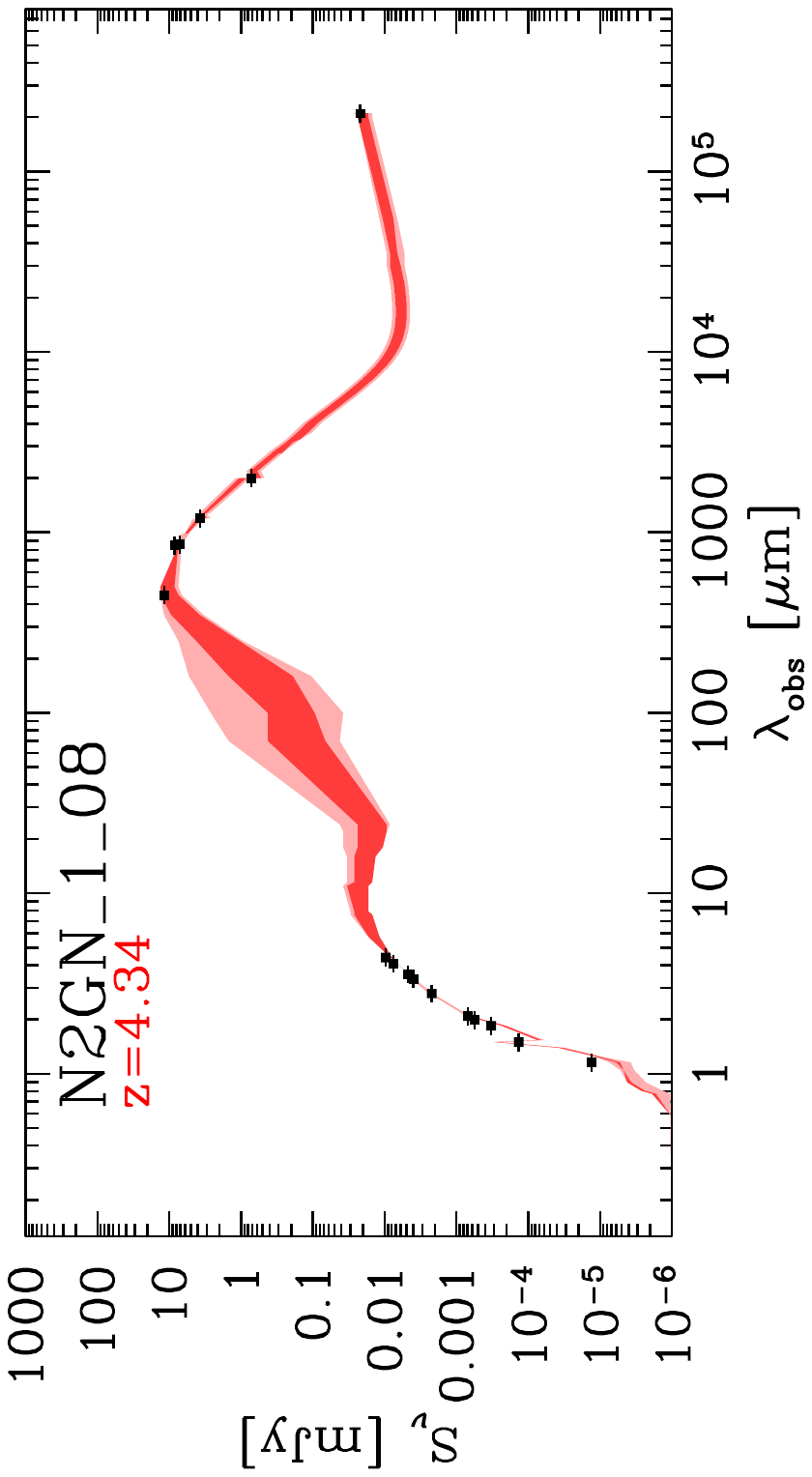}}
\rotatebox{-90}{\includegraphics[trim=0 0 0.22\imagewidth{} 0, clip, height=0.47\textwidth]{figs1/N2GN_1_08_sed3fit_hz_pho_land2_no_bf.pdf}}
\caption{Example of fit uncertainty when no NIR and MIR data are available and a big gap in wavelength exists between the rest-frame optical and FIR spectral domains. The two different tones of red shaded areas represent the 2.5\%, 16\% 84\% and 97.5\% percentiles of the models, as computed in the photometric bands of the input catalog; the corresponding $M_\star$ range is as large as 0.25\,dex (2.5th to 97.5th percentiles).}
\label{fig:no_NIR}
\end{figure}

The N2GN data set can arguably be considered the best photometric collection of an extragalactic blank field to date, including observations obtained with all major space-borne facilities available and covering the whole electromagnetic spectrum. Despite this opulence, it turns out that the majority of these $M_\star>10^{11}\, M_\odot$ high-$z$ galaxies are affected by a large, uncovered gap between their rest-frame optical and FIR data. One example is shown in Fig. \ref{fig:no_NIR}, where the 2.5\%, 16\% 84\% and 97.5\% percentiles of the modeled photometry are shown. Very deep MIR observations (in the observed frame, for example with JWST/MIRI) are necessary to constrain the NIR emission of these sources and their stellar mass.

%%%%%%%%%%%%%%%%%%%%%%%%%%%%%%%%%%%%%%%%%%%%%%%%%%%%%%%%%%%%%%%

\subsection{Blue excess}\label{sect:blue_excess}

A few N2GN sources are characterized by a blue excess detected in the short-wavelength JWST and HST bands, with respect to the extinguished stellar models fitting the optical-NIR SED and providing the power to heat the bright FIR dust emission. 
At the redshift of the N2GN galaxies, this excess lies in the rest-frame UV or blue optical domain. For these dusty powerful star forming galaxies -- hence excluding the case of passive galaxies with UV emission from evolved stellar populations -- two possible components might contribute to this excess: a type-1 AGN component, or a young stellar population with no (or low) extinction. 

As mentioned before, in absence of any other evidence of AGN activity, such as a MIR excess, optical/blue point-like morphology, (hard) X-ray emission or a radio excess with respect to the radio-FIR correlation, we opted to reproduce the observed blue data with a young SSP. Six N2GN sources were treated in this way, namely N2GN\_1\_06 and 13 (at $z>5$) and 15, 49, 53, and 56b (at ``cosmic noon''; Fig. \ref{fig:indiv_glxs}). 

A similar excess of emission at short wavelengths was found for ``little red dots'' observed with JWST: compact red galaxies at $z>5$, with F277W-F444W$>$1, and F150W-F200W$<$0.5\,mag and F444W$\le$28\,mag \citep{labbe2023,perezgonzalez2024,matthee2024,kocevski2024}. Based on the rest-frame optical-NIR JWST photometry, the very red color testifies the presence of a large amount of dust, while the blue excess can be explained by either
a QSO, a clumpy ISM surrounding the star forming regions, allowing us to see unobscured very young star formation, or a gray attenuation law, typically linked to significant scattering \citep{perezgonzalez2024}. The two $z>5$ such cases present in N2CLS GOODS-N (N2GN\_1\_06 and 13) benefit from a very complete multi-wavelength coverage and do not show any hints of AGN activity over the whole X-rays to radio broad-band SED.

\section{Comoving number density}\label{sect:comoving_N}

The comoving number density of the sources in the N2GN survey, in intervals of $M_\textrm{dust}$, also known as DMF, is given by
\begin{equation}\label{eq:phi_M}
\Phi(M_\textrm{dust})\Delta M_\textrm{dust} = \sum_i \frac{1}{V_a^i}\Delta M_\textrm{dust}\textrm{,}
\end{equation}
where the sum is computed over all galaxies in the given mass bin of width $\Delta M_\textrm{dust}$.
The volume within which each source is accessible to the N2CLS GOODS-N survey is a spherical shell: 
\begin{equation}\label{eq:Va}
V_a = \frac{\Omega}{4\pi} \int_{z_\textrm{min}^\textrm{bin}}^{\textrm{min}\left(z_\textrm{max},z_\textrm{max}^\textrm{bin}\right)}{\frac{\textrm{d}V}{\textrm{d}z}}\textrm{,}
\end{equation}
where the minimum redshift is given by the lower boundary of the given redshift bin considered and the maximum redshift is the minimum value between the upper boundary of the bin and the maximum accessible redshift. The latter is the highest redshift at which a galaxy would be observable in the survey \citep{schmidt1968}, given the N2GN 1.2 mm flux limit of 0.7 mJy. 

The total area of the N2GN field is 159\,arcmin$^2$ \citep{bing2023}. Nevertheless, the depth of the NIKA2 map is not uniform, but it varies by up to a factor of three across the N2GN field. As a consequence, $\Omega$ is the effective area associated to each galaxy, as computed by \citet{bing2023}. 

Because of the strong negative $k$-correction along the steep RJ tail of the dust SED, $z_\textrm{max}$ turns out to be very large (typically $z_\textrm{max}>20$). Hence, the accessible volume $V_a$ is limited by the upper boundary of the given redshift bin.

\subsection{Dust mass completeness}\label{sect:mass_compl}

The N2GN sample selection is based on a 1.2 mm flux cut in the observed frame. 
Flux completeness has already been taken into account by using the effective area of each source in the computation of the accessible volume. 

The dust mass budget is dominated by the rest-frame sub-millimeter emission. Therefore, there is an almost-direct link between the flux selection and the derived quantity $M_\textrm{dust}$.
Nevertheless, the relation between $S(1.2 \textrm{mm})$ and $M_\textrm{dust}$ is not univocal, because the latter is derived from the former by means of a proper SED fitting of each galaxy, and therefore the dust mass to sub-millimeter light ratio varies in the sample (Fig. \ref{fig:ML850} and associated text). 
Consequently, it is not possible to define a sharp mass limit encompassing the whole sample.

The bottom panel of Fig. \ref{fig:L_z} shows the distribution of $M_\textrm{dust}$ as a function of redshift. The dashed and dotted lines represent the minimum dust mass and maximal-completeness dust mass obtained rescaling the $L_{850}$ versus $z$ line adopting the minimum and maximum $M_\textrm{dust}/L_{850}$ ratio in the sample. The shaded areas are based on the best fit SED models of all N2GN sources.

\begin{figure}[t]
\centering
\settoheight{\imageheight}{\includegraphics{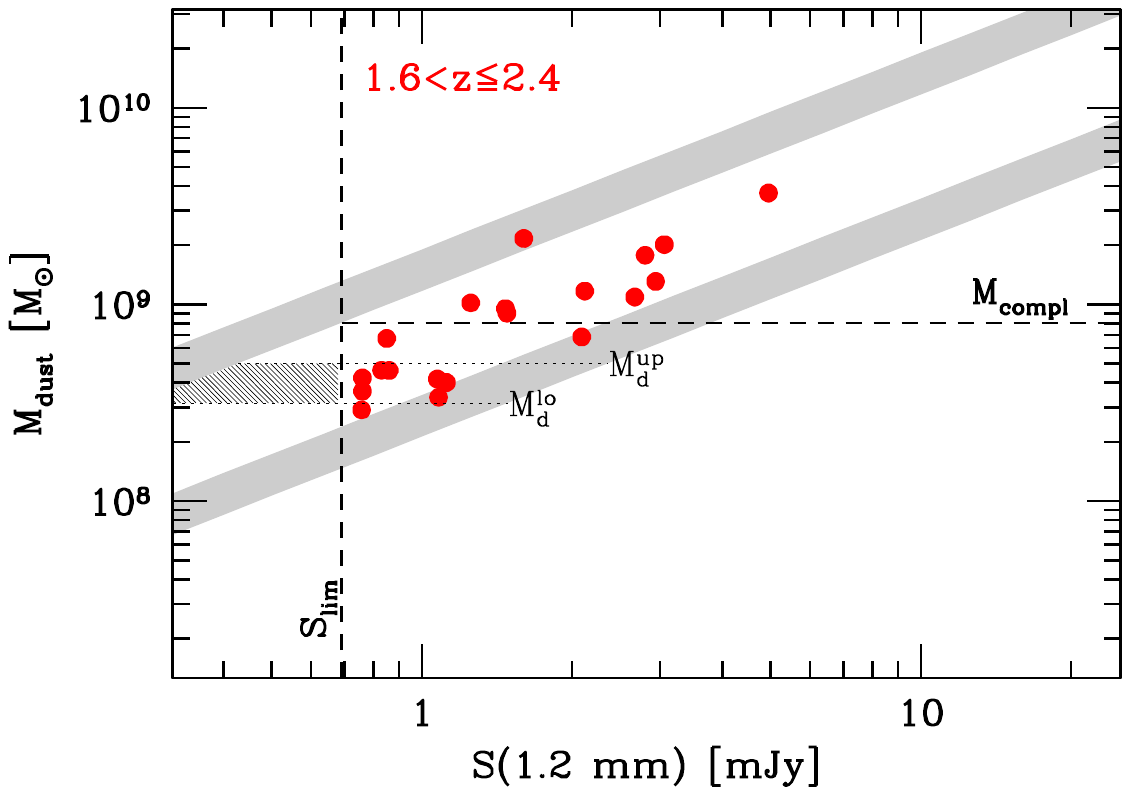}}
\includegraphics[trim=0 0.39\imageheight{} 0 0.10\imageheight{}, clip, width=0.47\textwidth]{figs1/example_completeness_1p6z2p4.pdf}
\caption{Exemplification of how the dust mass completeness is computed in the redshift bin $1.6<z\le2.4$ (Sect. \ref{sect:mass_compl}). The diagonal shaded areas represent tracks of the N2GN SED models, given  the minimum and maximum $M_\textrm{dust}/L_{850}$ ratios in the sample. The vertical dashed line is the 1.2~mm flux limit of the survey, and the horizontal dashed line is the corresponding mass maximal completeness threshold. The dotted horizontal lines mark the boundaries of an example mass bin, in which the hatched area would contain the sources missed by the flux cut.}
\label{fig:Mdust_compl}
\end{figure}

The minimal area represents the absolute minimum $M_\textrm{dust}$ that objects in the sample could have, at the given redshift, if they had the minimum ``observed'' $M_\textrm{dust}/L_{850}$ and a flux of 0.7\,mJy at 1.2\,mm, accounting for the scatter due to the different SED shapes of the sample.
The maximal area, instead, represents the $M_\textrm{dust}$ that an object would have, if its flux were 0.7\,mJy at 1.2\,mm, and if it were characterized by the maximum ``observed'' $M_\textrm{dust}/L_{850}$ ratio  in the N2GN sample, given the scatter of SED shapes. These maximal $M/L$ tracks do {\em not} indicate a maximum $M_\textrm{dust}$ limit. Schematically, above this area/line the sample is maximally complete in terms of dust mass. Below it, dust mass incompleteness must be corrected.

This effect has been first examined in the past for the stellar mass function \citep[e.g.,][]{dickinson2003,fontana2004,fontana2006,berta2007} and later on for the gas mass function and the DMF \citep[e.g.,][]{berta2013b,pozzi2020}. Here we applied the recipe by \citet{fontana2004} to the case of dust masses and the 1.2 mm flux cut. 

Figure \ref{fig:Mdust_compl} shows the distribution of dust masses as a function of the observed 1.2~mm flux for the N2GN galaxies in the $1.6<z\le2.4$ redshift bin, as an example. A similar procedure was applied also to the other redshift bins. The vertical line represents the adopted 1.2~mm flux limit. The diagonal shaded areas are the tracks described by the minimum and maximum $M_\textrm{dust}/L_{850}$ ratios in the sample, at the central redshift, taking into account the scatter due to the different SED shapes of the N2GN sources. The horizontal dashed line represents the dust mass level above which the sample is definitely complete, for a given SED. 

In a given mass bin, $M_\textrm{dust}^\textrm{lo}<M_\textrm{dust}<M_\textrm{dust}^\textrm{up}$, below the maximal completeness mass, the observed 1.2~mm flux corresponding to a given mass is encompassed between the maximal and minimum diagonal areas ($S^\textrm{lo}$, $S^\textrm{up}$), but the flux limit of the survey cuts the distribution. The sources in the shaded area are missed by the survey. The fraction of source actually observed, with respect to the total in that bin (i.e., the mass completeness) is
\begin{equation}
f_\textrm{compl} = \frac{\int_{S_{lim}}^{S_{up}} N(M,S^\prime)dS^\prime}{\int_{S_{lo}}^{S_{up}}N(M,S^\prime)dS^\prime}\textrm{.}
\end{equation}
Since the actual distribution of the galaxies below the flux limit is not known, we assumed that it does not depend of $M_\textrm{dust}$, implying that the distribution observed above the maximal mass threshold still holds below. 

We computed the observed DMF only in those bins with dust mass completeness $\ge0.5$. Table \ref{tab:dust_MF} lists the completeness values for each mass bin considered. We defer the reader to \citet{fontana2004,berta2007,pozzi2020} for further details.

\subsection{Dust mass function}\label{sect:DMF}

The comoving number density of galaxies as a function of dust mass (Eq.~\ref{eq:phi_M}) was corrected for mass incompleteness as described above. The incompleteness due to the non homogeneous map coverage and the limited S/N of the survey was already taken into account in the computation of $V_a$, when accounting for the effective area associated to each source.

\begin{figure*}[!ht]
\centering
\settowidth{\imagewidth}{\includegraphics{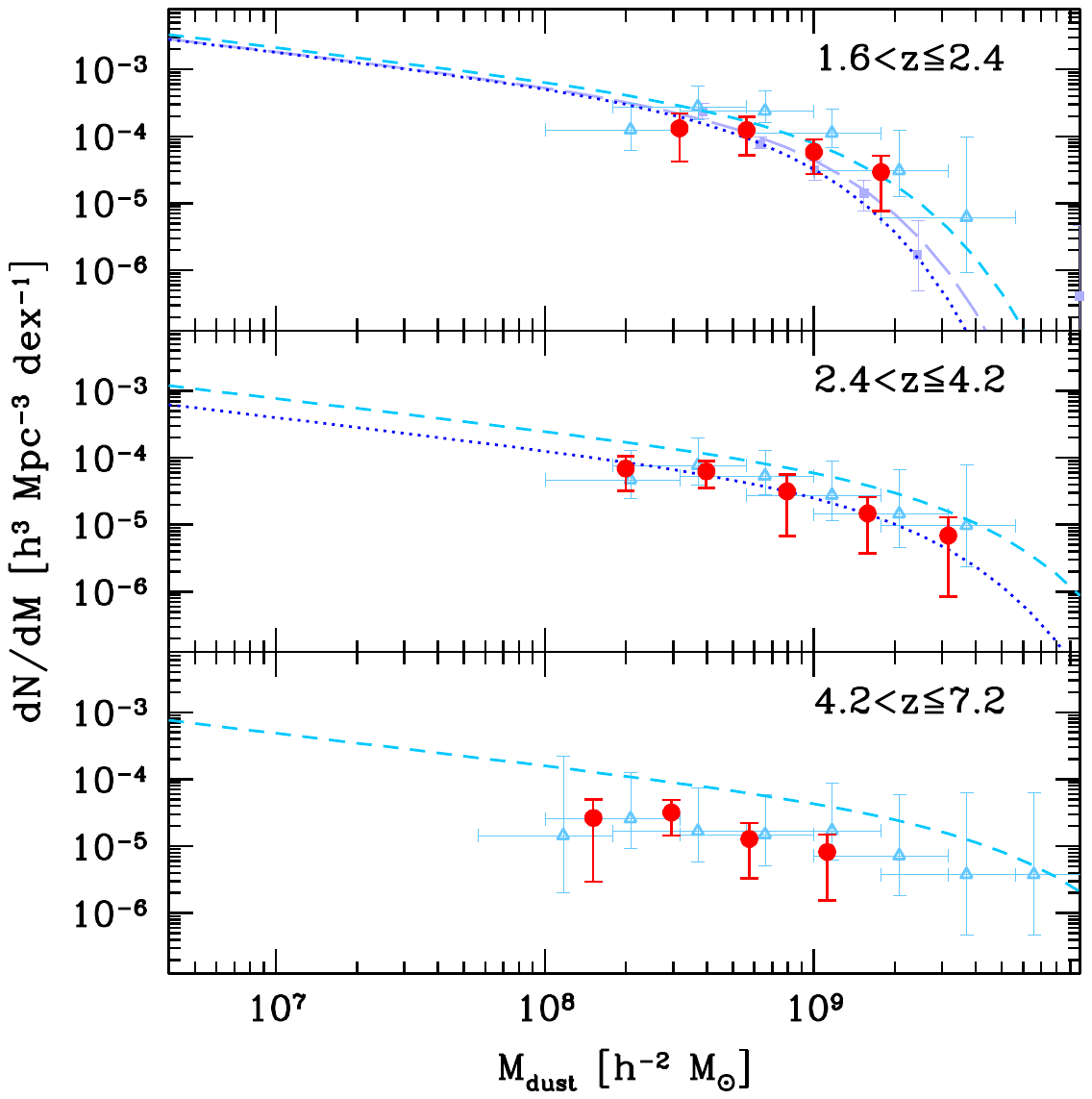}}
\includegraphics[trim=0 0.22\imagewidth{} 0 0.10\imagewidth{}, clip, height=0.47\textwidth]{figs1/pl_MFdust_9bb_l.pdf}
\includegraphics[trim=0 0.22\imagewidth{} 0 0.10\imagewidth{}, clip, height=0.47\textwidth]{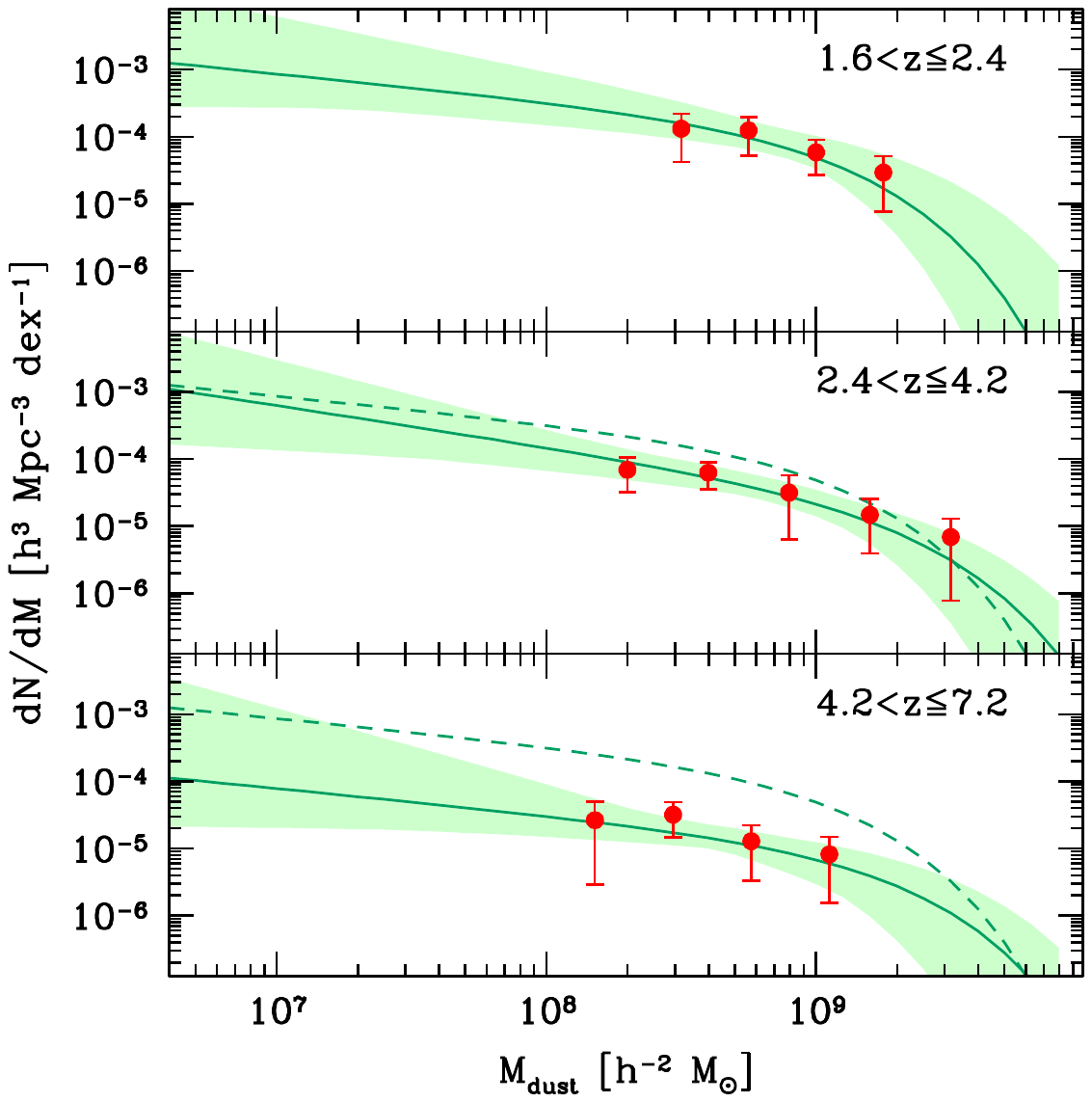}
\caption{Dust mass function of the N2GN sources (red symbols). Only the mass bins containing at least two galaxies and with a completeness $\ge50$\% are plotted. Uncertainties take into account $\Delta M_\textrm{dust}$, including SED fitting parameters sampling and $\Delta z$ contributions, and Poisson errors. {\em Left panel}: comparison to literature results by \citet[][blue filled squares and long-dashed line]{pozzi2020} and \citet[light blue triangles and short-dashed lines, fiducial results in the $2.0<z\le2.5$, $3.5<z\le4.5$ and $4.5<z\le6.0$ redshift bins]{traina2024}.  
The blue dotted lines are the same \citet{pozzi2020} DMF, evolved to $z=2.0$ and 3.3 (see Sect. \ref{sect:DMF}). 
{\em Right panel}: results of the STY analysis. The green solid lines and shaded areas are the result of the STY parametric analysis: most probable model (solid lines) and 1$\sigma$ uncertainty (shaded areas). The dashed green lines represent the $1.6<z\le2.4$ result, repeated in the other two redshift bins for comparison.}
\label{fig:dust_MF}
\end{figure*}

Figure \ref{fig:dust_MF} and Tab. \ref{tab:dust_MF} present the resulting DMF, in three redshift bins: $1.6<z\le2.4$, $2.4<z\le4.2$, and $4.2<z\le7.2$. The choice of these redshift bins comes from an optimization of the DMF signal and maximizes the number of sources in each bin. Only a few mass bins are populated because of the overall relatively small number of sources. 
Error bars account for $M_\textrm{dust}$ uncertainties, based on the probability distribution function (PDF) produced by SED fitting, redshift uncertainties, and Poisson errors.

\citet{pozzi2020} derived the DMF of 5546 galaxies detected by {\it Herschel} in the COSMOS field, in the redshift range $0.1<z<2.5$. They adopted a dust $\kappa_\nu(250\mu\textrm{m})=4.0$ cm$^2$ g$^{-1}$ \citep{bianchi2013}, which corresponds to the \citet{draine2003} dust models normalization.
Therefore, we rescaled their DMF to our $\kappa_\nu$ of reference \citep{draine2014}. The result is compared to the N2GN DMF at $1.8<z\le2.5$ in the left-hand panel of Fig. \ref{fig:dust_MF}. The two estimates are consistent within the errors, the {\it Herschel}-based DMF being systematically lower than the NIKA2 one.

\citet{traina2024} computed the DMF of 189 galaxies detected at millimeter wavelengths by ALMA in the COSMOS field as part of the A$^3$~COSMOS data collection \citep{liu2019b,adscheid2024,traina2024a}.  Their analysis covers the range from $z=0.5$ to $z=6.0$, split in eight redshift bins. In the left panel of Fig. \ref{fig:dust_MF}, the data by \citet{traina2024} are shown for only three $z$ bins, intersecting those of N2GN. The A$^3$~COSMOS DMF of the other intersecting bins is very similar and we omit it for sake of Figure readability. The DMF determined by \citet{traina2024} is very similar to our results, and extends to larger dust masses.

The DMF can be reproduced with a \citet{schechter1976} function. For logarithmic mass bins, this is parametrized as
\begin{equation}\label{eq:schechter_log}
\Phi(M_\textrm{dust})\,\textrm{d}\log M_\textrm{dust} = \frac{\Phi^\ast_\textrm{dust}}{M^\ast_\textrm{dust}}\left( \frac{M_\textrm{dust}}{M^\ast_\textrm{dust}} \right)^{\alpha+1} e^{-\frac{M_\textrm{dust}}{M^\ast_\textrm{dust}}} \ln(10)\,\textrm{d}\log M_\textrm{dust} \textrm{.}
\end{equation}

\citet{pozzi2020}  fitted the observed DMF with a non-linear least squares approach to determine the Schechter parameters over six redshift bins. At $z<0.25$ they sampled the DMF over a mass range sufficiently large to constrain the three parameters $\alpha$, $M^\ast_\textrm{dust}$ and $\Phi^\ast_\textrm{dust}$. At higher redshift, given the bright flux cut of their survey, they sampled only the massive tail of the DMF, and they needed to fix $\alpha$ to the local value of 1.48, in order to derive the evolution of $M^\ast_\textrm{dust}$ and $\Phi^\ast_\textrm{dust}$.

\begin{table}[t]
\centering
\small
\caption{\label{tab:dust_MF} The N2GN DMF.}
\begin{tabular}{ccc D{,}{\pm}{-1} cc}
%\begin{tabular}{ccc r@{$\pm$}l c}  %%%% for some unknown reason it's not aligning the heading properly (it always worked!!!), therefore I use an alternative method
\hline
\hline
\multicolumn{3}{c}{$\log\left(M_\textrm{dust} [M_\odot] \right)$} & \multicolumn{1}{c}{$dN/dM_\textrm{dust}$} & \multicolumn{1}{c}{N} & $M_\textrm{dust}$\\
Center & \multicolumn{2}{c}{Range} & \multicolumn{1}{c}{[$10^{-5}$ h$^3$ Mpc$^{-3}$ dex$^{-1}$]} & srcs. & Compl. \\ 
\hline
\multicolumn{6}{c}{$1.6<z\le2.4$}\\
8.50 & 8.38 & 8.62 & 13.18,8.98 & 5 & 0.55 \\
8.75 & 8.63 & 8.88 & 12.41,7.21 & 5 & 0.78 \\
9.00 & 8.88 & 9.12 &  5.85,3.12 & 6 & 1.00 \\
9.25 & 9.12 & 9.38 &  2.93,2.17 & 3 & 1.00 \\
\hline
\multicolumn{6}{c}{$2.4<z\le4.2$}\\
8.30 & 8.15 & 8.45 & 6.91,3.71 &  5 & 0.54 \\ 
8.60 & 8.45 & 8.75 & 6.27,2.73 & 10 & 0.76 \\
8.90 & 8.75 & 9.05 & 3.16,2.47 &  8 & 0.98 \\
9.20 & 9.05 & 9.35 & 1.48,1.10 &  4 & 1.00 \\
9.50 & 9.35 & 9.65 & 0.69,0.61 &  2 & 1.00 \\
\hline
\multicolumn{6}{c}{$4.2<z\le7.2$}\\
8.18 & 8.04 & 8.32 & 2.64,2.34 & 2 & 0.52 \\ 
8.47 & 8.32 & 8.61 & 3.18,1.73 & 6 & 0.86 \\
8.76 & 8.61 & 9.91 & 1.28,0.95 & 3 & 1.00 \\ 
9.05 & 8.81 & 9.20 & 0.82,0.55 & 3 & 1.00 \\ 
\hline
\end{tabular}
\tablefoot{The columns include: dust mass logarithmic bins (center, lower and upper boundaries); comoving number density; number of sources; and $M_\textrm{dust}$ fractional completeness (Sect. \ref{sect:mass_compl}). The DMF uncertainties take into account the contribution of the $M_\textrm{dust}$ probability distribution function produced by the SED fitting, as well as redshift uncertainties and Poisson errors.}
\end{table}

We studied the variation of $M^\ast_\textrm{dust}$ as a function of cosmic age, based on \citet{pozzi2020} results. By performing a simple least square fit to their $M^\ast_\textrm{dust}$ values as a function of cosmic time, we found the relation $\log M^\ast_\textrm{dust} = (-2.2 \pm 0.1) \times \log t_\textrm{Univ} + (30.1 \pm 1.0)$. The units are years for the age of the Universe and $M_\odot$ for $M^\ast_\textrm{dust}$.  While deriving this equation, we rescaled their values of $M^\ast_\textrm{dust}$ to the $\kappa_\nu$ normalization by \citep{draine2014}.

Using the derived equation, we evolved the \citet{pozzi2020} $M^\ast_\textrm{dust}$ to the central redshift of our two lower redshift bins ($z=2.0,\ 3.3$), but we did not extrapolate it to the third ($z=5.7$), because it is too far from the range of redshift actually covered by \citet{pozzi2020}.  
The light-blue long-dashed line in the top panel of Fig. \ref{fig:dust_MF} is the original $1.8<z<2.5$ DMF by \citet{pozzi2020}. The blue dotted lines in each panel  represent the \citet{pozzi2020} DMF evolved as described above, rescaling their value of $\Phi^\ast_\textrm{dust}$ to match the observed N2GN DMF.  

In both $1.6<z\le2.5$ and $2.5<z\le4.2$ redshift bins, the result (dotted lines in Fig. \ref{fig:dust_MF}) is consistent with the N2GN observed DMF. Notably so is also the low-mass end slope $\alpha=1.48$ derived locally by \citet{pozzi2020}, suggesting that the $M^\ast_\textrm{dust}$ derived from their data is a good approximation at least up to $z\sim 4.2$.

\citet{traina2024} performed a Markov Chain Monte Carlo (MCMC) fitting of the observed DMF. Their best fiducial model was obtained by fitting a combination of the A$^3$~COSMOS data and the {\it Herschel} results by \citet{pozzi2020}, simultaneously over all redshift bins. While doing so, they fixed the low-mass slope to the value determined locally by \citet{pozzi2020}. 
Their result is overplotted to the N2GN observed DMF in Fig. \ref{fig:dust_MF}. Their observed DMF is well consistent with the NIKA2 DMF at all redshifts, over the mass ranges in common, but at $z>2.4$ their Schechter parametrization differs significantly from ours (Sect. \ref{sect:STY}).

\subsection{STY analysis}\label{sect:STY}

We applied the STY method \citep*{sandage1979} to the N2GN galaxies, with the aim to derive new Schechter parameters describing the evolution of the DMF, based on the N2GN data themselves. It is important to remind that this is {\em not} a fit to the $1/V_a$ mass density using a Schechter function, but a statistical method taking into account the dust properties of each individual galaxy that contributes to the mass function. In this way, the mathod naturally includes all sources and account for the Poisson noise and the $M_\textrm{dust}$ uncertainty (including the errors on $z$).

The objects without a known redshift are included in this piece of analysis such that they contribute to the uncertainty of the Schechter parameters and of the integrated dust mass density (Sect. \ref{sect:rho}) in each redshift bin considered. Nevertheless, they have a very small impact ($\le2-3$\%) on the mass function and on the value of $\rho$. 

Following \citet{berta2007,berta2013b}, we used a MCMC sampling of the parameter space to explore the posterior probability function of the Schechter model, with parameters comprising both the three Schechter $M^\ast_\textrm{dust}$, $\Phi^\ast_\textrm{dust}$ and $\alpha$, and additional hyper-parameters to represent the dust mass of each individual galaxy dust masses. The MCMC engine was used with a Metropolis-Hastings sampling algorithm to explore the posterior probability function of the model. 
The hyper-parameters are constrained solely by the prior knowledge of the dust mass probability distribution function of each galaxy (produced by the SED fitting). Therefore, their marginalization automatically accounts for the mass uncertainties. We defer to \citet{berta2007} for a thorough description of the method and of the adopted algorithm.

Figure \ref{fig:dust_MF} shows the results of the STY analysis and Tab. \ref{tab:STY} summarizes the values found for the Schechter parameters. Despite the range of dust masses covered by the data is limited, we are able to set a constraint on all three Schechter parameters and their uncertainty. 

% --------------------------------------------------------------------

\section{Discussion}\label{sect:discussion}

The variation of $M^\ast$ and $\Phi^\ast$, obtained with the STY study of the DMF (Sect. \ref{sect:DMF}, Tab. \ref{tab:STY}), reveals a rapid evolution of the comoving density of DSFGs, roughly doubling in the first Gyr from $z=4.2-7.2$ to $z=2.4-4.2$, and increasing by more than a factor of 5 during the next two Gyr. Based on previous works, at more recent epochs, the comoving density evolution slows down and practically ceases by $z\sim0.5-1.0$ \citep{pozzi2020}.

The characteristic dust mass, $M^\ast_\textrm{dust}$, does not evolve at least until redshift $z\sim4$ and then is halved by $z\sim2$. The evolution continues down to the local Universe, decreasing $M^\ast_\textrm{dust}$ by more than an order of magnitude until $z\sim 0$.

These trends indicate a scenario in which DSFGs at ``cosmic noon'' are more numerous than at earlier epochs and contain more dust than their low-$z$ cousins. This two-fold evolution can be explained with dust being rapidly produced in the distant Universe, heating up and emitting at FIR-millimeter wavelengths, thus peaking at $z=2-3$, and finally being slowly and gradually consumed by star formation or expelled by galactic winds in more recent epochs, when dust production by AGB stars and Supernovae does not balance the ``losses'' anymore.

\subsection{Evolution of the dust mass cosmic density}\label{sect:rho}

\begin{figure*}[!ht]
\centering
\settowidth{\imageheight}{\includegraphics[height=0.73\textwidth]{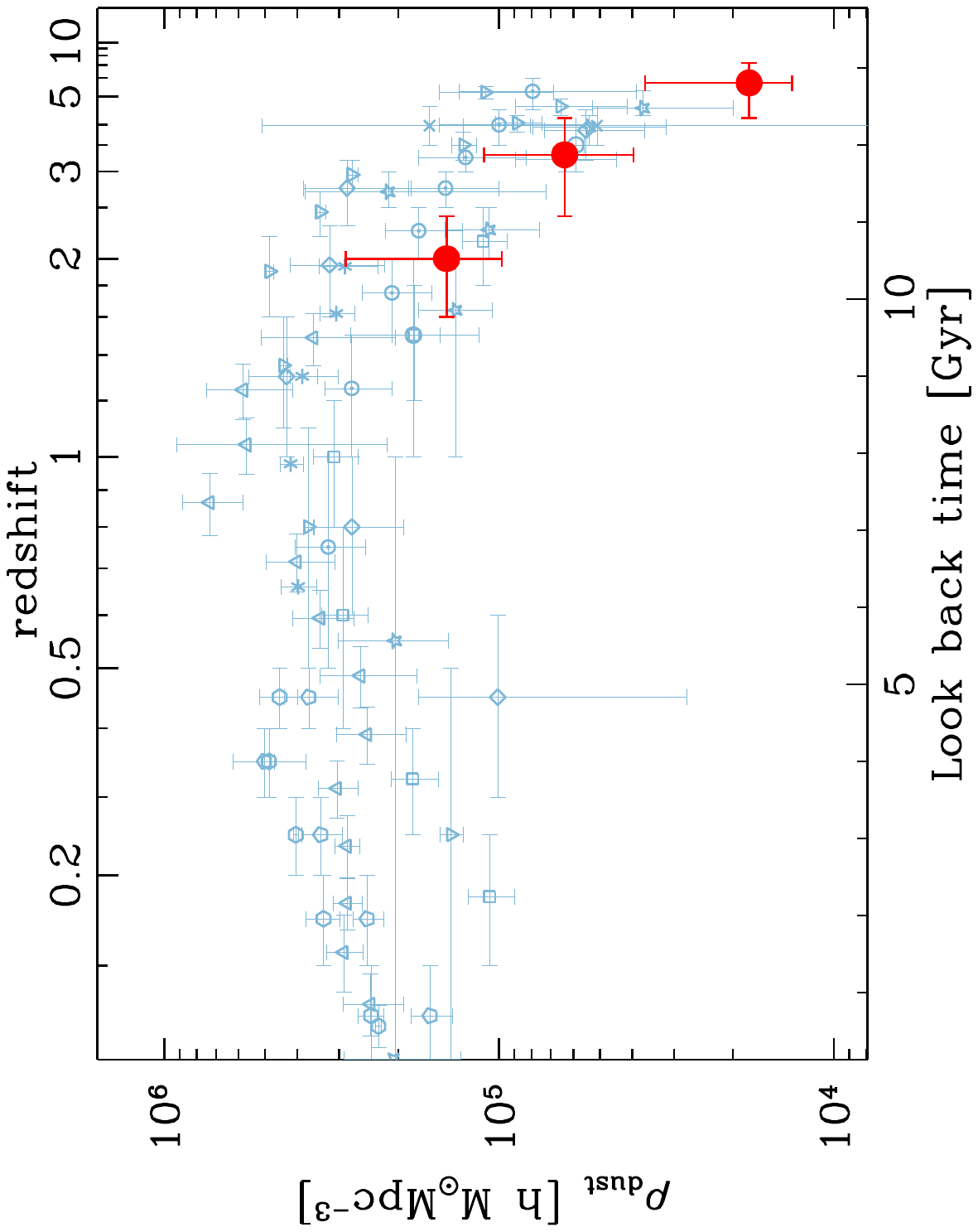}}
\settowidth{\imagewidth}{\includegraphics{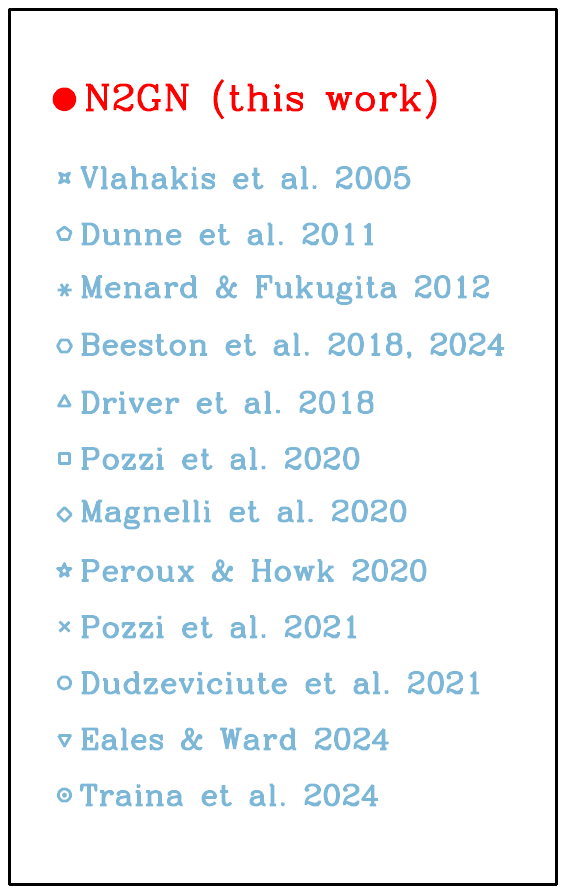}}
\begin{minipage}{0.74\textwidth}
\rotatebox{-90}{\includegraphics[width=\imageheight]{figs1/rho_z2_no_mod_20250130.pdf}}
\end{minipage}
\begin{minipage}{0.24\textwidth}
\includegraphics[trim=0.3\imagewidth{} 0.37\imagewidth{} 0.25\imagewidth{} 0.10\imagewidth{}, clip, height=0.8\imageheight]{figs1/legenda_rho4_20250130.pdf}
\end{minipage}
\caption{Evolution of the dust mass density as a function of look-back time. The N2GN results are depicted as red filled circles. All other symbols belong to literature data, as indicated in the right-hand panel. Local estimates are by \citet{vlahakis2005} and \citet{beeston2018}; combinations of optical-NIR and FIR-sub-millimeter data are by \citet{eales2024} and \citet{driver2018}; {\it Herschel}-based studies are by \citet{beeston2024}, \citet{pozzi2020} and \citet{dunne2011}; sub-millimeter and millimeter driven results are by \citet{pozzi2021},  \citet{dudzeviciute2021}, and \citet{magnelli2020}; and finally derivations of $\rho_\textrm{dust}$ are based on $\Omega_\textrm{gas}$ \citep{peroux2020} and Mg{\sc ii} absorbers \citep{menard2012}.
All data have been rescaled to our chosen cosmology and $\kappa_\nu$ \citep{draine2014}. Figure \ref{fig:kappa_rho} presents a full-color version and Fig. \ref{fig:rho_zzz} shows $\rho_\textrm{dust}$ as a function of redshift.}
\label{fig:rho_z}
\end{figure*}

The integral of the Schechter function is given by the upper incomplete Gamma function:
\begin{equation}
\begin{array}{rcl}
\rho_\textrm{dust}\left(M_\textrm{dust}>M_\textrm{inf} \right) &=& \int_{M_\textrm{inf}}^\infty{\Phi\left(M_\textrm{dust}\right)\textrm{d}\log M_\textrm{dust}}\\
&=&M^\ast_\textrm{dust}\Phi^\ast_\textrm{dust}\Gamma\left(\alpha+2,\frac{M_\textrm{dust}}{M^\ast_\textrm{dust}}\right)\textrm{.}\\
\end{array}
\end{equation}
In its calculation, we adopted $M_\textrm{inf}=10^4$ M$_\odot$. The resulting dust mass density of the N2GN survey is shown in Fig.\,\ref{fig:rho_z} and compared to data found in the literature.  % and models

All literature data have been rescaled to the \citet{draine2014} value of $\kappa_{850}=0.047$ m$^2$ kg$^{-1}$.  Appendix \ref{app:rho_NO_rescaling} lists the original assumptions of the different authors and shows the importance of rescaling all results to the same $\kappa_\nu$ in order to properly compare them to each other.

The local estimates of $\rho_\textrm{dust}$ shown in Fig.\,\ref{fig:rho_z} were established by \citet{vlahakis2005} and \citet{beeston2018}. The former observed a sample of 81 optically selected galaxies with SCUBA at 450 and 850\,$\mu$m. The latter studied the local DMF of 15\,750 galaxies drawn from the 145 deg$^2$ intersection of the Galaxy And Mass Assembly \citep[GAMA;][]{driver2009,liske2015} and the {\it Herschel} Astrophysical Terahertz Large Area Survey \citep[H-ATLAS;][]{eales2010} surveys, detected in the optical $r$ band and in the {\it Herschel} 250\,$\mu$m SPIRE band.

In this collection, FIR {\it Herschel}-based studies are included: the $z<0.5$ \citet{dunne2011} result, comprising 1867 H-ATLAS 250\,$\mu$m sources with a reliable SDSS counterpart; the $z<0.5$ \citet{beeston2024} data, based on a sample of 29\,241 galaxies in H-ATLAS; and \citet{pozzi2020}, already mentioned earlier, including 5546 PEP 160\,$\mu$m selected galaxies with multi-wavelength counterparts, in the redshift range $z=0.2-2.5$. 

Combining optically selected galaxies and long wavelength data, \citet{driver2018} brought together three complementary data sets, GAMA \citep{driver2011,liske2015}, G10-COSMOS \citep{davies2015,andrews2017}, and 3D-HST \citep{momcheva2016} for a total of more than 500\,000 optical and NIR selected galaxies up to $z\le1.75$.
\citet{eales2024} combined the stellar mass function by \citet{davidzon2017} in the COSMOS field \citep{scoville2007} with the SCUBA-2 850\,$\mu$m stacking analysis by \citet{millard2020} in the same field to study the evolution of the cosmic density of dust from $z=0.1$ to $z=5.5$.

\begin{table*}[t]
\centering
\caption{\label{tab:STY}Results of the STY analysis of the N2GN DMF using a Schechter function. }
\begin{tabular}{c|cc|cc|cc|cc}
\hline
\hline
$z$ & $\log M^\ast_\textrm{dust}$ & $1\sigma$ range & $\alpha$ & $1\sigma$ range & $\Phi^\ast_\textrm{dust}$ & $1\sigma$ range & $\rho_\textrm{dust}$ & $1\sigma$ range\\
& \multicolumn{2}{c|}{$\log M_\odot$} & & & \multicolumn{2}{c|}{$\rm 10^{-5} h\ Mpc^{-3} dex^{-1}$} & \multicolumn{2}{c}{10$^5$ h $M_\odot$Mpc$^{-3}$}\\ 
\hline
1.6-2.4 & 9.35 & 9.22 to 9.64 & -1.40 & -1.74 to -1.26 & 4.36 & 1.35 to 7.47 & 1.43 & 0.98 to 2.87\\	
2.4-4.2 & 9.61 & 9.46 to 9.73 & -1.61 & -1.83 to -1.41 & 0.69 & 0.37 to 1.39 & 0.64 & 0.40 to 1.11\\	
4.2-7.2 & 9.56 & 9.31 to 9.79 & -1.39 & -1.91 to -1.26 & 0.34 & 0.07 to 1.15 & 0.18 & 0.13 to 0.37\\	
\hline
\end{tabular}
\end{table*}

Few sub-millimeter-driven studies are included, based on SCUBA-2 and ALMA data. \citet{dudzeviciute2021} derived $\rho_\textrm{dust}$ by fitting the UV-to-radio SEDs of 450\,$\mu$m galaxies selected from the SCUBA-2 STUDIES survey \citep{wang2017,lim2020}. They compared and combined the results to a similar modeling carried out on 850\,$\mu$m ALMA/SCUBA-2 AS2UDS \citep{stach2019,dudzeviciute2020} galaxies.
\citet{magnelli2020} used the ALMA spectroscopic survey (ASPECS) in the Hubble Ultra Deep Field (HUDF) to study the contribution of $H$ band-selected galaxies to the dust cosmic density. These authors performed stacking on the 1.2\,mm ALMA maps of the selected galaxies in bins of stellar mass, over the redshift range $z=0.3-3.2$. The resulting $\rho_\textrm{dust}$ is limited by the lower limit of the stellar mass selection and can be considered a lower limit.
Also applying a stacking technique, \citet{pozzi2021} derived the contribution to $\rho_\textrm{dust}$ of 113 UV-selected galaxies at redshift $z=4.4-5.9$ in the ALMA/ALPINE survey \citep{lefevre2020}. They also estimated the dust density given by additional 23 serendipitously detected sub-millimeter galaxies.
\citet{traina2024} measured the evolutionary $\rho_\textrm{dust}$ from $z=6.0$ to $z=0.5$ analyzing the ALMA-selected A$^3$\,COSMOS sample mentioned in Sect. \ref{sect:DMF}.  In Fig.\,\ref{fig:rho_z}, we show their fiducial results, based on CIGALE SED fitting and the Schechter fit to ALMA plus {\it Herschel} data.
Finally, \citet{menard2012} computed the dust density studying Mg{\sc ii} absorbing clouds in the lines of sight of quasars, and deriving the amount of intervening dust by fitting the corresponding extinction curve, and \citet{peroux2020} derived the dust density $\Omega_\textrm{dust}=\rho_\textrm{dust}/\rho_\textrm{crit,0}$ from the gas density $\Omega_\textrm{gas}$, making an assumption on the dust to gas ratio.

The N2GN and A$^3$\,COSMOS samples are practically the only ones consisting of galaxies directly selected at observed millimeter wavelengths, that translate into the rest-frame sub-millimeter domain. As pointed out in Sect. \ref{sect:properties},  selecting along the RJ tail of the dust emission is paramount to have as unbiased a $M_\textrm{dust}$ estimate as possible. Moreover these  galaxies benefit from a very rich multi-wavelength coverage, that allowed us to avoid any strong assumption in the SED fitting that lead to the $M_\textrm{dust}$ measurements (e.g., fixing the dust temperature). The two samples lead to different results: in this work we detect a strong evolution of $\rho_\textrm{dust}$ from $z\sim7$ to $z>2$, while \citet{traina2024} find a  smoother increase of the dust density from $z=6$ to $z=0.5$. The A$^3$\,COSMOS values of $\rho_\textrm{dust}$ are larger than those derived here for N2GN by a factor $\sim 5$ in the largest redshift bin (i.e., $\sim 3.3\sigma$ higher) and by a factor 1.5 to 2.5 at $2.4<z\le4.2$ (i.e., 2-3$\sigma$).

The large spread in the observed values of $\rho_\textrm{dust}$ in Fig.\,\ref{fig:rho_z}, and the apparently discordant trends of $\rho_\textrm{dust}$ at $z<1.5$ in the different literature works shown here, are likely to be (at least in part) ascribed to differences in the selection of the samples and in the methods adopted to derive $M_\textrm{dust}$ among different authors. More systematic and extensive measurements, possibly carried out self-consistently over the whole $z=0-7$ redshift range, and over larger sky areas (e.g., COSMOS), are needed to be more conclusive about the actual cosmic evolution of dust. 

A larger survey area will also give the chance to detect rare objects at even higher redshift than in N2GN. The analysis of $z_\textrm{max}$ in Sect. \ref{sect:DMF} has in fact shown that -- if dust already existed in such amounts -- at the N2CLS depth a dusty star forming galaxy would be detected at $z\gg7$.

\subsection{Comparison to models predictions}

\begin{figure}[t]
\centering
\rotatebox{-90}{\includegraphics[height=0.47\textwidth]{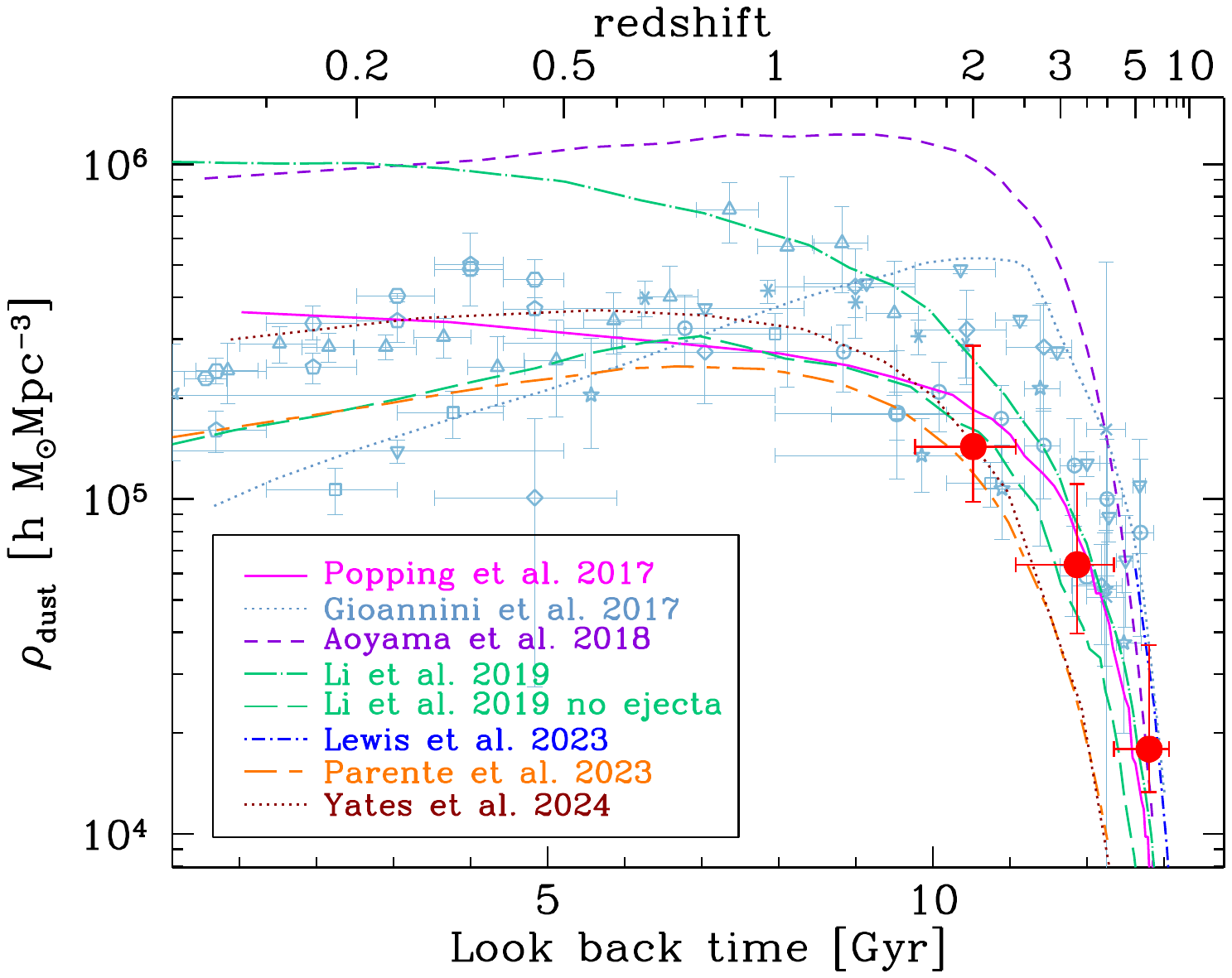}}
\caption{Comparison of the dust mass density to model predictions. The N2GN results are depicted as red filled circles.  The different lines represent the semi-analytical models by \citet{popping2017}, \citet{parente2023} and \citet{yates2024}, the chemical evolution model by \citet{gioannini2017}, and the cosmological hydrodynamical simulations by \citet{lewis2023}, \citet{li2019} and \citet{aoyama2018}. Figure \ref{fig:rho_zzz} presents a version of this Figure as a function of redshift.}
\label{fig:rho_z_mod}
\end{figure}

In Fig.\,\ref{fig:rho_z_mod}, the N2GN data and the collection of literature $\rho_\textrm{dust}$ results are compared to model predictions by \citet{popping2017}, \citet{gioannini2017}, \citet{aoyama2018}, \citet{li2019}, \citet{lewis2023}, \citet{parente2023}, and \citet{yates2024}.

\citet{popping2017} included dust production and destruction in a semi-analytic model of galaxy formation, taking into account condensation of dust in stellar ejecta, growth of dust in the ISM, the destruction of dust by supernovae and in hot halos, and dusty winds and inflows.  Their model reproduces the relation between stellar mass and dust mass in the local and high-redshift Universe, as well as the dust-to-gas ratio of local galaxies as a function of their stellar mass.

\citet{gioannini2017} studied the evolution of interstellar dust with chemical evolution models of galaxies of different morphological types, including dust production from supernovae and asymptotic giant branch stars, dust accretion and destruction processes. They explored different cosmological scenarios of galaxy formation: pure luminosity evolution and number density evolution. The latter is plotted in Fig. \ref{fig:rho_z_mod}.

\citet{aoyama2018} investigated the evolution of dust by means of hydrodynamic simulations, in which the enrichment of metals and dust is treated self-consistently with star formation and stellar feedback. Dust evolution is driven by production in stellar ejecta, destruction by sputtering, grain growth by accretion and coagulation, and grain disruption by shattering. The model is successful at reproducing the local relation between dust-to-gas ratio and metallicity of galaxies. In Fig. \ref{fig:rho_z_mod} only the dust component locked in galaxies is plotted.

\citet{li2019} implemented the production, growth, and destruction of dust grains into the SIMBA cosmological hydrodynamic galaxy formation simulation \citep{dave2019}. Dust forms in stellar ejecta, grows by the accretion of metals, and is destroyed by thermal sputtering and supernovae. This simulation reproduces the observed local DMF but under-predicts the DMF by a factor $\sim 3$ in the redshift range $z=1$ to 2. In Fig. \ref{fig:rho_z_mod} we show the prediction obtained including and excluding the dust ejected out of the galaxies via galactic winds.

The \citet{lewis2023} model, called DUSTiER (DUST in the Epoch of Reionization),
couples a physical model for dust production to the fully radiation-hydrodynamics cosmological simulation code RAMSES-CUDATON \citep{ocvirk2016}, with the aim to study how dust affects the escape of ionizing photons into the IGM during the Epoch of Reionization. In Fig. \ref{fig:rho_z_mod} the fiducial model is shown (at $z\ge5$); the uncertainty range is $\sim 0.3$ dex (not shown). 

\citet{parente2023} implemented some standard prescriptions for dust production and evolution on top of the L-GALAXIES2020 semi-analytic  model \citep{henriques2020}. Their approach adopts a new disc instability criterion to trigger bulge and central black hole growth. As for the dust properties, the model includes two populations of large and small grains \citep[e.g.,][]{granato2021}, production by stars, shattering and coagulation, grain growth in molecular clouds, destruction in SNe shocks, sputtering in the hot phase and ejection from the host galaxy. The model is successful at reproducing the local stellar mass function and DMF, but it is known to lack dusty highly star forming galaxies at $z>1$ \citep{traina2024}.
The \citet{yates2024} simulation is a further implementation of the L-GALAXIES model, taking into account binary stars evolution and dust production and destruction. This implementation reproduces observed dust-to-metal and dust-to-gas ratios at $z=0$ to 4 but systematically underpredicts the dust masses of galaxies at $z>4$.
In Fig. \ref{fig:rho_z_mod} the fiducial model of \citet{parente2023} and the total ISM dust component of \citet{yates2024} are shown.

All models predict a steep fall of the dust mass cosmic density above $z=2-3$, despite significant differences. The models by \citet{popping2017} and \citet{li2019} are the most successful at predicting the $z=2.4-7.2$ $\rho_\textrm{dust}$ as determined by N2GN. The model by \citet{gioannini2017} is consistent with higher measurements of $\rho_\textrm{dust}$ as those by \citet{eales2024}. 

At lower redshift, the differences between the different simulations become critical. In the \citet{popping2017} model, the dust density is roughly constant from $z\sim 2$ to the local Universe, suggesting a quasi-equilibrium between the production and destruction of dust grains \citep{magnelli2020}. This trend is consistent with the observed density by \citet{driver2018}, based on optically selected galaxies. On the other hand, a FIR selection leads to an evolving $\rho_\textrm{dust}(z)$, steadily decreasing from $z\sim1$ to $\sim0.2$ \citep{pozzi2020,eales2024}. The model by \citet{gioannini2017} fairly reproduces this decrease.

The hydrodynamical simulations by \citet{aoyama2018} and \citet{li2019} significantly over-predict the low-redshift dust density. Notably, when excluding the fraction of dust ejected from the galaxies into the IGM, the \citet{li2019} model predicts a decrease of $\rho_\textrm{dust}$ similar to the one observed by \citet{pozzi2020}. 
Arguably, the expelled dust residing in the IGM is too faint to contribute to the observed (sub-)mm flux of the N2GN galaxies, and therefore it is not included in the measured DMF and $\rho_\textrm{dust}$.

Given the large scatter in the data, after re-scaling all estimates to the same $\kappa_\nu$ assumptions, it is difficult to be conclusive on which model is most successful in reproducing the observed evolution of $\rho_\textrm{dust}$. Once again, selection effects and differences in the assumptions governing the dust mass derivation produce too large of a scatter to allow either of the models to be favored.

% --------------------------------------------------------------------

\section{Summary and conclusions}\label{sect:summary}

We have conducted a panchromatic study of the sources detected by NIKA2 in the GOODS-N field, combining multi-wavelength data obtained with NIKA2 at the IRAM 30~m telescope, NOEMA, SMA, SCUBA2, VLA, {\it Herschel}, {\it Spitzer}, the {\it Hubble} and {\it James Webb} space telescopes, and ground-based optical facilities. The sample is named N2GN.   

Out of the 71 individual galaxies associated with the N2GN detections, 68 have a detailed SED coverage and a redshift measurement. 
Redshifts span the range $0.6 < z < 7.2$, with a median of 2.819.
We performed SED fitting of these sources with eight different methods, including a MBB in the optically thin approximation and in its general form, the \citet{DL07} models, and UV-to-radio fits with MAGPHYS \citep{dacunha2008,dacunha2015,battisti2020}, CIGALE \citep{burgarella2005,noll2009,boquien2019}, and SED3FIT \citep{berta2013}. The end product of this analysis is a robust estimate of their dust masses and of the evolution of the DMF and the dust cosmic density as a function of redshift. The main results of this analysis are the following.

\begin{itemize}
\item Once the differences in their underlying assumptions are taken into account (e.g., in the reference value of the dust absorption coefficient $\kappa_\nu$), the eight methods produce consistent estimates of $M_\textrm{dust}$. The only exception is the MBB in general form, which tends to underestimate the dust mass. This model is known to fail when no information about the size of the dust emitting region, or the frequency at which the medium becomes optically thick, are known \citep{ismail2023}. We also found a similar good consistency between the different methods for $L_\textrm{IR}$. As for $M_\star$, the different codes produce results in general agreement, but significant differences exist for individual sources, especially those with very red optical SEDs.
\item Combining the SFR derived from $L_\textrm{IR}$ and the stellar mass, we compared the N2GN galaxies to the ``main sequence'' locus between redshift 1.6 and 7.2. When using the parametrization of the MS given by \citet{popesso2023}, the majority of the sources turn out to be outliers that are likely powered by an intense starburst. 
\item The stellar mass of several $z>4$ N2GN galaxies is of the order of a few $10^{11}\,M_\odot$. If confirmed, such a high M$_\star$ would challenge galaxy formation models and imply very high star formation efficiencies \citep{xiao2024}. Nevertheless, most of these sources have a very broad gap in their observed SEDs, as they lack data between rest-frame optical and FIR wavelengths. Moreover significant differences exist between the $M_\star$ estimates obtained by different codes, which is possibly driven by a different SFH or attenuation law details. Only future very deep MIR observations (e.g., with JWST/MIRI) will be able to constrain the rest-frame NIR emission and the stellar mass of these galaxies.
\item A few sources were detected in the X-rays or showed other evident signs of AGN activity (power-law-like MIR SED, radio excess, optical point-like morphology). These sources required an AGN-torus component in the SED fitting \citep{berta2013}. In some other cases, the rest-frame UV-blue optical SED had a strong excess with respect to the extinguished stellar component. This excess can be explained with the contribution of a non-extinguished young stellar population to the SED. Two of these blue-excess galaxies lie at $z>5$, and their SEDs are similar to those of the "little red dots" recently identified by JWST \citep{perezgonzalez2024}.
\item The millimeter selection, possible thanks to NIKA2, closely resembles a dust mass selection modulo the spread in redshift and in the dust mass to light ratio of the sample. 
The strong correlation between the dust mass and the rest-frame 850 $\mu$m luminosity of the N2GN galaxies is described by the equation: $\log(M_\textrm{dust}) = (1.24 \pm 0.08) \log(L_{850}) + (12.11 \pm 0.20)$. This relation can be useful for galaxies lacking an SED coverage detailed enough to derive $M_\textrm{dust}$.
\item We derived the DMF of the N2GN galaxies over the redshift range between $z=1.6$ and $z=7.2$, and we reproduced it with a Schechter function using the STY method. Combined with literature results \citep[e.g.,][]{pozzi2020,eales2024}, our data reveal a two-fold evolution of the DMF of millimeter galaxies: {\em (a)} a rapid evolution of the characteristic comoving density $\Phi^\ast$, that doubles from $z\sim7$ to $z\sim2.5$, increases by a further factor of five at later epochs and ceases at $z\sim0.5-1.0$; {\em (b)} a constant characteristic dust mass, $M^\ast_\textrm{dust}$, from $z\sim7$ to $z\sim4$ and then a smooth decrease by more than an order of magnitude down to the local Universe.
\item By integrating the DMF, we derived the comoving dust mass density of millimeter galaxies and compared it with previous studies. In this comparison, we showed that it is of uttermost importance to rescale all $\rho_\textrm{dust}$ determinations to the same $\kappa_\nu$ assumption. The comoving $\rho_\textrm{dust}$ increased by at least one order of magnitude from $z\sim7$ to $z\sim1.5$. At lower redshift, a different data set leads to different trends, with $\rho_\textrm{dust}$ possibly remaining constant down to $z\sim0$ or decreasing by a factor of approximately three. The N2GN sample leads to results systematically different from the recent ALMA A$^3$\,COSMOS determination of $\rho_\textrm{dust}$ \citep{traina2024}, with the latter showing a smoother evolution and a larger dust mass density at the highest redshifts.
A comparison with simulations of dust and galaxy evolution favors semi-analytical and chemical evolution models that take into account dust growth and destruction \citep{popping2017,gioannini2017}. Hydrodynamical simulations are also successful when excluding the fraction expelled into the IGM from the dust budget \citep{li2019}.
\end{itemize}

The deep NIKA2 maps of GOODS-N observed by the N2CLS survey have enabled for the first time the determination of the dust mass comoving density to be extended up to $\sim 7$, that is, an epoch when the Universe was less than 1 Gyr old. 
A very large spread in the determinations of $\rho_\textrm{dust}$ at all redshifts ($z=0-7$) still exists.
The discrepancies between different works likely arise from the differences in sample selections, in the methods employed to derive $M_\textrm{dust}$, and possibly in their underlying assumptions. A systematic and self-consistent study of $M_\textrm{dust}$ and $\rho_\textrm{dust}$ over as broad a redshift range as possible, from the local Universe to remote epochs, is envisaged and recommended in order to reach a consensus about the actual cosmic evolution of dust. 

Future studies will need to be carried out on larger areas in order to cover larger volumes and be sensitive to both lower redshift regimes and rare very distant dusty galaxies. Thanks to the negative $k$-correction at millimeter wavelengths, at the unprecedented depth of N2CLS, a dusty star forming galaxy would be easily detected at $z\gg7$ if dust already existed in galaxies in amounts similar to those in the N2GN sample. The natural next step in the exploration of $\rho_\textrm{dust}$ will exploit the N2CLS COSMOS field (N2CLS team, in prep.).

% --------------------------------------------------------------------

\section*{Data availability}

The N2CLS final maps and catalogs, the NOEMA follow-up data, and the multi-wavelength catalog are available on line on the survey's home page: \url{https://data.lam.fr/n2cls/home}. 
%
%Figure \href{https://doi.org/10.5281/zenodo.14965423}{E.1} is available on Zenodo, at the address \url{https://doi.org/10.5281/zenodo.14965423}.

% --------------------------------------------------------------------

\begin{acknowledgements}

We thank the anonymous referee for their insightful comments that improved the quality and the presentation of this work. We recognize the contribution of Mael Voyer, Jean Anquetil and Aymeric Garnier to the early identification process of the N2CLS sources. We are grateful to J. Lewis for providing the DUSTIER model.
This work is based on observations carried out under project numbers 192-16 with the IRAM 30~m telescope, and projects W16EE, E16AI, W21CV and W23CX with NOEMA. 
IRAM is supported by INSU/CNRS (France), MPG (Germany) and IGN (Spain). 
We would like to thank the IRAM staff for their support during the observation campaigns. 
The NIKA2 dilution cryostat has been designed and built at the Institut N\'eel. In particular, we acknowledge the crucial contribution of the Cryogenics Group, and in particular Gregory Garde, Henri Rodenas, Jean-Paul Leggeri, Philippe Camus. 
We acknowledge financial support from the ``Programme National de Cosmologie and Galaxies'' (PNCG) funded by CNRS/INSU-IN2P3-INP, CEA and CNES, France, from the European Research Council (ERC) under the European Union’s Horizon 2020 research and innovation programme (project CONCERTO, grant agreement No 788212) and from the Excellence Initiative of Aix-Marseille University-A*Midex, a French ``Investissements d’Avenir'' programme.
This work has been partially funded by the Foundation Nanoscience Grenoble and the LabEx FOCUS ANR-11-LABX-0013. This work is supported by the French National Research Agency under the contracts ``MKIDS'', ``NIKA'' and ANR-15-CE31-0017 and in the framework of the ``Investissements d’avenir'' program (ANR-15-IDEX-02). This work has been supported by the GIS KIDs. This work has benefited from the support of the European Research Council Advanced Grant ORISTARS under the European Union’s Seventh Framework Programme (Grant Agreement no. 291294). 
A.~M. acknowledges support the funding from the European Research Council (ERC) under the European Union’s Horizon 2020 research and innovation programme (Grant agreement No. 101098309 - PEBBLES)
A.~R. acknowledges financial support from the Italian Ministry of University and Research -- Project Proposal CIR01\_00010. 
E.~A. acknowledges funding from the French Programme ``Investissements d’avenir'' through the Enigmass Labex. 
M.~M.~E. acknowledges the support of the French Agence Nationale de la Recherche (ANR), under grant ANR-22-CE31-0010.
R.~A. acknowledges support from the Programme National Cosmology et Galaxies (PNCG) of CNRS/INSU with INP and IN2P3, co-funded by CEA and CNES.
R.~A. was supported by the French government through the France 2030 investment plan managed by the National Research Agency (ANR), as part of the Initiative of Excellence of Universit\'e C\^ote d'Azur under reference number ANR-15-IDEX-01.
The NIKA2 data were processed using the Pointing and Imaging In Continuum software \citep[PIIC;][]{zylka2013,piic2024}, developed by Robert Zylka at the Institut de Radioastronomie Millimetrique (IRAM) and distributed by IRAM via the GILDAS pages. PIIC is the extension of the MOPSIC data reduction software to the case of NIKA2 data.
This work made use of Astropy:\footnote{http://www.astropy.org} a community-developed core Python package and an ecosystem of tools and resources for astronomy \citep{astropy:2013, astropy:2018, astropy:2022}.

\end{acknowledgements}

\bibliographystyle{aa} 		% style aa.bst
\bibliography{n2cls_multiwave} 	% needs n2cls_multiwave.bib

\begin{thebibliography}{261}
\expandafter\ifx\csname natexlab\endcsname\relax\def\natexlab#1{#1}\fi

\bibitem[{{Adam} {et~al.}(2014){Adam}, {Adane}, {Ade}, {Andr{\'e}e}, {Beelen},
  {Belier}, {Beno{\^\i}t}, {Bideaud}, {Billot}, {Boudou}, {Bourrion}, {Calvo},
  {Catalano}, {Coiffard}, {Comis}, {D'Addabbo}, {D{\'e}sert}, {Doyle}, {Goupy},
  {Kramer}, {Leclercq}, {Mac{\'\i}as-P{\'e}rez}, {Martino}, {Mauskopf},
  {Mayet}, {Monfardini}, {Pajot}, {Pascale}, {Perotto}, {Pointecouteau},
  {Ponthieu}, {Rev{\'e}ret}, {Rodriguez}, {Ruppin}, {Savini}, {Schuster},
  {Sievers}, {Tucker}, \& {Zylka}}]{Adam2014}
{Adam}, R., {Adane}, A., {Ade}, P., {et~al.} 2014, arXiv e-prints,
  arXiv:1409.1137, {Proceedings of the 49th Rencontres de Moriond on Cosmology:
  La Thuile, Italy, March 15-22, 2014, 73-76}

\bibitem[{{Adam} {et~al.}(2018){Adam}, {Adane}, {Ade}, {Andr{\'e}},
  {Andrianasolo}, {Aussel}, {Beelen}, {Beno{\^\i}t}, {Bideaud}, {Billot},
  {Bourrion}, {Bracco}, {Calvo}, {Catalano}, {Coiffard}, {Comis}, {De Petris},
  {D{\'e}sert}, {Doyle}, {Driessen}, {Evans}, {Goupy}, {Kramer}, {Lagache},
  {Leclercq}, {Leggeri}, {Lestrade}, {Mac{\'\i}as-P{\'e}rez}, {Mauskopf},
  {Mayet}, {Maury}, {Monfardini}, {Navarro}, {Pascale}, {Perotto}, {Pisano},
  {Ponthieu}, {Rev{\'e}ret}, {Rigby}, {Ritacco}, {Romero}, {Roussel}, {Ruppin},
  {Schuster}, {Sievers}, {Triqueneaux}, {Tucker}, \& {Zylka}}]{adam2018}
{Adam}, R., {Adane}, A., {Ade}, P.~A.~R., {et~al.} 2018, \aap, 609, A115

\bibitem[{{Adscheid} {et~al.}(2024){Adscheid}, {Magnelli}, {Liu}, {Bertoldi},
  {Delvecchio}, {Gruppioni}, {Schinnerer}, {Traina}, {B{\'e}thermin}, \&
  {Gkogkou}}]{adscheid2024}
{Adscheid}, S., {Magnelli}, B., {Liu}, D., {et~al.} 2024, \aap, 685, A1

\bibitem[{{Alaghband-Zadeh} {et~al.}(2013){Alaghband-Zadeh}, {Chapman},
  {Swinbank}, {Smail}, {Danielson}, {Decarli}, {Ivison}, {Meijerink}, {Weiss},
  \& {van der Werf}}]{alaghbandzadeh2013}
{Alaghband-Zadeh}, S., {Chapman}, S.~C., {Swinbank}, A.~M., {et~al.} 2013,
  \mnras, 435, 1493

\bibitem[{{Alexander} {et~al.}(2003){Alexander}, {Bauer}, {Brandt},
  {Schneider}, {Hornschemeier}, {Vignali}, {Barger}, {Broos}, {Cowie},
  {Garmire}, {Townsley}, {Bautz}, {Chartas}, \& {Sargent}}]{alexander2003}
{Alexander}, D.~M., {Bauer}, F.~E., {Brandt}, W.~N., {et~al.} 2003, \aj, 126,
  539

\bibitem[{{Andrews} {et~al.}(2017){Andrews}, {Driver}, {Davies}, {Kafle},
  {Robotham}, \& {Wright}}]{andrews2017}
{Andrews}, S.~K., {Driver}, S.~P., {Davies}, L.~J.~M., {et~al.} 2017, \mnras,
  464, 1569

\bibitem[{{Aoyama} {et~al.}(2018){Aoyama}, {Hou}, {Hirashita}, {Nagamine}, \&
  {Shimizu}}]{aoyama2018}
{Aoyama}, S., {Hou}, K.-C., {Hirashita}, H., {Nagamine}, K., \& {Shimizu}, I.
  2018, \mnras, 478, 4905

\bibitem[{{Aravena} {et~al.}(2014){Aravena}, {Hodge}, {Wagg}, {Carilli},
  {Daddi}, {Dannerbauer}, {Lentati}, {Riechers}, {Sargent}, \&
  {Walter}}]{aravena2014}
{Aravena}, M., {Hodge}, J.~A., {Wagg}, J., {et~al.} 2014, \mnras, 442, 558

\bibitem[{{Aravena} {et~al.}(2013){Aravena}, {Murphy}, {Aguirre}, {Ashby},
  {Benson}, {Bothwell}, {Brodwin}, {Carlstrom}, {Chapman}, {Crawford}, {de
  Breuck}, {Fassnacht}, {Gonzalez}, {Greve}, {Gullberg}, {Hezaveh}, {Holder},
  {Holzapfel}, {Keisler}, {Malkan}, {Marrone}, {McIntyre}, {Reichardt},
  {Sharon}, {Spilker}, {Stalder}, {Stark}, {Vieira}, \&
  {Wei{\ss}}}]{aravena2013}
{Aravena}, M., {Murphy}, E.~J., {Aguirre}, J.~E., {et~al.} 2013, \mnras, 433,
  498

\bibitem[{{Aravena} {et~al.}(2016){Aravena}, {Spilker}, {Bethermin},
  {Bothwell}, {Chapman}, {de Breuck}, {Furstenau}, {G{\'o}nzalez-L{\'o}pez},
  {Greve}, {Litke}, {Ma}, {Malkan}, {Marrone}, {Murphy}, {Stark}, {Strandet},
  {Vieira}, {Weiss}, {Welikala}, {Wong}, \& {Collier}}]{aravena2016}
{Aravena}, M., {Spilker}, J.~S., {Bethermin}, M., {et~al.} 2016, \mnras, 457,
  4406

\bibitem[{{Arrabal Haro} {et~al.}(2018){Arrabal Haro}, {Rodr{\'\i}guez
  Espinosa}, {Mu{\~n}oz-Tu{\~n}{\'o}n}, {P{\'e}rez-Gonz{\'a}lez},
  {Dannerbauer}, {Bongiovanni}, {Barro}, {Cava}, {Lumbreras-Calle},
  {Hern{\'a}n-Caballero}, {Eliche-Moral}, {Dom{\'\i}nguez S{\'a}nchez},
  {Conselice}, {Tresse}, {Alcalde Pampliega}, {Balcells}, {Daddi}, \&
  {Rodighiero}}]{arrabal_haro2018}
{Arrabal Haro}, P., {Rodr{\'\i}guez Espinosa}, J.~M.,
  {Mu{\~n}oz-Tu{\~n}{\'o}n}, C., {et~al.} 2018, \mnras, 478, 3740

\bibitem[{{Astropy Collaboration} {et~al.}(2022){Astropy Collaboration},
  {Price-Whelan}, {Lim}, {Earl}, {Starkman}, {Bradley}, {Shupe}, {Patil},
  {Corrales}, {Brasseur}, {N{"o}the}, {Donath}, {Tollerud}, {Morris},
  {Ginsburg}, {Vaher}, {Weaver}, {Tocknell}, {Jamieson}, {van Kerkwijk},
  {Robitaille}, {Merry}, {Bachetti}, {G{"u}nther}, {Aldcroft},
  {Alvarado-Montes}, {Archibald}, {B{'o}di}, {Bapat}, {Barentsen}, {Baz{'a}n},
  {Biswas}, {Boquien}, {Burke}, {Cara}, {Cara}, {Conroy}, {Conseil}, {Craig},
  {Cross}, {Cruz}, {D'Eugenio}, {Dencheva}, {Devillepoix}, {Dietrich},
  {Eigenbrot}, {Erben}, {Ferreira}, {Foreman-Mackey}, {Fox}, {Freij}, {Garg},
  {Geda}, {Glattly}, {Gondhalekar}, {Gordon}, {Grant}, {Greenfield}, {Groener},
  {Guest}, {Gurovich}, {Handberg}, {Hart}, {Hatfield-Dodds}, {Homeier},
  {Hosseinzadeh}, {Jenness}, {Jones}, {Joseph}, {Kalmbach}, {Karamehmetoglu},
  {Ka{l}uszy{'n}ski}, {Kelley}, {Kern}, {Kerzendorf}, {Koch}, {Kulumani},
  {Lee}, {Ly}, {Ma}, {MacBride}, {Maljaars}, {Muna}, {Murphy}, {Norman},
  {O'Steen}, {Oman}, {Pacifici}, {Pascual}, {Pascual-Granado}, {Patil},
  {Perren}, {Pickering}, {Rastogi}, {Roulston}, {Ryan}, {Rykoff}, {Sabater},
  {Sakurikar}, {Salgado}, {Sanghi}, {Saunders}, {Savchenko}, {Schwardt},
  {Seifert-Eckert}, {Shih}, {Jain}, {Shukla}, {Sick}, {Simpson},
  {Singanamalla}, {Singer}, {Singhal}, {Sinha}, {Sip{H{o}}cz}, {Spitler},
  {Stansby}, {Streicher}, {{{S}}umak}, {Swinbank}, {Taranu}, {Tewary},
  {Tremblay}, {Val-Borro}, {Van Kooten}, {Vasovi{'c}}, {Verma}, {de Miranda
  Cardoso}, {Williams}, {Wilson}, {Winkel}, {Wood-Vasey}, {Xue}, {Yoachim},
  {Zhang}, {Zonca}, \& {Astropy Project Contributors}}]{astropy:2022}
{Astropy Collaboration}, {Price-Whelan}, A.~M., {Lim}, P.~L., {et~al.} 2022,
  ApJ, 935, 167

\bibitem[{{Astropy Collaboration} {et~al.}(2018){Astropy Collaboration},
  {Price-Whelan}, {Sip{\H{o}}cz}, {G{\"u}nther}, {Lim}, {Crawford}, {Conseil},
  {Shupe}, {Craig}, {Dencheva}, {Ginsburg}, {Vand erPlas}, {Bradley},
  {P{\'e}rez-Su{\'a}rez}, {de Val-Borro}, {Aldcroft}, {Cruz}, {Robitaille},
  {Tollerud}, {Ardelean}, {Babej}, {Bach}, {Bachetti}, {Bakanov}, {Bamford},
  {Barentsen}, {Barmby}, {Baumbach}, {Berry}, {Biscani}, {Boquien}, {Bostroem},
  {Bouma}, {Brammer}, {Bray}, {Breytenbach}, {Buddelmeijer}, {Burke},
  {Calderone}, {Cano Rodr{\'\i}guez}, {Cara}, {Cardoso}, {Cheedella}, {Copin},
  {Corrales}, {Crichton}, {D'Avella}, {Deil}, {Depagne}, {Dietrich}, {Donath},
  {Droettboom}, {Earl}, {Erben}, {Fabbro}, {Ferreira}, {Finethy}, {Fox},
  {Garrison}, {Gibbons}, {Goldstein}, {Gommers}, {Greco}, {Greenfield},
  {Groener}, {Grollier}, {Hagen}, {Hirst}, {Homeier}, {Horton}, {Hosseinzadeh},
  {Hu}, {Hunkeler}, {Ivezi{\'c}}, {Jain}, {Jenness}, {Kanarek}, {Kendrew},
  {Kern}, {Kerzendorf}, {Khvalko}, {King}, {Kirkby}, {Kulkarni}, {Kumar},
  {Lee}, {Lenz}, {Littlefair}, {Ma}, {Macleod}, {Mastropietro}, {McCully},
  {Montagnac}, {Morris}, {Mueller}, {Mumford}, {Muna}, {Murphy}, {Nelson},
  {Nguyen}, {Ninan}, {N{\"o}the}, {Ogaz}, {Oh}, {Parejko}, {Parley}, {Pascual},
  {Patil}, {Patil}, {Plunkett}, {Prochaska}, {Rastogi}, {Reddy Janga},
  {Sabater}, {Sakurikar}, {Seifert}, {Sherbert}, {Sherwood-Taylor}, {Shih},
  {Sick}, {Silbiger}, {Singanamalla}, {Singer}, {Sladen}, {Sooley},
  {Sornarajah}, {Streicher}, {Teuben}, {Thomas}, {Tremblay}, {Turner},
  {Terr{\'o}n}, {van Kerkwijk}, {de la Vega}, {Watkins}, {Weaver}, {Whitmore},
  {Woillez}, {Zabalza}, \& {Astropy Contributors}}]{astropy:2018}
{Astropy Collaboration}, {Price-Whelan}, A.~M., {Sip{\H{o}}cz}, B.~M., {et~al.}
  2018, \aj, 156, 123

\bibitem[{{Astropy Collaboration} {et~al.}(2013){Astropy Collaboration},
  {Robitaille}, {Tollerud}, {Greenfield}, {Droettboom}, {Bray}, {Aldcroft},
  {Davis}, {Ginsburg}, {Price-Whelan}, {Kerzendorf}, {Conley}, {Crighton},
  {Barbary}, {Muna}, {Ferguson}, {Grollier}, {Parikh}, {Nair}, {Unther},
  {Deil}, {Woillez}, {Conseil}, {Kramer}, {Turner}, {Singer}, {Fox}, {Weaver},
  {Zabalza}, {Edwards}, {Azalee Bostroem}, {Burke}, {Casey}, {Crawford},
  {Dencheva}, {Ely}, {Jenness}, {Labrie}, {Lim}, {Pierfederici}, {Pontzen},
  {Ptak}, {Refsdal}, {Servillat}, \& {Streicher}}]{astropy:2013}
{Astropy Collaboration}, {Robitaille}, T.~P., {Tollerud}, E.~J., {et~al.} 2013,
  \aap, 558, A33

\bibitem[{{Bakx} {et~al.}(2020){Bakx}, {Dannerbauer}, {Frayer}, {Eales},
  {P{\'e}rez-Fournon}, {Cai}, {Clements}, {De Zotti}, {Gonz{\'a}lez-Nuevo},
  {Ivison}, {Lapi}, {Micha{\l}owski}, {Negrello}, {Serjeant}, {Smith}, {Temi},
  {Urquhart}, \& {van der Werf}}]{bakx2020}
{Bakx}, T.~J.~L.~C., {Dannerbauer}, H., {Frayer}, D., {et~al.} 2020, \mnras,
  496, 2372

\bibitem[{{Barger} {et~al.}(2022){Barger}, {Cowie}, {Blair}, \&
  {Jones}}]{barger2022}
{Barger}, A.~J., {Cowie}, L.~L., {Blair}, A.~H., \& {Jones}, L.~H. 2022, \apj,
  934, 56

\bibitem[{{Barger} {et~al.}(2008){Barger}, {Cowie}, \& {Wang}}]{barger2008}
{Barger}, A.~J., {Cowie}, L.~L., \& {Wang}, W.~H. 2008, \apj, 689, 687

\bibitem[{{Barger} {et~al.}(2012){Barger}, {Wang}, {Cowie}, {Owen}, {Chen}, \&
  {Williams}}]{barger2012}
{Barger}, A.~J., {Wang}, W.~H., {Cowie}, L.~L., {et~al.} 2012, \apj, 761, 89

\bibitem[{{Barro} {et~al.}(2019){Barro}, {P{\'e}rez-Gonz{\'a}lez}, {Cava},
  {Brammer}, {Pandya}, {Eliche Moral}, {Esquej}, {Dom{\'\i}nguez-S{\'a}nchez},
  {Alcalde Pampliega}, {Guo}, {Koekemoer}, {Trump}, {Ashby}, {Cardiel},
  {Castellano}, {Conselice}, {Dickinson}, {Dolch}, {Donley}, {Espino Briones},
  {Faber}, {Fazio}, {Ferguson}, {Finkelstein}, {Fontana}, {Galametz},
  {Gardner}, {Gawiser}, {Giavalisco}, {Grazian}, {Grogin}, {Hathi}, {Hemmati},
  {Hern{\'a}n-Caballero}, {Kocevski}, {Koo}, {Kodra}, {Lee}, {Lin}, {Lucas},
  {Mobasher}, {McGrath}, {Nandra}, {Nayyeri}, {Newman}, {Pforr}, {Peth},
  {Rafelski}, {Rodr{\'\i}guez-Munoz}, {Salvato}, {Stefanon}, {van der Wel},
  {Willner}, {Wiklind}, \& {Wuyts}}]{barro2019}
{Barro}, G., {P{\'e}rez-Gonz{\'a}lez}, P.~G., {Cava}, A., {et~al.} 2019, \apjs,
  243, 22

\bibitem[{{Battisti} {et~al.}(2020){Battisti}, {Cunha}, {Shivaei}, \&
  {Calzetti}}]{battisti2020}
{Battisti}, A.~J., {Cunha}, E.~d., {Shivaei}, I., \& {Calzetti}, D. 2020, \apj,
  888, 108

\bibitem[{{Beeston} {et~al.}(2024){Beeston}, {Gomez}, {Dunne}, {Maddox},
  {Eales}, \& {Smith}}]{beeston2024}
{Beeston}, R.~A., {Gomez}, H.~L., {Dunne}, L., {et~al.} 2024, \mnras, 535, 3162

\bibitem[{{Beeston} {et~al.}(2018){Beeston}, {Wright}, {Maddox}, {Gomez},
  {Dunne}, {Driver}, {Robotham}, {Clark}, {Vinsen}, {Takeuchi}, {Popping},
  {Bourne}, {Bremer}, {Phillipps}, {Moffett}, {Baes}, {Bland-Hawthorn},
  {Brough}, {De Vis}, {Eales}, {Holwerda}, {Loveday}, {Liske}, {Smith},
  {Smith}, {Valiante}, {Vlahakis}, \& {Wang}}]{beeston2018}
{Beeston}, R.~A., {Wright}, A.~H., {Maddox}, S., {et~al.} 2018, \mnras, 479,
  1077

\bibitem[{{Berta} {et~al.}(2007){Berta}, {Lonsdale}, {Polletta}, {Savage},
  {Franceschini}, {Buttery}, {Cimatti}, {Dias}, {Feruglio}, {Fiore}, {Held},
  {La Franca}, {Maiolino}, {Marconi}, {Matute}, {Oliver}, {Ricciardelli},
  {Rubele}, {Sacchi}, {Shupe}, \& {Surace}}]{berta2007}
{Berta}, S., {Lonsdale}, C.~J., {Polletta}, M., {et~al.} 2007, \aap, 476, 151

\bibitem[{{Berta} {et~al.}(2016){Berta}, {Lutz}, {Genzel},
  {F{\"o}rster-Schreiber}, \& {Tacconi}}]{berta2016}
{Berta}, S., {Lutz}, D., {Genzel}, R., {F{\"o}rster-Schreiber}, N.~M., \&
  {Tacconi}, L.~J. 2016, \aap, 587, A73

\bibitem[{{Berta} {et~al.}(2013{\natexlab{a}}){Berta}, {Lutz}, {Nordon},
  {Genzel}, {Magnelli}, {Popesso}, {Rosario}, {Saintonge}, {Wuyts}, \&
  {Tacconi}}]{berta2013b}
{Berta}, S., {Lutz}, D., {Nordon}, R., {et~al.} 2013{\natexlab{a}}, \aap, 555,
  L8

\bibitem[{{Berta} {et~al.}(2013{\natexlab{b}}){Berta}, {Lutz}, {Santini},
  {Wuyts}, {Rosario}, {Brisbin}, {Cooray}, {Franceschini}, {Gruppioni},
  {Hatziminaoglou}, {Hwang}, {Le Floc'h}, {Magnelli}, {Nordon}, {Oliver},
  {Page}, {Popesso}, {Pozzetti}, {Pozzi}, {Riguccini}, {Rodighiero},
  {Roseboom}, {Scott}, {Symeonidis}, {Valtchanov}, {Viero}, \&
  {Wang}}]{berta2013}
{Berta}, S., {Lutz}, D., {Santini}, P., {et~al.} 2013{\natexlab{b}}, \aap, 551,
  A100

\bibitem[{{Berta} {et~al.}(2023){Berta}, {Stanley}, {Ismail}, {Cox}, {Neri},
  {Yang}, {Young}, {Jin}, {Dannerbauer}, {Bakx}, {Beelen}, {Wei{\ss}}, {Nanni},
  {Omont}, {van der Werf}, {Krips}, {Baker}, {Bendo}, {Borsato}, {Buat},
  {Butler}, {Chartab}, {Cooray}, {Dye}, {Eales}, {Gavazzi}, {Hughes}, {Ivison},
  {Jones}, {Lehnert}, {Marchetti}, {Messias}, {Negrello}, {Perez-Fournon},
  {Riechers}, {Serjeant}, {Urquhart}, \& {Vlahakis}}]{berta2023}
{Berta}, S., {Stanley}, F., {Ismail}, D., {et~al.} 2023, \aap, 678, A28

\bibitem[{{Berta} \& {Zylka}({2019-2024})}]{piic2024}
{Berta}, S. \& {Zylka}, R. {2019-2024}, {Welcome to the PIIC},
  {\url{https://www.iram.fr/~gildas/dist/piic.pdf}}

\bibitem[{{Bertin} \& {Arnouts}(1996)}]{bertin1996}
{Bertin}, E. \& {Arnouts}, S. 1996, \aaps, 117, 393

\bibitem[{{B{\'e}thermin} {et~al.}(2015){B{\'e}thermin}, {De Breuck},
  {Sargent}, \& {Daddi}}]{bethermin2015b}
{B{\'e}thermin}, M., {De Breuck}, C., {Sargent}, M., \& {Daddi}, E. 2015, \aap,
  576, L9

\bibitem[{{B{\'e}thermin} {et~al.}(2017){B{\'e}thermin}, {Wu}, {Lagache},
  {Davidzon}, {Ponthieu}, {Cousin}, {Wang}, {Dor{\'e}}, {Daddi}, \&
  {Lapi}}]{bethermin2017}
{B{\'e}thermin}, M., {Wu}, H.-Y., {Lagache}, G., {et~al.} 2017, \aap, 607, A89

\bibitem[{{Bianchi}(2013)}]{bianchi2013}
{Bianchi}, S. 2013, \aap, 552, A89

\bibitem[{{Bianchi} \& {Ferrara}(2005)}]{bianchi2005}
{Bianchi}, S. \& {Ferrara}, A. 2005, \mnras, 358, 379

\bibitem[{{Bing} {et~al.}(2023){Bing}, {B{\'e}thermin}, {Lagache}, {Adam},
  {Ade}, {Ajeddig}, {Andr{\'e}}, {Artis}, {Aussel}, {Beelen}, {Beno{\^\i}t},
  {Berta}, {Billot}, {Bourrion}, {Calvo}, {Catalano}, {De Petris},
  {D{\'e}sert}, {Doyle}, {Driessen}, {Elbaz}, {Gkogkou}, {Gomez}, {Goupy},
  {Hanser}, {K{\'e}ruzor{\'e}}, {Kramer}, {Ladjelate}, {Liu}, {Leclercq},
  {Lestrade}, {Lustig}, {Mac{\'\i}as-P{\'e}rez}, {Maury}, {Mauskopf}, {Mayet},
  {Monfardini}, {Mu{\~n}oz-Echeverr{\'\i}a}, {Perotto}, {Pisano}, {Ponthieu},
  {Rev{\'e}ret}, {Rigby}, {Ritacco}, {Romero}, {Roussel}, {Ruppin}, {Schuster},
  {Sievers}, {Tucker}, \& {Zylka}}]{bing2023}
{Bing}, L., {B{\'e}thermin}, M., {Lagache}, G., {et~al.} 2023, \aap, 677, A66

\bibitem[{{Bing}({2022})}]{bingthesis}
{Bing}, L.~J. {2022}, {PhD thesis, "Mapping the dusty star formation at high
  redshift with the NIKA2 cosmological surveys"},
  {\url{https://theses.fr/2022AIXM0560}}

\bibitem[{{Blain} {et~al.}(2002){Blain}, {Smail}, {Ivison}, {Kneib}, \&
  {Frayer}}]{blain2002}
{Blain}, A.~W., {Smail}, I., {Ivison}, R.~J., {Kneib}, J.~P., \& {Frayer},
  D.~T. 2002, \physrep, 369, 111

\bibitem[{{Boquien} {et~al.}(2019){Boquien}, {Burgarella}, {Roehlly}, {Buat},
  {Ciesla}, {Corre}, {Inoue}, \& {Salas}}]{boquien2019}
{Boquien}, M., {Burgarella}, D., {Roehlly}, Y., {et~al.} 2019, \aap, 622, A103

\bibitem[{{Borys} {et~al.}(2004){Borys}, {Scott}, {Chapman}, {Halpern},
  {Nandra}, \& {Pope}}]{borys2004}
{Borys}, C., {Scott}, D., {Chapman}, S., {et~al.} 2004, \mnras, 355, 485

\bibitem[{{Bothwell} {et~al.}(2017){Bothwell}, {Aguirre}, {Aravena},
  {Bethermin}, {Bisbas}, {Chapman}, {De Breuck}, {Gonzalez}, {Greve},
  {Hezaveh}, {Ma}, {Malkan}, {Marrone}, {Murphy}, {Spilker}, {Strandet},
  {Vieira}, \& {Wei{\ss}}}]{bothwell2017}
{Bothwell}, M.~S., {Aguirre}, J.~E., {Aravena}, M., {et~al.} 2017, \mnras, 466,
  2825

\bibitem[{{Bothwell} {et~al.}(2010){Bothwell}, {Chapman}, {Tacconi}, {Smail},
  {Ivison}, {Casey}, {Bertoldi}, {Beswick}, {Biggs}, {Blain}, {Cox}, {Genzel},
  {Greve}, {Kennicutt}, {Muxlow}, {Neri}, \& {Omont}}]{bothwell2010}
{Bothwell}, M.~S., {Chapman}, S.~C., {Tacconi}, L., {et~al.} 2010, \mnras, 405,
  219

\bibitem[{{Bothwell} {et~al.}(2013){Bothwell}, {Smail}, {Chapman}, {Genzel},
  {Ivison}, {Tacconi}, {Alaghband-Zadeh}, {Bertoldi}, {Blain}, {Casey}, {Cox},
  {Greve}, {Lutz}, {Neri}, {Omont}, \& {Swinbank}}]{bothwell2013}
{Bothwell}, M.~S., {Smail}, I., {Chapman}, S.~C., {et~al.} 2013, \mnras, 429,
  3047

\bibitem[{{Bouch{\'e}} {et~al.}(2007){Bouch{\'e}}, {Lehnert}, {Aguirre},
  {P{\'e}roux}, \& {Bergeron}}]{bouche2007}
{Bouch{\'e}}, N., {Lehnert}, M.~D., {Aguirre}, A., {P{\'e}roux}, C., \&
  {Bergeron}, J. 2007, \mnras, 378, 525

\bibitem[{{Bourrion} {et~al.}(2016){Bourrion}, {Benoit}, {Bouly}, {Bouvier},
  {Bosson}, {Calvo}, {Catalano}, {Goupy}, {Li}, {Mac{\'\i}as-P{\'e}rez},
  {Monfardini}, {Tourres}, {Ponchant}, \& {Vescovi}}]{bourrion2016}
{Bourrion}, O., {Benoit}, A., {Bouly}, J.~L., {et~al.} 2016, Journal of
  Instrumentation, 11, P11001

\bibitem[{{Bouwens} {et~al.}(2015){Bouwens}, {Illingworth}, {Oesch}, {Trenti},
  {Labb{\'e}}, {Bradley}, {Carollo}, {van Dokkum}, {Gonzalez}, {Holwerda},
  {Franx}, {Spitler}, {Smit}, \& {Magee}}]{bouwens2015}
{Bouwens}, R.~J., {Illingworth}, G.~D., {Oesch}, P.~A., {et~al.} 2015, \apj,
  803, 34

\bibitem[{{Brandt} {et~al.}(2001){Brandt}, {Alexander}, {Hornschemeier},
  {Garmire}, {Schneider}, {Barger}, {Bauer}, {Broos}, {Cowie}, {Townsley},
  {Burrows}, {Chartas}, {Feigelson}, {Griffiths}, {Nousek}, \&
  {Sargent}}]{brandt2001}
{Brandt}, W.~N., {Alexander}, D.~M., {Hornschemeier}, A.~E., {et~al.} 2001,
  \aj, 122, 2810

\bibitem[{{Bruzual} \& {Charlot}(2003)}]{bc03}
{Bruzual}, G. \& {Charlot}, S. 2003, \mnras, 344, 1000

\bibitem[{{Burgarella} {et~al.}(2005){Burgarella}, {Buat}, \&
  {Iglesias-P{\'a}ramo}}]{burgarella2005}
{Burgarella}, D., {Buat}, V., \& {Iglesias-P{\'a}ramo}, J. 2005, \mnras, 360,
  1413

\bibitem[{{Calvo} {et~al.}(2016){Calvo}, {Beno{\^\i}t}, {Catalano}, {Goupy},
  {Monfardini}, {Ponthieu}, {Barria}, {Bres}, {Grollier}, {Garde}, {Leggeri},
  {Pont}, {Triqueneaux}, {Adam}, {Bourrion}, {Mac{\'\i}as-P{\'e}rez}, {Rebolo},
  {Ritacco}, {Scordilis}, {Tourres}, {Adane}, {Coiffard}, {Leclercq},
  {D{\'e}sert}, {Doyle}, {Mauskopf}, {Tucker}, {Ade}, {Andr{\'e}}, {Beelen},
  {Belier}, {Bideaud}, {Billot}, {Comis}, {D'Addabbo}, {Kramer}, {Martino},
  {Mayet}, {Pajot}, {Pascale}, {Perotto}, {Rev{\'e}ret}, {Ritacco},
  {Rodriguez}, {Savini}, {Schuster}, {Sievers}, \& {Zylka}}]{calvo2016}
{Calvo}, M., {Beno{\^\i}t}, A., {Catalano}, A., {et~al.} 2016, Journal of Low
  Temperature Physics, 184, 816

\bibitem[{{Calzetti} {et~al.}(2000){Calzetti}, {Armus}, {Bohlin}, {Kinney},
  {Koornneef}, \& {Storchi-Bergmann}}]{calzetti2000}
{Calzetti}, D., {Armus}, L., {Bohlin}, R.~C., {et~al.} 2000, \apj, 533, 682

\bibitem[{{Carilli} {et~al.}(2010){Carilli}, {Daddi}, {Riechers}, {Walter},
  {Weiss}, {Dannerbauer}, {Morrison}, {Wagg}, {Dav{\'e}}, {Elbaz}, {Stern},
  {Dickinson}, {Krips}, \& {Aravena}}]{carilli2010}
{Carilli}, C.~L., {Daddi}, E., {Riechers}, D., {et~al.} 2010, \apj, 714, 1407

\bibitem[{{Catalano} {et~al.}(2014){Catalano}, {Calvo}, {Ponthieu}, {Adam},
  {Adane}, {Ade}, {Andr{\'e}}, {Beelen}, {Belier}, {Beno{\^\i}t}, {Bideaud},
  {Billot}, {Boudou}, {Bourrion}, {Coiffard}, {Comis}, {D'Addabbo},
  {D{\'e}sert}, {Doyle}, {Goupy}, {Kramer}, {Leclercq},
  {Mac{\'\i}as-P{\'e}rez}, {Martino}, {Mauskopf}, {Mayet}, {Monfardini},
  {Pajot}, {Pascale}, {Perotto}, {Rev{\'e}ret}, {Rodriguez}, {Savini},
  {Schuster}, {Sievers}, {Tucker}, \& {Zylka}}]{Catalano2014}
{Catalano}, A., {Calvo}, M., {Ponthieu}, N., {et~al.} 2014, \aap, 569, A9

\bibitem[{{Chabrier}(2003)}]{chabrier2003}
{Chabrier}, G. 2003, \apjl, 586, L133

\bibitem[{{Chapin} {et~al.}(2009){Chapin}, {Pope}, {Scott}, {Aretxaga},
  {Austermann}, {Chary}, {Coppin}, {Halpern}, {Hughes}, {Lowenthal},
  {Morrison}, {Perera}, {Scott}, {Wilson}, \& {Yun}}]{chapin2009}
{Chapin}, E.~L., {Pope}, A., {Scott}, D., {et~al.} 2009, \mnras, 398, 1793

\bibitem[{{Chapman} {et~al.}(2005){Chapman}, {Blain}, {Smail}, \&
  {Ivison}}]{chapman2005}
{Chapman}, S.~C., {Blain}, A.~W., {Smail}, I., \& {Ivison}, R.~J. 2005, \apj,
  622, 772

\bibitem[{{Charlot} \& {Fall}(2000)}]{charlot2000}
{Charlot}, S. \& {Fall}, S.~M. 2000, \apj, 539, 718

\bibitem[{{Chung} {et~al.}(2009){Chung}, {Rhee}, {Kim}, {Yun}, {Heyer}, \&
  {Young}}]{chung2009}
{Chung}, E.~J., {Rhee}, M.-H., {Kim}, H., {et~al.} 2009, \apjs, 184, 199

\bibitem[{{Ciesla} {et~al.}(2015){Ciesla}, {Charmandaris}, {Georgakakis},
  {Bernhard}, {Mitchell}, {Buat}, {Elbaz}, {LeFloc'h}, {Lacey}, {Magdis}, \&
  {Xilouris}}]{ciesla2015}
{Ciesla}, L., {Charmandaris}, V., {Georgakakis}, A., {et~al.} 2015, \aap, 576,
  A10

\bibitem[{{Colina} {et~al.}(2023){Colina}, {Crespo G{\'o}mez},
  {{\'A}lvarez-M{\'a}rquez}, {Bik}, {Walter}, {Boogaard}, {Labiano},
  {Peissker}, {P{\'e}rez-Gonz{\'a}lez}, {{\"O}stlin}, {Greve},
  {N{\o}rgaard-Nielsen}, {Wright}, {Alonso-Herrero}, {Azollini}, {Caputi},
  {Dicken}, {Garc{\'\i}a-Mar{\'\i}n}, {Hjorth}, {Ilbert}, {Kendrew}, {Pye},
  {Tikkanen}, {van der Werf}, {Costantin}, {Iani}, {Gillman}, {Jermann},
  {Langeroodi}, {Moutard}, {Rinaldi}, {Topinka}, {van Dishoeck}, {G{\"u}del},
  {Henning}, {Lagage}, {Ray}, \& {Vandenbussche}}]{colina2023}
{Colina}, L., {Crespo G{\'o}mez}, A., {{\'A}lvarez-M{\'a}rquez}, J., {et~al.}
  2023, \aap, 673, L6

\bibitem[{{Combes} {et~al.}(2011){Combes}, {Garc{\'\i}a-Burillo}, {Braine},
  {Schinnerer}, {Walter}, \& {Colina}}]{combes2011}
{Combes}, F., {Garc{\'\i}a-Burillo}, S., {Braine}, J., {et~al.} 2011, \aap,
  528, A124

\bibitem[{{Combes} {et~al.}(2013){Combes}, {Garc{\'\i}a-Burillo}, {Braine},
  {Schinnerer}, {Walter}, \& {Colina}}]{combes2013}
{Combes}, F., {Garc{\'\i}a-Burillo}, S., {Braine}, J., {et~al.} 2013, \aap,
  550, A41

\bibitem[{{Cortzen} {et~al.}(2020){Cortzen}, {Magdis}, {Valentino}, {Daddi},
  {Liu}, {Rigopoulou}, {Sargent}, {Riechers}, {Cormier}, {Hodge}, {Walter},
  {Elbaz}, {B{\'e}thermin}, {Greve}, {Kokorev}, \& {Toft}}]{cortzen2020}
{Cortzen}, I., {Magdis}, G.~E., {Valentino}, F., {et~al.} 2020, \aap, 634, L14

\bibitem[{{Cowie} {et~al.}(2017){Cowie}, {Barger}, {Hsu}, {Chen}, {Owen}, \&
  {Wang}}]{cowie2017}
{Cowie}, L.~L., {Barger}, A.~J., {Hsu}, L.~Y., {et~al.} 2017, \apj, 837, 139

\bibitem[{{Cowie} {et~al.}(2004){Cowie}, {Barger}, {Hu}, {Capak}, \&
  {Songaila}}]{cowie2004}
{Cowie}, L.~L., {Barger}, A.~J., {Hu}, E.~M., {Capak}, P., \& {Songaila}, A.
  2004, \aj, 127, 3137

\bibitem[{{Cox} {et~al.}(2023){Cox}, {Neri}, {Berta}, {Ismail}, {Stanley},
  {Young}, {Jin}, {Bakx}, {Beelen}, {Dannerbauer}, {Krips}, {Lehnert}, {Omont},
  {Riechers}, {Baker}, {Bendo}, {Borsato}, {Buat}, {Butler}, {Chartab},
  {Cooray}, {Dye}, {Eales}, {Gavazzi}, {Hughes}, {Ivison}, {Jones},
  {Marchetti}, {Messias}, {Nanni}, {Negrello}, {Perez-Fournon}, {Serjeant},
  {Urquhart}, {Vlahakis}, {Wei{\ss}}, {van der Werf}, \& {Yang}}]{cox2023}
{Cox}, P., {Neri}, R., {Berta}, S., {et~al.} 2023, \aap, 678, A26

\bibitem[{{Crespo G{\'o}mez} {et~al.}(2024){Crespo G{\'o}mez}, {Colina},
  {{\'A}lvarez-M{\'a}rquez}, {Bik}, {Boogaard}, {{\"O}stlin}, {Pei{\ss}ker},
  {Walter}, {Labiano}, {P{\'e}rez-Gonz{\'a}lez}, {Greve}, {Wright},
  {Alonso-Herrero}, {Caputi}, {Costantin}, {Eckart}, {Garc{\'\i}a-Mar{\'\i}n},
  {Gillman}, {Hjorth}, {Iani}, {Langeroodi}, {Pye}, {Rinaldi}, {Tikkanen}, {van
  der Werf}, {Lagage}, \& {van Dishoeck}}]{crespogomez2024}
{Crespo G{\'o}mez}, A., {Colina}, L., {{\'A}lvarez-M{\'a}rquez}, J., {et~al.}
  2024, arXiv e-prints, arXiv:2402.18672

\bibitem[{{Cutri} {et~al.}(2013){Cutri}, {Wright}, {Conrow}, {Fowler},
  {Eisenhardt}, {Grillmair}, {Kirkpatrick}, {Masci}, {McCallon}, {Wheelock},
  {Fajardo-Acosta}, {Yan}, {Benford}, {Harbut}, {Jarrett}, {Lake}, {Leisawitz},
  {Ressler}, {Stanford}, {Tsai}, {Liu}, {Helou}, {Mainzer}, {Gettings},
  {Gonzalez}, {Hoffman}, {Marsh}, {Padgett}, {Skrutskie}, {Beck}, {Papin}, \&
  {Wittman}}]{wise2013}
{Cutri}, R.~M., {Wright}, E.~L., {Conrow}, T., {et~al.} 2013, {Explanatory
  Supplement to the AllWISE Data Release Products}, Explanatory Supplement to
  the AllWISE Data Release Products, by R. M. Cutri et al.

\bibitem[{{da Cunha} {et~al.}(2008){da Cunha}, {Charlot}, \&
  {Elbaz}}]{dacunha2008}
{da Cunha}, E., {Charlot}, S., \& {Elbaz}, D. 2008, \mnras, 388, 1595

\bibitem[{{da Cunha} {et~al.}(2013){da Cunha}, {Walter}, {Decarli}, {Bertoldi},
  {Carilli}, {Daddi}, {Elbaz}, {Ivison}, {Maiolino}, {Riechers}, {Rix},
  {Sargent}, {Smail}, \& {Weiss}}]{dacunha2013}
{da Cunha}, E., {Walter}, F., {Decarli}, R., {et~al.} 2013, \apj, 765, 9

\bibitem[{{da Cunha} {et~al.}(2015){da Cunha}, {Walter}, {Smail}, {Swinbank},
  {Simpson}, {Decarli}, {Hodge}, {Weiss}, {van der Werf}, {Bertoldi},
  {Chapman}, {Cox}, {Danielson}, {Dannerbauer}, {Greve}, {Ivison}, {Karim}, \&
  {Thomson}}]{dacunha2015}
{da Cunha}, E., {Walter}, F., {Smail}, I.~R., {et~al.} 2015, \apj, 806, 110

\bibitem[{{Daddi} {et~al.}(2009){Daddi}, {Dannerbauer}, {Stern}, {Dickinson},
  {Morrison}, {Elbaz}, {Giavalisco}, {Mancini}, {Pope}, \&
  {Spinrad}}]{daddi2009}
{Daddi}, E., {Dannerbauer}, H., {Stern}, D., {et~al.} 2009, \apj, 694, 1517

\bibitem[{{Dale} {et~al.}(2014){Dale}, {Helou}, {Magdis}, {Armus},
  {D{\'\i}az-Santos}, \& {Shi}}]{dale2014}
{Dale}, D.~A., {Helou}, G., {Magdis}, G.~E., {et~al.} 2014, \apj, 784, 83

\bibitem[{{Dannerbauer} {et~al.}(2019){Dannerbauer}, {Harrington},
  {D{\'\i}az-S{\'a}nchez}, {Iglesias-Groth}, {Rebolo}, {Genova-Santos}, \&
  {Krips}}]{dannerbauer2019}
{Dannerbauer}, H., {Harrington}, K., {D{\'\i}az-S{\'a}nchez}, A., {et~al.}
  2019, \aj, 158, 34

\bibitem[{{Dav{\'e}} {et~al.}(2019){Dav{\'e}}, {Angl{\'e}s-Alc{\'a}zar},
  {Narayanan}, {Li}, {Rafieferantsoa}, \& {Appleby}}]{dave2019}
{Dav{\'e}}, R., {Angl{\'e}s-Alc{\'a}zar}, D., {Narayanan}, D., {et~al.} 2019,
  \mnras, 486, 2827

\bibitem[{{Davidzon} {et~al.}(2017){Davidzon}, {Ilbert}, {Laigle}, {Coupon},
  {McCracken}, {Delvecchio}, {Masters}, {Capak}, {Hsieh}, {Le F{\`e}vre},
  {Tresse}, {Bethermin}, {Chang}, {Faisst}, {Le Floc'h}, {Steinhardt}, {Toft},
  {Aussel}, {Dubois}, {Hasinger}, {Salvato}, {Sanders}, {Scoville}, \&
  {Silverman}}]{davidzon2017}
{Davidzon}, I., {Ilbert}, O., {Laigle}, C., {et~al.} 2017, \aap, 605, A70

\bibitem[{{Davies} {et~al.}(2015){Davies}, {Driver}, {Robotham}, {Baldry},
  {Lange}, {Liske}, {Meyer}, {Popping}, {Wilkins}, \& {Wright}}]{davies2015}
{Davies}, L.~J.~M., {Driver}, S.~P., {Robotham}, A.~S.~G., {et~al.} 2015,
  \mnras, 447, 1014

\bibitem[{{Decarli} {et~al.}(2016){Decarli}, {Walter}, {Aravena}, {Carilli},
  {Bouwens}, {da Cunha}, {Daddi}, {Elbaz}, {Riechers}, {Smail}, {Swinbank},
  {Weiss}, {Bacon}, {Bauer}, {Bell}, {Bertoldi}, {Chapman}, {Colina}, {Cortes},
  {Cox}, {G{\'o}nzalez-L{\'o}pez}, {Inami}, {Ivison}, {Hodge}, {Karim},
  {Magnelli}, {Ota}, {Popping}, {Rix}, {Sargent}, {van der Wel}, \& {van der
  Werf}}]{decarli2016}
{Decarli}, R., {Walter}, F., {Aravena}, M., {et~al.} 2016, \apj, 833, 70

\bibitem[{{Decarli} {et~al.}(2019){Decarli}, {Walter},
  {G{\'o}nzalez-L{\'o}pez}, {Aravena}, {Boogaard}, {Carilli}, {Cox}, {Daddi},
  {Popping}, {Riechers}, {Uzgil}, {Weiss}, {Assef}, {Bacon}, {Bauer},
  {Bertoldi}, {Bouwens}, {Contini}, {Cortes}, {da Cunha}, {D{\'\i}az-Santos},
  {Elbaz}, {Inami}, {Hodge}, {Ivison}, {Le F{\`e}vre}, {Magnelli}, {Novak},
  {Oesch}, {Rix}, {Sargent}, {Smail}, {Swinbank}, {Somerville}, {van der Werf},
  {Wagg}, \& {Wisotzki}}]{decarli2019}
{Decarli}, R., {Walter}, F., {G{\'o}nzalez-L{\'o}pez}, J., {et~al.} 2019, \apj,
  882, 138

\bibitem[{{Delhaize} {et~al.}(2017){Delhaize}, {Smol{\v{c}}i{\'c}},
  {Delvecchio}, {Novak}, {Sargent}, {Baran}, {Magnelli}, {Zamorani},
  {Schinnerer}, {Murphy}, {Aravena}, {Berta}, {Bondi}, {Capak}, {Carilli},
  {Ciliegi}, {Civano}, {Ilbert}, {Karim}, {Laigle}, {Le F{\`e}vre}, {Marchesi},
  {McCracken}, {Salvato}, {Seymour}, \& {Tasca}}]{delhaize2017}
{Delhaize}, J., {Smol{\v{c}}i{\'c}}, V., {Delvecchio}, I., {et~al.} 2017, \aap,
  602, A4

\bibitem[{{Dickinson} {et~al.}({2003}){Dickinson}, {Bergeron}, {Casertano}, \&
  et~al.}]{dickinson2003b}
{Dickinson}, M., {Bergeron}, J., {Casertano}, S., \& et~al. {2003}, {Great
  Observatories Origins Deep Survey (GOODS), Validation Observations, Spitzer
  Proposal}

\bibitem[{{Dickinson} \& {GOODS Legacy Team}(2001)}]{dickinson2001}
{Dickinson}, M. \& {GOODS Legacy Team}. 2001, in American Astronomical Society
  Meeting Abstracts, Vol. 198, American Astronomical Society Meeting Abstracts
  \#198, 25.01

\bibitem[{{Dickinson} {et~al.}(2003){Dickinson}, {Papovich}, {Ferguson}, \&
  {Budav{\'a}ri}}]{dickinson2003}
{Dickinson}, M., {Papovich}, C., {Ferguson}, H.~C., \& {Budav{\'a}ri}, T. 2003,
  \apj, 587, 25

\bibitem[{{Dole} {et~al.}(2006){Dole}, {Lagache}, {Puget}, {Caputi},
  {Fern{\'a}ndez-Conde}, {Le Floc'h}, {Papovich}, {P{\'e}rez-Gonz{\'a}lez},
  {Rieke}, \& {Blaylock}}]{dole2006}
{Dole}, H., {Lagache}, G., {Puget}, J.~L., {et~al.} 2006, \aap, 451, 417

\bibitem[{{Draine}(2003)}]{draine2003}
{Draine}, B.~T. 2003, \araa, 41, 241

\bibitem[{{Draine} {et~al.}(2014){Draine}, {Aniano}, {Krause}, {Groves},
  {Sandstrom}, {Braun}, {Leroy}, {Klaas}, {Linz}, {Rix}, {Schinnerer},
  {Schmiedeke}, \& {Walter}}]{draine2014}
{Draine}, B.~T., {Aniano}, G., {Krause}, O., {et~al.} 2014, \apj, 780, 172

\bibitem[{{Draine} {et~al.}(2007){Draine}, {Dale}, {Bendo}, {Gordon}, {Smith},
  {Armus}, {Engelbracht}, {Helou}, {Kennicutt}, {Li}, {Roussel}, {Walter},
  {Calzetti}, {Moustakas}, {Murphy}, {Rieke}, {Bot}, {Hollenbach}, {Sheth}, \&
  {Teplitz}}]{draine2007}
{Draine}, B.~T., {Dale}, D.~A., {Bendo}, G., {et~al.} 2007, \apj, 663, 866

\bibitem[{{Draine} \& {Li}(2001)}]{DL01}
{Draine}, B.~T. \& {Li}, A. 2001, \apj, 551, 807

\bibitem[{{Draine} \& {Li}(2007)}]{DL07}
{Draine}, B.~T. \& {Li}, A. 2007, \apj, 657, 810

\bibitem[{{Draine} \& {Salpeter}(1979)}]{draine1979}
{Draine}, B.~T. \& {Salpeter}, E.~E. 1979, \apj, 231, 438

\bibitem[{{Driver} {et~al.}(2018){Driver}, {Andrews}, {da Cunha}, {Davies},
  {Lagos}, {Robotham}, {Vinsen}, {Wright}, {Alpaslan}, {Bland-Hawthorn},
  {Bourne}, {Brough}, {Bremer}, {Cluver}, {Colless}, {Conselice}, {Dunne},
  {Eales}, {Gomez}, {Holwerda}, {Hopkins}, {Kafle}, {Kelvin}, {Loveday},
  {Liske}, {Maddox}, {Phillipps}, {Pimbblet}, {Rowlands}, {Sansom}, {Taylor},
  {Wang}, \& {Wilkins}}]{driver2018}
{Driver}, S.~P., {Andrews}, S.~K., {da Cunha}, E., {et~al.} 2018, \mnras, 475,
  2891

\bibitem[{{Driver} {et~al.}(2011){Driver}, {Hill}, {Kelvin}, {Robotham},
  {Liske}, {Norberg}, {Baldry}, {Bamford}, {Hopkins}, {Loveday}, {Peacock},
  {Andrae}, {Bland-Hawthorn}, {Brough}, {Brown}, {Cameron}, {Ching}, {Colless},
  {Conselice}, {Croom}, {Cross}, {de Propris}, {Dye}, {Drinkwater}, {Ellis},
  {Graham}, {Grootes}, {Gunawardhana}, {Jones}, {van Kampen}, {Maraston},
  {Nichol}, {Parkinson}, {Phillipps}, {Pimbblet}, {Popescu}, {Prescott},
  {Roseboom}, {Sadler}, {Sansom}, {Sharp}, {Smith}, {Taylor}, {Thomas},
  {Tuffs}, {Wijesinghe}, {Dunne}, {Frenk}, {Jarvis}, {Madore}, {Meyer},
  {Seibert}, {Staveley-Smith}, {Sutherland}, \& {Warren}}]{driver2011}
{Driver}, S.~P., {Hill}, D.~T., {Kelvin}, L.~S., {et~al.} 2011, \mnras, 413,
  971

\bibitem[{{Driver} {et~al.}(2009){Driver}, {Norberg}, {Baldry}, {Bamford},
  {Hopkins}, {Liske}, {Loveday}, {Peacock}, {Hill}, {Kelvin}, {Robotham},
  {Cross}, {Parkinson}, {Prescott}, {Conselice}, {Dunne}, {Brough}, {Jones},
  {Sharp}, {van Kampen}, {Oliver}, {Roseboom}, {Bland-Hawthorn}, {Croom},
  {Ellis}, {Cameron}, {Cole}, {Frenk}, {Couch}, {Graham}, {Proctor}, {De
  Propris}, {Doyle}, {Edmondson}, {Nichol}, {Thomas}, {Eales}, {Jarvis},
  {Kuijken}, {Lahav}, {Madore}, {Seibert}, {Meyer}, {Staveley-Smith},
  {Phillipps}, {Popescu}, {Sansom}, {Sutherland}, {Tuffs}, \&
  {Warren}}]{driver2009}
{Driver}, S.~P., {Norberg}, P., {Baldry}, I.~K., {et~al.} 2009, Astronomy and
  Geophysics, 50, 5.12

\bibitem[{{Driver} {et~al.}(2008){Driver}, {Popescu}, {Tuffs}, {Graham},
  {Liske}, \& {Baldry}}]{driver2008}
{Driver}, S.~P., {Popescu}, C.~C., {Tuffs}, R.~J., {et~al.} 2008, \apjl, 678,
  L101

\bibitem[{{Dudzevi{\v{c}}i{\={u}}t{\.{e}}}
  {et~al.}(2021){Dudzevi{\v{c}}i{\={u}}t{\.{e}}}, {Smail}, {Swinbank}, {Lim},
  {Wang}, {Simpson}, {Ao}, {Chapman}, {Chen}, {Clements}, {Dannerbauer}, {Ho},
  {Hwang}, {Koprowski}, {Lee}, {Scott}, {Shim}, {Shirley}, \&
  {Toba}}]{dudzeviciute2021}
{Dudzevi{\v{c}}i{\={u}}t{\.{e}}}, U., {Smail}, I., {Swinbank}, A.~M., {et~al.}
  2021, \mnras, 500, 942

\bibitem[{{Dudzevi{\v{c}}i{\={u}}t{\.{e}}}
  {et~al.}(2020){Dudzevi{\v{c}}i{\={u}}t{\.{e}}}, {Smail}, {Swinbank}, {Stach},
  {Almaini}, {da Cunha}, {An}, {Arumugam}, {Birkin}, {Blain}, {Chapman},
  {Chen}, {Conselice}, {Coppin}, {Dunlop}, {Farrah}, {Geach}, {Gullberg},
  {Hartley}, {Hodge}, {Ivison}, {Maltby}, {Scott}, {Simpson}, {Simpson},
  {Thomson}, {Walter}, {Wardlow}, {Weiss}, \& {van der
  Werf}}]{dudzeviciute2020}
{Dudzevi{\v{c}}i{\={u}}t{\.{e}}}, U., {Smail}, I., {Swinbank}, A.~M., {et~al.}
  2020, \mnras, 494, 3828

\bibitem[{{Dunne} {et~al.}(2020){Dunne}, {Bonavera}, {Gonzalez-Nuevo},
  {Maddox}, \& {Vlahakis}}]{dunne2020b}
{Dunne}, L., {Bonavera}, L., {Gonzalez-Nuevo}, J., {Maddox}, S.~J., \&
  {Vlahakis}, C. 2020, \mnras, 498, 4635

\bibitem[{{Dunne} {et~al.}(2000){Dunne}, {Eales}, {Edmunds}, {Ivison},
  {Alexander}, \& {Clements}}]{dunne2000}
{Dunne}, L., {Eales}, S., {Edmunds}, M., {et~al.} 2000, \mnras, 315, 115

\bibitem[{{Dunne} \& {Eales}(2001)}]{dunne2001}
{Dunne}, L. \& {Eales}, S.~A. 2001, \mnras, 327, 697

\bibitem[{{Dunne} {et~al.}(2011){Dunne}, {Gomez}, {da Cunha}, {Charlot}, {Dye},
  {Eales}, {Maddox}, {Rowlands}, {Smith}, {Auld}, {Baes}, {Bonfield}, {Bourne},
  {Buttiglione}, {Cava}, {Clements}, {Coppin}, {Cooray}, {Dariush}, {de Zotti},
  {Driver}, {Fritz}, {Geach}, {Hopwood}, {Ibar}, {Ivison}, {Jarvis}, {Kelvin},
  {Pascale}, {Pohlen}, {Popescu}, {Rigby}, {Robotham}, {Rodighiero}, {Sansom},
  {Serjeant}, {Temi}, {Thompson}, {Tuffs}, {van der Werf}, \&
  {Vlahakis}}]{dunne2011}
{Dunne}, L., {Gomez}, H.~L., {da Cunha}, E., {et~al.} 2011, \mnras, 417, 1510

\bibitem[{{Dunne} {et~al.}(2009){Dunne}, {Maddox}, {Ivison}, {Rudnick},
  {Delaney}, {Matthews}, {Crowe}, {Gomez}, {Eales}, \& {Dye}}]{dunne2009}
{Dunne}, L., {Maddox}, S.~J., {Ivison}, R.~J., {et~al.} 2009, \mnras, 394, 1307

\bibitem[{{Dunne} {et~al.}(2021){Dunne}, {Maddox}, {Vlahakis}, \&
  {Gomez}}]{dunne2021}
{Dunne}, L., {Maddox}, S.~J., {Vlahakis}, C., \& {Gomez}, H.~L. 2021, \mnras,
  501, 2573

\bibitem[{{Eales} {et~al.}(2010){Eales}, {Dunne}, {Clements}, {Cooray}, {de
  Zotti}, {Dye}, {Ivison}, {Jarvis}, {Lagache}, {Maddox}, {Negrello},
  {Serjeant}, {Thompson}, {Kampen}, {Amblard}, {Andreani}, {Baes}, {Beelen},
  {Bendo}, {Benford}, {Bertoldi}, {Bock}, {Bonfield}, {Boselli}, {Bridge},
  {Buat}, {Burgarella}, {Carlberg}, {Cava}, {Chanial}, {Charlot},
  {Christopher}, {Coles}, {Cortese}, {Dariush}, {da Cunha}, {Dalton}, {Danese},
  {Dannerbauer}, {Driver}, {Dunlop}, {Fan}, {Farrah}, {Frayer}, {Frenk},
  {Geach}, {Gardner}, {Gomez}, {Gonz{\'a}lez-Nuevo}, {Gonz{\'a}lez-Solares},
  {Griffin}, {Hardcastle}, {Hatziminaoglou}, {Herranz}, {Hughes}, {Ibar},
  {Jeong}, {Lacey}, {Lapi}, {Lawrence}, {Lee}, {Leeuw}, {Liske},
  {L{\'o}pez-Caniego}, {M{\"u}ller}, {Nandra}, {Panuzzo}, {Papageorgiou},
  {Patanchon}, {Peacock}, {Pearson}, {Phillipps}, {Pohlen}, {Popescu},
  {Rawlings}, {Rigby}, {Rigopoulou}, {Robotham}, {Rodighiero}, {Sansom},
  {Schulz}, {Scott}, {Smith}, {Sibthorpe}, {Smail}, {Stevens}, {Sutherland},
  {Takeuchi}, {Tedds}, {Temi}, {Tuffs}, {Trichas}, {Vaccari}, {Valtchanov},
  {van der Werf}, {Verma}, {Vieria}, {Vlahakis}, \& {White}}]{eales2010}
{Eales}, S., {Dunne}, L., {Clements}, D., {et~al.} 2010, \pasp, 122, 499

\bibitem[{{Eales} \& {Ward}(2024)}]{eales2024}
{Eales}, S. \& {Ward}, B. 2024, \mnras, 529, 1130

\bibitem[{{Eisenstein} {et~al.}(2023){Eisenstein}, {Johnson}, {Robertson},
  {Tacchella}, {Hainline}, {Jakobsen}, {Maiolino}, {Bonaventura}, {Bunker},
  {Cameron}, {Cargile}, {Curtis-Lake}, {Hausen}, {Pusk{\'a}s}, {Rieke}, {Sun},
  {Willmer}, {Willott}, {Alberts}, {Arribas}, {Baker}, {Baum}, {Bhatawdekar},
  {Carniani}, {Charlot}, {Chen}, {Chevallard}, {Curti}, {DeCoursey},
  {D'Eugenio}, {de Graaff}, {Egami}, {Helton}, {Ji}, {Jones}, {Kumari},
  {L{\"u}tzgendorf}, {Laseter}, {Looser}, {Lyu}, {Maseda}, {Nelson},
  {Parlanti}, {Rauscher}, {Rawle}, {Rieke}, {Rix}, {Rujopakarn}, {Sandles},
  {Saxena}, {Scholtz}, {Sharpe}, {Shivaei}, {Simmonds}, {Smit}, {Topping},
  {{\"U}bler}, {Venturi}, {Williams}, {Witstok}, \& {Woodrum}}]{eisenstein2023}
{Eisenstein}, D.~J., {Johnson}, B.~D., {Robertson}, B., {et~al.} 2023, arXiv
  e-prints, arXiv:2310.12340, {Submitted to ApJ Supplement}

\bibitem[{{Elbaz} {et~al.}(2011){Elbaz}, {Dickinson}, {Hwang},
  {D{\'\i}az-Santos}, {Magdis}, {Magnelli}, {Le Borgne}, {Galliano},
  {Pannella}, {Chanial}, {Armus}, {Charmandaris}, {Daddi}, {Aussel}, {Popesso},
  {Kartaltepe}, {Altieri}, {Valtchanov}, {Coia}, {Dannerbauer}, {Dasyra},
  {Leiton}, {Mazzarella}, {Alexander}, {Buat}, {Burgarella}, {Chary}, {Gilli},
  {Ivison}, {Juneau}, {Le Floc'h}, {Lutz}, {Morrison}, {Mullaney}, {Murphy},
  {Pope}, {Scott}, {Brodwin}, {Calzetti}, {Cesarsky}, {Charlot}, {Dole},
  {Eisenhardt}, {Ferguson}, {F{\"o}rster Schreiber}, {Frayer}, {Giavalisco},
  {Huynh}, {Koekemoer}, {Papovich}, {Reddy}, {Surace}, {Teplitz}, {Yun}, \&
  {Wilson}}]{elbaz2011}
{Elbaz}, D., {Dickinson}, M., {Hwang}, H.~S., {et~al.} 2011, \aap, 533, A119

\bibitem[{{Enia} {et~al.}(2022){Enia}, {Talia}, {Pozzi}, {Cimatti},
  {Delvecchio}, {Zamorani}, {D'Amato}, {Bisigello}, {Gruppioni}, {Rodighiero},
  {Calura}, {Dallacasa}, {Giulietti}, {Barchiesi}, {Behiri}, \&
  {Romano}}]{enia2022}
{Enia}, A., {Talia}, M., {Pozzi}, F., {et~al.} 2022, \apj, 927, 204

\bibitem[{{Evans} {et~al.}(2024){Evans}, {Evans}, {Mart{\'\i}nez-Galarza},
  {Miller}, {Primini}, {Azadi}, {Burke}, {Civano}, {D'Abrusco}, {Fabbiano},
  {Graessle}, {Grier}, {Houck}, {Lauer}, {McCollough}, {Nowak}, {Plummer},
  {Rots}, {Siemiginowska}, \& {Tibbetts}}]{evans2024}
{Evans}, I.~N., {Evans}, J.~D., {Mart{\'\i}nez-Galarza}, J.~R., {et~al.} 2024,
  \apjs, 274, 22

\bibitem[{{Feltre} {et~al.}(2012){Feltre}, {Hatziminaoglou}, {Fritz}, \&
  {Franceschini}}]{feltre2012}
{Feltre}, A., {Hatziminaoglou}, E., {Fritz}, J., \& {Franceschini}, A. 2012,
  \mnras, 426, 120

\bibitem[{{Ferrarotti} \& {Gail}(2006)}]{ferrarotti2006}
{Ferrarotti}, A.~S. \& {Gail}, H.~P. 2006, \aap, 447, 553

\bibitem[{{Finkelstein} {et~al.}(2015){Finkelstein}, {Ryan}, {Papovich},
  {Dickinson}, {Song}, {Somerville}, {Ferguson}, {Salmon}, {Giavalisco},
  {Koekemoer}, {Ashby}, {Behroozi}, {Castellano}, {Dunlop}, {Faber}, {Fazio},
  {Fontana}, {Grogin}, {Hathi}, {Jaacks}, {Kocevski}, {Livermore}, {McLure},
  {Merlin}, {Mobasher}, {Newman}, {Rafelski}, {Tilvi}, \&
  {Willner}}]{finkelstein2015}
{Finkelstein}, S.~L., {Ryan}, Russell~E., J., {Papovich}, C., {et~al.} 2015,
  \apj, 810, 71

\bibitem[{{Fixsen} {et~al.}(1998){Fixsen}, {Dwek}, {Mather}, {Bennett}, \&
  {Shafer}}]{fixsen1998}
{Fixsen}, D.~J., {Dwek}, E., {Mather}, J.~C., {Bennett}, C.~L., \& {Shafer},
  R.~A. 1998, \apj, 508, 123

\bibitem[{{Fontana} {et~al.}(2004){Fontana}, {Pozzetti}, {Donnarumma},
  {Renzini}, {Cimatti}, {Zamorani}, {Menci}, {Daddi}, {Giallongo}, {Mignoli},
  {Perna}, {Salimbeni}, {Saracco}, {Broadhurst}, {Cristiani}, {D'Odorico}, \&
  {Gilmozzi}}]{fontana2004}
{Fontana}, A., {Pozzetti}, L., {Donnarumma}, I., {et~al.} 2004, \aap, 424, 23

\bibitem[{{Fontana} {et~al.}(2006){Fontana}, {Salimbeni}, {Grazian},
  {Giallongo}, {Pentericci}, {Nonino}, {Fontanot}, {Menci}, {Monaco},
  {Cristiani}, {Vanzella}, {de Santis}, \& {Gallozzi}}]{fontana2006}
{Fontana}, A., {Salimbeni}, S., {Grazian}, A., {et~al.} 2006, \aap, 459, 745

\bibitem[{{Frayer} {et~al.}(2006){Frayer}, {Huynh}, {Chary}, {Dickinson},
  {Elbaz}, {Fadda}, {Surace}, {Teplitz}, {Yan}, \& {Mobasher}}]{frayer2006}
{Frayer}, D.~T., {Huynh}, M.~T., {Chary}, R., {et~al.} 2006, \apjl, 647, L9

\bibitem[{{Freundlich} {et~al.}(2019){Freundlich}, {Combes}, {Tacconi},
  {Genzel}, {Garcia-Burillo}, {Neri}, {Contini}, {Bolatto}, {Lilly},
  {Salom{\'e}}, {Bicalho}, {Boissier}, {Boone}, {Bouch{\'e}}, {Bournaud},
  {Burkert}, {Carollo}, {Cooper}, {Cox}, {Feruglio}, {F{\"o}rster Schreiber},
  {Juneau}, {Lippa}, {Lutz}, {Naab}, {Renzini}, {Saintonge}, {Sternberg},
  {Walter}, {Weiner}, {Wei{\ss}}, \& {Wuyts}}]{freundlich2019}
{Freundlich}, J., {Combes}, F., {Tacconi}, L.~J., {et~al.} 2019, \aap, 622,
  A105

\bibitem[{{Fritz} {et~al.}(2006){Fritz}, {Franceschini}, \&
  {Hatziminaoglou}}]{fritz2006}
{Fritz}, J., {Franceschini}, A., \& {Hatziminaoglou}, E. 2006, \mnras, 366, 767

\bibitem[{{Fujimoto} {et~al.}(2022){Fujimoto}, {Brammer}, {Watson}, {Magdis},
  {Kokorev}, {Greve}, {Toft}, {Walter}, {Valiante}, {Ginolfi}, {Schneider},
  {Valentino}, {Colina}, {Vestergaard}, {Marques-Chaves}, {Fynbo}, {Krips},
  {Steinhardt}, {Cortzen}, {Rizzo}, \& {Oesch}}]{fujimoto2022}
{Fujimoto}, S., {Brammer}, G.~B., {Watson}, D., {et~al.} 2022, \nat, 604, 261

\bibitem[{{Fujimoto} {et~al.}(2017){Fujimoto}, {Ouchi}, {Shibuya}, \&
  {Nagai}}]{fujimoto2017}
{Fujimoto}, S., {Ouchi}, M., {Shibuya}, T., \& {Nagai}, H. 2017, \apj, 850, 83

\bibitem[{{Fukugita}(2011)}]{fukugita2011}
{Fukugita}, M. 2011, arXiv e-prints, arXiv:1103.4191

\bibitem[{{Geach} {et~al.}(2011){Geach}, {Smail}, {Moran}, {MacArthur},
  {Lagos}, \& {Edge}}]{geach2011}
{Geach}, J.~E., {Smail}, I., {Moran}, S.~M., {et~al.} 2011, \apjl, 730, L19

\bibitem[{{Gehrz}(1989)}]{gehrz1989}
{Gehrz}, R. 1989, in IAU Symposium, Vol. 135, Interstellar Dust, ed. L.~J.
  {Allamandola} \& A.~G.~G.~M. {Tielens}, 445

\bibitem[{{Genzel} {et~al.}(2003){Genzel}, {Baker}, {Tacconi}, {Lutz}, {Cox},
  {Guilloteau}, \& {Omont}}]{genzel2003}
{Genzel}, R., {Baker}, A.~J., {Tacconi}, L.~J., {et~al.} 2003, \apj, 584, 633

\bibitem[{{Genzel} {et~al.}(2015){Genzel}, {Tacconi}, {Lutz}, {Saintonge},
  {Berta}, {Magnelli}, {Combes}, {Garc{\'\i}a-Burillo}, {Neri}, {Bolatto},
  {Contini}, {Lilly}, {Boissier}, {Boone}, {Bouch{\'e}}, {Bournaud}, {Burkert},
  {Carollo}, {Colina}, {Cooper}, {Cox}, {Feruglio}, {F{\"o}rster Schreiber},
  {Freundlich}, {Gracia-Carpio}, {Juneau}, {Kovac}, {Lippa}, {Naab}, {Salome},
  {Renzini}, {Sternberg}, {Walter}, {Weiner}, {Weiss}, \& {Wuyts}}]{genzel2015}
{Genzel}, R., {Tacconi}, L.~J., {Lutz}, D., {et~al.} 2015, \apj, 800, 20

\bibitem[{{George} {et~al.}(2013){George}, {Ivison}, {Hopwood}, {Riechers},
  {Bussmann}, {Cox}, {Dye}, {Krips}, {Negrello}, {Neri}, {Serjeant},
  {Valtchanov}, {Baes}, {Bourne}, {Clements}, {de Zotti}, {Dunne}, {Eales},
  {Ibar}, {Maddox}, {Smith}, {Valiante}, \& {van der Werf}}]{george2013}
{George}, R.~D., {Ivison}, R.~J., {Hopwood}, R., {et~al.} 2013, \mnras, 436,
  L99

\bibitem[{{Gioannini} {et~al.}(2017){Gioannini}, {Matteucci}, \&
  {Calura}}]{gioannini2017}
{Gioannini}, L., {Matteucci}, F., \& {Calura}, F. 2017, \mnras, 471, 4615

\bibitem[{{Granato} {et~al.}(2021){Granato}, {Ragone-Figueroa}, {Taverna},
  {Silva}, {Valentini}, {Borgani}, {Monaco}, {Murante}, \&
  {Tornatore}}]{granato2021}
{Granato}, G.~L., {Ragone-Figueroa}, C., {Taverna}, A., {et~al.} 2021, \mnras,
  503, 511

\bibitem[{{Greve} {et~al.}(2008){Greve}, {Pope}, {Scott}, {Ivison}, {Borys},
  {Conselice}, \& {Bertoldi}}]{greve2008}
{Greve}, T.~R., {Pope}, A., {Scott}, D., {et~al.} 2008, \mnras, 389, 1489

\bibitem[{{Griffin} {et~al.}(2010){Griffin}, {Abergel}, {Abreu}, {Ade},
  {Andr{\'e}}, {Augueres}, {Babbedge}, {Bae}, {Baillie}, {Baluteau}, {Barlow},
  {Bendo}, {Benielli}, {Bock}, {Bonhomme}, {Brisbin}, {Brockley-Blatt},
  {Caldwell}, {Cara}, {Castro-Rodriguez}, {Cerulli}, {Chanial}, {Chen},
  {Clark}, {Clements}, {Clerc}, {Coker}, {Communal}, {Conversi}, {Cox},
  {Crumb}, {Cunningham}, {Daly}, {Davis}, {de Antoni}, {Delderfield}, {Devin},
  {di Giorgio}, {Didschuns}, {Dohlen}, {Donati}, {Dowell}, {Dowell}, {Duband},
  {Dumaye}, {Emery}, {Ferlet}, {Ferrand}, {Fontignie}, {Fox}, {Franceschini},
  {Frerking}, {Fulton}, {Garcia}, {Gastaud}, {Gear}, {Glenn}, {Goizel},
  {Griffin}, {Grundy}, {Guest}, {Guillemet}, {Hargrave}, {Harwit}, {Hastings},
  {Hatziminaoglou}, {Herman}, {Hinde}, {Hristov}, {Huang}, {Imhof}, {Isaak},
  {Israelsson}, {Ivison}, {Jennings}, {Kiernan}, {King}, {Lange}, {Latter},
  {Laurent}, {Laurent}, {Leeks}, {Lellouch}, {Levenson}, {Li}, {Li},
  {Lilienthal}, {Lim}, {Liu}, {Lu}, {Madden}, {Mainetti}, {Marliani}, {McKay},
  {Mercier}, {Molinari}, {Morris}, {Moseley}, {Mulder}, {Mur}, {Naylor},
  {Nguyen}, {O'Halloran}, {Oliver}, {Olofsson}, {Olofsson}, {Orfei}, {Page},
  {Pain}, {Panuzzo}, {Papageorgiou}, {Parks}, {Parr-Burman}, {Pearce},
  {Pearson}, {P{\'e}rez-Fournon}, {Pinsard}, {Pisano}, {Podosek}, {Pohlen},
  {Polehampton}, {Pouliquen}, {Rigopoulou}, {Rizzo}, {Roseboom}, {Roussel},
  {Rowan-Robinson}, {Rownd}, {Saraceno}, {Sauvage}, {Savage}, {Savini},
  {Sawyer}, {Scharmberg}, {Schmitt}, {Schneider}, {Schulz}, {Schwartz},
  {Shafer}, {Shupe}, {Sibthorpe}, {Sidher}, {Smith}, {Smith}, {Smith},
  {Spencer}, {Stobie}, {Sudiwala}, {Sukhatme}, {Surace}, {Stevens}, {Swinyard},
  {Trichas}, {Tourette}, {Triou}, {Tseng}, {Tucker}, {Turner}, {Vaccari},
  {Valtchanov}, {Vigroux}, {Virique}, {Voellmer}, {Walker}, {Ward}, {Waskett},
  {Weilert}, {Wesson}, {White}, {Whitehouse}, {Wilson}, {Winter}, {Woodcraft},
  {Wright}, {Xu}, {Zavagno}, {Zemcov}, {Zhang}, \& {Zonca}}]{griffin2010}
{Griffin}, M.~J., {Abergel}, A., {Abreu}, A., {et~al.} 2010, \aap, 518, L3

\bibitem[{{Guidetti} {et~al.}(2017){Guidetti}, {Bondi}, {Prandoni}, {Muxlow},
  {Beswick}, {Wrigley}, {Smail}, {McHardy}, {Thomson}, {Radcliffe}, \&
  {Argo}}]{guidetti2017}
{Guidetti}, D., {Bondi}, M., {Prandoni}, I., {et~al.} 2017, \mnras, 471, 210

\bibitem[{{Hagimoto} {et~al.}(2023){Hagimoto}, {Bakx}, {Serjeant}, {Bendo},
  {Urquhart}, {Eales}, {Harrington}, {Tamura}, {Umehata}, {Berta}, {Cooray},
  {Cox}, {De Zotti}, {Lehnert}, {Riechers}, {Scott}, {Temi}, {van der Werf},
  {Yang}, {Amvrosiadis}, {Andreani}, {Baker}, {Beelen}, {Borsato}, {Buat},
  {Butler}, {Dannerbauer}, {Dunne}, {Dye}, {Enia}, {Fan}, {Gavazzi},
  {Gonz{\'a}lez-Nuevo}, {Harris}, {Herrera}, {Hughes}, {Ismail}, {Ivison},
  {Jones}, {Kohno}, {Krips}, {Lagache}, {Marchetti}, {Massardi}, {Messias},
  {Negrello}, {Neri}, {Omont}, {Perez-Fournon}, {Sedgwick}, {Smith}, {Stanley},
  {Verma}, {Vlahakis}, {Ward}, {Weiner}, {Wei{\ss}}, \& {Young}}]{hagimoto2023}
{Hagimoto}, M., {Bakx}, T.~J.~L.~C., {Serjeant}, S., {et~al.} 2023, \mnras,
  521, 5508

\bibitem[{{Harris} {et~al.}(2012){Harris}, {Baker}, {Frayer}, {Smail},
  {Swinbank}, {Riechers}, {van der Werf}, {Auld}, {Baes}, {Bussmann},
  {Buttiglione}, {Cava}, {Clements}, {Cooray}, {Dannerbauer}, {Dariush}, {De
  Zotti}, {Dunne}, {Dye}, {Eales}, {Fritz}, {Gonz{\'a}lez-Nuevo}, {Hopwood},
  {Ibar}, {Ivison}, {Jarvis}, {Maddox}, {Negrello}, {Rigby}, {Smith}, {Temi},
  \& {Wardlow}}]{harris2012}
{Harris}, A.~I., {Baker}, A.~J., {Frayer}, D.~T., {et~al.} 2012, \apj, 752, 152

\bibitem[{{Harris} {et~al.}(2010){Harris}, {Baker}, {Zonak}, {Sharon},
  {Genzel}, {Rauch}, {Watts}, \& {Creager}}]{harris2010}
{Harris}, A.~I., {Baker}, A.~J., {Zonak}, S.~G., {et~al.} 2010, \apj, 723, 1139

\bibitem[{{Henriques} {et~al.}(2020){Henriques}, {Yates}, {Fu}, {Guo},
  {Kauffmann}, {Srisawat}, {Thomas}, \& {White}}]{henriques2020}
{Henriques}, B. M.~B., {Yates}, R.~M., {Fu}, J., {et~al.} 2020, \mnras, 491,
  5795

\bibitem[{{Hezaveh} {et~al.}(2013){Hezaveh}, {Marrone}, {Fassnacht}, {Spilker},
  {Vieira}, {Aguirre}, {Aird}, {Aravena}, {Ashby}, {Bayliss}, {Benson},
  {Bleem}, {Bothwell}, {Brodwin}, {Carlstrom}, {Chang}, {Chapman}, {Crawford},
  {Crites}, {De Breuck}, {de Haan}, {Dobbs}, {Fomalont}, {George}, {Gladders},
  {Gonzalez}, {Greve}, {Halverson}, {High}, {Holder}, {Holzapfel}, {Hoover},
  {Hrubes}, {Husband}, {Hunter}, {Keisler}, {Lee}, {Leitch}, {Lueker},
  {Luong-Van}, {Malkan}, {McIntyre}, {McMahon}, {Mehl}, {Menten}, {Meyer},
  {Mocanu}, {Murphy}, {Natoli}, {Padin}, {Plagge}, {Reichardt}, {Rest}, {Ruel},
  {Ruhl}, {Sharon}, {Schaffer}, {Shaw}, {Shirokoff}, {Stalder}, {Staniszewski},
  {Stark}, {Story}, {Vanderlinde}, {Wei{\ss}}, {Welikala}, \&
  {Williamson}}]{hezaveh2013}
{Hezaveh}, Y.~D., {Marrone}, D.~P., {Fassnacht}, C.~D., {et~al.} 2013, \apj,
  767, 132

\bibitem[{{Hollenbach} \& {Salpeter}(1971)}]{hollenbach1971}
{Hollenbach}, D. \& {Salpeter}, E.~E. 1971, \apj, 163, 155

\bibitem[{{Hughes} {et~al.}(1998){Hughes}, {Serjeant}, {Dunlop},
  {Rowan-Robinson}, {Blain}, {Mann}, {Ivison}, {Peacock}, {Efstathiou}, {Gear},
  {Oliver}, {Lawrence}, {Longair}, {Goldschmidt}, \& {Jenness}}]{hughes1998}
{Hughes}, D.~H., {Serjeant}, S., {Dunlop}, J., {et~al.} 1998, \nat, 394, 241

\bibitem[{{Inoue} {et~al.}(2014){Inoue}, {Shimizu}, {Iwata}, \&
  {Tanaka}}]{inoue2014}
{Inoue}, A.~K., {Shimizu}, I., {Iwata}, I., \& {Tanaka}, M. 2014, \mnras, 442,
  1805

\bibitem[{{Ismail} {et~al.}(2023){Ismail}, {Beelen}, {Buat}, {Berta}, {Cox},
  {Stanley}, {Young}, {Jin}, {Neri}, {Bakx}, {Dannerbauer}, {Butler}, {Cooray},
  {Nanni}, {Omont}, {Serjeant}, {van der Werf}, {Vlahakis}, {Wei{\ss}}, {Yang},
  {Baker}, {Bendo}, {Borsato}, {Chartab}, {Dye}, {Eales}, {Gavazzi}, {Hughes},
  {Ivison}, {Jones}, {Krips}, {Lehnert}, {Marchetti}, {Messias}, {Negrello},
  {Perez-Fournon}, {Riechers}, \& {Urquhart}}]{ismail2023}
{Ismail}, D., {Beelen}, A., {Buat}, V., {et~al.} 2023, \aap, 678, A27

\bibitem[{{Ivison} {et~al.}(2007){Ivison}, {Greve}, {Dunlop}, {Peacock},
  {Egami}, {Smail}, {Ibar}, {van Kampen}, {Aretxaga}, {Babbedge}, {Biggs},
  {Blain}, {Chapman}, {Clements}, {Coppin}, {Farrah}, {Halpern}, {Hughes},
  {Jarvis}, {Jenness}, {Jones}, {Mortier}, {Oliver}, {Papovich},
  {P{\'e}rez-Gonz{\'a}lez}, {Pope}, {Rawlings}, {Rieke}, {Rowan-Robinson},
  {Savage}, {Scott}, {Seigar}, {Serjeant}, {Simpson}, {Stevens}, {Vaccari},
  {Wagg}, \& {Willott}}]{ivison2007}
{Ivison}, R.~J., {Greve}, T.~R., {Dunlop}, J.~S., {et~al.} 2007, \mnras, 380,
  199

\bibitem[{{Ivison} {et~al.}(2002){Ivison}, {Greve}, {Smail}, {Dunlop}, {Roche},
  {Scott}, {Page}, {Stevens}, {Almaini}, {Blain}, {Willott}, {Fox}, {Gilbank},
  {Serjeant}, \& {Hughes}}]{ivison2002}
{Ivison}, R.~J., {Greve}, T.~R., {Smail}, I., {et~al.} 2002, \mnras, 337, 1

\bibitem[{{Ivison} {et~al.}(2011){Ivison}, {Papadopoulos}, {Smail}, {Greve},
  {Thomson}, {Xilouris}, \& {Chapman}}]{ivison2011}
{Ivison}, R.~J., {Papadopoulos}, P.~P., {Smail}, I., {et~al.} 2011, \mnras,
  412, 1913

\bibitem[{{Ivison} {et~al.}(2013){Ivison}, {Swinbank}, {Smail}, {Harris},
  {Bussmann}, {Cooray}, {Cox}, {Fu}, {Kov{\'a}cs}, {Krips}, {Narayanan},
  {Negrello}, {Neri}, {Pe{\~n}arrubia}, {Richard}, {Riechers}, {Rowlands},
  {Staguhn}, {Targett}, {Amber}, {Baker}, {Bourne}, {Bertoldi}, {Bremer},
  {Calanog}, {Clements}, {Dannerbauer}, {Dariush}, {De Zotti}, {Dunne},
  {Eales}, {Farrah}, {Fleuren}, {Franceschini}, {Geach}, {George}, {Helly},
  {Hopwood}, {Ibar}, {Jarvis}, {Kneib}, {Maddox}, {Omont}, {Scott}, {Serjeant},
  {Smith}, {Thompson}, {Valiante}, {Valtchanov}, {Vieira}, \& {van der
  Werf}}]{ivison2013}
{Ivison}, R.~J., {Swinbank}, A.~M., {Smail}, I., {et~al.} 2013, \apj, 772, 137

\bibitem[{{Ivison} {et~al.}(2010){Ivison}, {Swinbank}, {Swinyard}, {Smail},
  {Pearson}, {Rigopoulou}, {Polehampton}, {Baluteau}, {Barlow}, {Blain},
  {Bock}, {Clements}, {Coppin}, {Cooray}, {Danielson}, {Dwek}, {Edge},
  {Franceschini}, {Fulton}, {Glenn}, {Griffin}, {Isaak}, {Leeks}, {Lim},
  {Naylor}, {Oliver}, {Page}, {P{\'e}rez Fournon}, {Rowan-Robinson}, {Savini},
  {Scott}, {Spencer}, {Valtchanov}, {Vigroux}, \& {Wright}}]{ivison2010}
{Ivison}, R.~J., {Swinbank}, A.~M., {Swinyard}, B., {et~al.} 2010, \aap, 518,
  L35

\bibitem[{{Jin} {et~al.}(2022){Jin}, {Daddi}, {Magdis}, {Liu}, {Weaver}, {Tan},
  {Valentino}, {Gao}, {Schinnerer}, {Calabr{\`o}}, {Gu}, \& {Sese}}]{jin2022}
{Jin}, S., {Daddi}, E., {Magdis}, G.~E., {et~al.} 2022, \aap, 665, A3

\bibitem[{{Jones} {et~al.}(1994){Jones}, {Tielens}, {Hollenbach}, \&
  {McKee}}]{jones1994}
{Jones}, A.~P., {Tielens}, A.~G.~G.~M., {Hollenbach}, D.~J., \& {McKee}, C.~F.
  1994, \apj, 433, 797

\bibitem[{{Kocevski} {et~al.}(2024){Kocevski}, {Finkelstein}, {Barro},
  {Taylor}, {Calabr{\`o}}, {Laloux}, {Buchner}, {Trump}, {Leung}, {Yang},
  {Dickinson}, {P{\'e}rez-Gonz{\'a}lez}, {Pacucci}, {Inayoshi}, {Somerville},
  {McGrath}, {Akins}, {Bagley}, {Bisigello}, {Bowler}, {Carnall}, {Casey},
  {Cheng}, {Cleri}, {Costantin}, {Cullen}, {Davis}, {Donnan}, {Dunlop},
  {Ellis}, {Ferguson}, {Fujimoto}, {Fontana}, {Giavalisco}, {Grazian},
  {Grogin}, {Hathi}, {Hirschmann}, {Huertas-Company}, {Holwerda},
  {Illingworth}, {Juneau}, {Kartaltepe}, {Koekemoer}, {Li}, {Lucas}, {Magee},
  {Mason}, {McLeod}, {McLure}, {Napolitano}, {Papovich}, {Pirzkal},
  {Rodighiero}, {Santini}, {Wilkins}, \& {Yung}}]{kocevski2024}
{Kocevski}, D.~D., {Finkelstein}, S.~L., {Barro}, G., {et~al.} 2024, arXiv
  e-prints, arXiv:2404.03576, {accepted or publication in ApJ}

\bibitem[{{Kodra} {et~al.}(2023){Kodra}, {Andrews}, {Newman}, {Finkelstein},
  {Fontana}, {Hathi}, {Salvato}, {Wiklind}, {Wuyts}, {Broussard}, {Chartab},
  {Conselice}, {Cooper}, {Dekel}, {Dickinson}, {Ferguson}, {Gawiser}, {Grogin},
  {Iyer}, {Kartaltepe}, {Kassin}, {Koekemoer}, {Koo}, {Lucas}, {Mantha},
  {McIntosh}, {Mobasher}, {Pacifici}, {P{\'e}rez-Gonz{\'a}lez}, \&
  {Santini}}]{kodra2023}
{Kodra}, D., {Andrews}, B.~H., {Newman}, J.~A., {et~al.} 2023, \apj, 942, 36

\bibitem[{{Labbe} {et~al.}(2025){Labbe}, {Greene}, {Bezanson}, {Fujimoto},
  {Furtak}, {Goulding}, {Matthee}, {Naidu}, {Oesch}, {Atek}, {Brammer},
  {Chemerynska}, {Coe}, {Cutler}, {Dayal}, {Feldmann}, {Franx}, {Glazebrook},
  {Leja}, {Maseda}, {Marchesini}, {Nanayakkara}, {Nelson}, {Pan}, {Papovich},
  {Price}, {Suess}, {Wang}, {Weaver}, {Whitaker}, {Williams}, \&
  {Zitrin}}]{labbe2023}
{Labbe}, I., {Greene}, J.~E., {Bezanson}, R., {et~al.} 2025, \apj, 978, 92

\bibitem[{{Lagache} {et~al.}(2005){Lagache}, {Puget}, \& {Dole}}]{lagache2005}
{Lagache}, G., {Puget}, J.-L., \& {Dole}, H. 2005, \araa, 43, 727

\bibitem[{{Le F{\`e}vre} {et~al.}(2020){Le F{\`e}vre}, {B{\'e}thermin},
  {Faisst}, {Jones}, {Capak}, {Cassata}, {Silverman}, {Schaerer}, {Yan},
  {Amorin}, {Bardelli}, {Boquien}, {Cimatti}, {Dessauges-Zavadsky},
  {Giavalisco}, {Hathi}, {Fudamoto}, {Fujimoto}, {Ginolfi}, {Gruppioni},
  {Hemmati}, {Ibar}, {Koekemoer}, {Khusanova}, {Lagache}, {Lemaux}, {Loiacono},
  {Maiolino}, {Mancini}, {Narayanan}, {Morselli}, {M{\'e}ndez-Hern{\`a}ndez},
  {Oesch}, {Pozzi}, {Romano}, {Riechers}, {Scoville}, {Talia}, {Tasca},
  {Thomas}, {Toft}, {Vallini}, {Vergani}, {Walter}, {Zamorani}, \&
  {Zucca}}]{lefevre2020}
{Le F{\`e}vre}, O., {B{\'e}thermin}, M., {Faisst}, A., {et~al.} 2020, \aap,
  643, A1

\bibitem[{{Lewis} {et~al.}(2023){Lewis}, {Ocvirk}, {Dubois}, {Aubert},
  {Chardin}, {Gillet}, \& {Th{\'e}lie}}]{lewis2023}
{Lewis}, J. S.~W., {Ocvirk}, P., {Dubois}, Y., {et~al.} 2023, \mnras, 519, 5987

\bibitem[{{Li} \& {Draine}(2001)}]{LD01}
{Li}, A. \& {Draine}, B.~T. 2001, \apj, 554, 778

\bibitem[{{Li} {et~al.}(2019){Li}, {Narayanan}, \& {Dav{\'e}}}]{li2019}
{Li}, Q., {Narayanan}, D., \& {Dav{\'e}}, R. 2019, \mnras, 490, 1425

\bibitem[{{Lim} {et~al.}(2020){Lim}, {Wang}, {Smail}, {Scott}, {Chen}, {Chang},
  {Simpson}, {Toba}, {Shu}, {Clements}, {Greenslade}, {Ao}, {Babul}, {Birkin},
  {Chapman}, {Cheng}, {Cho}, {Dannerbauer}, {Dudzevi{\v{c}}i{\={u}}t{\.{e}}},
  {Dunlop}, {Gao}, {Goto}, {Ho}, {Hsu}, {Hwang}, {Jeong}, {Koprowski}, {Lee},
  {Lin}, {Lin}, {Micha{\l}owski}, {Parsons}, {Sawicki}, {Shirley}, {Shim},
  {Urquhart}, {Wang}, \& {Wang}}]{lim2020}
{Lim}, C.-F., {Wang}, W.-H., {Smail}, I., {et~al.} 2020, \apj, 889, 80

\bibitem[{{Liske} {et~al.}(2015){Liske}, {Baldry}, {Driver}, {Tuffs},
  {Alpaslan}, {Andrae}, {Brough}, {Cluver}, {Grootes}, {Gunawardhana},
  {Kelvin}, {Loveday}, {Robotham}, {Taylor}, {Bamford}, {Bland-Hawthorn},
  {Brown}, {Drinkwater}, {Hopkins}, {Meyer}, {Norberg}, {Peacock}, {Agius},
  {Andrews}, {Bauer}, {Ching}, {Colless}, {Conselice}, {Croom}, {Davies}, {De
  Propris}, {Dunne}, {Eardley}, {Ellis}, {Foster}, {Frenk}, {H{\"a}u{\ss}ler},
  {Holwerda}, {Howlett}, {Ibarra}, {Jarvis}, {Jones}, {Kafle}, {Lacey},
  {Lange}, {Lara-L{\'o}pez}, {L{\'o}pez-S{\'a}nchez}, {Maddox}, {Madore},
  {McNaught-Roberts}, {Moffett}, {Nichol}, {Owers}, {Palamara}, {Penny},
  {Phillipps}, {Pimbblet}, {Popescu}, {Prescott}, {Proctor}, {Sadler},
  {Sansom}, {Seibert}, {Sharp}, {Sutherland}, {V{\'a}zquez-Mata}, {van Kampen},
  {Wilkins}, {Williams}, \& {Wright}}]{liske2015}
{Liske}, J., {Baldry}, I.~K., {Driver}, S.~P., {et~al.} 2015, \mnras, 452, 2087

\bibitem[{{Liu} {et~al.}(2018){Liu}, {Daddi}, {Dickinson}, {Owen}, {Pannella},
  {Sargent}, {B{\'e}thermin}, {Magdis}, {Gao}, {Shu}, {Wang}, {Jin}, \&
  {Inami}}]{liu2018}
{Liu}, D., {Daddi}, E., {Dickinson}, M., {et~al.} 2018, \apj, 853, 172

\bibitem[{{Liu} {et~al.}(2019){Liu}, {Schinnerer}, {Groves}, {Magnelli},
  {Lang}, {Leslie}, {Jim{\'e}nez-Andrade}, {Riechers}, {Popping}, {Magdis},
  {Daddi}, {Sargent}, {Gao}, {Fudamoto}, {Oesch}, \& {Bertoldi}}]{liu2019b}
{Liu}, D., {Schinnerer}, E., {Groves}, B., {et~al.} 2019, \apj, 887, 235

\bibitem[{{Lutz}(2014)}]{lutz2014}
{Lutz}, D. 2014, \araa, 52, 373

\bibitem[{{Lutz} {et~al.}(2011){Lutz}, {Poglitsch}, {Altieri}, {Andreani},
  {Aussel}, {Berta}, {Bongiovanni}, {Brisbin}, {Cava}, {Cepa}, {Cimatti},
  {Daddi}, {Dominguez-Sanchez}, {Elbaz}, {F{\"o}rster Schreiber}, {Genzel},
  {Grazian}, {Gruppioni}, {Harwit}, {Le Floc'h}, {Magdis}, {Magnelli},
  {Maiolino}, {Nordon}, {P{\'e}rez Garc{\'\i}a}, {Popesso}, {Pozzi},
  {Riguccini}, {Rodighiero}, {Saintonge}, {Sanchez Portal}, {Santini}, {Shao},
  {Sturm}, {Tacconi}, {Valtchanov}, {Wetzstein}, \& {Wieprecht}}]{lutz2011}
{Lutz}, D., {Poglitsch}, A., {Altieri}, B., {et~al.} 2011, \aap, 532, A90

\bibitem[{{Madau}(1995)}]{madau1995}
{Madau}, P. 1995, \apj, 441, 18

\bibitem[{{Magdis} {et~al.}(2012){Magdis}, {Daddi}, {B{\'e}thermin}, {Sargent},
  {Elbaz}, {Pannella}, {Dickinson}, {Dannerbauer}, {da Cunha}, {Walter},
  {Rigopoulou}, {Charmandaris}, {Hwang}, \& {Kartaltepe}}]{magdis2012}
{Magdis}, G.~E., {Daddi}, E., {B{\'e}thermin}, M., {et~al.} 2012, \apj, 760, 6

\bibitem[{{Magnelli} {et~al.}(2020){Magnelli}, {Boogaard}, {Decarli},
  {G{\'o}nzalez-L{\'o}pez}, {Novak}, {Popping}, {Smail}, {Walter}, {Aravena},
  {Assef}, {Bauer}, {Bertoldi}, {Carilli}, {Cortes}, {Cunha}, {Daddi},
  {D{\'\i}az-Santos}, {Inami}, {Ivison}, {F{\`e}vre}, {Oesch}, {Riechers},
  {Rix}, {Sargent}, {Werf}, {Wagg}, \& {Weiss}}]{magnelli2020}
{Magnelli}, B., {Boogaard}, L., {Decarli}, R., {et~al.} 2020, \apj, 892, 66

\bibitem[{{Magnelli} {et~al.}(2015){Magnelli}, {Ivison}, {Lutz}, {Valtchanov},
  {Farrah}, {Berta}, {Bertoldi}, {Bock}, {Cooray}, {Ibar}, {Karim}, {Le
  Floc'h}, {Nordon}, {Oliver}, {Page}, {Popesso}, {Pozzi}, {Rigopoulou},
  {Riguccini}, {Rodighiero}, {Rosario}, {Roseboom}, {Wang}, \&
  {Wuyts}}]{magnelli2015}
{Magnelli}, B., {Ivison}, R.~J., {Lutz}, D., {et~al.} 2015, \aap, 573, A45

\bibitem[{{Magnelli} {et~al.}(2012){Magnelli}, {Lutz}, {Santini}, {Saintonge},
  {Berta}, {Albrecht}, {Altieri}, {Andreani}, {Aussel}, {Bertoldi},
  {B{\'e}thermin}, {Bongiovanni}, {Capak}, {Chapman}, {Cepa}, {Cimatti},
  {Cooray}, {Daddi}, {Danielson}, {Dannerbauer}, {Dunlop}, {Elbaz}, {Farrah},
  {F{\"o}rster Schreiber}, {Genzel}, {Hwang}, {Ibar}, {Ivison}, {Le Floc'h},
  {Magdis}, {Maiolino}, {Nordon}, {Oliver}, {P{\'e}rez Garc{\'\i}a},
  {Poglitsch}, {Popesso}, {Pozzi}, {Riguccini}, {Rodighiero}, {Rosario},
  {Roseboom}, {Salvato}, {Sanchez-Portal}, {Scott}, {Smail}, {Sturm},
  {Swinbank}, {Tacconi}, {Valtchanov}, {Wang}, \& {Wuyts}}]{magnelli2012}
{Magnelli}, B., {Lutz}, D., {Santini}, P., {et~al.} 2012, \aap, 539, A155

\bibitem[{{Magnelli} {et~al.}(2013){Magnelli}, {Popesso}, {Berta}, {Pozzi},
  {Elbaz}, {Lutz}, {Dickinson}, {Altieri}, {Andreani}, {Aussel},
  {B{\'e}thermin}, {Bongiovanni}, {Cepa}, {Charmandaris}, {Chary}, {Cimatti},
  {Daddi}, {F{\"o}rster Schreiber}, {Genzel}, {Gruppioni}, {Harwit}, {Hwang},
  {Ivison}, {Magdis}, {Maiolino}, {Murphy}, {Nordon}, {Pannella}, {P{\'e}rez
  Garc{\'\i}a}, {Poglitsch}, {Rosario}, {Sanchez-Portal}, {Santini}, {Scott},
  {Sturm}, {Tacconi}, \& {Valtchanov}}]{magnelli2013}
{Magnelli}, B., {Popesso}, P., {Berta}, S., {et~al.} 2013, \aap, 553, A132

\bibitem[{{Mancini} {et~al.}(2009){Mancini}, {Matute}, {Cimatti}, {Daddi},
  {Dickinson}, {Rodighiero}, {Bolzonella}, \& {Pozzetti}}]{mancini2009}
{Mancini}, C., {Matute}, I., {Cimatti}, A., {et~al.} 2009, \aap, 500, 705

\bibitem[{{Maraston}(2005)}]{maraston2005}
{Maraston}, C. 2005, \mnras, 362, 799

\bibitem[{{Maseda} {et~al.}(2024){Maseda}, {de Graaff}, {Franx}, {Rix},
  {Carniani}, {Laseter}, {Dudzevi{\v{c}}i{\={u}}t{\.{e}}}, {Rawle}, {Parlanti},
  {Arribas}, {Bunker}, {Cameron}, {Charlot}, {Curti}, {D'Eugenio}, {Jones},
  {Kumari}, {Maiolino}, {{\"U}bler}, {Saxena}, {Smit}, {Willott}, \&
  {Witstok}}]{maseda2024}
{Maseda}, M.~V., {de Graaff}, A., {Franx}, M., {et~al.} 2024, \aap, 689, A73

\bibitem[{{Mathis}(1990)}]{mathis1990}
{Mathis}, J.~S. 1990, \araa, 28, 37

\bibitem[{{Matthee} {et~al.}(2024){Matthee}, {Naidu}, {Brammer}, {Chisholm},
  {Eilers}, {Goulding}, {Greene}, {Kashino}, {Labbe}, {Lilly}, {Mackenzie},
  {Oesch}, {Weibel}, {Wuyts}, {Xiao}, {Bordoloi}, {Bouwens}, {van Dokkum},
  {Illingworth}, {Kramarenko}, {Maseda}, {Mason}, {Meyer}, {Nelson}, {Reddy},
  {Shivaei}, {Simcoe}, \& {Yue}}]{matthee2024}
{Matthee}, J., {Naidu}, R.~P., {Brammer}, G., {et~al.} 2024, \apj, 963, 129

\bibitem[{{McKee} {et~al.}(1987){McKee}, {Hollenbach}, {Seab}, \&
  {Tielens}}]{mckee1987}
{McKee}, C.~F., {Hollenbach}, D.~J., {Seab}, G.~C., \& {Tielens}, A.~G.~G.~M.
  1987, \apj, 318, 674

\bibitem[{{M{\'e}nard} \& {Fukugita}(2012)}]{menard2012}
{M{\'e}nard}, B. \& {Fukugita}, M. 2012, \apj, 754, 116

\bibitem[{{Millard} {et~al.}(2020){Millard}, {Eales}, {Smith}, {Gomez},
  {Ma{\l}ek}, {Simpson}, {Peng}, {Sawicki}, {Beeston}, {Bunker}, {Ao}, {Babul},
  {Ho}, {Hwang}, {Micha{\l}owski}, {Scoville}, {Shim}, \& {Toba}}]{millard2020}
{Millard}, J.~S., {Eales}, S.~A., {Smith}, M.~W.~L., {et~al.} 2020, \mnras,
  494, 293

\bibitem[{{Momcheva} {et~al.}(2016){Momcheva}, {Brammer}, {van Dokkum},
  {Skelton}, {Whitaker}, {Nelson}, {Fumagalli}, {Maseda}, {Leja}, {Franx},
  {Rix}, {Bezanson}, {Da Cunha}, {Dickey}, {F{\"o}rster Schreiber},
  {Illingworth}, {Kriek}, {Labb{\'e}}, {Ulf Lange}, {Lundgren}, {Magee},
  {Marchesini}, {Oesch}, {Pacifici}, {Patel}, {Price}, {Tal}, {Wake}, {van der
  Wel}, \& {Wuyts}}]{momcheva2016}
{Momcheva}, I.~G., {Brammer}, G.~B., {van Dokkum}, P.~G., {et~al.} 2016, \apjs,
  225, 27

\bibitem[{{Monfardini} {et~al.}(2014){Monfardini}, {Adam}, {Adane}, {Ade},
  {Andr{\'e}}, {Beelen}, {Belier}, {Benoit}, {Bideaud}, {Billot}, {Bourrion},
  {Calvo}, {Catalano}, {Coiffard}, {Comis}, {D'Addabbo}, {D{\'e}sert}, {Doyle},
  {Goupy}, {Kramer}, {Leclercq}, {Macias-Perez}, {Martino}, {Mauskopf},
  {Mayet}, {Pajot}, {Pascale}, {Ponthieu}, {Rev{\'e}ret}, {Rodriguez},
  {Savini}, {Schuster}, {Sievers}, {Tucker}, \& {Zylka}}]{monfardini2014}
{Monfardini}, A., {Adam}, R., {Adane}, A., {et~al.} 2014, Journal of Low
  Temperature Physics, 176, 787

\bibitem[{{Nanni} {et~al.}(2013){Nanni}, {Bressan}, {Marigo}, \&
  {Girardi}}]{nanni2013}
{Nanni}, A., {Bressan}, A., {Marigo}, P., \& {Girardi}, L. 2013, \mnras, 434,
  2390

\bibitem[{{Neri} {et~al.}(2014){Neri}, {Downes}, {Cox}, \& {Walter}}]{neri2014}
{Neri}, R., {Downes}, D., {Cox}, P., \& {Walter}, F. 2014, \aap, 562, A35

\bibitem[{{Noll} {et~al.}(2009){Noll}, {Burgarella}, {Giovannoli}, {Buat},
  {Marcillac}, \& {Mu{\~n}oz-Mateos}}]{noll2009}
{Noll}, S., {Burgarella}, D., {Giovannoli}, E., {et~al.} 2009, \aap, 507, 1793

\bibitem[{{Nozawa} {et~al.}(2007){Nozawa}, {Kozasa}, {Habe}, {Dwek}, {Umeda},
  {Tominaga}, {Maeda}, \& {Nomoto}}]{nozawa2007}
{Nozawa}, T., {Kozasa}, T., {Habe}, A., {et~al.} 2007, \apj, 666, 955

\bibitem[{{Ocvirk} {et~al.}(2016){Ocvirk}, {Gillet}, {Shapiro}, {Aubert},
  {Iliev}, {Teyssier}, {Yepes}, {Choi}, {Sullivan}, {Knebe}, {Gottl{\"o}ber},
  {D'Aloisio}, {Park}, {Hoffman}, \& {Stranex}}]{ocvirk2016}
{Ocvirk}, P., {Gillet}, N., {Shapiro}, P.~R., {et~al.} 2016, \mnras, 463, 1462

\bibitem[{{Oesch} {et~al.}(2023){Oesch}, {Brammer}, {Naidu}, {Bouwens},
  {Chisholm}, {Illingworth}, {Matthee}, {Nelson}, {Qin}, {Reddy}, {Shapley},
  {Shivaei}, {van Dokkum}, {Weibel}, {Whitaker}, {Wuyts}, {Covelo-Paz},
  {Endsley}, {Fudamoto}, {Giovinazzo}, {Herard-Demanche}, {Kerutt},
  {Kramarenko}, {Labbe}, {Leonova}, {Lin}, {Magee}, {Marchesini}, {Maseda},
  {Mason}, {Matharu}, {Meyer}, {Neufeld}, {Prieto Lyon}, {Schaerer}, {Sharma},
  {Shuntov}, {Smit}, {Stefanon}, {Wyithe}, \& {Xiao}}]{oesch2023}
{Oesch}, P.~A., {Brammer}, G., {Naidu}, R.~P., {et~al.} 2023, \mnras, 525, 2864

\bibitem[{{Oliver} {et~al.}(2012){Oliver}, {Bock}, {Altieri}, {Amblard},
  {Arumugam}, {Aussel}, {Babbedge}, {Beelen}, {B{\'e}thermin}, {Blain},
  {Boselli}, {Bridge}, {Brisbin}, {Buat}, {Burgarella},
  {Castro-Rodr{\'\i}guez}, {Cava}, {Chanial}, {Cirasuolo}, {Clements},
  {Conley}, {Conversi}, {Cooray}, {Dowell}, {Dubois}, {Dwek}, {Dye}, {Eales},
  {Elbaz}, {Farrah}, {Feltre}, {Ferrero}, {Fiolet}, {Fox}, {Franceschini},
  {Gear}, {Giovannoli}, {Glenn}, {Gong}, {Gonz{\'a}lez Solares}, {Griffin},
  {Halpern}, {Harwit}, {Hatziminaoglou}, {Heinis}, {Hurley}, {Hwang}, {Hyde},
  {Ibar}, {Ilbert}, {Isaak}, {Ivison}, {Lagache}, {Le Floc'h}, {Levenson},
  {Faro}, {Lu}, {Madden}, {Maffei}, {Magdis}, {Mainetti}, {Marchetti},
  {Marsden}, {Marshall}, {Mortier}, {Nguyen}, {O'Halloran}, {Omont}, {Page},
  {Panuzzo}, {Papageorgiou}, {Patel}, {Pearson}, {P{\'e}rez-Fournon}, {Pohlen},
  {Rawlings}, {Raymond}, {Rigopoulou}, {Riguccini}, {Rizzo}, {Rodighiero},
  {Roseboom}, {Rowan-Robinson}, {S{\'a}nchez Portal}, {Schulz}, {Scott},
  {Seymour}, {Shupe}, {Smith}, {Stevens}, {Symeonidis}, {Trichas}, {Tugwell},
  {Vaccari}, {Valtchanov}, {Vieira}, {Viero}, {Vigroux}, {Wang}, {Ward},
  {Wardlow}, {Wright}, {Xu}, \& {Zemcov}}]{oliver2012}
{Oliver}, S.~J., {Bock}, J., {Altieri}, B., {et~al.} 2012, \mnras, 424, 1614

\bibitem[{{Ostriker} \& {Silk}(1973)}]{ostriker1973}
{Ostriker}, J. \& {Silk}, J. 1973, \apjl, 184, L113

\bibitem[{{Owen}(2018)}]{owen2018}
{Owen}, F.~N. 2018, \apjs, 235, 34

\bibitem[{{Parente} {et~al.}(2023){Parente}, {Ragone-Figueroa}, {Granato}, \&
  {Lapi}}]{parente2023}
{Parente}, M., {Ragone-Figueroa}, C., {Granato}, G.~L., \& {Lapi}, A. 2023,
  \mnras, 521, 6105

\bibitem[{{Pearson} {et~al.}(2010){Pearson}, {Serjeant}, {Negrello}, {Takagi},
  {Jeong}, {Matsuhara}, {Wada}, {Oyabu}, {Lee}, \& {Im}}]{pearson2010}
{Pearson}, C.~P., {Serjeant}, S., {Negrello}, M., {et~al.} 2010, \aap, 514, A9

\bibitem[{{Penner} {et~al.}(2011){Penner}, {Pope}, {Chapin}, {Greve},
  {Bertoldi}, {Brodwin}, {Chary}, {Conselice}, {Coppin}, {Giavalisco},
  {Hughes}, {Ivison}, {Perera}, {Scott}, {Scott}, \& {Wilson}}]{Penner2011}
{Penner}, K., {Pope}, A., {Chapin}, E.~L., {et~al.} 2011, \mnras, 410, 2749

\bibitem[{{Penney} {et~al.}(2020){Penney}, {Blain}, {Assef}, {Diaz-Santos},
  {Gonz{\'a}lez-L{\'o}pez}, {Tsai}, {Aravena}, {Eisenhardt}, {Jones}, {Jun},
  {Kim}, {Stern}, \& {Wu}}]{penney2020}
{Penney}, J.~I., {Blain}, A.~W., {Assef}, R.~J., {et~al.} 2020, \mnras, 496,
  1565

\bibitem[{{Perera} {et~al.}(2008){Perera}, {Chapin}, {Austermann}, {Scott},
  {Wilson}, {Halpern}, {Pope}, {Scott}, {Yun}, {Lowenthal}, {Morrison},
  {Aretxaga}, {Bock}, {Coppin}, {Crowe}, {Frey}, {Hughes}, {Kang}, {Kim}, \&
  {Mauskopf}}]{perera2008}
{Perera}, T.~A., {Chapin}, E.~L., {Austermann}, J.~E., {et~al.} 2008, \mnras,
  391, 1227

\bibitem[{{P{\'e}rez-Gonz{\'a}lez} {et~al.}(2024){P{\'e}rez-Gonz{\'a}lez},
  {Barro}, {Rieke}, {Lyu}, {Rieke}, {Alberts}, {Williams}, {Hainline}, {Sun},
  {Pusk{\'a}s}, {Annunziatella}, {Baker}, {Bunker}, {Egami}, {Ji}, {Johnson},
  {Robertson}, {Rodr{\'\i}guez Del Pino}, {Rujopakarn}, {Shivaei}, {Tacchella},
  {Willmer}, \& {Willott}}]{perezgonzalez2024}
{P{\'e}rez-Gonz{\'a}lez}, P.~G., {Barro}, G., {Rieke}, G.~H., {et~al.} 2024,
  \apj, 968, 4

\bibitem[{{P{\'e}rez-Gonz{\'a}lez} {et~al.}(2005){P{\'e}rez-Gonz{\'a}lez},
  {Rieke}, {Egami}, {Alonso-Herrero}, {Dole}, {Papovich}, {Blaylock}, {Jones},
  {Rieke}, {Rigby}, {Barmby}, {Fazio}, {Huang}, \& {Martin}}]{pg2005}
{P{\'e}rez-Gonz{\'a}lez}, P.~G., {Rieke}, G.~H., {Egami}, E., {et~al.} 2005,
  \apj, 630, 82

\bibitem[{{P{\'e}rez-Gonz{\'a}lez} {et~al.}(2008){P{\'e}rez-Gonz{\'a}lez},
  {Rieke}, {Villar}, {Barro}, {Blaylock}, {Egami}, {Gallego}, {Gil de Paz},
  {Pascual}, {Zamorano}, \& {Donley}}]{pg2008}
{P{\'e}rez-Gonz{\'a}lez}, P.~G., {Rieke}, G.~H., {Villar}, V., {et~al.} 2008,
  \apj, 675, 234

\bibitem[{{Perotto} {et~al.}(2020){Perotto}, {Ponthieu},
  {Mac{\'\i}as-P{\'e}rez}, {Adam}, {Ade}, {Andr{\'e}}, {Andrianasolo},
  {Aussel}, {Beelen}, {Beno{\^\i}t}, {Berta}, {Bideaud}, {Bourrion}, {Calvo},
  {Catalano}, {Comis}, {De Petris}, {D{\'e}sert}, {Doyle}, {Driessen},
  {Garc{\'\i}a}, {Gomez}, {Goupy}, {John}, {K{\'e}ruzor{\'e}}, {Kramer},
  {Ladjelate}, {Lagache}, {Leclercq}, {Lestrade}, {Maury}, {Mauskopf}, {Mayet},
  {Monfardini}, {Navarro}, {Pe{\~n}alver}, {Pierfederici}, {Pisano},
  {Rev{\'e}ret}, {Ritacco}, {Romero}, {Roussel}, {Ruppin}, {Schuster}, {Shu},
  {Sievers}, {Tucker}, \& {Zylka}}]{perotto2020}
{Perotto}, L., {Ponthieu}, N., {Mac{\'\i}as-P{\'e}rez}, J.~F., {et~al.} 2020,
  \aap, 637, A71

\bibitem[{{P{\'e}roux} \& {Howk}(2020)}]{peroux2020}
{P{\'e}roux}, C. \& {Howk}, J.~C. 2020, \araa, 58, 363

\bibitem[{{Pilbratt} {et~al.}(2010){Pilbratt}, {Riedinger}, {Passvogel},
  {Crone}, {Doyle}, {Gageur}, {Heras}, {Jewell}, {Metcalfe}, {Ott}, \&
  {Schmidt}}]{pilbratt2010}
{Pilbratt}, G.~L., {Riedinger}, J.~R., {Passvogel}, T., {et~al.} 2010, \aap,
  518, L1

\bibitem[{{Planck Collaboration} {et~al.}(2020){Planck Collaboration},
  {Aghanim}, {Akrami}, {Ashdown}, {Aumont}, {Baccigalupi}, {Ballardini},
  {Banday}, {Barreiro}, {Bartolo}, {Basak}, {Battye}, {Benabed}, {Bernard},
  {Bersanelli}, {Bielewicz}, {Bock}, {Bond}, {Borrill}, {Bouchet}, {Boulanger},
  {Bucher}, {Burigana}, {Butler}, {Calabrese}, {Cardoso}, {Carron},
  {Challinor}, {Chiang}, {Chluba}, {Colombo}, {Combet}, {Contreras}, {Crill},
  {Cuttaia}, {de Bernardis}, {de Zotti}, {Delabrouille}, {Delouis}, {Di
  Valentino}, {Diego}, {Dor{\'e}}, {Douspis}, {Ducout}, {Dupac}, {Dusini},
  {Efstathiou}, {Elsner}, {En{\ss}lin}, {Eriksen}, {Fantaye}, {Farhang},
  {Fergusson}, {Fernandez-Cobos}, {Finelli}, {Forastieri}, {Frailis},
  {Fraisse}, {Franceschi}, {Frolov}, {Galeotta}, {Galli}, {Ganga},
  {G{\'e}nova-Santos}, {Gerbino}, {Ghosh}, {Gonz{\'a}lez-Nuevo}, {G{\'o}rski},
  {Gratton}, {Gruppuso}, {Gudmundsson}, {Hamann}, {Handley}, {Hansen},
  {Herranz}, {Hildebrandt}, {Hivon}, {Huang}, {Jaffe}, {Jones}, {Karakci},
  {Keih{\"a}nen}, {Keskitalo}, {Kiiveri}, {Kim}, {Kisner}, {Knox},
  {Krachmalnicoff}, {Kunz}, {Kurki-Suonio}, {Lagache}, {Lamarre}, {Lasenby},
  {Lattanzi}, {Lawrence}, {Le Jeune}, {Lemos}, {Lesgourgues}, {Levrier},
  {Lewis}, {Liguori}, {Lilje}, {Lilley}, {Lindholm}, {L{\'o}pez-Caniego},
  {Lubin}, {Ma}, {Mac{\'\i}as-P{\'e}rez}, {Maggio}, {Maino}, {Mandolesi},
  {Mangilli}, {Marcos-Caballero}, {Maris}, {Martin}, {Martinelli},
  {Mart{\'\i}nez-Gonz{\'a}lez}, {Matarrese}, {Mauri}, {McEwen}, {Meinhold},
  {Melchiorri}, {Mennella}, {Migliaccio}, {Millea}, {Mitra},
  {Miville-Desch{\^e}nes}, {Molinari}, {Montier}, {Morgante}, {Moss}, {Natoli},
  {N{\o}rgaard-Nielsen}, {Pagano}, {Paoletti}, {Partridge}, {Patanchon},
  {Peiris}, {Perrotta}, {Pettorino}, {Piacentini}, {Polastri}, {Polenta},
  {Puget}, {Rachen}, {Reinecke}, {Remazeilles}, {Renzi}, {Rocha}, {Rosset},
  {Roudier}, {Rubi{\~n}o-Mart{\'\i}n}, {Ruiz-Granados}, {Salvati}, {Sandri},
  {Savelainen}, {Scott}, {Shellard}, {Sirignano}, {Sirri}, {Spencer},
  {Sunyaev}, {Suur-Uski}, {Tauber}, {Tavagnacco}, {Tenti}, {Toffolatti},
  {Tomasi}, {Trombetti}, {Valenziano}, {Valiviita}, {Van Tent}, {Vibert},
  {Vielva}, {Villa}, {Vittorio}, {Wandelt}, {Wehus}, {White}, {White},
  {Zacchei}, \& {Zonca}}]{planck2020_cosmo}
{Planck Collaboration}, {Aghanim}, N., {Akrami}, Y., {et~al.} 2020, \aap, 641,
  A6

\bibitem[{{Poglitsch} {et~al.}(2010){Poglitsch}, {Waelkens}, {Geis},
  {Feuchtgruber}, {Vandenbussche}, {Rodriguez}, {Krause}, {Renotte}, {van
  Hoof}, {Saraceno}, {Cepa}, {Kerschbaum}, {Agn{\`e}se}, {Ali}, {Altieri},
  {Andreani}, {Augueres}, {Balog}, {Barl}, {Bauer}, {Belbachir}, {Benedettini},
  {Billot}, {Boulade}, {Bischof}, {Blommaert}, {Callut}, {Cara}, {Cerulli},
  {Cesarsky}, {Contursi}, {Creten}, {De Meester}, {Doublier}, {Doumayrou},
  {Duband}, {Exter}, {Genzel}, {Gillis}, {Gr{\"o}zinger}, {Henning},
  {Herreros}, {Huygen}, {Inguscio}, {Jakob}, {Jamar}, {Jean}, {de Jong},
  {Katterloher}, {Kiss}, {Klaas}, {Lemke}, {Lutz}, {Madden}, {Marquet},
  {Martignac}, {Mazy}, {Merken}, {Montfort}, {Morbidelli}, {M{\"u}ller},
  {Nielbock}, {Okumura}, {Orfei}, {Ottensamer}, {Pezzuto}, {Popesso},
  {Putzeys}, {Regibo}, {Reveret}, {Royer}, {Sauvage}, {Schreiber}, {Stegmaier},
  {Schmitt}, {Schubert}, {Sturm}, {Thiel}, {Tofani}, {Vavrek}, {Wetzstein},
  {Wieprecht}, \& {Wiezorrek}}]{poglitsch2010}
{Poglitsch}, A., {Waelkens}, C., {Geis}, N., {et~al.} 2010, \aap, 518, L2

\bibitem[{{Pope} {et~al.}(2005){Pope}, {Borys}, {Scott}, {Conselice},
  {Dickinson}, \& {Mobasher}}]{pope2005}
{Pope}, A., {Borys}, C., {Scott}, D., {et~al.} 2005, \mnras, 358, 149

\bibitem[{{Pope} {et~al.}(2008){Pope}, {Chary}, {Alexander}, {Armus},
  {Dickinson}, {Elbaz}, {Frayer}, {Scott}, \& {Teplitz}}]{pope2008}
{Pope}, A., {Chary}, R.-R., {Alexander}, D.~M., {et~al.} 2008, \apj, 675, 1171

\bibitem[{{Pope} {et~al.}(2006){Pope}, {Scott}, {Dickinson}, {Chary},
  {Morrison}, {Borys}, {Sajina}, {Alexander}, {Daddi}, {Frayer}, {MacDonald},
  \& {Stern}}]{pope2006}
{Pope}, A., {Scott}, D., {Dickinson}, M., {et~al.} 2006, \mnras, 370, 1185

\bibitem[{{Popesso} {et~al.}(2023){Popesso}, {Concas}, {Cresci}, {Belli},
  {Rodighiero}, {Inami}, {Dickinson}, {Ilbert}, {Pannella}, \&
  {Elbaz}}]{popesso2023}
{Popesso}, P., {Concas}, A., {Cresci}, G., {et~al.} 2023, \mnras, 519, 1526

\bibitem[{{Popping} {et~al.}(2017){Popping}, {Somerville}, \&
  {Galametz}}]{popping2017}
{Popping}, G., {Somerville}, R.~S., \& {Galametz}, M. 2017, \mnras, 471, 3152

\bibitem[{{Pozzi} {et~al.}(2021){Pozzi}, {Calura}, {Fudamoto},
  {Dessauges-Zavadsky}, {Gruppioni}, {Talia}, {Zamorani}, {Bethermin},
  {Cimatti}, {Enia}, {Khusanova}, {Decarli}, {Le F{\`e}vre}, {Capak},
  {Cassata}, {Faisst}, {Yan}, {Schaerer}, {Silverman}, {Bardelli}, {Boquien},
  {Enia}, {Narayanan}, {Ginolfi}, {Hathi}, {Jones}, {Koekemoer}, {Lemaux},
  {Loiacono}, {Maiolino}, {Riechers}, {Rodighiero}, {Romano}, {Vallini},
  {Vergani}, \& {Zucca}}]{pozzi2021}
{Pozzi}, F., {Calura}, F., {Fudamoto}, Y., {et~al.} 2021, \aap, 653, A84

\bibitem[{{Pozzi} {et~al.}(2020){Pozzi}, {Calura}, {Zamorani}, {Delvecchio},
  {Gruppioni}, \& {Santini}}]{pozzi2020}
{Pozzi}, F., {Calura}, F., {Zamorani}, G., {et~al.} 2020, \mnras, 491, 5073

\bibitem[{{Puget} {et~al.}(1996){Puget}, {Abergel}, {Bernard}, {Boulanger},
  {Burton}, {Desert}, \& {Hartmann}}]{puget1996}
{Puget}, J.~L., {Abergel}, A., {Bernard}, J.~P., {et~al.} 1996, \aap, 308, L5

\bibitem[{{Reuter} {et~al.}(2020){Reuter}, {Vieira}, {Spilker}, {Weiss},
  {Aravena}, {Archipley}, {B{\'e}thermin}, {Chapman}, {De Breuck}, {Dong},
  {Everett}, {Fu}, {Greve}, {Hayward}, {Hill}, {Hezaveh}, {Jarugula}, {Litke},
  {Malkan}, {Marrone}, {Narayanan}, {Phadke}, {Stark}, \&
  {Strandet}}]{reuter2020}
{Reuter}, C., {Vieira}, J.~D., {Spilker}, J.~S., {et~al.} 2020, \apj, 902, 78

\bibitem[{{Rho} {et~al.}(2008){Rho}, {Kozasa}, {Reach}, {Smith}, {Rudnick},
  {DeLaney}, {Ennis}, {Gomez}, \& {Tappe}}]{rho2008}
{Rho}, J., {Kozasa}, T., {Reach}, W.~T., {et~al.} 2008, \apj, 673, 271

\bibitem[{{Richards} {et~al.}(1998){Richards}, {Kellermann}, {Fomalont},
  {Windhorst}, \& {Partridge}}]{richards1998}
{Richards}, E.~A., {Kellermann}, K.~I., {Fomalont}, E.~B., {Windhorst}, R.~A.,
  \& {Partridge}, R.~B. 1998, \aj, 116, 1039

\bibitem[{{Riechers} {et~al.}(2011){Riechers}, {Hodge}, {Walter}, {Carilli}, \&
  {Bertoldi}}]{riechers2011}
{Riechers}, D.~A., {Hodge}, J., {Walter}, F., {Carilli}, C.~L., \& {Bertoldi},
  F. 2011, \apjl, 739, L31

\bibitem[{{Riechers} {et~al.}(2020){Riechers}, {Hodge}, {Pavesi}, {Daddi},
  {Decarli}, {Ivison}, {Sharon}, {Smail}, {Walter}, {Aravena}, {Capak},
  {Carilli}, {Cox}, {Cunha}, {Dannerbauer}, {Dickinson}, {Neri}, \&
  {Wagg}}]{riechers2020}
{Riechers}, D.~A., {Hodge}, J.~A., {Pavesi}, R., {et~al.} 2020, \apj, 895, 81

\bibitem[{{Rovilos} {et~al.}(2010){Rovilos}, {Georgantopoulos}, {Akylas}, \&
  {Fotopoulou}}]{rovilos2010}
{Rovilos}, E., {Georgantopoulos}, I., {Akylas}, A., \& {Fotopoulou}, S. 2010,
  \aap, 522, A11

\bibitem[{{Rudnick} {et~al.}(2017){Rudnick}, {Hodge}, {Walter}, {Momcheva},
  {Tran}, {Papovich}, {da Cunha}, {Decarli}, {Saintonge}, {Willmer}, {Lotz}, \&
  {Lentati}}]{rudnick2017}
{Rudnick}, G., {Hodge}, J., {Walter}, F., {et~al.} 2017, \apj, 849, 27

\bibitem[{{Saintonge} {et~al.}(2013){Saintonge}, {Lutz}, {Genzel}, {Magnelli},
  {Nordon}, {Tacconi}, {Baker}, {Bandara}, {Berta}, {F{\"o}rster Schreiber},
  {Poglitsch}, {Sturm}, {Wuyts}, \& {Wuyts}}]{saintonge2013}
{Saintonge}, A., {Lutz}, D., {Genzel}, R., {et~al.} 2013, \apj, 778, 2

\bibitem[{{Sandage} {et~al.}(1979){Sandage}, {Tammann}, \&
  {Yahil}}]{sandage1979}
{Sandage}, A., {Tammann}, G.~A., \& {Yahil}, A. 1979, \apj, 232, 352

\bibitem[{{Santini} {et~al.}(2014){Santini}, {Maiolino}, {Magnelli}, {Lutz},
  {Lamastra}, {Li Causi}, {Eales}, {Andreani}, {Berta}, {Buat}, {Cooray},
  {Cresci}, {Daddi}, {Farrah}, {Fontana}, {Franceschini}, {Genzel}, {Granato},
  {Grazian}, {Le Floc'h}, {Magdis}, {Magliocchetti}, {Mannucci}, {Menci},
  {Nordon}, {Oliver}, {Popesso}, {Pozzi}, {Riguccini}, {Rodighiero}, {Rosario},
  {Salvato}, {Scott}, {Silva}, {Tacconi}, {Viero}, {Wang}, {Wuyts}, \&
  {Xu}}]{santini2014}
{Santini}, P., {Maiolino}, R., {Magnelli}, B., {et~al.} 2014, \aap, 562, A30

\bibitem[{{Sargent} {et~al.}(2010){Sargent}, {Srinivasan}, {Meixner}, {Kemper},
  {Tielens}, {Speck}, {Matsuura}, {Bernard}, {Hony}, {Gordon}, {Indebetouw},
  {Marengo}, {Sloan}, \& {Woods}}]{sargent2010}
{Sargent}, B.~A., {Srinivasan}, S., {Meixner}, M., {et~al.} 2010, \apj, 716,
  878

\bibitem[{{Schechter}(1976)}]{schechter1976}
{Schechter}, P. 1976, \apj, 203, 297

\bibitem[{{Schmidt}(1968)}]{schmidt1968}
{Schmidt}, M. 1968, \apj, 151, 393

\bibitem[{{Schneider} {et~al.}(2014){Schneider}, {Valiante}, {Ventura},
  {dell'Agli}, {Di Criscienzo}, {Hirashita}, \& {Kemper}}]{schneider2014}
{Schneider}, R., {Valiante}, R., {Ventura}, P., {et~al.} 2014, \mnras, 442,
  1440

\bibitem[{{Scoville} {et~al.}(2007){Scoville}, {Aussel}, {Brusa}, {Capak},
  {Carollo}, {Elvis}, {Giavalisco}, {Guzzo}, {Hasinger}, {Impey}, {Kneib},
  {LeFevre}, {Lilly}, {Mobasher}, {Renzini}, {Rich}, {Sanders}, {Schinnerer},
  {Schminovich}, {Shopbell}, {Taniguchi}, \& {Tyson}}]{scoville2007}
{Scoville}, N., {Aussel}, H., {Brusa}, M., {et~al.} 2007, \apjs, 172, 1

\bibitem[{{Sharon} {et~al.}(2016){Sharon}, {Riechers}, {Hodge}, {Carilli},
  {Walter}, {Wei{\ss}}, {Knudsen}, \& {Wagg}}]{sharon2016}
{Sharon}, C.~E., {Riechers}, D.~A., {Hodge}, J., {et~al.} 2016, \apj, 827, 18

\bibitem[{{Simpson} {et~al.}(2020){Simpson}, {Smail},
  {Dudzevi{\v{c}}i{\={u}}t{\.{e}}}, {Matsuda}, {Hsieh}, {Wang}, {Swinbank},
  {Stach}, {An}, {Birkin}, {Ao}, {Bunker}, {Chapman}, {Chen}, {Coppin},
  {Ikarashi}, {Ivison}, {Mitsuhashi}, {Saito}, {Umehata}, {Wang}, \&
  {Zhao}}]{simpson2020}
{Simpson}, J.~M., {Smail}, I., {Dudzevi{\v{c}}i{\={u}}t{\.{e}}}, U., {et~al.}
  2020, \mnras, 495, 3409

\bibitem[{{Simpson} {et~al.}(2017){Simpson}, {Smail}, {Swinbank}, {Ivison},
  {Dunlop}, {Geach}, {Almaini}, {Arumugam}, {Bremer}, {Chen}, {Conselice},
  {Coppin}, {Farrah}, {Ibar}, {Hartley}, {Ma}, {Micha{\l}owski}, {Scott},
  {Spaans}, {Thomson}, \& {van der Werf}}]{simpson2017}
{Simpson}, J.~M., {Smail}, I., {Swinbank}, A.~M., {et~al.} 2017, \apj, 839, 58

\bibitem[{{Skelton} {et~al.}(2014){Skelton}, {Whitaker}, {Momcheva}, {Brammer},
  {van Dokkum}, {Labb{\'e}}, {Franx}, {van der Wel}, {Bezanson}, {Da Cunha},
  {Fumagalli}, {F{\"o}rster Schreiber}, {Kriek}, {Leja}, {Lundgren}, {Magee},
  {Marchesini}, {Maseda}, {Nelson}, {Oesch}, {Pacifici}, {Patel}, {Price},
  {Rix}, {Tal}, {Wake}, \& {Wuyts}}]{skelton2014}
{Skelton}, R.~E., {Whitaker}, K.~E., {Momcheva}, I.~G., {et~al.} 2014, \apjs,
  214, 24

\bibitem[{{Smail} {et~al.}(2000){Smail}, {Ivison}, {Owen}, {Blain}, \&
  {Kneib}}]{smail2000}
{Smail}, I., {Ivison}, R.~J., {Owen}, F.~N., {Blain}, A.~W., \& {Kneib}, J.~P.
  2000, \apj, 528, 612

\bibitem[{{Solomon} {et~al.}(1997){Solomon}, {Downes}, {Radford}, \&
  {Barrett}}]{solomon1997}
{Solomon}, P.~M., {Downes}, D., {Radford}, S.~J.~E., \& {Barrett}, J.~W. 1997,
  \apj, 478, 144

\bibitem[{{Speagle} {et~al.}(2014){Speagle}, {Steinhardt}, {Capak}, \&
  {Silverman}}]{speagle2014}
{Speagle}, J.~S., {Steinhardt}, C.~L., {Capak}, P.~L., \& {Silverman}, J.~D.
  2014, \apjs, 214, 15

\bibitem[{{Stach} {et~al.}(2019){Stach}, {Dudzevi{\v{c}}i{\={u}}t{\.{e}}},
  {Smail}, {Swinbank}, {Geach}, {Simpson}, {An}, {Almaini}, {Arumugam},
  {Blain}, {Chapman}, {Chen}, {Conselice}, {Cooke}, {Coppin}, {da Cunha},
  {Dunlop}, {Farrah}, {Gullberg}, {Hodge}, {Ivison}, {Kocevski},
  {Micha{\l}owski}, {Miyaji}, {Scott}, {Thomson}, {Wardlow}, {Weiss}, \& {van
  der Werf}}]{stach2019}
{Stach}, S.~M., {Dudzevi{\v{c}}i{\={u}}t{\.{e}}}, U., {Smail}, I., {et~al.}
  2019, \mnras, 487, 4648

\bibitem[{{Staguhn} {et~al.}(2014){Staguhn}, {Kov{\'a}cs}, {Arendt}, {Benford},
  {Decarli}, {Dwek}, {Fixsen}, {Hilton}, {Irwin}, {Jhabvala}, {Karim},
  {Leclercq}, {Maher}, {Miller}, {Moseley}, {Sharp}, {Walter}, \&
  {Wollack}}]{staguhn2014}
{Staguhn}, J.~G., {Kov{\'a}cs}, A., {Arendt}, R.~G., {et~al.} 2014, \apj, 790,
  77

\bibitem[{{Stalevski} {et~al.}(2012){Stalevski}, {Fritz}, {Baes}, {Nakos}, \&
  {Popovi{\'c}}}]{stalevski2012}
{Stalevski}, M., {Fritz}, J., {Baes}, M., {Nakos}, T., \& {Popovi{\'c}},
  L.~{\v{C}}. 2012, \mnras, 420, 2756

\bibitem[{{Stalevski} {et~al.}(2016){Stalevski}, {Ricci}, {Ueda}, {Lira},
  {Fritz}, \& {Baes}}]{stalevski2016}
{Stalevski}, M., {Ricci}, C., {Ueda}, Y., {et~al.} 2016, \mnras, 458, 2288

\bibitem[{{Steidel} {et~al.}(2003){Steidel}, {Adelberger}, {Shapley},
  {Pettini}, {Dickinson}, \& {Giavalisco}}]{steidel2003}
{Steidel}, C.~C., {Adelberger}, K.~L., {Shapley}, A.~E., {et~al.} 2003, \apj,
  592, 728

\bibitem[{{Sun} {et~al.}(2024){Sun}, {Helton}, {Egami}, {Hainline}, {Rieke},
  {Willmer}, {Eisenstein}, {Johnson}, {Rieke}, {Robertson}, {Tacchella},
  {Alberts}, {Baker}, {Bhatawdekar}, {Boyett}, {Bunker}, {Charlot}, {Chen},
  {Chevallard}, {Curtis-Lake}, {Danhaive}, {DeCoursey}, {Ji}, {Lyu},
  {Maiolino}, {Rujopakarn}, {Sandles}, {Shivaei}, {{\"U}bler}, {Willott}, \&
  {Witstok}}]{sun2024}
{Sun}, F., {Helton}, J.~M., {Egami}, E., {et~al.} 2024, \apj, 961, 69

\bibitem[{{Swinbank} {et~al.}(2004){Swinbank}, {Smail}, {Chapman}, {Blain},
  {Ivison}, \& {Keel}}]{swinbank2004}
{Swinbank}, A.~M., {Smail}, I., {Chapman}, S.~C., {et~al.} 2004, \apj, 617, 64

\bibitem[{{Tacconi} {et~al.}(2018){Tacconi}, {Genzel}, {Saintonge}, {Combes},
  {Garc{\'\i}a-Burillo}, {Neri}, {Bolatto}, {Contini}, {F{\"o}rster Schreiber},
  {Lilly}, {Lutz}, {Wuyts}, {Accurso}, {Boissier}, {Boone}, {Bouch{\'e}},
  {Bournaud}, {Burkert}, {Carollo}, {Cooper}, {Cox}, {Feruglio}, {Freundlich},
  {Herrera-Camus}, {Juneau}, {Lippa}, {Naab}, {Renzini}, {Salome}, {Sternberg},
  {Tadaki}, {{\"U}bler}, {Walter}, {Weiner}, \& {Weiss}}]{tacconi2018}
{Tacconi}, L.~J., {Genzel}, R., {Saintonge}, A., {et~al.} 2018, \apj, 853, 179

\bibitem[{{Tacconi} {et~al.}(2020){Tacconi}, {Genzel}, \&
  {Sternberg}}]{tacconi2020}
{Tacconi}, L.~J., {Genzel}, R., \& {Sternberg}, A. 2020, \araa, 58, 157

\bibitem[{{Tacconi} {et~al.}(2013){Tacconi}, {Neri}, {Genzel}, {Combes},
  {Bolatto}, {Cooper}, {Wuyts}, {Bournaud}, {Burkert}, {Comerford}, {Cox},
  {Davis}, {F{\"o}rster Schreiber}, {Garc{\'\i}a-Burillo}, {Gracia-Carpio},
  {Lutz}, {Naab}, {Newman}, {Omont}, {Saintonge}, {Shapiro Griffin}, {Shapley},
  {Sternberg}, \& {Weiner}}]{tacconi2013}
{Tacconi}, L.~J., {Neri}, R., {Genzel}, R., {et~al.} 2013, \apj, 768, 74

\bibitem[{{Teplitz} {et~al.}(2011){Teplitz}, {Chary}, {Elbaz}, {Dickinson},
  {Bridge}, {Colbert}, {Le Floc'h}, {Frayer}, {Howell}, {Koo}, {Papovich},
  {Phillips}, {Scarlata}, {Siana}, {Spinrad}, \& {Stern}}]{teplitz2011}
{Teplitz}, H.~I., {Chary}, R., {Elbaz}, D., {et~al.} 2011, \aj, 141, 1

\bibitem[{{Thomson} {et~al.}(2012){Thomson}, {Ivison}, {Smail}, {Swinbank},
  {Weiss}, {Kneib}, {Papadopoulos}, {Baker}, {Sharon}, \& {van
  Moorsel}}]{thomson2012}
{Thomson}, A.~P., {Ivison}, R.~J., {Smail}, I., {et~al.} 2012, \mnras, 425,
  2203

\bibitem[{{Tielens}(1998)}]{tielens1998}
{Tielens}, A.~G.~G.~M. 1998, \apj, 499, 267

\bibitem[{{Todini} \& {Ferrara}(2001)}]{todini2001}
{Todini}, P. \& {Ferrara}, A. 2001, \mnras, 325, 726

\bibitem[{{Traina} {et~al.}(2024{\natexlab{a}}){Traina}, {Gruppioni},
  {Delvecchio}, {Calura}, {Bisigello}, {Feltre}, {Magnelli}, {Schinnerer},
  {Liu}, {Adscheid}, {Behiri}, {Gentile}, {Pozzi}, {Talia}, {Zamorani},
  {Algera}, {Gillman}, {Lambrides}, \& {Symeonidis}}]{traina2024a}
{Traina}, A., {Gruppioni}, C., {Delvecchio}, I., {et~al.} 2024{\natexlab{a}},
  \aap, 681, A118

\bibitem[{{Traina} {et~al.}(2024{\natexlab{b}}){Traina}, {Magnelli},
  {Gruppioni}, {Delvecchio}, {Parente}, {Calura}, {Bisigello}, {Feltre},
  {Pozzi}, \& {Vallini}}]{traina2024}
{Traina}, A., {Magnelli}, B., {Gruppioni}, C., {et~al.} 2024{\natexlab{b}},
  \aap, 690, A84

\bibitem[{{{\"U}bler} {et~al.}(2024){{\"U}bler}, {D'Eugenio}, {Perna},
  {Arribas}, {Jones}, {Bunker}, {Carniani}, {Charlot}, {Maiolino},
  {Rodr{\'\i}guez del Pino}, {Willott}, {B{\"o}ker}, {Cresci}, {Kumari},
  {Lamperti}, {Parlanti}, {Scholtz}, \& {Venturi}}]{uebler2024}
{{\"U}bler}, H., {D'Eugenio}, F., {Perna}, M., {et~al.} 2024, \mnras, 533, 4287

\bibitem[{{Valentino} {et~al.}(2018){Valentino}, {Magdis}, {Daddi}, {Liu},
  {Aravena}, {Bournaud}, {Cibinel}, {Cormier}, {Dickinson}, {Gao}, {Jin},
  {Juneau}, {Kartaltepe}, {Lee}, {Madden}, {Puglisi}, {Sanders}, \&
  {Silverman}}]{valentino2018}
{Valentino}, F., {Magdis}, G.~E., {Daddi}, E., {et~al.} 2018, \apj, 869, 27

\bibitem[{{Villanueva} {et~al.}(2017){Villanueva}, {Ibar}, {Hughes},
  {Lara-L{\'o}pez}, {Dunne}, {Eales}, {Ivison}, {Aravena}, {Baes}, {Bourne},
  {Cassata}, {Cooray}, {Dannerbauer}, {Davies}, {Driver}, {Dye}, {Furlanetto},
  {Herrera-Camus}, {Maddox}, {Micha{\l}owski}, {Molina}, {Riechers}, {Sansom},
  {Smith}, {Rodighiero}, {Valiante}, \& {van der Werf}}]{villanueva2017}
{Villanueva}, V., {Ibar}, E., {Hughes}, T.~M., {et~al.} 2017, \mnras, 470, 3775

\bibitem[{{Vlahakis} {et~al.}(2005){Vlahakis}, {Dunne}, \&
  {Eales}}]{vlahakis2005}
{Vlahakis}, C., {Dunne}, L., \& {Eales}, S. 2005, \mnras, 364, 1253

\bibitem[{{Walter} {et~al.}(2012){Walter}, {Decarli}, {Carilli}, {Bertoldi},
  {Cox}, {da Cunha}, {Daddi}, {Dickinson}, {Downes}, {Elbaz}, {Ellis}, {Hodge},
  {Neri}, {Riechers}, {Weiss}, {Bell}, {Dannerbauer}, {Krips}, {Krumholz},
  {Lentati}, {Maiolino}, {Menten}, {Rix}, {Robertson}, {Spinrad}, {Stark}, \&
  {Stern}}]{walter2012}
{Walter}, F., {Decarli}, R., {Carilli}, C., {et~al.} 2012, \nat, 486, 233

\bibitem[{{Wang} {et~al.}(2016{\natexlab{a}}){Wang}, {Liu}, {Qiu}, {Bai},
  {Yang}, {Guo}, \& {Zhang}}]{wang2016}
{Wang}, S., {Liu}, J., {Qiu}, Y., {et~al.} 2016{\natexlab{a}}, \apjs, 224, 40

\bibitem[{{Wang} {et~al.}(2018){Wang}, {Elbaz}, {Daddi}, {Liu}, {Kodama},
  {Tanaka}, {Schreiber}, {Zanella}, {Valentino}, {Sargent}, {Kohno}, {Xiao},
  {Pannella}, {Ciesla}, {Gobat}, \& {Koyama}}]{wang2018}
{Wang}, T., {Elbaz}, D., {Daddi}, E., {et~al.} 2018, \apjl, 867, L29

\bibitem[{{Wang} {et~al.}(2016{\natexlab{b}}){Wang}, {Elbaz}, {Schreiber},
  {Pannella}, {Shu}, {Willner}, {Ashby}, {Huang}, {Fontana}, {Dekel}, {Daddi},
  {Ferguson}, {Dunlop}, {Ciesla}, {Koekemoer}, {Giavalisco}, {Boutsia},
  {Finkelstein}, {Juneau}, {Barro}, {Koo}, {Micha{\l}owski}, {Orellana}, {Lu},
  {Castellano}, {Bourne}, {Buitrago}, {Santini}, {Faber}, {Hathi}, {Lucas}, \&
  {P{\'e}rez-Gonz{\'a}lez}}]{wang2016b}
{Wang}, T., {Elbaz}, D., {Schreiber}, C., {et~al.} 2016{\natexlab{b}}, \apj,
  816, 84

\bibitem[{{Wang} {et~al.}(2004){Wang}, {Cowie}, \& {Barger}}]{wang2004}
{Wang}, W.~H., {Cowie}, L.~L., \& {Barger}, A.~J. 2004, \apj, 613, 655

\bibitem[{{Wang} {et~al.}(2017){Wang}, {Lin}, {Lim}, {Smail}, {Chapman},
  {Zheng}, {Shim}, {Kodama}, {Almaini}, {Ao}, {Blain}, {Bourne}, {Bunker},
  {Chang}, {Chao}, {Chen}, {Clements}, {Conselice}, {Cowley}, {Dannerbauer},
  {Dunlop}, {Geach}, {Goto}, {Jiang}, {Ivison}, {Jeong}, {Kohno}, {Kong},
  {Lee}, {Lee}, {Lee}, {Micha{\l}owski}, {Oteo}, {Sawicki}, {Scott}, {Shu},
  {Simpson}, {Tee}, {Toba}, {Valiante}, {Wang}, {Wang}, \&
  {Wardlow}}]{wang2017}
{Wang}, W.-H., {Lin}, W.-C., {Lim}, C.-F., {et~al.} 2017, \apj, 850, 37

\bibitem[{{Weibel} {et~al.}(2024){Weibel}, {Oesch}, {Barrufet}, {Gottumukkala},
  {Ellis}, {Santini}, {Weaver}, {Allen}, {Bouwens}, {Bowler}, {Brammer},
  {Carnall}, {Cullen}, {Dayal}, {Dickinson}, {Donnan}, {Dunlop}, {Giavalisco},
  {Grogin}, {Illingworth}, {Koekemoer}, {Labbe}, {Marchesini}, {McLeod},
  {McLure}, {Naidu}, {P{\'e}rez-Gonz{\'a}lez}, {Shuntov}, {Stefanon}, {Toft},
  \& {Xiao}}]{weibel2024}
{Weibel}, A., {Oesch}, P.~A., {Barrufet}, L., {et~al.} 2024, \mnras, 533, 1808

\bibitem[{{Weingartner} \& {Draine}(2001)}]{wd01}
{Weingartner}, J.~C. \& {Draine}, B.~T. 2001, \apj, 548, 296

\bibitem[{{Williams} {et~al.}(1996){Williams}, {Blacker}, {Dickinson}, {Dixon},
  {Ferguson}, {Fruchter}, {Giavalisco}, {Gilliland}, {Heyer}, {Katsanis},
  {Levay}, {Lucas}, {McElroy}, {Petro}, {Postman}, {Adorf}, \&
  {Hook}}]{williams1996}
{Williams}, R.~E., {Blacker}, B., {Dickinson}, M., {et~al.} 1996, \aj, 112,
  1335

\bibitem[{{Wolfire} {et~al.}(1995){Wolfire}, {Hollenbach}, {McKee}, {Tielens},
  \& {Bakes}}]{wolfire1995}
{Wolfire}, M.~G., {Hollenbach}, D., {McKee}, C.~F., {Tielens}, A.~G.~G.~M., \&
  {Bakes}, E.~L.~O. 1995, \apj, 443, 152

\bibitem[{{Xiao} {et~al.}(2023){Xiao}, {Oesch}, {Elbaz}, {Bing}, {Nelson},
  {Weibel}, {Naidu}, {Daddi}, {Bouwens}, {Matthee}, {Wuyts}, {Chisholm},
  {Brammer}, {Dickinson}, {Magnelli}, {Leroy}, {van Dokkum}, {Schaerer},
  {Herard-Demanche}, {Barrufet}, {Endsley}, {Fudamoto}, {G{\'o}mez-Guijarro},
  {Gottumukkala}, {Illingworth}, {Labbe}, {Magee}, {Marchesini}, {Maseda},
  {Qin}, {Reddy}, {Shapley}, {Shivaei}, {Shuntov}, {Stefanon}, {Whitaker}, \&
  {Wyithe}}]{xiao2024}
{Xiao}, M., {Oesch}, P., {Elbaz}, D., {et~al.} 2023, arXiv e-prints,
  arXiv:2309.02492

\bibitem[{{Yang} {et~al.}(2017){Yang}, {Omont}, {Beelen}, {Gao}, {van der
  Werf}, {Gavazzi}, {Zhang}, {Ivison}, {Lehnert}, {Liu}, {Oteo},
  {Gonz{\'a}lez-Alfonso}, {Dannerbauer}, {Cox}, {Krips}, {Neri}, {Riechers},
  {Baker}, {Micha{\l}owski}, {Cooray}, \& {Smail}}]{yang2017}
{Yang}, C., {Omont}, A., {Beelen}, A., {et~al.} 2017, \aap, 608, A144

\bibitem[{{Yates} {et~al.}(2024){Yates}, {Hendriks}, {Vijayan}, {Izzard},
  {Thomas}, \& {Das}}]{yates2024}
{Yates}, R.~M., {Hendriks}, D., {Vijayan}, A.~P., {et~al.} 2024, \mnras, 527,
  6292

\bibitem[{{Zhukovska} {et~al.}(2008){Zhukovska}, {Gail}, \&
  {Trieloff}}]{zhukovska2008}
{Zhukovska}, S., {Gail}, H.~P., \& {Trieloff}, M. 2008, \aap, 479, 453

\bibitem[{{Zylka}(2013)}]{zylka2013}
{Zylka}, R. 2013, {MOPSIC: Extended Version of MOPSI, Astrophysics Source Code
  Library, record ascl:1303.011},
  \url{https://ui.adsabs.harvard.edu/abs/2013ascl.soft03011Z}

\end{thebibliography}

% --------------------------------------------------------------------

% --------------------------------------------------------------------

\begin{appendix}

\onecolumn
\section{Comparison between the different fitting methods}\label{app:comparison_codes}

The eight SED fitting codes used in this work are here compared in terms of their $M_\textrm{dust}$, $M_\star$ and $L_\textrm{IR}(8-1000\mu\textrm{m})$ outputs.
For each of these quantities, we directly compare the results of each method to those of each other approach. 

For each object, we compute the dispersion of all eight determinations around their median value (i.e., computing the median absolute deviation, M.A.D., per object).
To quantify the scatter in each combination of two methods, we compute the distribution of the median absolute deviation of the quantity $(y-x)/x$, computed for each pair of codes using all sources. The quantities  $x$ and $y$ simply represent the $M_\textrm{dust}$ or $L_\textrm{IR}(8-1000\mu\textrm{m})$ values computed with methods in the abscissa or ordinate axis in the comparison diagrams.

\subsection{Dust mass}

In order to produce a meaningful comparison, all dust masses have been re-scaled to the $\kappa_\nu$ normalization by \citet{draine2014}.  We remind that the first DL07 modeling came with the parametrization of the dust absorption coefficient $\kappa_\nu$ by \citet{LD01}, later revised by \citet{draine2003} ($\kappa_{850\mu\textrm{m}}=0.038$ m$^2$ kg$^{-1}$). \citet{draine2014} re-normalized their assumption on dust properties such that $\kappa_{850\mu\textrm{m}}=0.047$ m$^2$ kg$^{-1}$. On the other hand, \citet{dacunha2008,dacunha2015} adopted the normalization by \citet{dunne2000}, that is $\kappa_{850\mu\textrm{m}}=0.077$ m$^2$ kg$^{-1}$. 

\begin{figure*}[!t]	%[!ht]
\centering
\settowidth{\imagewidth}{\includegraphics{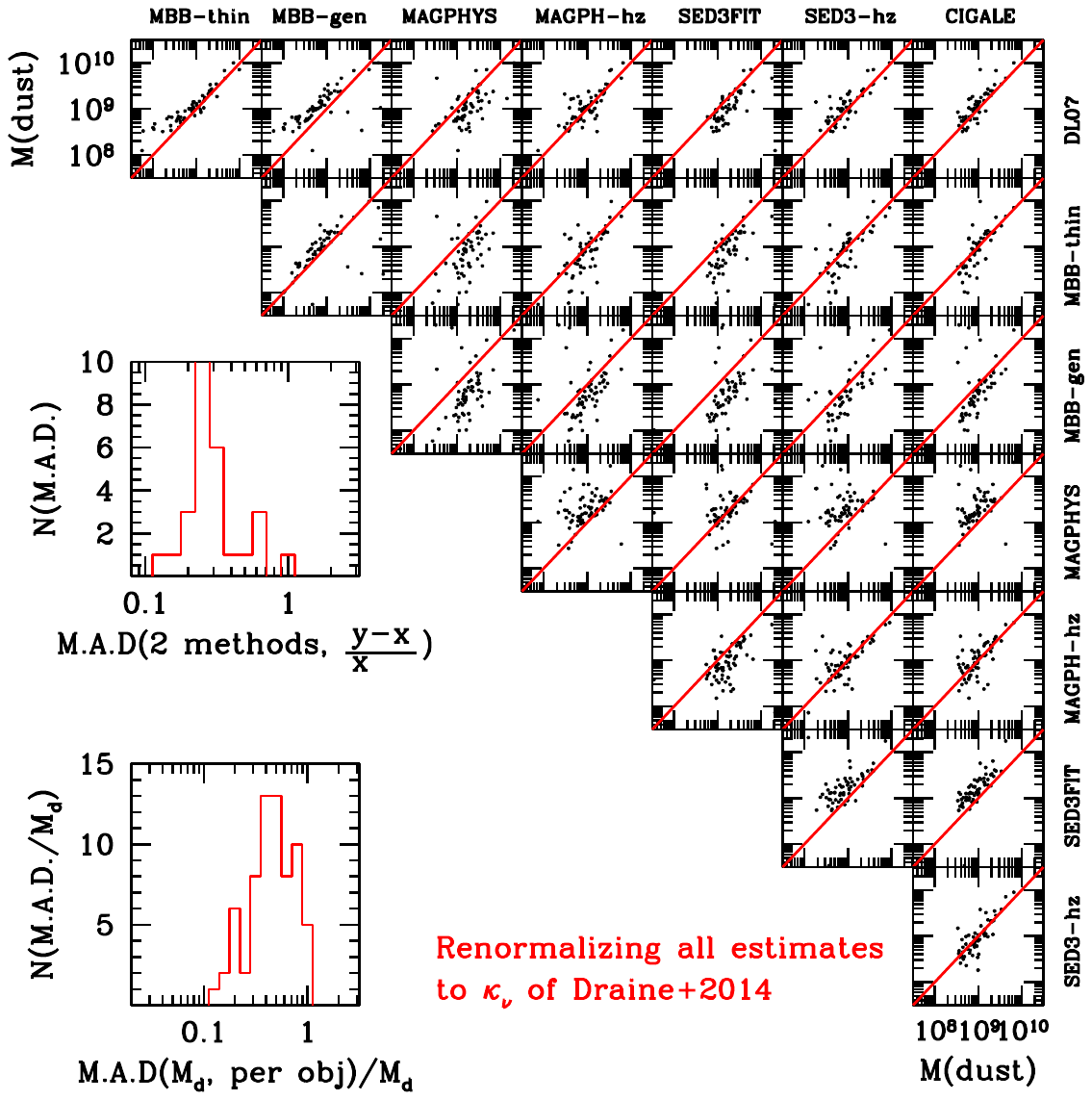}}
\includegraphics[trim=0 0.22\imagewidth{} 0 0.10\imagewidth{}, clip, width=0.7\textwidth]{figs1/comparison_Mdust_renorm_202410.pdf}
\caption{Comparison of $M_\textrm{dust}$ obtained with the eight different SED fitting methods considered here. 
The top histogram shows the distribution of the median absolute deviation (M.A.D.) of the quantity $(y-x)/x$, computed for each pair of codes using all sources.
The bottom histogram depicts the dispersion of all eight determinations around their median value (i.e., computing the M.A.D. per object). 
}
\label{fig:compare_Mdust}
\end{figure*}

Figure \ref{fig:compare_Mdust} compares the $M_\textrm{dust}$ results of each method to those of each other approach. 
The comparison shows a general agreement between the different methods, modulo their natural limitations (especially those of the MBB models). 

The most striking exception is the MBB-gen model, which systematically underestimates $M_\textrm{dust}$ with respect to the other methods. The effect varies between 0.2 and 0.5 dex (median offset over all sources, with respect to different codes), corresponding to up to $4\times$ the M.A.D. of the quantity $(y-x)/x$. 
This confirms the results by \citet{ismail2023}, who showed that a more thorough knowledge of the source properties (specifically the size of the dusty IR-emitting region) is needed to properly apply this method (Sect. \ref{sect:MBB}).

The MBB-thin model shows a milder systematic under-estimation of $M_\textrm{dust}$, to be at least partially ascribed to the missing warmer dust component that dominates the $\lambda_\textrm{rest}<50$  $\mu$m emission.
A systematic offset between the original and the high-$z$ versions of MAGPHYS and SED3FIT exists, most likely caused by the broader range of dust properties allowed in the latter, which leads to a better fit to the observed SEDs than in the original version. 
Moreover, MAGPHYS and SED3FIT (which are tightly related to each other) show the largest scatter when compared to the other methods. This drives the preference to use the result of the high-$z$ version of these codes for the N2CLS sample.

\subsection{Stellar mass}

For the stellar component, the MAGPHYS-based codes adopt \citet[][BC03]{bc03} stellar populations with a \citet{chabrier2003} IMF and apply the \citet[][CF00]{charlot2000} two-components attenuation law. The CIGALE fits have been performed adopting a similar setup: the BC03 stellar populations with the \citet{chabrier2003} IMF and the same CF00 attenuation law.

\begin{figure*}[!t]	%[!ht]
\centering
\settowidth{\imagewidth}{\includegraphics{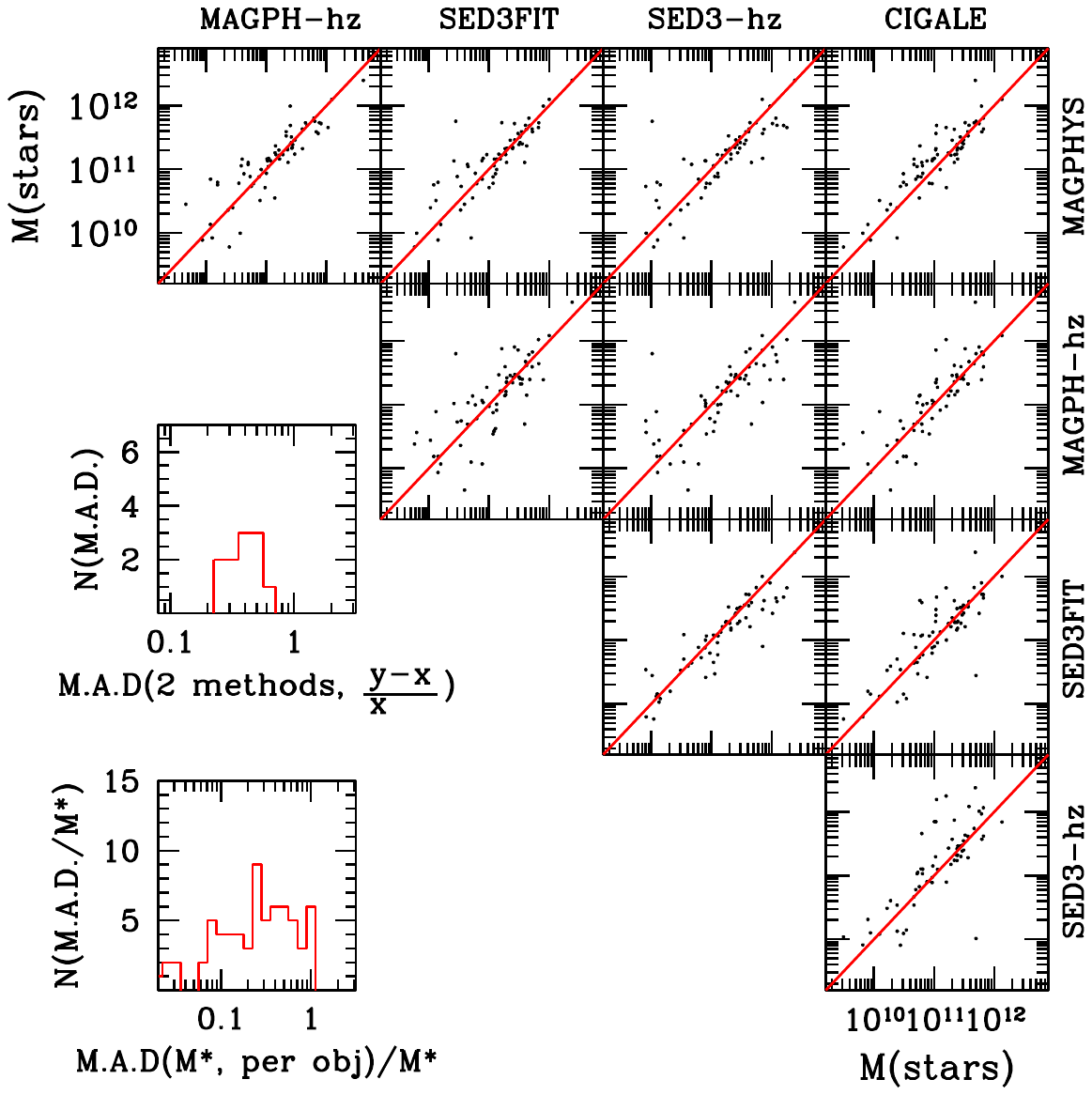}}
\includegraphics[trim=0 0.22\imagewidth{} 0 0.10\imagewidth{}, clip, width=0.5\textwidth]{figs1/comparison_Ms_202410.pdf}
\caption{Comparison of $M_\star$ obtained with the five different UV-to-radio SED fitting methods considered here.}
\label{fig:compare_Mstar}
\end{figure*}

Figure \ref{fig:compare_Mstar} presents the comparison of stellar masses and shows a very good agreement, although a few important outliers exist, especially in the high-$z$ versions of MAGPHYS and SED3FIT with respect to CIGALE. Section \ref{sect:starbursty} show the consequence of these differences on the position of the N2GN galaxies in the $M_\star$-SFR plane. Appendix \ref{app:taudep_change} investigates the effect on depletion timescales. The most striking cases are very red galaxies at high redshift, with $M_\star$ differences up to 1\,dex between MAGPHYS and CIGALE. In these cases, despite adopting the same stellar models and the same dust attenuation law, MAGPHYS converges to best fit models with NIR luminosities much brighter than those found by CIGALE. The SEDs of the majority of these cases miss data points to constrain the rest-frame NIR emission. The most likely culprits are differences in the adopted SFH and in the retrieved $A_V$. For these very red outliers, $A_V>3$ and $A_V(\textrm{MAGP})-A_V(\textrm{CIG})>1.3$. These differences testify how critical and uncertain can the $M_\star$ derivation be in the case of extreme, high-$z$ SMGs.

%%%%%%%%%%

\subsection{Infrared luminosity}

As for the infrared luminosity, the single-temperature MBB models do not include any warm dust component and are therefore limited to data points at $\lambda_\textrm{rest}>50\ \mu$m. Consequently, the integrated luminosity of these models is $L_\textrm{FIR}(50-1000\mu\textrm{m})$, rather than $L_\textrm{IR}$. By integrating the templates family of {\it Herschel} star forming galaxies defined by \citet{berta2013}, we recovered a typical ratio $L_\textrm{FIR}/L_\textrm{IR}\simeq0.7$ for this kind of galaxies \citep[][]{berta2023}. We therefore rescaled the MBB $L_\textrm{FIR}$ estimate to the galaxies' $L_\textrm{IR}$ using this ratio.

\begin{figure*}[!t]	%[!ht]
\centering
\settowidth{\imagewidth}{\includegraphics{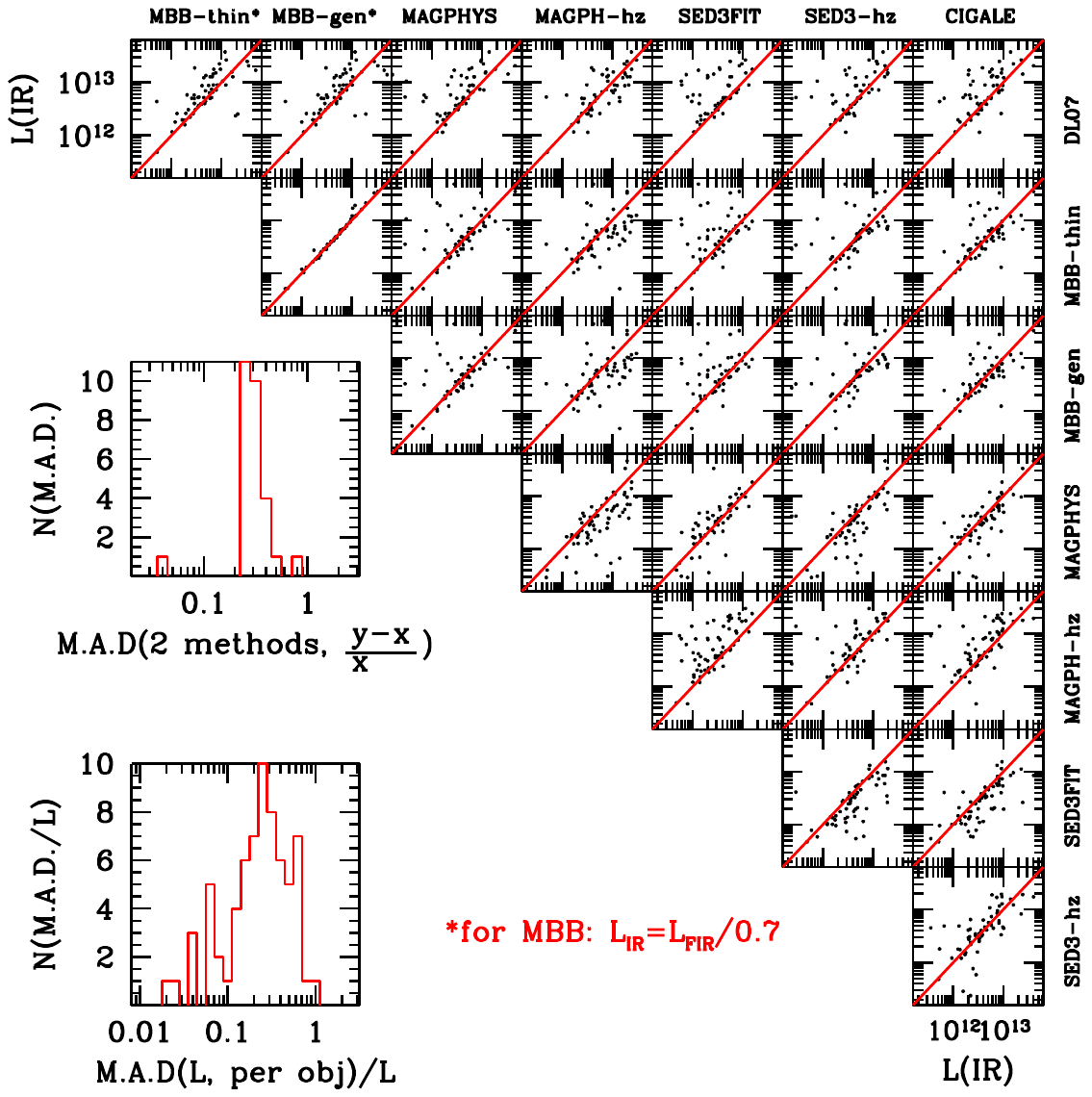}}
\includegraphics[trim=0 0.22\imagewidth{} 0 0.10\imagewidth{}, clip, width=0.6\textwidth]{figs1/comparison_LIR_202410.pdf}
\caption{Comparison of $L_\textrm{IR}$ obtained with the eight different SED fitting methods considered here.}
\label{fig:compare_LIR}
\end{figure*}

Figure \ref{fig:compare_LIR} compares the $L_\textrm{IR}$ results. In the cases when an AGN-torus was necessary in the fit, its contribution to the $L_\textrm{IR}$ has been excluded. A good overall agreement between the different approaches is found. CIGALE behaves naturally similarly to DL07 and is in good agreement with SED3FIT-h$z$. The extreme outliers in some of the panels are those sources for which the codes diverge because not enough data are available to constrain all modeled components.

%\clearpage

%\clearpage
\newpage

\section{Effect of $M_\star$ on the $\tau_\textrm{dep}$ estimate}\label{app:taudep_change}

In Sect. \ref{sect:starbursty}, the results of SED fitting have been used to derive the molecular gas depletion timescale of the N2GN galaxies, by applying the \citet{tacconi2020} relation that links $M_\star$, redshift and distance from the MS to $\tau_\textrm{dep}$. This approach takes as reference the MS parametrization by \citet{speagle2014}, shown in Fig. \ref{fig:MS}.
The $\tau_\textrm{dep}$ thus obtained depends on the value of $M_\star$. In the same Sect. \ref{sect:starbursty}, it has also been shown that for a few extreme galaxies at $z>4$, with very red rest-frame optical colors, MAGPHYS/SED3FIT gives a significantly different estimate of $M_\star$ with respect to CIGALE. These sources are outliers in Fig. \ref{fig:compare_Mstar}.

\begin{figure*}[!h]
\centering
\rotatebox{-90}{\includegraphics[height=0.47\textwidth]{figs1/pl_taudep_z4a.pdf}}
\rotatebox{-90}{\includegraphics[height=0.47\textwidth]{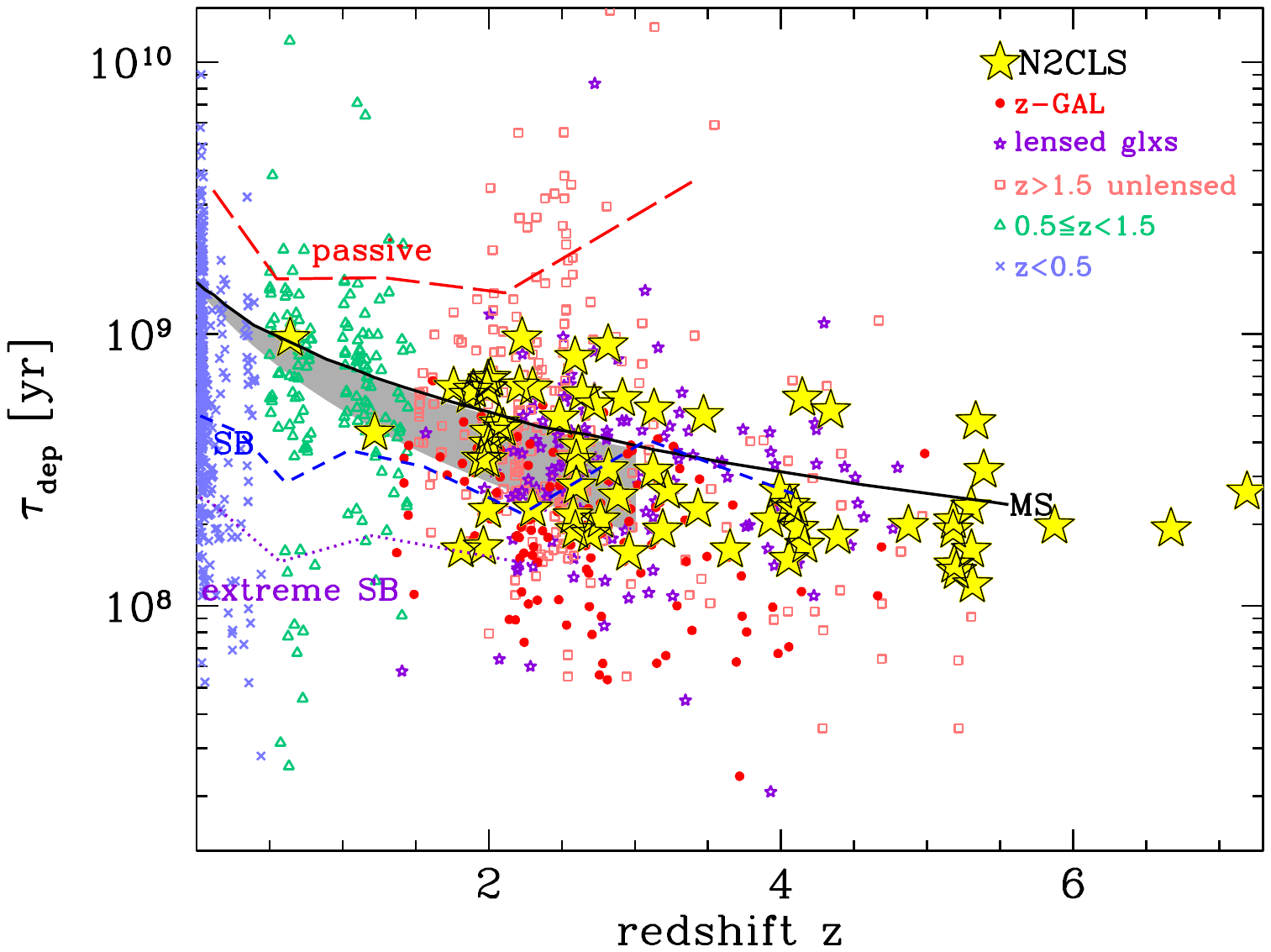}}
\caption{Consequence of changing $M_\star$ when computing $\tau_\textrm{dep}$ with the \citet{tacconi2020} scaling relations. {\em Left panel}: computation performed adopting the MAGPHYS/SED3FIT fiducial results. {\em Right panel}: same result obtained adopting the CIGALE estimate of $M_\star$. The references of the literature data and lines are listed in the caption of Fig. \ref{fig:taudep}}
\label{fig:taudep_Mstar_dependence}
\end{figure*}

Here we show in Fig.\,\ref{fig:taudep_Mstar_dependence} the consequence of using the value of $M_\star$ obtained with MAGPHYS/SED3FIT or with CIGALE. Few sources migrate to lower values of $\tau_\textrm{dep}$, and from the locus of the MS to the region occupied by starbursts. This happens specifically for the group of galaxies at $z\sim5$, that is going to be the subject of Lagache et al. (in prep.). 

%\clearpage
\newpage

\section{The importance of $\kappa_\nu$ when comparing different works}\label{app:rho_NO_rescaling}

Section \ref{sect:rho} presented the evolution of the dust cosmic density computed by integrating the STY results of the N2GN DMF in the redshift range $z=1.6$ to 7.2. The N2GN results have been there compared to literature data, pointing out the need to reconcile all data to the same dust model normalization, namely to the same $\kappa_\nu$. Our choice of reference is $\kappa_{850\mu\textrm{m}}=0.047$ m$^2$ kg$^{-1}$ \citep{draine2014}. 

If a different dust model was adopted, a re-normalization of dust masses is needed, because $M_\textrm{dust}$ depends inversely on $\kappa_\nu$. This is easy to visualize in the simple case of the MBB (Eq.~\ref{eq:mbb_mass}). The re-normalization factor is simply the ratio between the two different $\kappa_\nu$ values of reference (at the same reference frequency). Consequently, the observed DMF is shifted (in log space) by the same amount along the mass axis and so does its integral.

For example, if a work found in the literature adopted the \citet{draine2003} $\kappa_{850}=0.038$ m$^2$ kg$^{-1}$ value, in order to compare the literature dust masses to N2GN's, they would need to be rescaled by a factor $\kappa_{850}(D03)/\kappa_{850}(D14)$, because dust mass is inversely proportional to $\kappa_\nu$. In other words, if the adopted $\kappa_{850}$ is smaller than the \citet{draine2014} value, the corresponding dust masses need to be scaled down, in order to be compared to those obtained with the \citet{draine2014} normalization. Vice versa if the adopted $\kappa_{850}$ is larger than the \citet{draine2014} value, they will need to be scaled up. This holds if all adopted dust models differ only by their normalization but have the same functional dependence on wavelength (e.g. the $R_V=3.1$ MW models by Draine and collaborators). On the other hand, if more complex differences exist in the dust model assumptions, further transformations might be needed.
%
%%%% vedi macro test_scaling_rho.m per conferma (ho provato semplicemente a farlo)
%
The specific $\kappa_\nu$ assumptions of the different works shown in Fig. \ref{fig:rho_z} are as follows.

The SED fitting code MAGPHYS \citep{dacunha2008} was used by \citet[][local galaxies]{beeston2018}, \citet[][$0.0<z<0.5$]{beeston2024}, \citet[][optically selected galaxies at $z<1.75$]{driver2018}, and \citet[][sub-millimeter galaxies at $z=1-2$ and $z=3-4$]{dudzeviciute2021}. MAGPHYS adopts $\kappa_{850}=0.077$ m$^2$ kg$^{-1}$. Other works in the literature collection shown in Fig. \ref{fig:rho_z} also adopted the same value, namely \citet{vlahakis2005} and \citet{eales2024}. \citet{dunne2011} adopted $\kappa_{250}=0.89$ m$^2$ kg$^{-1}$, which they state corresponds to $\kappa_{850}=0.077$ m$^2$ kg$^{-1}$.

\citet{pozzi2020,pozzi2021} used the value $\kappa_{1.2\textrm{THz}}=4.0$ cm$^2$ g$^{-1}$, corresponding to \citet[][$\kappa_{850}=0.038$ m$^2$ kg$^{-1}$]{draine2003}. 
\citet{magnelli2020} adopted $\kappa_{850}=0.043$ m$^2$ kg$^{-1}$.    %%%, misquoted as \citet{LD01}. 
\citet{traina2024} estimated the dust mass of their 189 galaxies in three ways: a multi-wavelength SED fit with CIGALE, and a RJ flux scaling based on two different $T_\textrm{dust}$ assumptions of 25 and 35 K. In all three cases, they adopted the \citet{draine2014} $\kappa_\nu$ re-normalization.

\citet{menard2012} computed the dust density studying Mg{\sc ii} absorbing clouds in the lines of sight of quasars, applying the \citet{wd01} SMC (Small Magellanic Cloud) dust model. Transforming to the MW dust model implies multiplying by a factor 1.8 (see their Sect. 4). A further rescaling by 0.93 brings the value to the \citet{draine2003} normalization and finally this needs to be transformed to the \citet{draine2014} assumption.
Finally, \citet{peroux2020} derived the dust cosmic density by applying the dust to gas ratio by \citet{draine2003} -- with the correction prescribed by \citet{draine2014} -- to the gas cosmic density. Therefore, their data do not need any further correction.

The consequence of rescaling some results to a value of $\kappa_\nu$ different from the one originally adopted is a shift of the DMF in logarithmic space to larger or smaller masses. The net result on $\rho_\textrm{dust}$ (the integral of the DMF) is therefore a simple scaling by the same amount as for $M_\textrm{dust}$.

\begin{figure*}[!h]
\centering
\settowidth{\imageheight}{\includegraphics[height=0.43\textwidth]{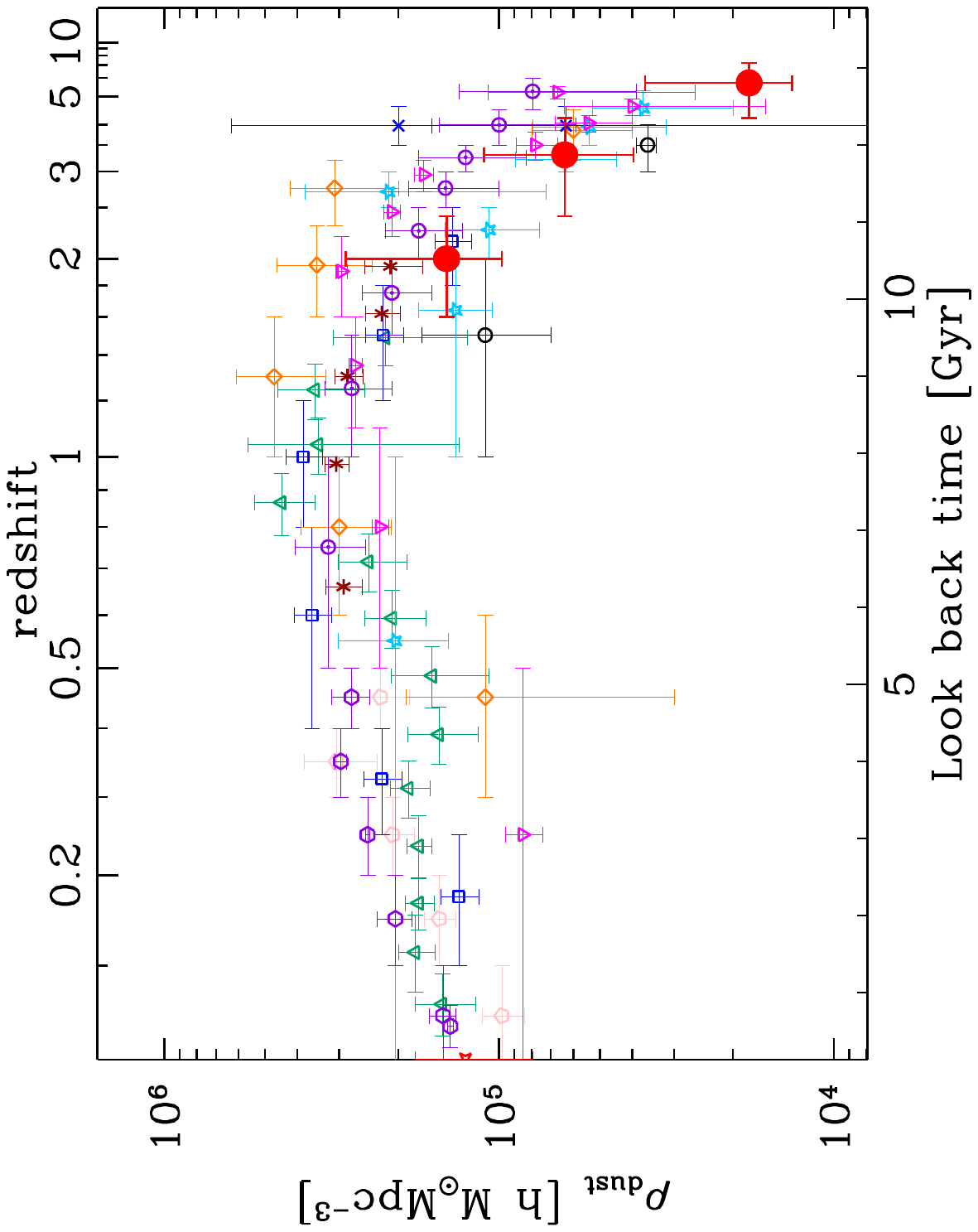}}
\settowidth{\imagewidth}{\includegraphics{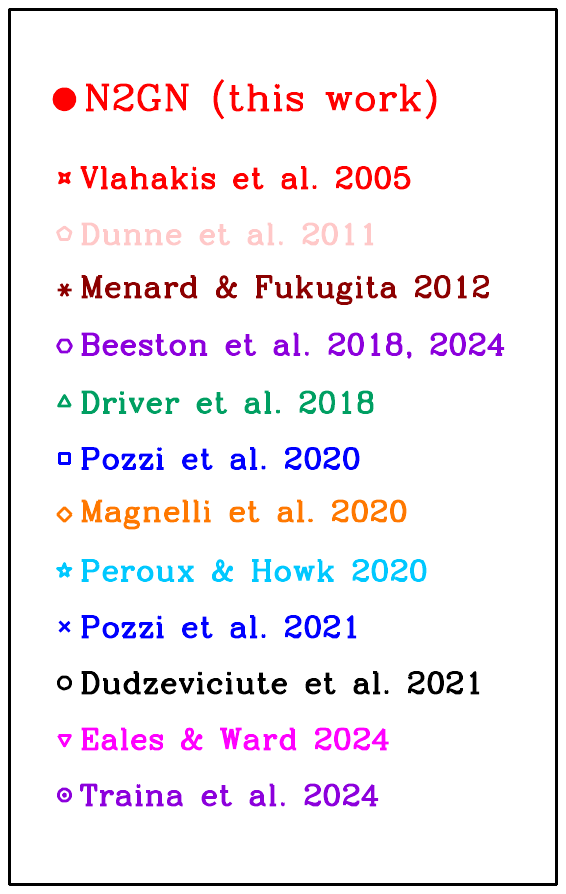}}
\begin{minipage}{0.4\textwidth}
\rotatebox{-90}{\includegraphics[width=\imageheight]{figs1/rho_z_no_mod_NO_renorm_20250130.pdf}}
\end{minipage}
\begin{minipage}{0.4\textwidth}
\rotatebox{-90}{\includegraphics[width=\imageheight]{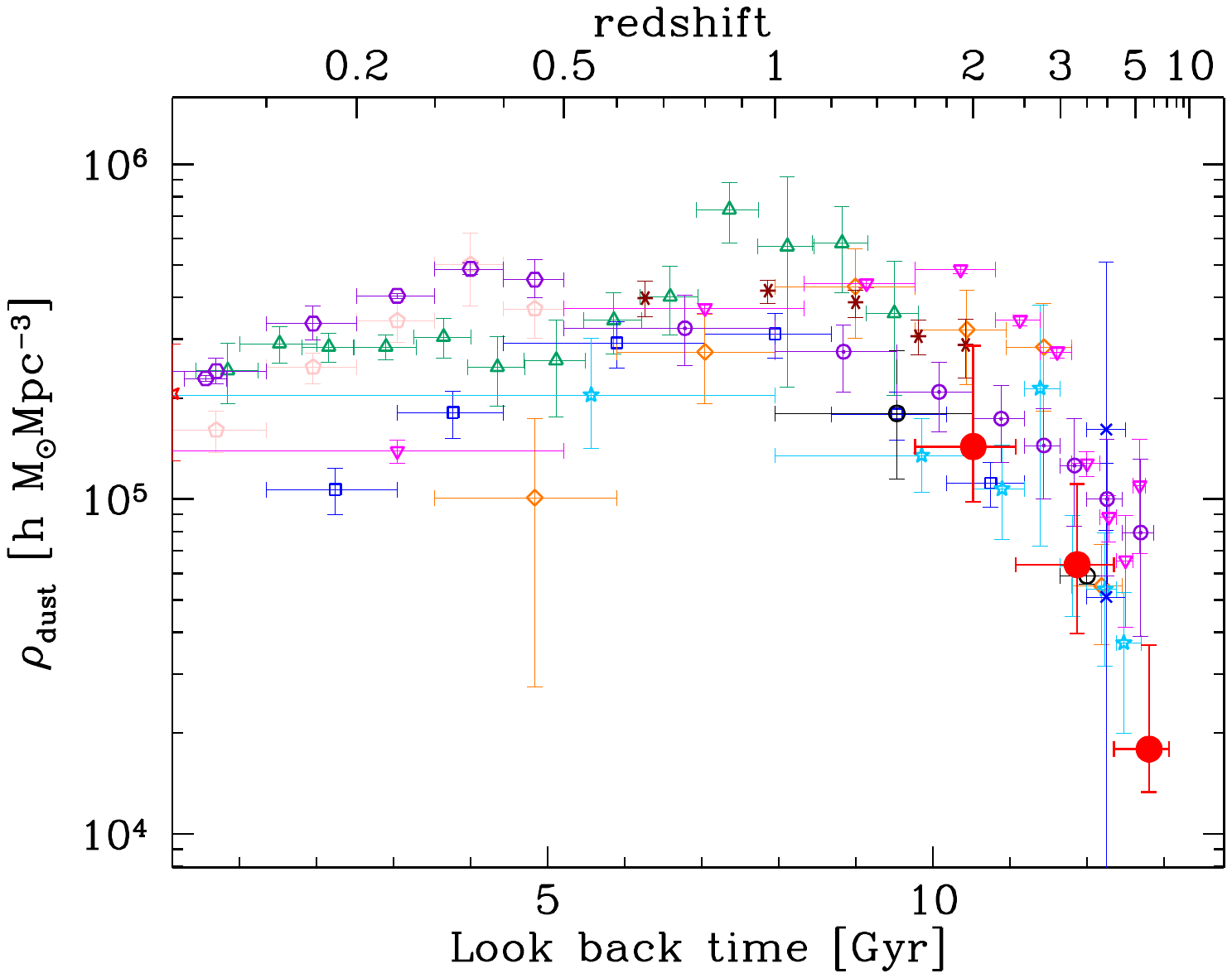}}
\end{minipage}
\begin{minipage}{0.19\textwidth}
\raggedleft
\includegraphics[trim=0.3\imagewidth{} 0.37\imagewidth{} 0.25\imagewidth{} 0.10\imagewidth{}, clip, height=0.9\imageheight]{figs1/legenda_rho5_20250130.pdf}
\end{minipage}
\caption{Effect of rescaling $\rho_\textrm{dust}$ to the same $\kappa_\nu$. {\em Left}: data without rescaling; each data set is shown as obtained with its $\kappa_\nu$ underlying assumption. {\em Right}: the same data after rescaling to $\kappa_{850}=0.047$ m$^2$ kg$^{-1}$ \citep{draine2014}. }
\label{fig:kappa_rho}
\end{figure*}

Table \ref{tab:kappa_summary} summarizes the values adopted in literature, as described above. Figure \ref{fig:kappa_rho} shows the importance of reconciling all the data to the same $\kappa_\nu$ assumptions. In the left panel the data are shown ``as they are'' computed by each work, each with its specific value of $\kappa_\nu$. The right panel is a repetition of Figs. \ref{fig:rho_z} and \ref{fig:rho_z_mod}, in full color and excluding the models, after rescaling all results to the \citet{draine2014} choice of reference.

\begin{table*}[!h]
\centering
\caption{\label{tab:kappa_summary}Summary of the $\kappa_\nu$ values adopted in the literature.}
\begin{tabular}{lll}
\hline
\hline
Adopted $\kappa_\nu$ & $\kappa_\nu$ reference & Work\\
$[$m$^2$ kg$^{-1}]$ & &\\
\hline
MW $\kappa_{850}=0.047$ & \citet{draine2014} & This work\\
&& \citet{peroux2020}\\
&& \citet{traina2024}\\
%&& \\
MW $\kappa_{850}=0.038$ & \citet{draine2003} & \citet{pozzi2020}\\
&& \citet{pozzi2021}\\
%&& \\
MW $\kappa_{850}=0.077$ & \citet{dunne2000} & MAGPHYS\\
&& \citet{beeston2018}\\
&& \citet{beeston2024}\\
&& \citet{driver2018}\\
&& \citet{dudzeviciute2021}\\
&& \citet{dunne2011}\\
&& \citet{eales2024}\\
&& \citet{vlahakis2005}\\
MW $\kappa_{850}=0.043$ & \citet{LD01} & \citet{magnelli2020}\\
SMC $\kappa_V=1.54\times10^3$& \citet{wd01} & \citet{menard2012}\\
\hline
\end{tabular}
\end{table*}

\section{$\rho_\textrm{dust}$ versus $z$}

In this brief Appendix, Figs. \ref{fig:rho_z} and \ref{fig:rho_z_mod} are presented with redshift on the main x-axis instead of look-back time, as is often found in the literature. 

\begin{figure*}[!h]
\centering
\settowidth{\imageheight}{\includegraphics[height=0.43\textwidth]{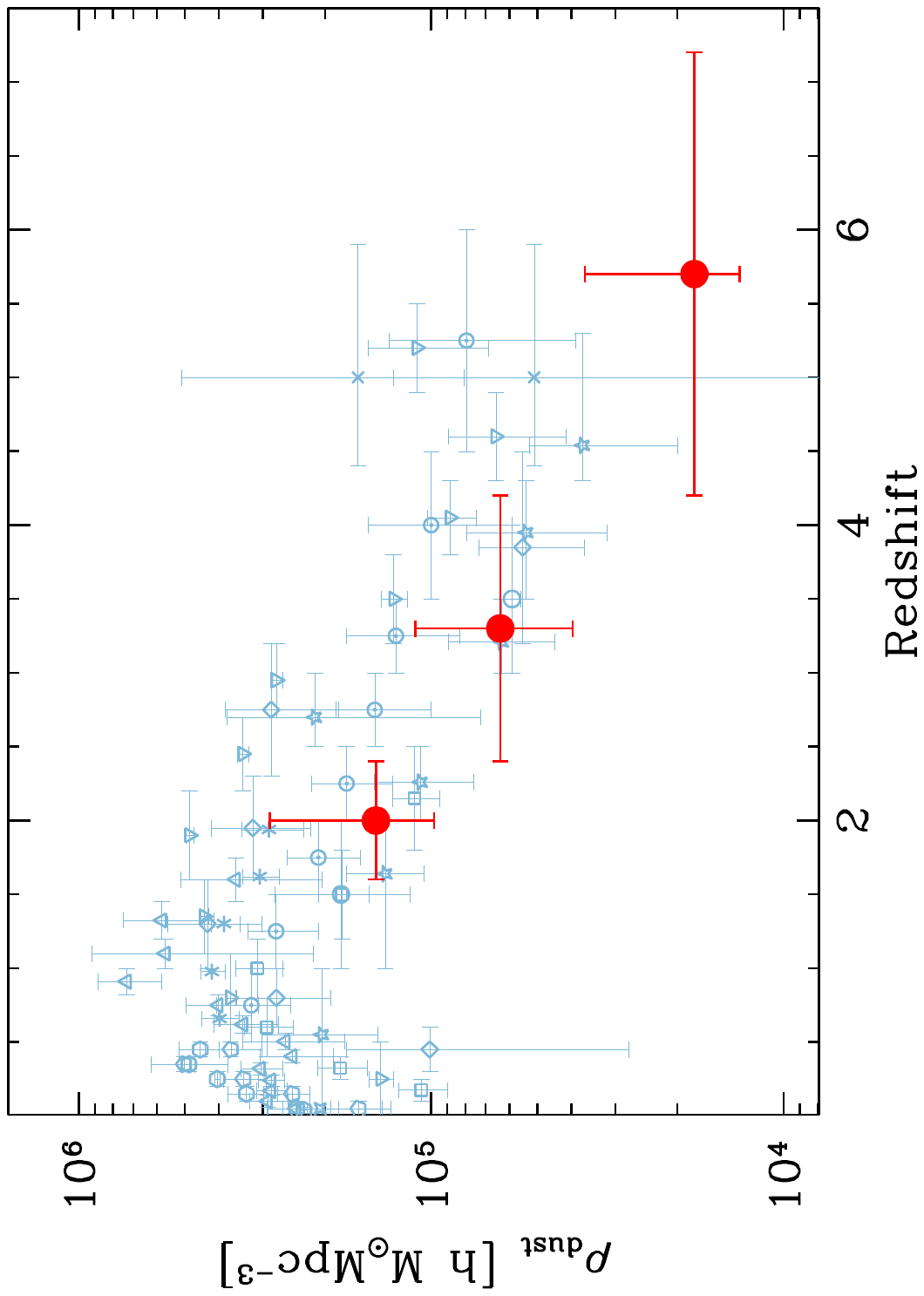}}
\settowidth{\imagewidth}{\includegraphics{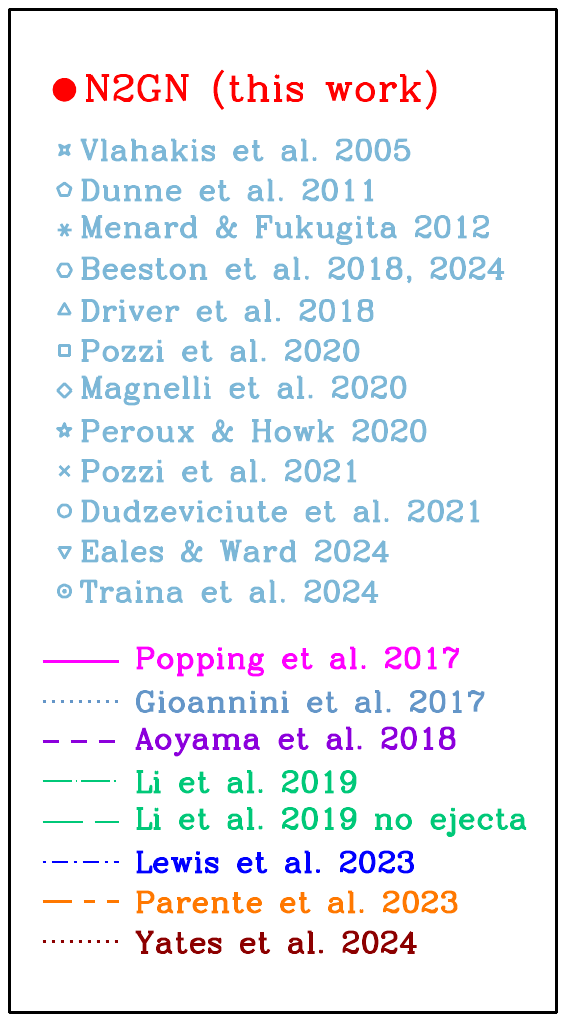}}
\begin{minipage}{0.4\textwidth}
\rotatebox{-90}{\includegraphics[width=\imageheight]{figs1/rho_zzz3_ash_no_mod_20250130.pdf}}
\end{minipage}
\begin{minipage}{0.4\textwidth}
\rotatebox{-90}{\includegraphics[width=\imageheight]{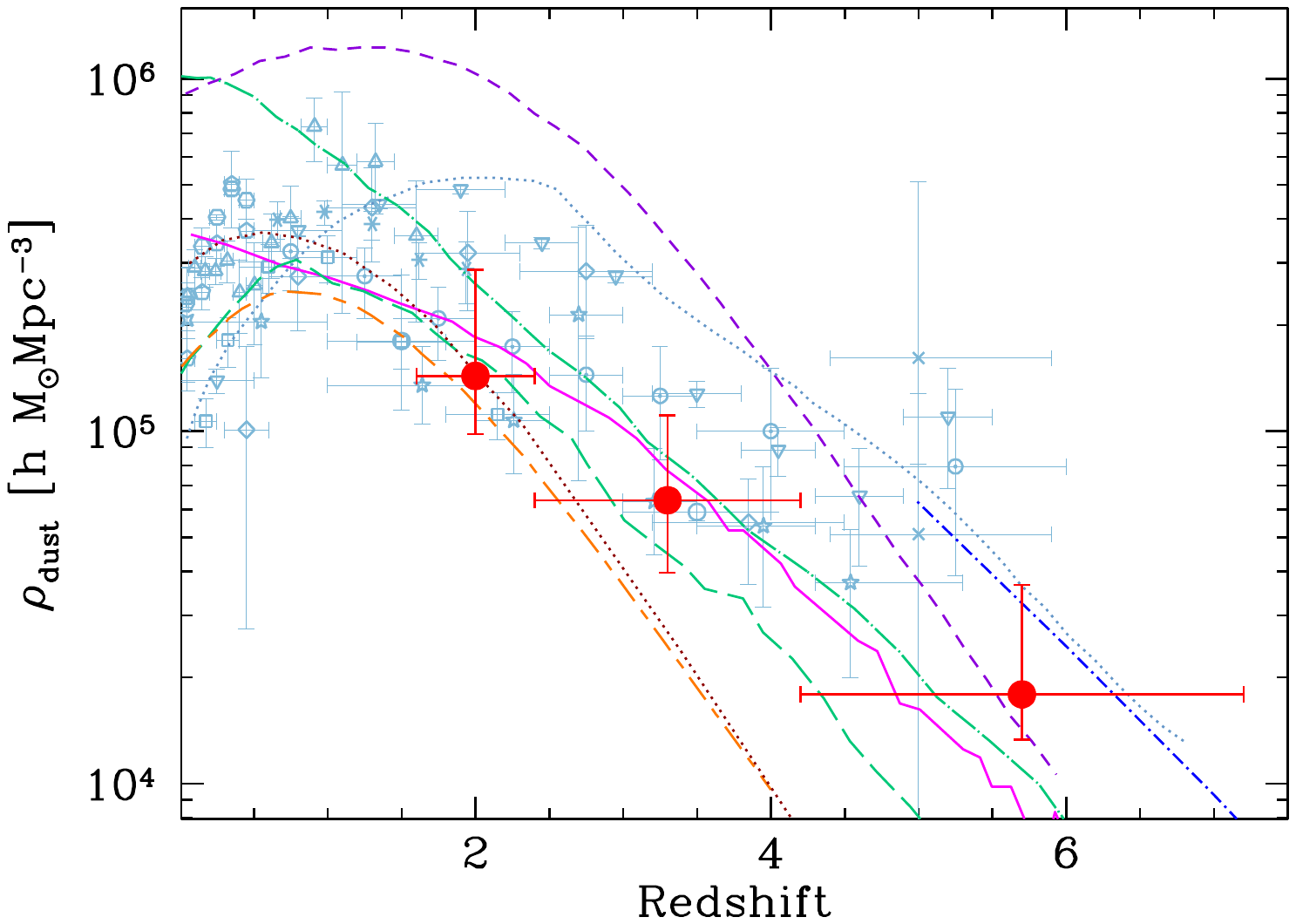}}
\end{minipage}
\begin{minipage}{0.19\textwidth}
\raggedleft
\includegraphics[trim=0.3\imagewidth{} 0.32\imagewidth{} 0.25\imagewidth{} 0.10\imagewidth{}, clip, height=0.9\imageheight]{figs1/legenda_rho3_20250130.pdf}
\end{minipage}
\caption{Cosmic density of dust in star forming galaxies as a function of redshift. Figures \ref{fig:rho_z} and \ref{fig:rho_z_mod} present the same data as a function of look-back time.}
\label{fig:rho_zzz}
\end{figure*}

\clearpage

\twocolumn
\section{Notes on individual sources}\label{app:indiv_srcs}

%%% ---------

%%% X-ray identifications (cfr also Sect \ref{sect:xray}).

%OK N2GN\_1\_03      in Barro+2019 (matched to Alexander+2003) + 2020 CDC, AGN fit
%OK N2GN\_1\_05      in Barro+2019 (matched to Alexander+2003), no need for AGN 
%OK N2GN\_1\_10      in Barro+2019 (matched to Alexander+2003) + 2020 CDC, AGN fit
%OK N2GN\_1\_11      in the 2020 CDC catalog, AGN fit
%OK N2GN\_1\_14      in Barro+2019 (matched to Alexander+2003), no need for AGN
%OK N2GN\_1\_19	    in Wang+2016, AGN fit
%OK N2GN\_1\_21      in Barro+2019 (matched to Alexander+2003), no need for AGN
%OK N2GN\_1\_25      in Barro+2019 (matched to Alexander+2003), no need for AGN
%OK N2GN\_1\_32      in Barro+2019 (matched to Alexander+2003) + 2020 CDC, AGN fit
%N2GN\_1\_37      in Barro+2019 (matched to Alexander+2003), no need for AGN
%OK N2GN\_1\_39      in Barro+2019 (matched to Alexander+2003), no need for AGN
%OK N2GN\_1\_45      in Barro+2019 (matched to Alexander+2003), no need for AGN
%OK N2GN\_1\_46	    Brandt+2001 + WISE + IRS 16um, AGN fit
%OK N2GN\_1\_47      in Barro+2019 (matched to Alexander+2003), no need for AGN
%OK N2GN\_1\_51      in Barro+2019 (matched to Alexander+2003), no need for AGN
%OK N2GN\_1\_54      in Barro+2019 (matched to Alexander+2003), no need for AGN
%OK N2GN\_1\_56_a    in Barro+2019 (matched to Alexander+2003), no need for AGN

%%% ---------

Figure \ref{fig:indiv_glxs} presents the multi-wavelength postage stamps, the FIR SED fitting and the panchromatic optical-to-radio modeling of all N2GN sources. In this Appendix, some notes about the N2GN individual sources are given.

{\bf N2GN\_1\_01} is an optically dark galaxy, identified with the GN10 SCUBA source \citep{pope2006} at $z=5.303$. It was detected at 2~mm by GISMO on the IRAM 30~m telescope \citep[GDF2000.6;][]{staguhn2014}. It is detected by the SMA \citep{cowie2017} and NOEMA \citep{riechers2020}. The latter authors studied the $\rm ^{12}CO$ spectral line energy distribution (SLED) of this source and its dynamical properties, by combining NOEMA and VLA observations of $\rm ^{12}CO$ and [C{\sc ii}] 158 $\mu$m. GN10 is a compact starburst ($\sim1.6$ kpc in diameter) hosted in a $\sim6.4$ kpc rotating cold gas disk, with a total dynamical mass of $8.6\times\,10^{10}\, M_\odot$. The observed SED includes data from JWST,  HST, {\it Spitzer, Herschel}, NIKA2, NOEMA and VLA.

{\bf N2GN\_1\_02}'s SMA position is $\sim2$ arcsec away from the SCUBA source GOODS 850.6 \citep{wang2004,barger2012} and 0.25 arcsec from AzGN5 \citep{chapin2009}. It was previously assigned a low photometric redshift \citep{cowie2017}, likely related to the nearby galaxy lying at only $\sim6$ arcsec away. We adopt the $z_\textrm{phot}=1.963$ reported by \citet{kodra2023} and based on optical-NIR SED fitting.

{\bf N2GN\_1\_03} was detected by SCUBA2 at 450 $\mu$m and corresponds to GOODS 850-9 and GN19 \citep{pope2006}, with a spectroscopic redshift $z=2.484$. It lies close to a pair of radio sources. The SMA position identifies it with the southwestern one. It was also detected by IRS at 16 $\mu$m with a flux of 32.6\,$\mu$Jy \citet{teplitz2011}. At its position, the 2 Ms {\it Chandra} map of the CDFN reveals a X-ray source with a hard spectrum and fluxes of 5.94 and $1.56\,10^{-5}$ $\mu$Jy in the hard (2-7 keV) and soft (1.2-2 keV) band, respectively \citep{alexander2003,evans2024}. We fit the SED including an AGN-torus component, which contributes to the UV (and X-rays) emission of the galaxy but is negligible in the FIR-millimeter domain. The VLA radio measurements are in agreement with the radio-FIR correlation, given the {\it Herschel}, SCUBA2 and NIKA2 flux densities.

{\bf N2GN\_1\_04} is identified with GN20 \citep{pope2005,daddi2009}, a well studied $z=4.055$ starburst system of two interacting galaxies. The NIKA2 source coincides with the brightest of the two (GN20 proper), while GN20.2 is too far for contributing to the NIKA2 flux density.  
\citet{cortzen2020} studied the [C{\sc i}](2-1) to (1-0) ratio, as measured with NOEMA, finding that its excitation temperature points to an optically thick dusty medium with optical depth reaching unity at $170\pm23$ $\mu$m. Recent JWST MIRI and NIRSpec observations reveal a conspicuous central source hosted by an extended envelope, faint stellar clumps associated to CO similar structures, a turbulent rotating disk (500 km/s) with additional non circular motions, and a very broad H$\alpha$ line in the central regions, suggestive of strong, possibly AGN driven, winds \citep{colina2023, uebler2024, crespogomez2024}.

{\bf N2GN\_1\_05} corresponds to the source HDF254 observed at 1.4 GHz at high resolution with MERLIN by \citet{bothwell2010}. The radio observations show five peaks within a 1.5 arcsec maximum distance from each other, with a large velocity dispersion ($249\pm18$ km/s) indicative of a possible merger. A strong MERLIN point source 0.5 arcsec to the west might be an AGN component, although the optical and MIR spectroscopy show a normal H$\alpha$ to [N{\sc ii}] line ratio and strong PAH features \citep{swinbank2004,pope2008}. Our W21CV NOEMA program detected an emission line at 154.2 GHz, consistent with being $\rm ^{12}CO$(4-3) at $z=1.996$, the same redshift measured by optical spectroscopy \citep[3D-HST;][and those previously mentioned]{skelton2014}.
The proxy position lies between the two GISMO 2\,mm sources GDF2000.3 and GDF2000.8 \citep{staguhn2014}.
Although the source is detected in the 2 Ms CDFN X-ray map \citep{barro2019, alexander2003, bothwell2010}, no AGN-torus component is needed to fit its optical to millimeter SED, nor the VLA continuum shows any excess with respect to the radio-FIR correlation.

{\bf N2GN\_1\_06} is HDF850.1, the first and brightest sub-millimeter galaxy discovered by SCUBA in the {\it Hubble} Deep Field \citep{hughes1998}. It was also  detected at 2~mm by GISMO \citep[GDF2000.1;][]{staguhn2014}. It is an optically dark galaxy and it took nearly 14 years to determine its redshift. \citet{walter2012} and \citet{neri2014} placed it at $z=5.184$, detecting CO and [C{\sc ii}] 158 $\mu$m at millimeter wavelengths with the PdBI, precursor of NOEMA. HDF850.1 is a starburst galaxy belonging to the known $z\sim5.2$ overdensity in the GOODS-N field. JWST detected its rest-frame UV-optical emission, solving the long-lasting puzzle of its identification \citep{sun2024}. The galaxy is mildly lensed by a $z=1.224$ elliptical galaxy, with a magnification factor $\mu\sim2.5$ 
The NIRCAM photometry reveals a rest-frame UV excess with respect to the light emitted from the extinguished stars that power its bright IR emission. We fit the SED of this galaxy adding a unextinguished young SSP to reproduce this excess (Sects.~\ref{sect:sed3fit} and~\ref{sect:blue_excess}).

{\bf N2GN\_1\_07} corresponds to GN12 \citep{pope2005,daddi2009,mancini2009}. It was detected at 2~mm also by GISMO \citep[GDF2000.11;][]{staguhn2014}. NIRSpec Wide GTO data give z$_{\rm spec}$=2.9601.
The source lies in a very crowded field, with the consequence that the {\it Herschel}/SPIRE photometry possibly includes the contribution of a few strongly blended sources and was therefore not used in our SED modeling. We kept instead the PACS photometry, that has a PSF size matching well the NIKA2 beam, as well as the 450\,$\mu$m SCUBA2 measurement \citep{barger2022}. The SED is well reproduced with a starburst model; the available VLA data are in slight excess with respect to the radio-FIR correlation. 

{\bf N2GN\_1\_08} is detected by SMA at a position 0.15 arcsec away from AzGN2 \citep{chapin2009} and $\sim 4$ arcsec away from  
JWST JADES-GN-189.13100+62.28684 \citep[$z=5.267$;][]{sun2024}. A new JWST extraction reveals a large obscured galaxy at a projected distance of 0.04 arcsec, composed of four ``blobs'' (M.~Xiao, priv. comm.). We used CIGALE to derive the photometric redshift of the latter, $z_\textrm{phot}=4.34\pm 0.16$. 

{\bf N2GN\_1\_09} is the ID12646 in the {\it Spitzer}-driven super-deblended catalog by \citet{liu2018}, and the AzGN10 source detected by Atzec at 1.1\,mm \citep{perera2008,chapin2009}, pinpointed by VLA. Its position is 0.465 arcsec away from the IRAC J123627.54+621218.4 object reported in \cite{enia2022}. It is an optically dark galaxy and was observed with NOEMA by \citet{jin2022} and \citet{bingthesis} at $z_\textrm{spec}=4.147$. Its optical counterpart is finally identified on JWST maps with a very red galaxy at $\sim0.3$ arcsec distance.

{\bf N2GN\_1\_10} is a known bright AGN at $z=2.005$, detected in the {\it Chandra} 2 Ms map with fluxes of 2.92, 3.66 and $18.10\times10^{-5}\,\mu$Jy in the 0.5-1.2, 1.2-2.0 and 2.0-7.0 keV bands, respectively \citep{evans2024}. It is also detected by WISE and {\it Akari} in the MIR and at 8.5, 5.5 and 1.4\,GHz in the radio \citep{wise2013, pearson2010, richards1998, guidetti2017, cowie2017}. The MIR SED is a well defined power-law and is reproduced with an important AGN-torus component in the SED fitting. The radio emission, nevertheless, is consistent with the radio-FIR correlation, indicating that the AGN does not contribute significantly in this spectral domain.

{\bf N2GN\_1\_11} is an optically dark source not present in the \citet{barro2019} catalogs. It lies at 0.223 arcsec from the IRAC J123713.86+621826.6 object reported by \citet{enia2022}. It has a bright radio counterpart, which is here used to pinpoint its precise position and corresponds to GN40 or AzGN26. To date, no optical photometry is available. The photometric redshift $z=3.99$ was estimated by \citet{wang2016} based on F160W, Ks and {\it Spitzer} photometry. The object is a bright X-ray source with fluxes of 5.1, 7.9 and $10.7\,10^{-5}\,\mu$Jy in the 0.2-0.5, 0.5-1.2 and 2.0-7.0 {\it Chandra} bands. The SED fitting includes a bright AGN-torus component that dominates the IRAC 8 $\mu$m, IRS 16 $\mu$m and MIPS 24 $\mu$m observed fluxes. The 21 cm emission is in excess with respect to the radio-FIR correlation by at least a factor $\sim5$.

{\bf N2GN\_1\_12} is a blend of two PdBI 255 GHz sources \citep{bingthesis}.
The two components, named 12a and 12b, correspond to ID13161 and ID13207 in \citet{liu2018} and are detected by both HST and JWST. The coordinates of 12a are 0.7\,arcsec away from those of GN11 \citep{pope2005,pope2006,mancini2009}. The high-resolution optical data show two peaks each, belonging to the same galaxies. The photometric redshift of 12a is determined here with CIGALE ($z=2.59\pm0.08$) and that of 12b was estimated by \citet[][$z=4.163$]{kodra2023}. Because of blending, the FIR-millimeter emission is defined by the NOEMA fluxes only. Source 12b is also detected at 21\,cm and is consistent with the radio-FIR correlation.

{\bf N2GN\_1\_13} is an optically dark galaxy. Its position and spectroscopic redshift has been determined by recent NOEMA observations with $\rm ^{12}CO$(4-3) and (5-4) transitions at $z=5.181$ (this work and Lagache et al. in prep.). It is now detected by JWST/FRESCO (M.~Xiao, priv. comm.). No MIR data are available to verify the possible presence of an AGN-torus component contributing to its MIR SED. The optical photometry reveals a rest-frame UV excess, that we reproduce by adding a unextinguished young SSP to the SED modeling (Sect.~\ref{sect:blue_excess}).

{\bf N2GN\_1\_14} is identified with a faint optical galaxy at $z_\textrm{spec}=1.76$ thanks to the SMA observations \citep{cowie2017}. Its coordinates match those of GN17. The SED is very complete and finely sampled with 27 data points, a well-defined 1.6\,$\mu$m (rest-frame) stellar ``bump'', and data on both sides of the FIR emission peak. It is detected in the X-rays \citep{barro2019, alexander2003}, but no AGN-torus component is needed to reproduce its optical-to-radio emission.

{\bf N2GN\_1\_15}, similarly to the previous one, is identified by SMA and has a very well defined SED at $z_\textrm{spec}=2.0$. It corresponds to GN6. The deep optical data show an excess that we model with an unextinguished young SSP. 

{\bf N2GN\_1\_16} is a high-redshift galaxy, identified by SMA and detected in the optical by HST \citep{kodra2023,liu2018}. It lies at the position of GN9. Including JWST data (one source at a distance of 0.23 arcsec, M.~Xiao priv. comm.) we obtain a photometric redshift of $z_\textrm{phot}=3.47\pm0.45$.

{\bf N2GN\_1\_17} is split in two components by NOEMA (Sect. \ref{sect:data_NOEMA}), named 17a and 17b. The component 17b lies very close to a bright star, and therefore no optical-IR photometry is available, nor a redshift estimate. Because of blending, the SPIRE, SCUBA2 and NIKA2 data are not considered and the FIR emission of 17a is defined by NOEMA alone. \citet{liu2018} deblended the MIPS and IRAC emission. Its position matches that of GN37. The spectroscopic redshift, $z_\textrm{spec}=3.190$, was measured by \citet{cowie2004}. 

{\bf N2GN\_1\_18} is an optically dark galaxy. \citet{cowie2017} associated it to a Keck/LRIS and DEIMOS redshift but it has no K band counterpart on the CFHT maps; therefore we ignored it. In the JWST grism spectra, it  has a bright neighboring source in the same dispersion direction. Therefore, these spectra were of no use either. Thanks to the NOEMA data, we identify it on the JWST maps and derived a new $z_\textrm{phot}=4.39\pm0.15$ with CIGALE. 

{\bf N2GN\_1\_19} is matched through SMA to a X-ray source at $z=2.578$. Its integrated 0.3-8.0\,keV flux is $1.68\times10^{-4}\,\mu$Jy \citep{wang2016}. It lies at the same position of GN4 and AzGN16 and shows an 8.0\,$\mu$m excess with respect to its stellar emission but no radio excess. We model the SED including an AGN-torus component.

{\bf N2GN\_1\_20} is another optically dark galaxy, detected by SMA. It lies at a distance of 1.3 arcsec from AzGN4 \citep{chapin2009}. It corresponds to ID14914 by \citet{liu2018} and has a Ks band flux of 0.75 $\mu$Jy, resulting in a very red Ks to 3.6 $\mu$m color, $S_{3.6}/S_\textrm{Ks}\sim 13$, interpreted as a deep Balmer break combined to a large extinction, that leads to a $z_\textrm{phot}=5.333$ \citep{liu2018}. In our SED fitting, the bright 24 $\mu$m flux is contributed by a strong 3.3\,$\mu$m PAH feature.

{\bf N2GN\_1\_21} is identified with a $z_\textrm{spec}=1.219$ radio source at a projected distance of 0.25\,arcsec from the N2CLS position, with a rich and very well sampled SED. \citet{barro2019} matched it to an X-ray source \citep{alexander2003}, but no AGN-torus component is needed in the SED fitting. A JADES $z=5.446$ source \citep{sun2024} lies at a projected distance of 1.05\,arcsec from the N2CLS position.

{\bf N2GN\_1\_22} lies in a crowded area. The SMA counterpart seems blended with a different bright 24 $\mu$m object. Hence, we adopted the MIPS and IRS fluxes deblended by \citet{liu2018} and we do not consider the {\it Herschel} photometry. \citet{kodra2023} give a spectroscopic redshift $z_\textrm{spec}=2.098$.

{\bf N2GN\_1\_23} is an optically dark galaxy detected by JWST/FRESCO at $z_\textrm{spec}$=5.179 \citep{xiao2024}. The NOEMA identification is at a distance of only 0.167\,arcsec. The MIR and FIR corresponding sources are blends of several objects, and therefore we ignored them for the SED. The NIKA2 position is 0.336 arcsec away from the GOODS J123656.5+621207 objects mentioned by \citet[][with $z_\textrm{phot}=3.03$]{enia2022}.

{\bf N2GN\_1\_24} is a blend of two millimeter sources detected by NOEMA at $\sim 10$ arcsec distance from each other and named 24a and 24b. In their SEDs, the FIR emission is defined by NOEMA alone because of blending. The NOEMA positions allow the sources to be matched with two optical galaxies at $z_\textrm{phot}=2.697$ and 2.817 \citep{kodra2023}.

{\bf N2GN\_1\_25} was observed with NOEMA and identified as an optical galaxy at $z_\textrm{spec}=2.914$. The exquisite SED shows clear Lyman and Balmer breaks. In the \citet{barro2019} catalog it is matched to an X-ray source, but neither the MIR or the radio emission is in excess with respect to the SED of a star forming galaxy.

{\bf N2GN\_1\_26} is identified with a high-$z$ HST galaxy thanks to SMA data. It corresponds to AzGN18 or GN38. \citet{kodra2023} give a photometric redshift $z_\textrm{phot}=5.31$. The FIR SED is well defined by the PACS, SCUBA2, SMA, and NIKA2 data. The 24\,$\mu$m flux is well matched by an intense 3.3 $\mu$m PAH feature.

{\bf N2GN\_1\_27} is the merger of two galaxies (27a and 27b), identified with NOEMA and at nearly the same redshift \citep[$z_\textrm{spec}=1.988,\,1.989$;][]{kodra2023}. The system lies at the position of GN7. It has been detected by IRS at 16\,$\mu$m \citep{teplitz2011}, for which we adopt the deblended fluxes by \citet{liu2018}. 

{\bf N2GN\_1\_28} is identified with VLA 21\,cm source \citep{owen2018} and has a spectroscopic redshift $z_\textrm{spec}$= 3.2223 from the NIRSpec Wide GTO. The very red optical SED is well fitted by an intense starburst model and the VLA flux is consistent with the radio-FIR correlation.

{\bf N2GN\_1\_29} is also identified through VLA data and is at the position of GN18. A JADES $z=5.432$ source lies at a projected distance of 3.9 arcsec \citep{sun2024}, but the $z=2.819$ radio source has a higher match probability. The SED is well sampled and is well fitted with a starburst model.

{\bf N2GN\_1\_30} is very close to N2GN\_1\_51; therefore we excluded the SPIRE photometry that is partially blended. An AGN-torus component is required to fit the optical blue slope, %much shallower: 
as a pure stellar model does not reproduce successfully both the 8\,$\mu$m and the observed U, F435W and F606W photometry at the same time; in the best fit solution, the AGN emission supplies the optical and the stellar component dominates in the NIR.

{\bf N2GN\_1\_31} has a bright VLA 1.4\,GHz counterpart, also detected by HST and JWST. It corresponds to AzGN20 and its redshift is $z_\textrm{phot}=2.638$ \citep{kodra2023}. In the SPIRE 500\,$\mu$m channel, the source is heavily blended with another, brighter, object; therefore we ignore this band.

{\bf N2GN\_1\_32} is a radio source at $z_\textrm{spec}=3.652$ \citep{owen2018,barger2008,kodra2023}. It is detected in the {\it Chandra} 2 Ms map of the CDFN with fluxes of 2.42, 2.23, 0.95, and $1.11\times\,10^{-4}\,\mu$Jy in the 0.2-0.5, 0.5-1.2, 1.2-2.0, and 2.0-7.0 keV bands, respectively \citep{evans2024}. The observed rest-frame NIR-MIR SED is a power law and is dominated by an AGN-torus component in the IRAC and MIPS bands. A stellar component contributes the bulk of the optical emission. The radio emission does not show any excess and is consistent with the radio-FIR correlation. 

{\bf N2GN\_1\_33} corresponds to the SMA GOODS-N source nr. 13 and SCUBA2 nr. 63 \citep{cowie2017}, as well as to GN25. Three sources lie inside the SMA beam: a bright $z_{\rm spec}=1.013$ optical galaxy, a very faint optical galaxy revealed by JWST, and a radio source which lies in the middle of the two optical sources. 
The bright and faint optical galaxies are respectively at 1.21 and 0.56\,arcsec from the SMA position. We chose the closest one as the identification. 
\citet{kodra2023} reports a redshift $z_\textrm{phot}=5.388$. Because of blending, we ignore the {\it Spitzer} and {\it Herschel} data in the SED fitting. Therefore, a large gap in the wavelength coverage exists between the rest-frame optical and sub-millimeter spectral domains.

{\bf N2GN\_1\_34} corresponds to AzGN28 \citep{chapin2009} but breaks in two components (34a and 34b), when observed with NOEMA. Both are optically dark sources. The 34a component lies very close to a bright nearby galaxy, it might be lensed, but no counterpart is found in any catalog or map. On the other hand, 34b is an isolated source detected in the IRAC, IRS 16\,$\mu$m, MIPS 24\,$\mu$m channels, and at 1.4\,GHz by VLA. \citet{enia2022} report an optically dark source at 0.233 arcsec from its position (GOODS J123644.0+621938). CIGALE gives a redshift $z_\textrm{phot}=4.12\pm 0.53$ taking into account the available photometry (Sect.~\ref{sect:redshift}). 

{\bf N2GN\_1\_35} is identified with VLA with an edge-on $z_\textrm{phot}$=1.95 galaxy. The SED is superb, with 26 bands covering the whole electromagnetic spectrum from 3500 \AA\ to 21 cm and allows a textbook fit.

{\bf N2GN\_1\_36} is an optically dark source that lies very close ($<5$ arcsec) to a bright star. The SMA data pinpoint its position. It lies at 0.233 arcsec from the GOODS J123658.5+620931 source discussed in \cite{enia2022}. \citet{liu2018} derive a redshift $z_\textrm{phot}=3.55$, but their photometry is affected by the nearby star. Here we adopt a new HST extraction by the FRESCO team (M.~Xiao, priv. comm.) and derive a new estimate of $z_\textrm{phot}=3.93\pm0.17$ with CIGALE. Similarly to the case of N2GN\_1\_30, an AGN-torus component is needed:  the bright NIR emission is dominated by stars and the blue excess by the AGN.

{\bf N2GN\_1\_37} is an isolated galaxy at $z_\textrm{spec}=2.302$ \citep{kodra2023}, whose accurate position is given by the VLA data. Its observed SED consists of 24 data points. \citet{barro2019} match it to an X-ray source in the 2 Ms {\it Chandra} map \citet{alexander2003}, but its optical-to-radio SED is well reproduced by a simple starburst model.

{\bf N2GN\_1\_38} lies in a crowded region of GOODS-N. The NOEMA position matches that of a $z_\textrm{phot}=3.126$ optical galaxy detected by HST and {\it Spitzer}/IRAC \citep{kodra2023, liu2018}. The FIR-millimeter emission is defined by the 850\,$\mu$m to 2.0\,mm data only, because of the heavy blending affecting the {\it Herschel} data.

{\bf N2GN\_1\_39} is a $z_\textrm{spec}=2.301$ galaxy, identified thanks to the NOEMA data. Its SED shows a clear 1.6 $\mu$m stellar peak and a well defined FIR dust emission sampled by PACS, SCUBA2, NIKA2 and NOEMA. It is detected in the X-rays \citep{barro2019,alexander2003} but no AGN torus is needed in the SED fitting. The radio measurement is consistent with being powered by star formation only.

{\bf N2GN\_1\_40} is a galaxy at $z_\textrm{phot}=2.764$ \citep{kodra2023}. In the 1.2\,mm NOEMA map, two millimeter sources are detected, but the fainter one lies 7.87\,arcsec apart and does not contribute significantly to the NIKA2 flux. The {\it Herschel} and MIPS photometry is affected by a significant blending and is therefore ignored. It is clearly detected in the HST and JWST maps at the same NOEMA position. The observed SED has a large wavelength gap between the observed NIR (IRAC) and sub-millimeter data, implying that the rest-frame MIR and the peak of the FIR emission are not well sampled.

{\bf N2GN\_1\_41} has a very red, but complex optical SED, with a slight flux offset between the HST and JWST measurements. The redshift $z_\textrm{phot}=4.101$ by \citet{kodra2023} is based on the HST, Ks  and IRAC photometry. Its position is pinpointed by our NOEMA data. It was not detected by {\it Herschel}, nor by MIPS. 

{\bf N2GN\_1\_42} is in a crowded area and the actual multi-wavelength counterparts are identified thanks to the NOEMA data. The IR deblended photometry by \citet{liu2018} is used. It is a $z_\textrm{phot}=2.228$ optically bright galaxy showing a distinct  break at $\sim 1.5$\,$\mu$m in the observed frame and a strong 1.6\,$\mu$m stellar peak in the IRAC 5.8\,$\mu$m band. The 1.4\,GHz measurement is in excess to the radio synchrotron powered by star formation by more than an order of magnitude, indicating the presence of an AGN, but no signs  are detected at MIR or optical wavelengths. 

{\bf N2GN\_1\_43} is an optically and K-band dark galaxy, detected by JWST at $\sim 10\sigma$ in the F277W band, $\sim3\sigma$ in F200W, and tentatively by HST at shorter wavelengths (M.~Xiao, priv. comm.). NOEMA matches it with a clear detection in the {\it Spitzer} IRAC bands, but the object is not visible in the MIPS and {\it Herschel} maps. CIGALE places it at $z_\textrm{phot}=5.87\pm0.38$ with the Balmer break between the two JWST measurements mentioned here.

{\bf N2GN\_1\_44} is identified as the GNz7q source studied by \citet{fujimoto2022}. These authors report on the results by NOEMA and HST, highlighting its compact morphology at rest-frame UV wavelength, a bright FIR-millimeter emission, and a MIR excess in the observed 24\,$\mu$m band. The object is not detected in the {\it Chandra} 2 Ms X-ray map. Its rest-frame UV spectrum does not show any bright, nor broad, line between Ly$\alpha$ and C{\sc iii}]$\lambda$1909. NOEMA detected the [C{\sc ii}] 158 $\mu$m line at redshift $z=7.1899$, consistent with the observed Lyman break.
Two possible interpretations of the emission of GNz7q are proposed by \citet{fujimoto2022}: a UV-compact star-forming object or (their favorite scenario) a red quasar explaining the MIR excess at odds with the lack of broad UV Mg{\sc ii} and C{\sc iv} lines.
Our SED fitting reproduces the source with a blue star forming galaxy model, characterized by a prominent 3.3 $\mu$m PAH feature that accounts for the observed MIR excess. On the other hand, SED3FIT and CIGALE failed at finding a viable solution including a type-1 AGN torus, as suggested by \citet[][their Fig.~2]{fujimoto2022}.

{\bf N2GN\_1\_45} is pinpointed by the VLA data \citep{owen2018} and matches a $z_\textrm{phot}\sim3$ galaxy detected by HST, JST, {\it Spitzer}, {\it Herschel} and SCUBA2. It corresponds to AzGN24 and GN23 \citep{chapin2009,mancini2009,pope2006}. \citet{barro2019} identify it with an X-ray source. No AGN torus is needed to describe the observed SED, because the 1.6 $\mu$m stellar peak falls in the {\it Spitzer}/IRAC 8.0 $\mu$m channel and the stellar component  fits well the IRAC monotonically increasing SED. The VLA radio flux is in significant excess to the radio-FIR correlation, as applied by MAGPHYS, but consistent with the \citet{delhaize2017} correlation scaled from the MBB fit.

{\bf N2GN\_1\_46} is matched by NOEMA to one of three tightly grouped optical sources. It was previously erroneously associated with a $z_\textrm{spec}=0.55$ galaxy \citep{kodra2023, barger2012, owen2018}, but the high-resolution HST and JWST optical photometry reveals that this is a nearby galaxy at $\sim1.5$\,arcsec distance and not the NOEMA source. Moreover this low redshift is not consistent with the observed 1.6\,$\mu$m stellar peak detected between the IRAC 4.5 and 5.8\,$\mu$m bands. Therefore we adopted the $z_\textrm{phot}=1.894$ redshift estimate by \citet{kodra2023}. The source is detected in the 2 Ms {\it Chandra} map with fluxes of 0.99 and $7.6\times10^{-5}\,\mu$Jy in the 0.5-2.0 and 2.0-8.0 keV bands. In the SED fitting we include an AGN-torus component, that turns out to contribute significantly only at  the bluest optical wavelengths.

{\bf N2GN\_1\_47} is a high-$z$ galaxy detected by HST and identified via the NOEMA observations. The observed photometry shows a break between the Ks and IRAC 3.6\,$\mu$m band, placing it at $z_\textrm{phot}=4.87$ \citep{kodra2023}. \citet{barro2019} associate it with an X-ray source \citep{alexander2003}, but no AGN component is required to fit its SED.

{\bf N2GN\_1\_48} is detected by HST and is identified by the SMA, It corresponds to GN21. JWST spectroscopy sets the redshift to $z_\textrm{spec}=2.6$ (JADES, DJA). The SED is reproduced by a very extinguished starburst model.

{\bf N2GN\_1\_49} lies close to a very bright star. Identified on the basis of its VLA radio counterpart \citep{owen2018}, is blended to another galaxy in the IRAC bands \citep{liu2018}. \citet{kodra2023} derive a $z_\textrm{phot}=2.592$. The optical SED is characterized by a blue excess, that we fit with an unextinguished young SSP (Sect.~\ref{sect:blue_excess}).

{\bf N2GN\_1\_50} corresponds to AzGN22, so far associated to a radio galaxy at $z=1.0662$. The VLA position is at 0.19\,arcsec from a JWST source at $z_\textrm{spec}=3.1312$, that nicely matches the observed 1.6\,$\mu$m stellar peak detected in the IRAC 5.8\,$\mu$m channel and the Balmer break in the F105W JWST band. The adopted IRAC photometry is from the super-deblended catalog by \citet{liu2018}, that removes the contribution of a close by ($\sim3$\,arcsec away) face-on spiral galaxy (see HST postage stamps, arguably the $z=1.0662$ galaxy). Closer to the N2CLS position, there is a JWST galaxy at $z_\textrm{spec}=5.186$ \citep{sun2024},

{\bf N2GN\_1\_51} lies roughly two NIKA2 HPBW away from another, much brighter 1.2\,mm source, in a crowded region with several MIR objects (on the IRAC and MIPS maps). NOEMA confirms the NIKA2 detection and identifies it with a HST object at $z_\textrm{phot}=2.013$ that becomes progressively more point-like in the bluest HST bands. \citet{barro2019} matches it to an X-ray source in the \citet{alexander2003} CDFN catalog. Despite the compact blue-optical morphology and the X-rays detection, no AGN component is needed to fit the optical-to-millimeter SED of this source.

{\bf N2GN\_1\_52} is an optically dark source and lies outside of the current JWST covered area. Only a very faint {\it Spitzer} photometry is available, identified with the NOEMA positioning. \citet{liu2018} give a redshift $z_\textrm{phot}=4.06$. The IRAC 3.6\,$\mu$m flux is brighter than the one at 4.6 $\mu$m, which could be explained with a bright H$\alpha$ line at this redshift, contributing to the broad-band photometry.

{\bf N2GN\_1\_53} is an isolated IRAC galaxy at $z_\textrm{phot}=1.994$, located by NOEMA. The rich optical SED has a remarkable blue excess, that we reproduce adding an unextinguished young SSP to the dusty starburst model. 

{\bf N2GN\_1\_54} is confirmed by NOEMA, that defined its precise position. It is detected on  the {\it Spitzer} IRAC and MIPS maps, as well as by HST and JWST. Its SED shows a remarkable Lyman break at $z_\textrm{phot}=2.885$. The \citet{barro2019} catalog includes a X-ray identification \citep{alexander2003}. A big gap between the NIR and sub-millimeter observed photometry still exists and does not allow us to verify if an AGN torus is needed to fit the MIR SED of this source.

{\bf N2GN\_1\_55} lies on top of three radio sources \citep{owen2018}: two very bright nearby galaxies (at 3.8 and 4.6\,arcsec from the NIKA2 position) and one optically and IRAC-dark object (at 0.08\,arcsec from the NIKA2 position). The SCUBA2 data were matched to one of the nearby galaxies because the SCUBA2 position is shifted to the north with respect to the NIKA2 position. We adopt the dark source as the counterpart. The {\it Spitzer} map being highly blended in this region, currently no photometry other than NIKA2 and VLA is available, nor any redshift estimate is at hand.

{\bf N2GN\_1\_56} splits in two NOEMA sources at a distance of 13.5 arcsec from each other, named 56a and 56b. The former corresponds to AzGN21 \citep{chapin2009}. Both sources have IRAC, MIPS and HST counterparts. Their redshifts are $z_\textrm{phot}=1.866$ and 2.598, respectively \citep{kodra2023}. Because of blending, the {\it Herschel} photometry is ignored. \citet{barro2019} matched the 56a counterpart to an X-ray source. The 56b galaxy shows a significant optical-blue excess that is modeled with a unextinguished, young SSP as usual. The radio emission is consistent with the radio-FIR correlation in both cases.

{\bf N2GN\_1\_57} is detected by SCUBA at 450 and 850\,$\mu$m \citep{barger2012}. The corresponding SCUBA2 source was matched to a $z=0.9$ radio galaxy. In the NIKA2 beam lies the high-$z$ candidate HRG14 J123731.66+621616.7 \citep{finkelstein2015}, also detected by the SHARDS middle-band survey at $z_\textrm{phot}=5.3$ \citep{kodra2023}. This is our counterpart choice. It is detected in the four IRAC channels. At longer wavelengths it is strongly blended to a couple of nearby objects, and therefore we ignored the MIPS and {\it Herschel} photometry. Summarizing: the observed SED includes HST and IRAC photometry, with a pronounced Balmer break at $\lambda\sim 2$\,$\mu$m, as well as SCUBA2 and NIKA2 data. 

{\bf N2GN\_1\_58} is identified by VLA with a $z_\textrm{phot}=2.723$ galaxy with an exceptional SED coverage. A total of 20 data points cover the wavelength range from 3500 \AA\ to 21\,cm, excluding the SPIRE data that are a blend of different objects.  

{\bf N2GN\_1\_59} is also pinpointed by VLA, although the radio source is at 5.74 arcsec from the N2CLS position. It was observed by NOEMA as part of our follow-up program, but only a $5\sigma$ source at 9.7\,arcsec distance from the NIKA2 peak was detected. The high-$z$ candidate HRG14 J123702.11+621733.4 \citep[$z_\textrm{phot}=3.434$;][]{arrabal_haro2018} lies at 0.79\,arcsec from the radio position, and CANDELS-GDN J123702.1+621734.2 is 0.3 arcsec away. The final, observed SED combines the HST, {\it Spitzer}, NIKA2 and VLA data and is best fitted with a very young starburst ($1.3\times10^7$ yr).

{\bf N2GN\_1\_60} is blended with a few large nearby galaxies and is identified by SMA at $z_\textrm{phot}=1.811$ \citep{kodra2023}. It is detected both in the Ks band and at 1.4\,GHz \citep{owen2018}. The IRAC photometry is finely deblended by \citet{liu2018}, but the MIPS, PACS and SPIRE data are not recoverable. HST constrains the optical emission of this red, dusty galaxy.

{\bf N2GN\_1\_61} is an optical- and IRAC-dark source, confirmed by NOEMA. JWST grism observations detected one emission line, consistent with H$\alpha$ at $z_\textrm{spec}=5.201$ (M.~Xiao, priv. comm.). A faint source is visible on the IRAC 3.6 $\mu$m and 4.5 $\mu$m maps at $\sim3$ arcsec distance. The observed SED consists of only three data points (NIKA2 1.2\,mm, SCUBA2 850\,$\mu$m and JWST F444W), but the availability of the redshift allowed us to perform a fit with the sole aim of deriving $M_\textrm{dust}$.

{\bf N2GN\_1\_62} is a nearby spiral galaxy at $z_\textrm{spec}=0.641$, confirmed by NOEMA. Its SED is well sampled by all available instruments.

{\bf N2GN\_1\_63} is identified by VLA with a galaxy at $z_\textrm{spec}=2.21$ \citep{kodra2023} and corresponds to GN5. The SED is defined by 23 bands from the optical to the radio domain and shows a clear 1.6\,$\mu$m stellar peak, a very red NIR-to-MIR color ($L_{7.5 \mu\textrm{m}}/L_{2.5\mu\textrm{m}}\sim12$ in the rest frame) and a well define FIR-millimeter emission with data on both the Wien and RJ tails.

{\bf N2GN\_2\_13} is a NIKA2 2\,mm source with no 1.2\,mm detection (in the blind catalog with S/N$>3\sigma$). It was observed by NOEMA and located in a crowded region with several optical sources. The NOEMA position corresponds to a $z_\textrm{phot}=2.071$ galaxy \citep{kodra2023} detected by HST. We use the super-deblended IRAC catalog by \citet{liu2018}, but MIPS and SPIRE fluxes are not recoverable. 

{\bf N2GN\_2\_20}, finally, is an optically dark source, identified by VLA with a $z_\textrm{phot}=6.67$ isolated galaxy. The photometric redshift is constrained  mainly by the Balmer break observed between the Ks band and the IRAC 3.6\,$\mu$m band. Its position corresponds to GN3 \citep{pope2006}.

%\twocolumn

%\end{appendix}

%\end{document}

%%%%%%% 2025/03/04: the long figure E.1 (here below) has been shifted to an online Zenodo entry,
%%%%%%% as requested by the A&A editorial office

\settoheight{\imageheight}{\includegraphics{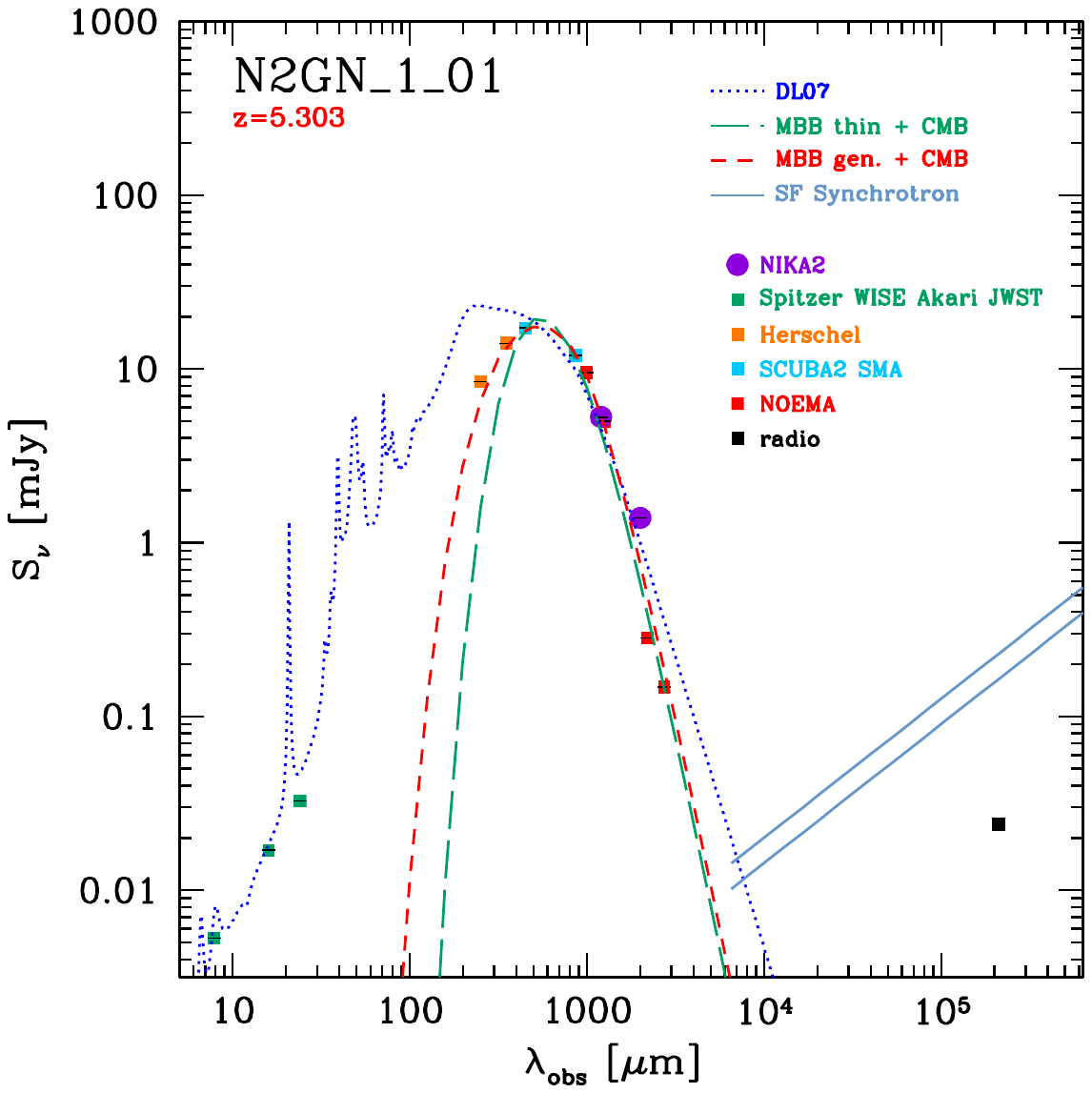}}

\begin{figure*}[t]
\centering
\includegraphics[align=c,width=0.4\textwidth]{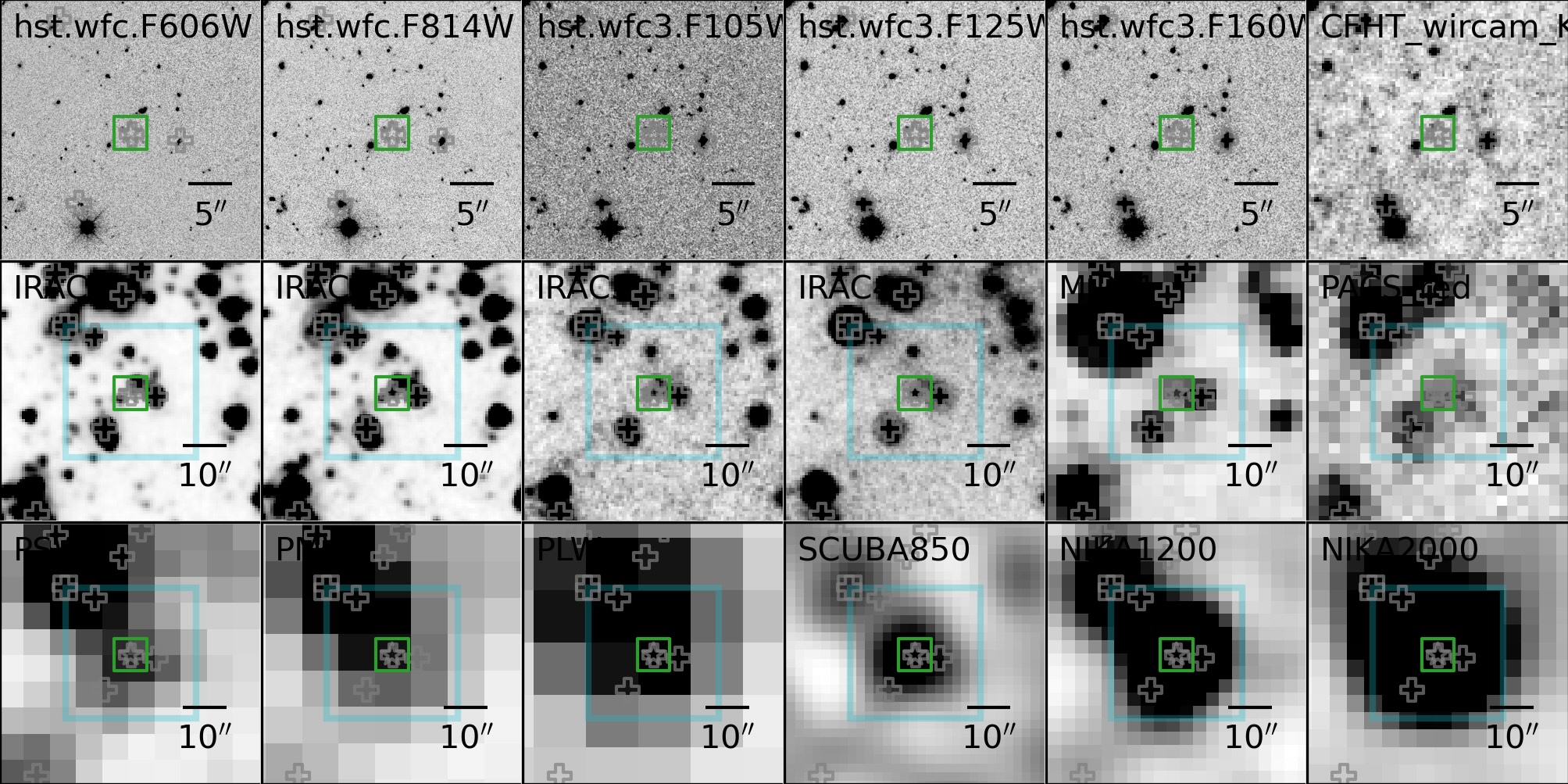}
\includegraphics[align=c,trim=0 0.18\imageheight{} 0 0.075\imageheight{}, clip, width=0.25\textwidth]{figs2_indiv_objs/sed_FIR_fit_N2GN_1_01.pdf}
\includegraphics[align=c,trim=0 0.18\imageheight{} 0 0.075\imageheight{}, clip, width=0.25\textwidth]{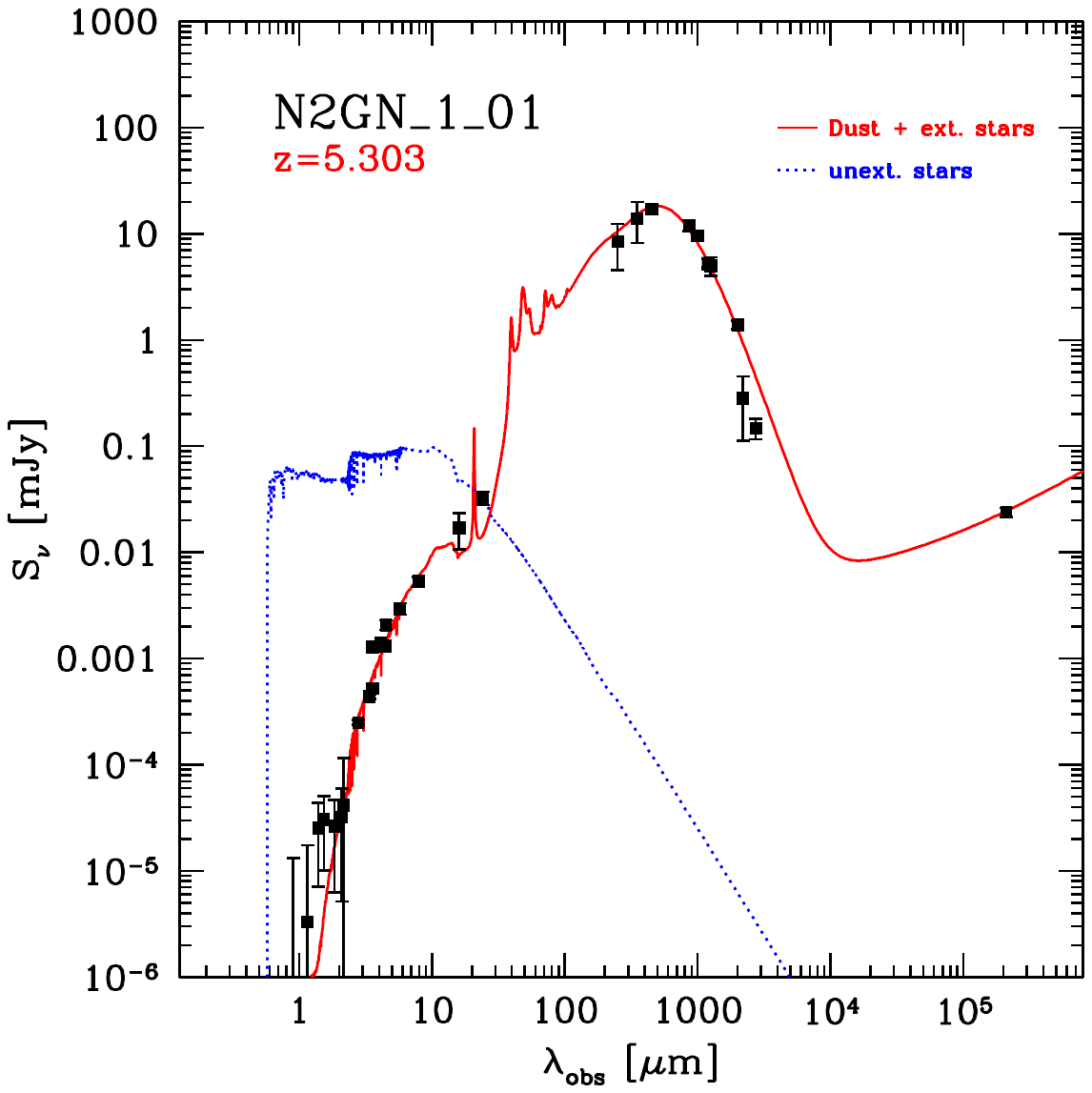}
\includegraphics[align=c,width=0.4\textwidth]{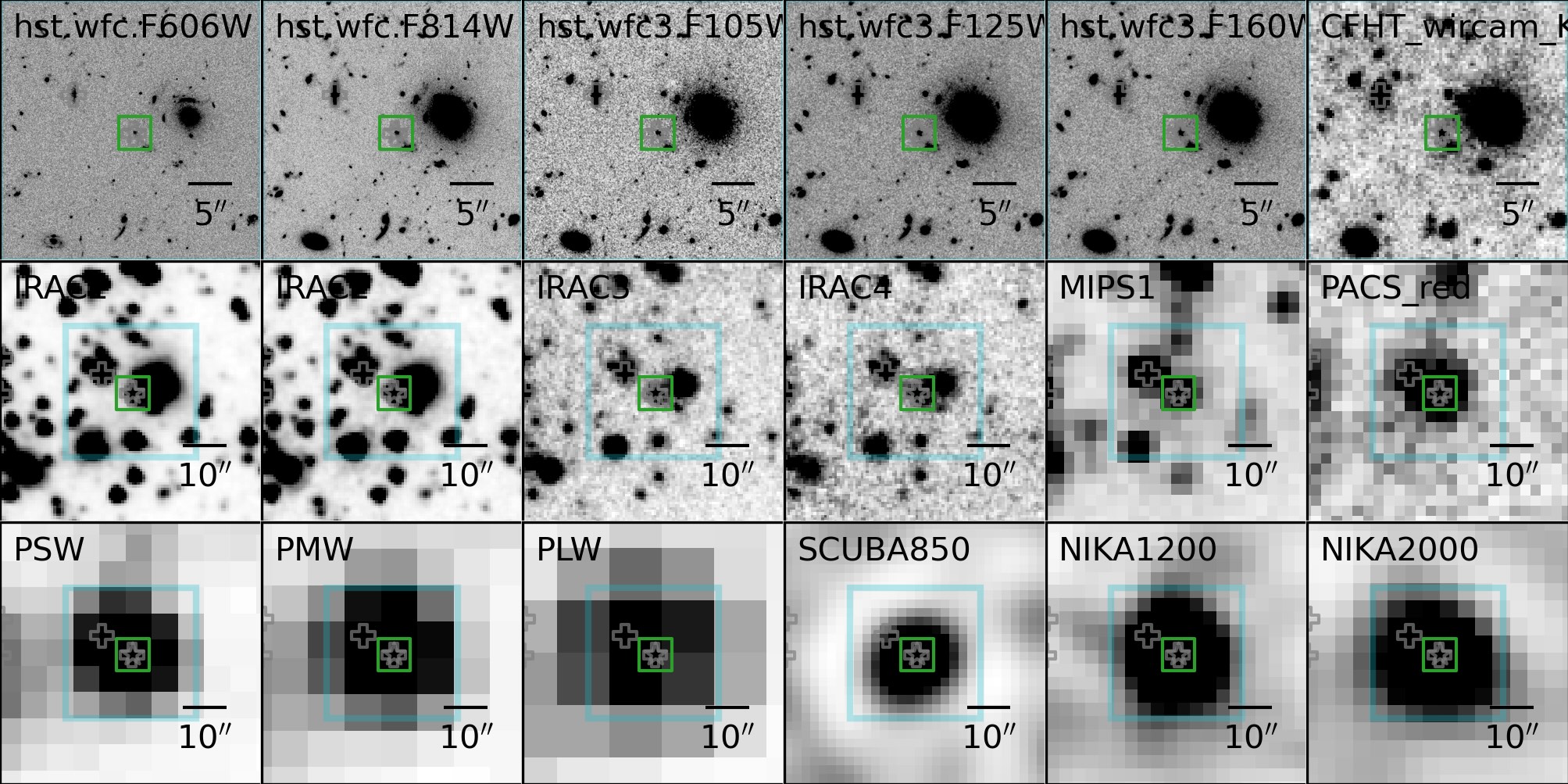}
\includegraphics[align=c,trim=0 0.18\imageheight{} 0 0.075\imageheight{}, clip, width=0.25\textwidth]{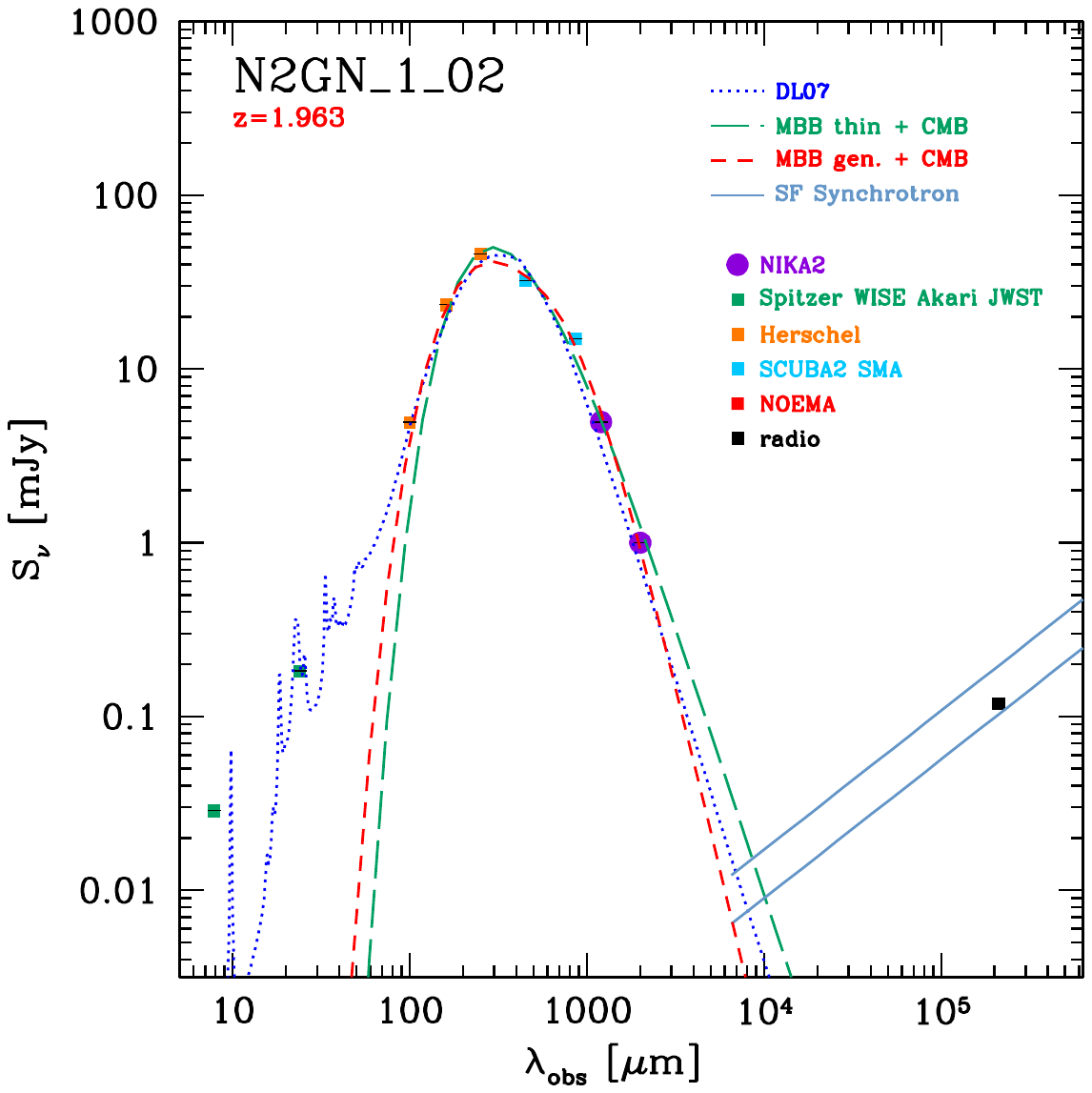}
\includegraphics[align=c,trim=0 0.18\imageheight{} 0 0.075\imageheight{}, clip, width=0.25\textwidth]{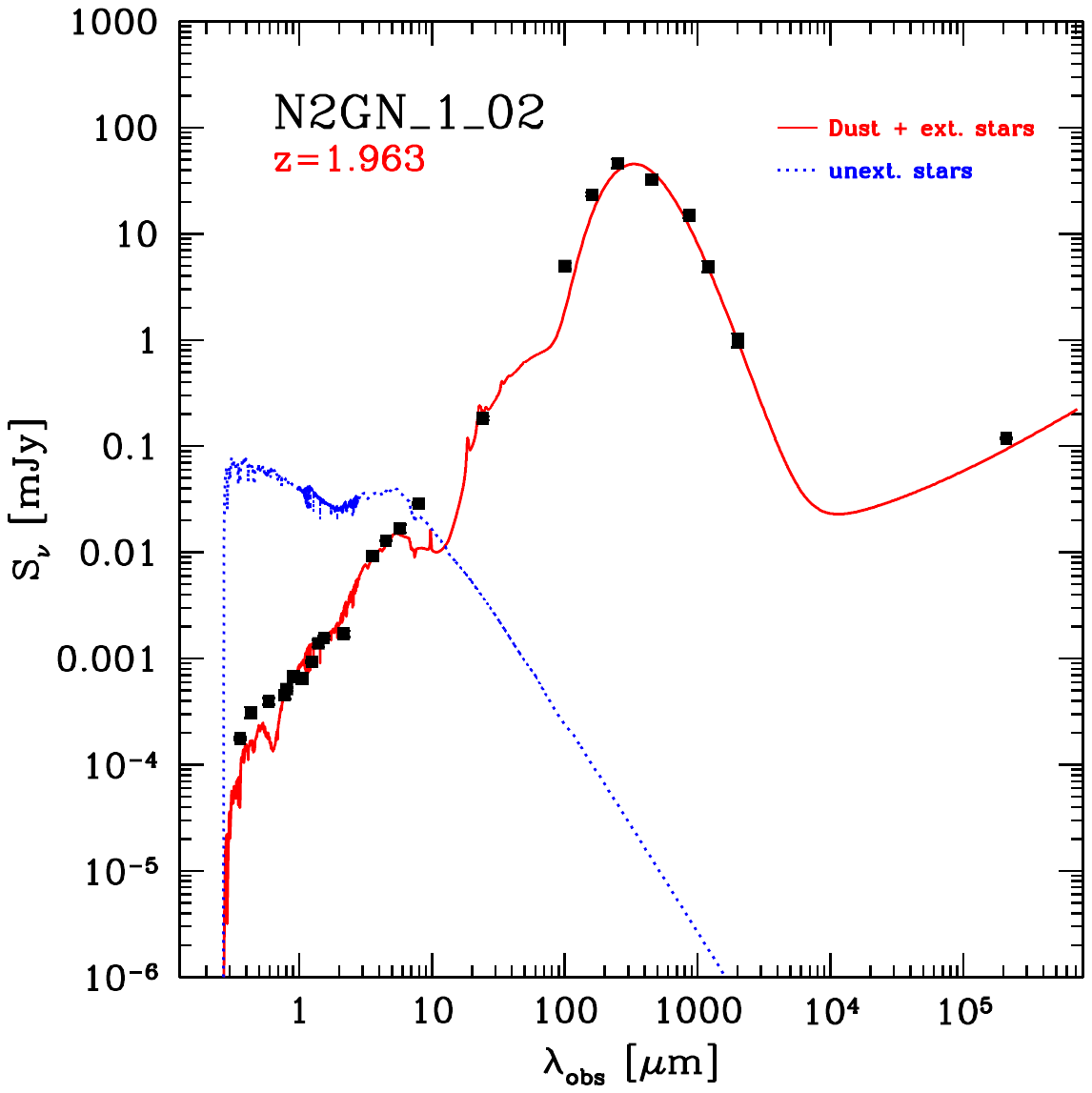}
\includegraphics[align=c,width=0.4\textwidth]{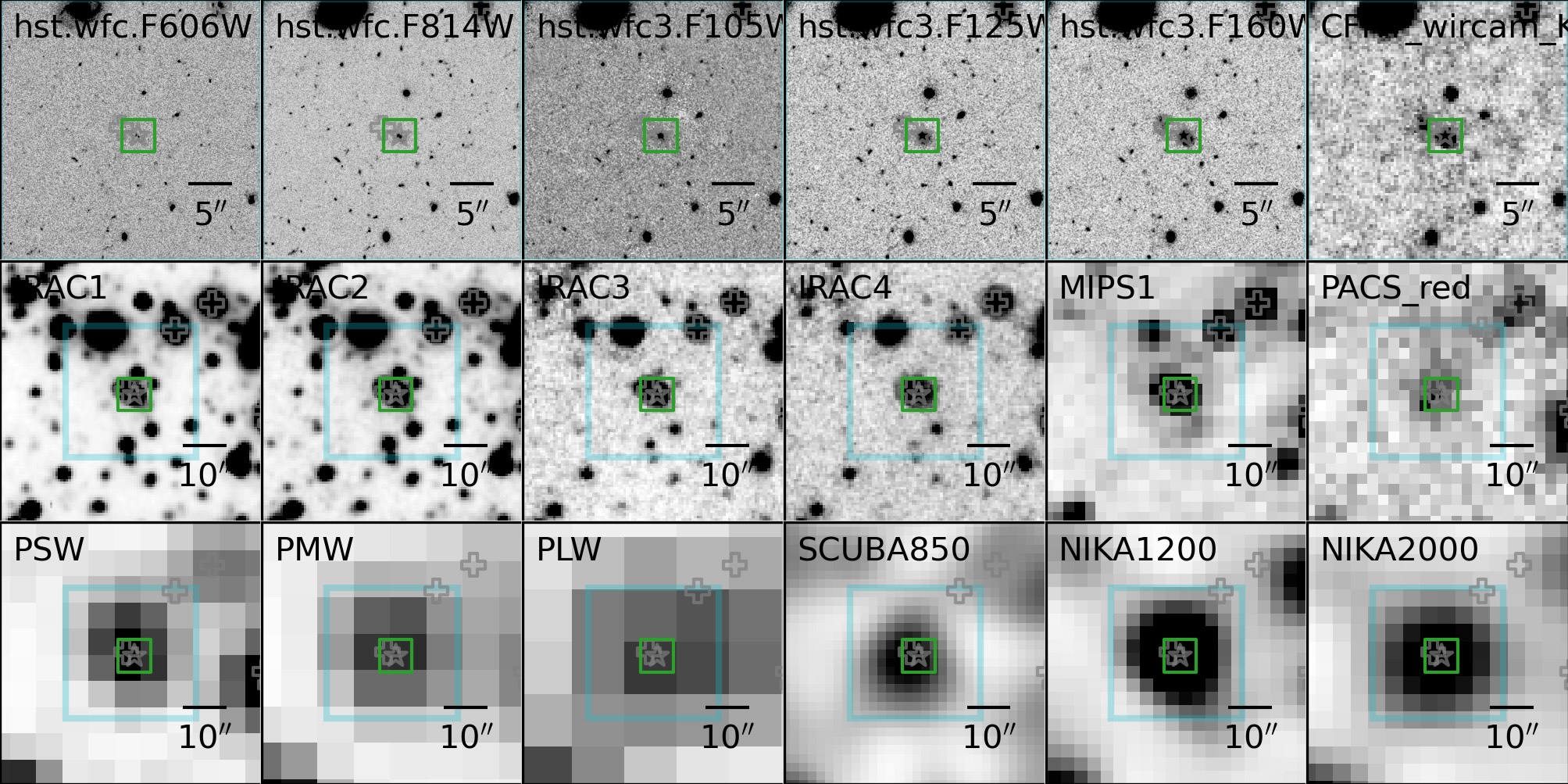}
\includegraphics[align=c,trim=0 0.18\imageheight{} 0 0.075\imageheight{}, clip, width=0.25\textwidth]{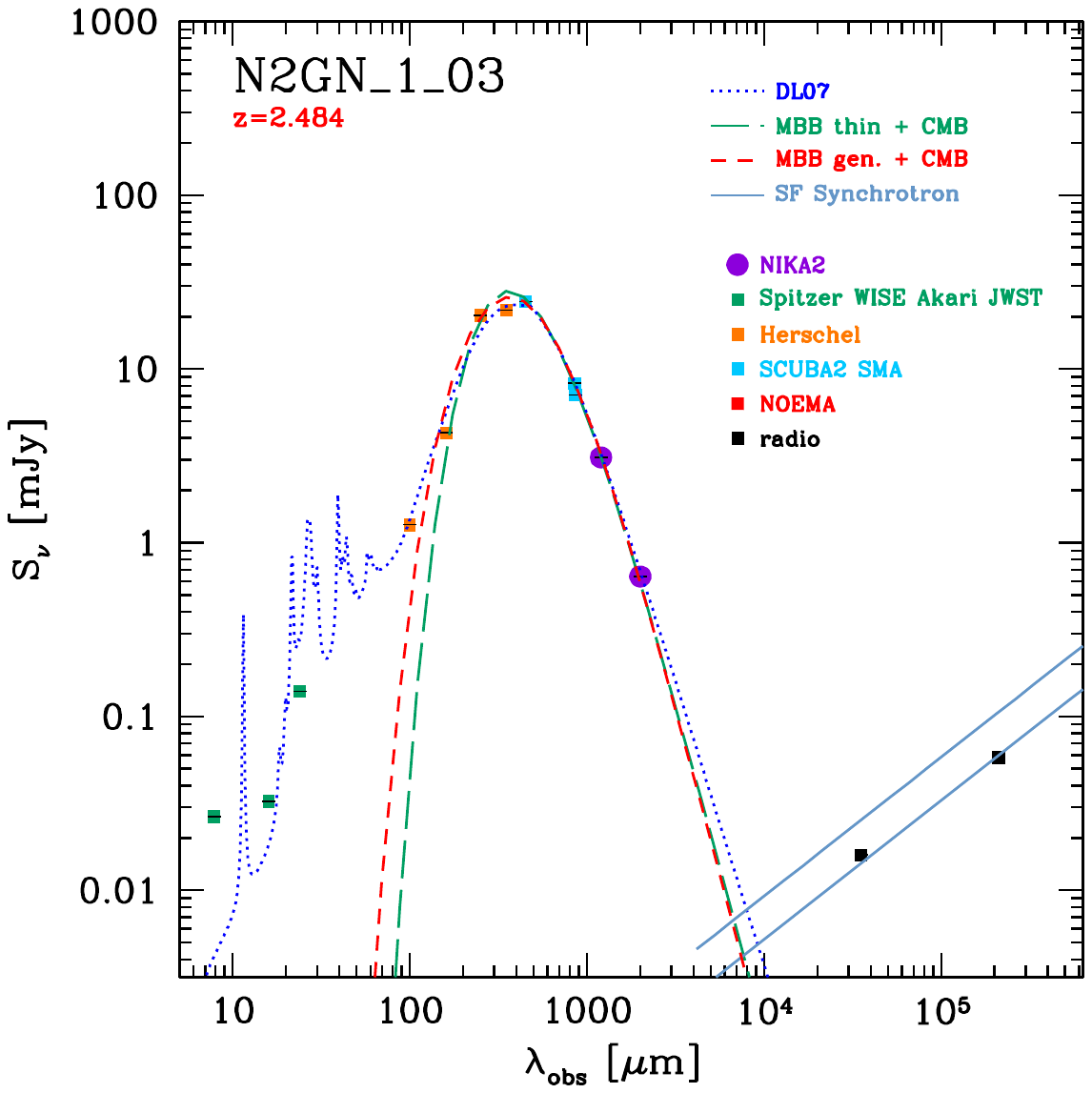}
\includegraphics[align=c,trim=0 0.18\imageheight{} 0 0.075\imageheight{}, clip, width=0.25\textwidth]{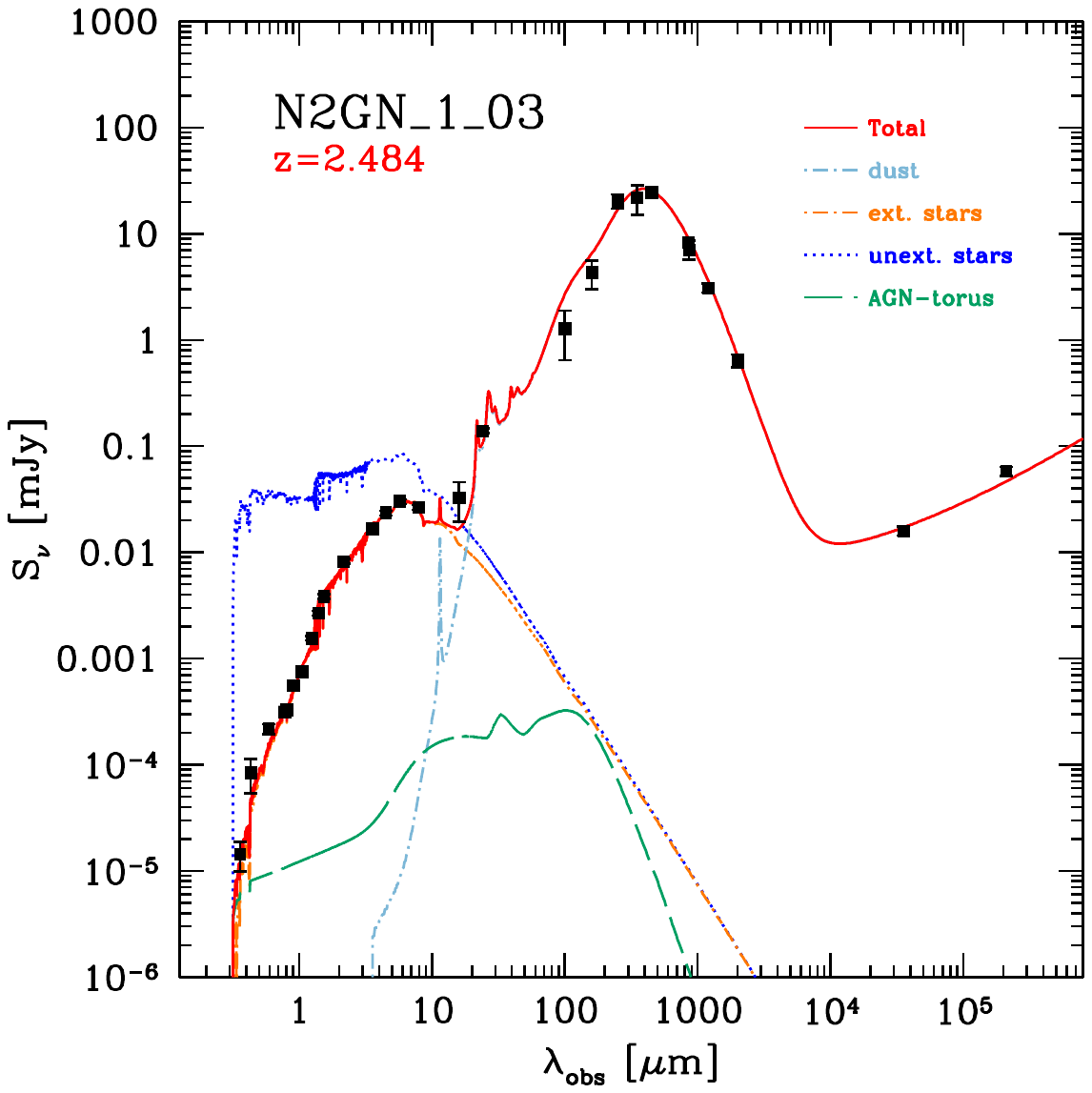}
\includegraphics[align=c,width=0.4\textwidth]{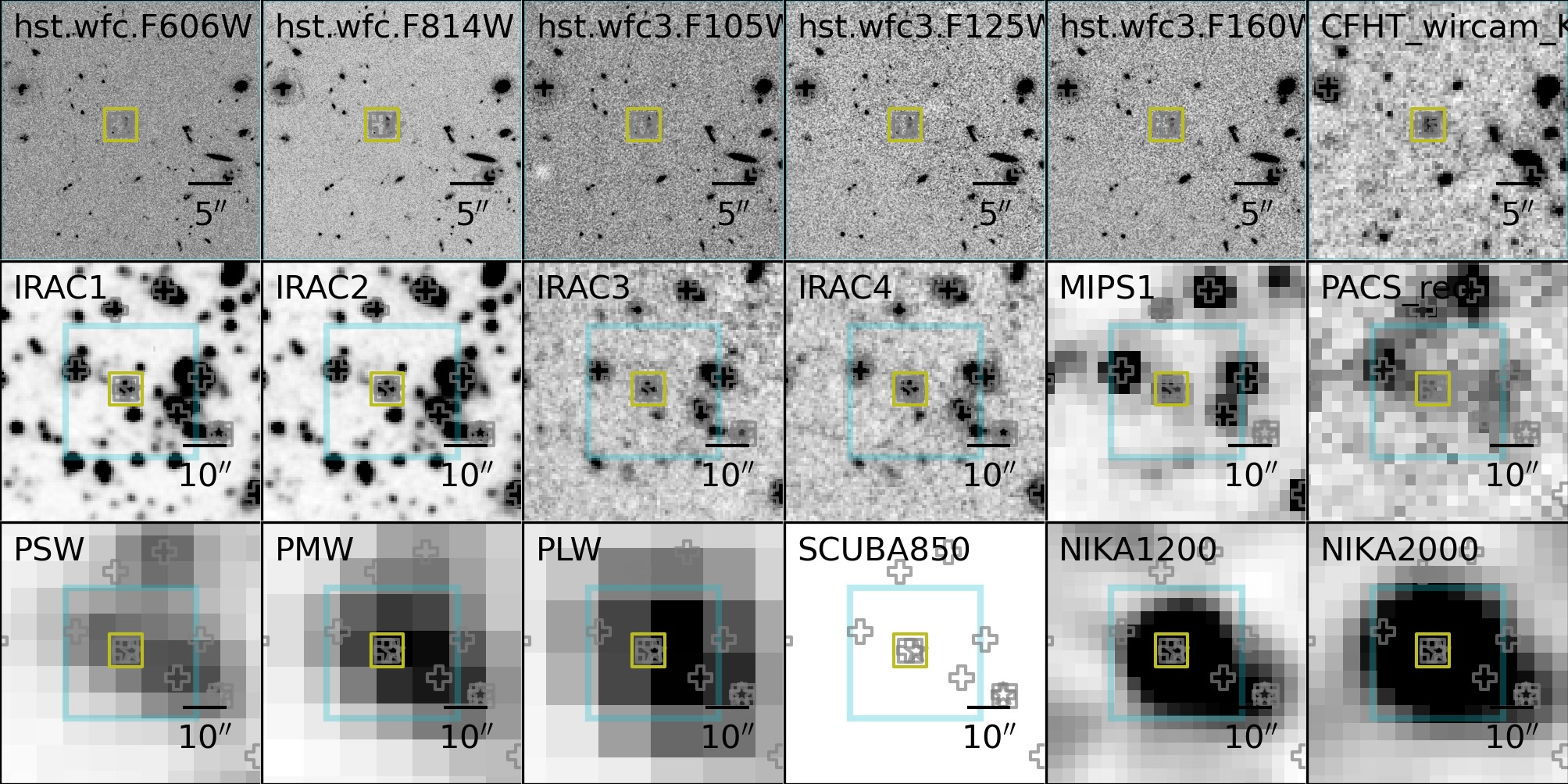}
\includegraphics[align=c,trim=0 0.18\imageheight{} 0 0.075\imageheight{}, clip, width=0.25\textwidth]{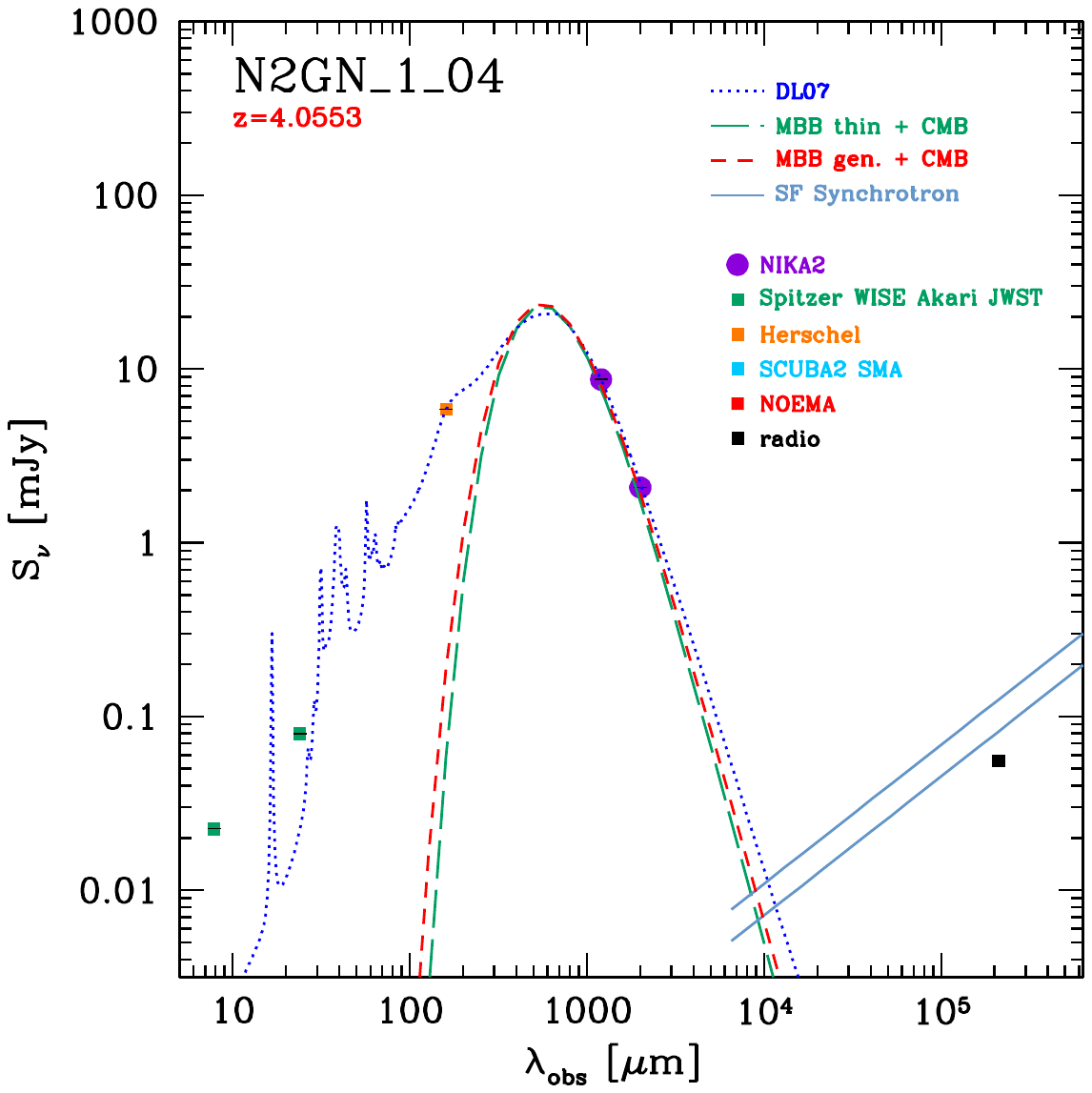}
\includegraphics[align=c,trim=0 0.18\imageheight{} 0 0.075\imageheight{}, clip, width=0.25\textwidth]{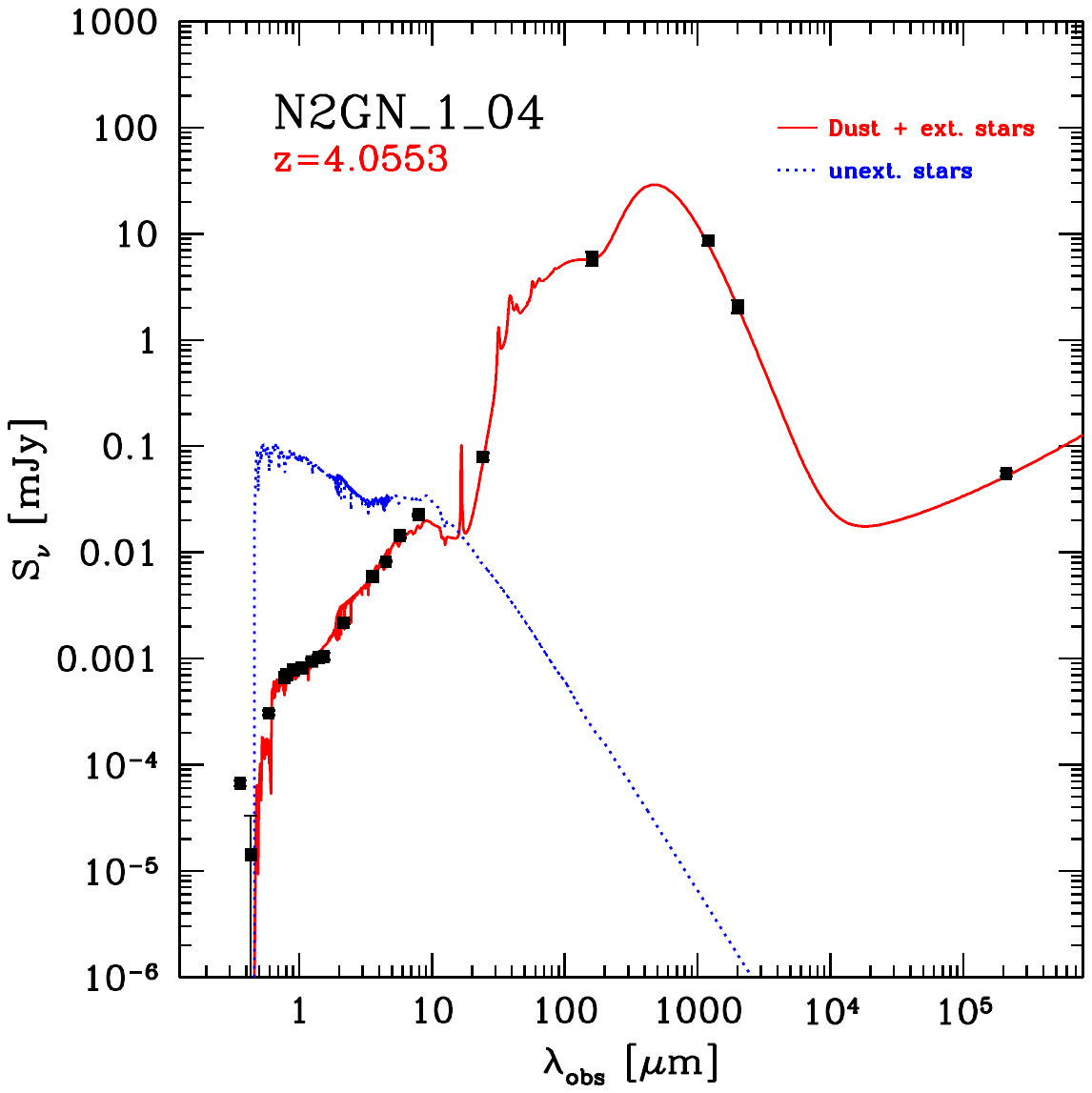}
\caption{N2GN individual sources: multi-wavelength postage stamps and SED fitting results.
{\em Postage stamps}: the bands from left to right and top to bottom are: HST F606W, F814W, F105W, F125W, F160W, CFHT Ks, {\it Spitzer} 3.6 $\mu$m, 4.5 $\mu$m, 5.8 $\mu$m, 8.0 $\mu$m, 24 $\mu$m, {\it Herschel} 160 $\mu$m, 250 $\mu$m, 350 $\mu$m, 500 $\mu$m, SCUBA2 850 $\mu$m, and NIKA2 1.2 mm, 2.0 mm. The identifications appear with squares of different colors: green for SMA, red for VLA, purple for NOEMA/PDBI and yellow for identification by hand. In addition: VLA sources are marked with gray crosses, SMA sources with gray stars, NOEMA sources with gray X. 
The chosen counterparts at all wavelengths are marked with circles: blue for SCUBA 850 $\mu$m; orange for SCUBA 450 $\mu$m;  red for IRAC, MIPS 24 $\mu$m \& PACS; cyan for CFHT {\em Ks}; pink for \citet{barro2019}; cyan for SPIRE \citep{liu2018}; red for SPIRE \citep{elbaz2011}. 
{\em Middle panels}: FIR SED fitting, obtained with DL07 (blue dotted lines), optically thin MBB (green long-dashed lines) and MBB in general form (red short-dashed lines) models. The light-blue solid lines in the radio domain represent a synchrotron power law linked to the FIR emission by the radio-FIR correlation, in the derivations by \citet[][upper line]{magnelli2015} and \citet[][lower line]{delhaize2017}.
{\em Right-hand panels}: best fit model obtained with MAGPHYS or SED3FIT (high-$z$ versions). The blue dotted lines represent the stellar un-absorbed emission and the orange dashed lines the extinguished model. In case an AGN-torus component is needed (SED3FIT), it is depicted with the dark green long-dashed lines.  In few cases where a UV excess is observed but there are no hints for an AGN, we used a young SSP (10 to 100 Myr, bright green long-dashed line) instead of the torus model. The red solid line represents the sum of all components, when needed.
}
\label{fig:indiv_glxs}
\end{figure*}

\addtocounter{figure}{-1}
\newpage

\begin{figure*}[t]
\centering
\includegraphics[align=c,width=0.4\textwidth]{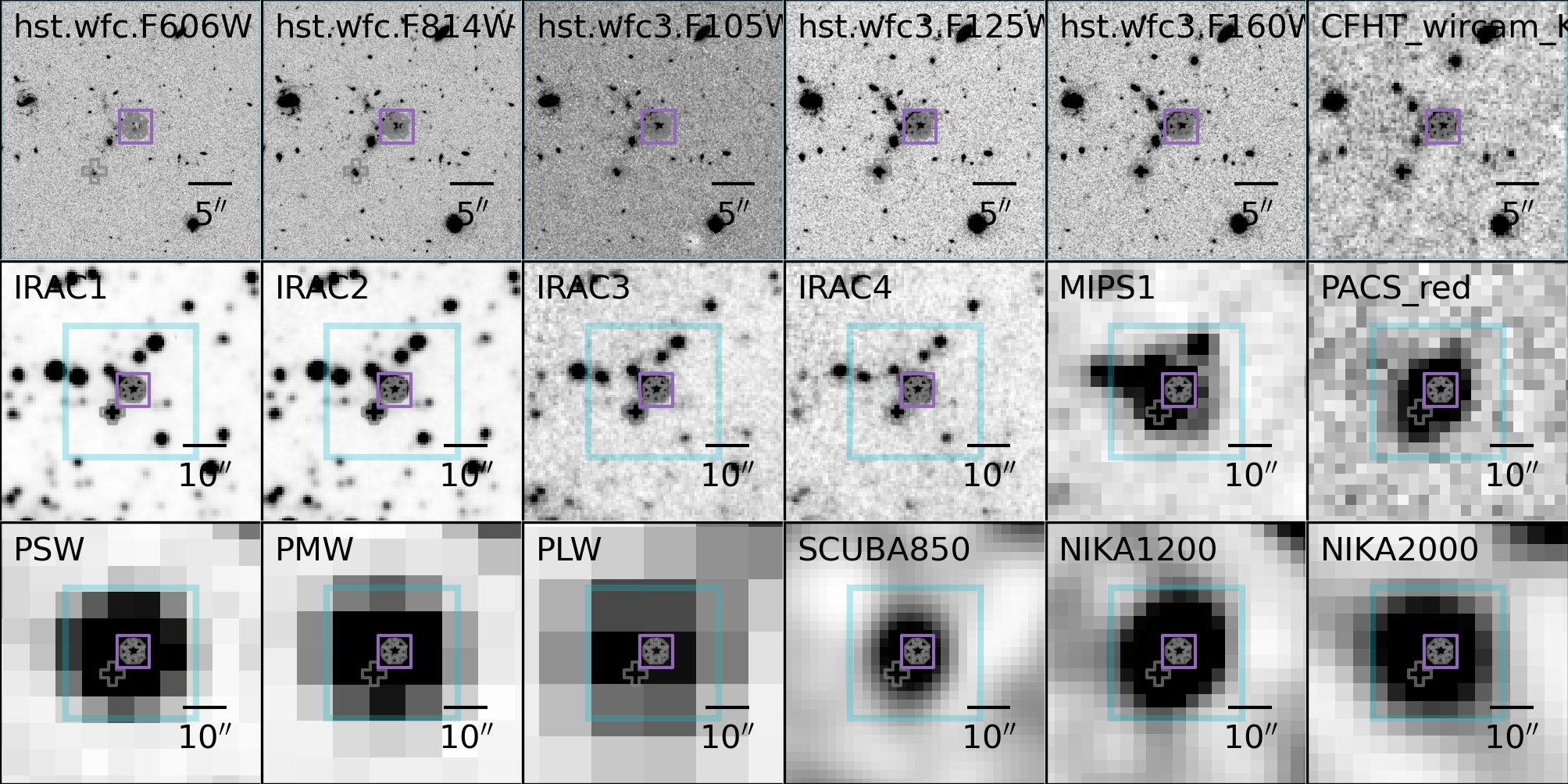}
\includegraphics[align=c,trim=0 0.18\imageheight{} 0 0.075\imageheight{}, clip, width=0.25\textwidth]{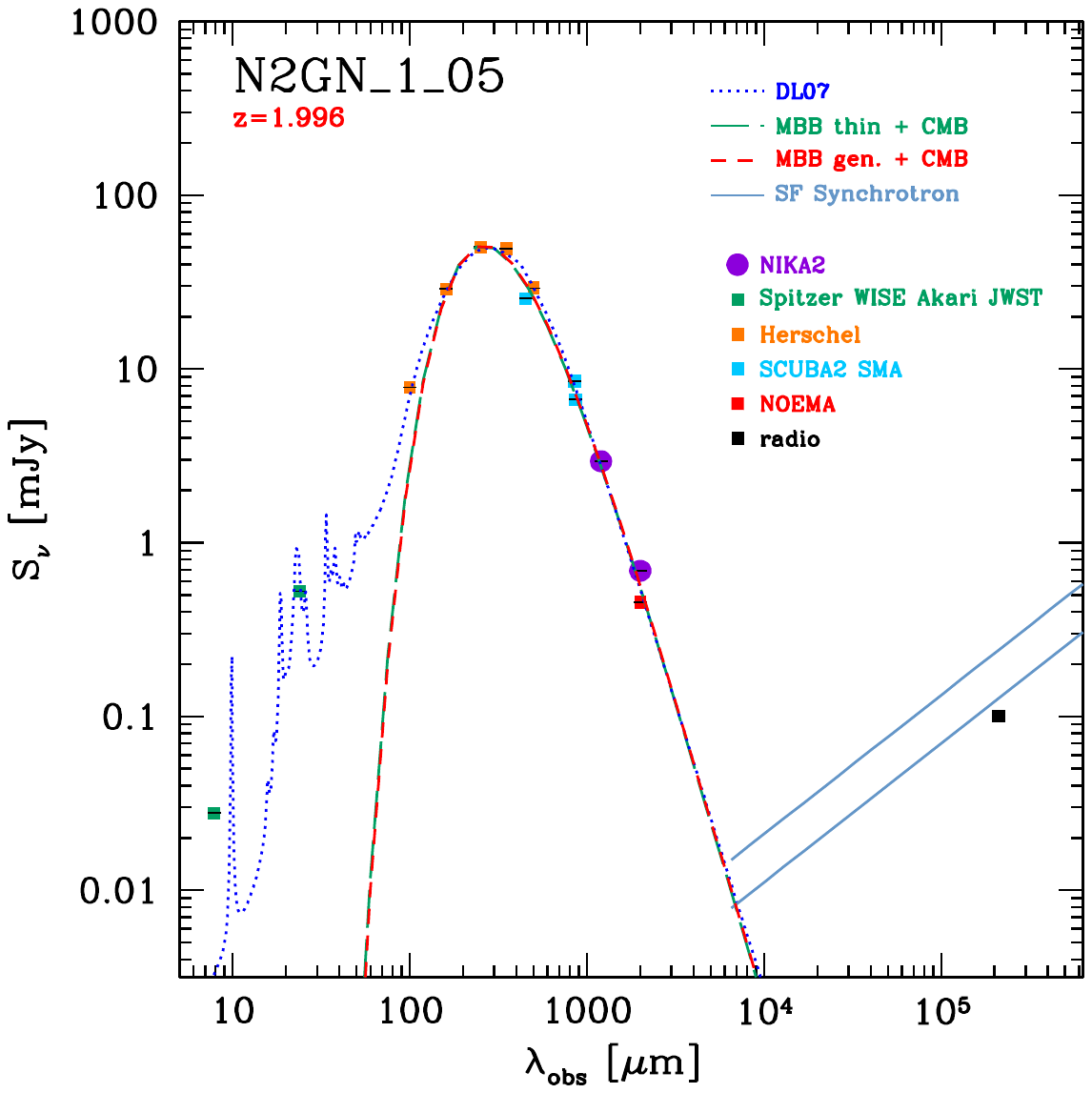}
\includegraphics[align=c,trim=0 0.18\imageheight{} 0 0.075\imageheight{}, clip, width=0.25\textwidth]{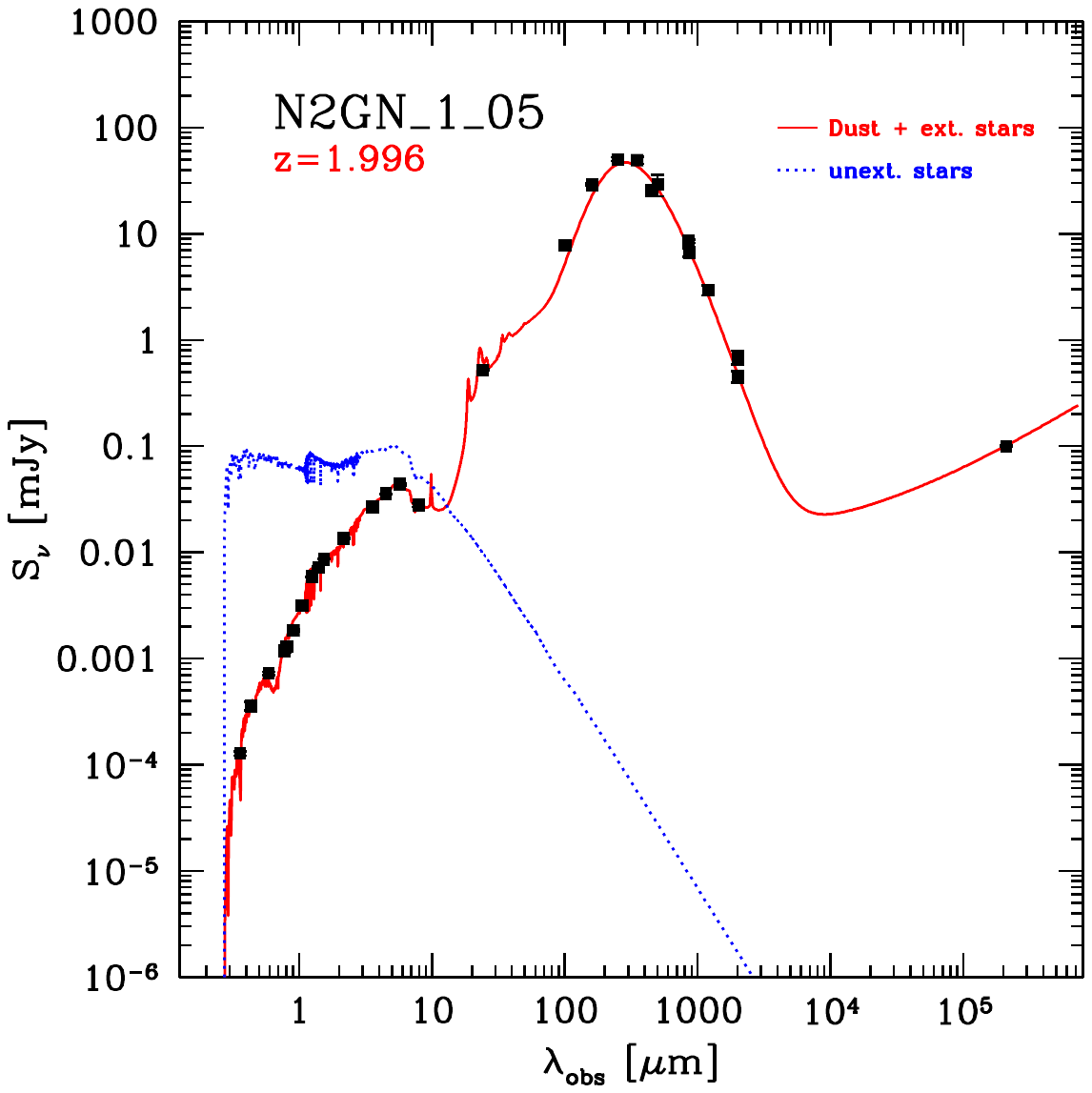}
\includegraphics[align=c,width=0.4\textwidth]{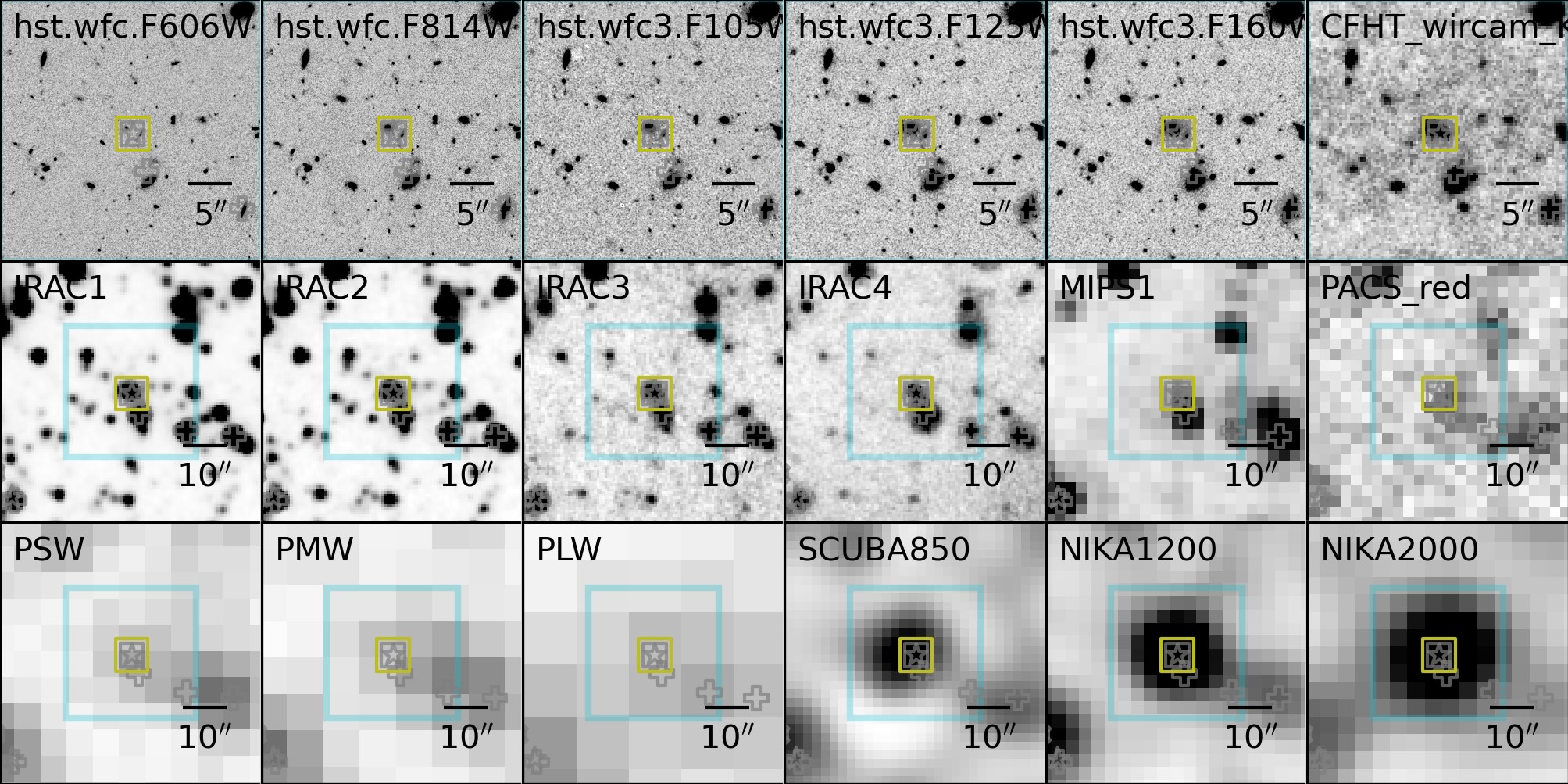}
\includegraphics[align=c,trim=0 0.18\imageheight{} 0 0.075\imageheight{}, clip, width=0.25\textwidth]{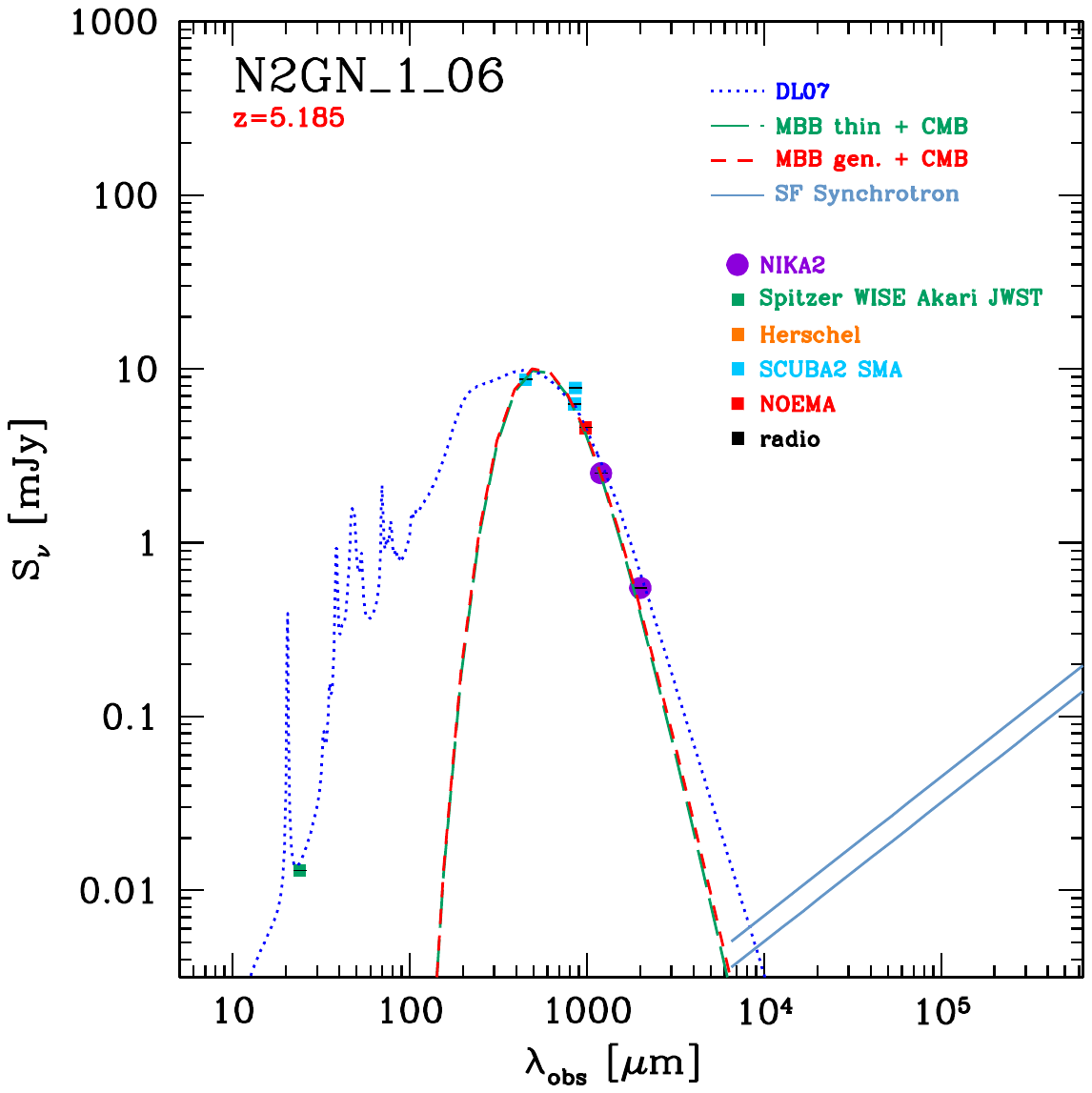}
\includegraphics[align=c,trim=0 0.18\imageheight{} 0 0.075\imageheight{}, clip, width=0.25\textwidth]{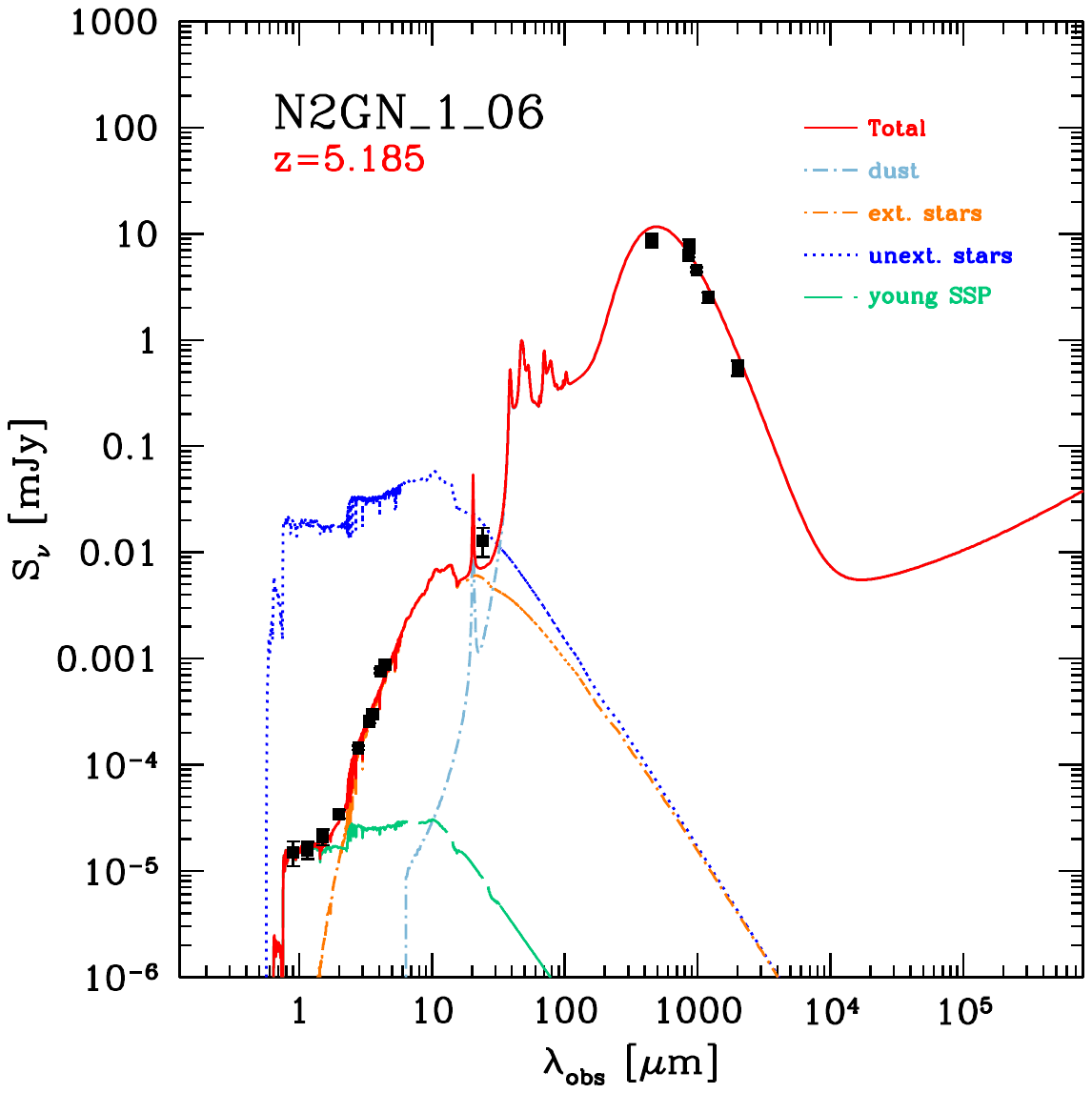}
\includegraphics[align=c,width=0.4\textwidth]{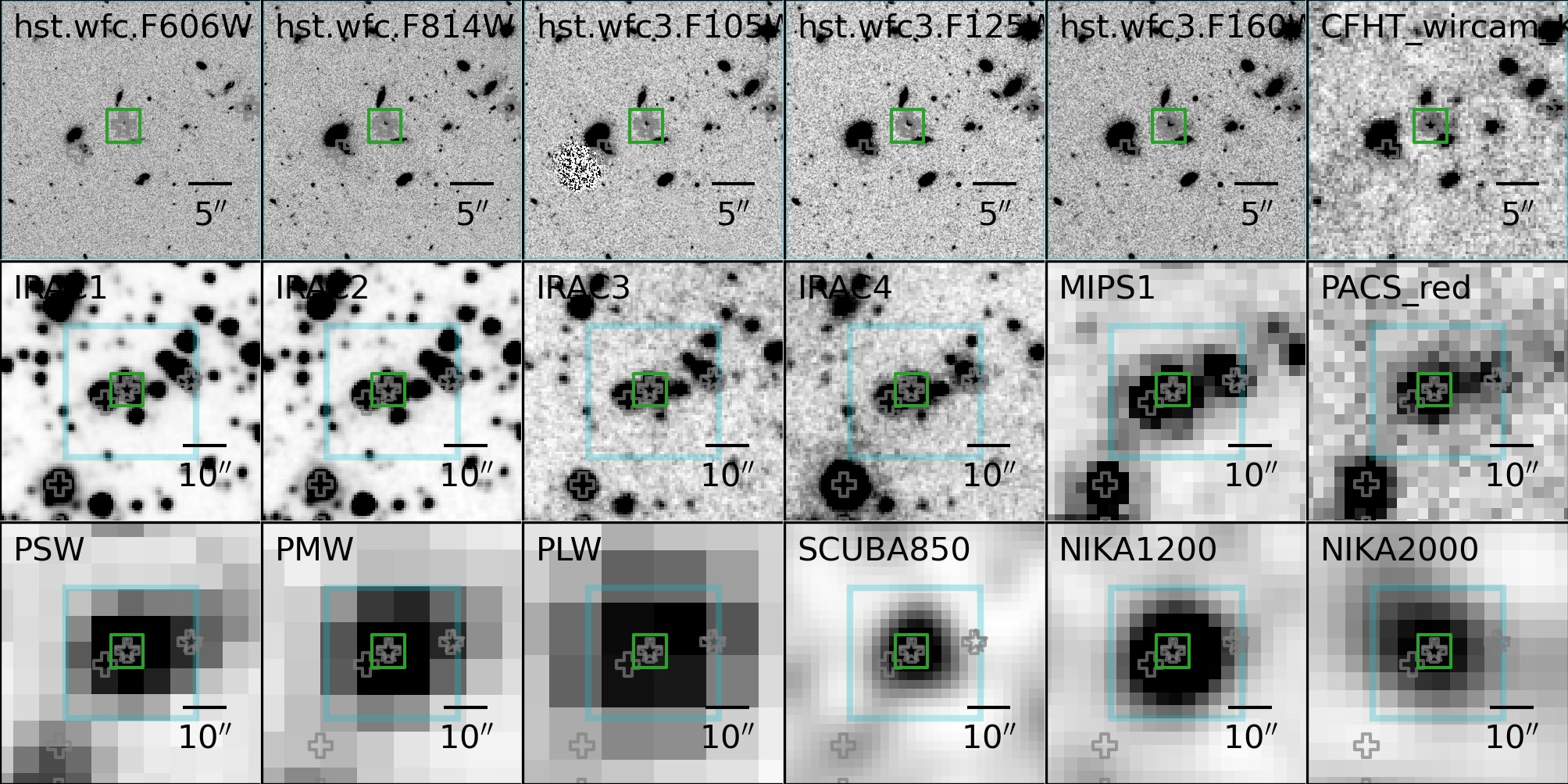}
\includegraphics[align=c,trim=0 0.18\imageheight{} 0 0.075\imageheight{}, clip, width=0.25\textwidth]{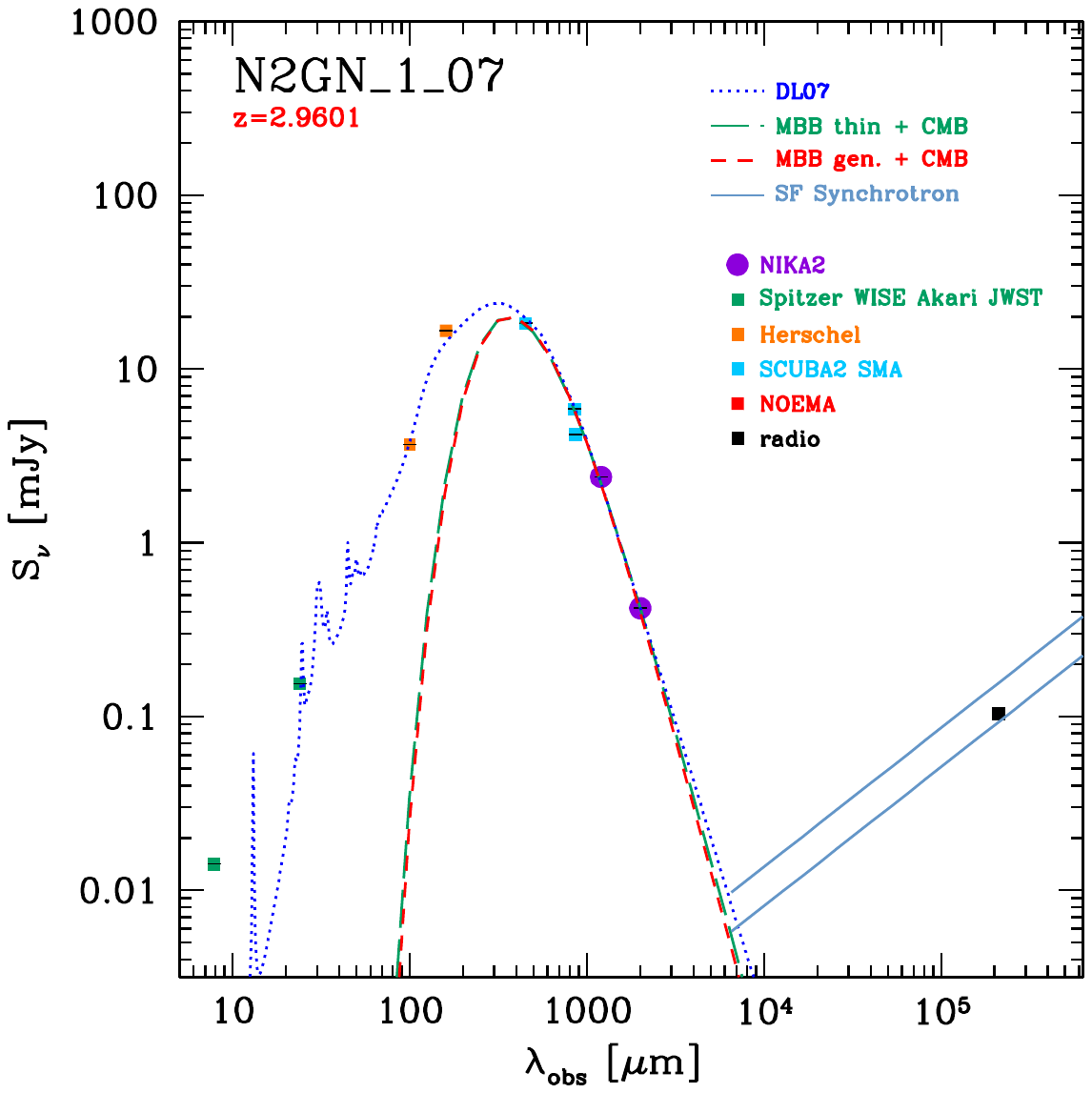}
\includegraphics[align=c,trim=0 0.18\imageheight{} 0 0.075\imageheight{}, clip, width=0.25\textwidth]{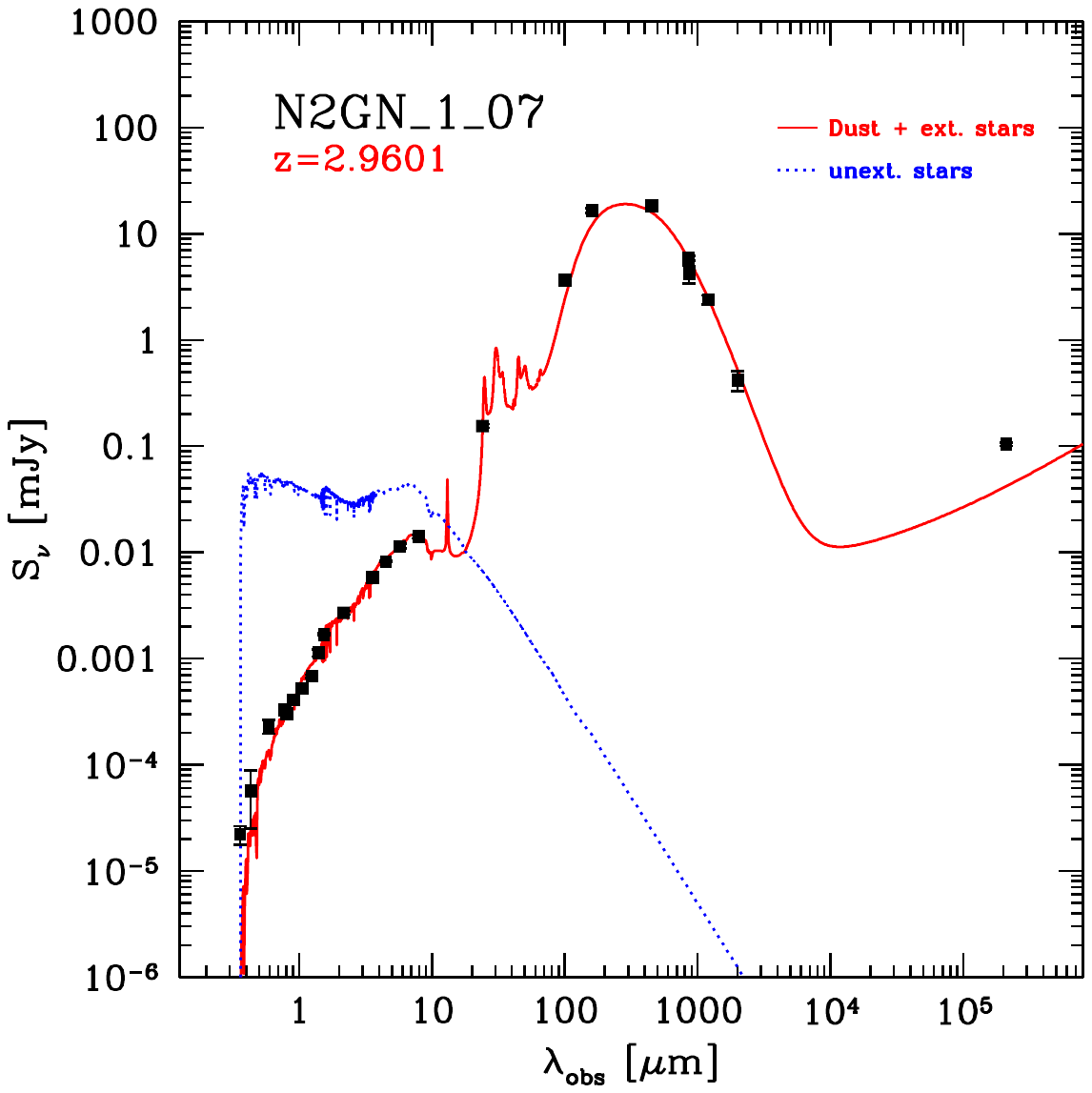}
\includegraphics[align=c,width=0.4\textwidth]{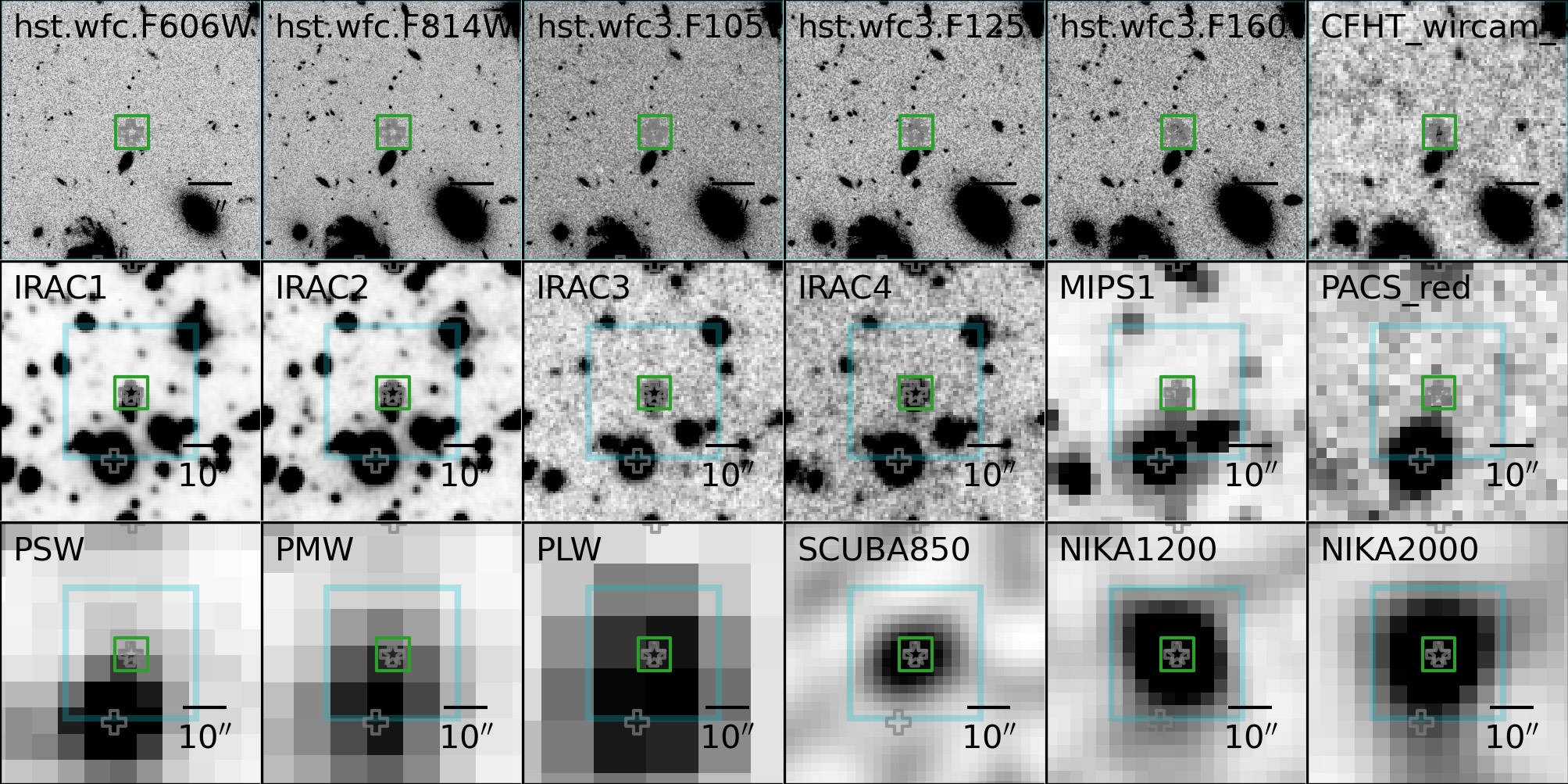}
\includegraphics[align=c,trim=0 0.18\imageheight{} 0 0.075\imageheight{}, clip, width=0.25\textwidth]{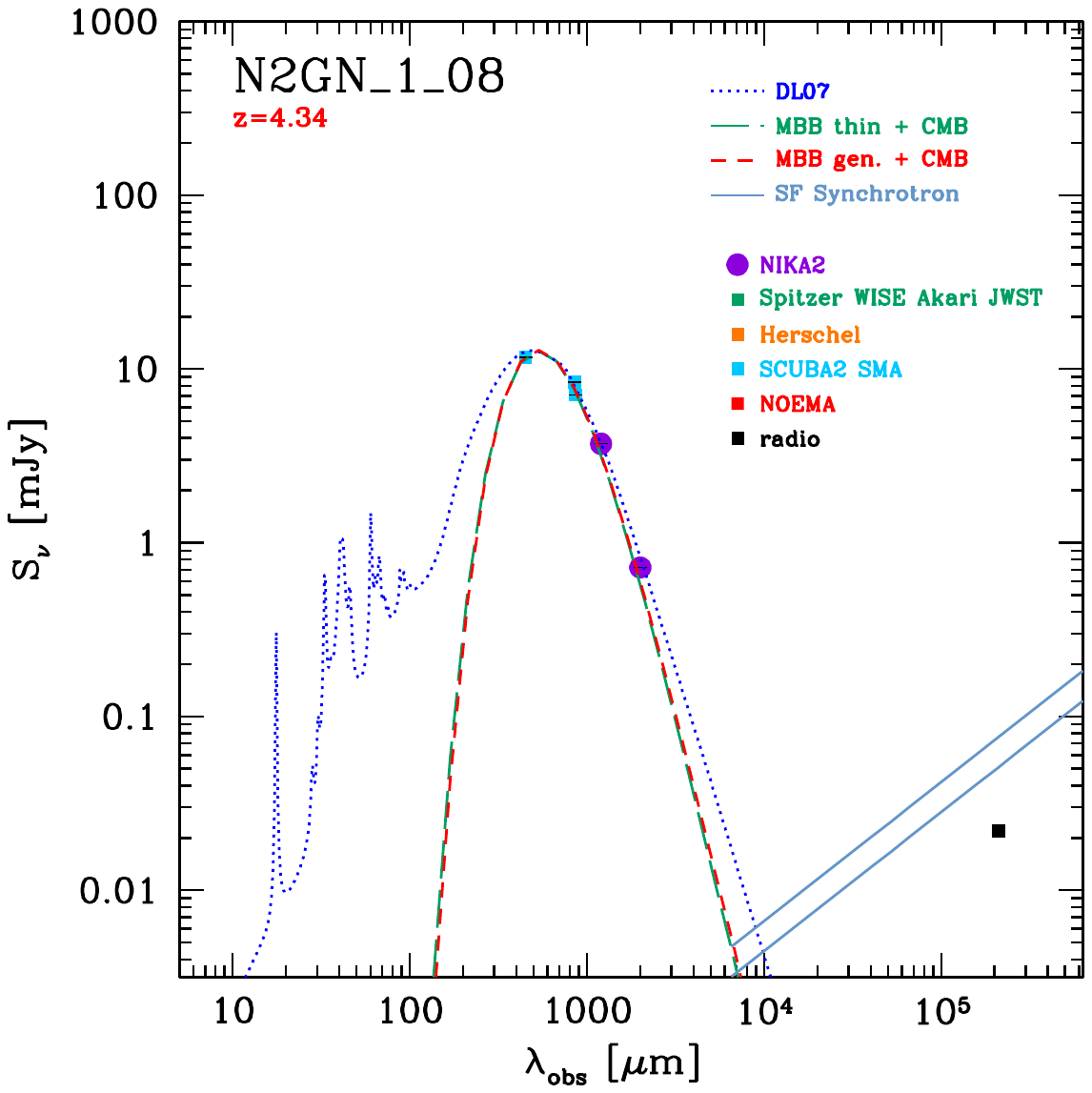}
\includegraphics[align=c,trim=0 0.18\imageheight{} 0 0.075\imageheight{}, clip, width=0.25\textwidth]{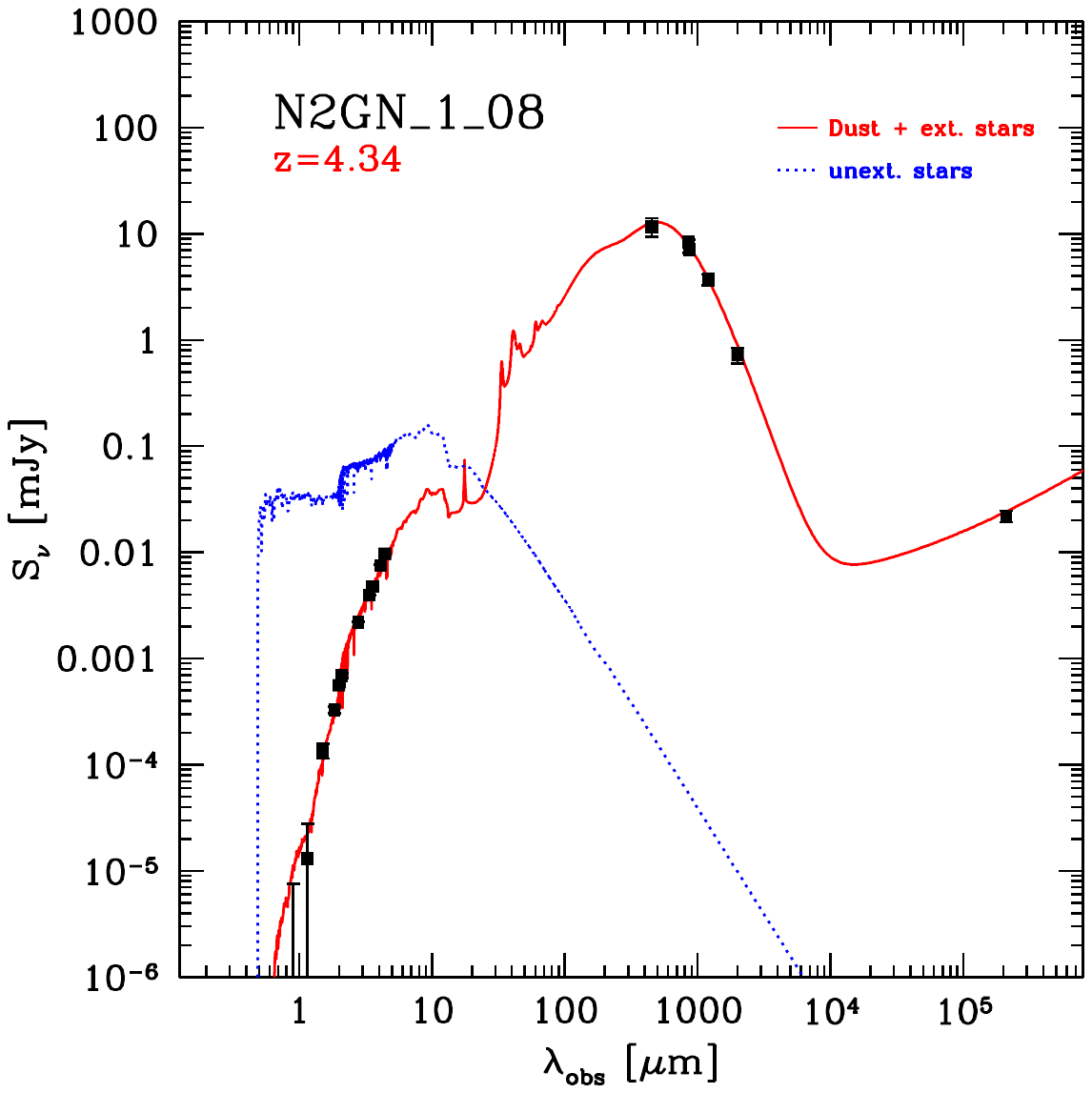}
\includegraphics[align=c,width=0.4\textwidth]{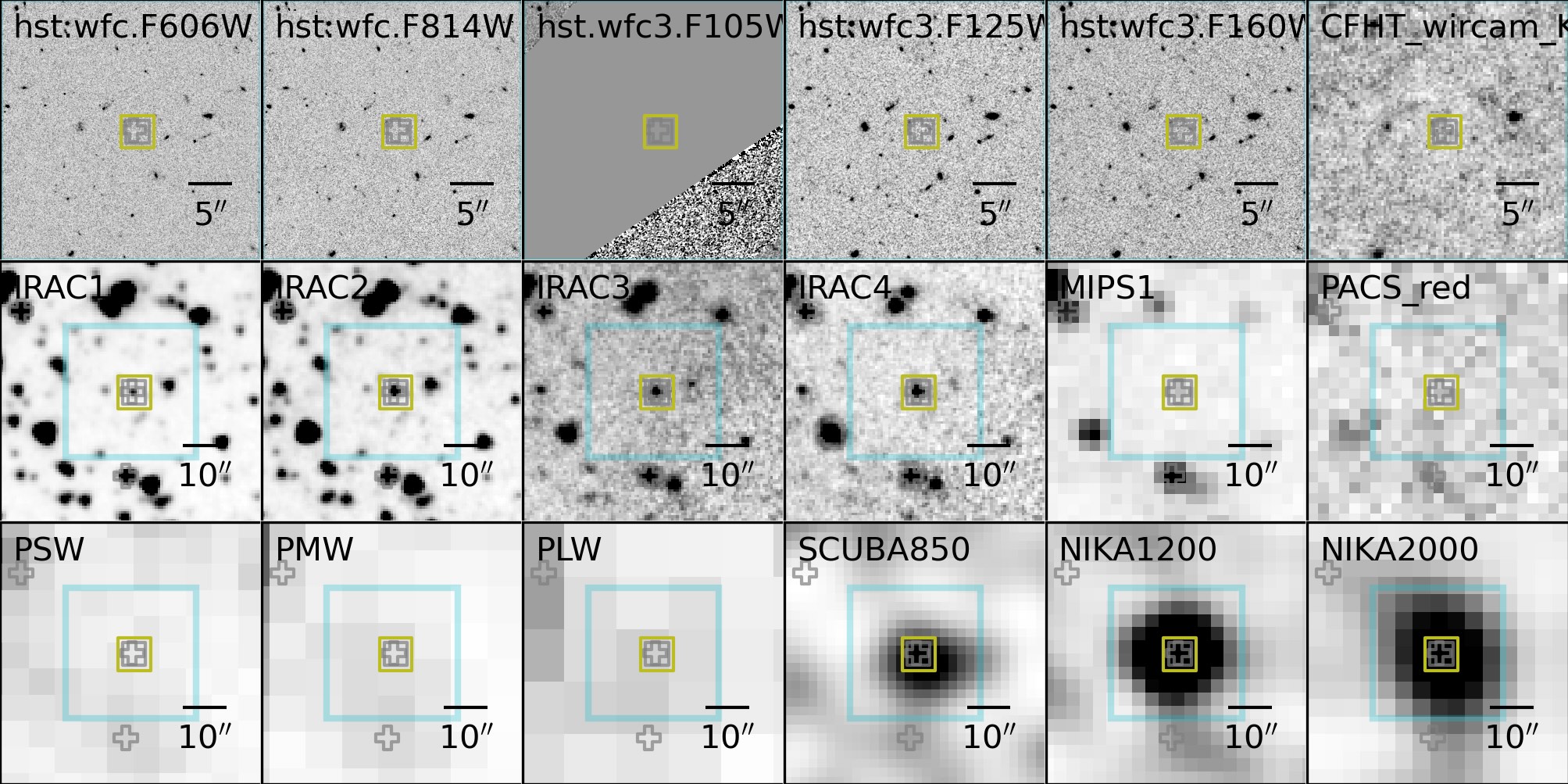}
\includegraphics[align=c,trim=0 0.18\imageheight{} 0 0.075\imageheight{}, clip, width=0.25\textwidth]{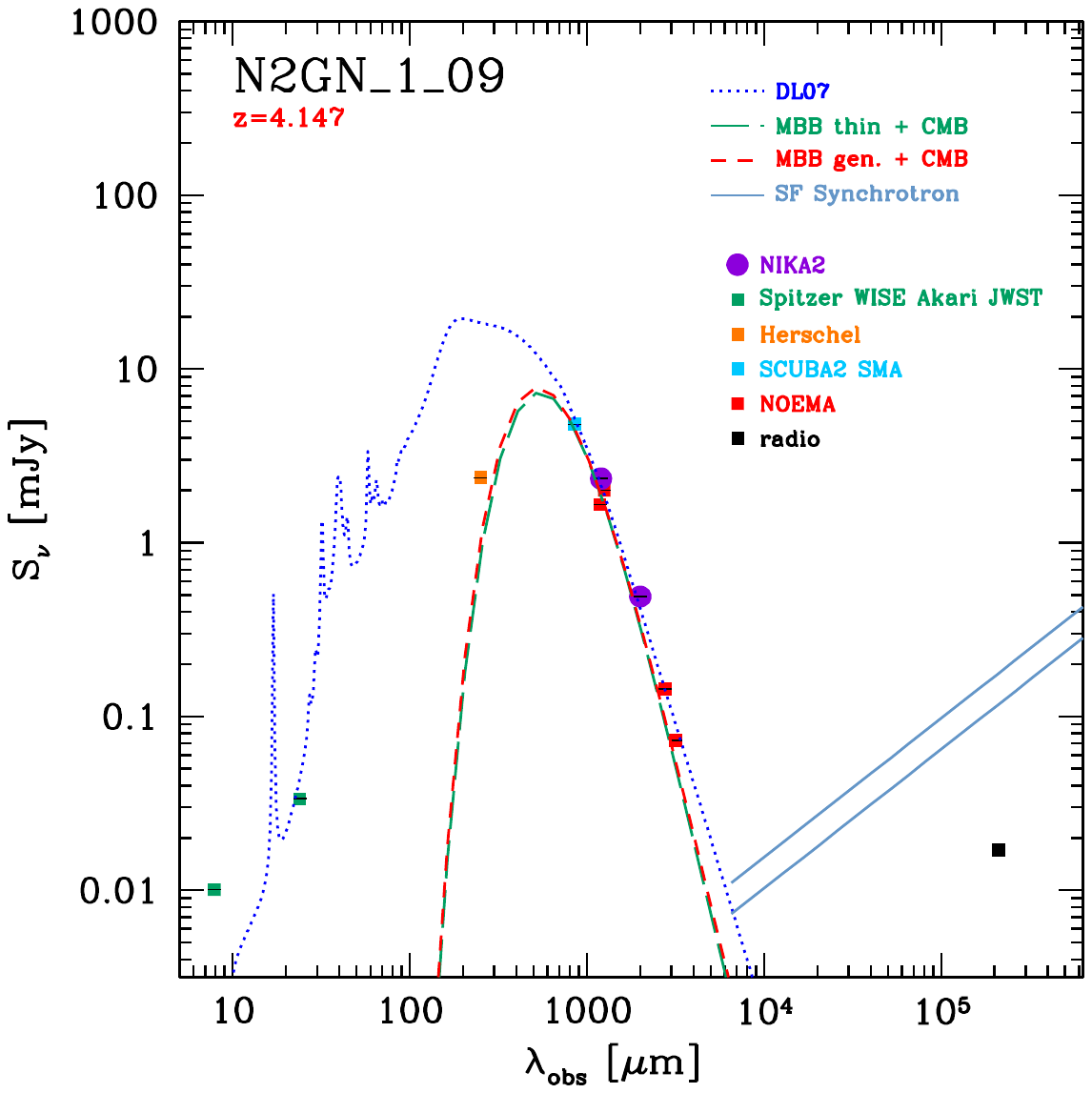}
\includegraphics[align=c,trim=0 0.18\imageheight{} 0 0.075\imageheight{}, clip, width=0.25\textwidth]{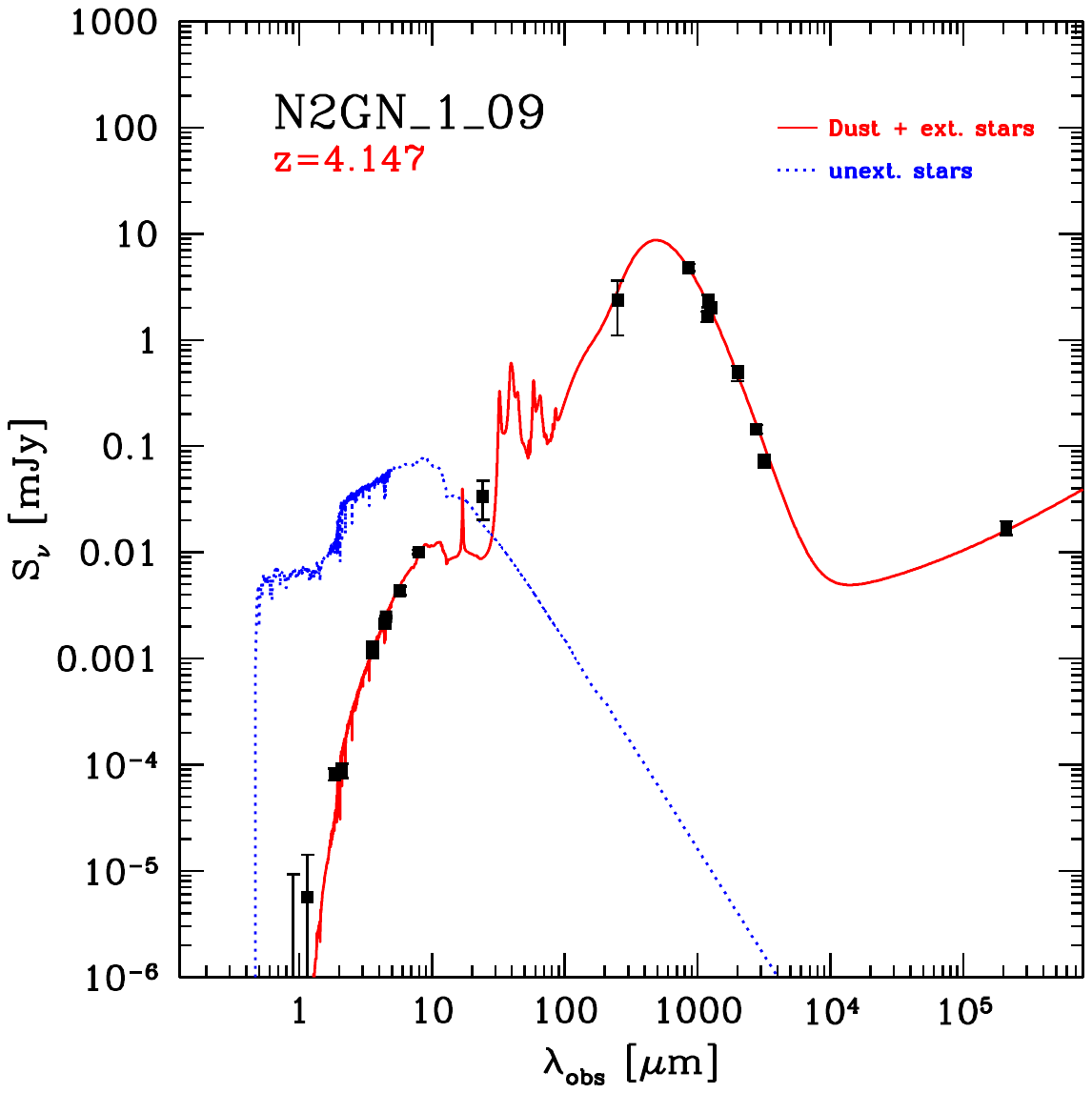}
\caption{continued.}
\end{figure*}

\addtocounter{figure}{-1}
\newpage

\begin{figure*}[t]
\centering
\includegraphics[align=c,width=0.4\textwidth]{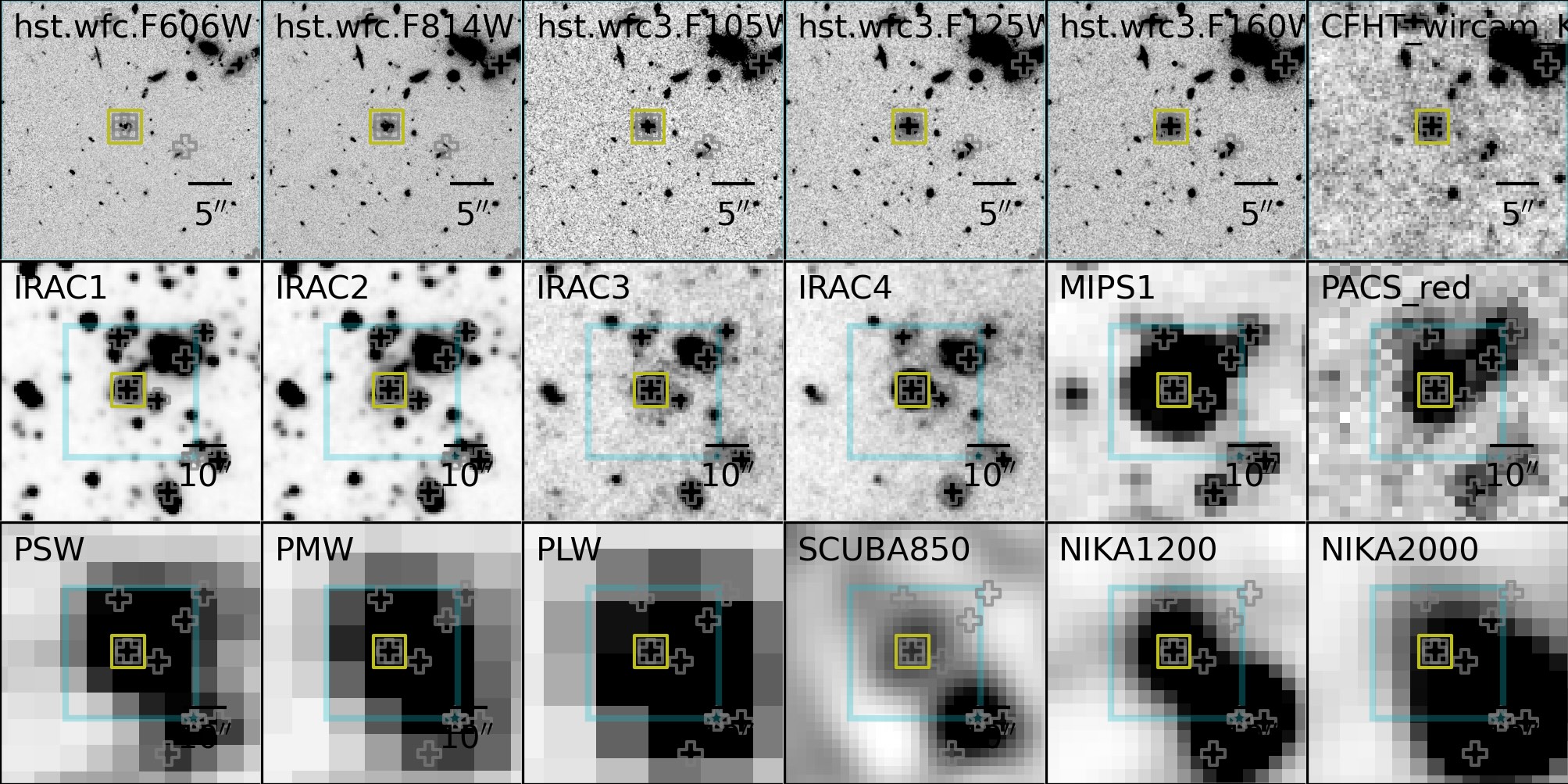}
\includegraphics[align=c,trim=0 0.18\imageheight{} 0 0.075\imageheight{}, clip, width=0.25\textwidth]{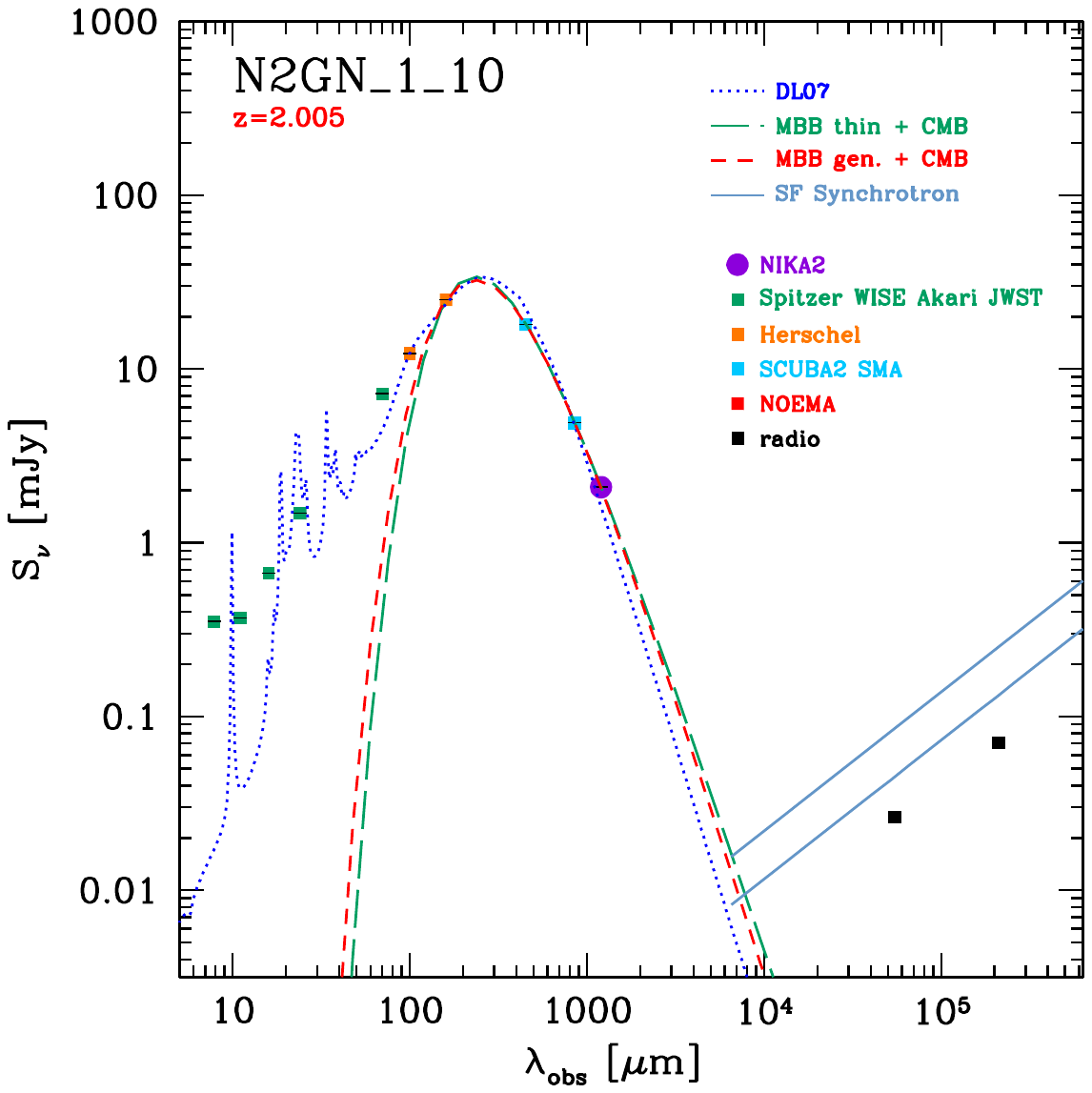}
\includegraphics[align=c,trim=0 0.18\imageheight{} 0 0.075\imageheight{}, clip, width=0.25\textwidth]{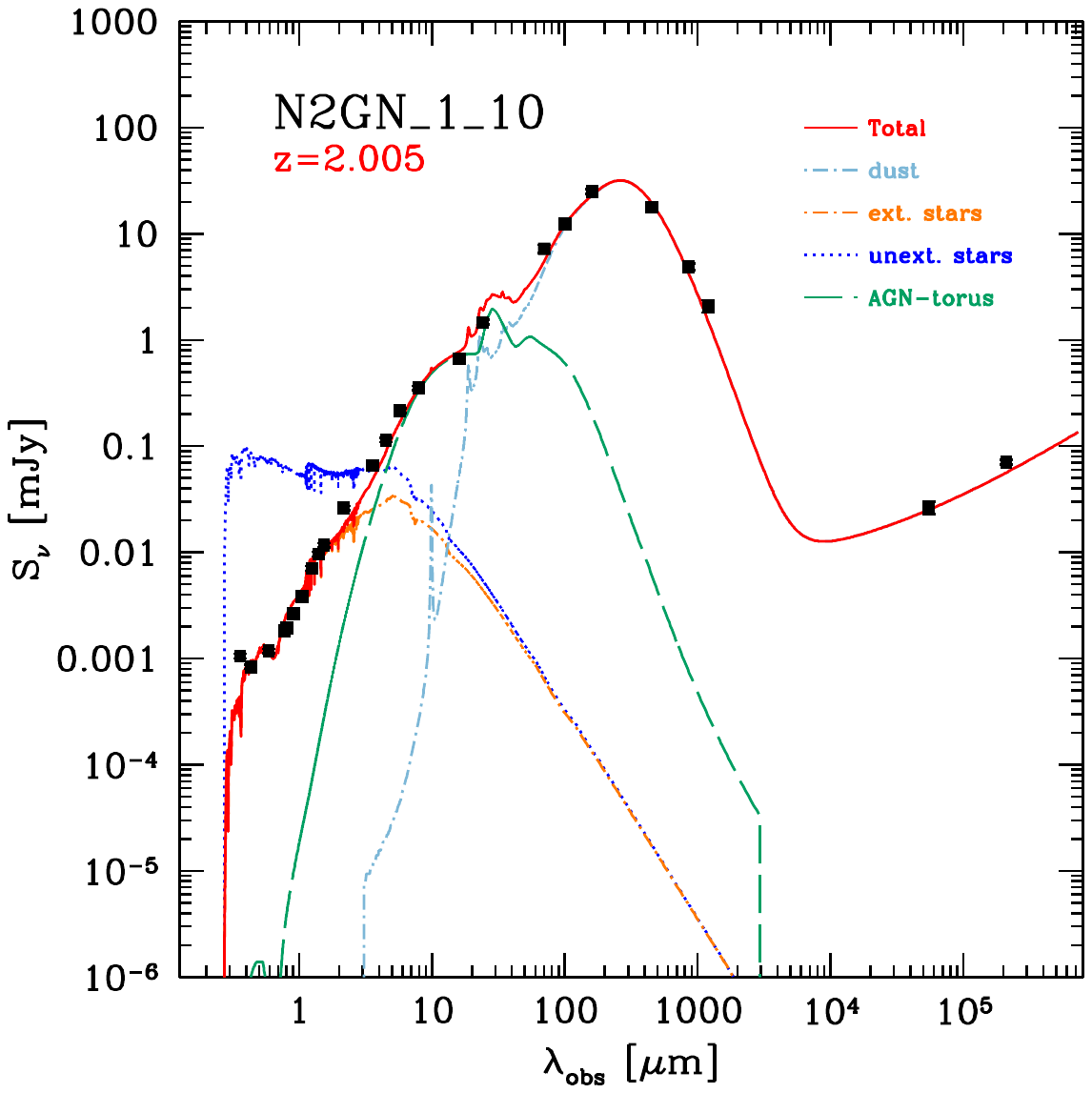}
\includegraphics[align=c,width=0.4\textwidth]{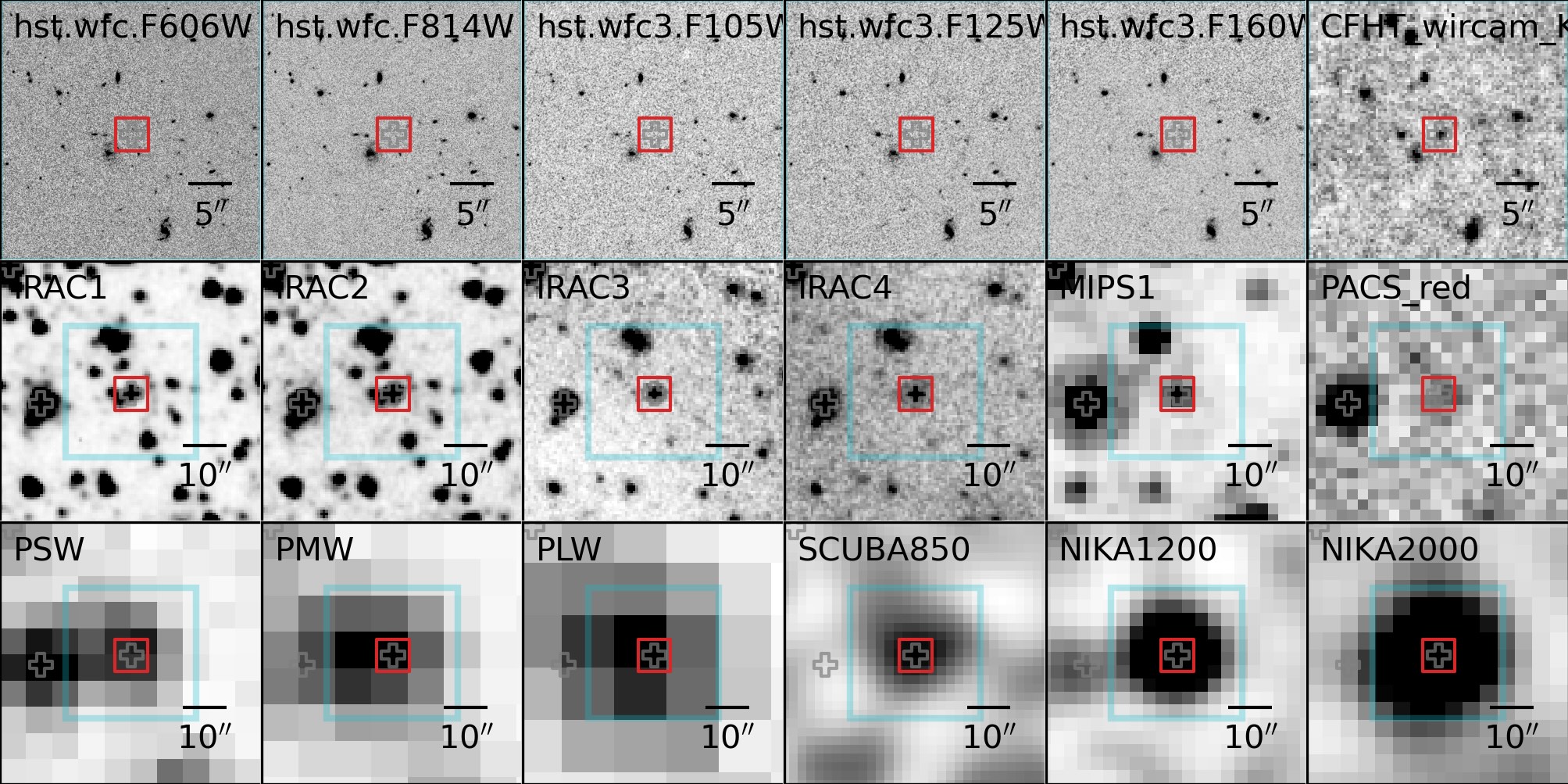}
\includegraphics[align=c,trim=0 0.18\imageheight{} 0 0.075\imageheight{}, clip, width=0.25\textwidth]{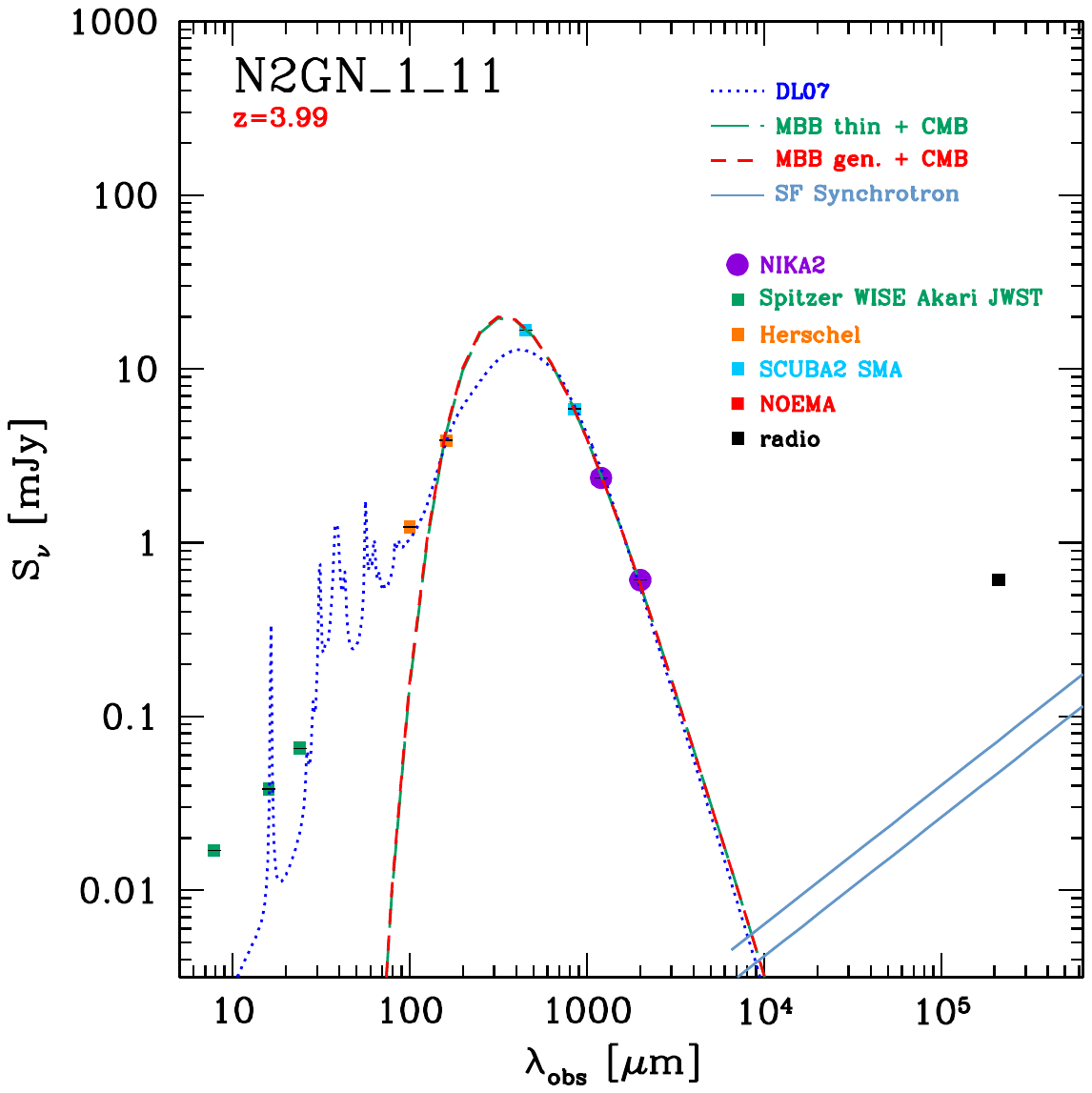}
\includegraphics[align=c,trim=0 0.18\imageheight{} 0 0.075\imageheight{}, clip, width=0.25\textwidth]{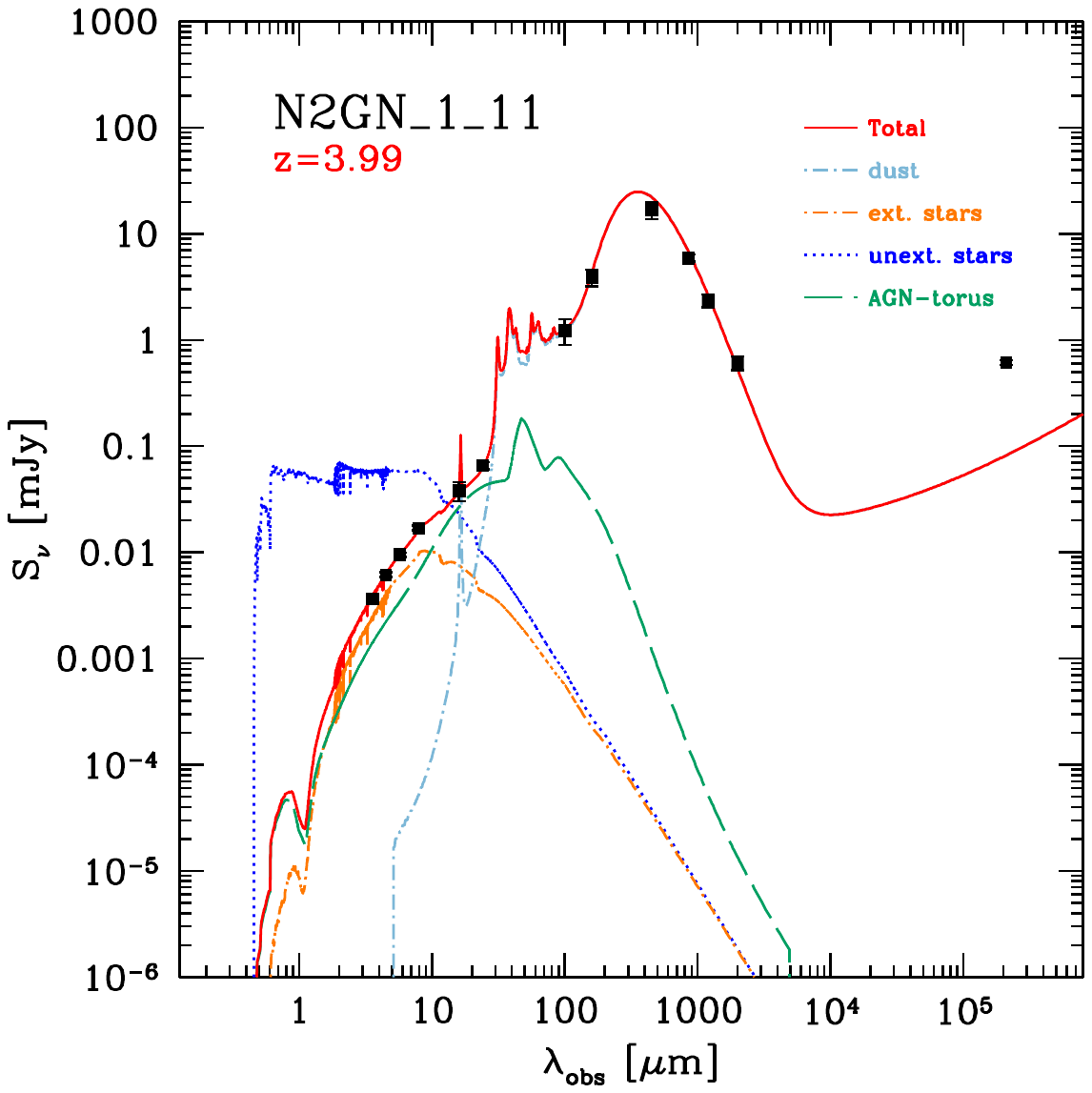}
\includegraphics[align=c,width=0.4\textwidth]{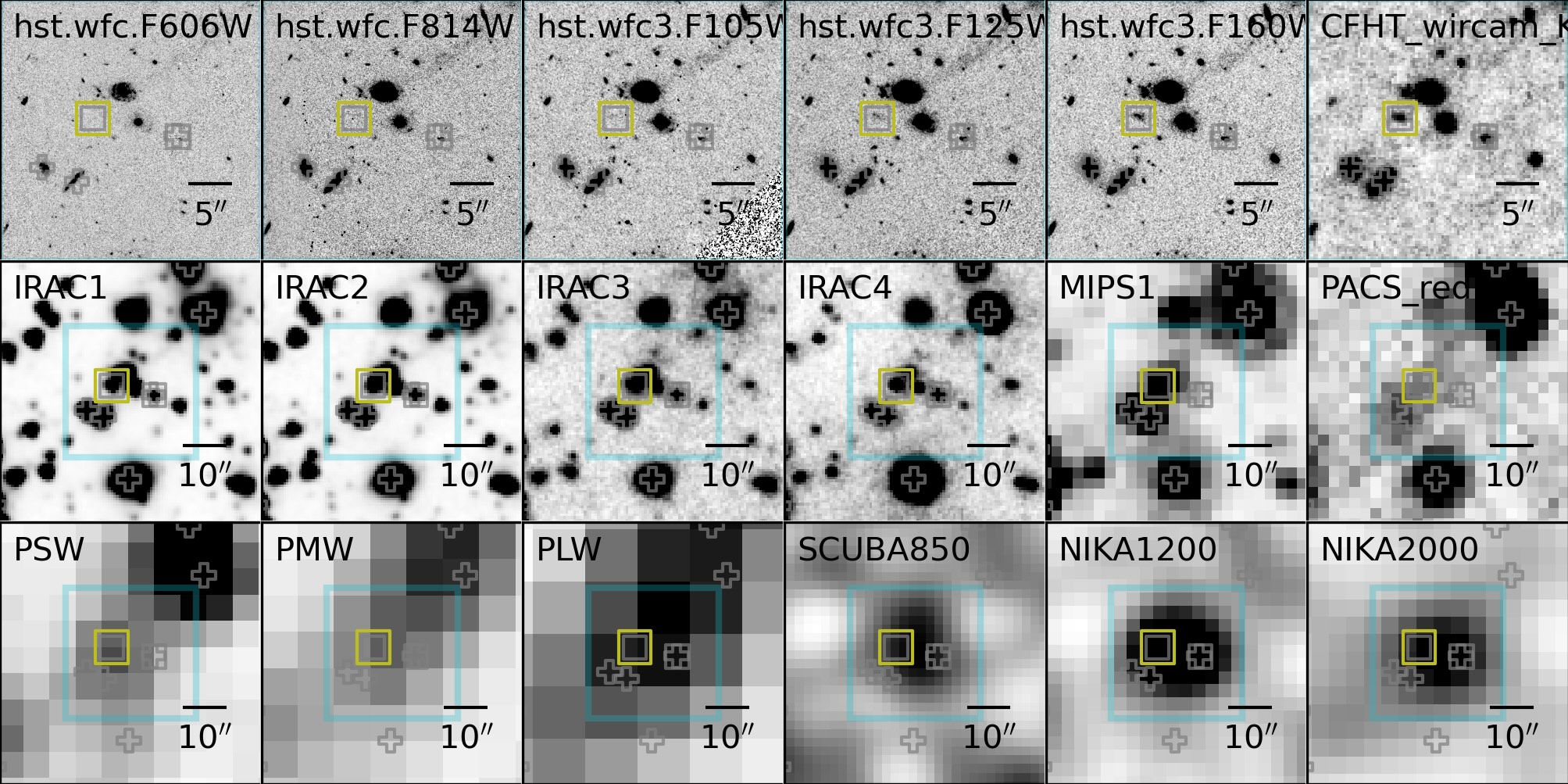}
\includegraphics[align=c,trim=0 0.18\imageheight{} 0 0.075\imageheight{}, clip, width=0.25\textwidth]{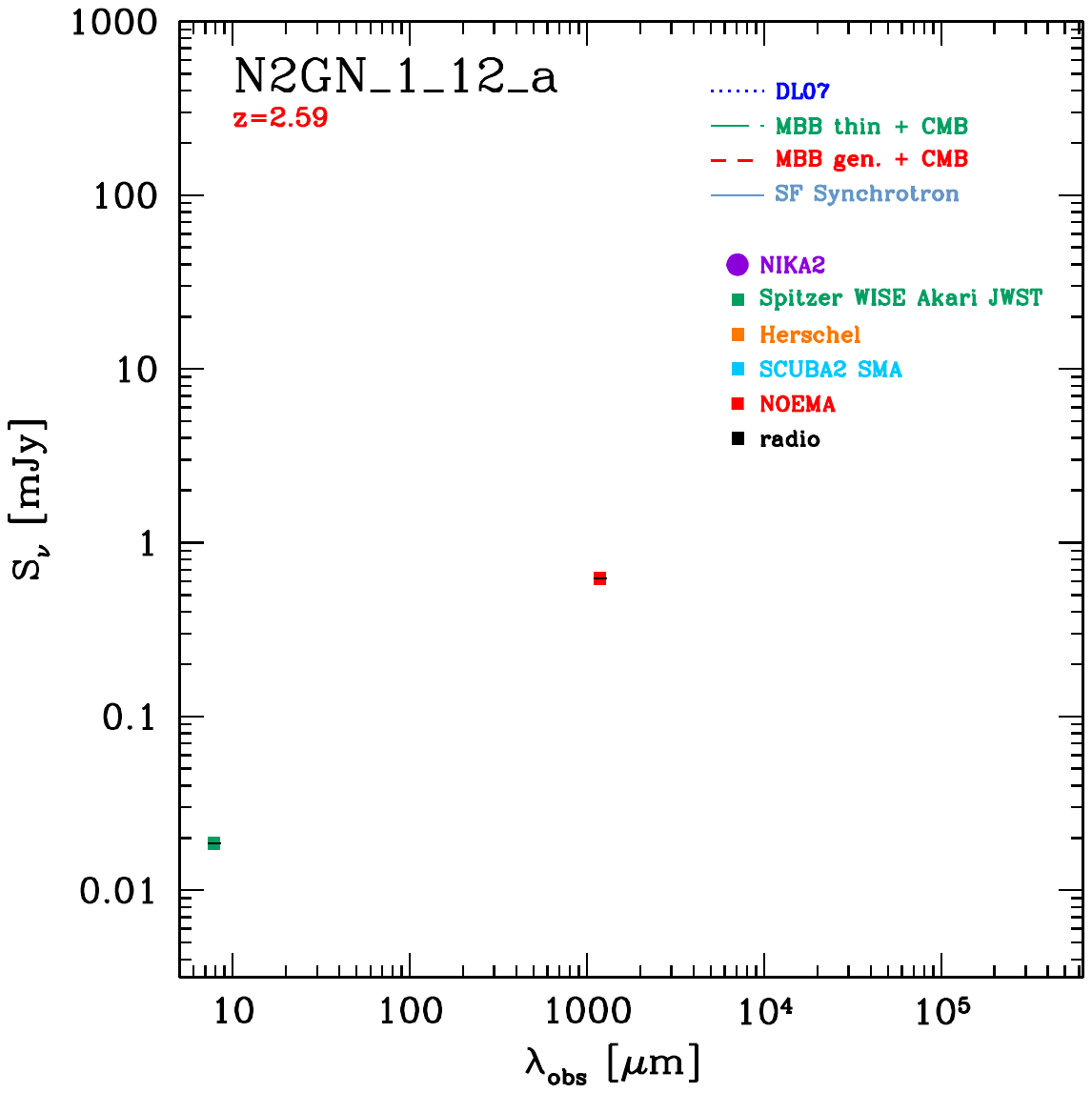}
\includegraphics[align=c,trim=0 0.18\imageheight{} 0 0.075\imageheight{}, clip, width=0.25\textwidth]{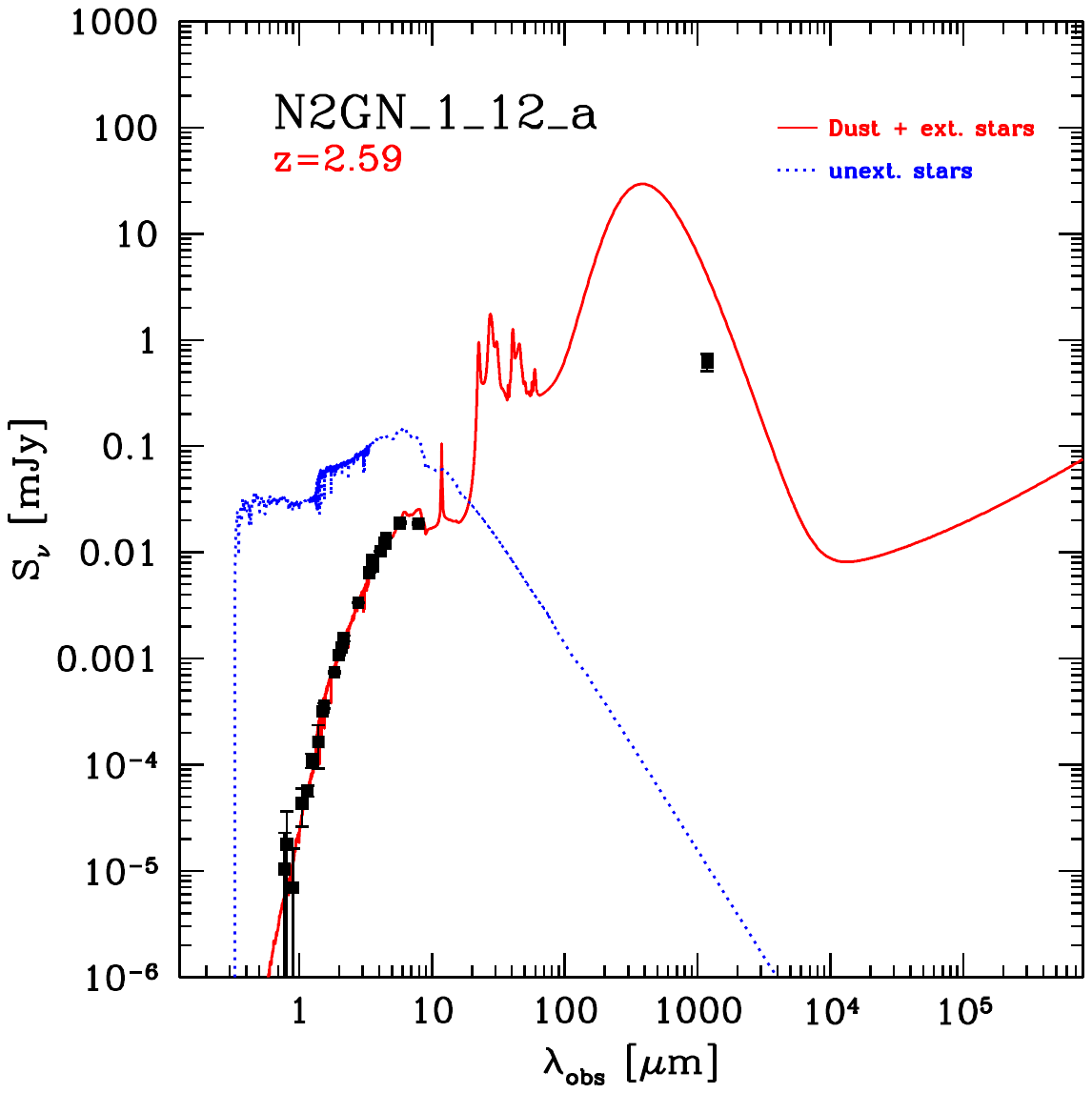}
\settoheight{\imageheight}{\includegraphics{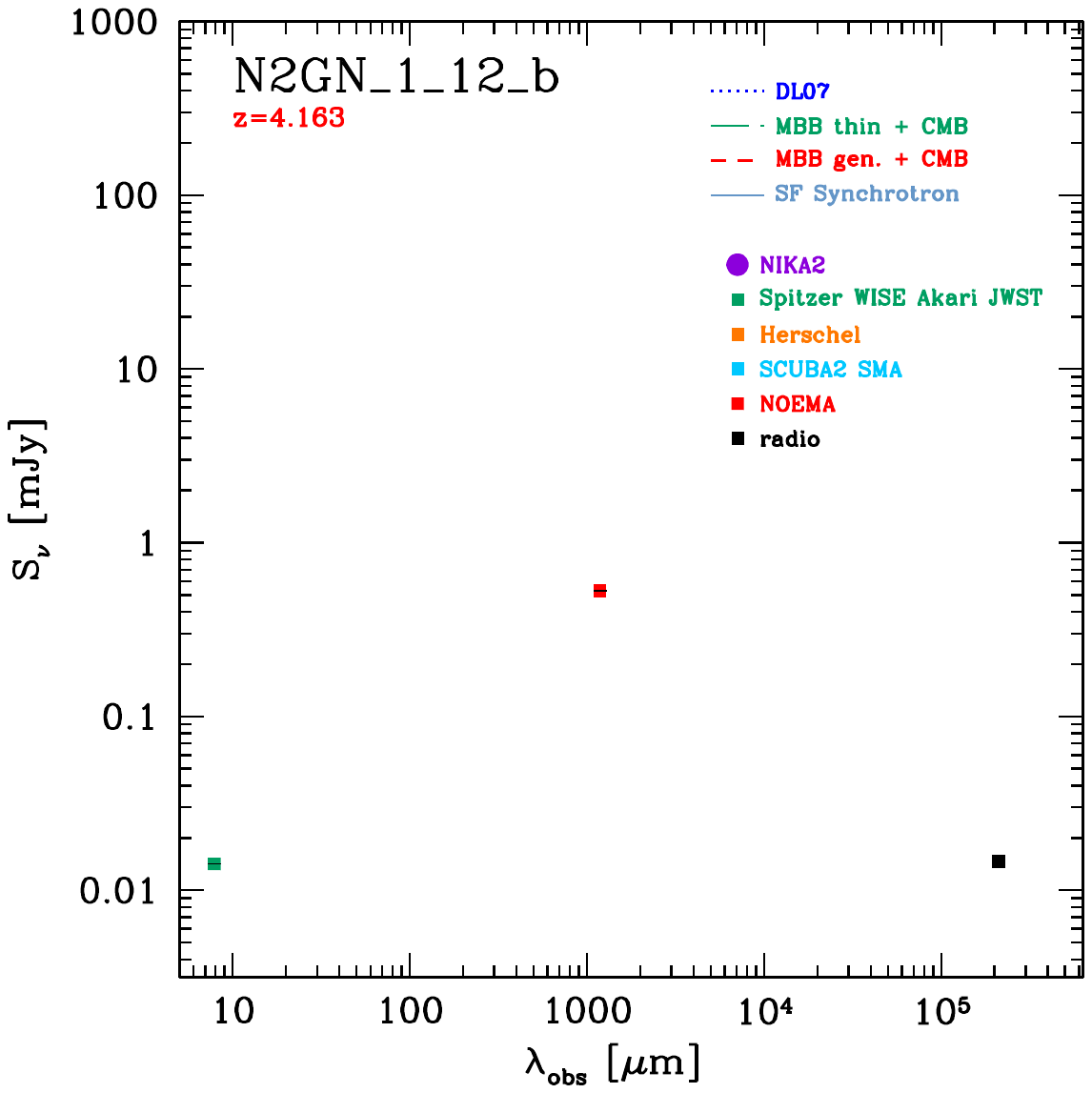}}
\includegraphics[align=c,width=0.4\textwidth]{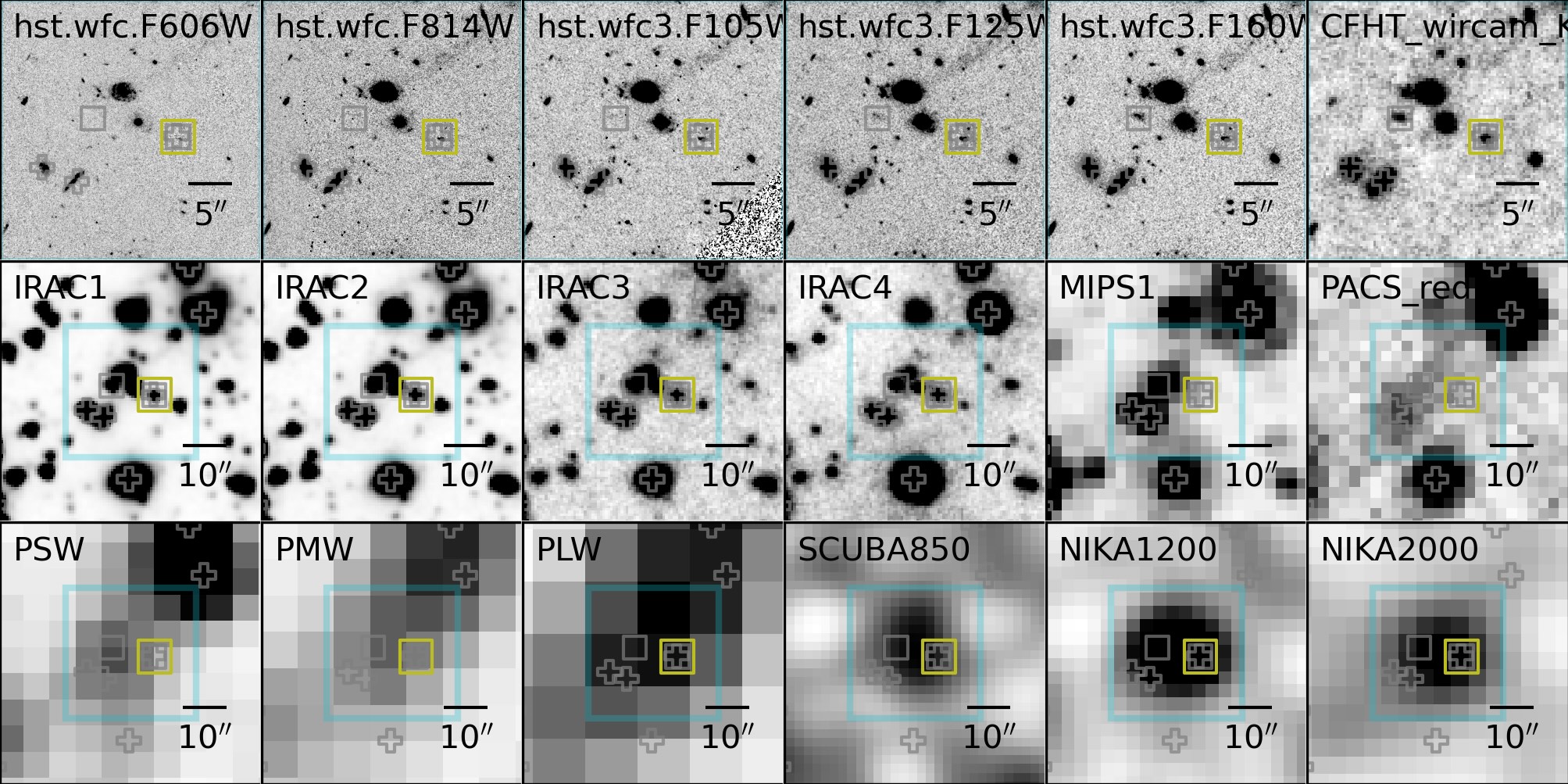}
\includegraphics[align=c,trim=0 0.18\imageheight{} 0 0.075\imageheight{}, clip, width=0.25\textwidth]{figs2_indiv_objs/sed_FIR_fit_N2GN_1_12_b.pdf}
\includegraphics[align=c,trim=0 0.18\imageheight{} 0 0.075\imageheight{}, clip, width=0.25\textwidth]{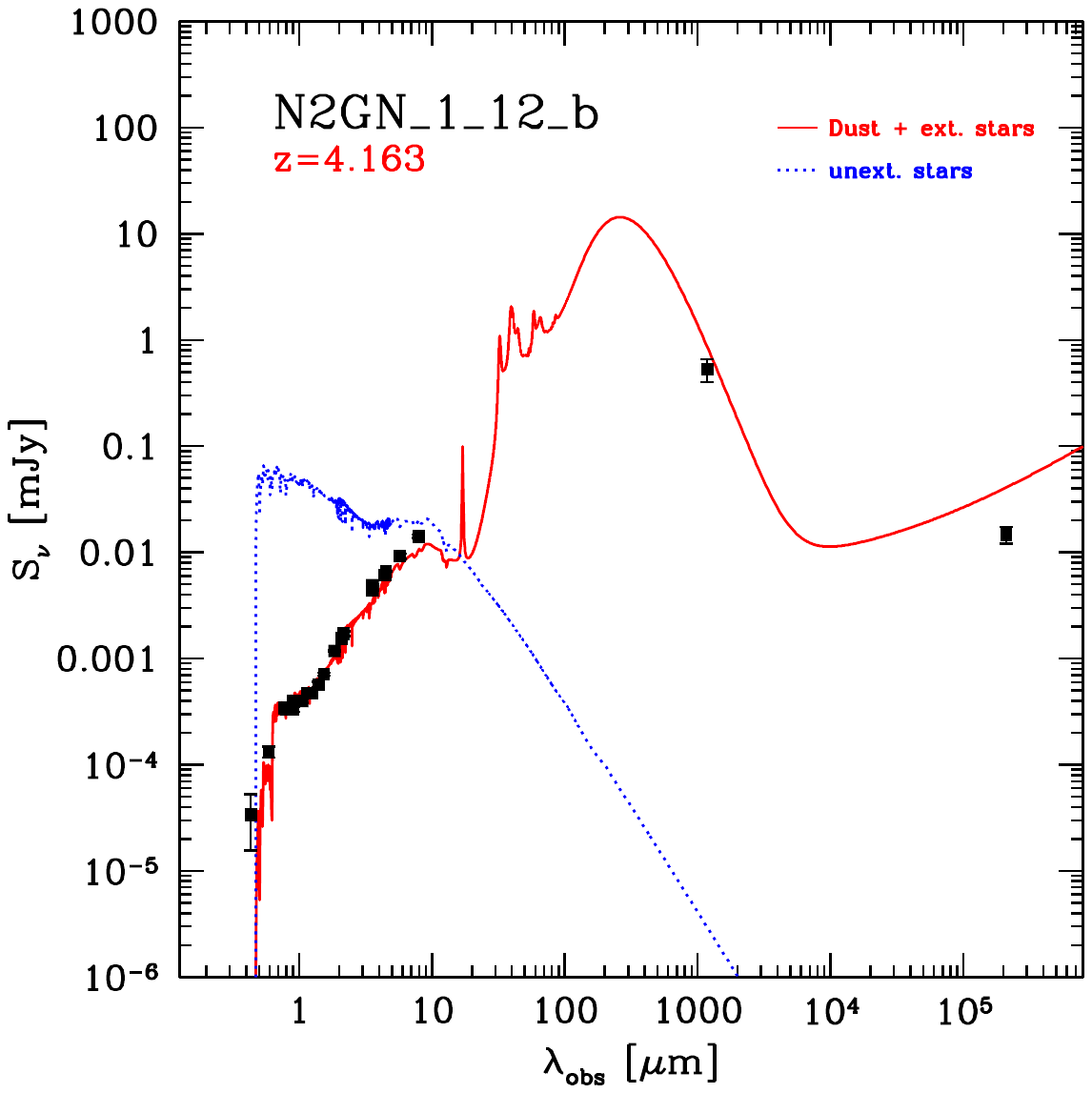}
\includegraphics[align=c,width=0.4\textwidth]{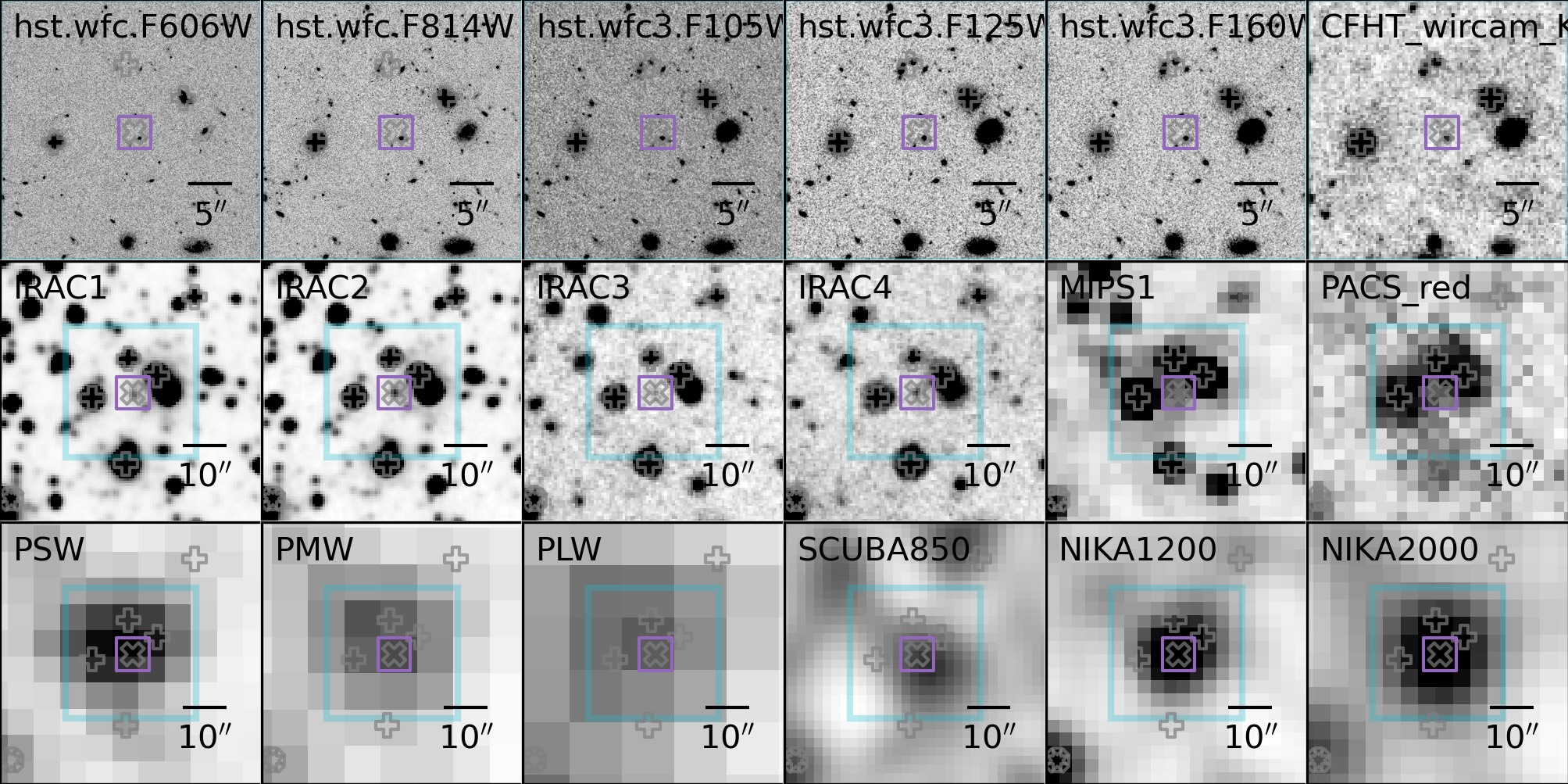}
\includegraphics[align=c,trim=0 0.18\imageheight{} 0 0.075\imageheight{}, clip, width=0.25\textwidth]{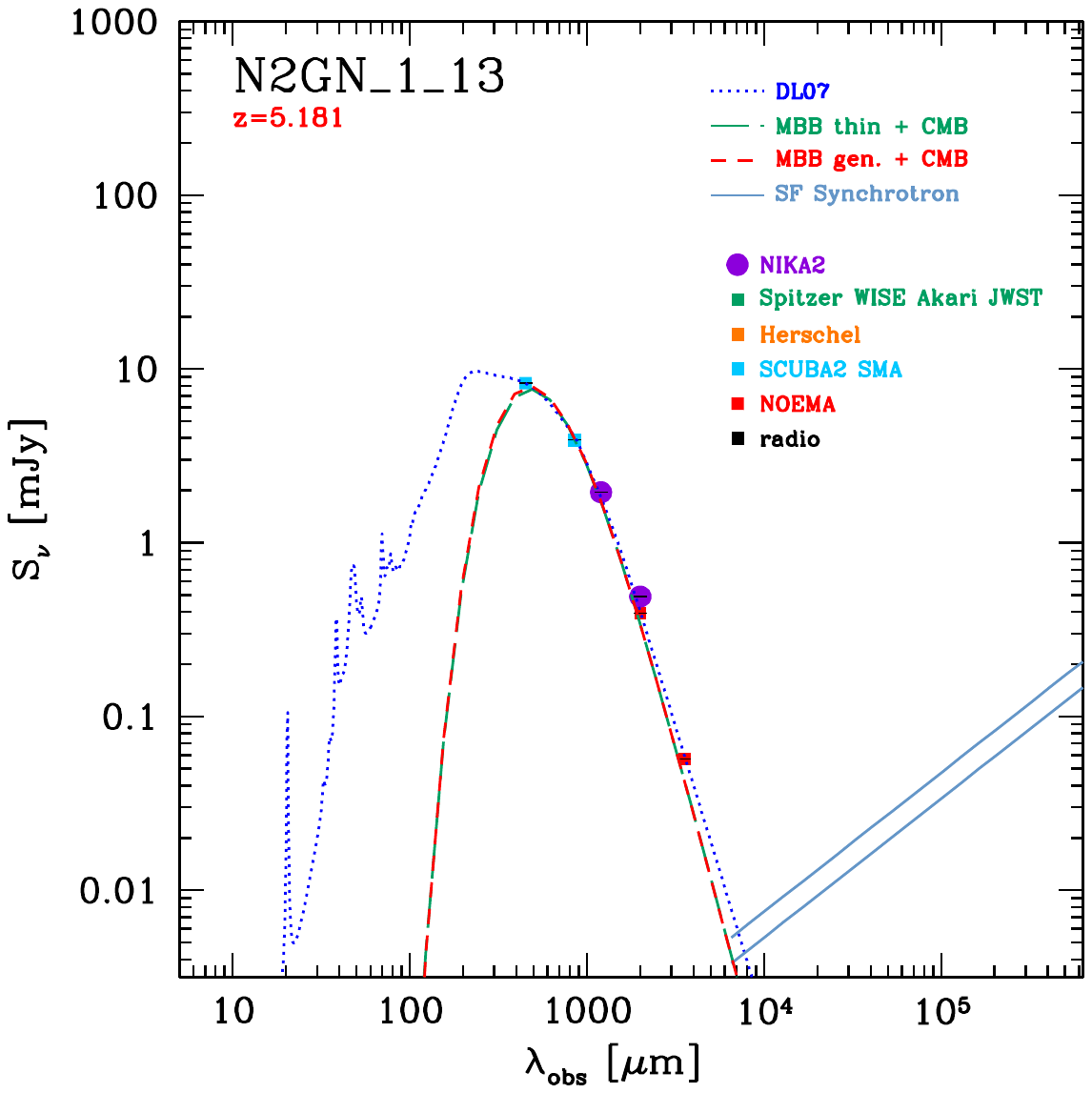}
\includegraphics[align=c,trim=0 0.18\imageheight{} 0 0.075\imageheight{}, clip, width=0.25\textwidth]{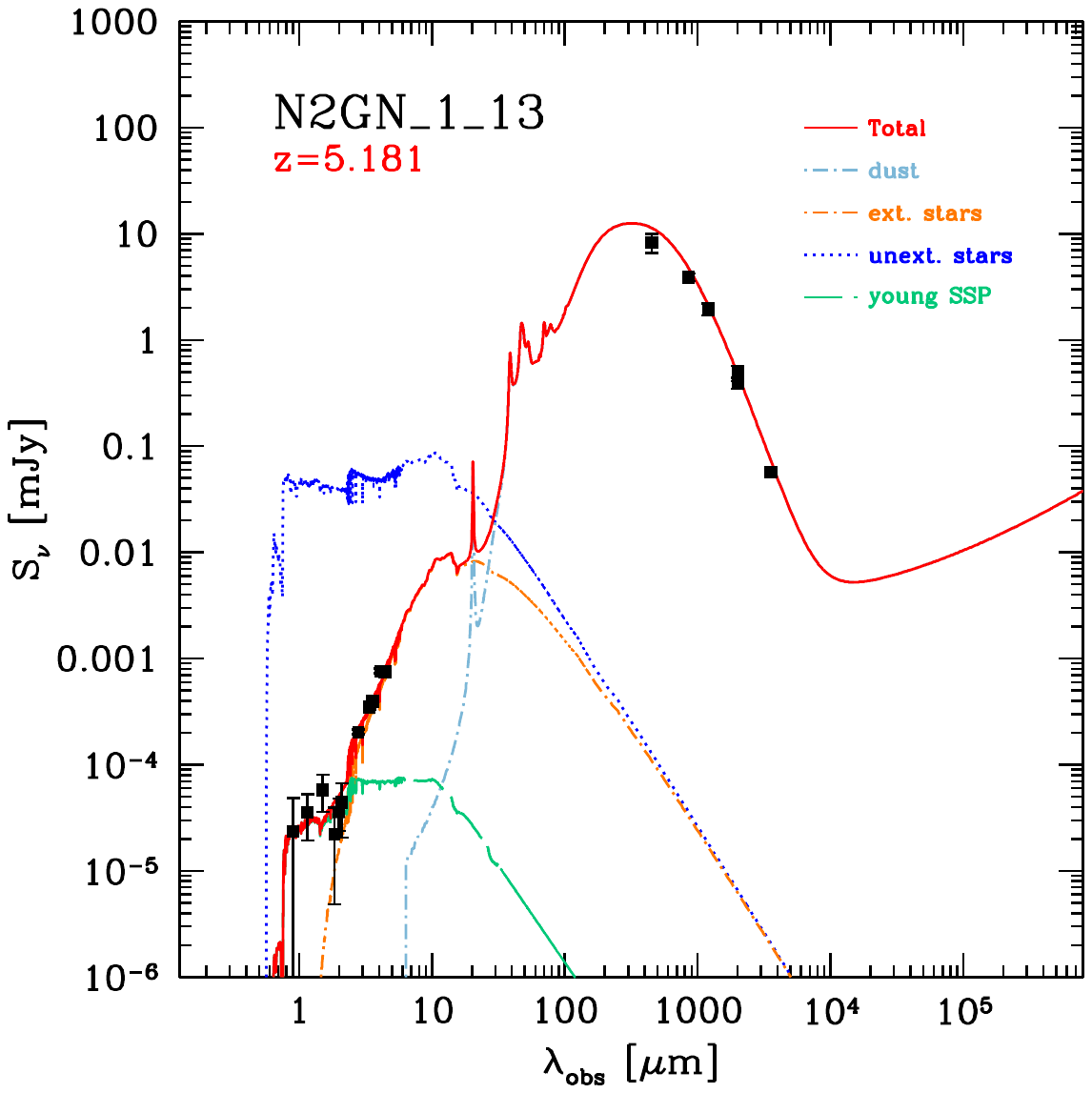}
\caption{continued.}
\end{figure*}

\addtocounter{figure}{-1}
\newpage

\begin{figure*}[t]
\centering
\includegraphics[align=c,width=0.4\textwidth]{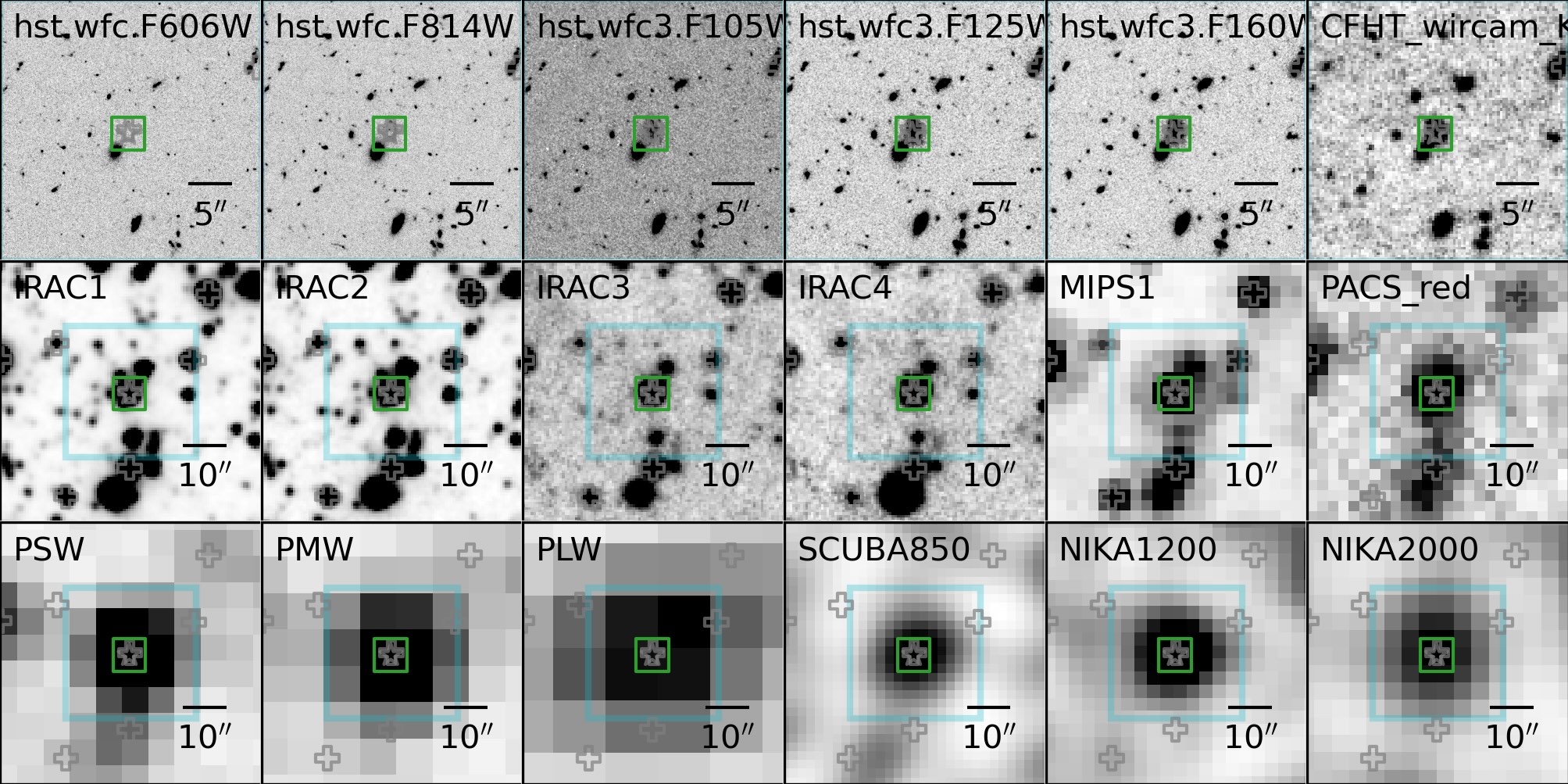}
\includegraphics[align=c,trim=0 0.18\imageheight{} 0 0.075\imageheight{}, clip, width=0.25\textwidth]{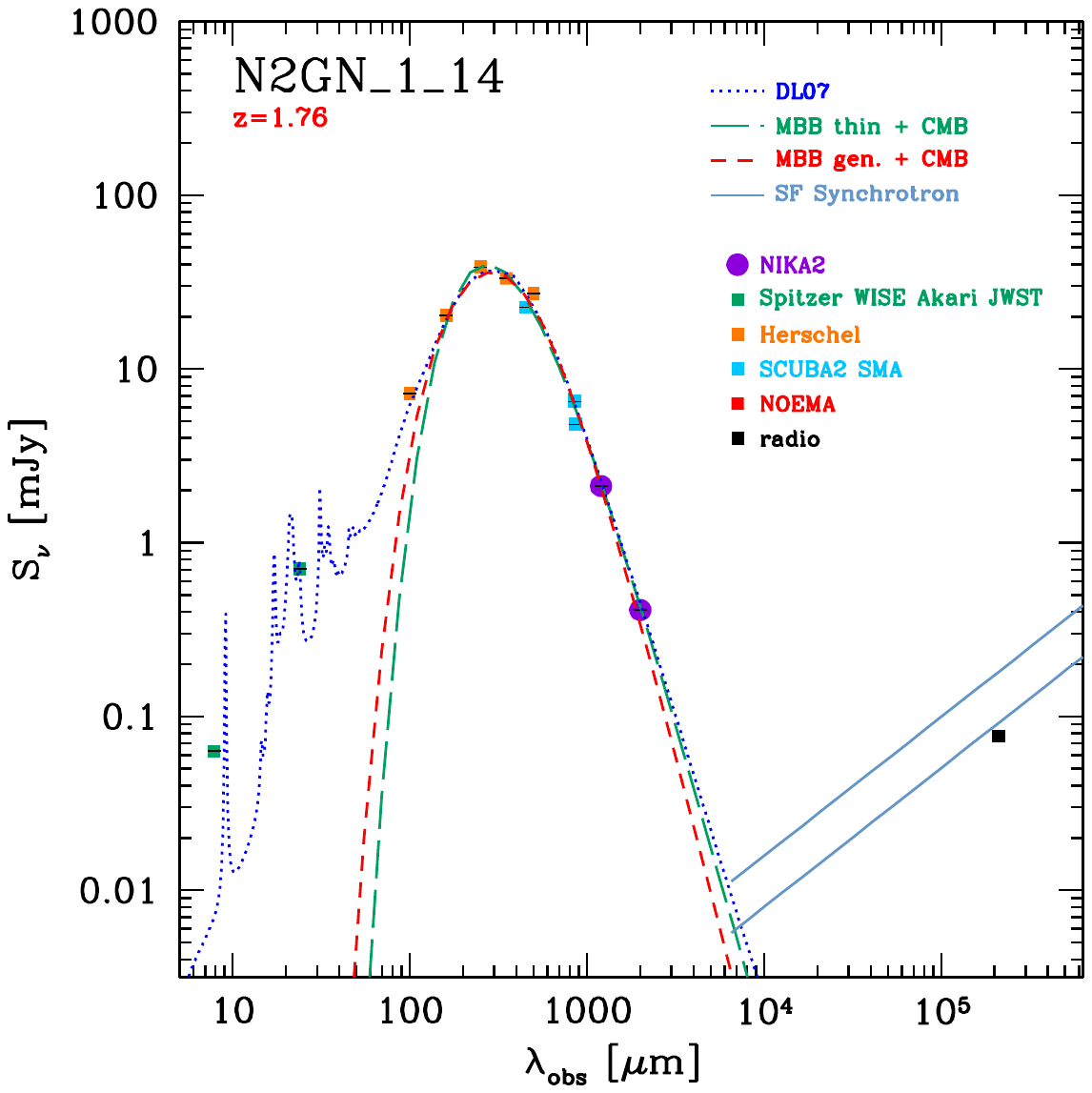}
\includegraphics[align=c,trim=0 0.18\imageheight{} 0 0.075\imageheight{}, clip, width=0.25\textwidth]{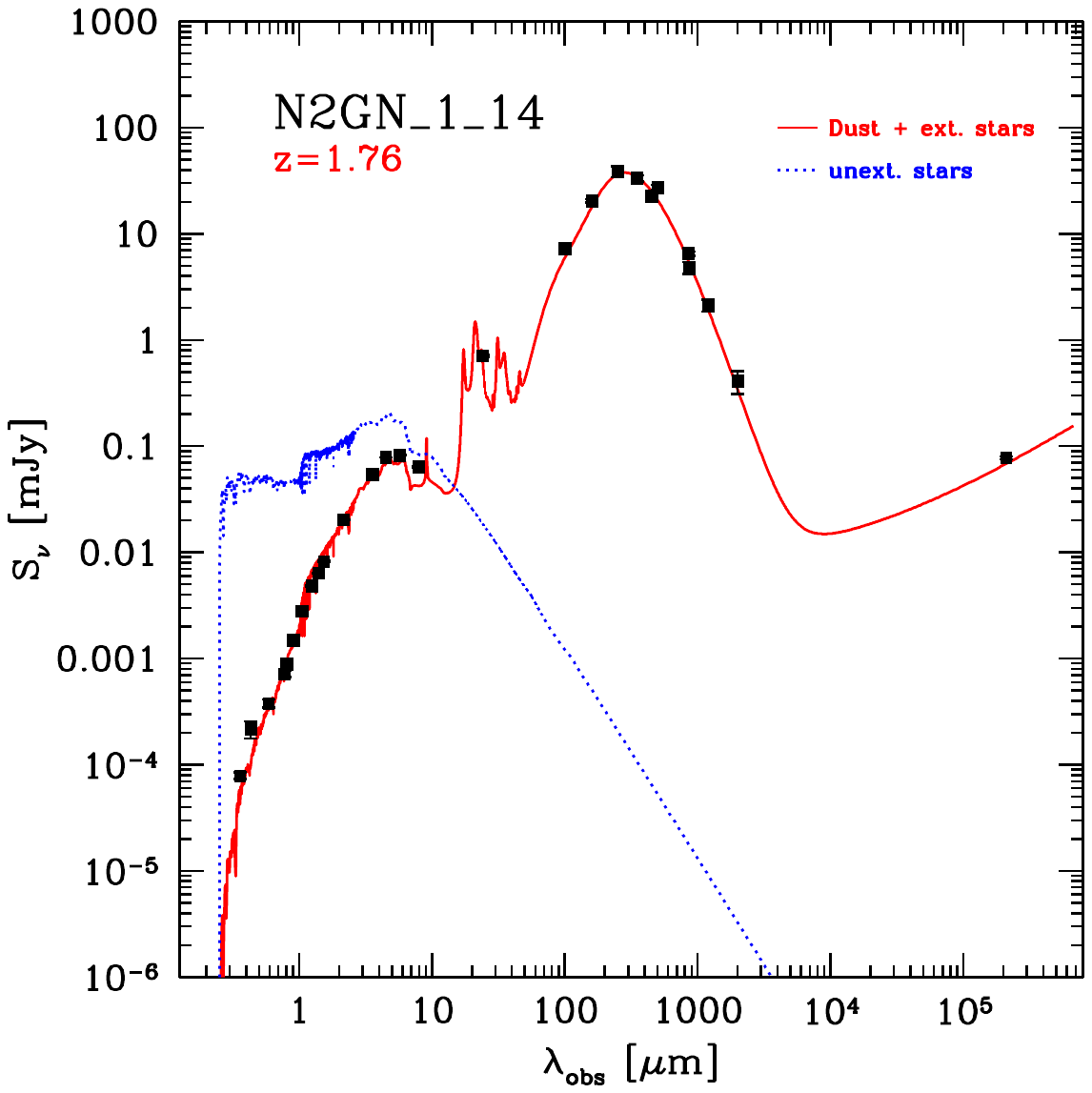}
\includegraphics[align=c,width=0.4\textwidth]{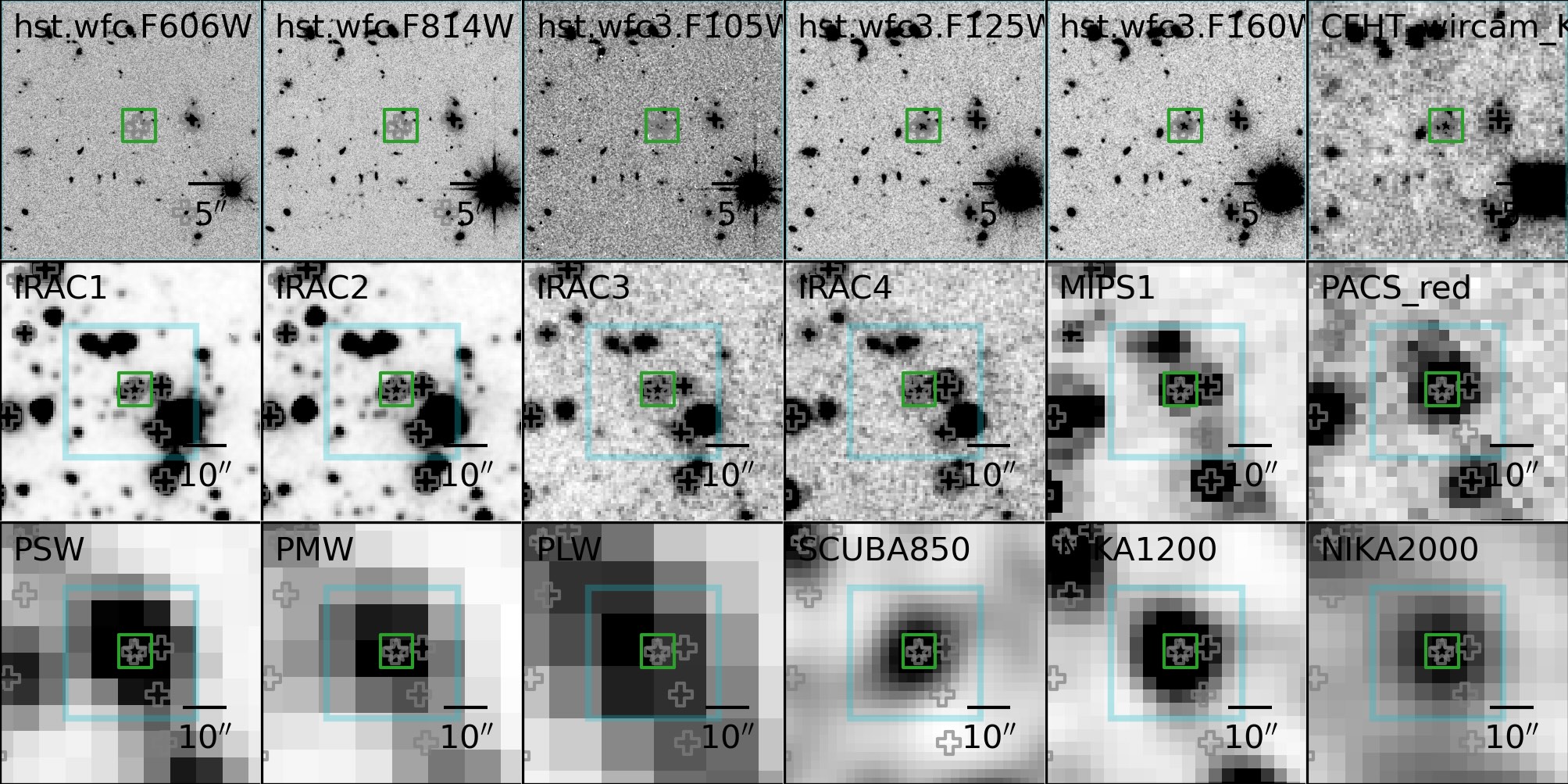}
\includegraphics[align=c,trim=0 0.18\imageheight{} 0 0.075\imageheight{}, clip, width=0.25\textwidth]{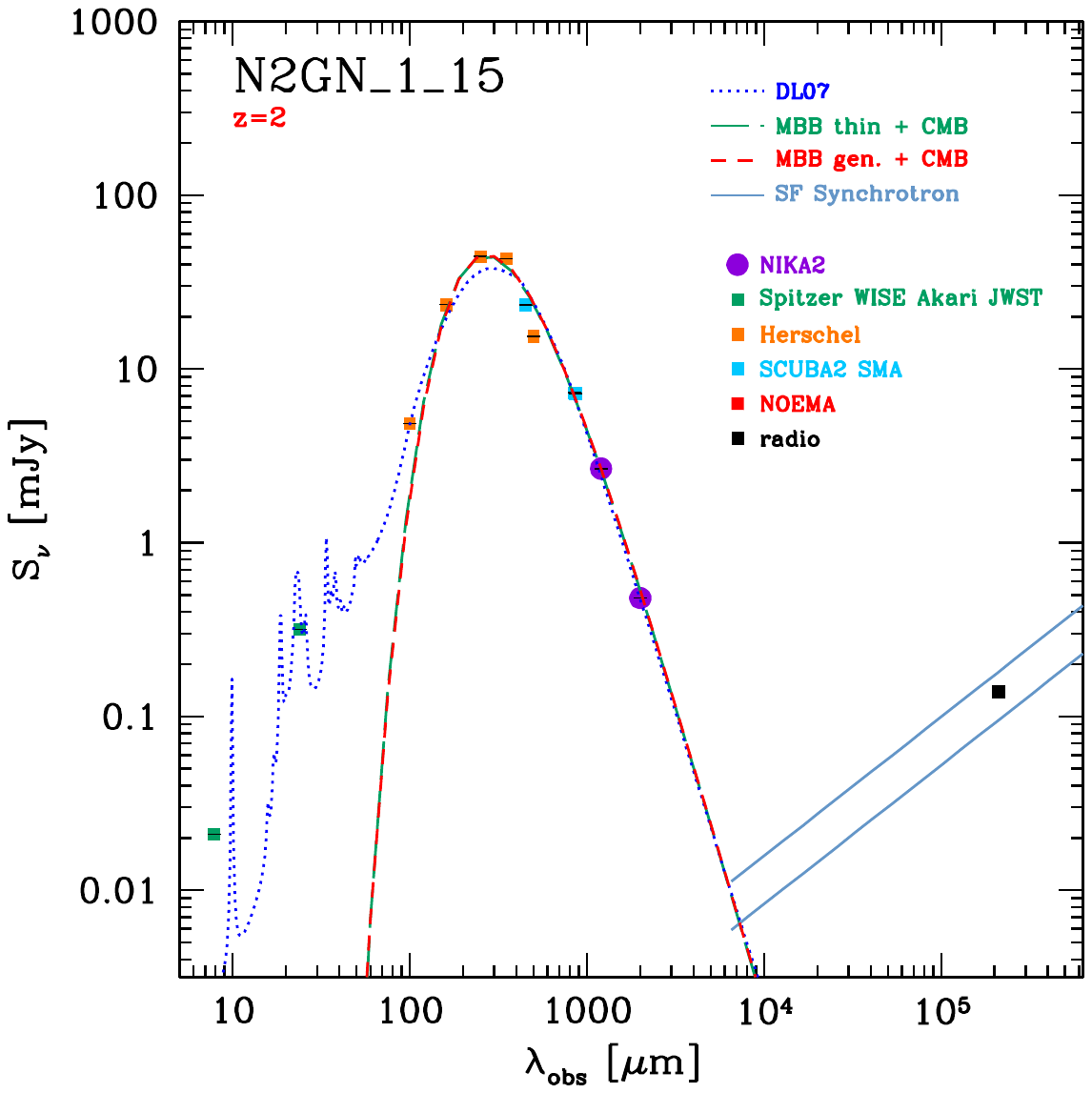}
\includegraphics[align=c,trim=0 0.18\imageheight{} 0 0.075\imageheight{}, clip, width=0.25\textwidth]{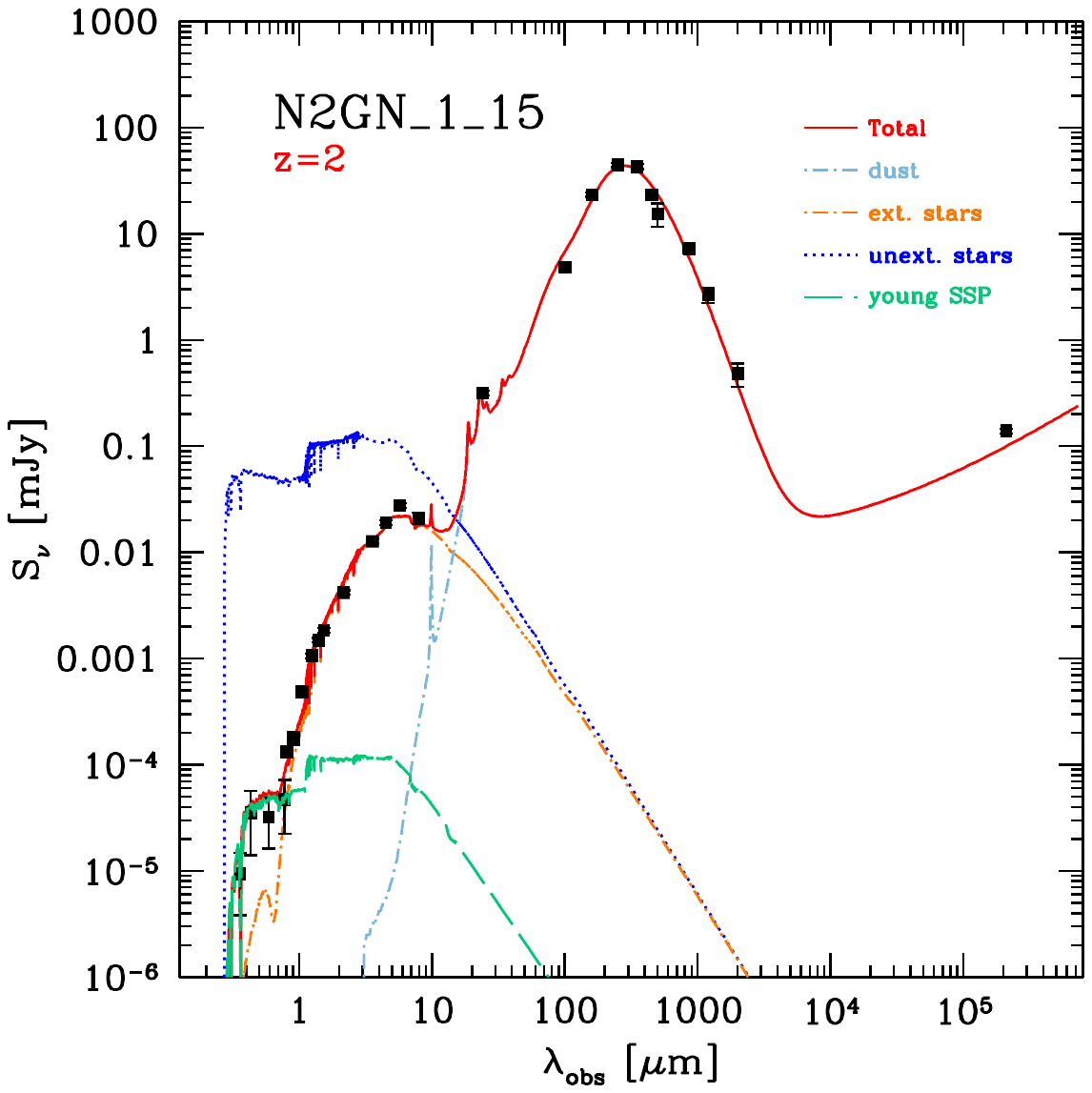}
\includegraphics[align=c,width=0.4\textwidth]{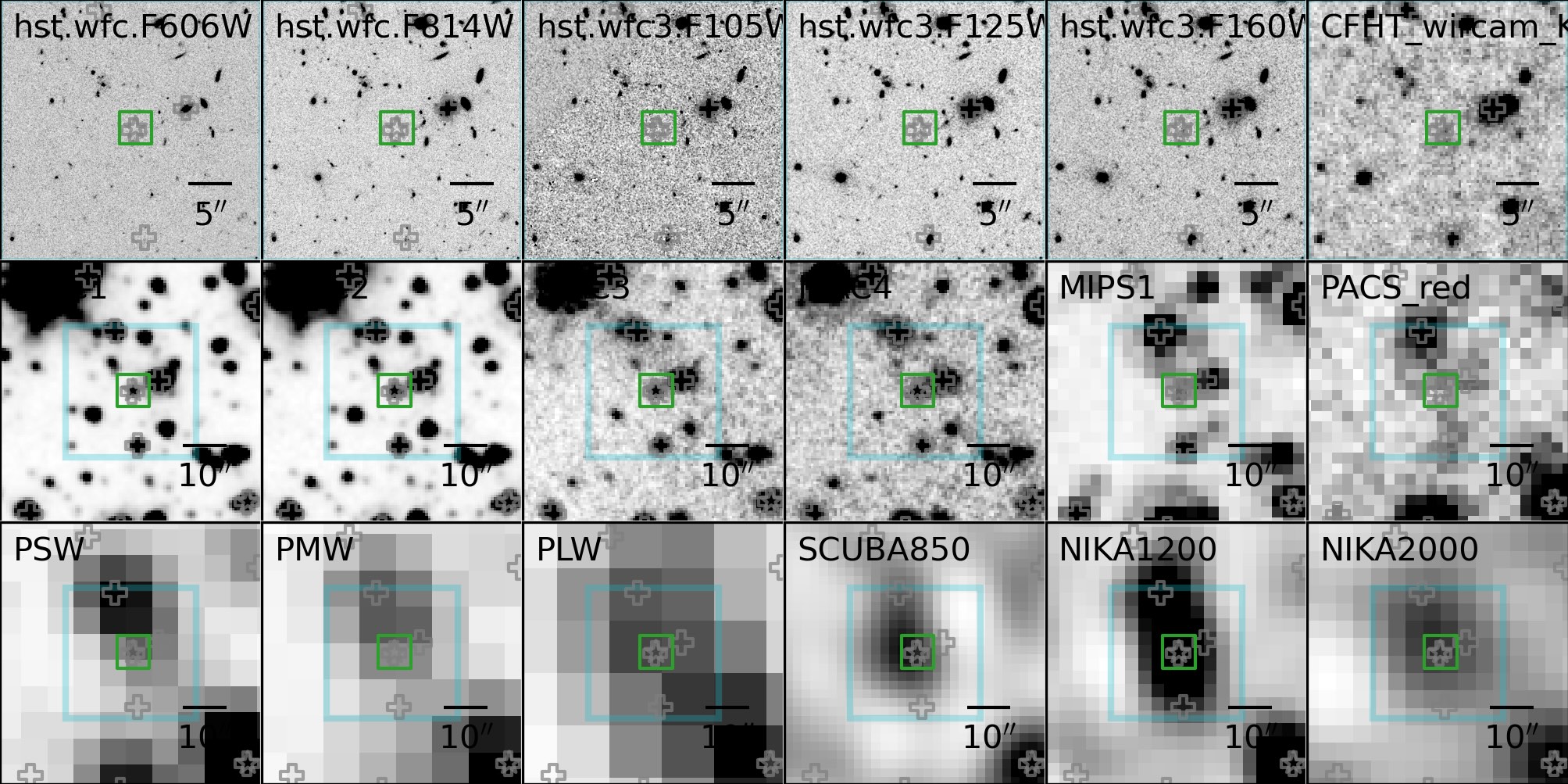}
\includegraphics[align=c,trim=0 0.18\imageheight{} 0 0.075\imageheight{}, clip, width=0.25\textwidth]{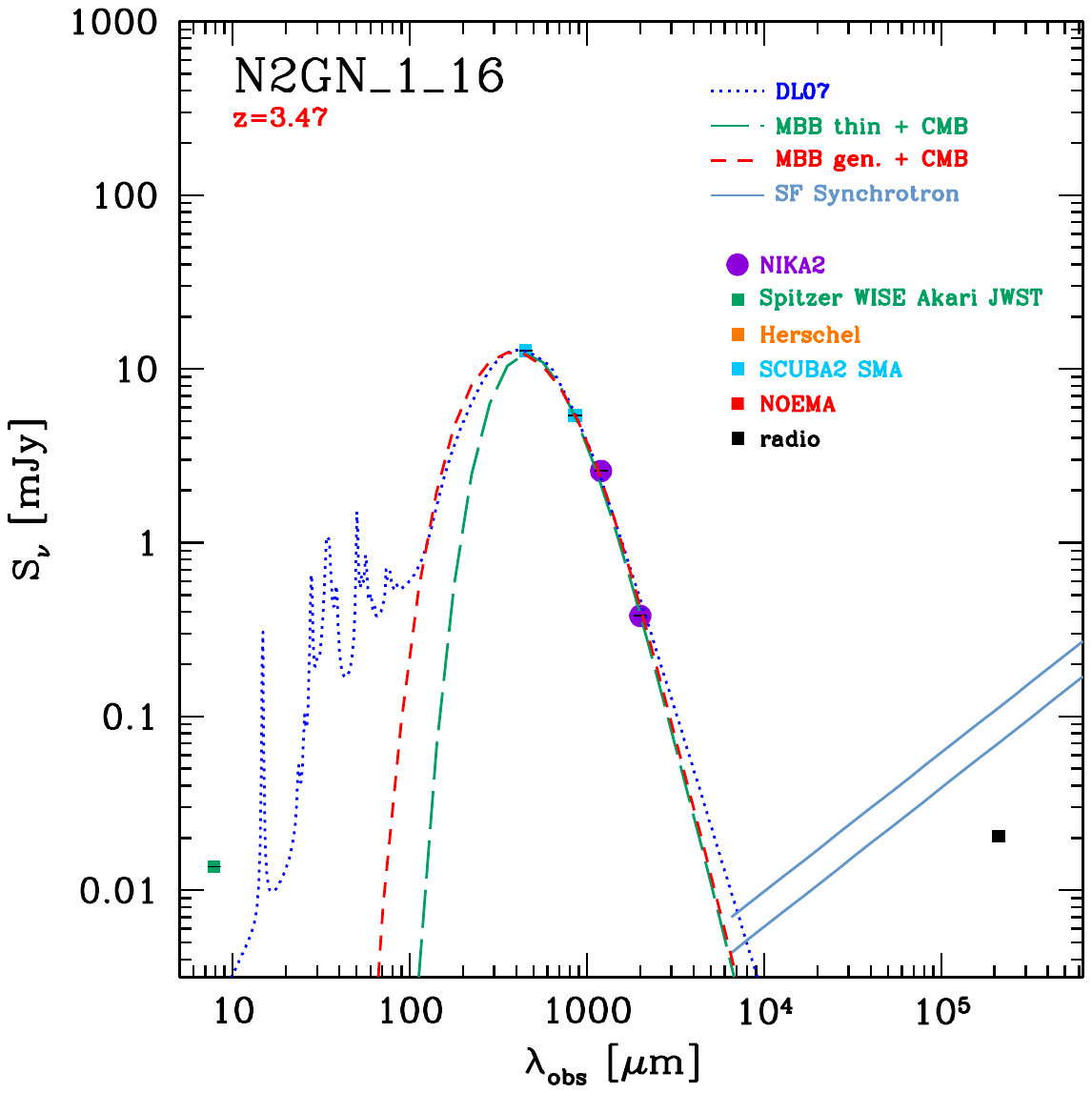}
\includegraphics[align=c,trim=0 0.18\imageheight{} 0 0.075\imageheight{}, clip, width=0.25\textwidth]{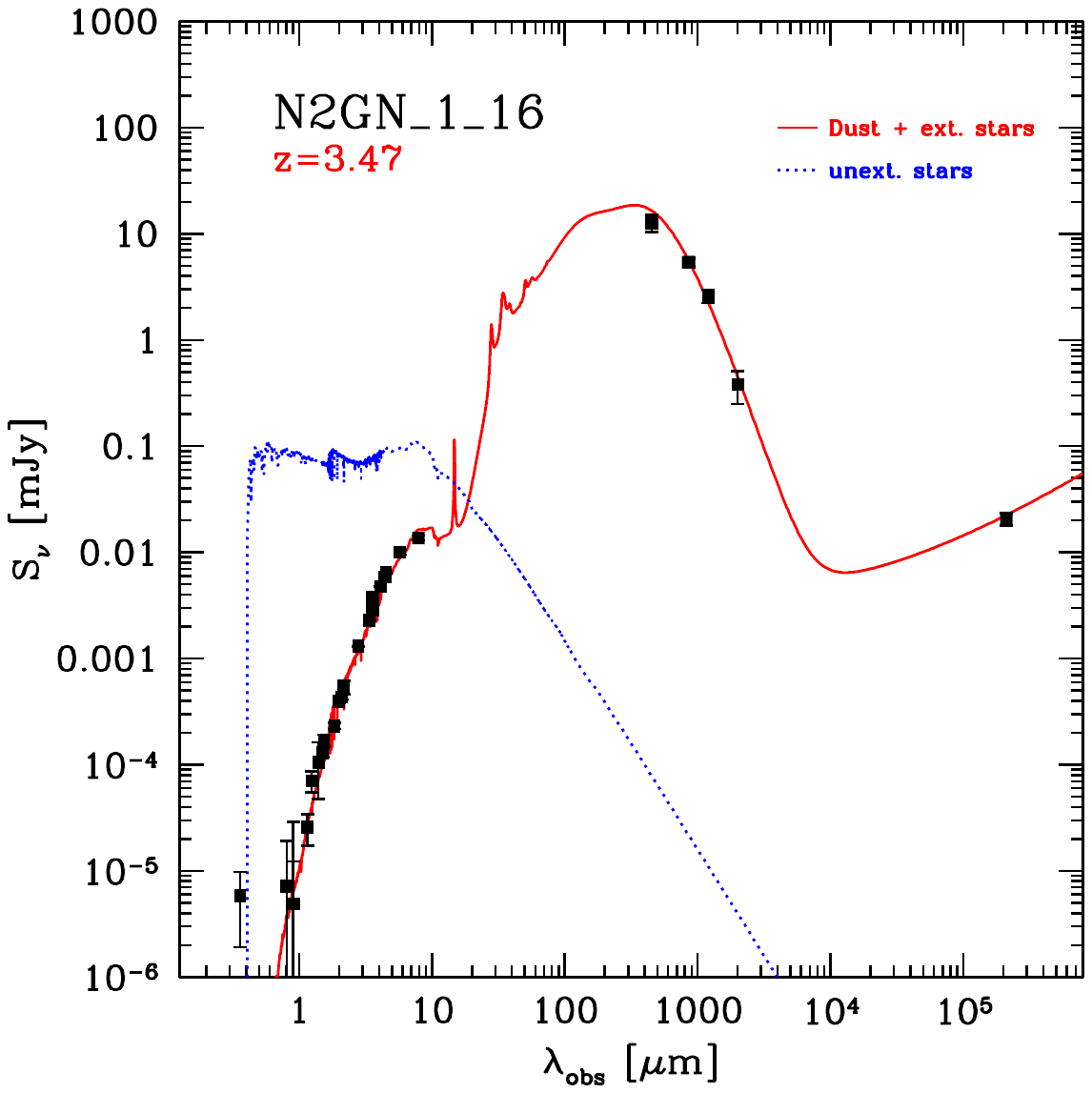}
\includegraphics[align=c,width=0.4\textwidth]{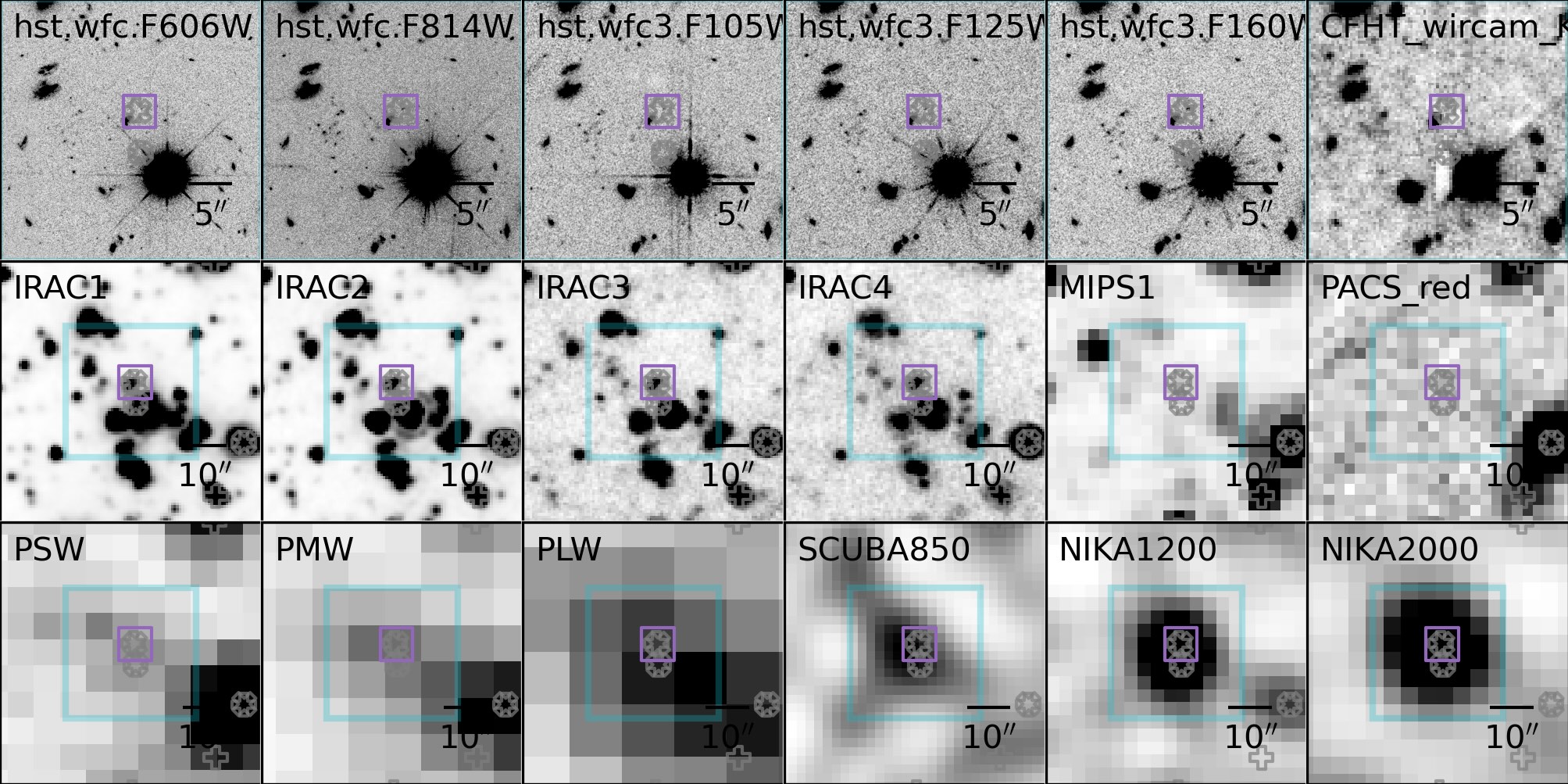}
\includegraphics[align=c,trim=0 0.18\imageheight{} 0 0.075\imageheight{}, clip, width=0.25\textwidth]{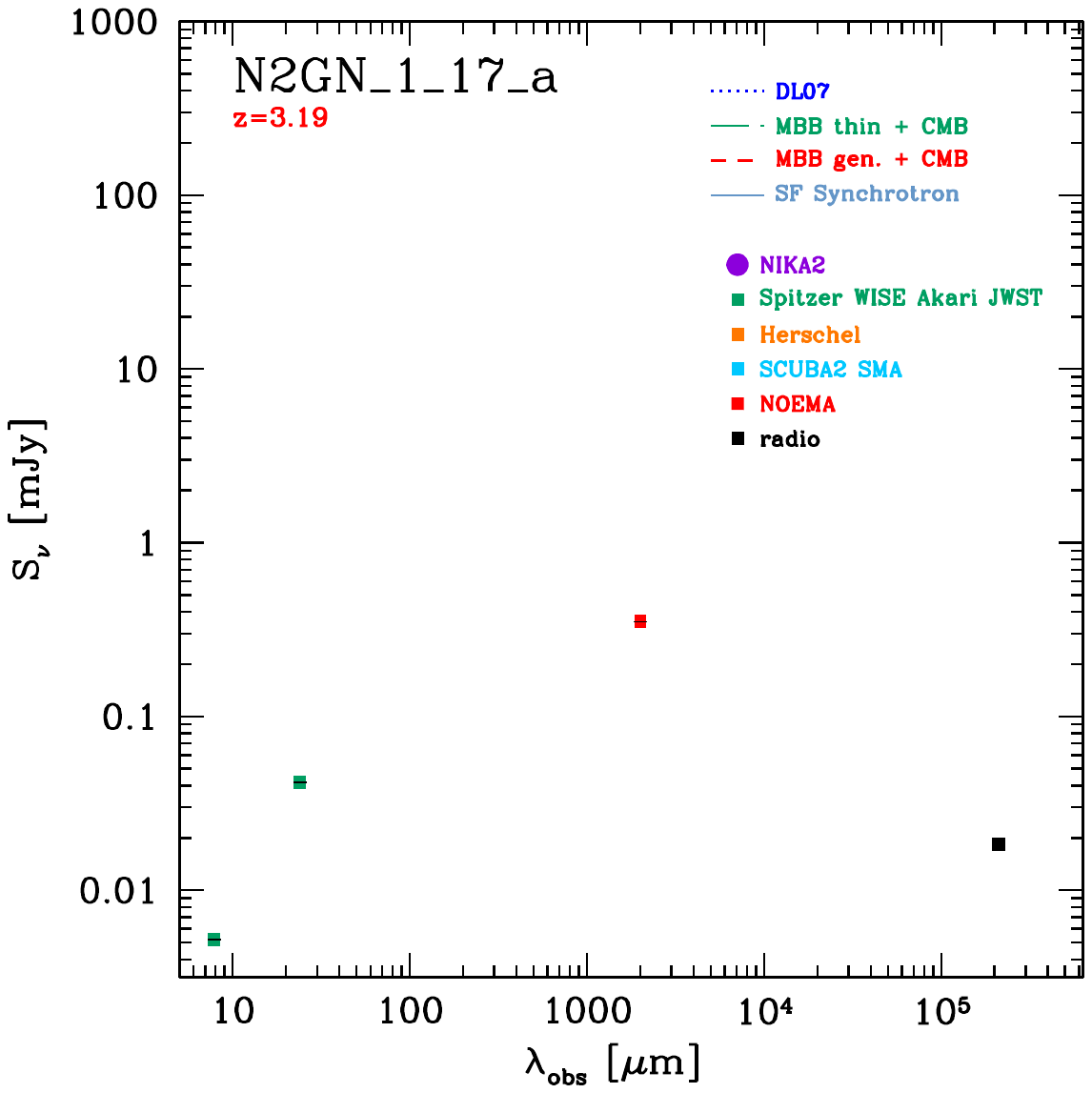}
\includegraphics[align=c,trim=0 0.18\imageheight{} 0 0.075\imageheight{}, clip, width=0.25\textwidth]{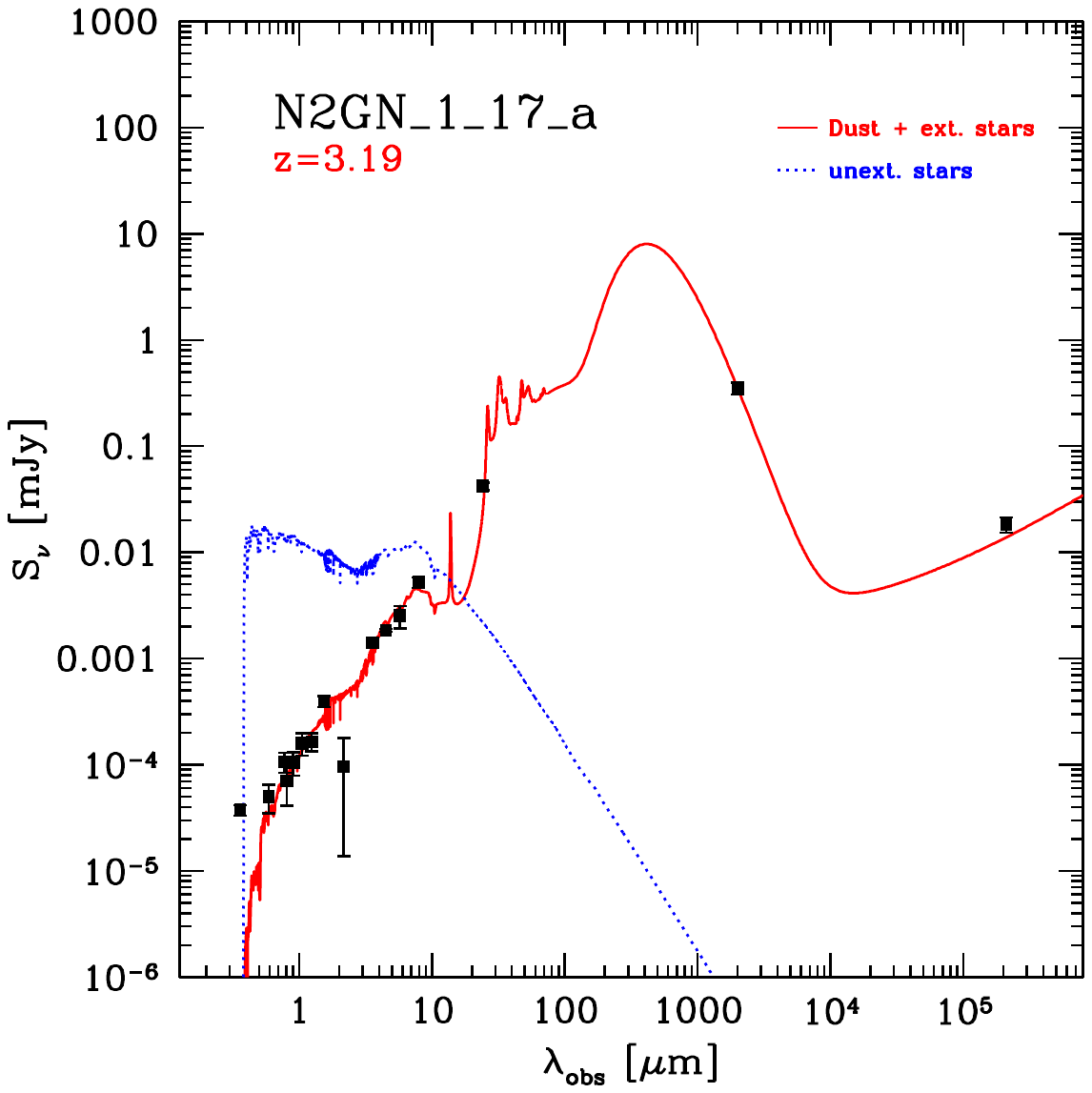}
\includegraphics[align=c,width=0.4\textwidth]{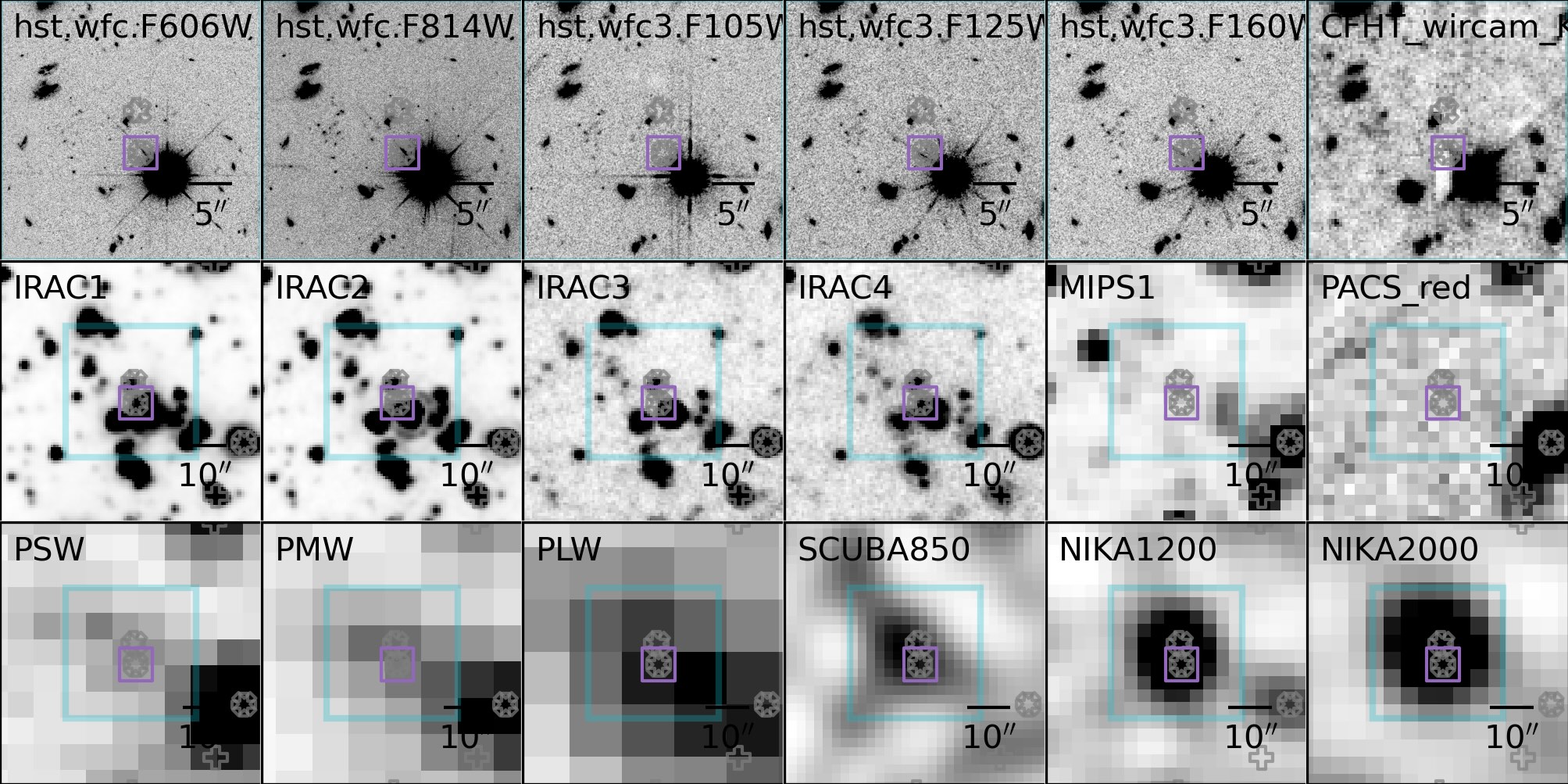}
\includegraphics[align=c,trim=0 0.18\imageheight{} 0 0.075\imageheight{}, clip, width=0.25\textwidth]{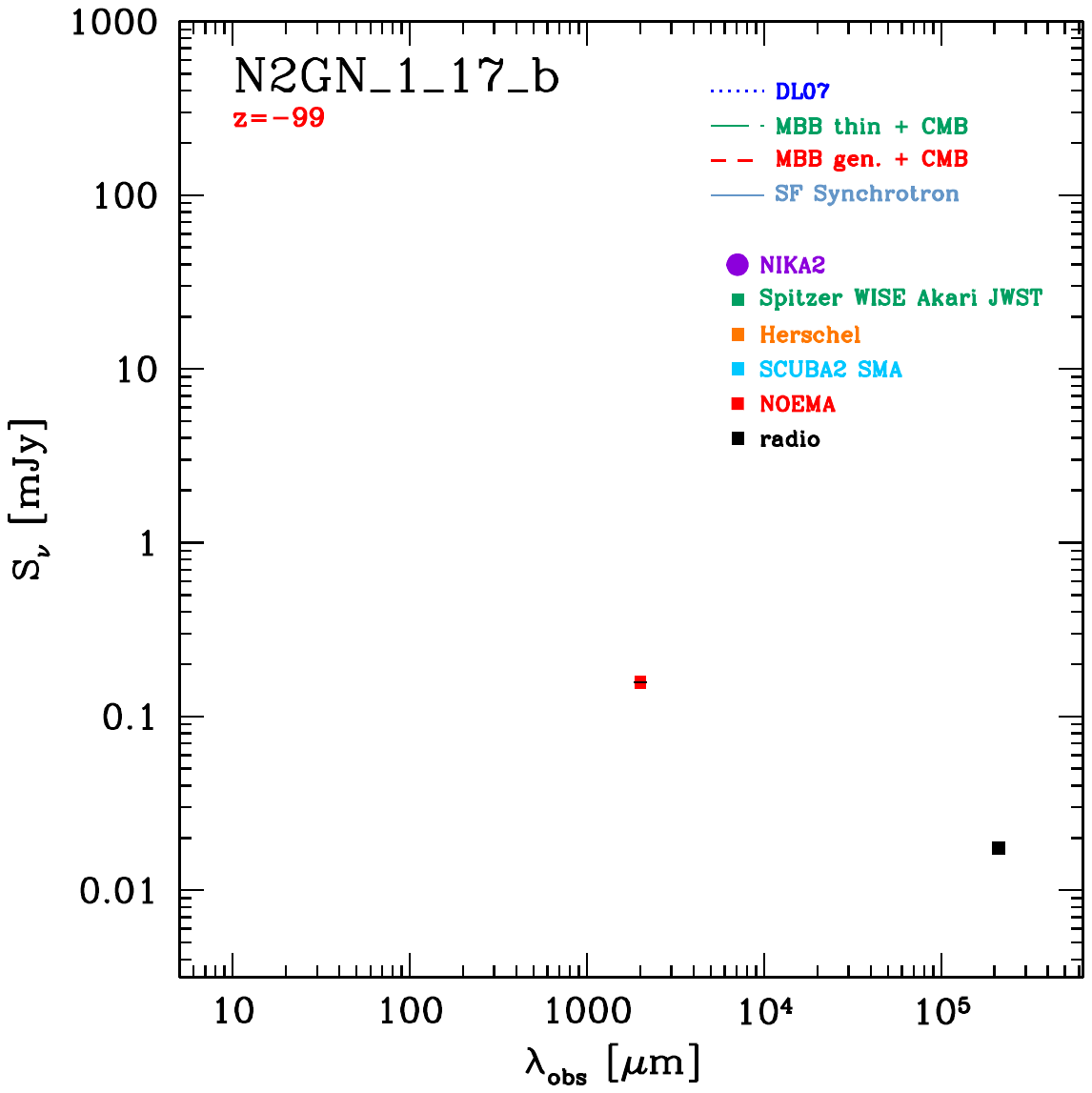}
\includegraphics[align=c,trim=0 0.18\imageheight{} 0 0.075\imageheight{}, clip, width=0.25\textwidth]{figs2_indiv_objs/white.pdf}
\caption{continued.}
\end{figure*}

\addtocounter{figure}{-1}
\newpage

\begin{figure*}[t]
\centering
\includegraphics[align=c,width=0.4\textwidth]{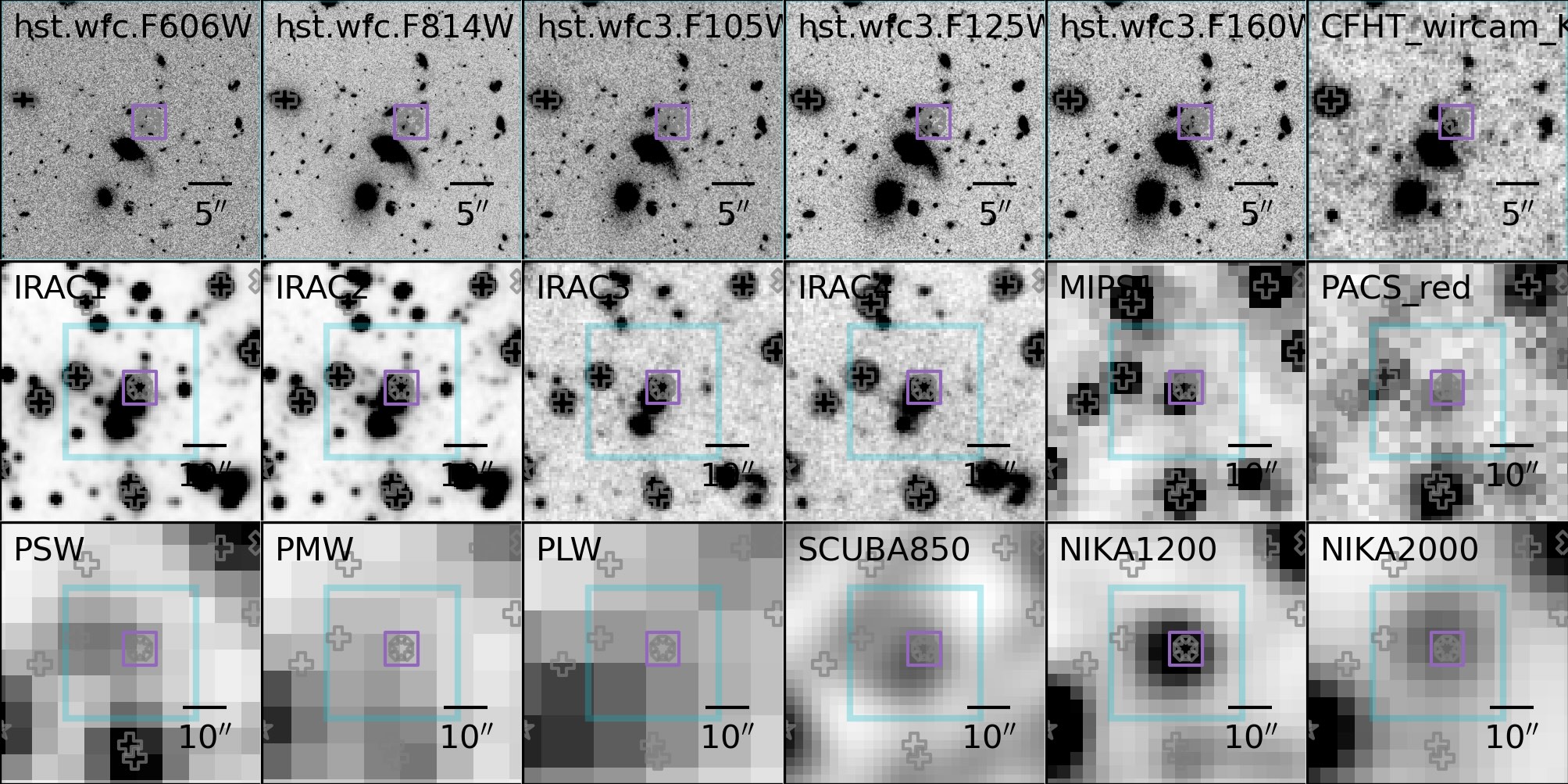}
\includegraphics[align=c,trim=0 0.18\imageheight{} 0 0.075\imageheight{}, clip, width=0.25\textwidth]{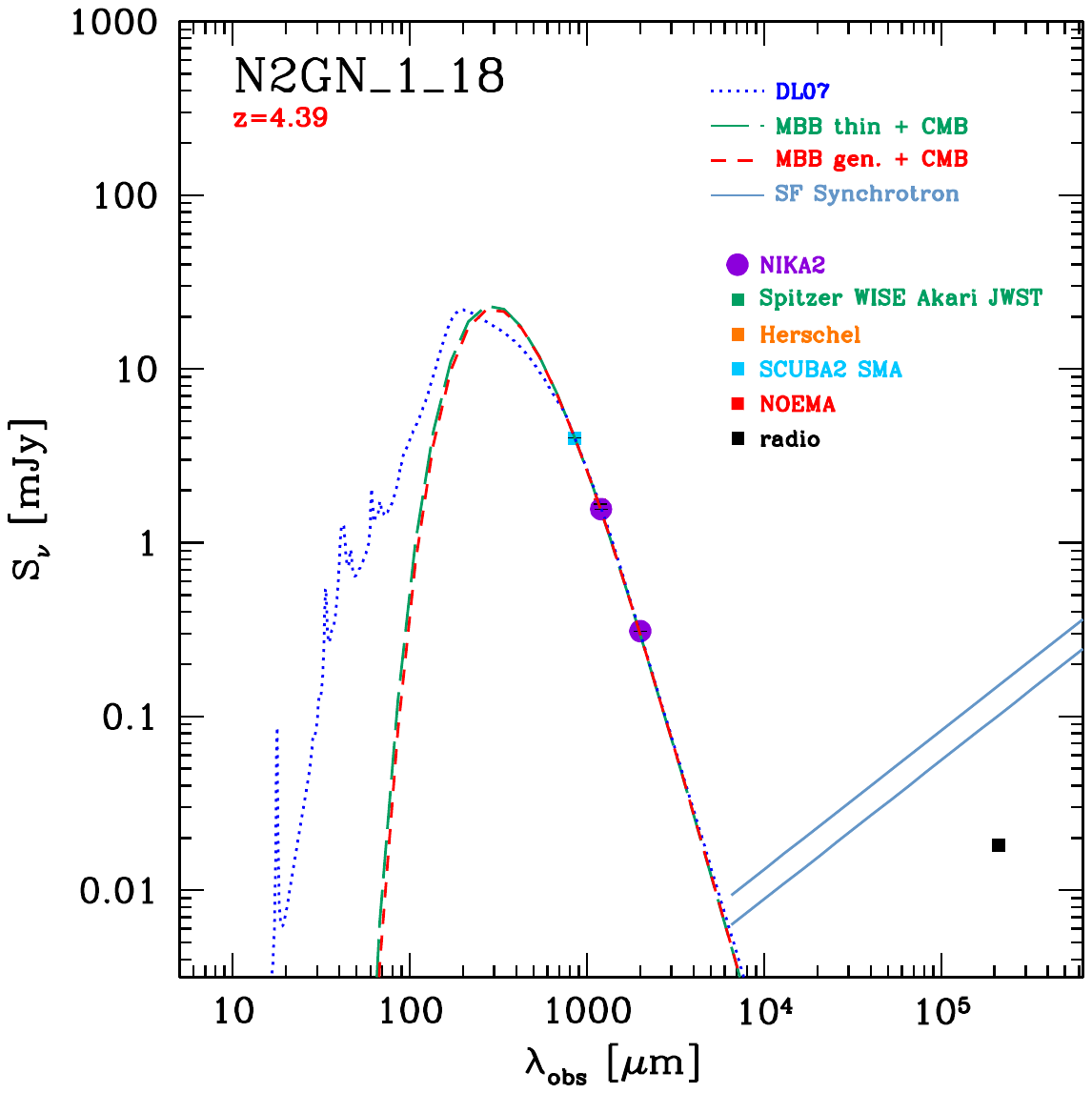}
\includegraphics[align=c,trim=0 0.18\imageheight{} 0 0.075\imageheight{}, clip, width=0.25\textwidth]{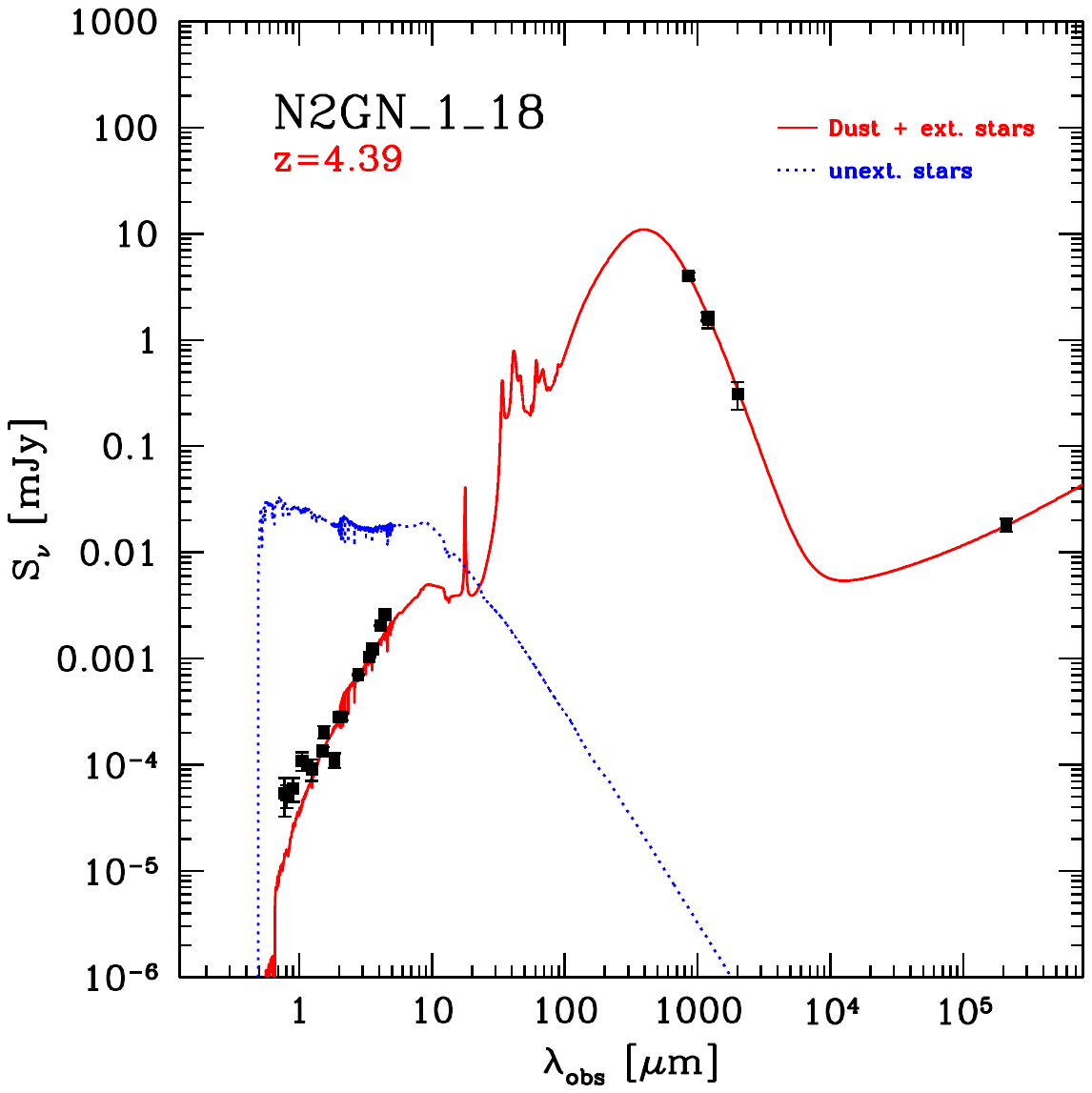}
\includegraphics[align=c,width=0.4\textwidth]{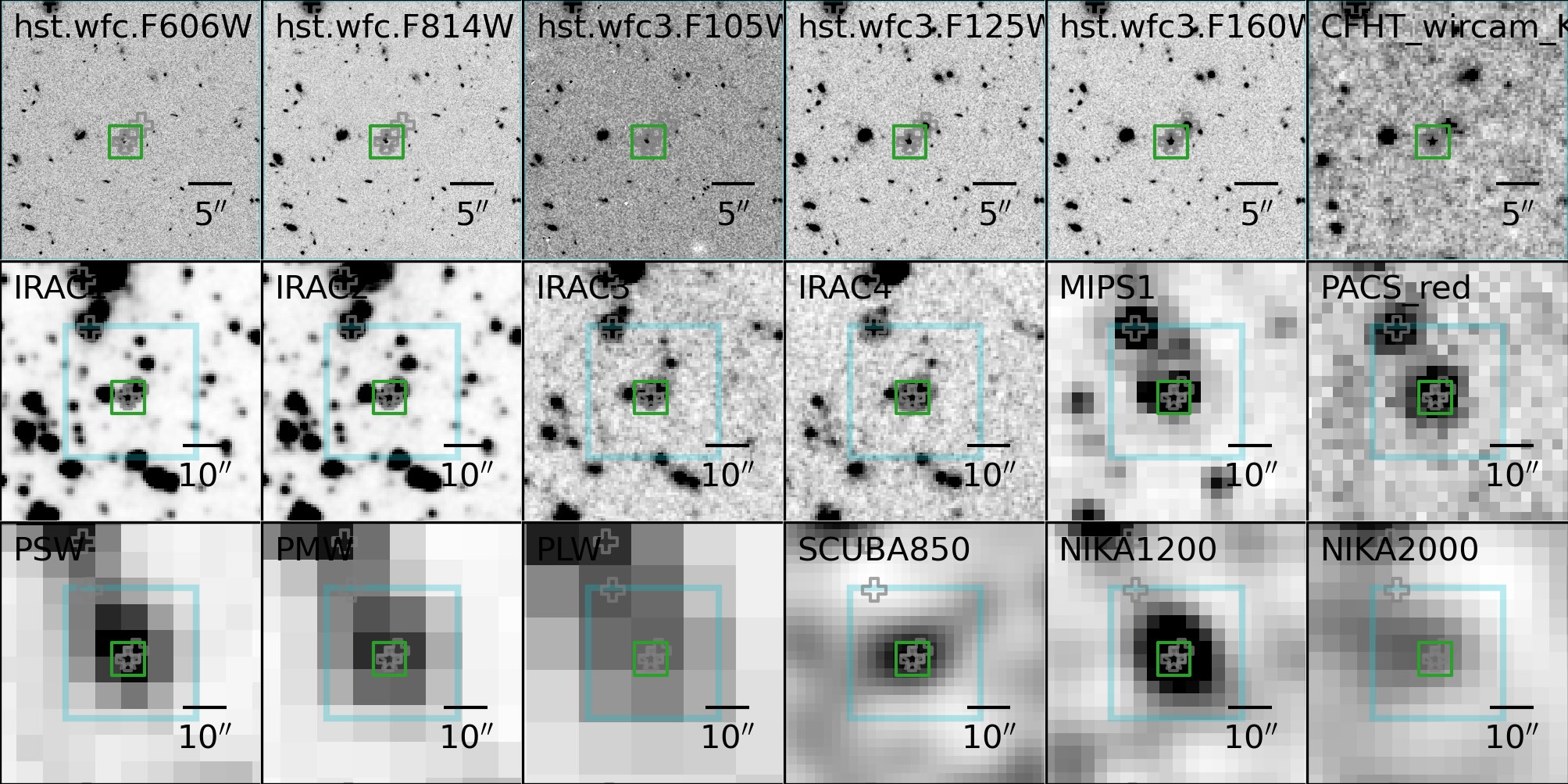}
\includegraphics[align=c,trim=0 0.18\imageheight{} 0 0.075\imageheight{}, clip, width=0.25\textwidth]{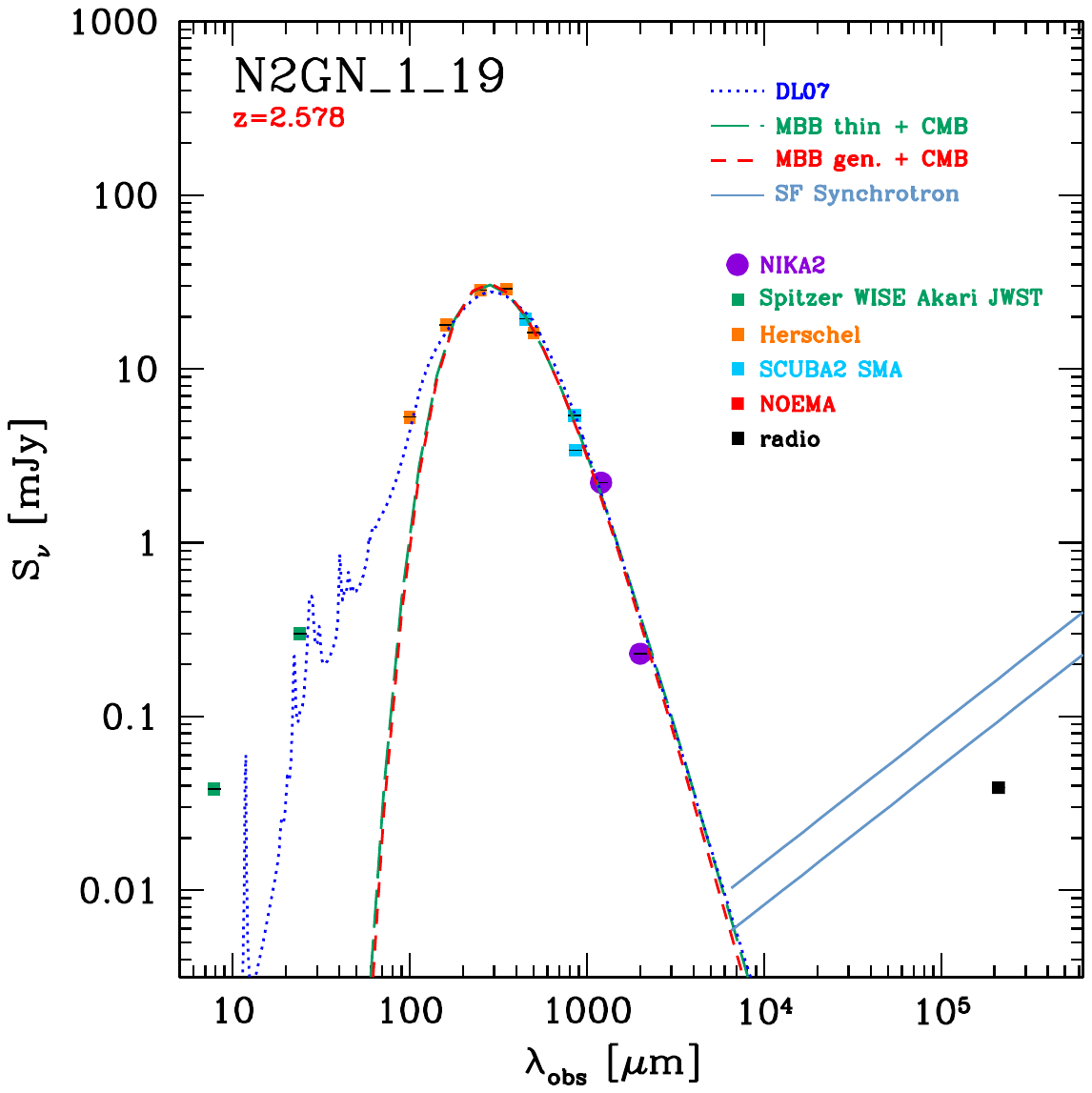}
\includegraphics[align=c,trim=0 0.18\imageheight{} 0 0.075\imageheight{}, clip, width=0.25\textwidth]{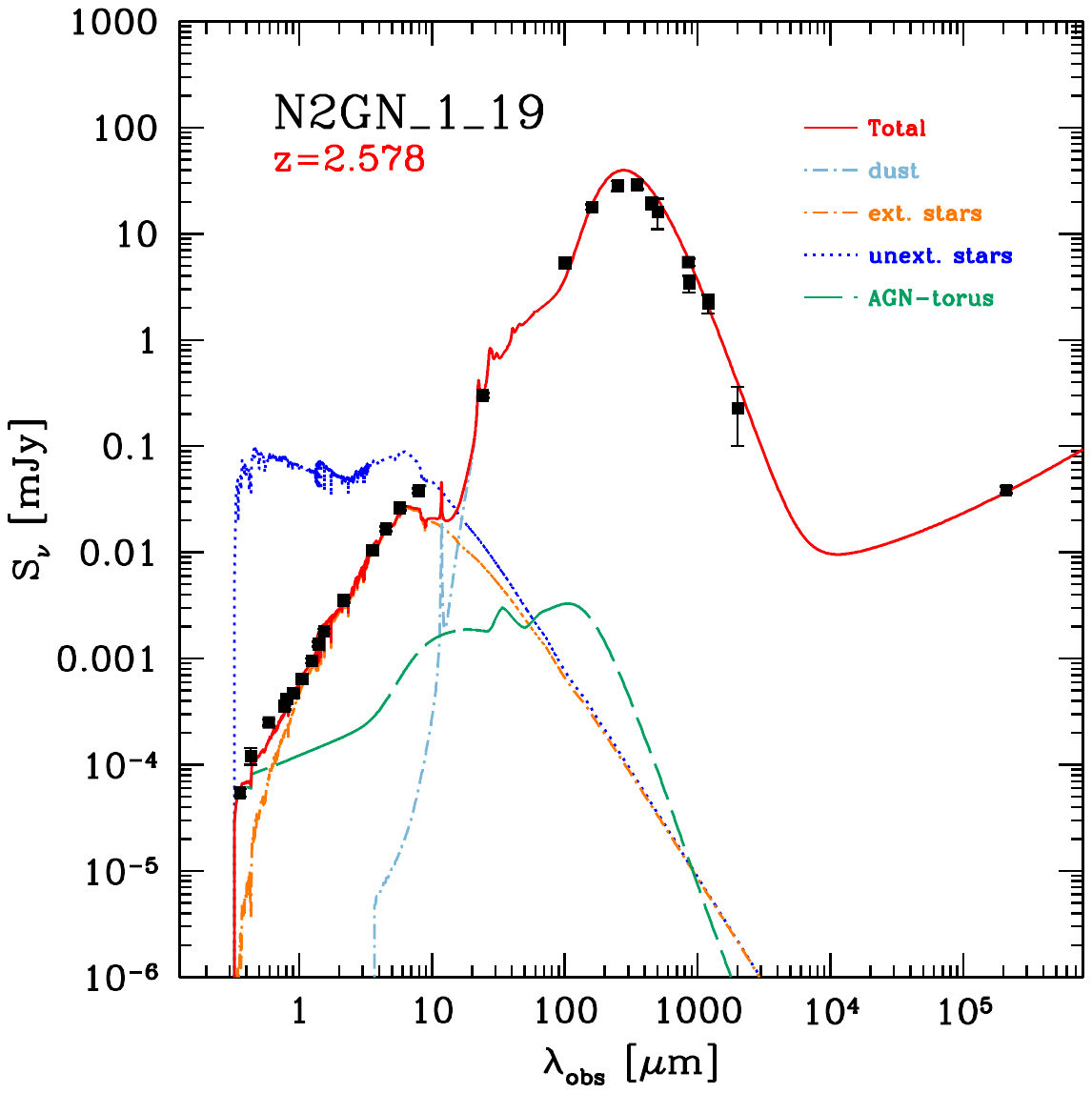}
\includegraphics[align=c,width=0.4\textwidth]{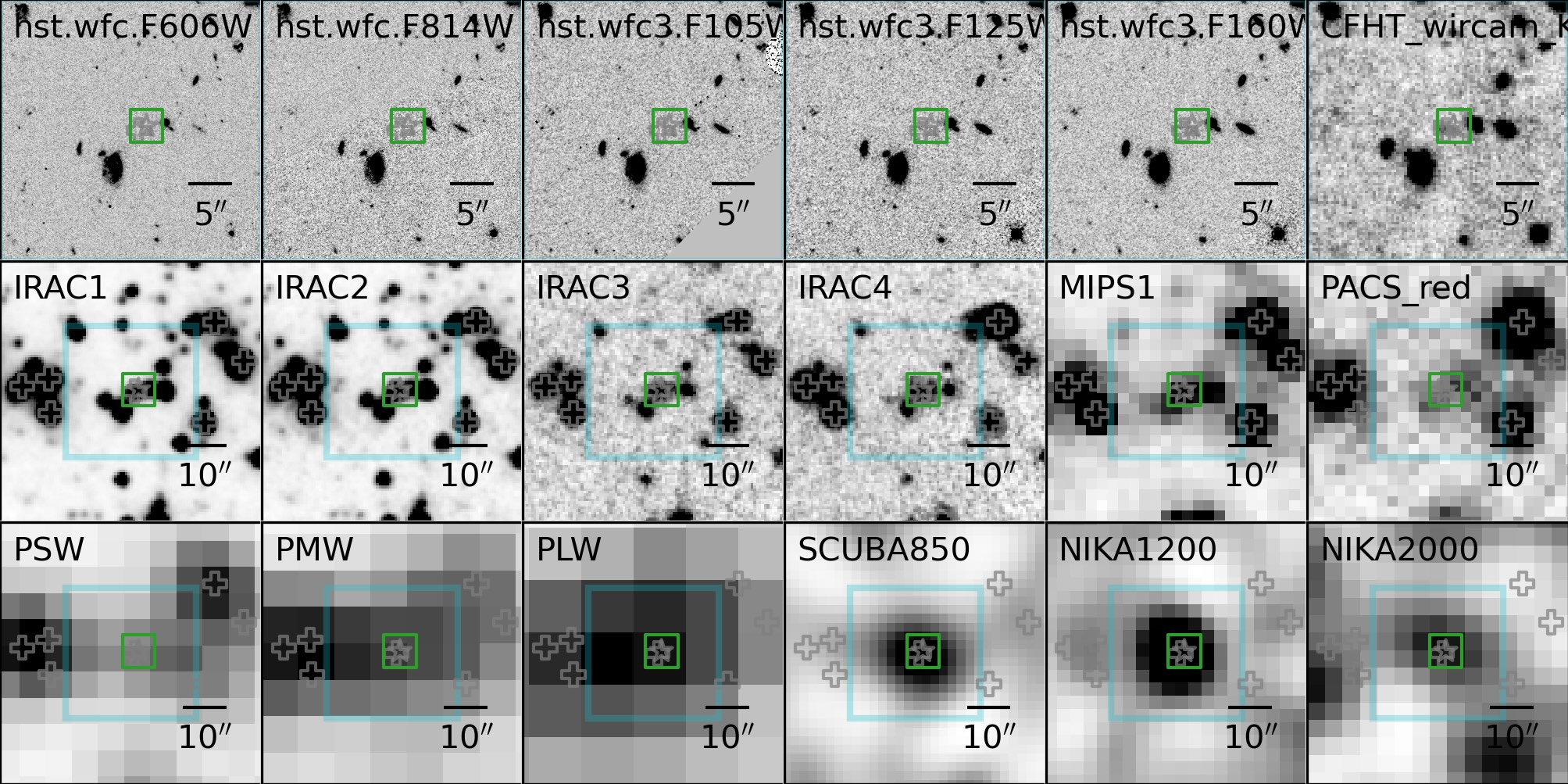}
\includegraphics[align=c,trim=0 0.18\imageheight{} 0 0.075\imageheight{}, clip, width=0.25\textwidth]{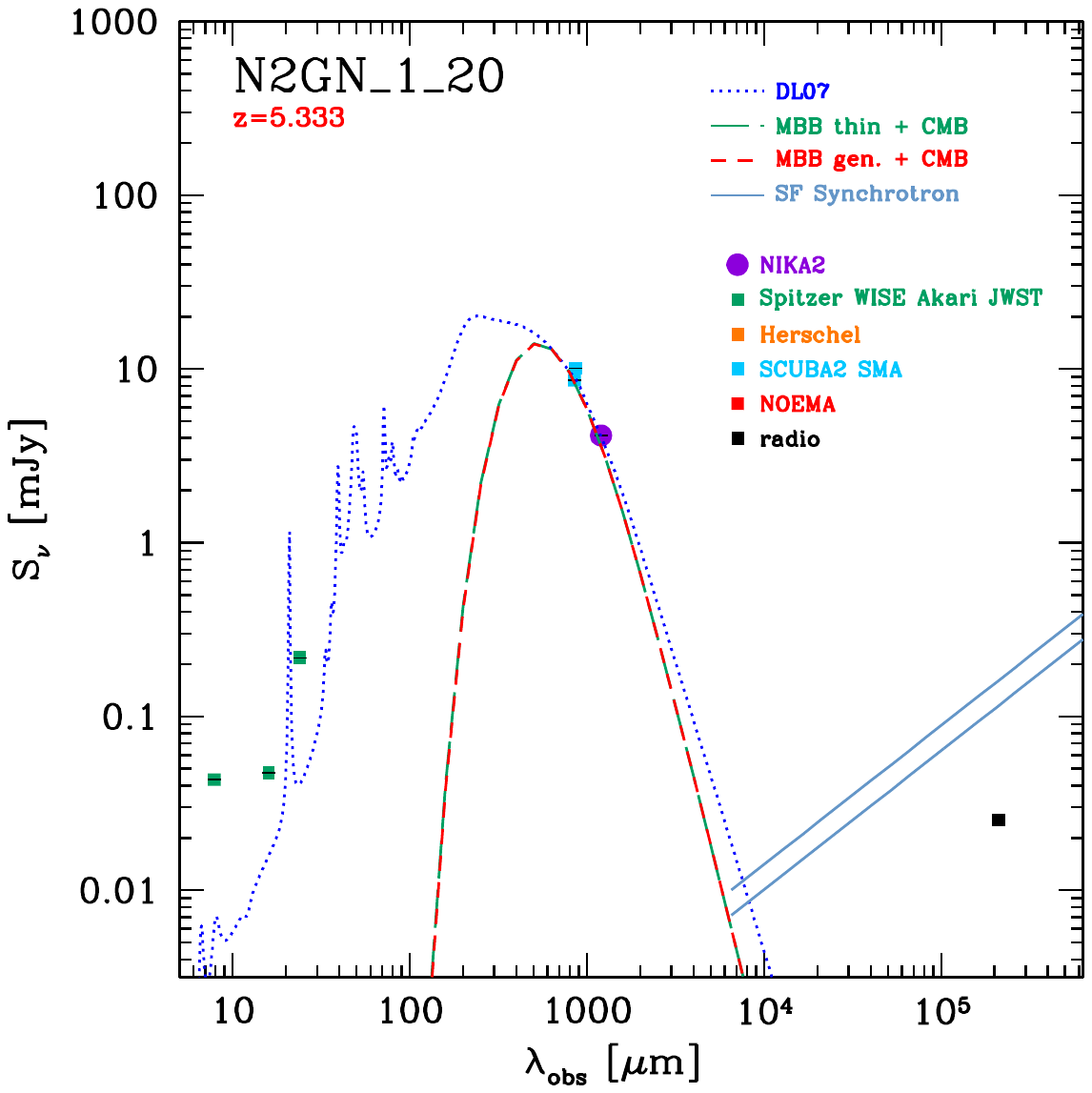}
\includegraphics[align=c,trim=0 0.18\imageheight{} 0 0.075\imageheight{}, clip, width=0.25\textwidth]{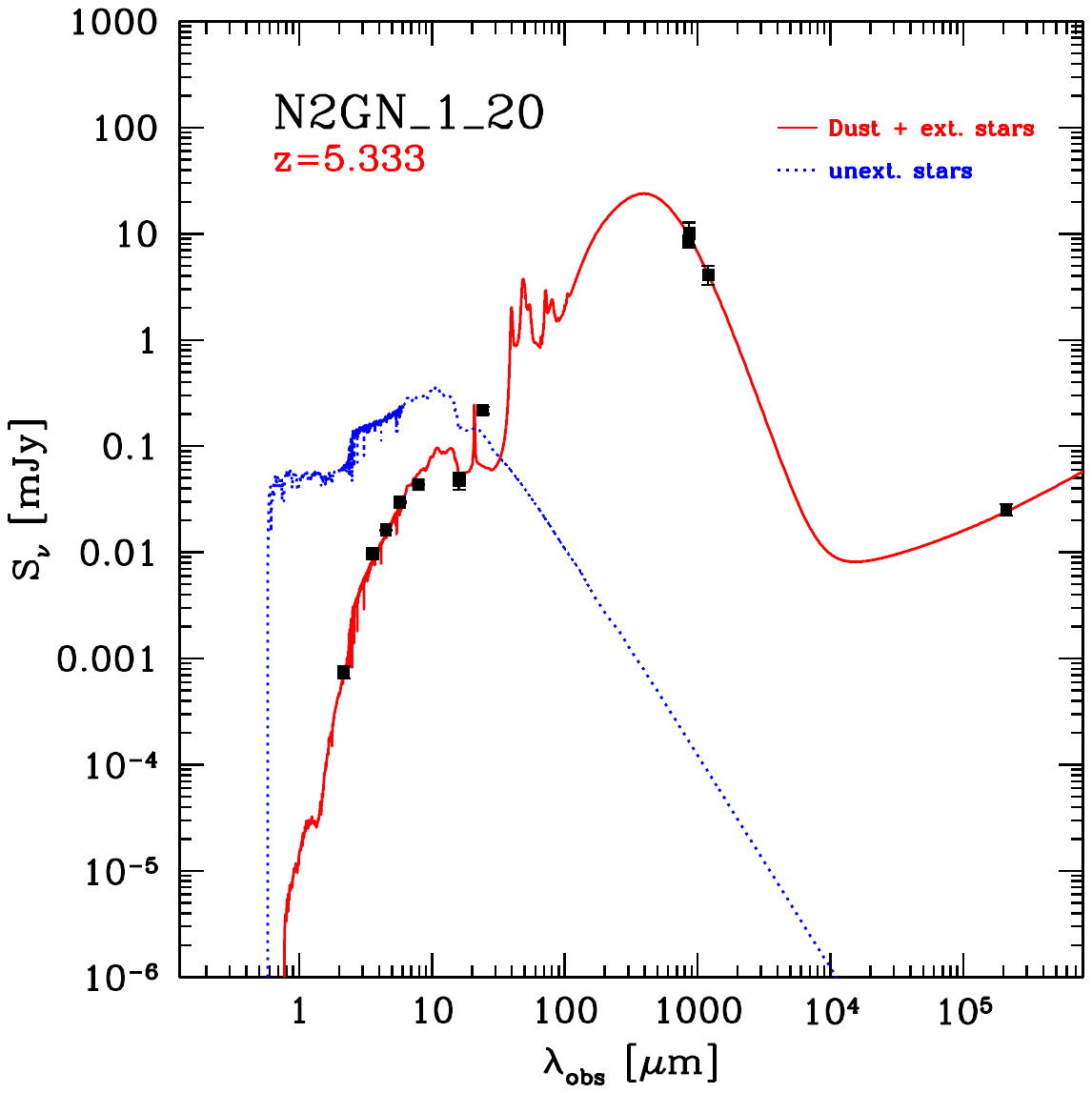}
\includegraphics[align=c,width=0.4\textwidth]{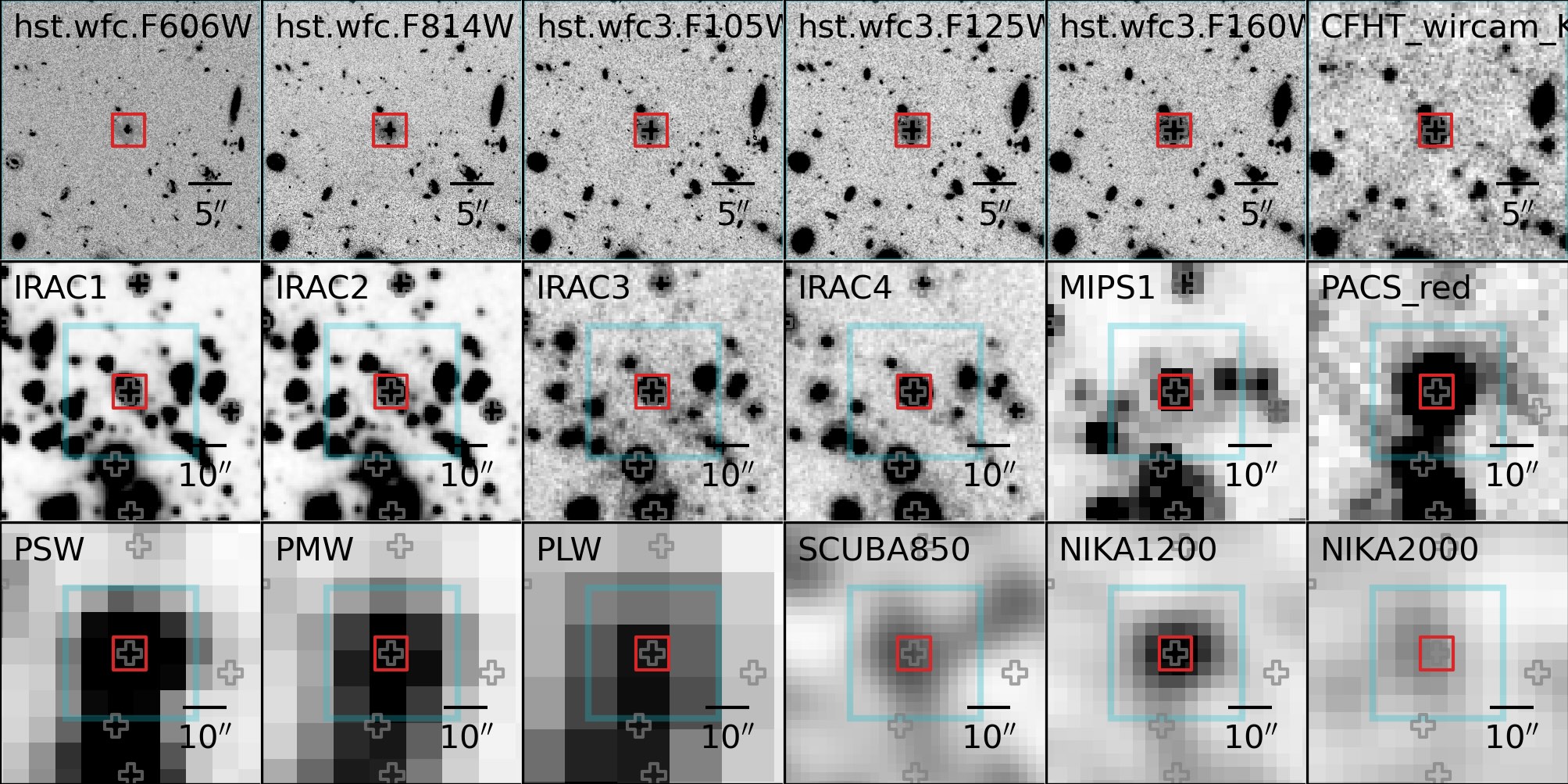}
\includegraphics[align=c,trim=0 0.18\imageheight{} 0 0.075\imageheight{}, clip, width=0.25\textwidth]{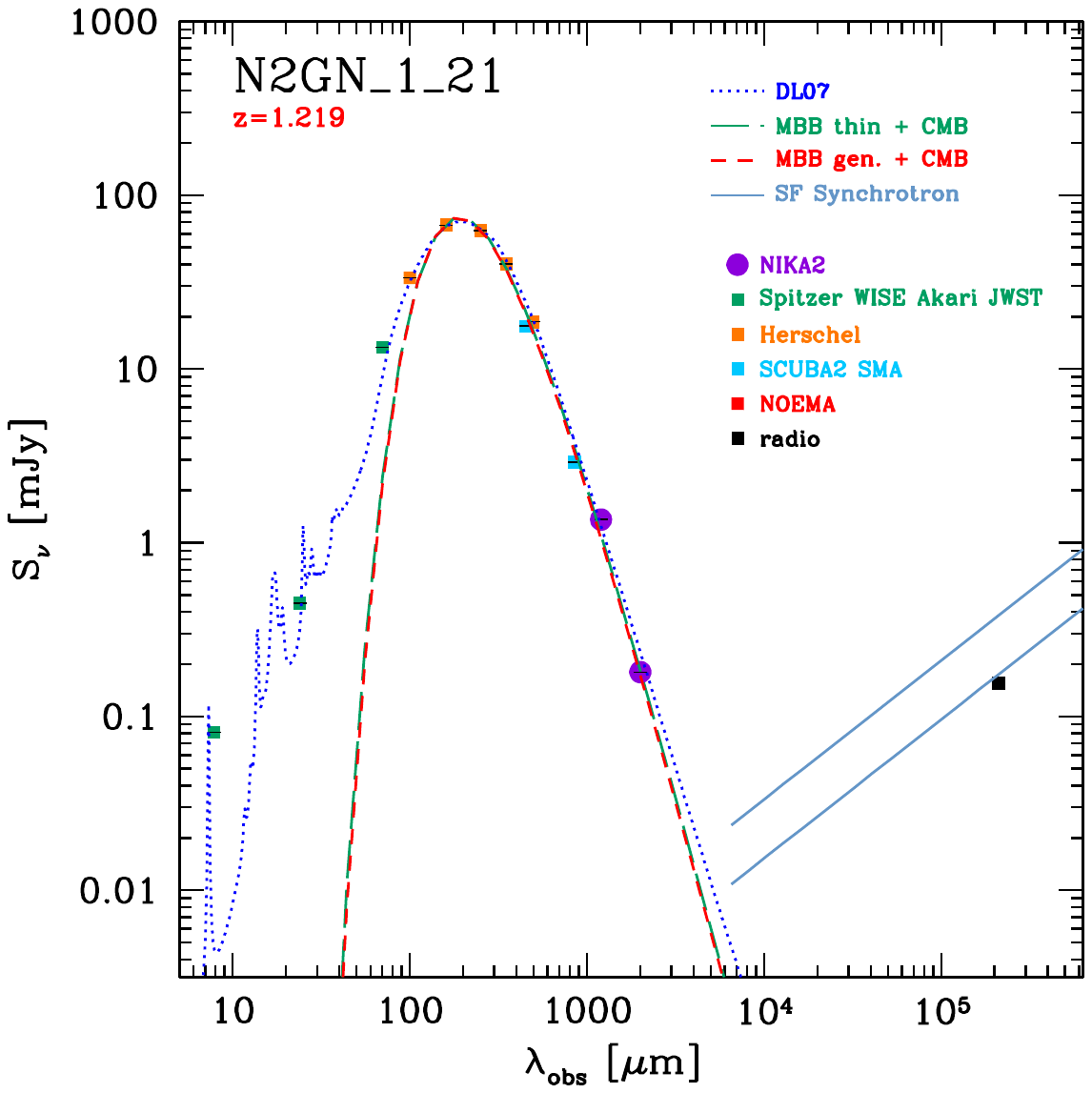}
\includegraphics[align=c,trim=0 0.18\imageheight{} 0 0.075\imageheight{}, clip, width=0.25\textwidth]{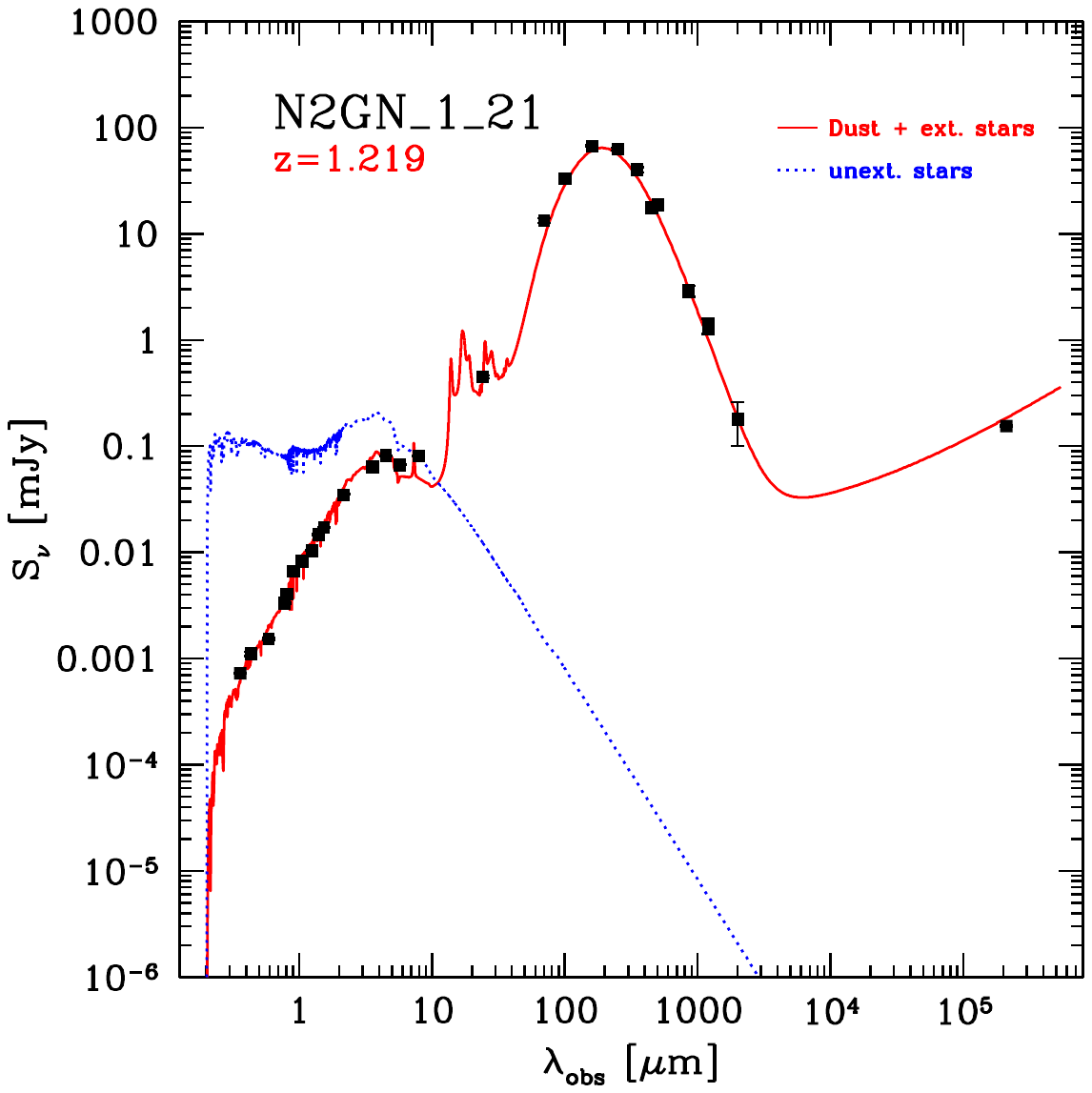}
\includegraphics[align=c,width=0.4\textwidth]{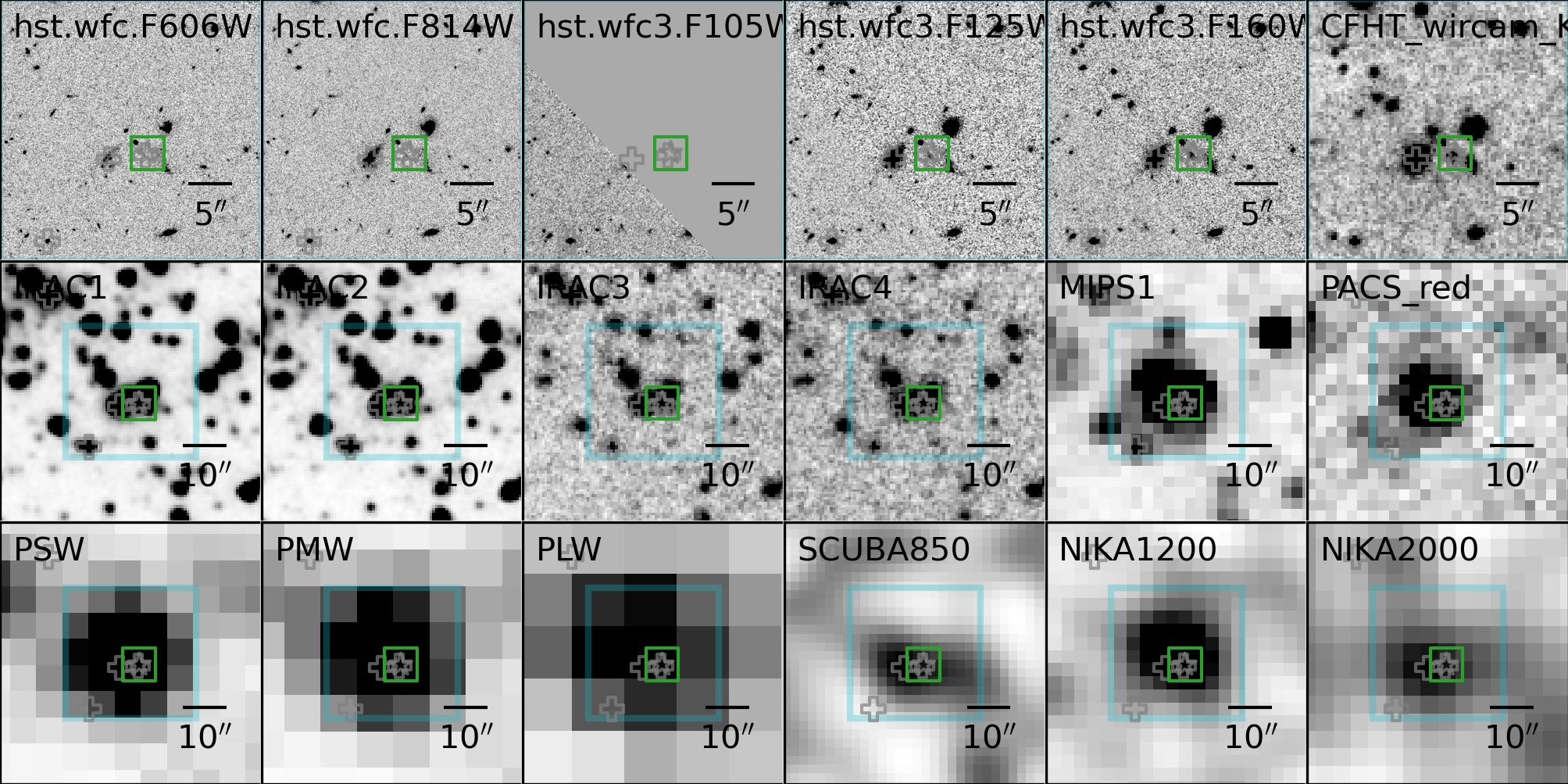}
\includegraphics[align=c,trim=0 0.18\imageheight{} 0 0.075\imageheight{}, clip, width=0.25\textwidth]{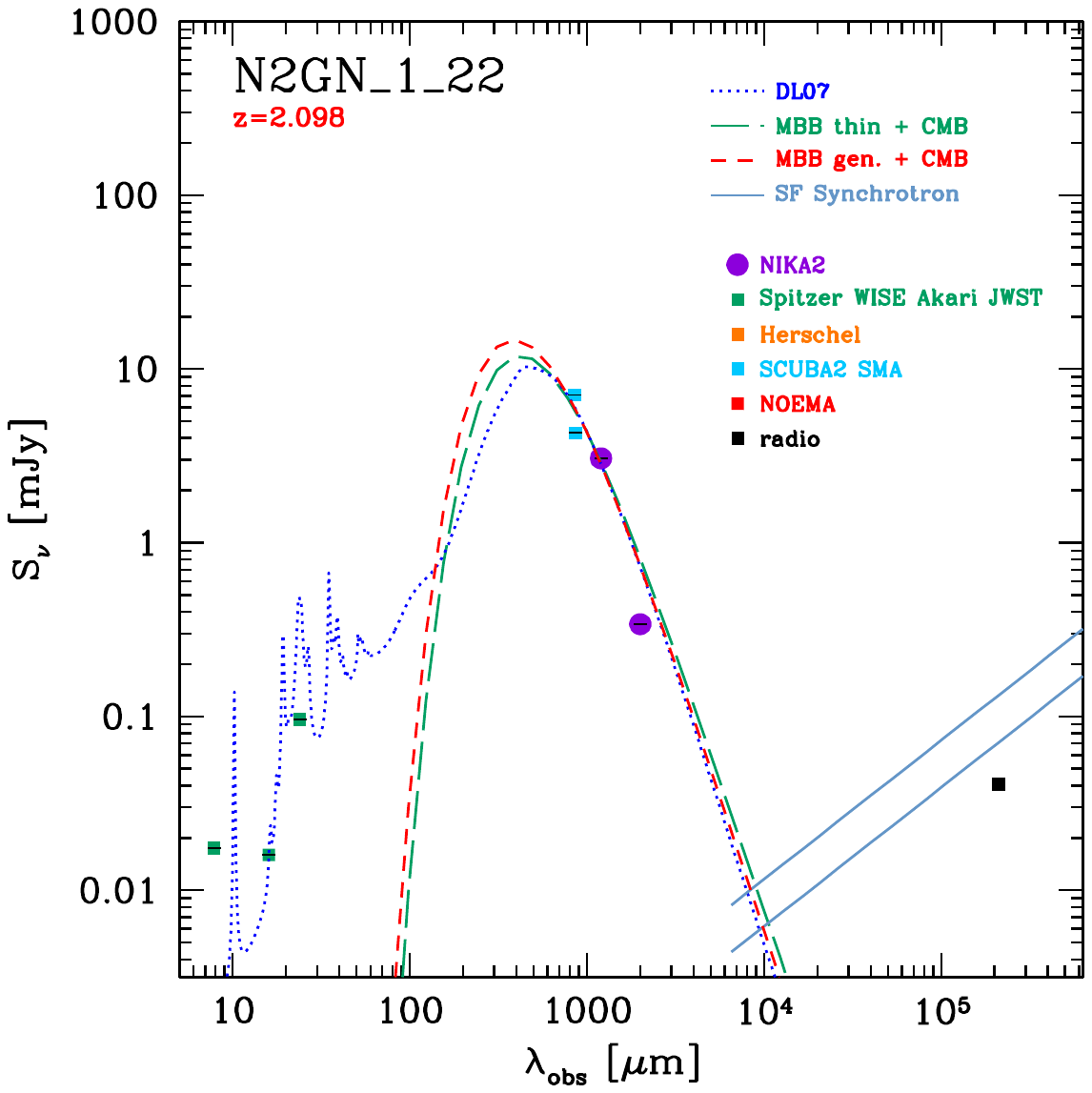}
\includegraphics[align=c,trim=0 0.18\imageheight{} 0 0.075\imageheight{}, clip, width=0.25\textwidth]{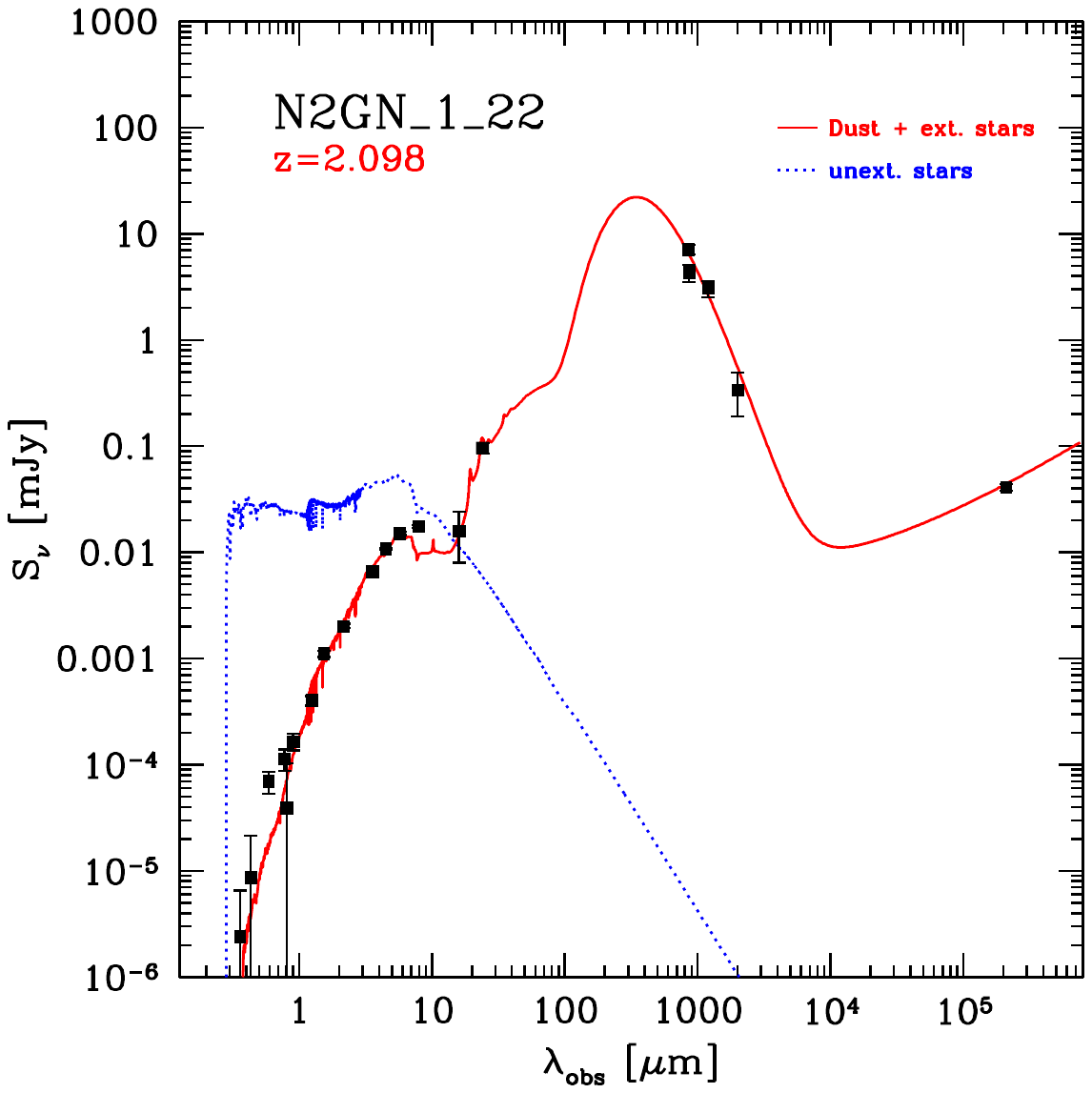}
\caption{continued.}
\end{figure*}

\addtocounter{figure}{-1}
\newpage

\begin{figure*}[t]
\centering
\includegraphics[align=c,width=0.4\textwidth]{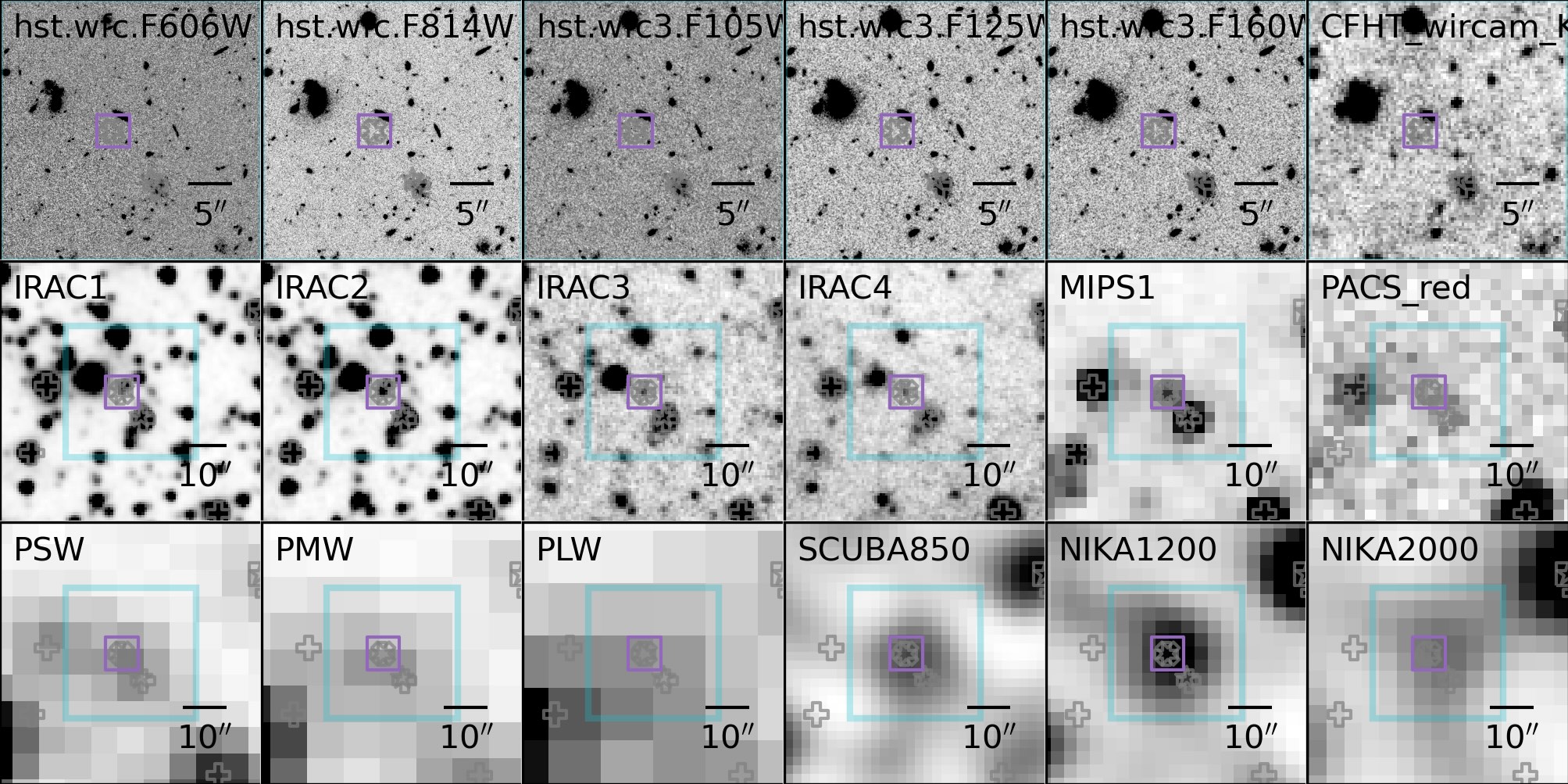}
\includegraphics[align=c,trim=0 0.18\imageheight{} 0 0.075\imageheight{}, clip, width=0.25\textwidth]{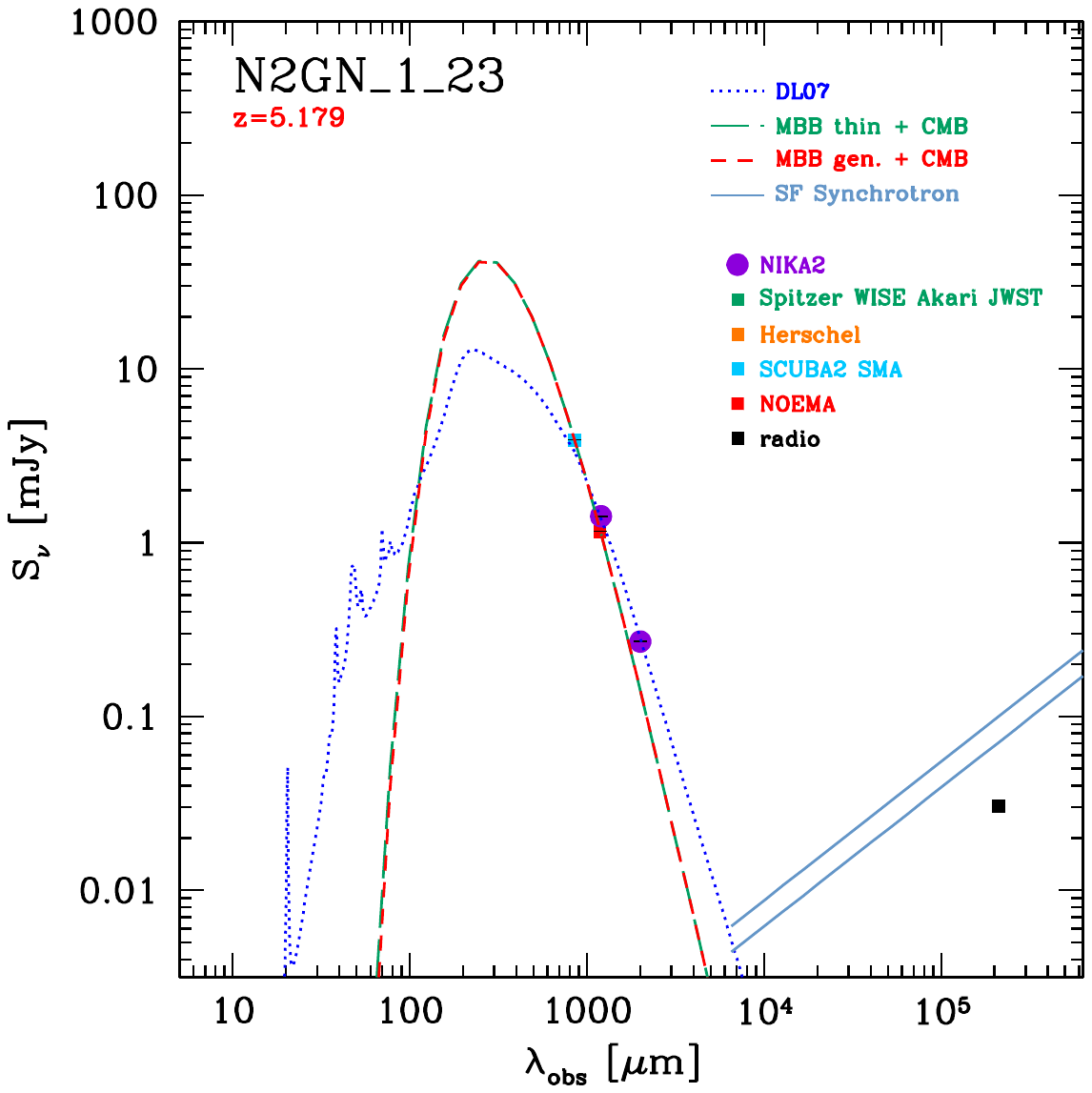}
\includegraphics[align=c,trim=0 0.18\imageheight{} 0 0.075\imageheight{}, clip, width=0.25\textwidth]{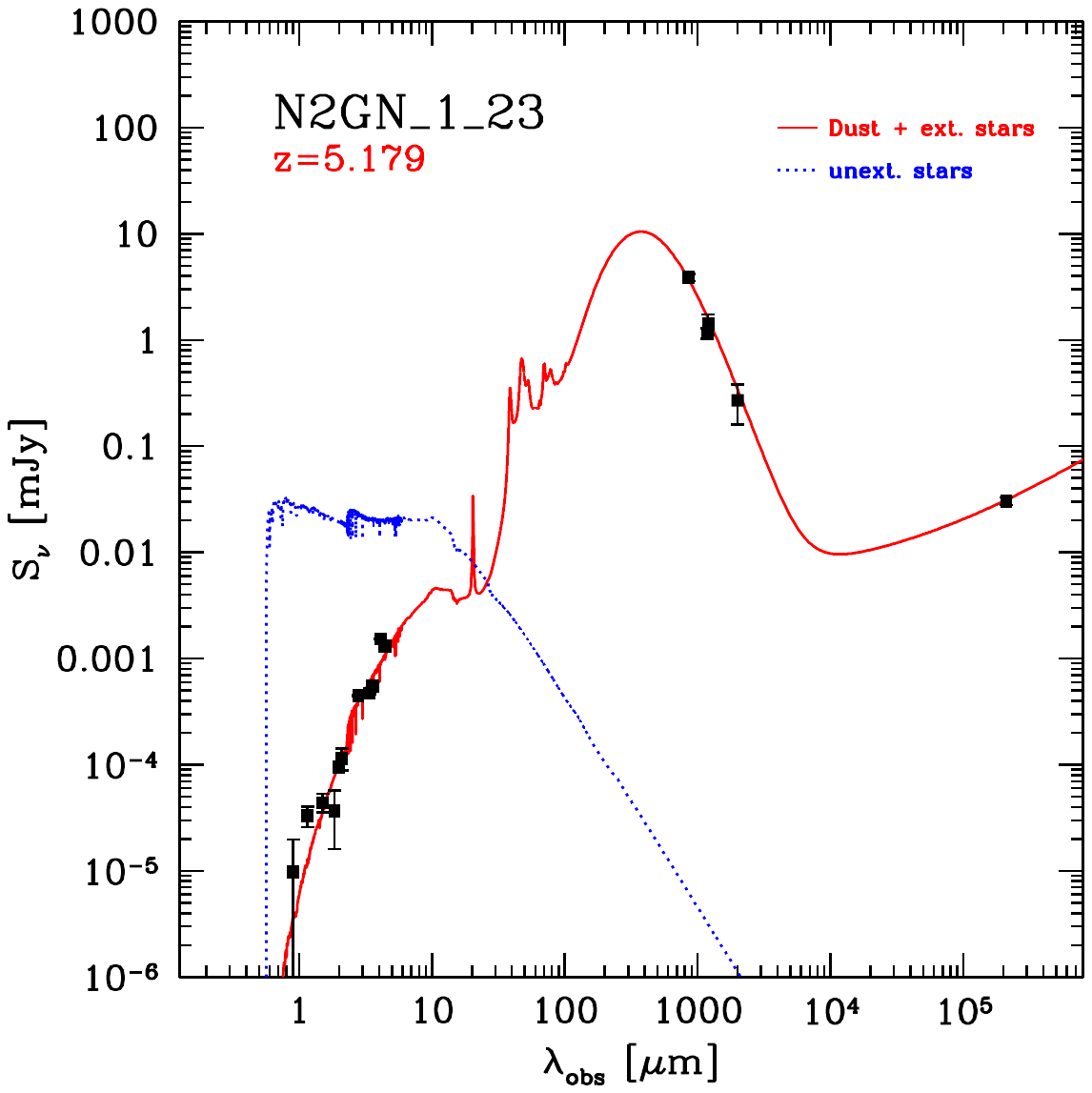}
\includegraphics[align=c,width=0.4\textwidth]{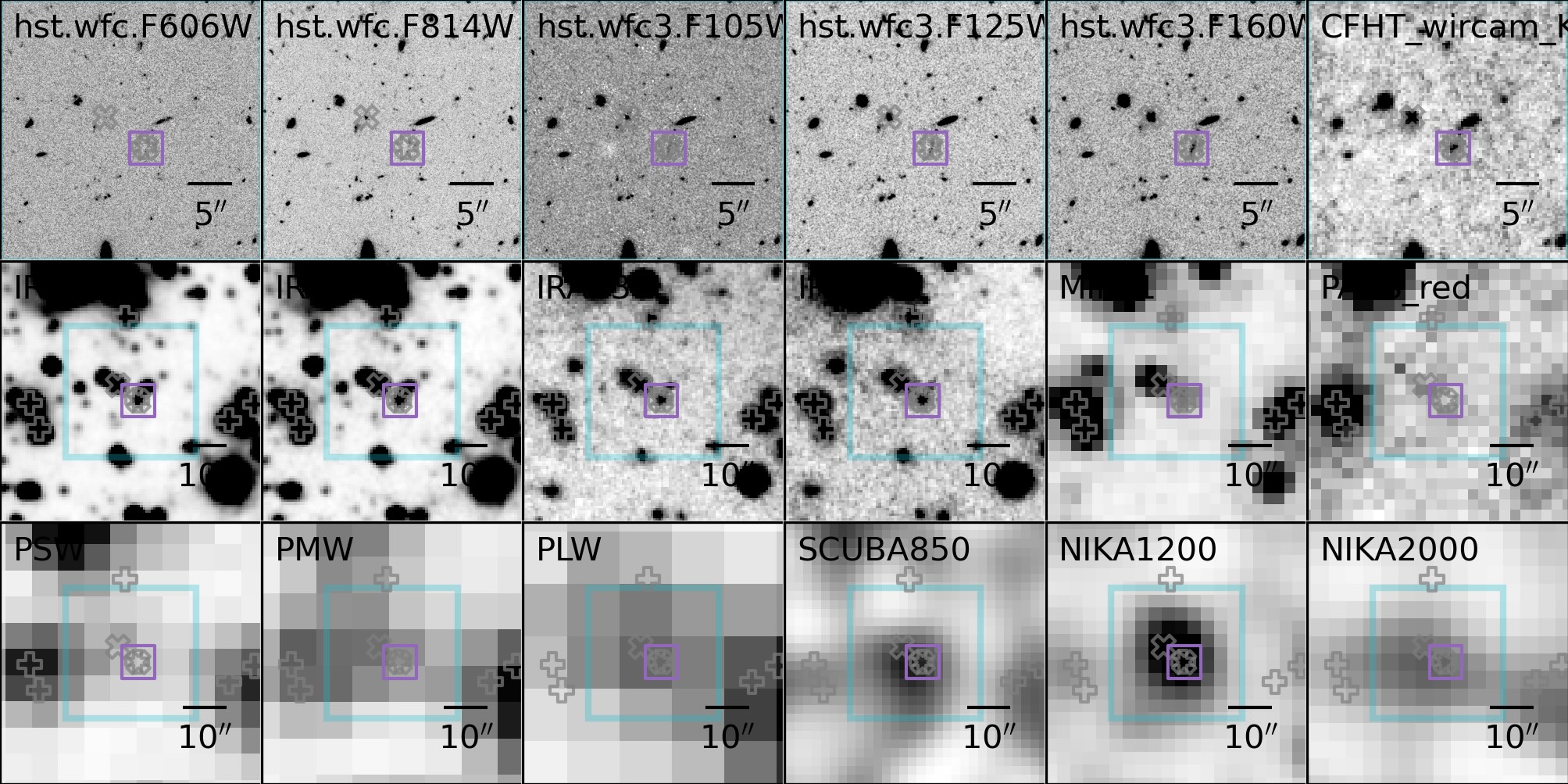} %\\
\includegraphics[align=c,trim=0 0.18\imageheight{} 0 0.075\imageheight{}, clip, width=0.25\textwidth]{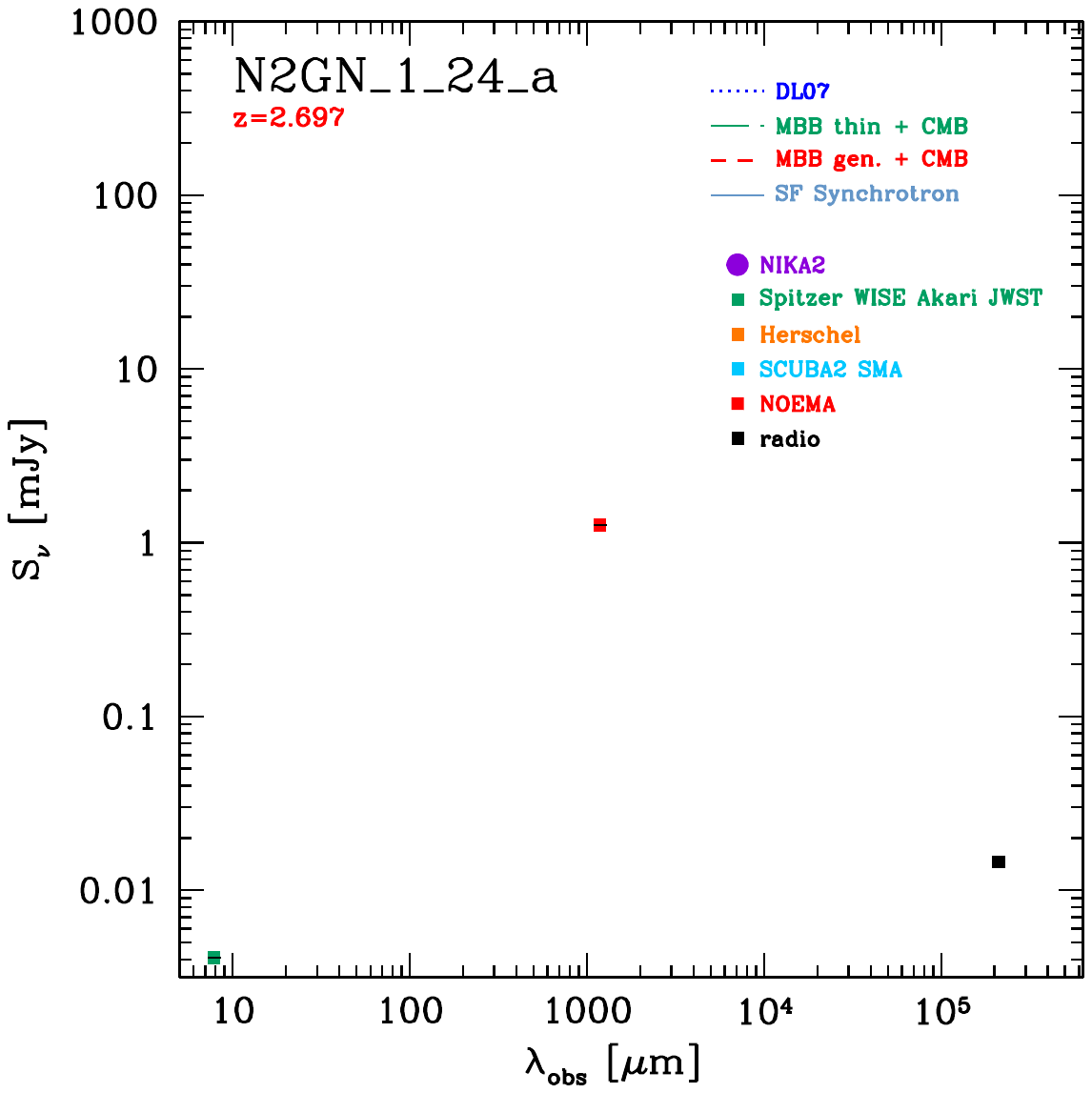}
\includegraphics[align=c,trim=0 0.18\imageheight{} 0 0.075\imageheight{}, clip, width=0.25\textwidth]{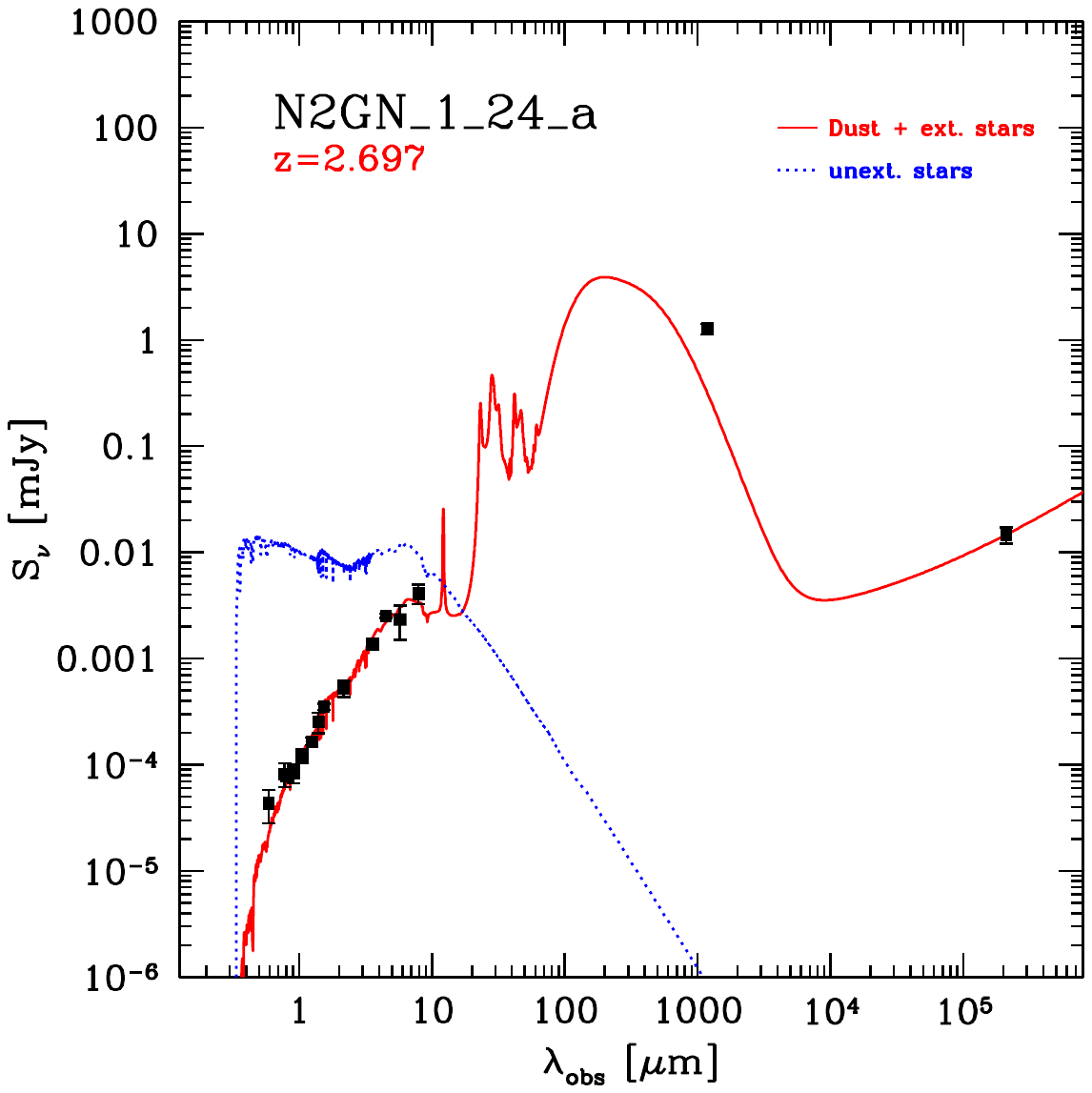}
\includegraphics[align=c,width=0.4\textwidth]{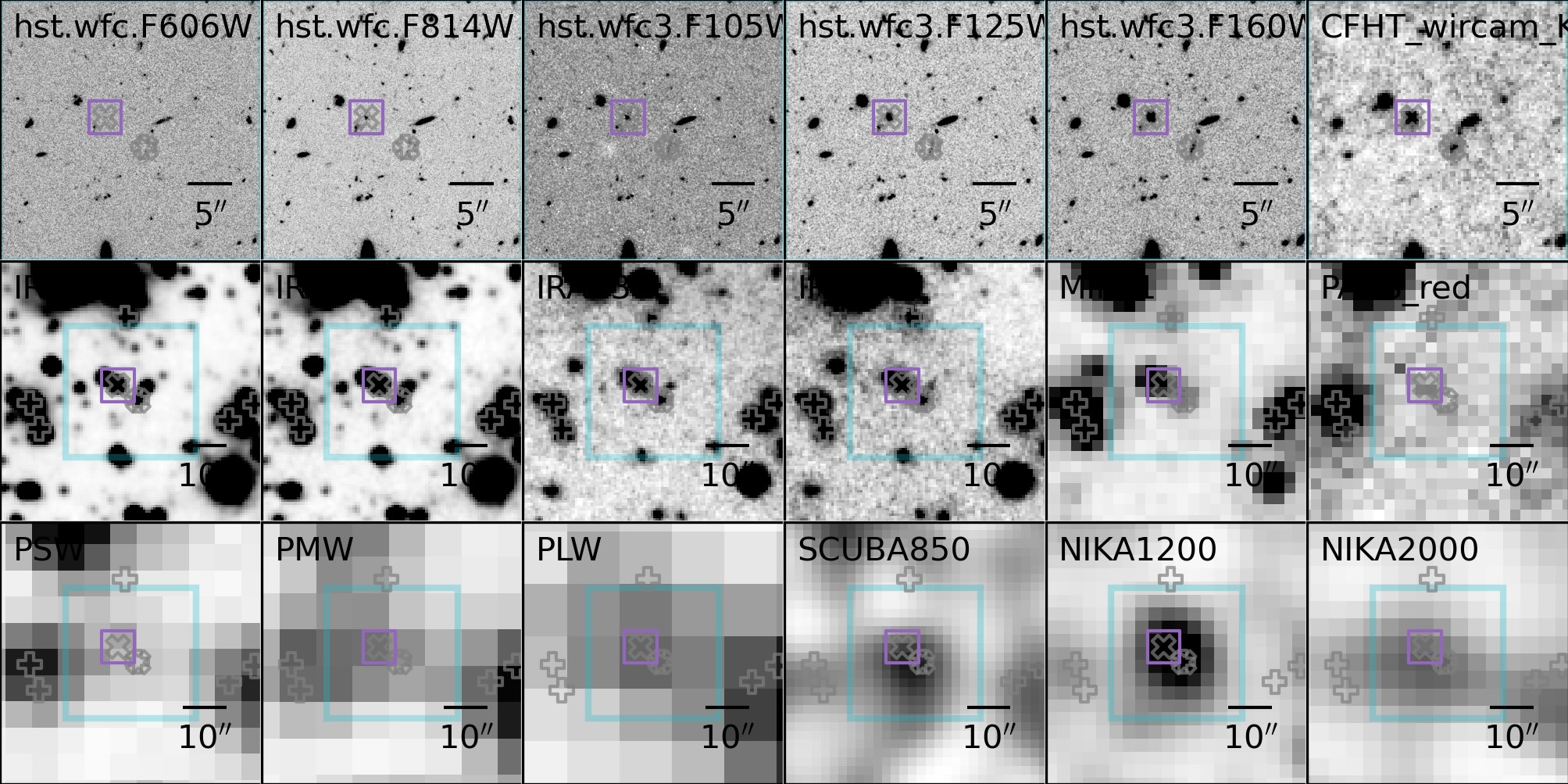}
\includegraphics[align=c,trim=0 0.18\imageheight{} 0 0.075\imageheight{}, clip, width=0.25\textwidth]{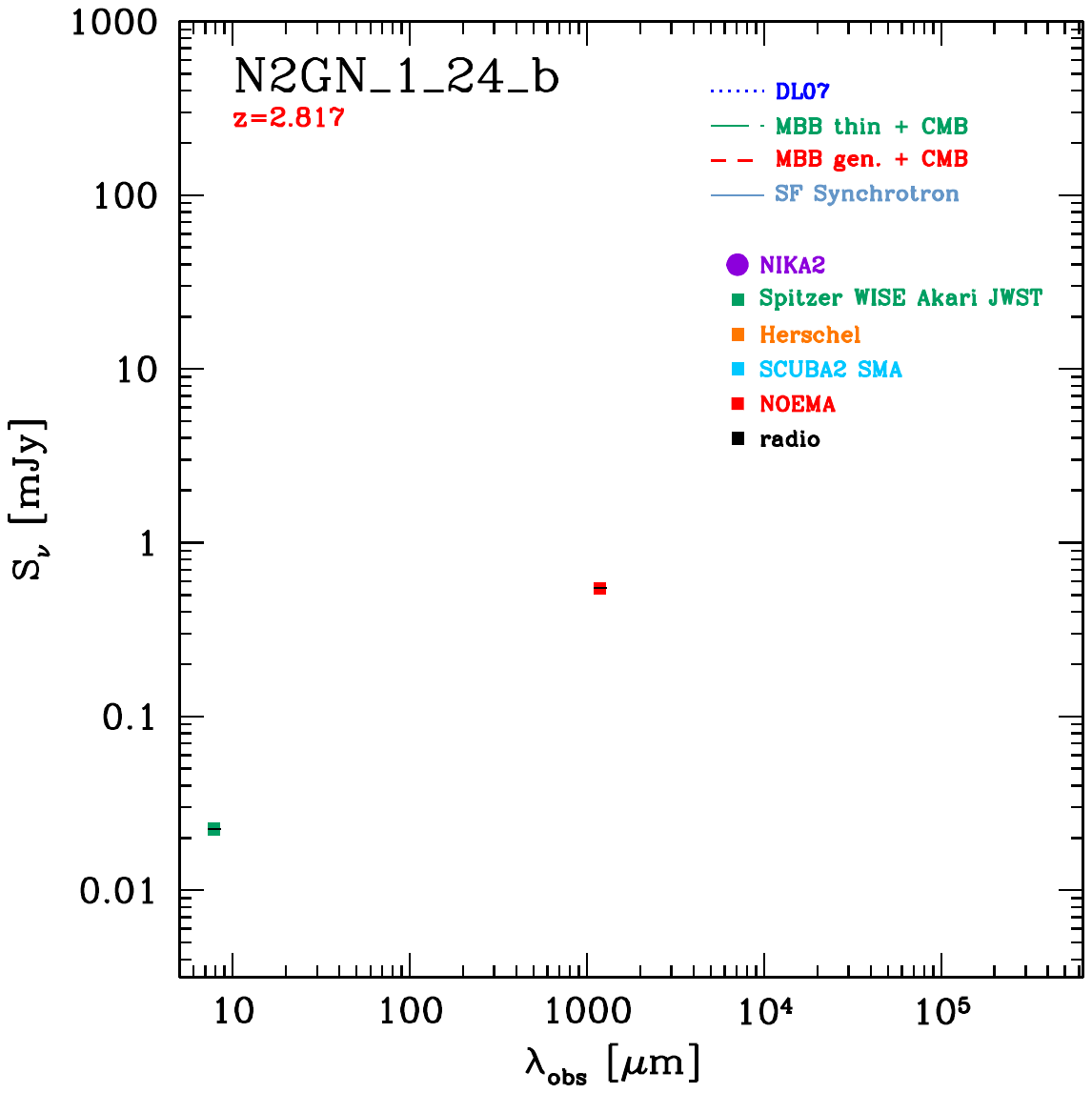}
\includegraphics[align=c,trim=0 0.18\imageheight{} 0 0.075\imageheight{}, clip, width=0.25\textwidth]{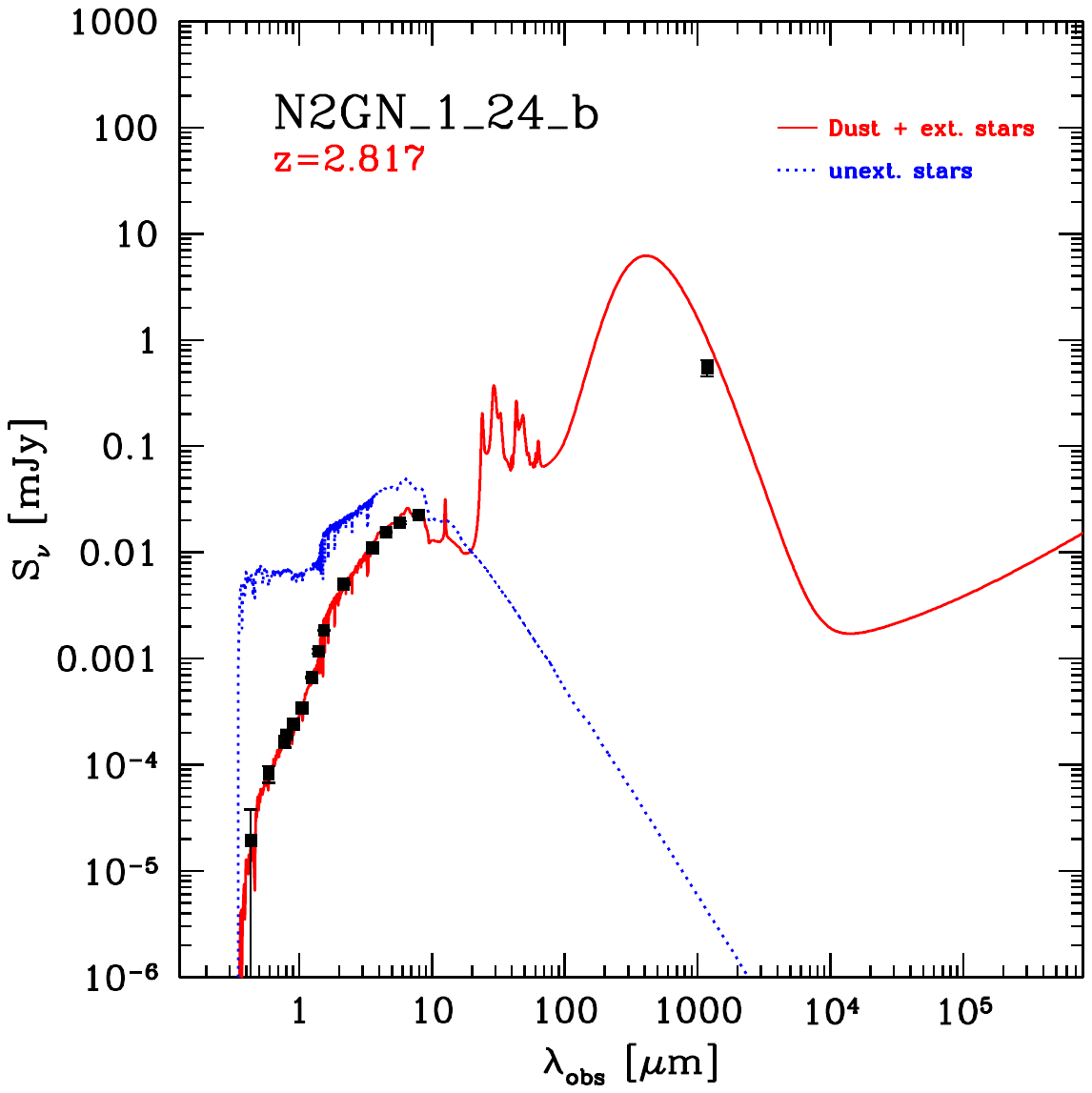}
\includegraphics[align=c,width=0.4\textwidth]{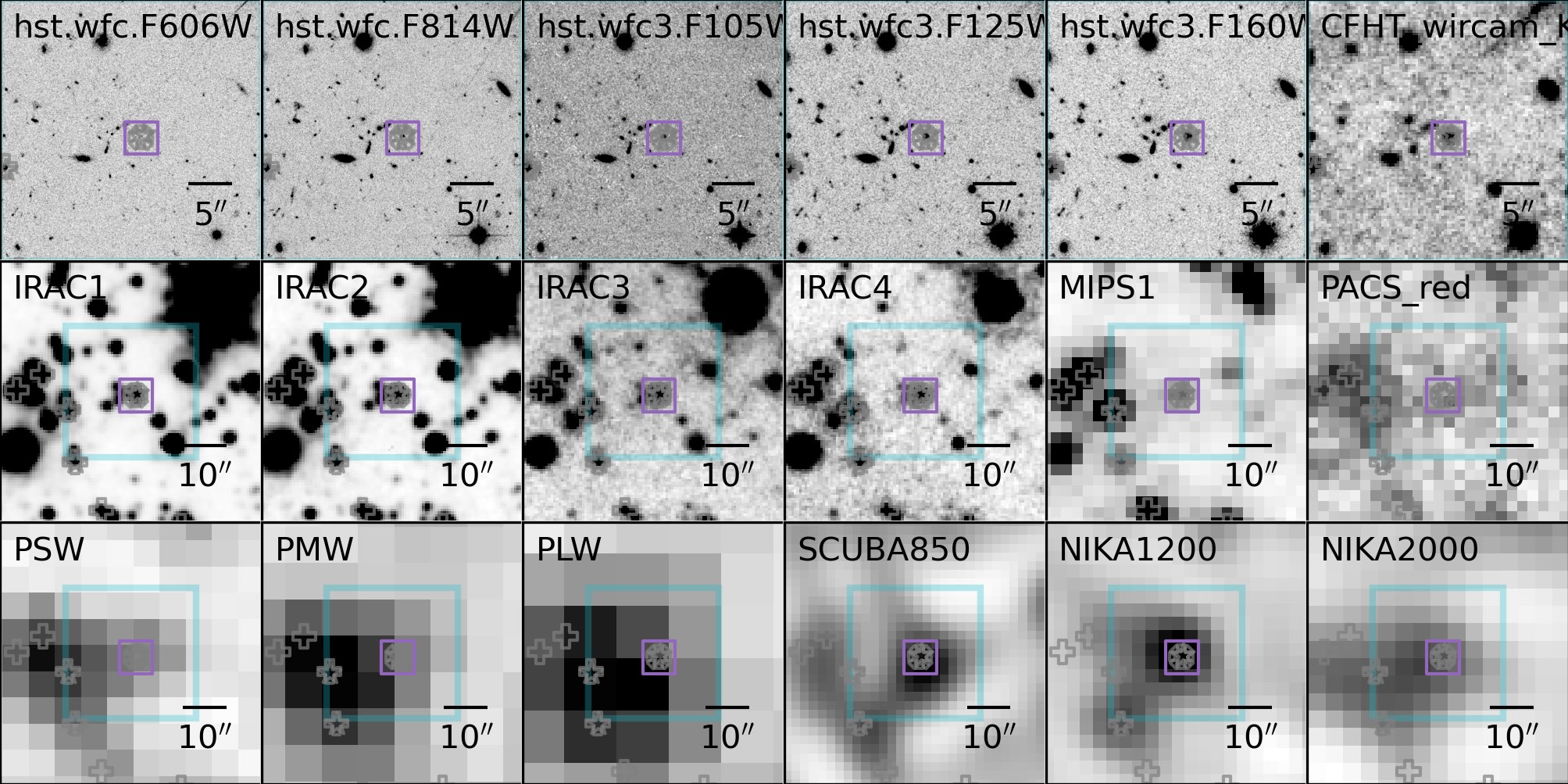}
\includegraphics[align=c,trim=0 0.18\imageheight{} 0 0.075\imageheight{}, clip, width=0.25\textwidth]{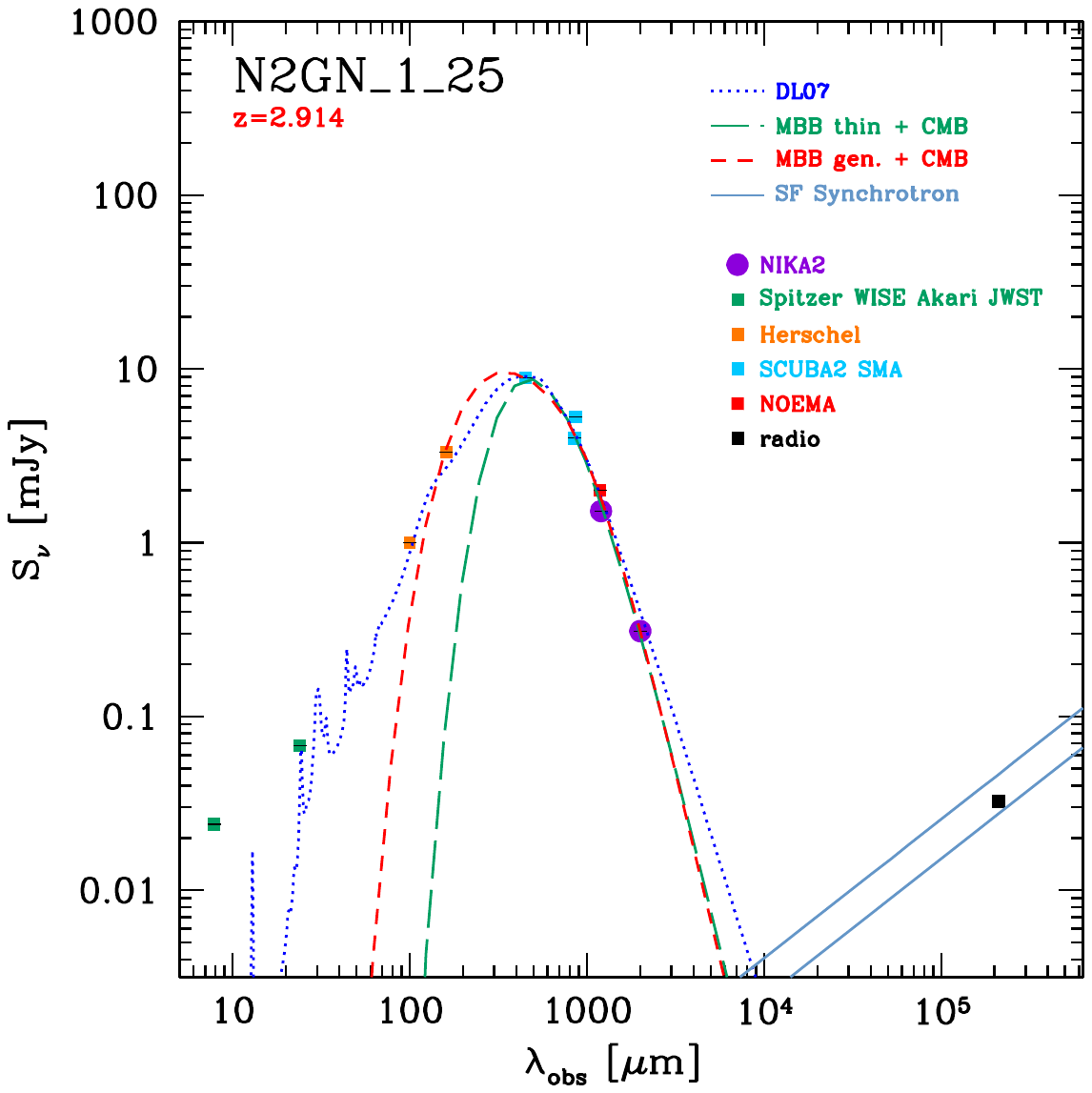}
\includegraphics[align=c,trim=0 0.18\imageheight{} 0 0.075\imageheight{}, clip, width=0.25\textwidth]{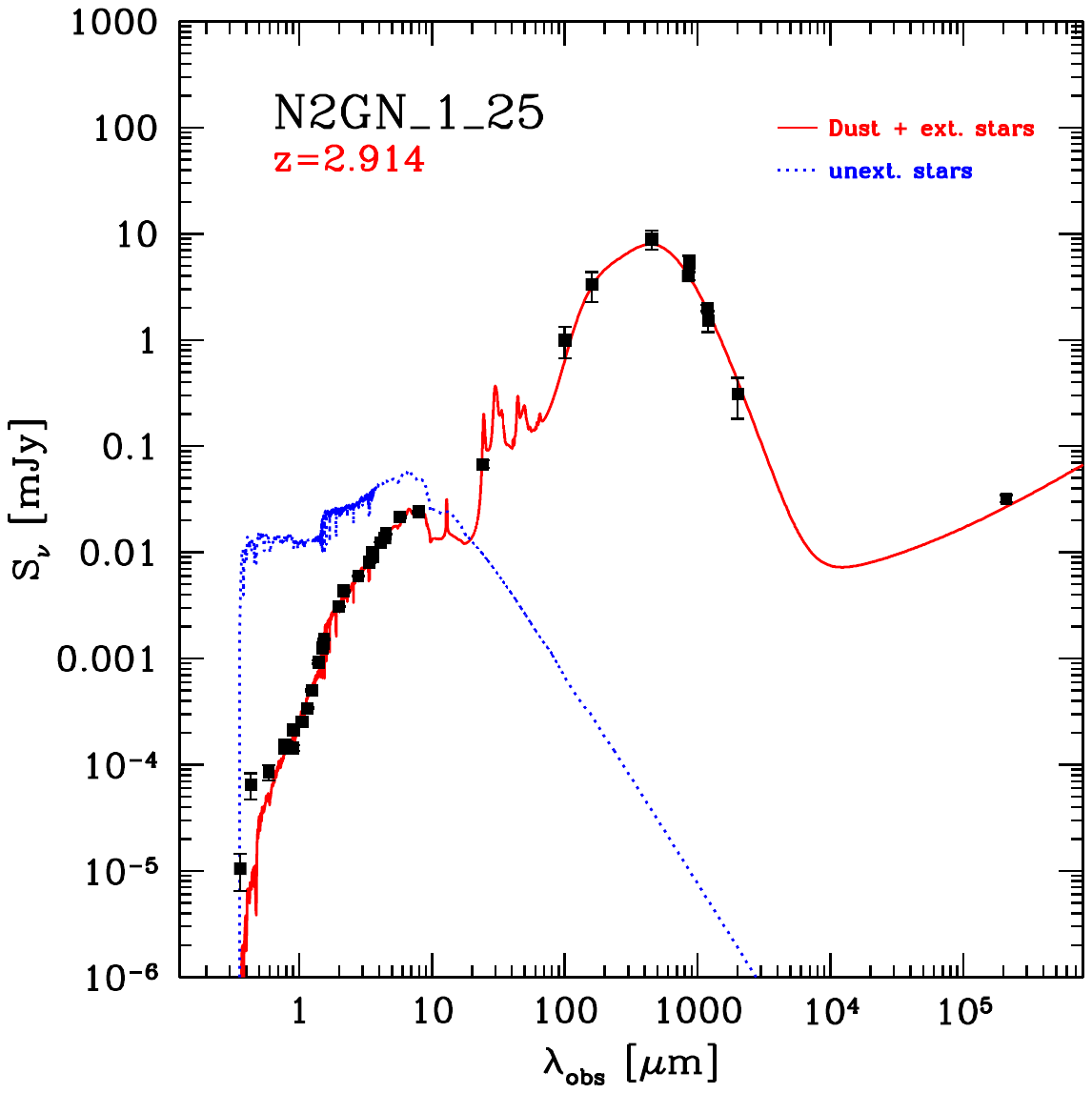}
\includegraphics[align=c,width=0.4\textwidth]{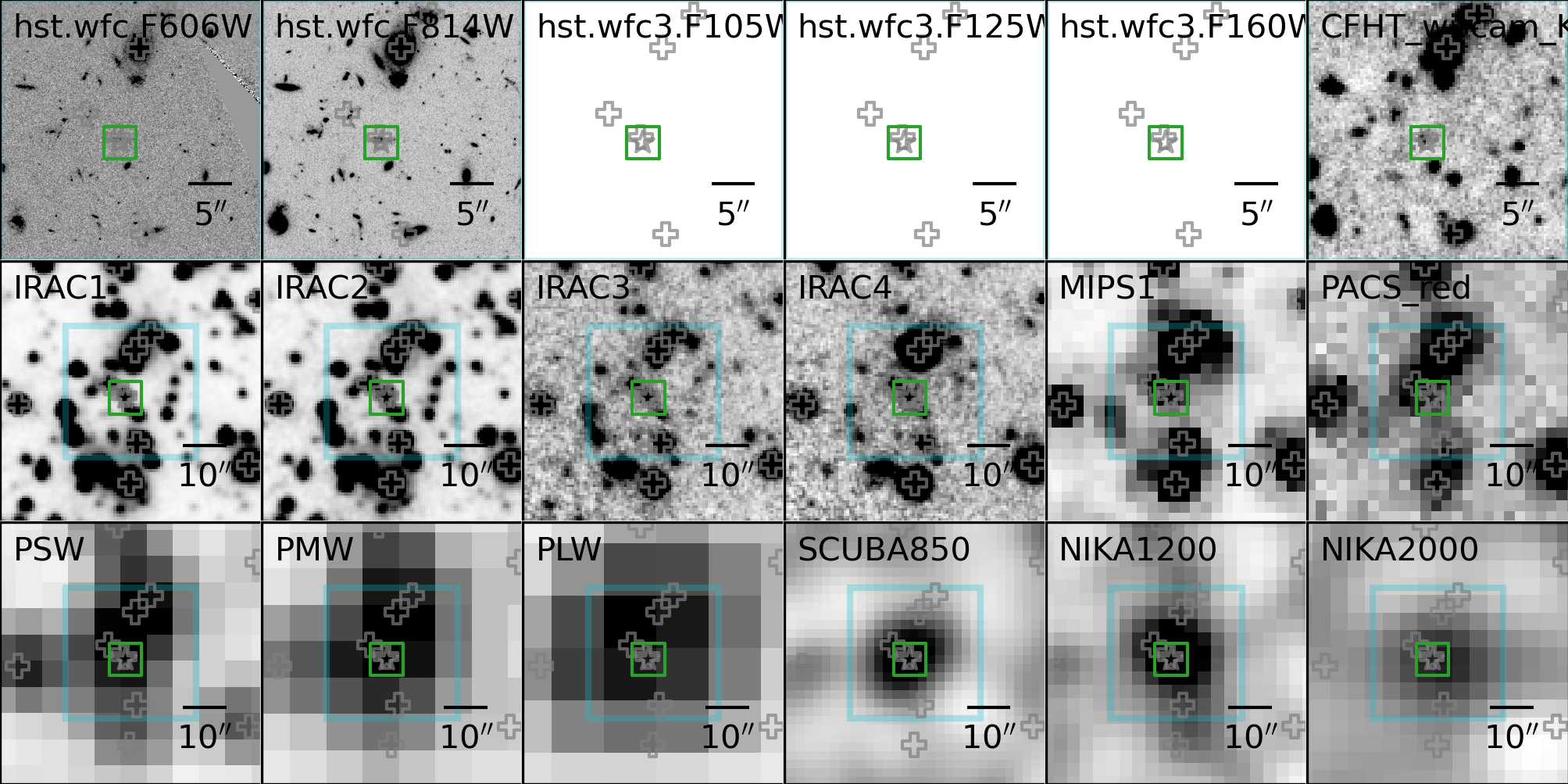}
\includegraphics[align=c,trim=0 0.18\imageheight{} 0 0.075\imageheight{}, clip, width=0.25\textwidth]{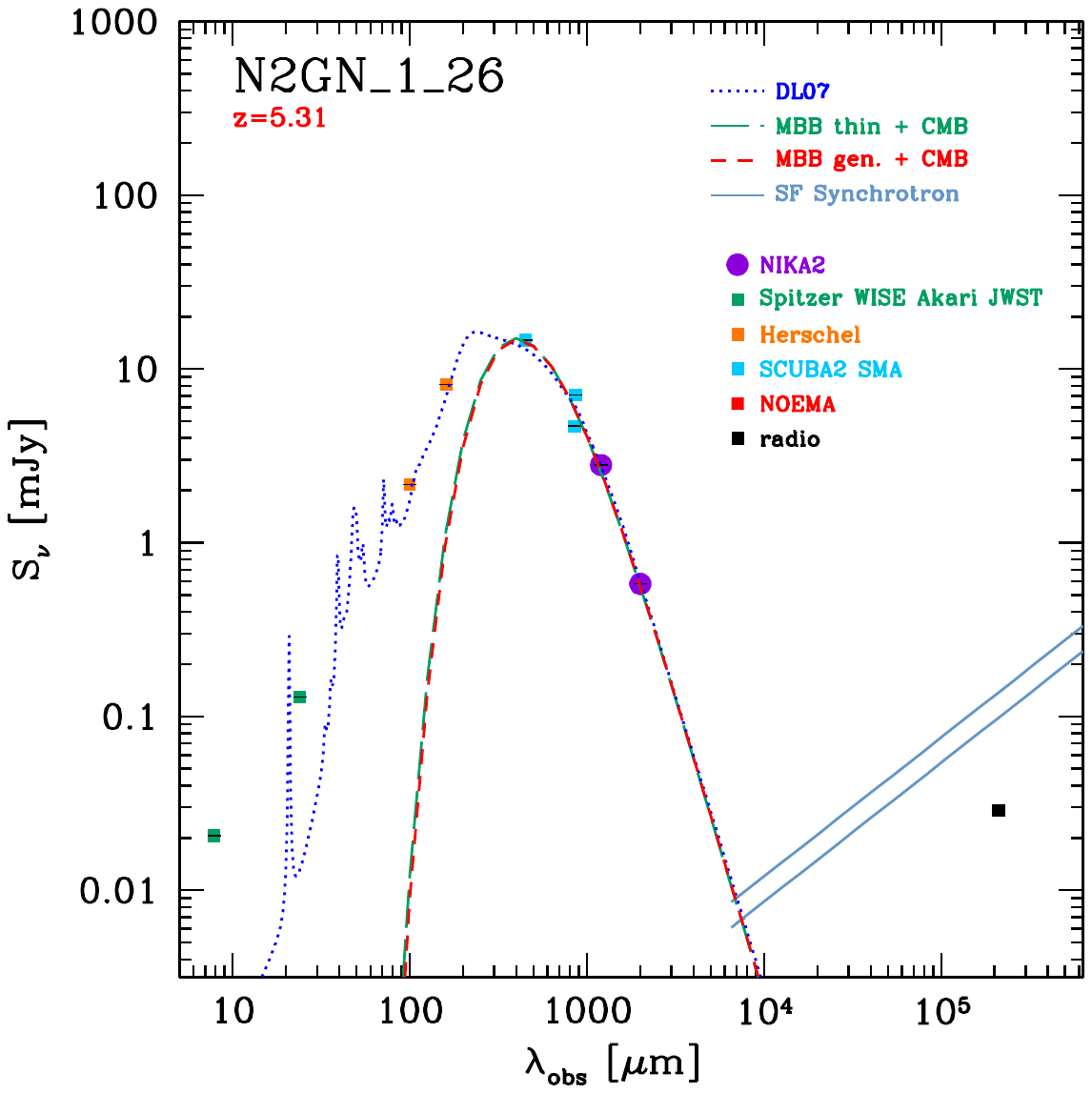}
\includegraphics[align=c,trim=0 0.18\imageheight{} 0 0.075\imageheight{}, clip, width=0.25\textwidth]{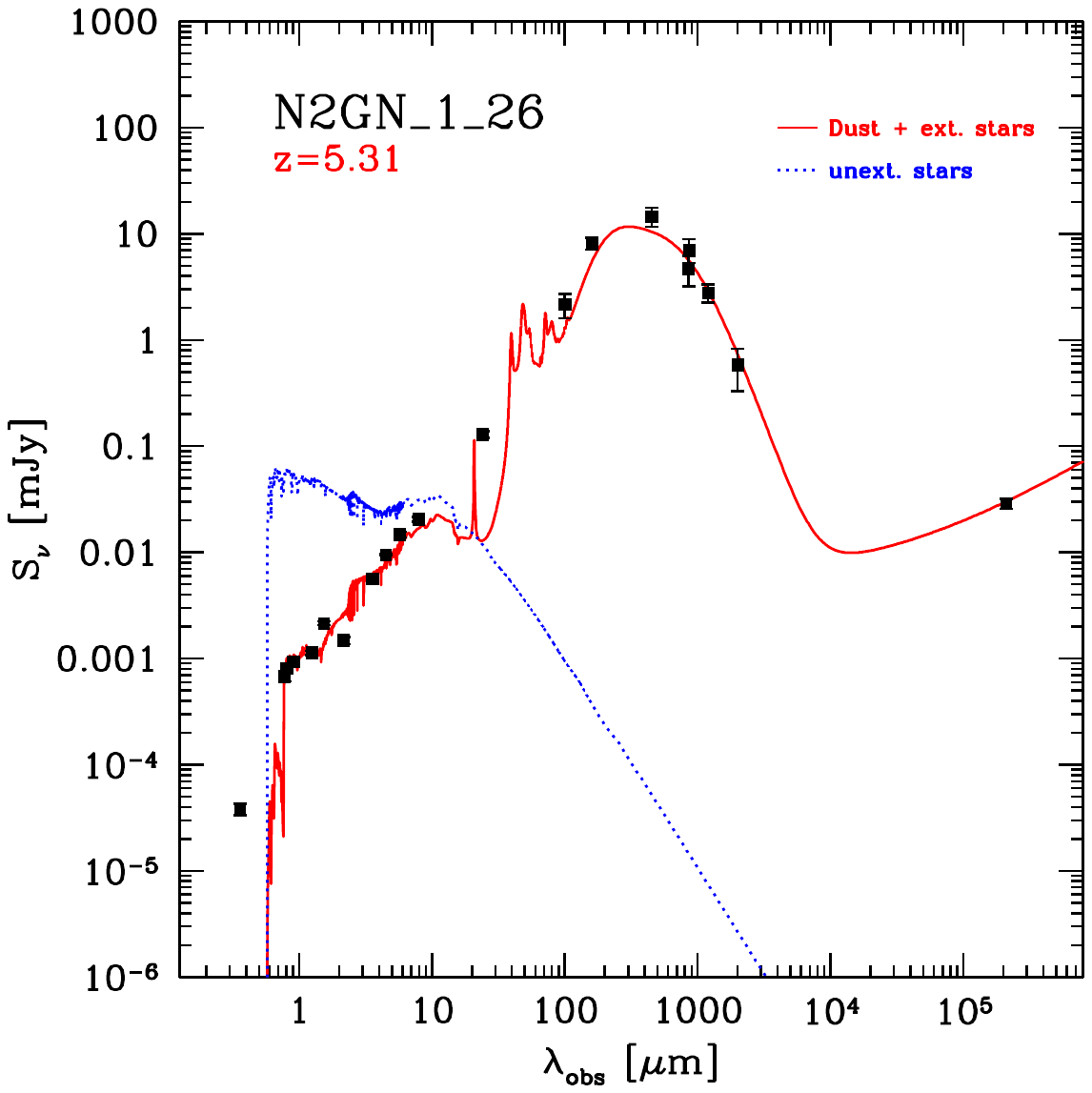}
\caption{continued.}
\end{figure*}

\addtocounter{figure}{-1}
\newpage

\begin{figure*}[t]
\centering
\includegraphics[align=c,width=0.4\textwidth]{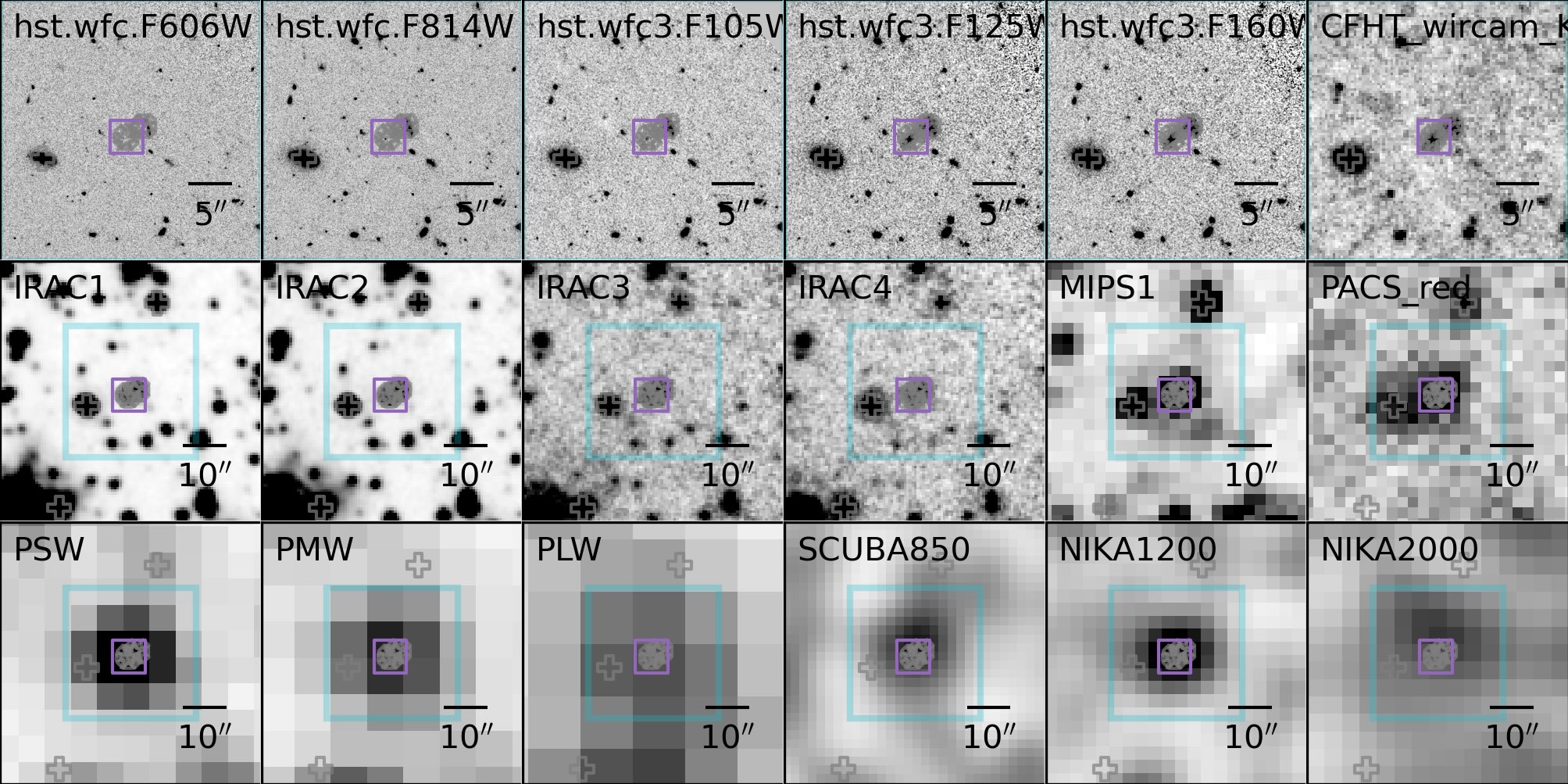}
\includegraphics[align=c,trim=0 0.18\imageheight{} 0 0.075\imageheight{}, clip, width=0.25\textwidth]{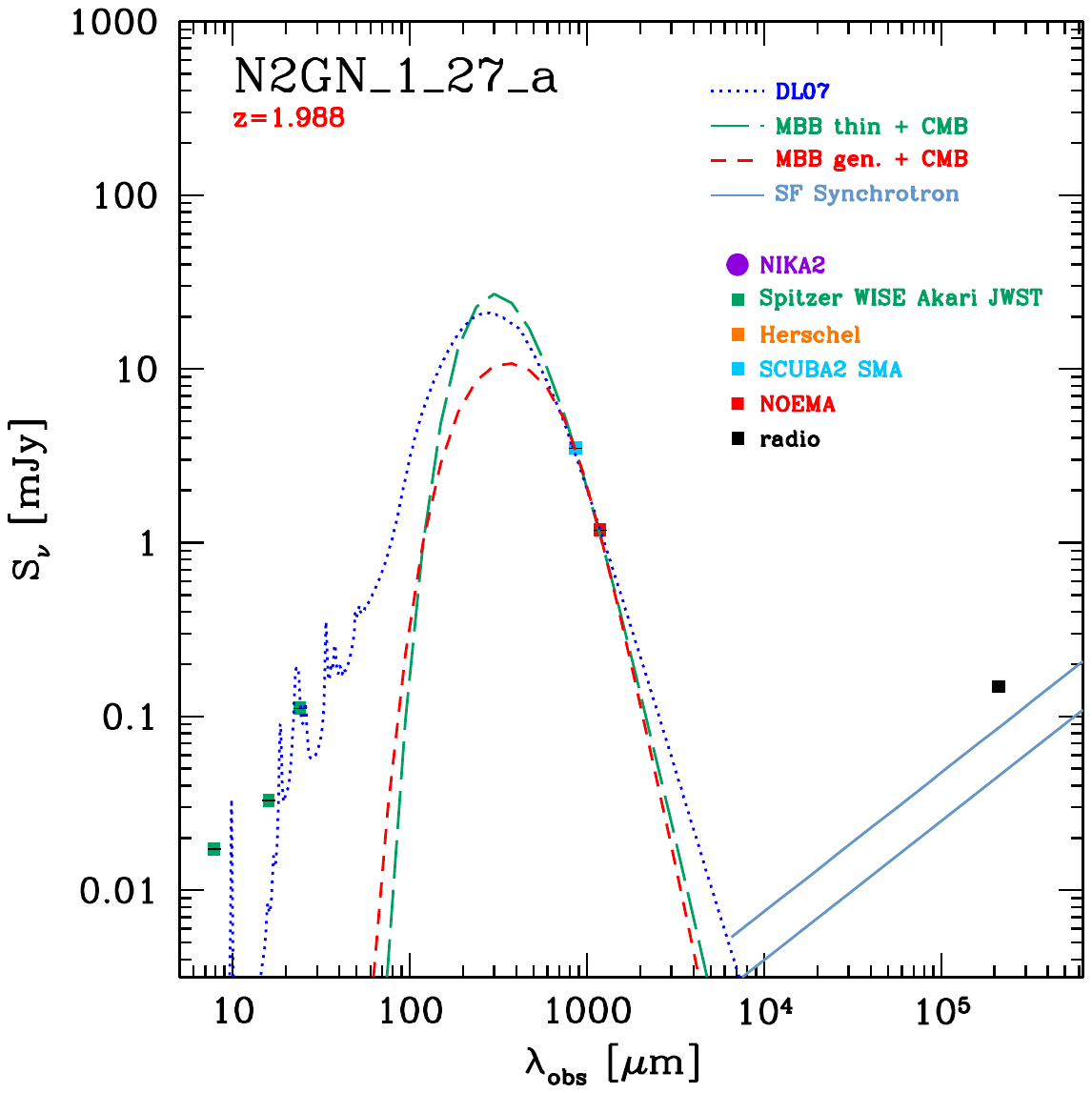}
\includegraphics[align=c,trim=0 0.18\imageheight{} 0 0.075\imageheight{}, clip, width=0.25\textwidth]{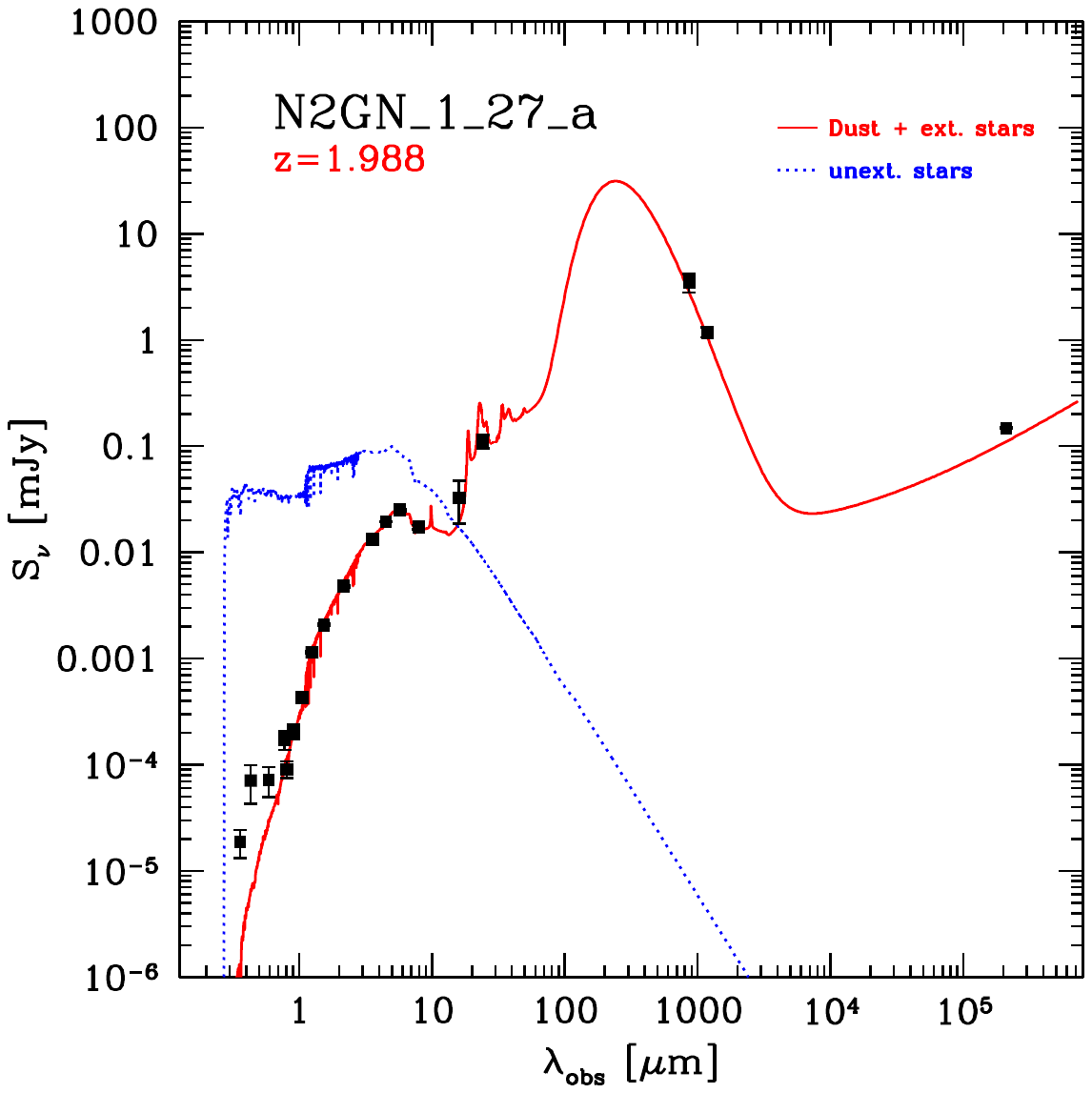}
\includegraphics[align=c,width=0.4\textwidth]{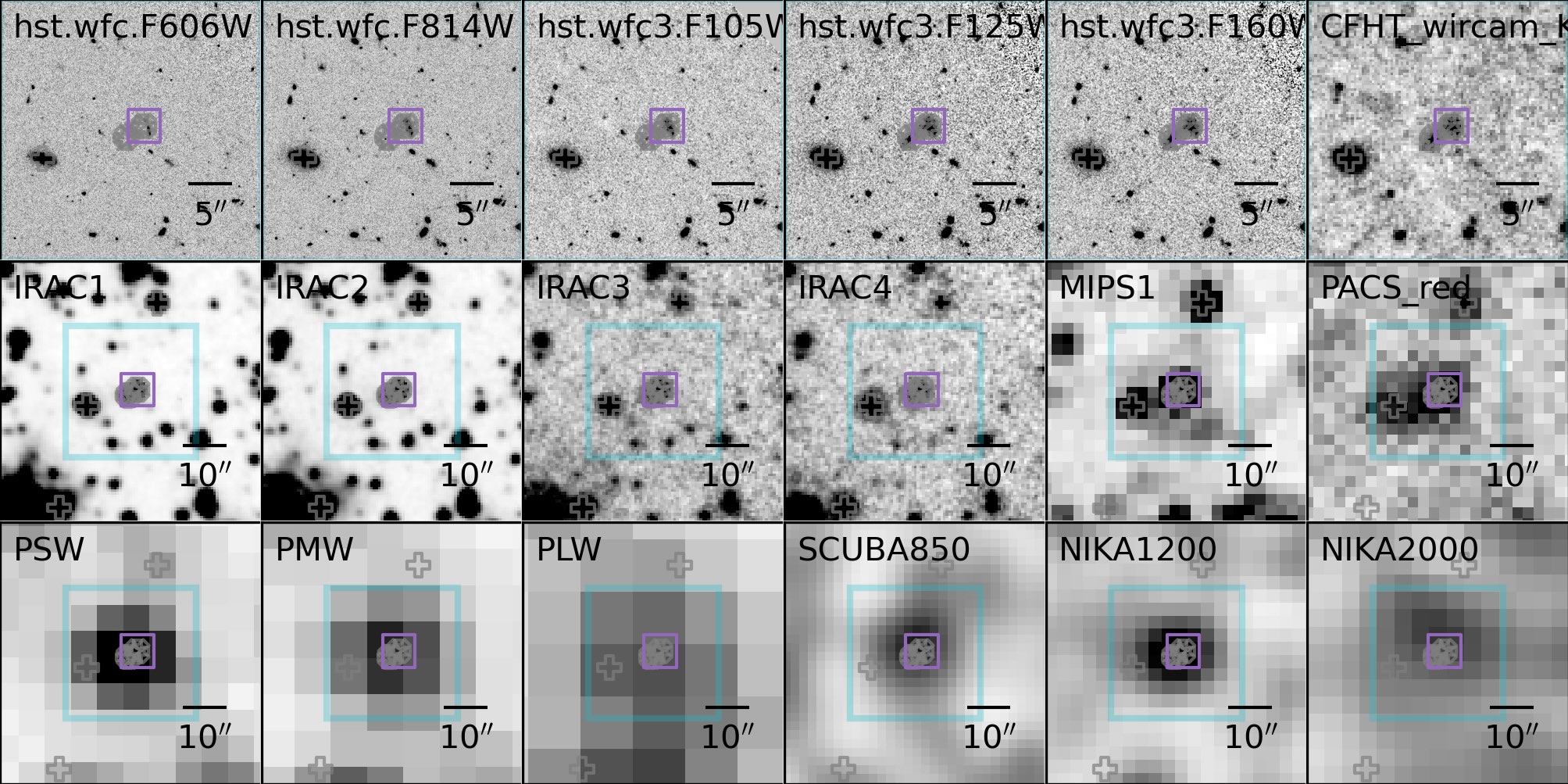}
\includegraphics[align=c,trim=0 0.18\imageheight{} 0 0.075\imageheight{}, clip, width=0.25\textwidth]{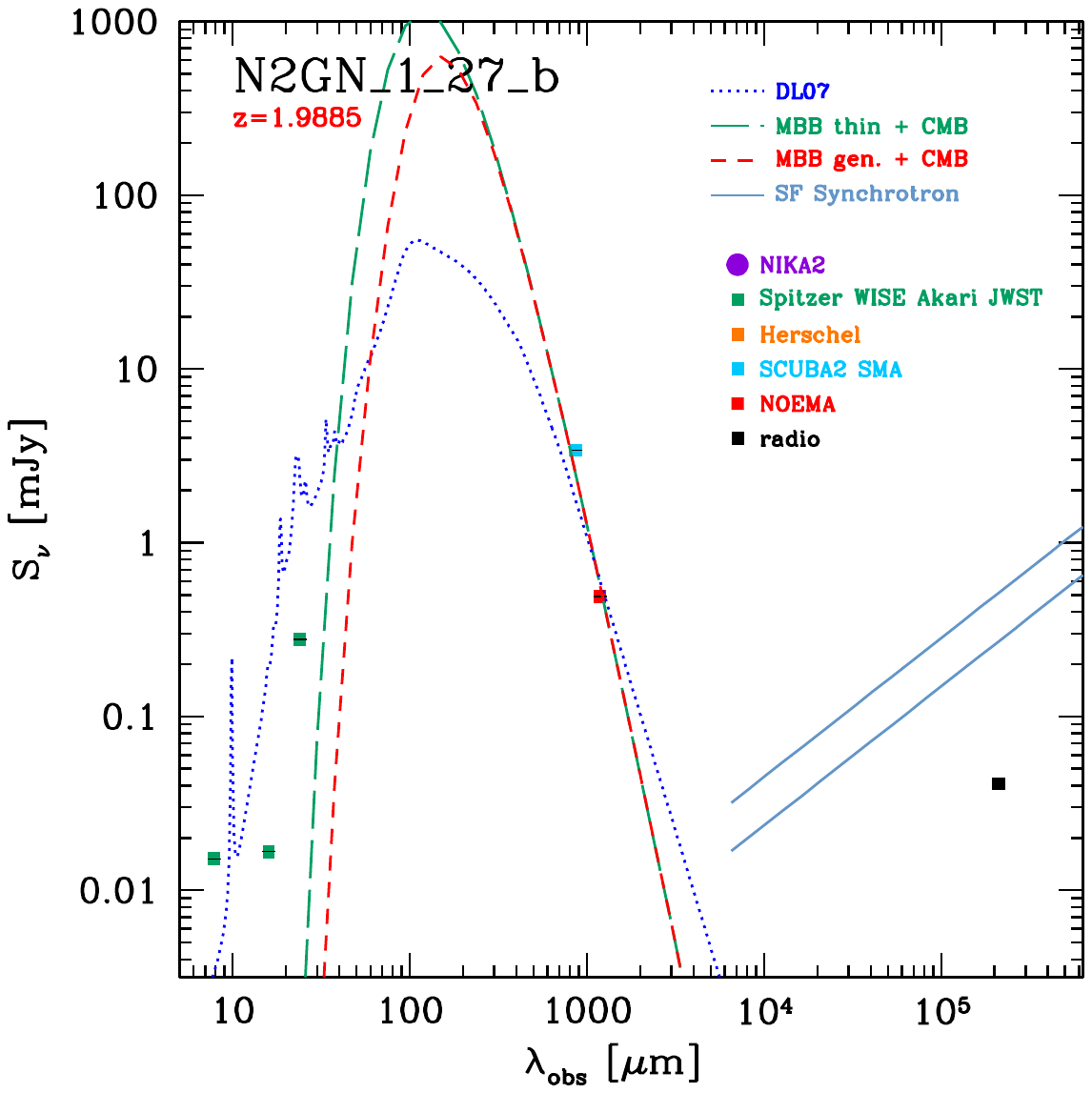}
\includegraphics[align=c,trim=0 0.18\imageheight{} 0 0.075\imageheight{}, clip, width=0.25\textwidth]{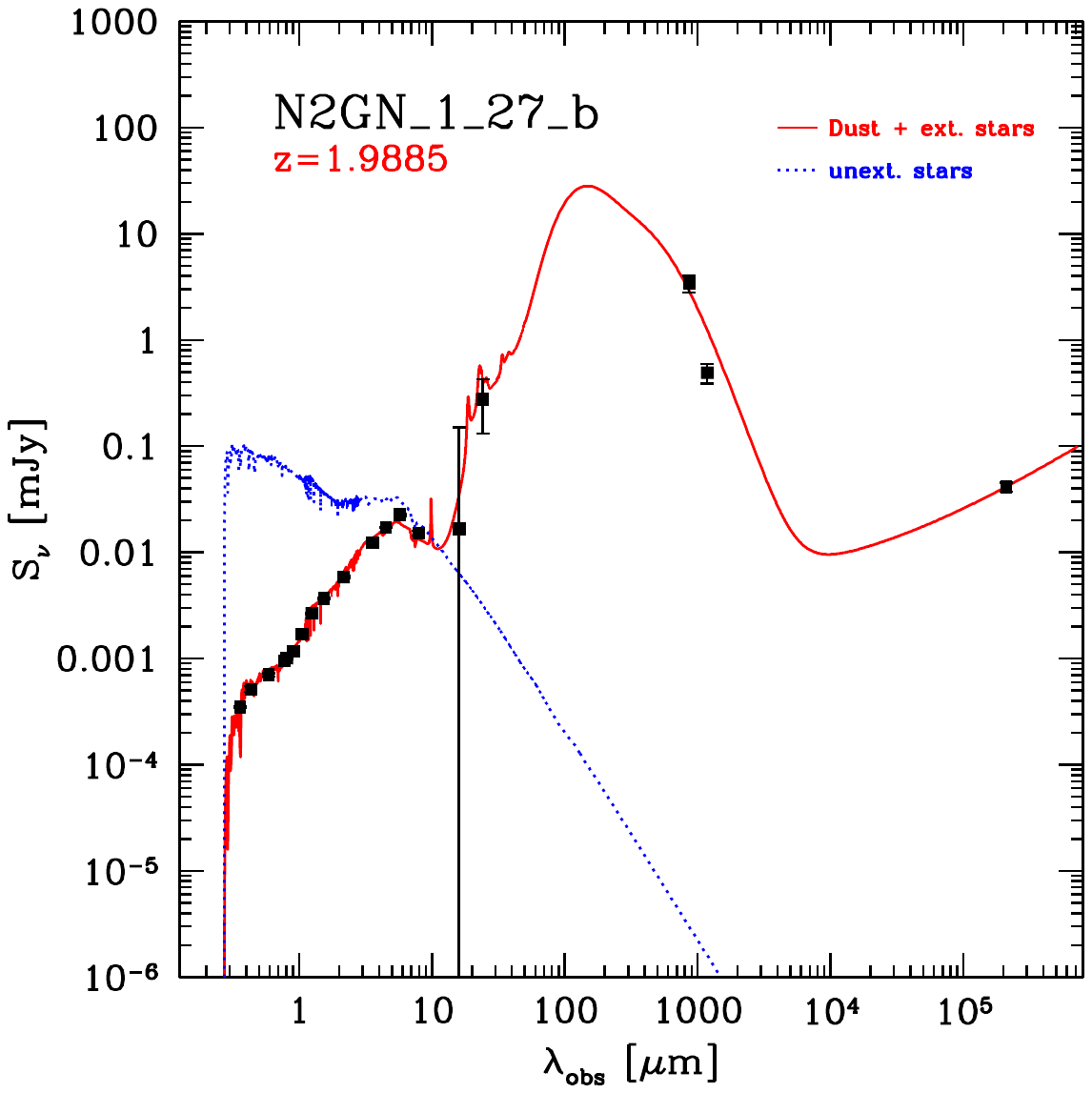}
\includegraphics[align=c,width=0.4\textwidth]{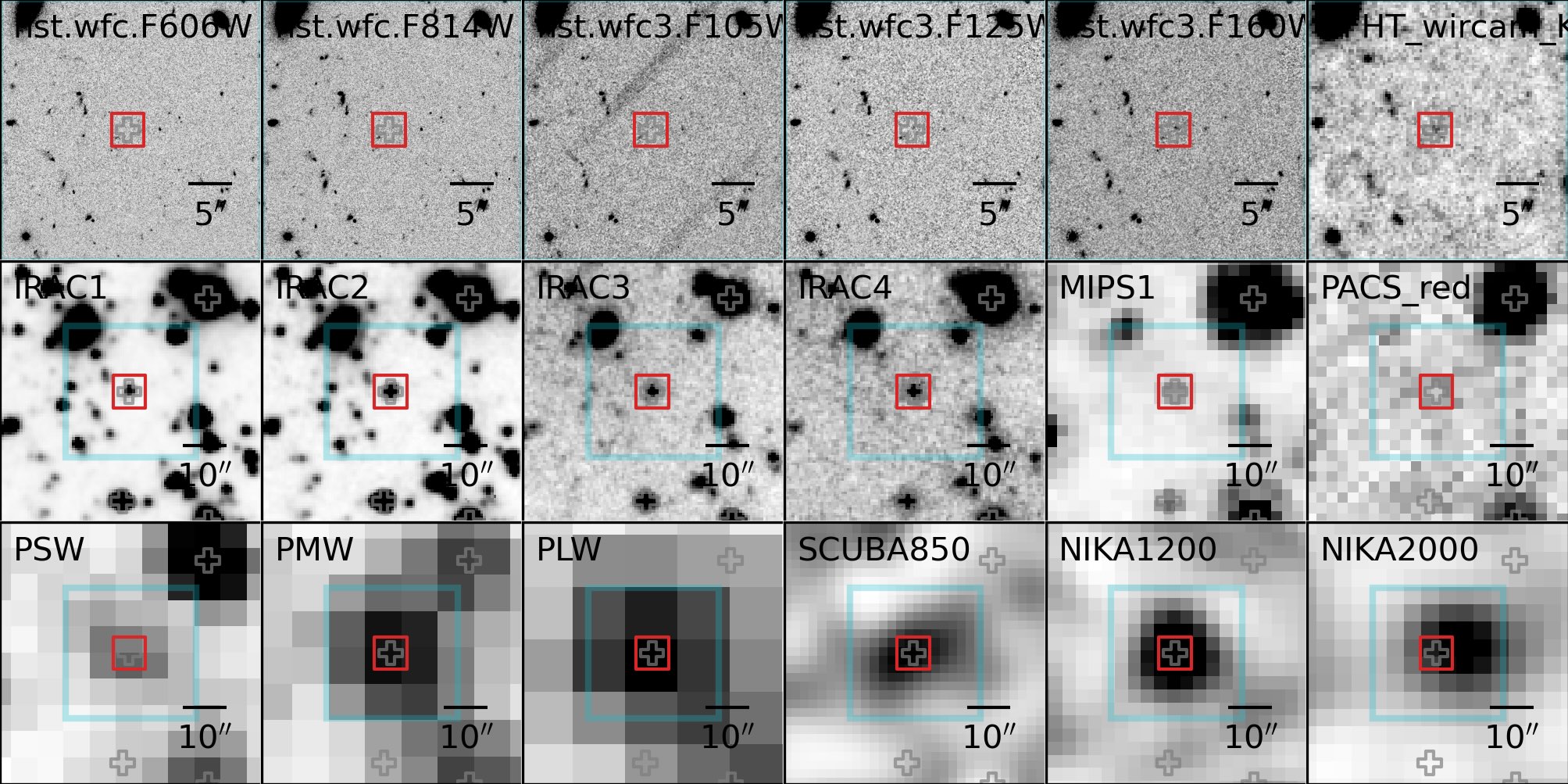}
\includegraphics[align=c,trim=0 0.18\imageheight{} 0 0.075\imageheight{}, clip, width=0.25\textwidth]{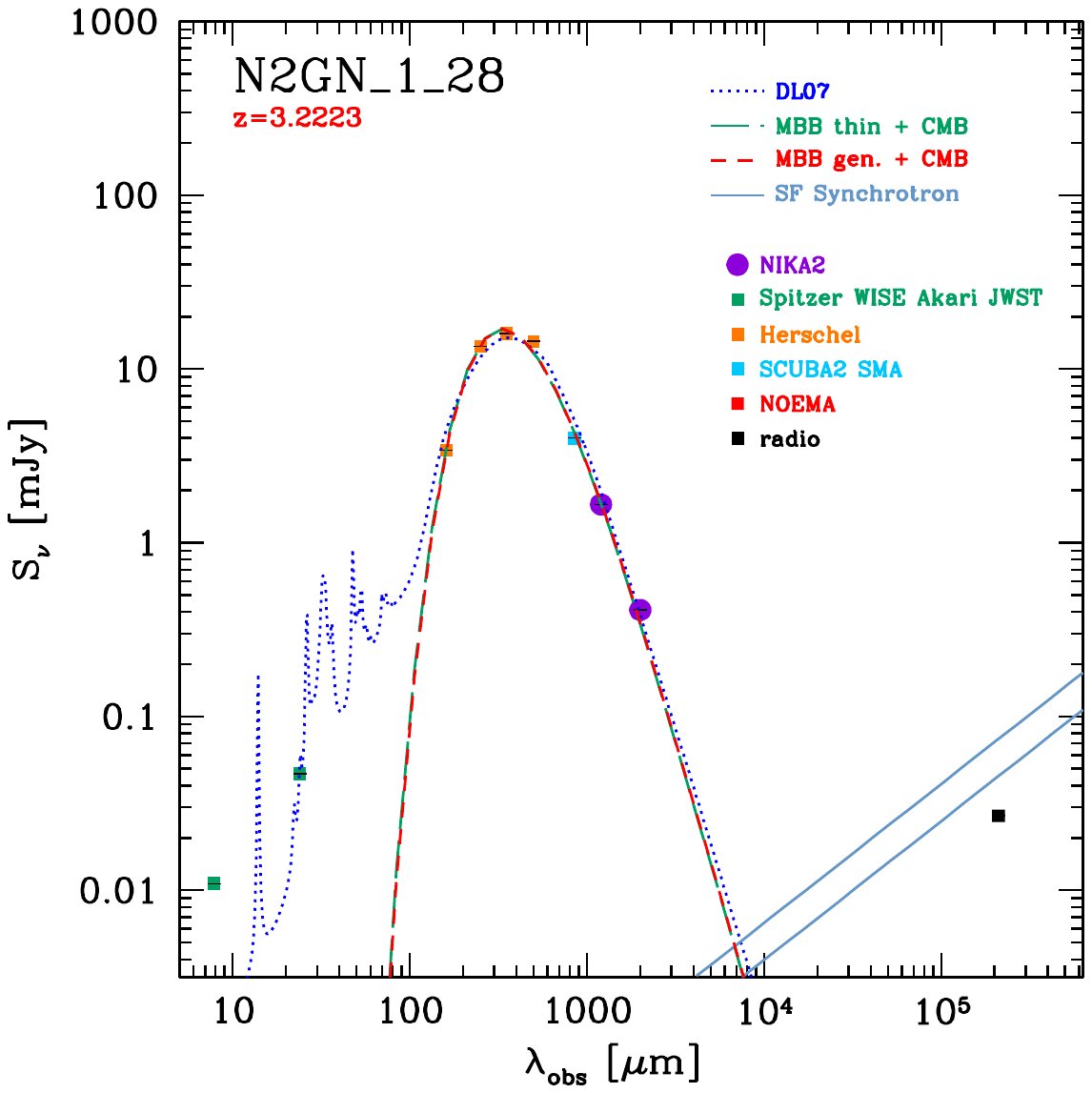}
\includegraphics[align=c,trim=0 0.18\imageheight{} 0 0.075\imageheight{}, clip, width=0.25\textwidth]{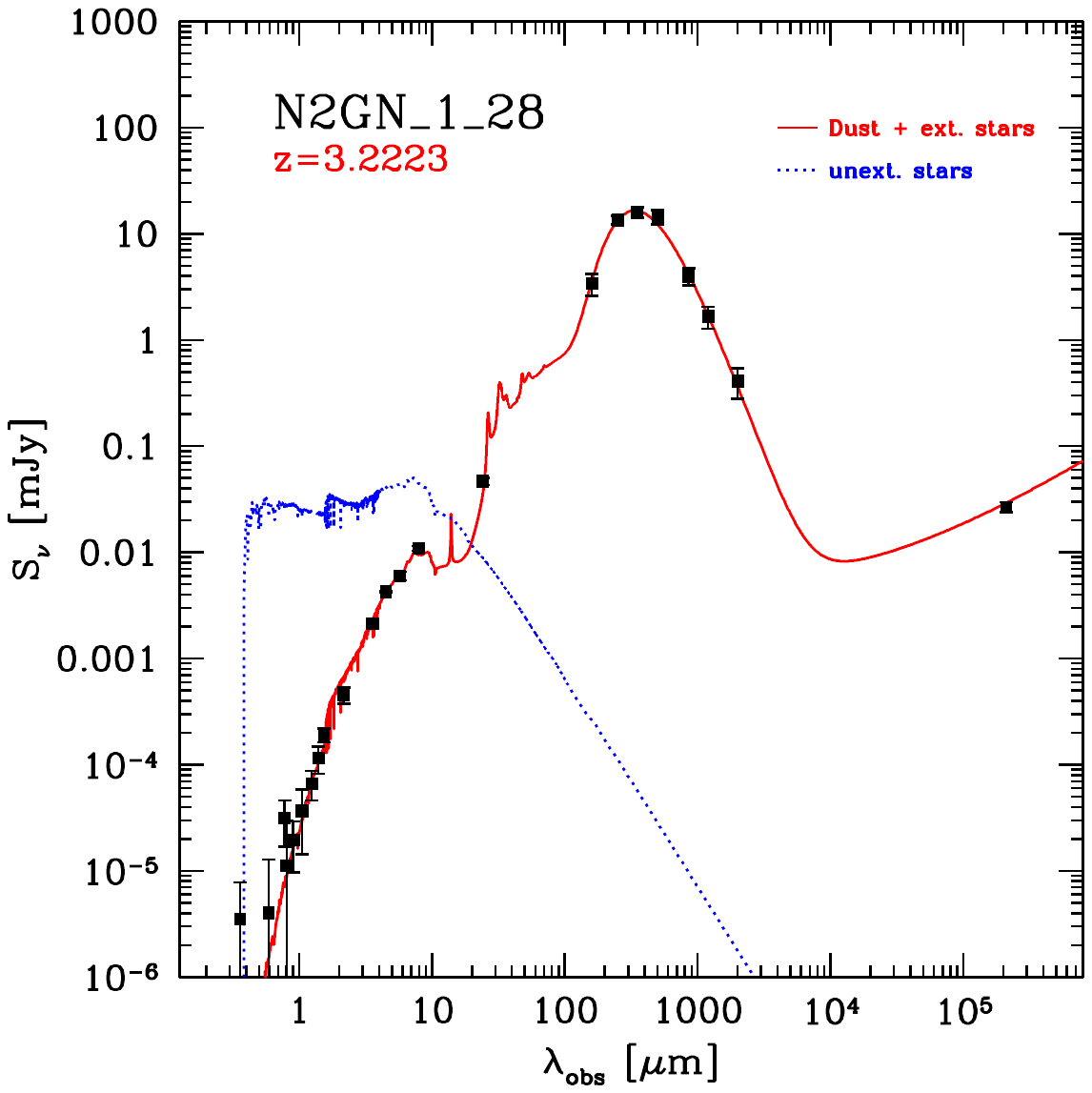}
\includegraphics[align=c,width=0.4\textwidth]{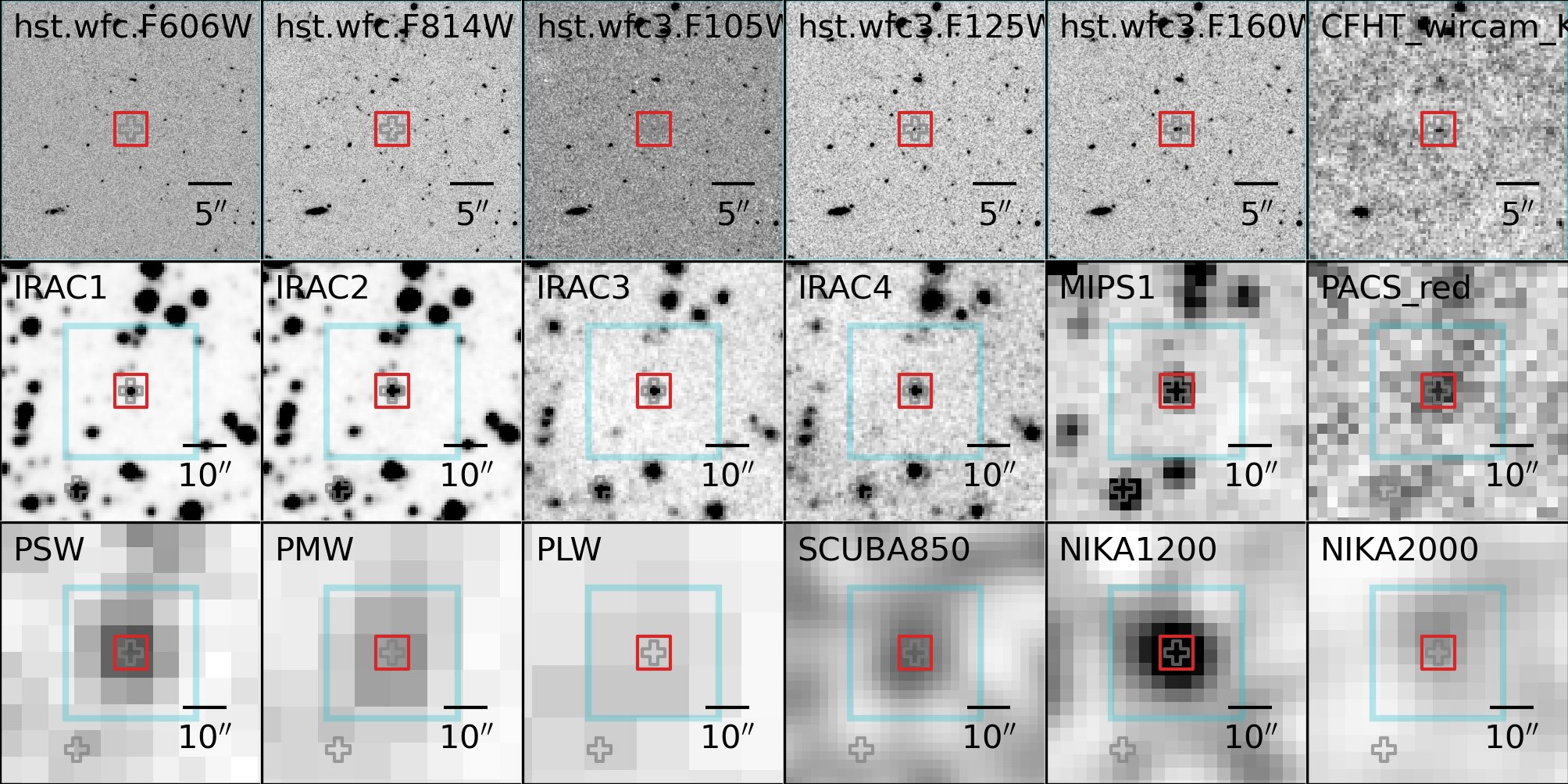}
\includegraphics[align=c,trim=0 0.18\imageheight{} 0 0.075\imageheight{}, clip, width=0.25\textwidth]{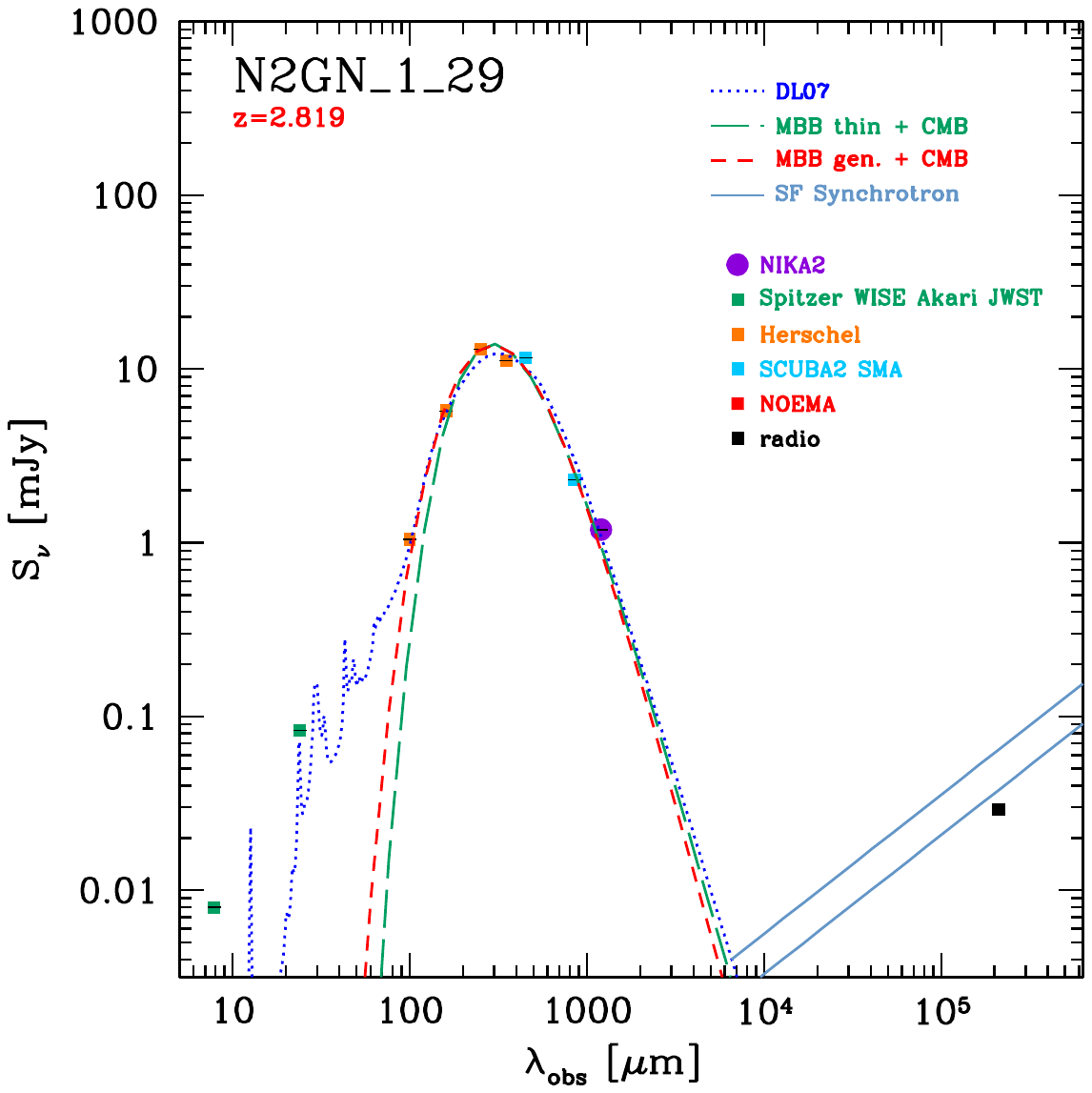}
\includegraphics[align=c,trim=0 0.18\imageheight{} 0 0.075\imageheight{}, clip, width=0.25\textwidth]{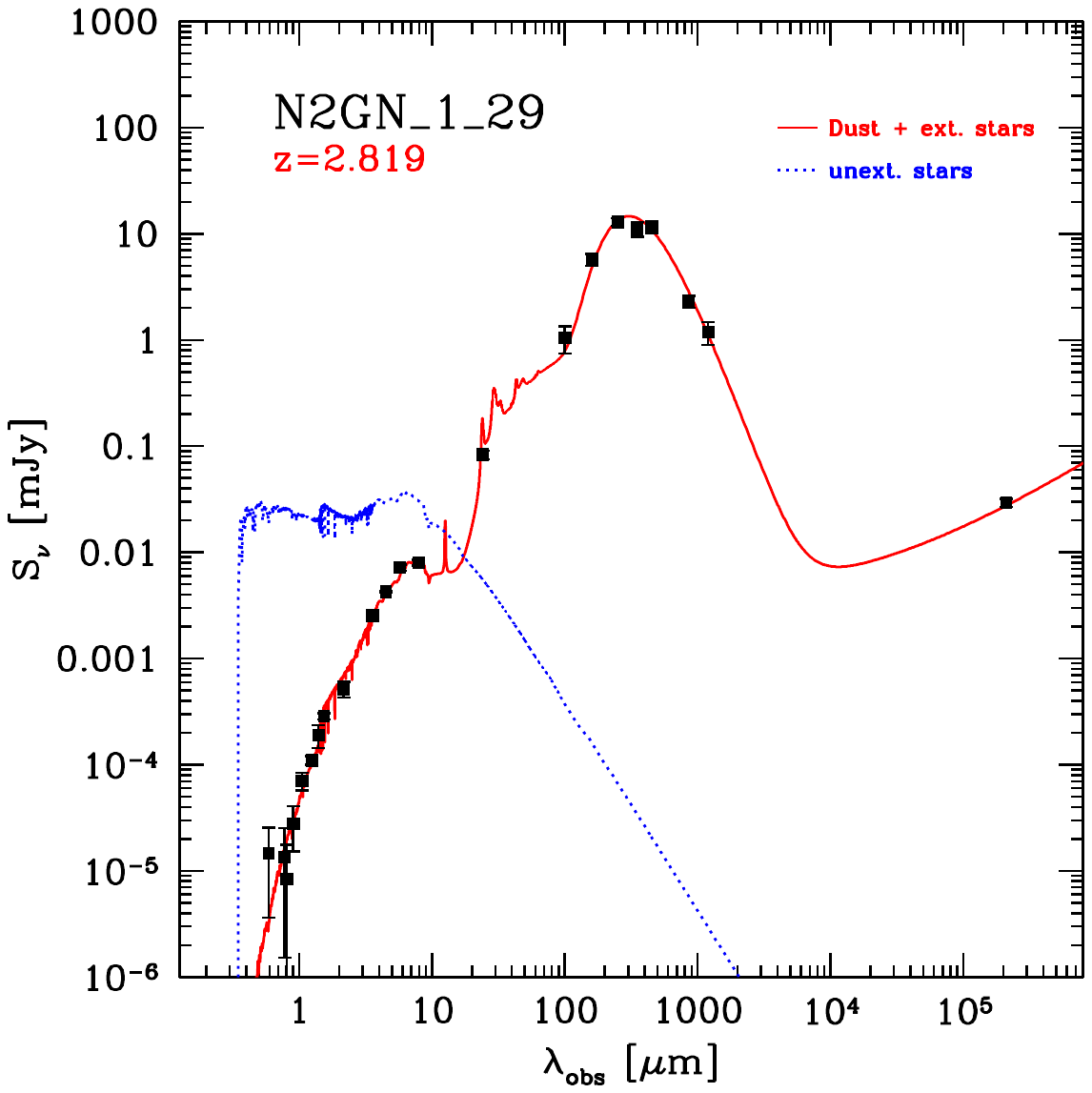}
\includegraphics[align=c,width=0.4\textwidth]{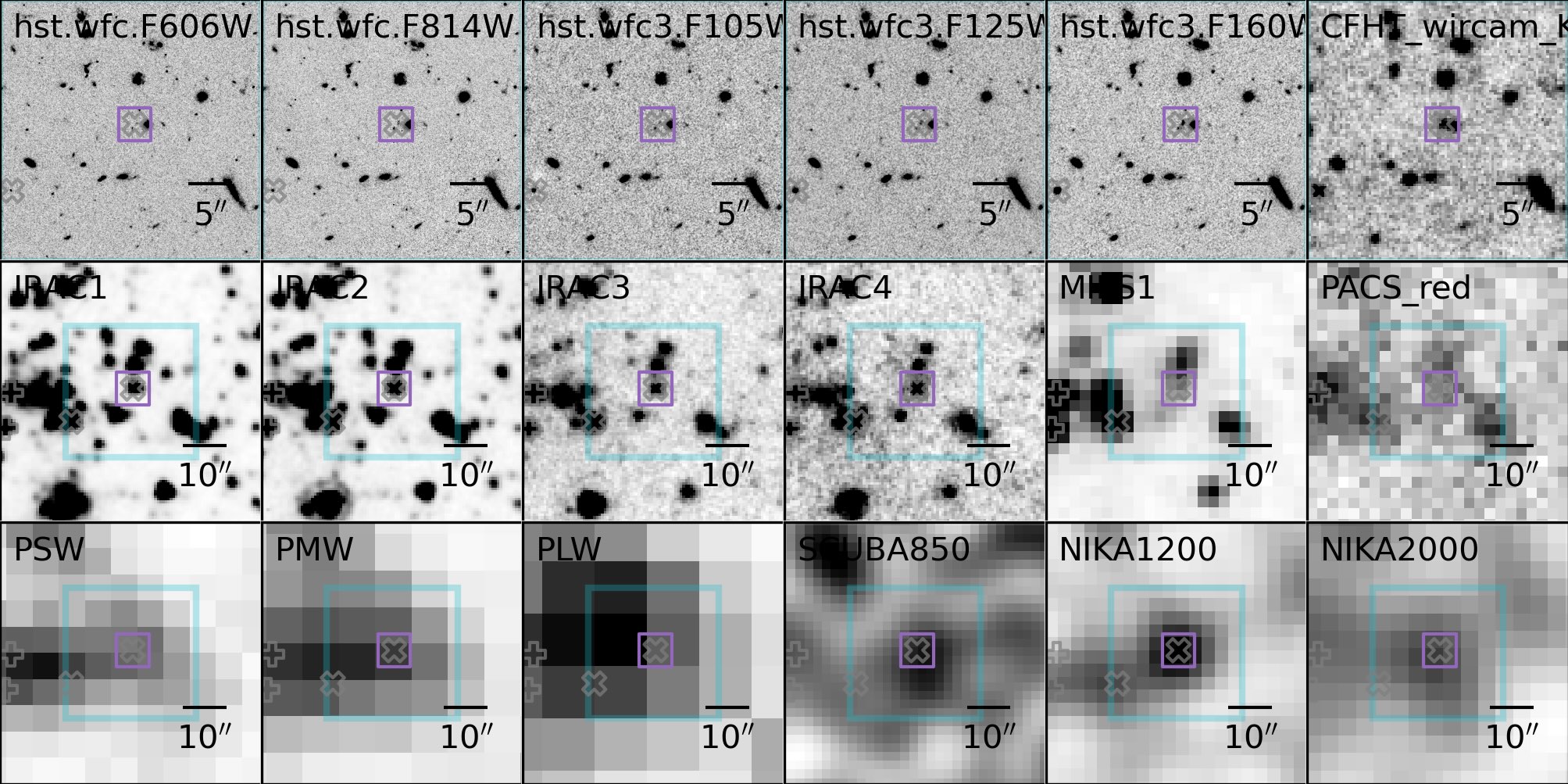}
\includegraphics[align=c,trim=0 0.18\imageheight{} 0 0.075\imageheight{}, clip, width=0.25\textwidth]{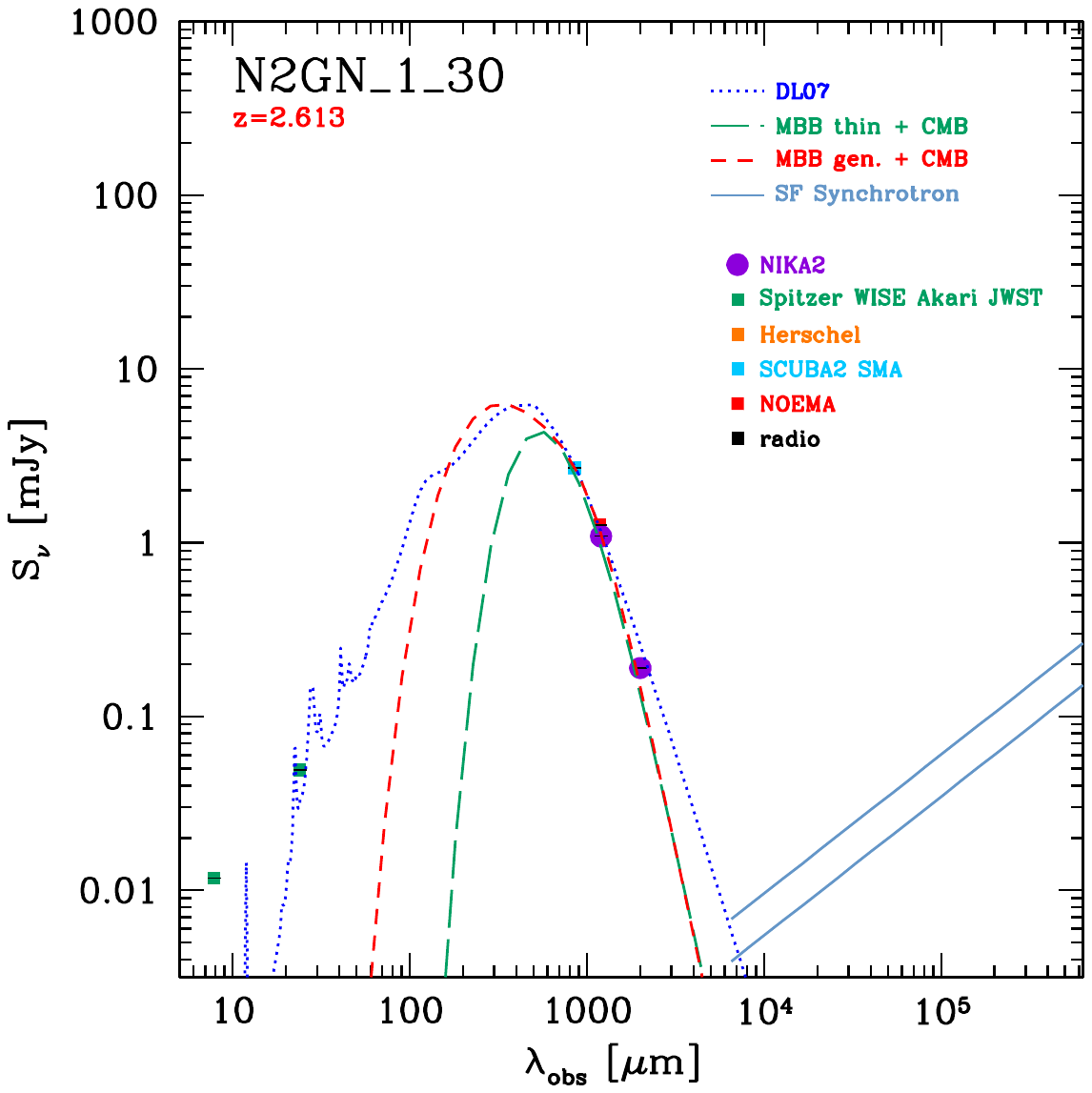}
\includegraphics[align=c,trim=0 0.18\imageheight{} 0 0.075\imageheight{}, clip, width=0.25\textwidth]{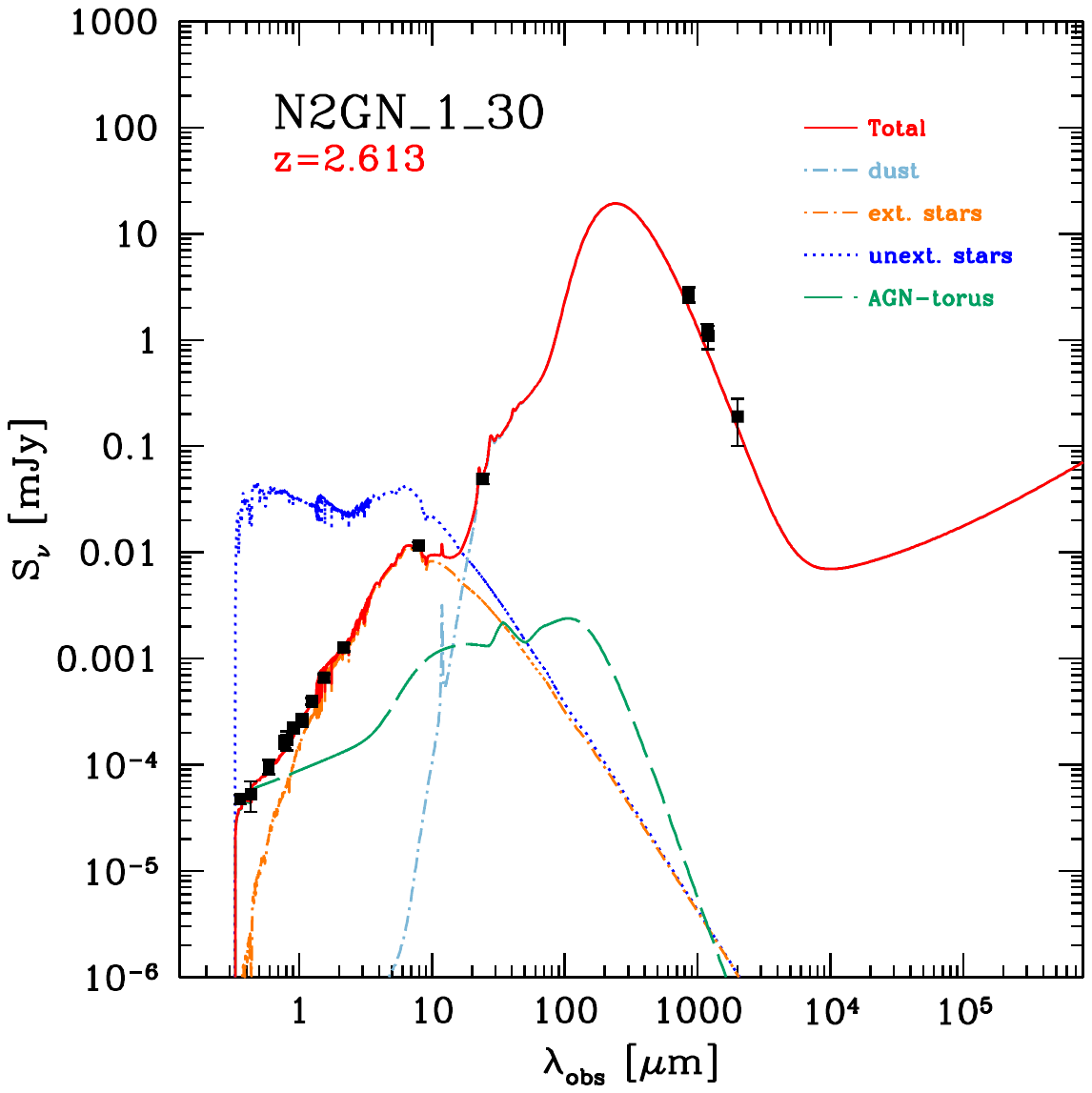}
\caption{continued.}
\end{figure*}

\addtocounter{figure}{-1}
\newpage

\begin{figure*}[t]
\centering
\includegraphics[align=c,width=0.4\textwidth]{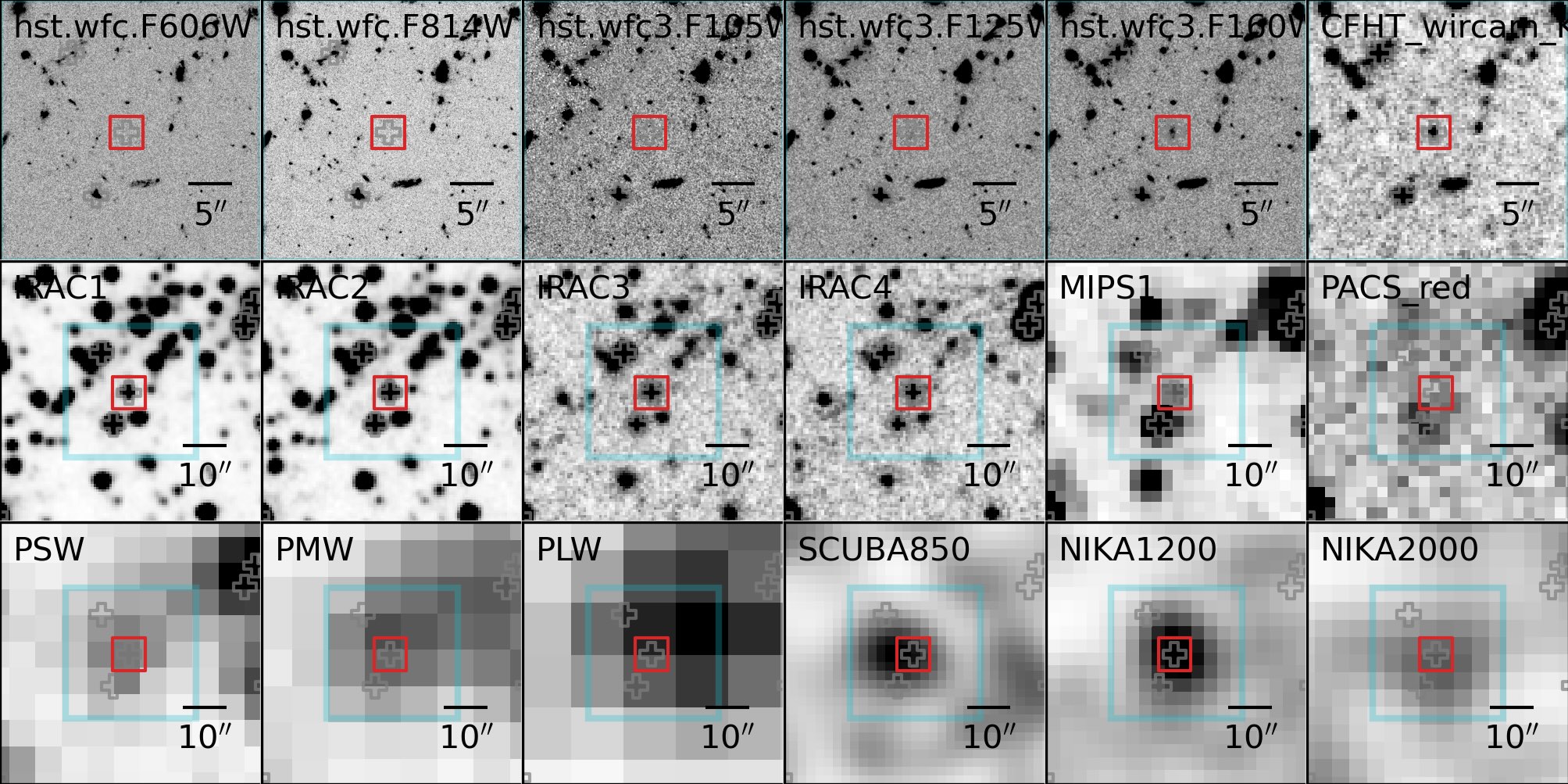}
\includegraphics[align=c,trim=0 0.18\imageheight{} 0 0.075\imageheight{}, clip, width=0.25\textwidth]{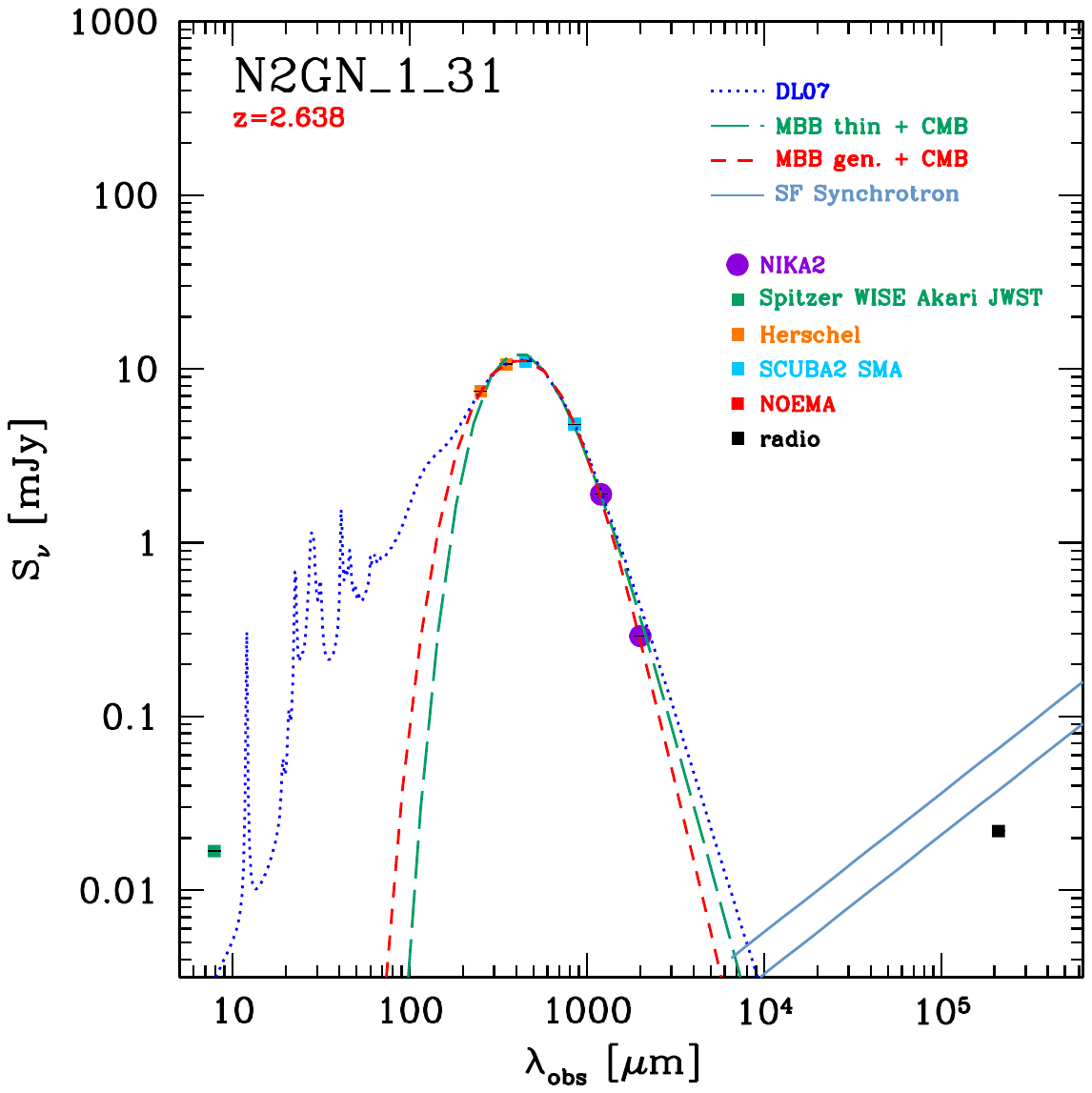}
\includegraphics[align=c,trim=0 0.18\imageheight{} 0 0.075\imageheight{}, clip, width=0.25\textwidth]{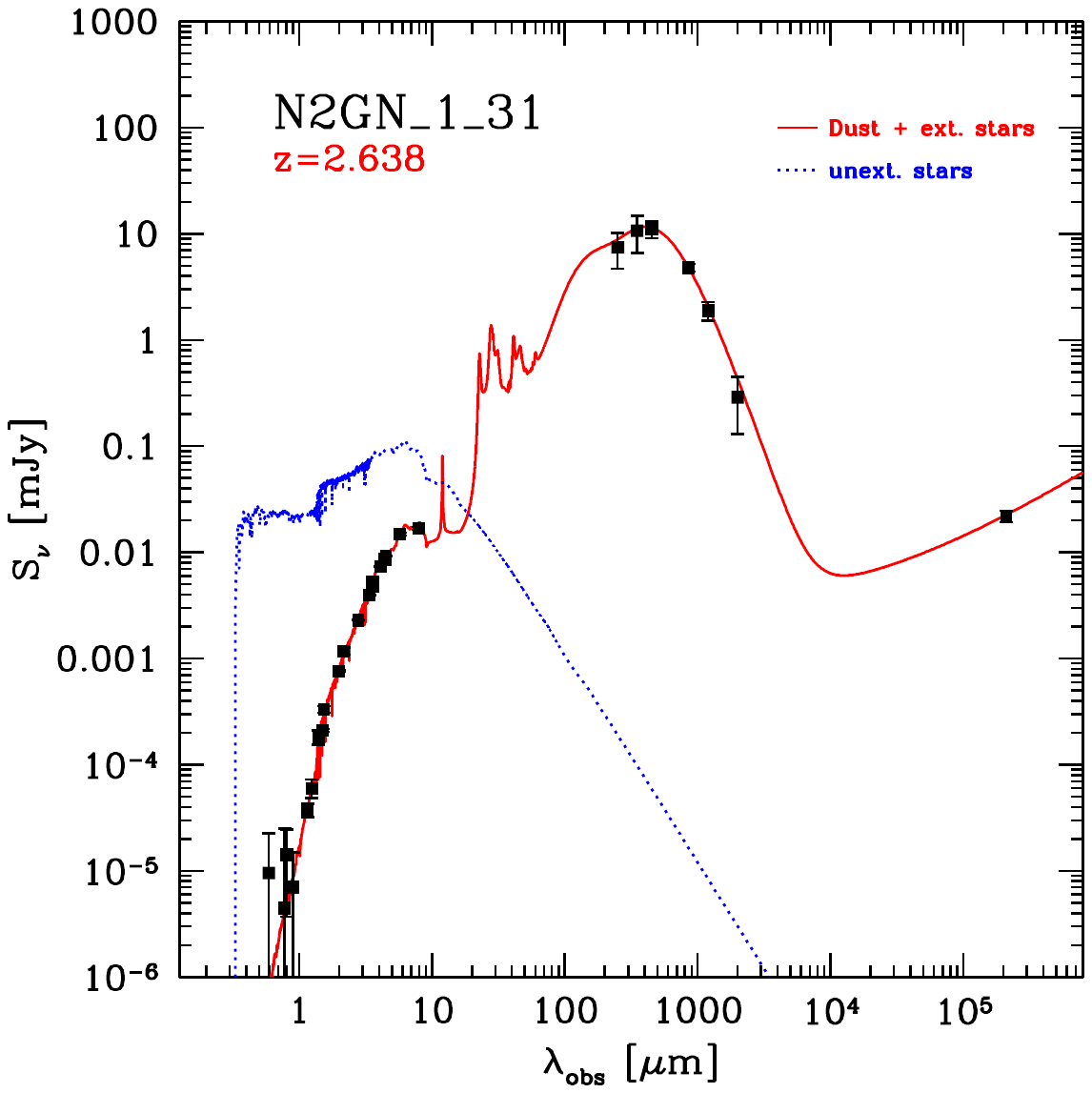}
\includegraphics[align=c,width=0.4\textwidth]{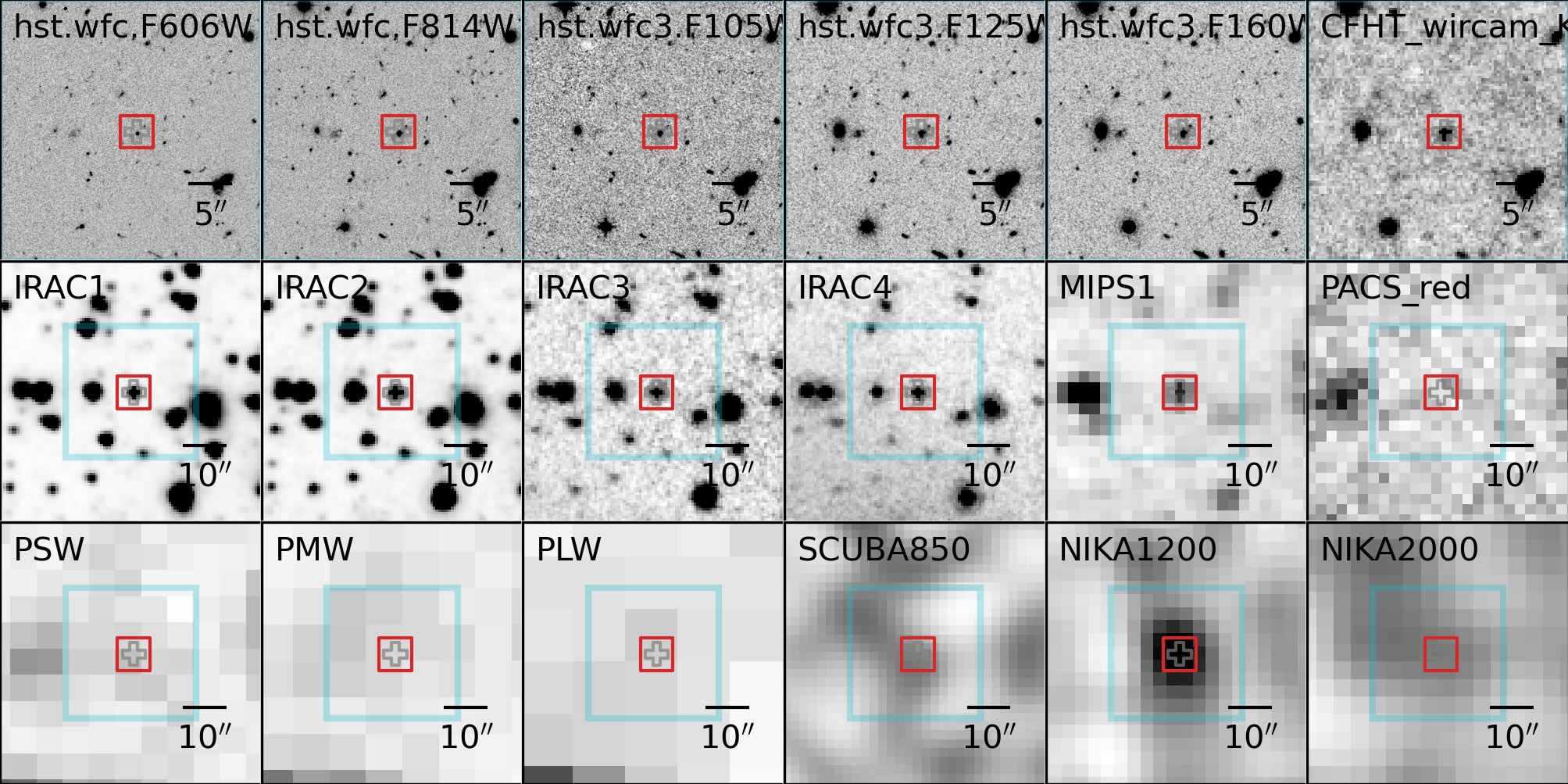}
\includegraphics[align=c,trim=0 0.18\imageheight{} 0 0.075\imageheight{}, clip, width=0.25\textwidth]{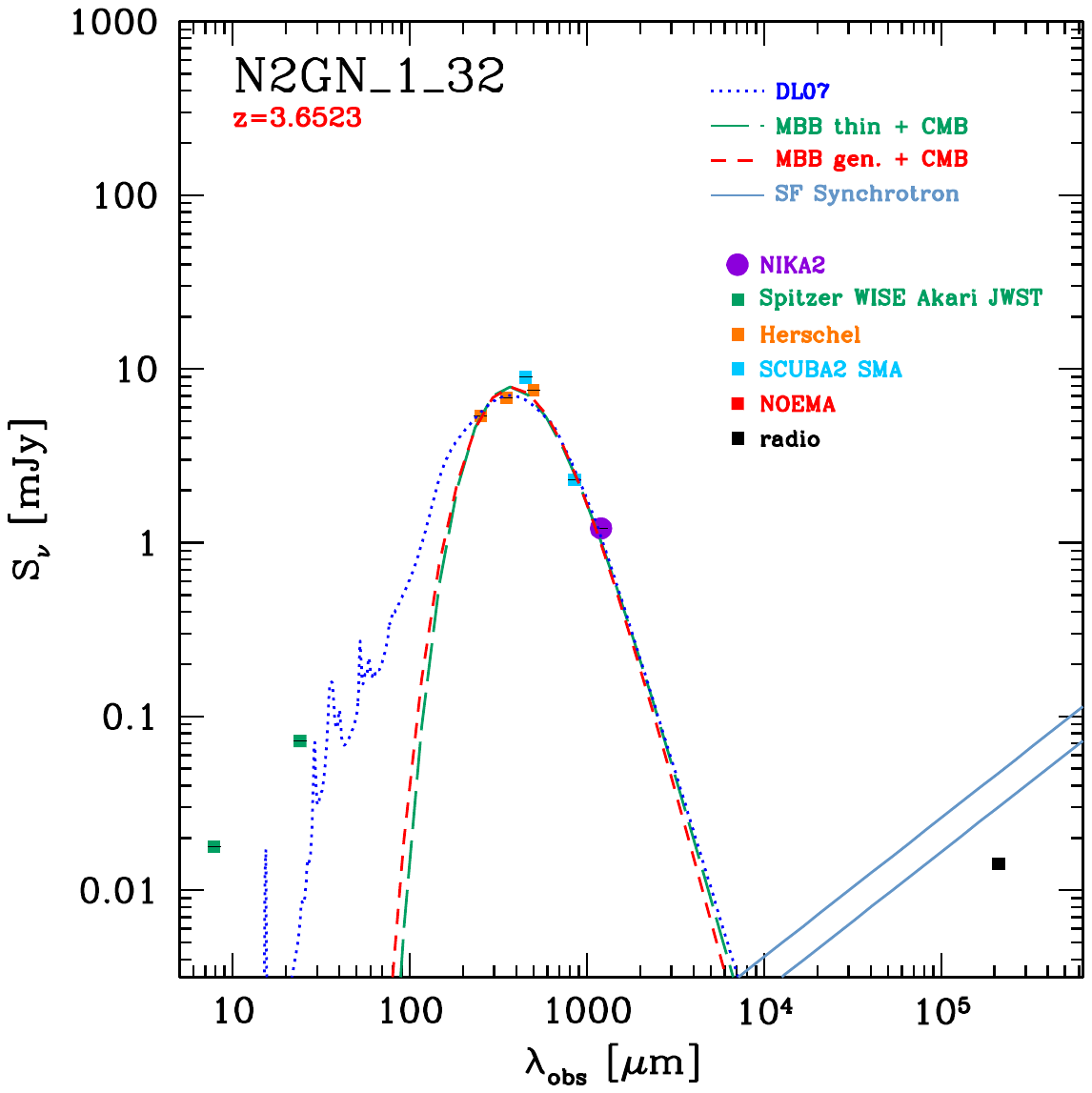}
\includegraphics[align=c,trim=0 0.18\imageheight{} 0 0.075\imageheight{}, clip, width=0.25\textwidth]{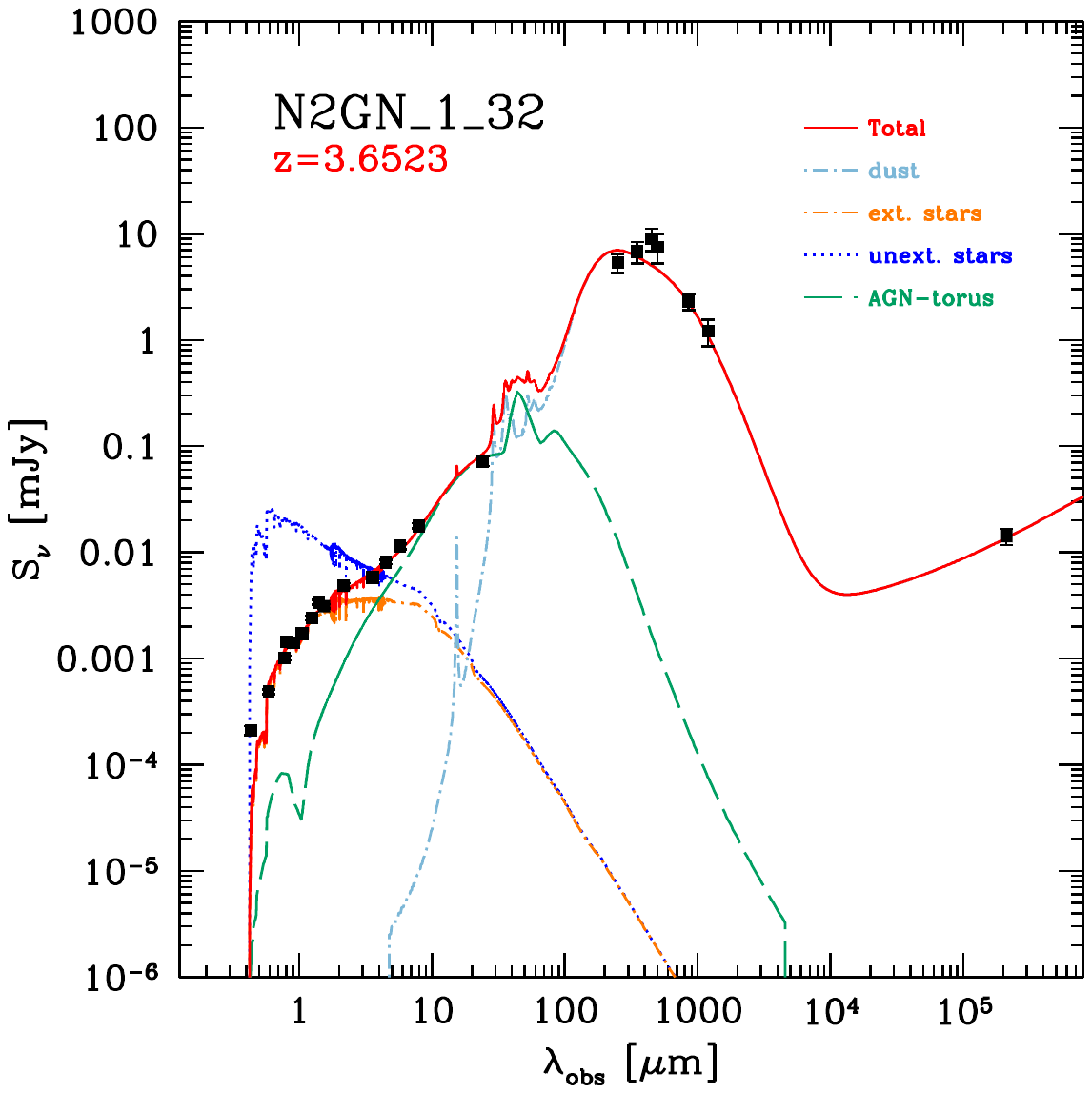}
\includegraphics[align=c,width=0.4\textwidth]{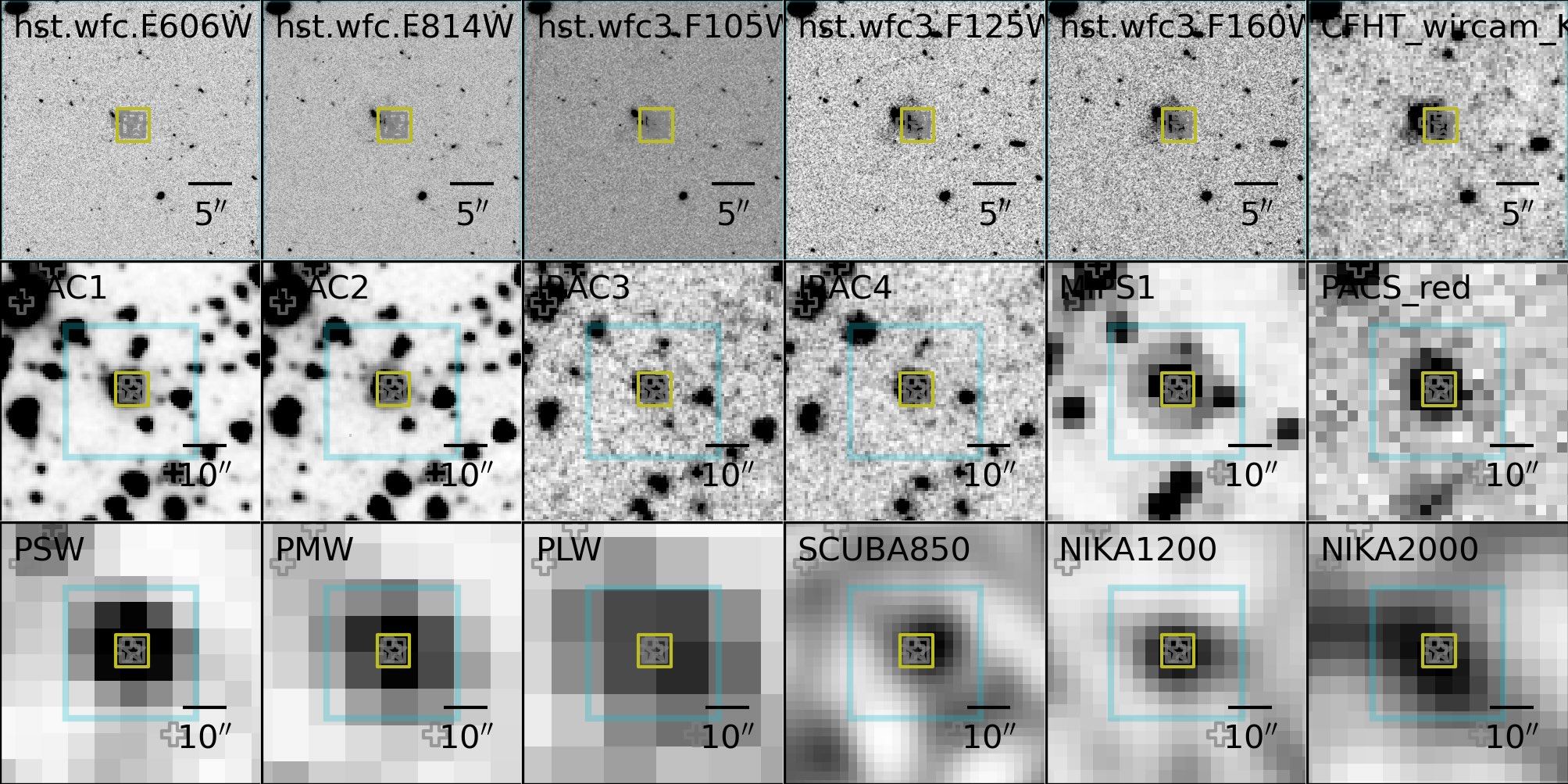}
\includegraphics[align=c,trim=0 0.18\imageheight{} 0 0.075\imageheight{}, clip, width=0.25\textwidth]{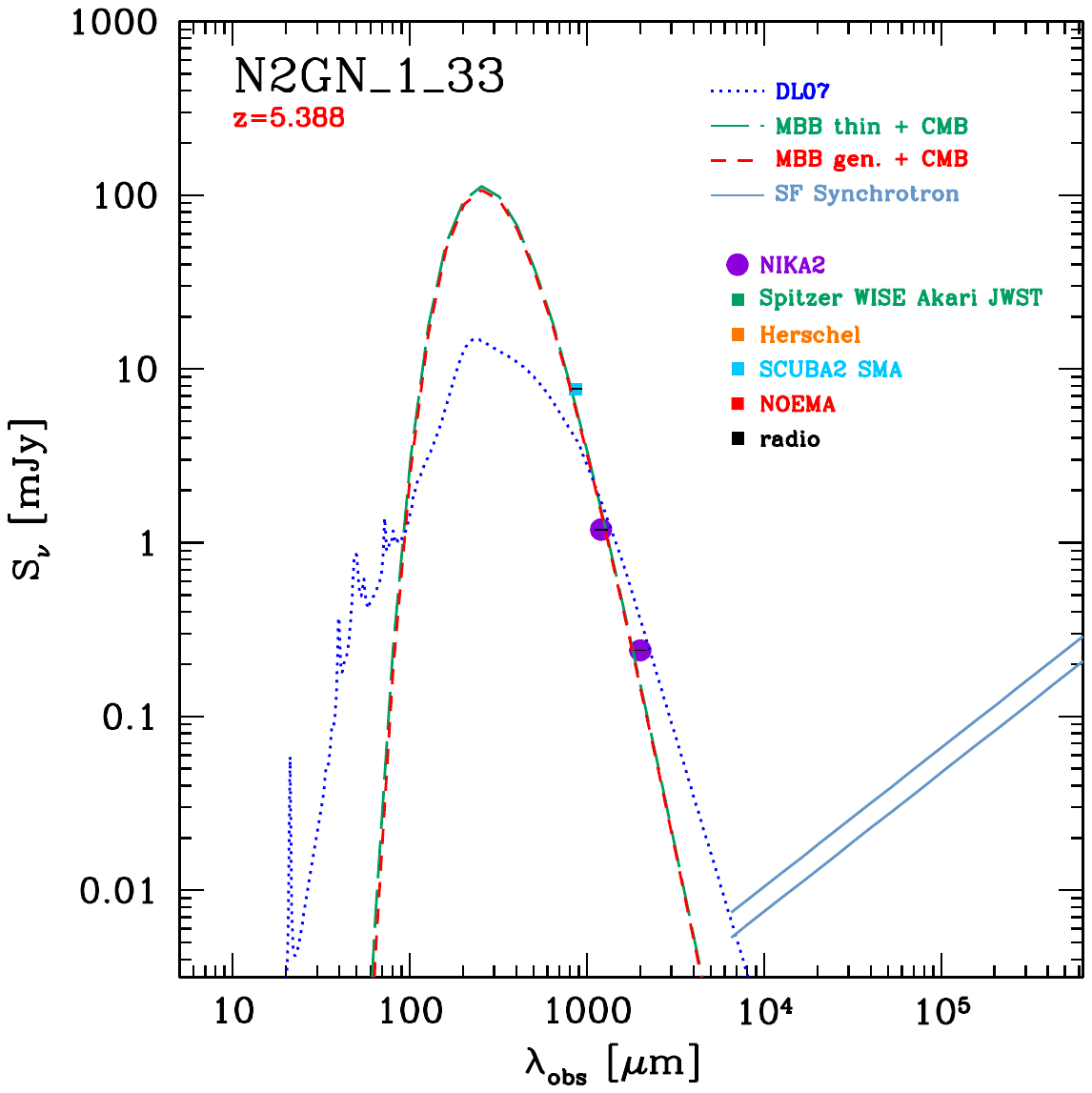}
\includegraphics[align=c,trim=0 0.18\imageheight{} 0 0.075\imageheight{}, clip, width=0.25\textwidth]{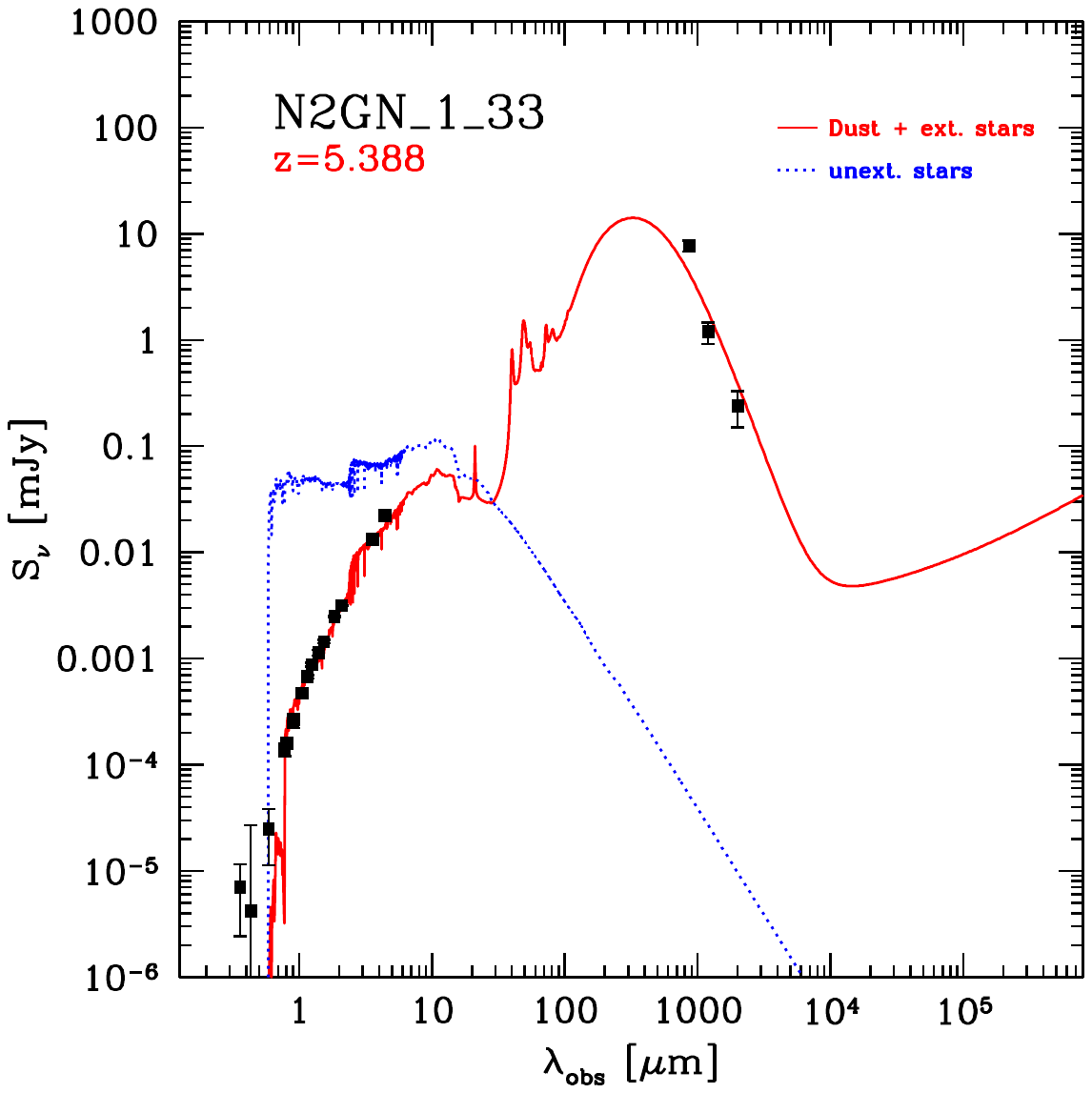}
\includegraphics[align=c,width=0.4\textwidth]{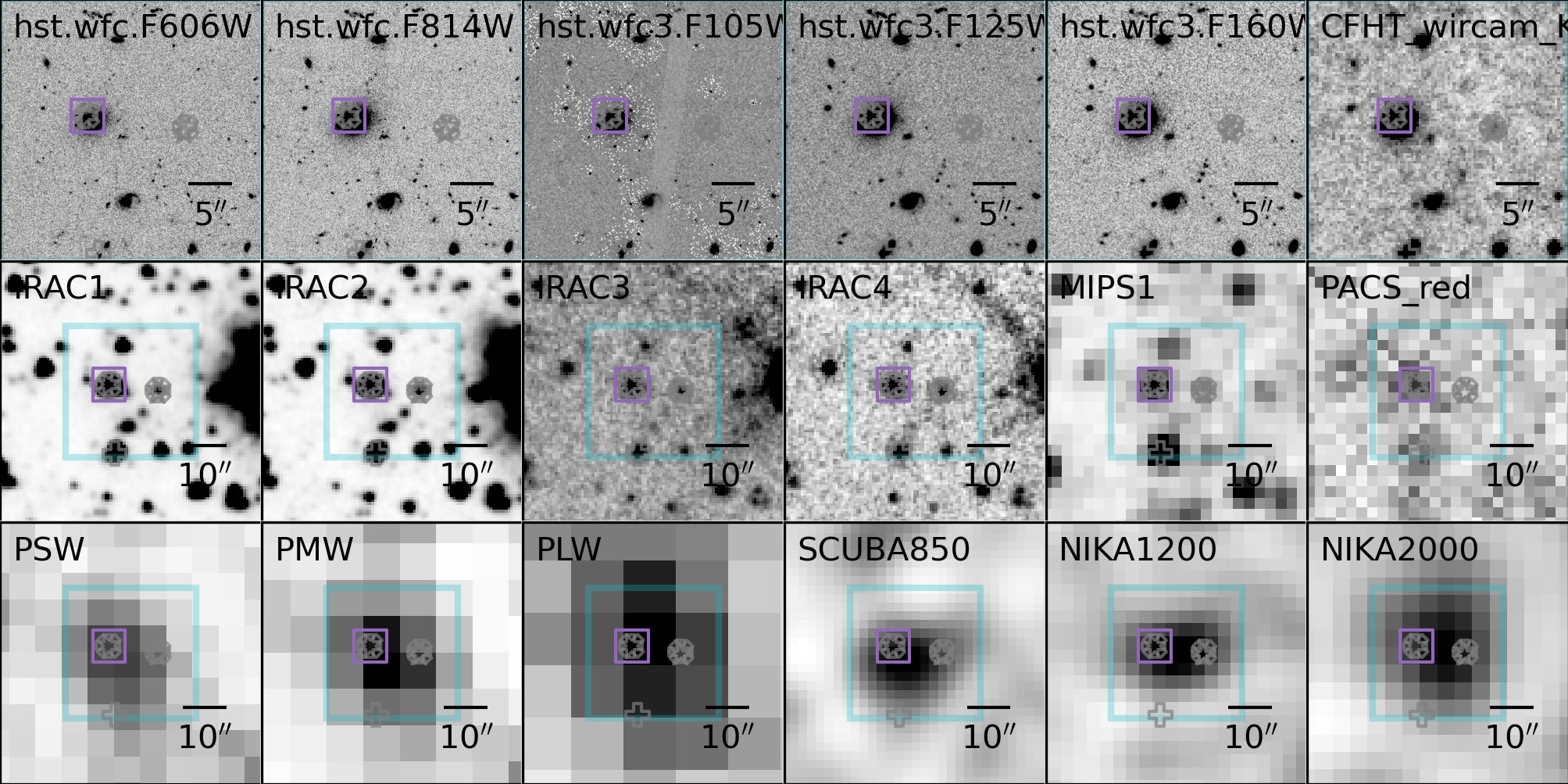}
\includegraphics[align=c,trim=0 0.18\imageheight{} 0 0.075\imageheight{}, clip, width=0.25\textwidth]{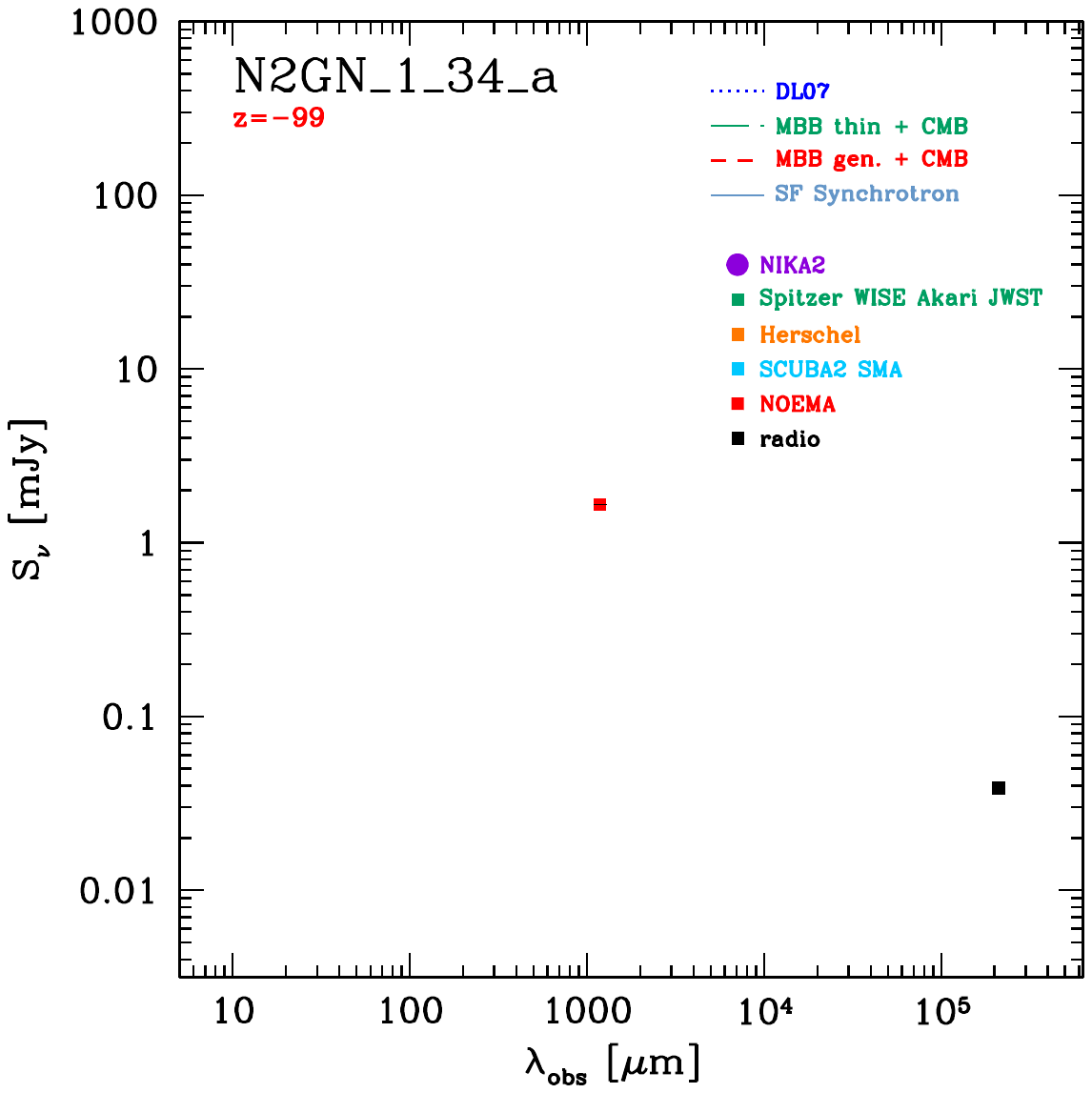}
\includegraphics[align=c,trim=0 0.18\imageheight{} 0 0.075\imageheight{}, clip, width=0.25\textwidth]{figs2_indiv_objs/white.pdf}
\includegraphics[align=c,width=0.4\textwidth]{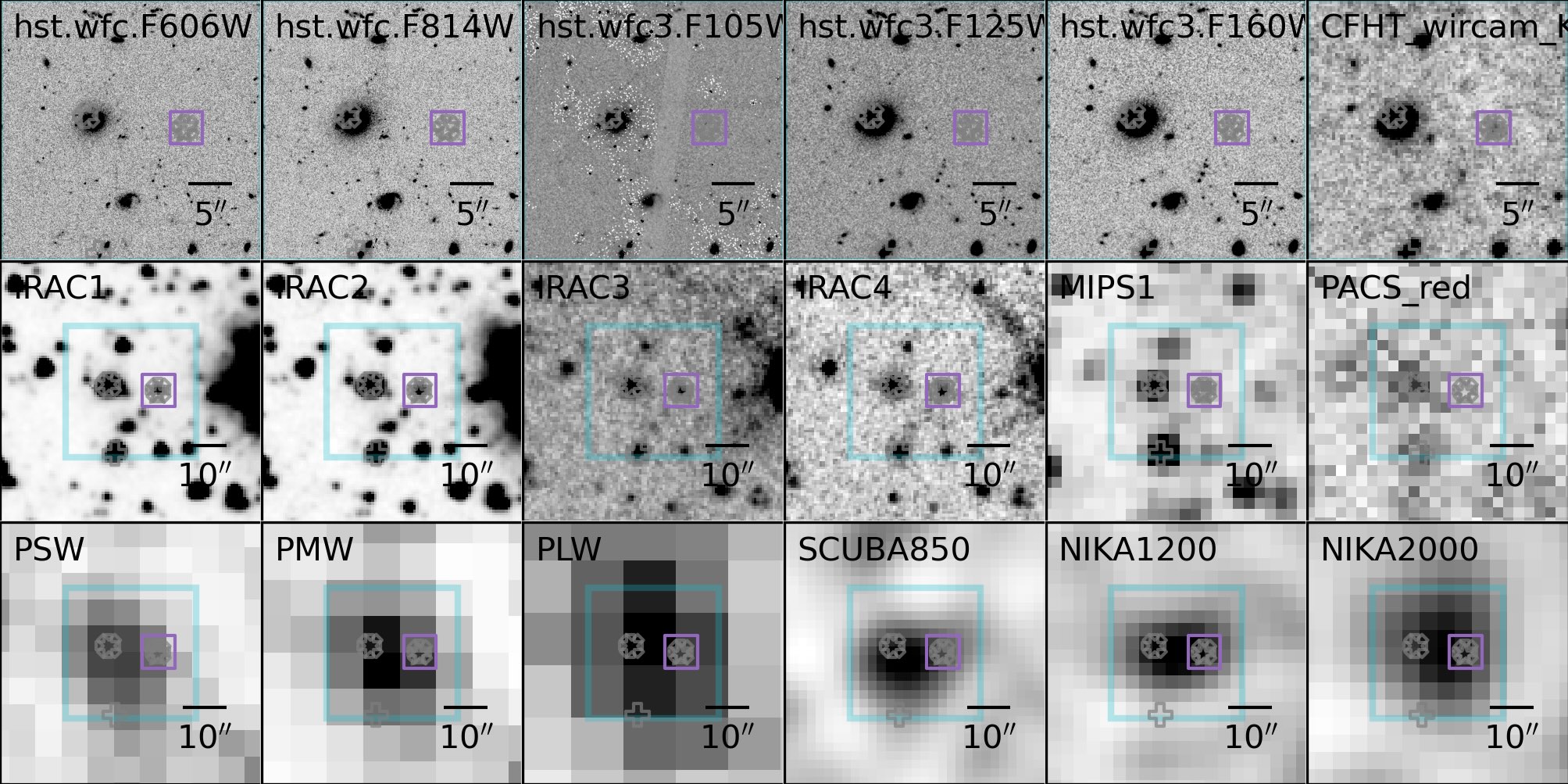}
\includegraphics[align=c,trim=0 0.18\imageheight{} 0 0.075\imageheight{}, clip, width=0.25\textwidth]{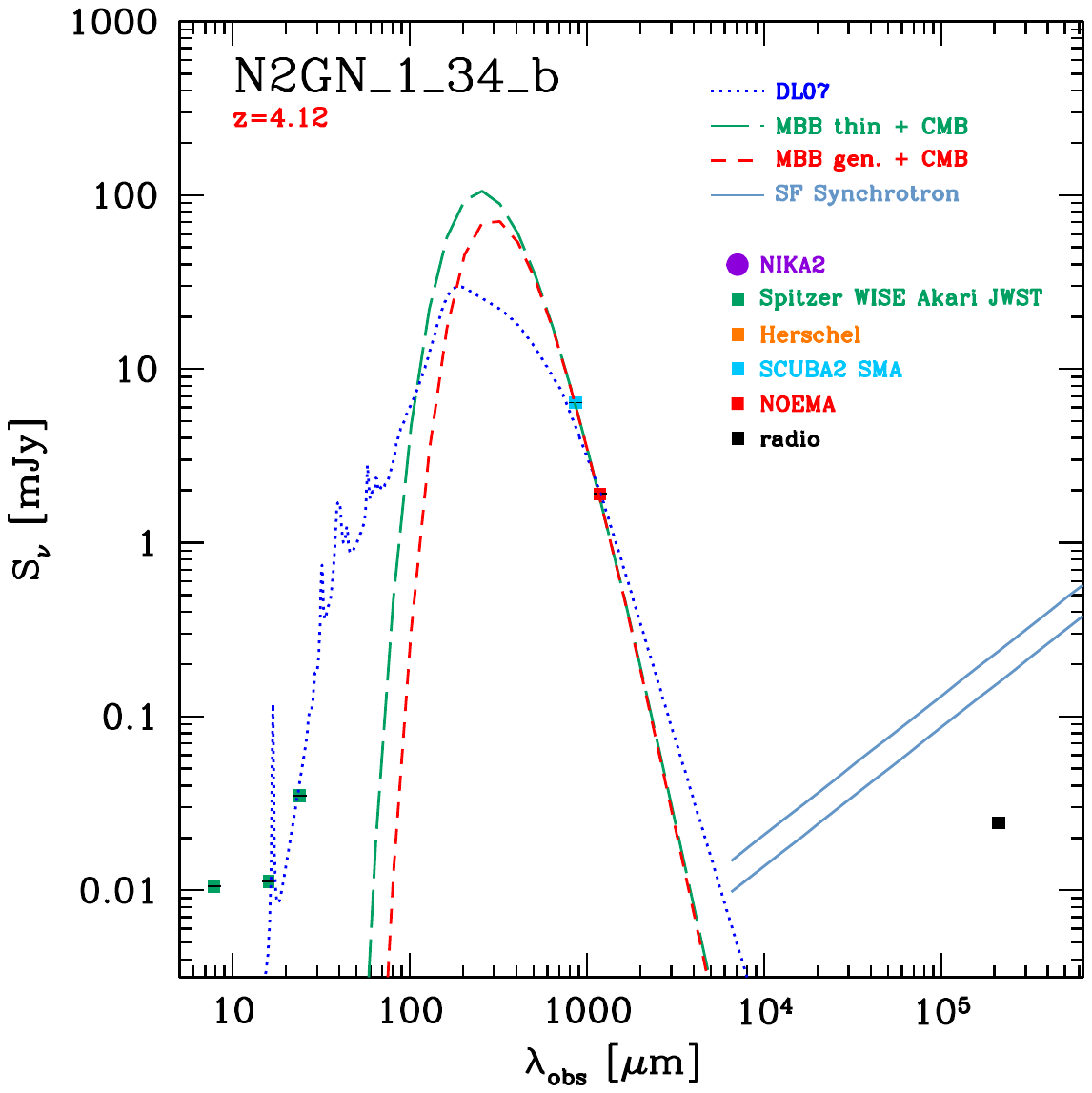}
\includegraphics[align=c,trim=0 0.18\imageheight{} 0 0.075\imageheight{}, clip, width=0.25\textwidth]{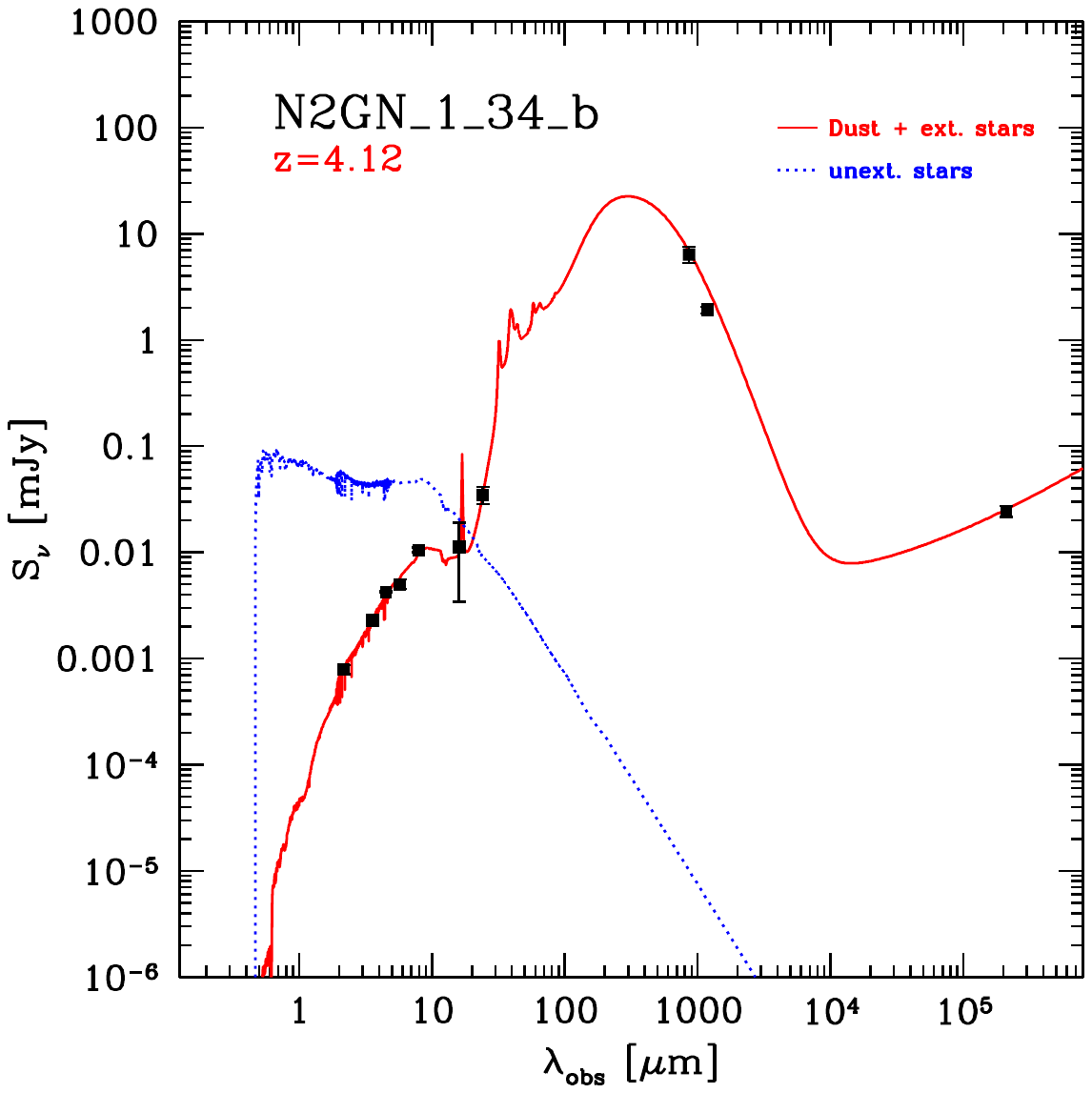}
\caption{continued.}
\end{figure*}

\addtocounter{figure}{-1}
\newpage

\begin{figure*}[t]
\centering
\includegraphics[align=c,width=0.4\textwidth]{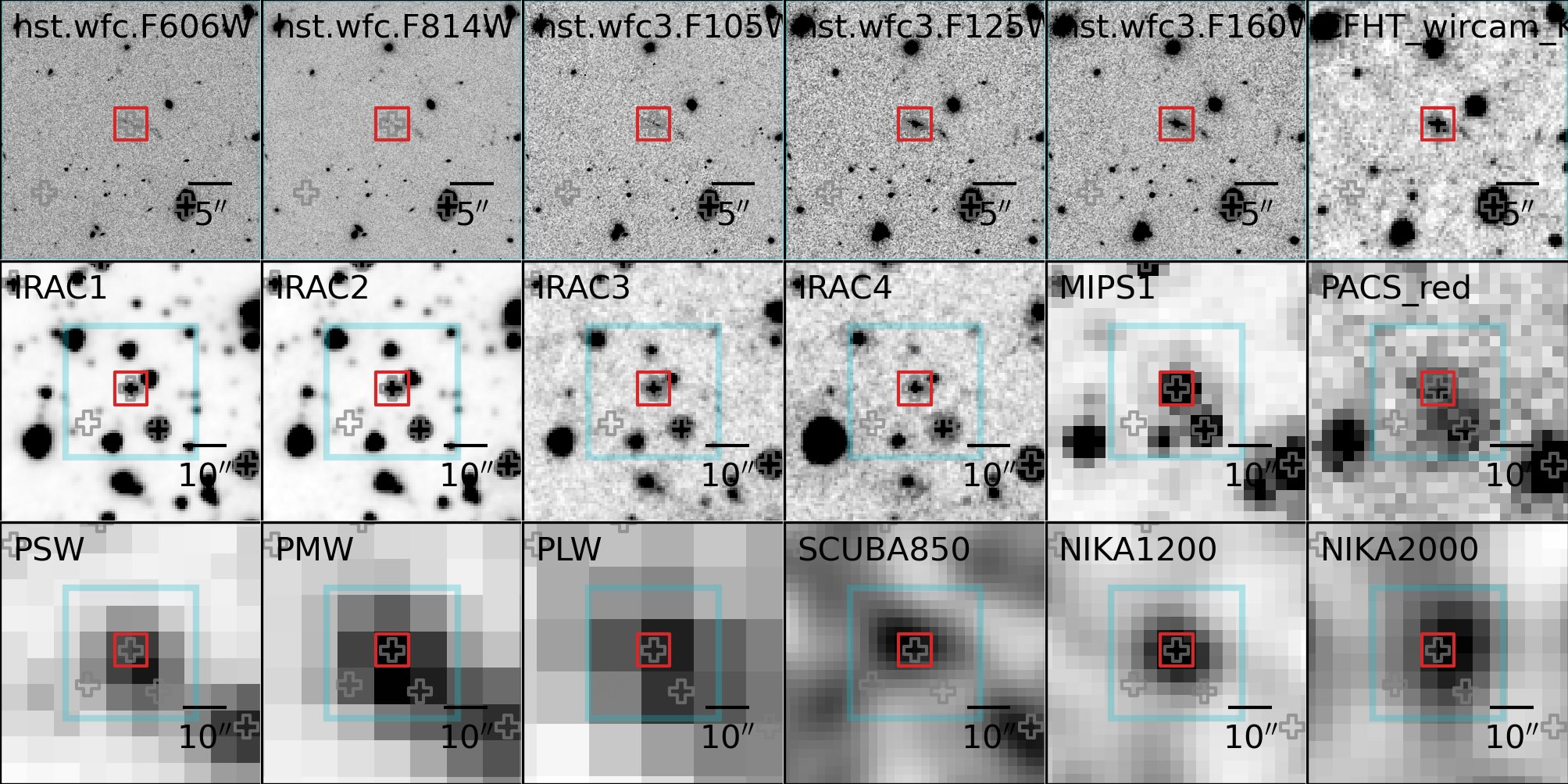}
\includegraphics[align=c,trim=0 0.18\imageheight{} 0 0.075\imageheight{}, clip, width=0.25\textwidth]{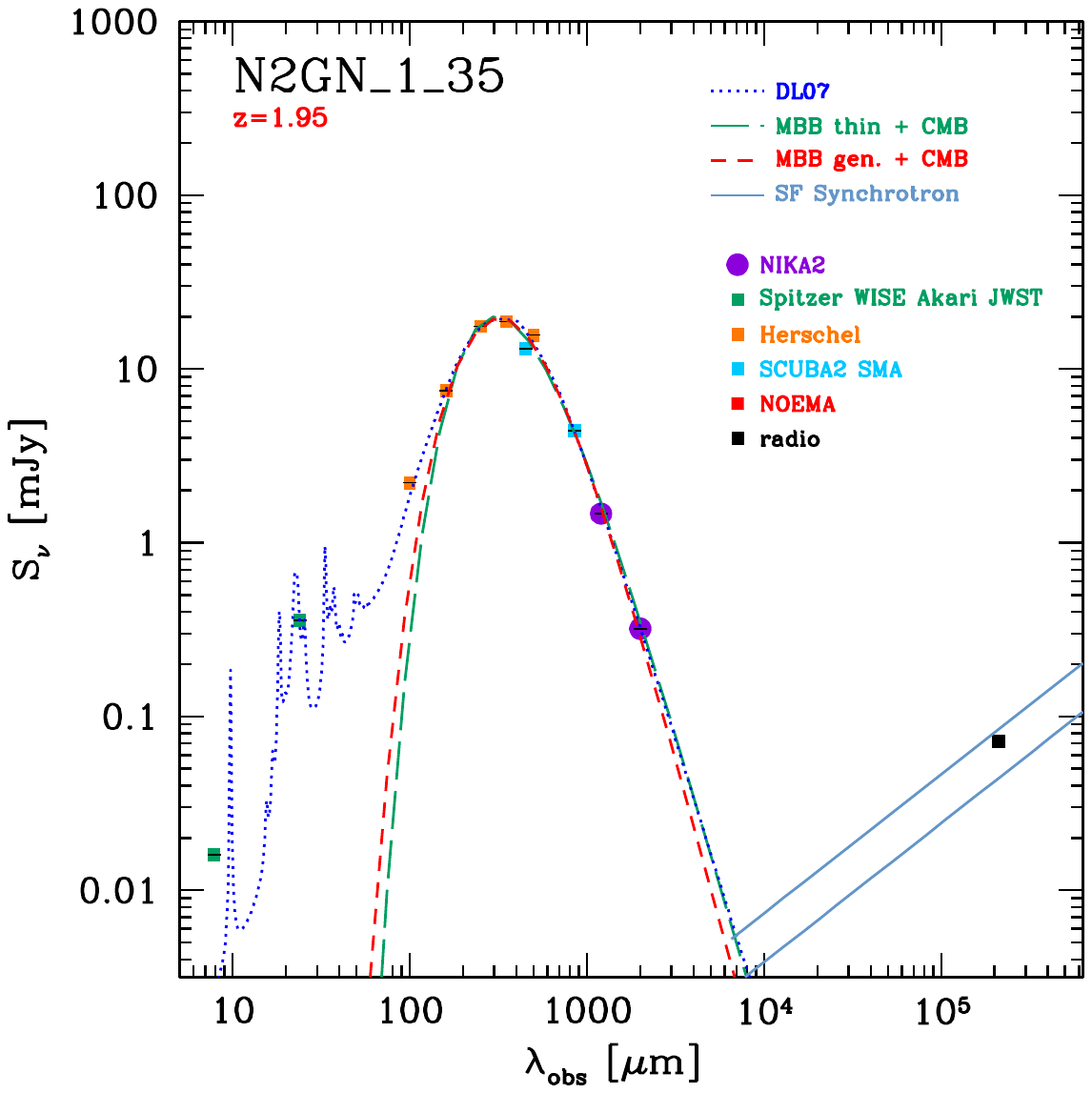}
\includegraphics[align=c,trim=0 0.18\imageheight{} 0 0.075\imageheight{}, clip, width=0.25\textwidth]{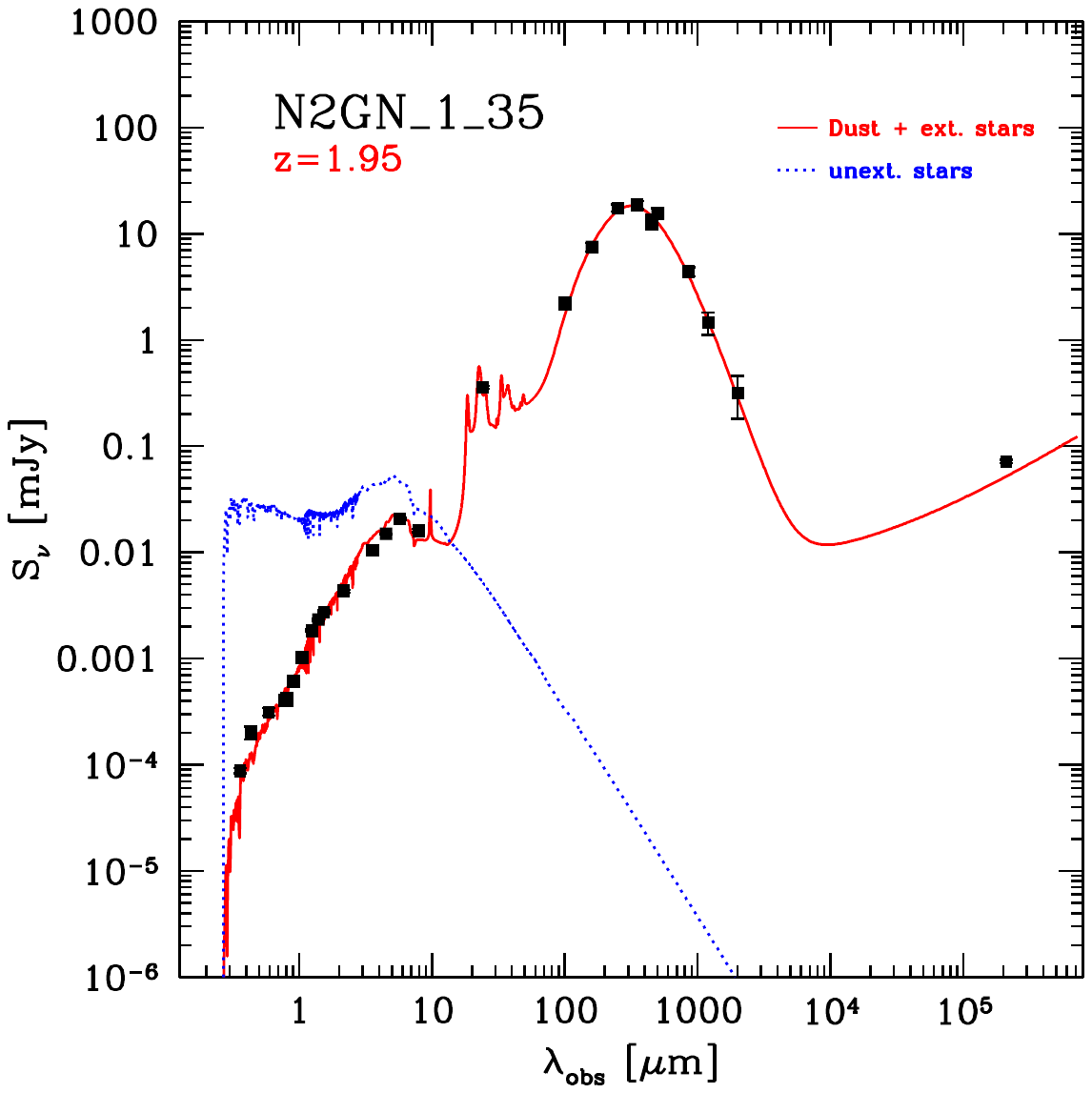}
\includegraphics[align=c,width=0.4\textwidth]{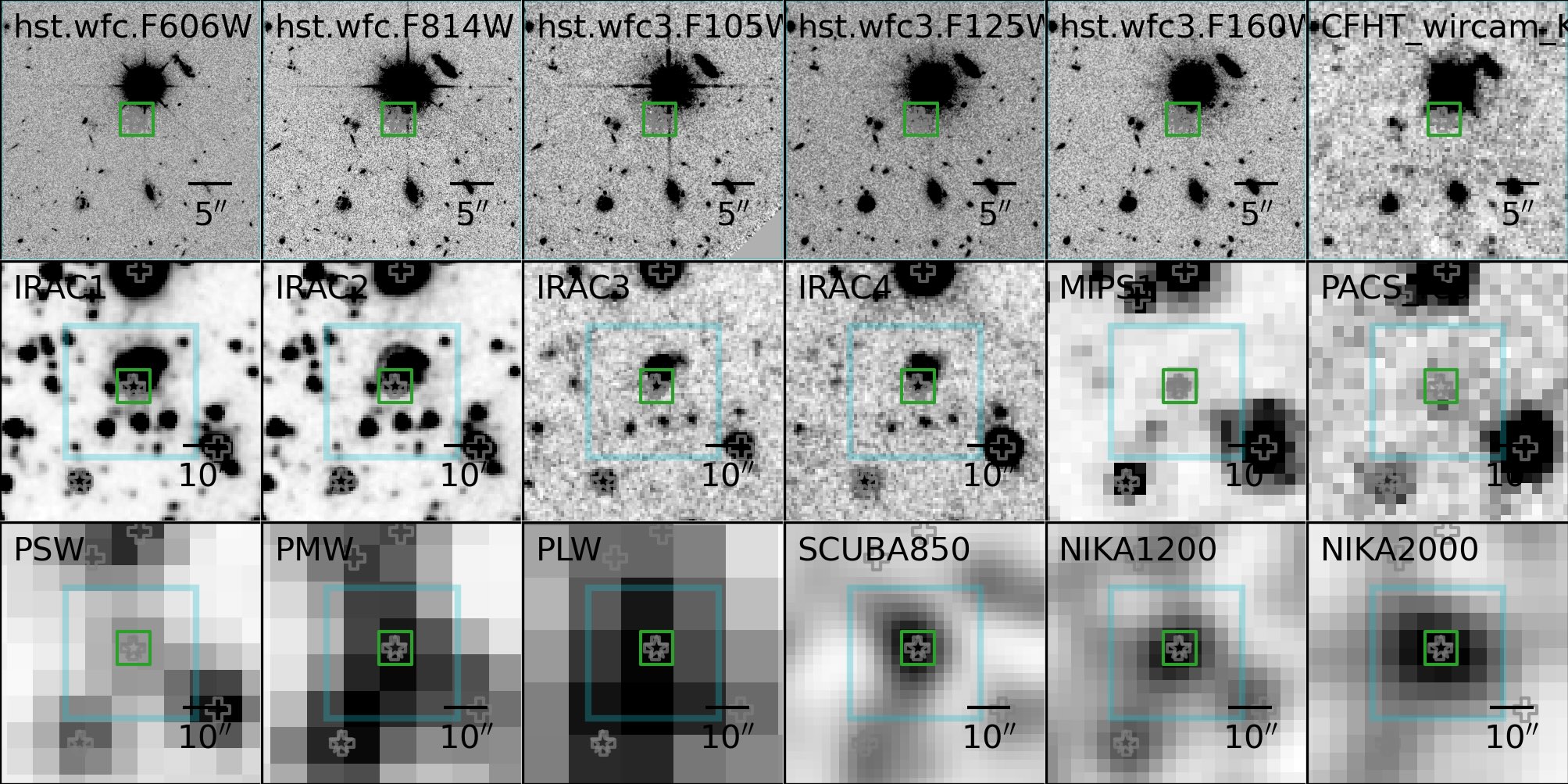}
\includegraphics[align=c,trim=0 0.18\imageheight{} 0 0.075\imageheight{}, clip, width=0.25\textwidth]{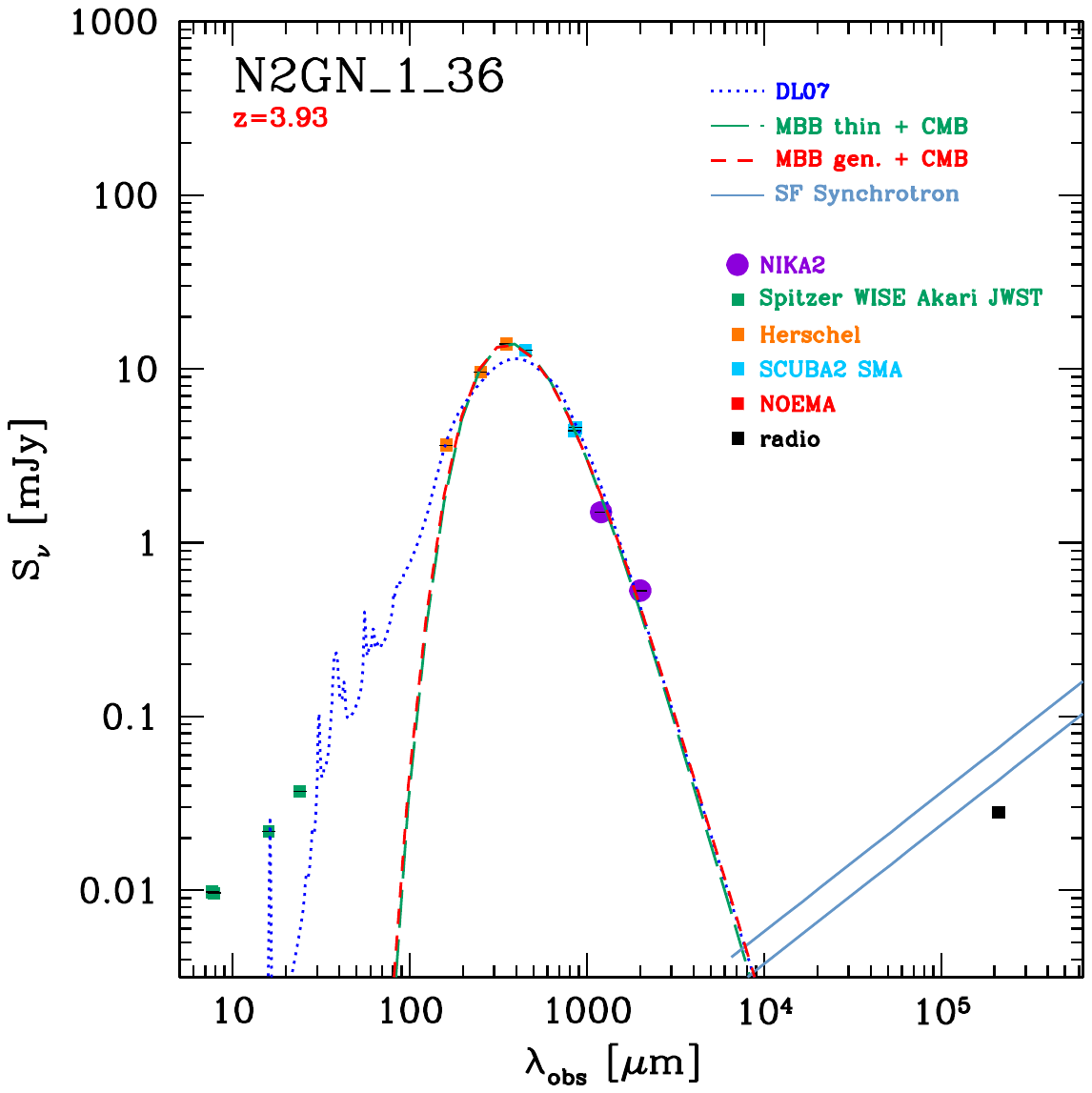}
\includegraphics[align=c,trim=0 0.18\imageheight{} 0 0.075\imageheight{}, clip, width=0.25\textwidth]{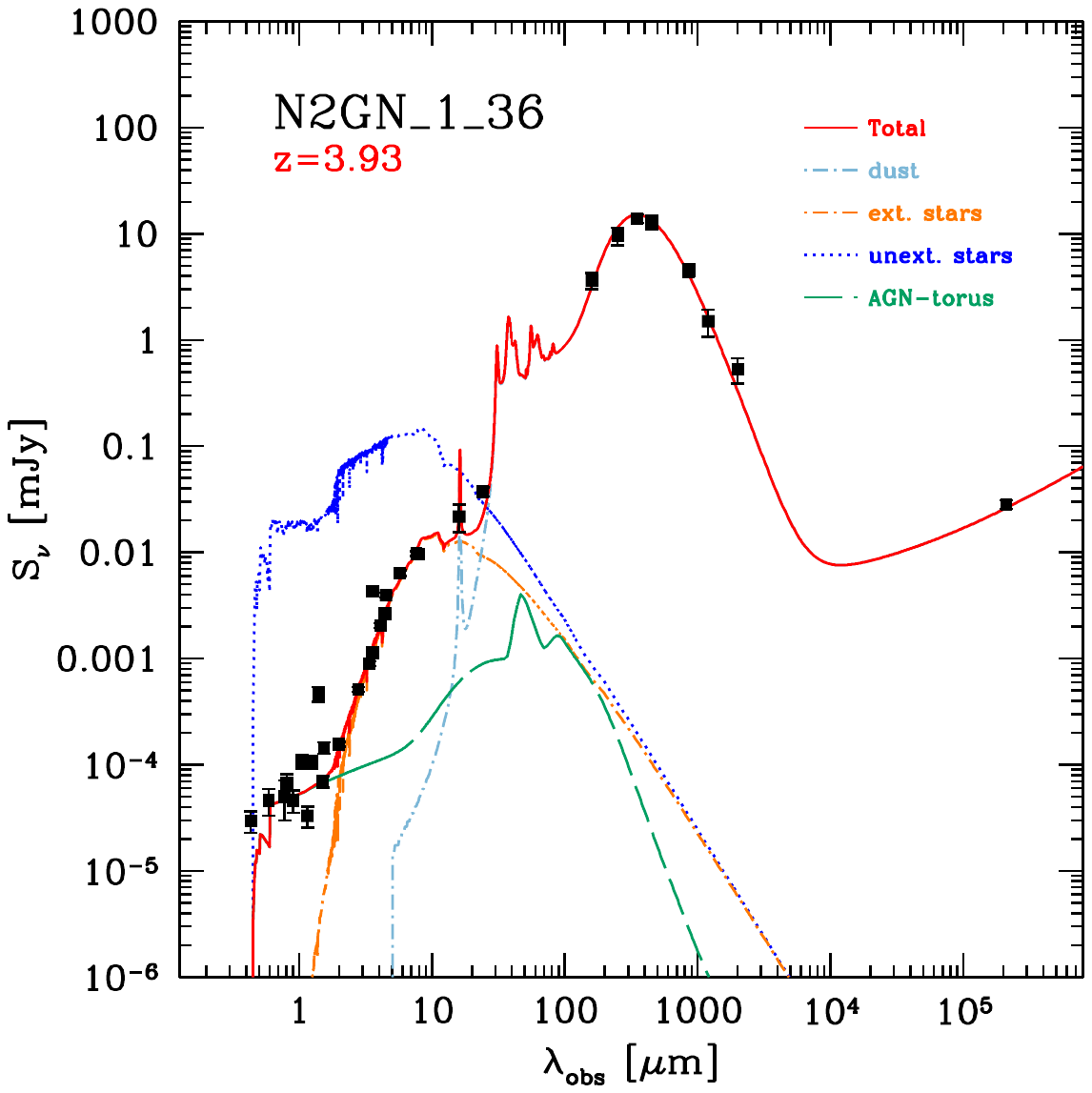}
\includegraphics[align=c,width=0.4\textwidth]{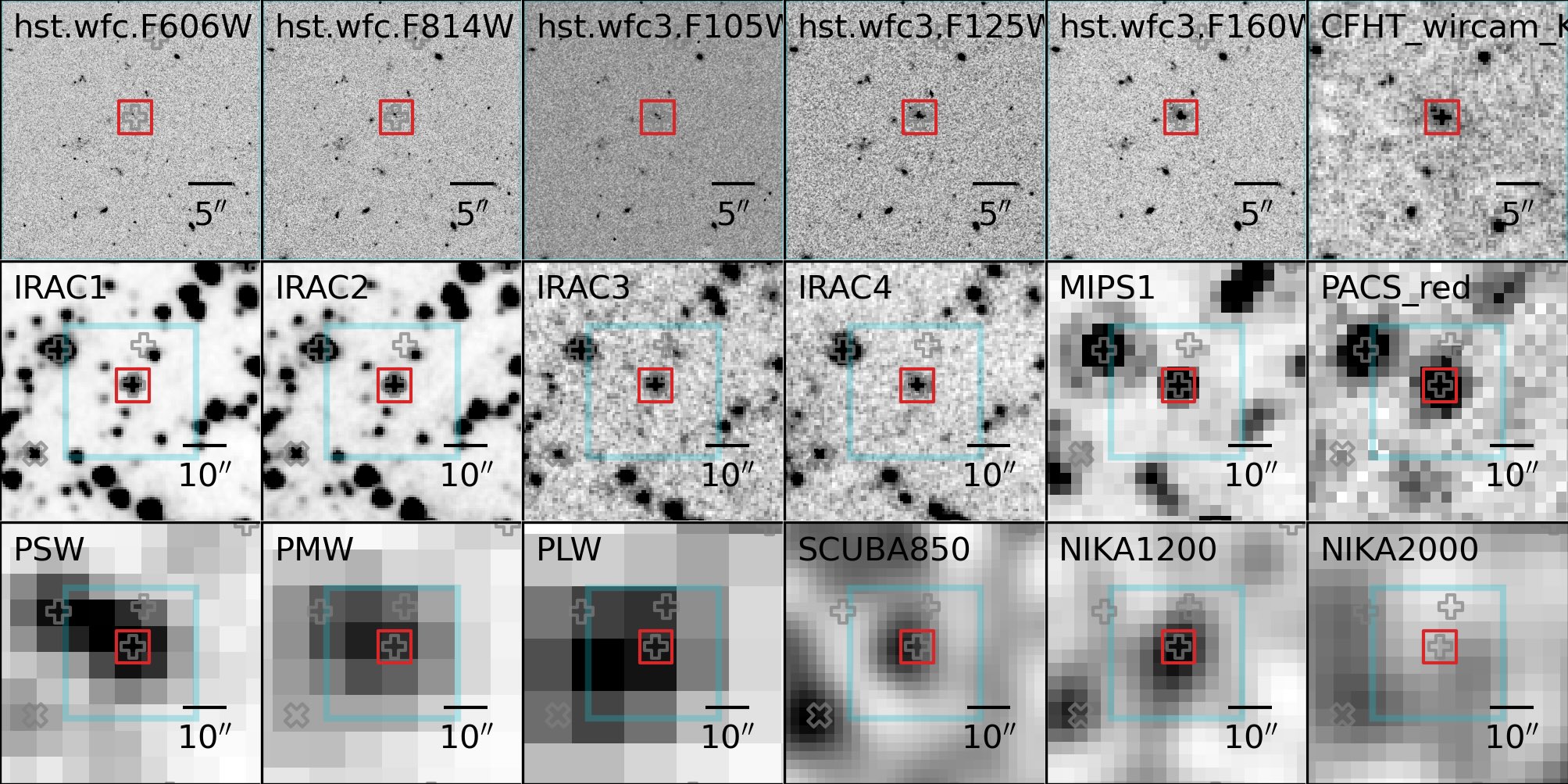}
\includegraphics[align=c,trim=0 0.18\imageheight{} 0 0.075\imageheight{}, clip, width=0.25\textwidth]{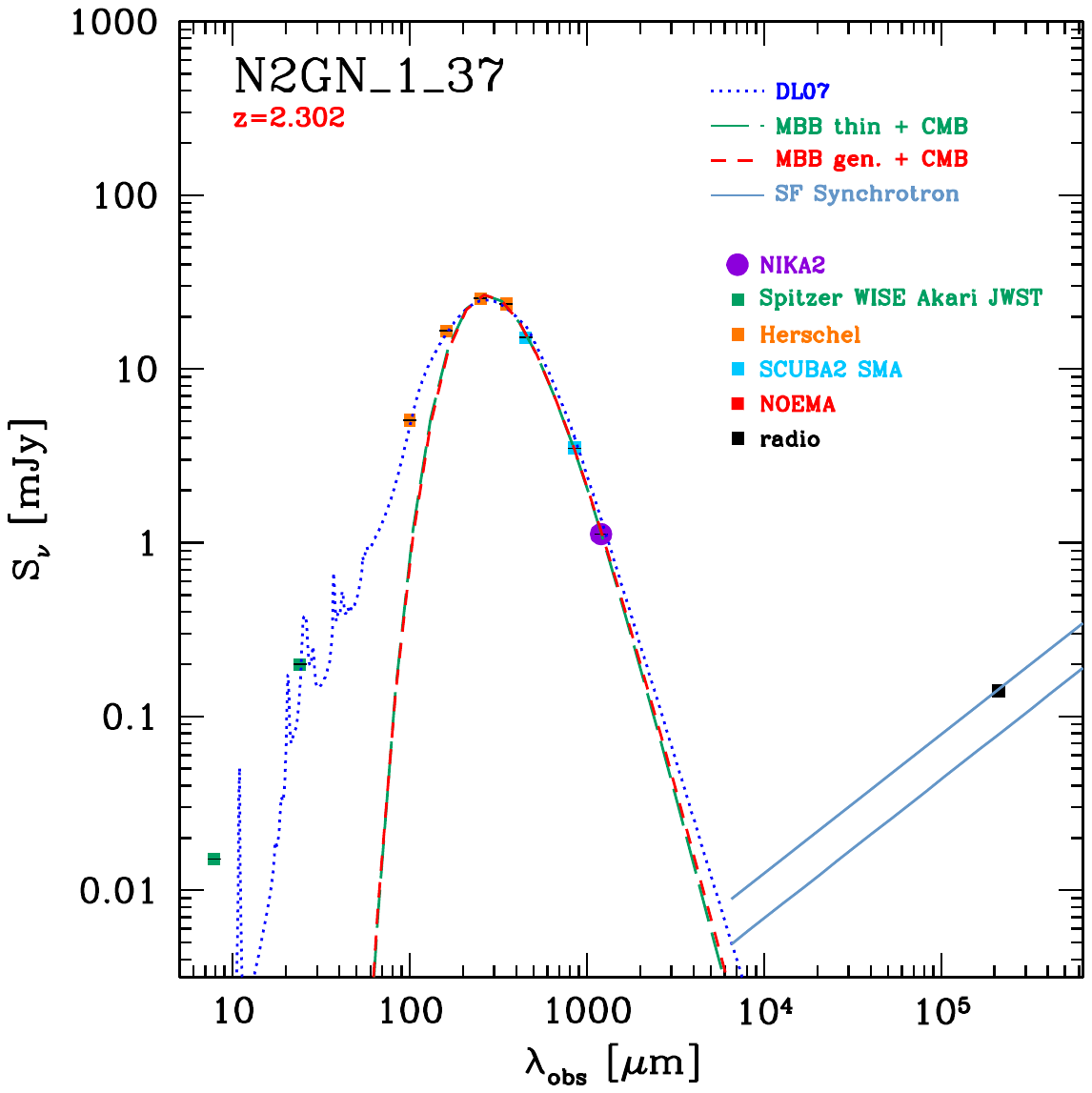}
\includegraphics[align=c,trim=0 0.18\imageheight{} 0 0.075\imageheight{}, clip, width=0.25\textwidth]{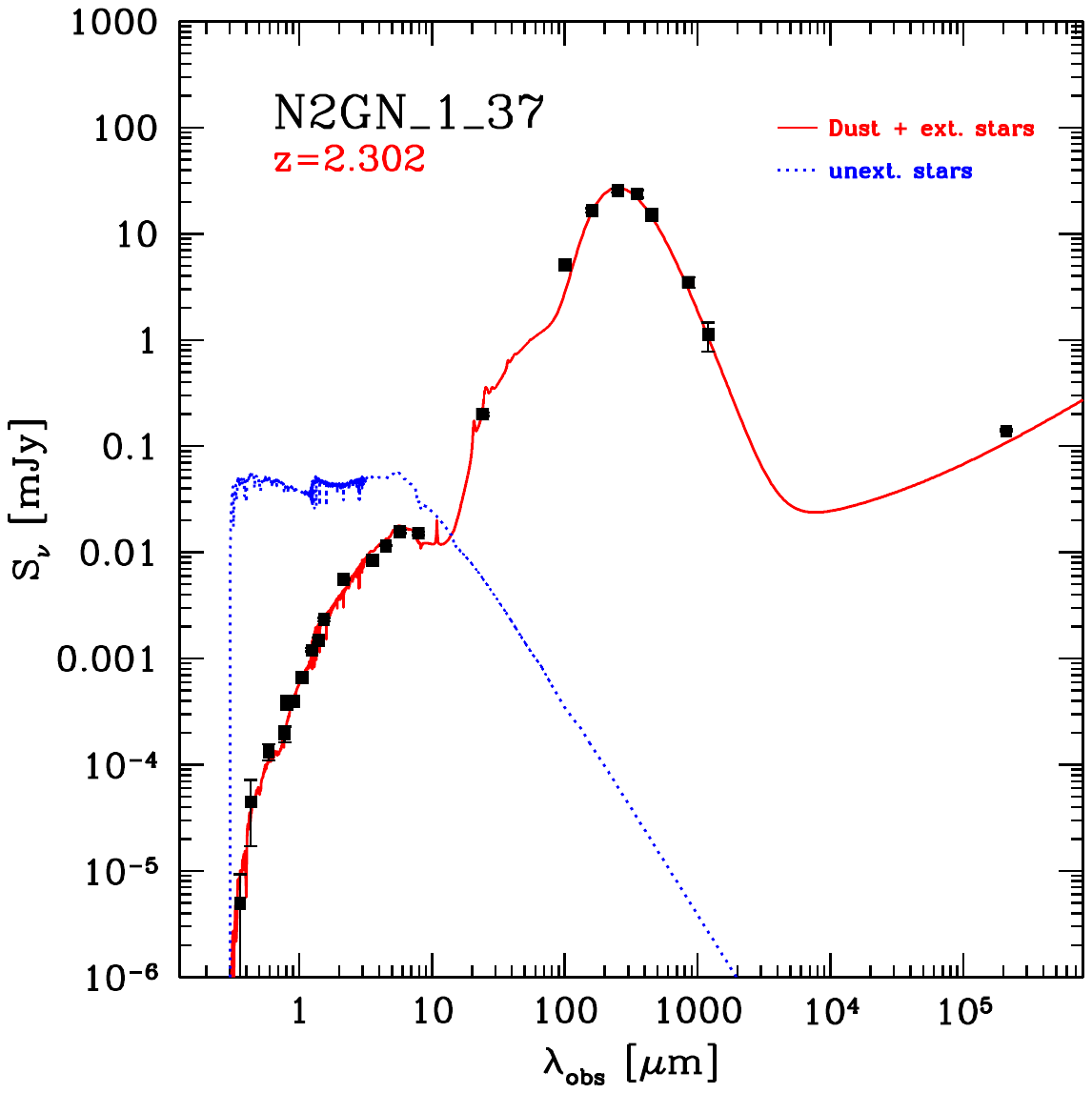}
\includegraphics[align=c,width=0.4\textwidth]{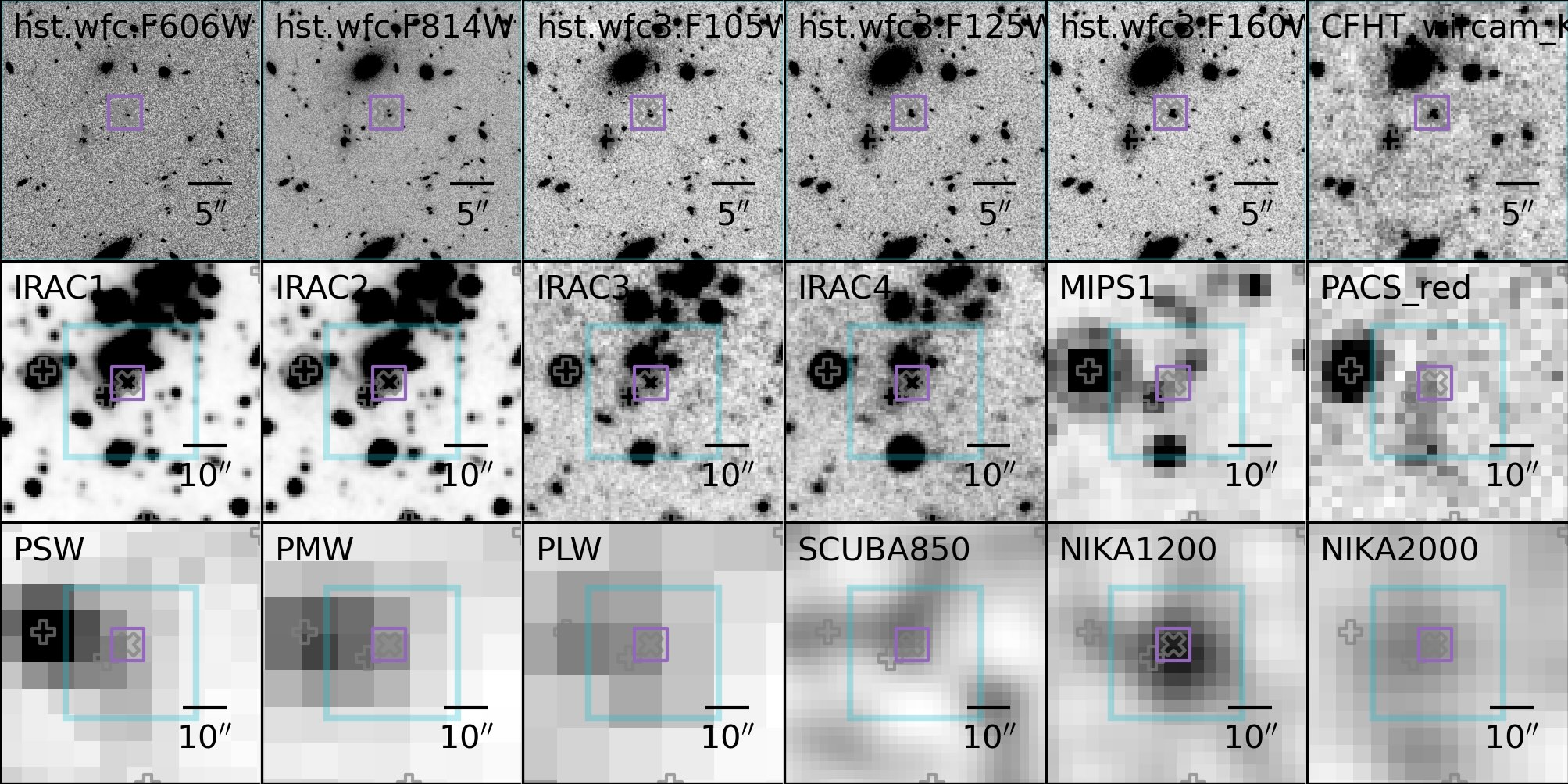}
\includegraphics[align=c,trim=0 0.18\imageheight{} 0 0.075\imageheight{}, clip, width=0.25\textwidth]{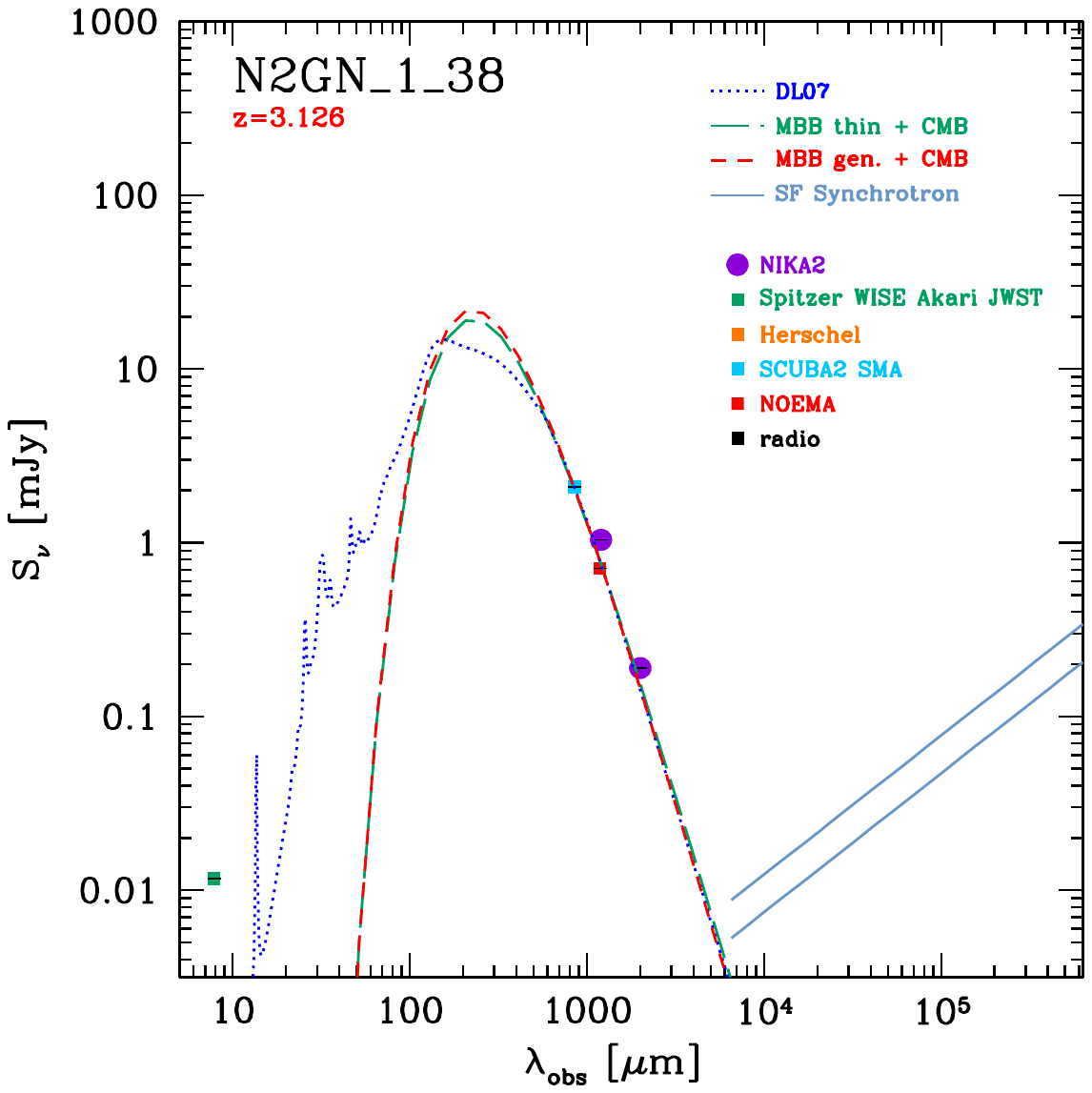}
\includegraphics[align=c,trim=0 0.18\imageheight{} 0 0.075\imageheight{}, clip, width=0.25\textwidth]{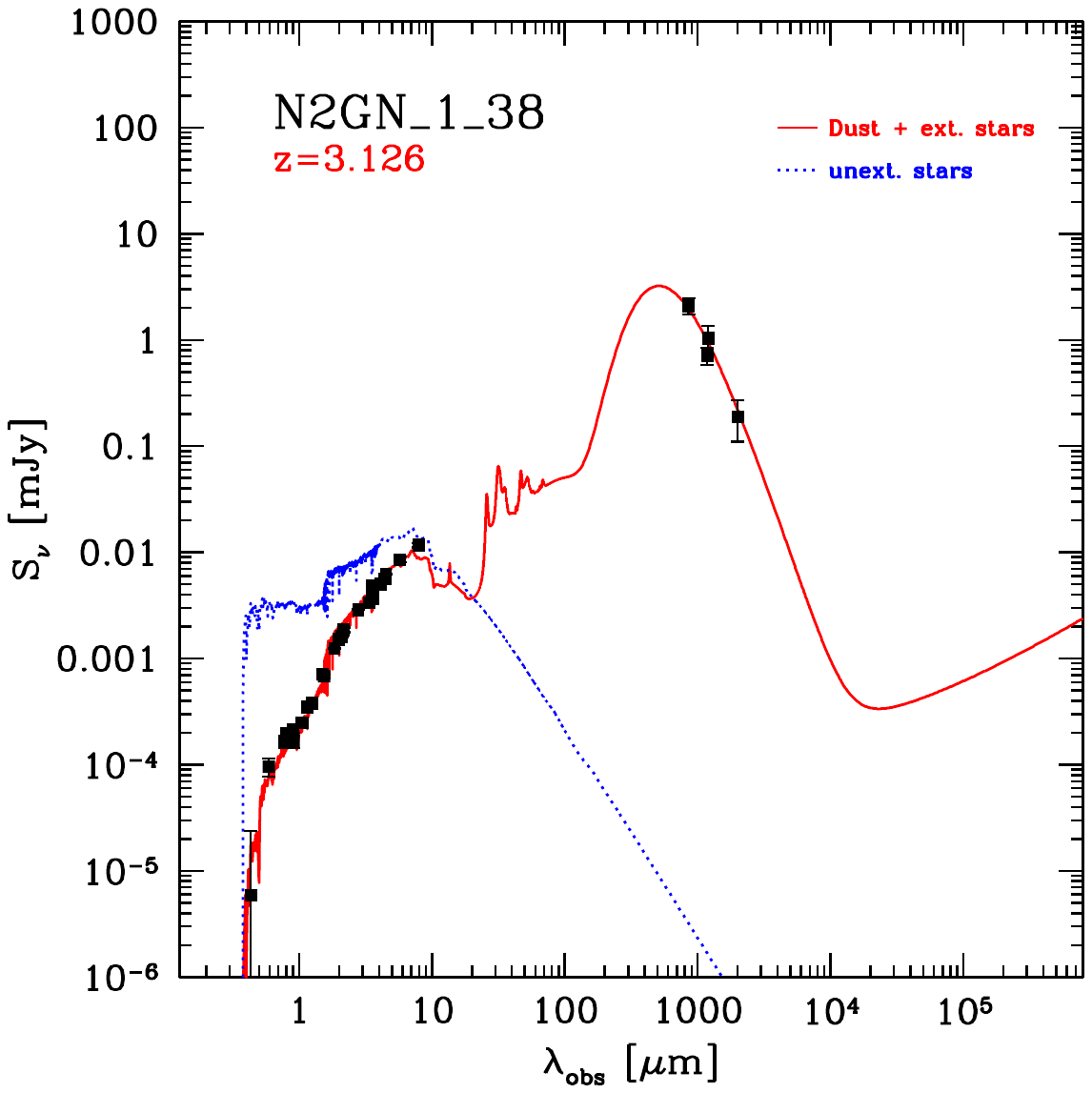}
\includegraphics[align=c,width=0.4\textwidth]{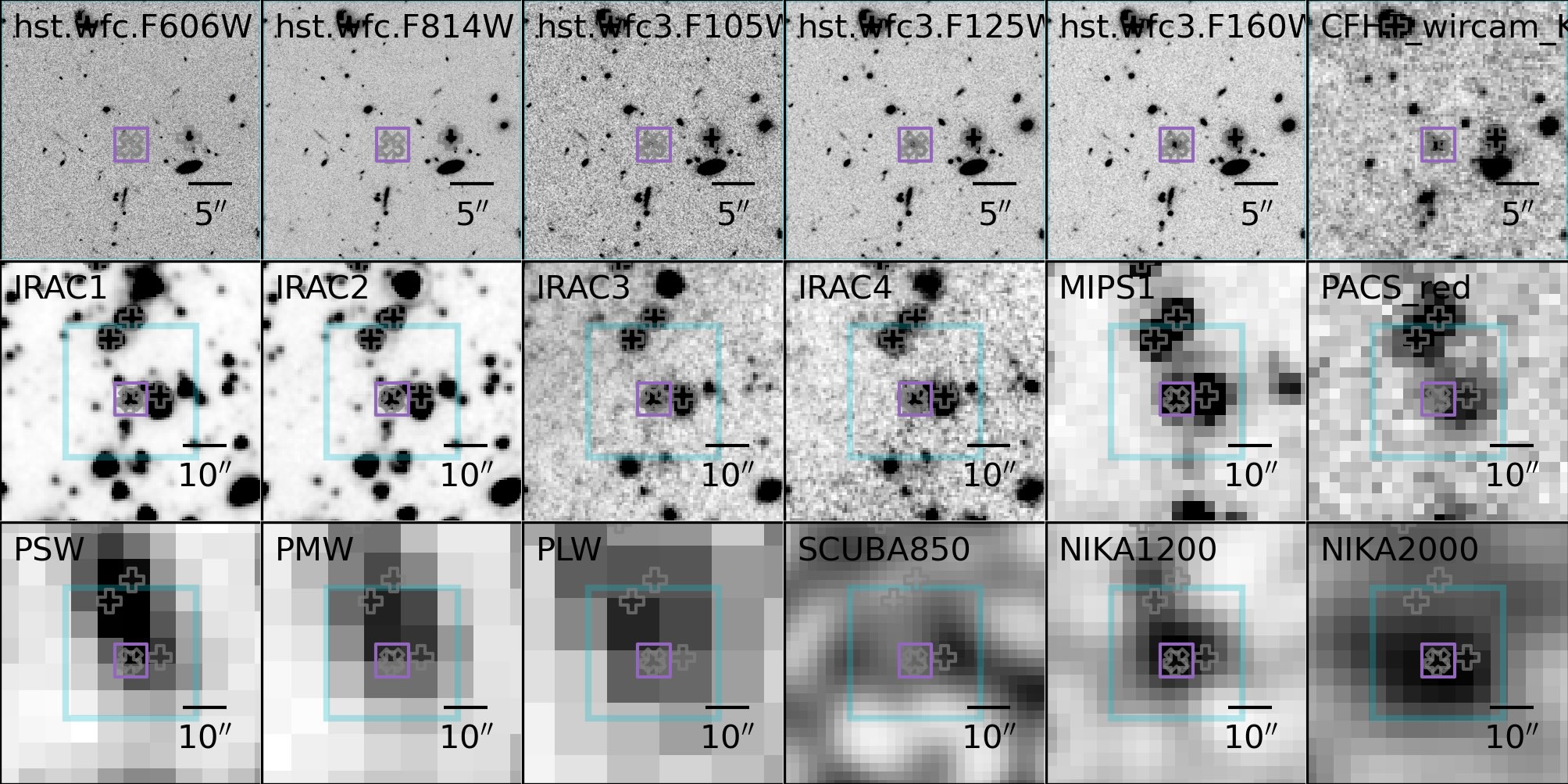}
\includegraphics[align=c,trim=0 0.18\imageheight{} 0 0.075\imageheight{}, clip, width=0.25\textwidth]{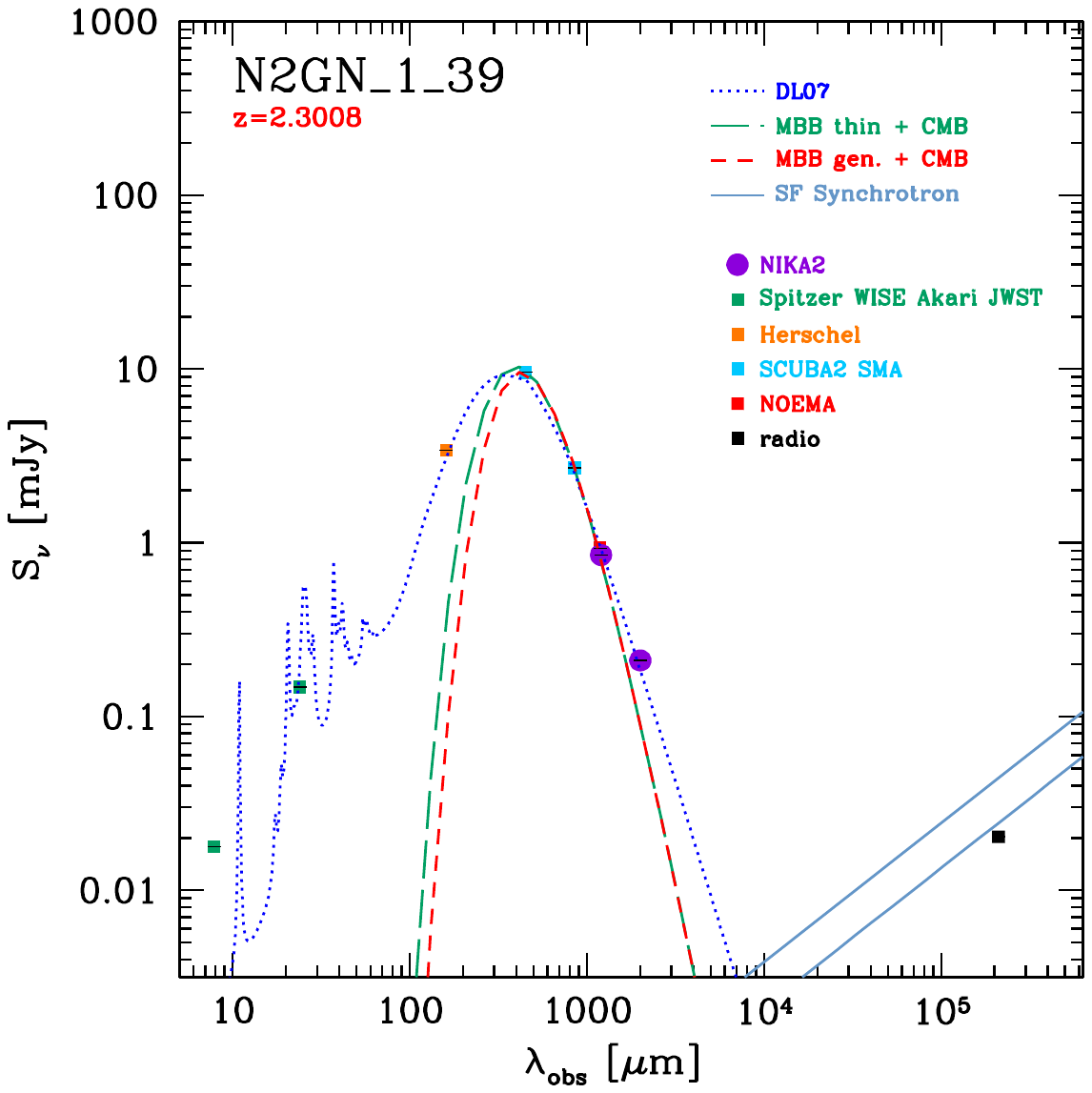}
\includegraphics[align=c,trim=0 0.18\imageheight{} 0 0.075\imageheight{}, clip, width=0.25\textwidth]{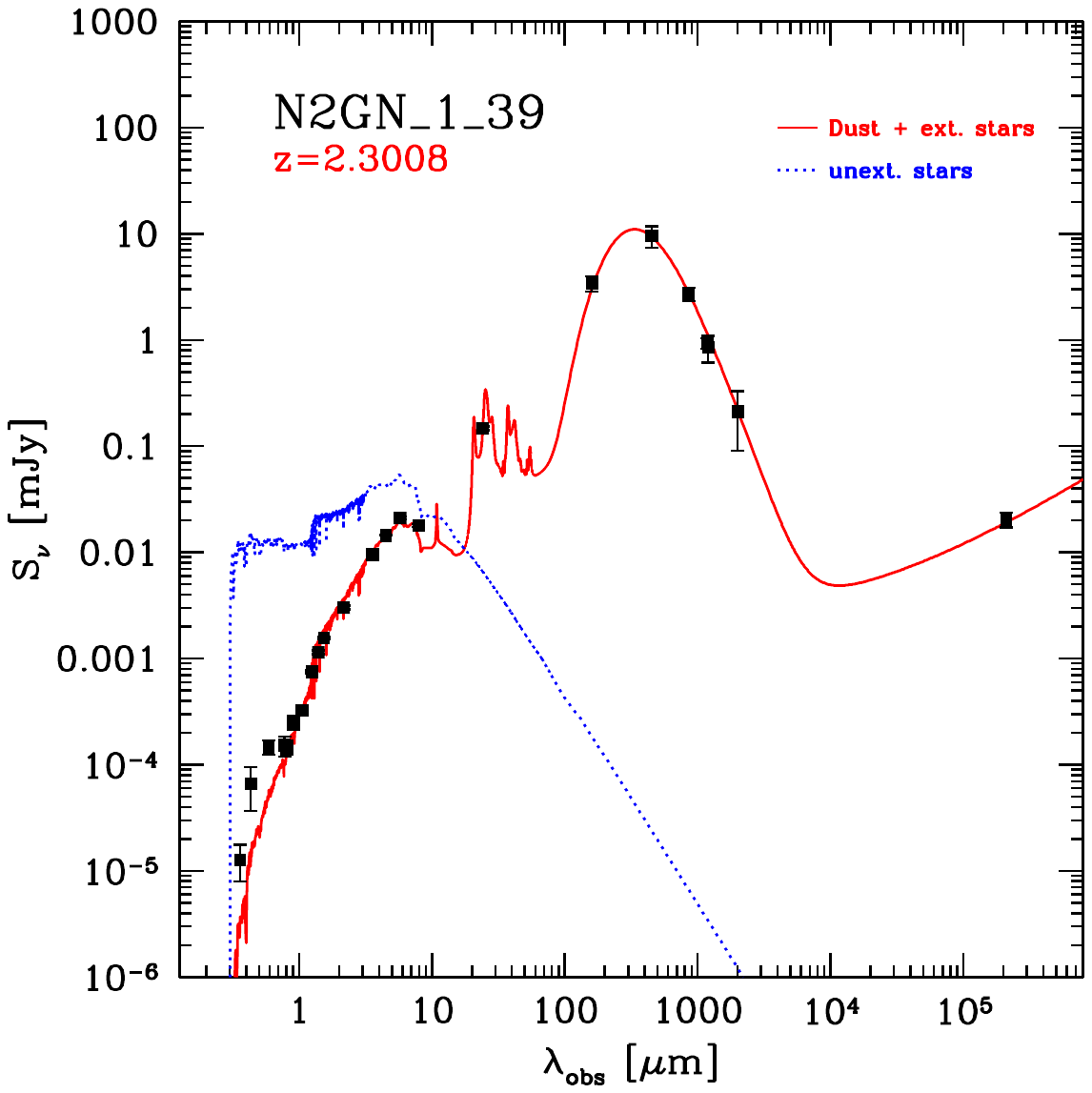}
\caption{continued.}
\end{figure*}

\addtocounter{figure}{-1}
\newpage

\begin{figure*}[t]
\centering
\includegraphics[align=c,width=0.4\textwidth]{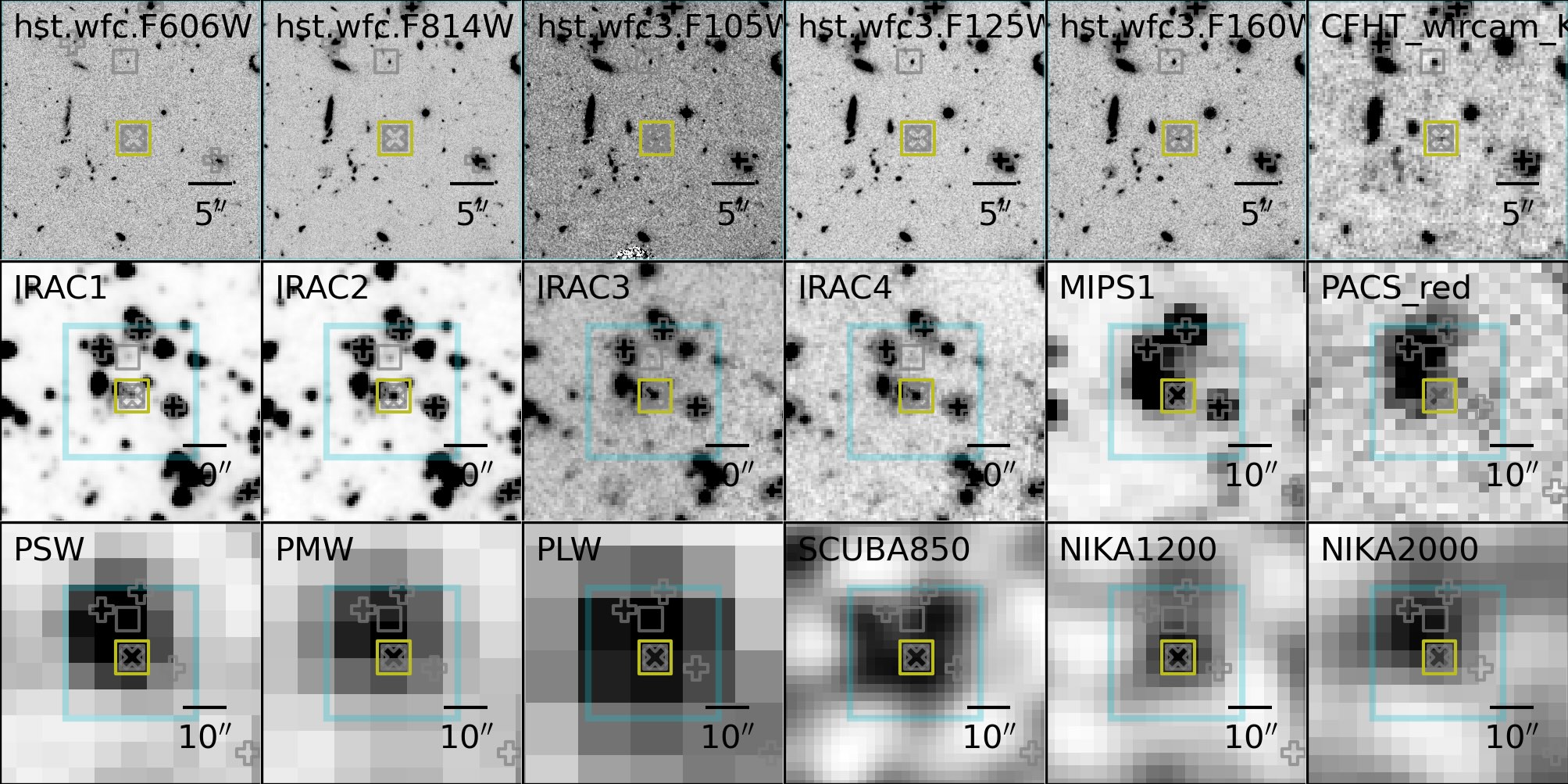}
\includegraphics[align=c,trim=0 0.18\imageheight{} 0 0.075\imageheight{}, clip, width=0.25\textwidth]{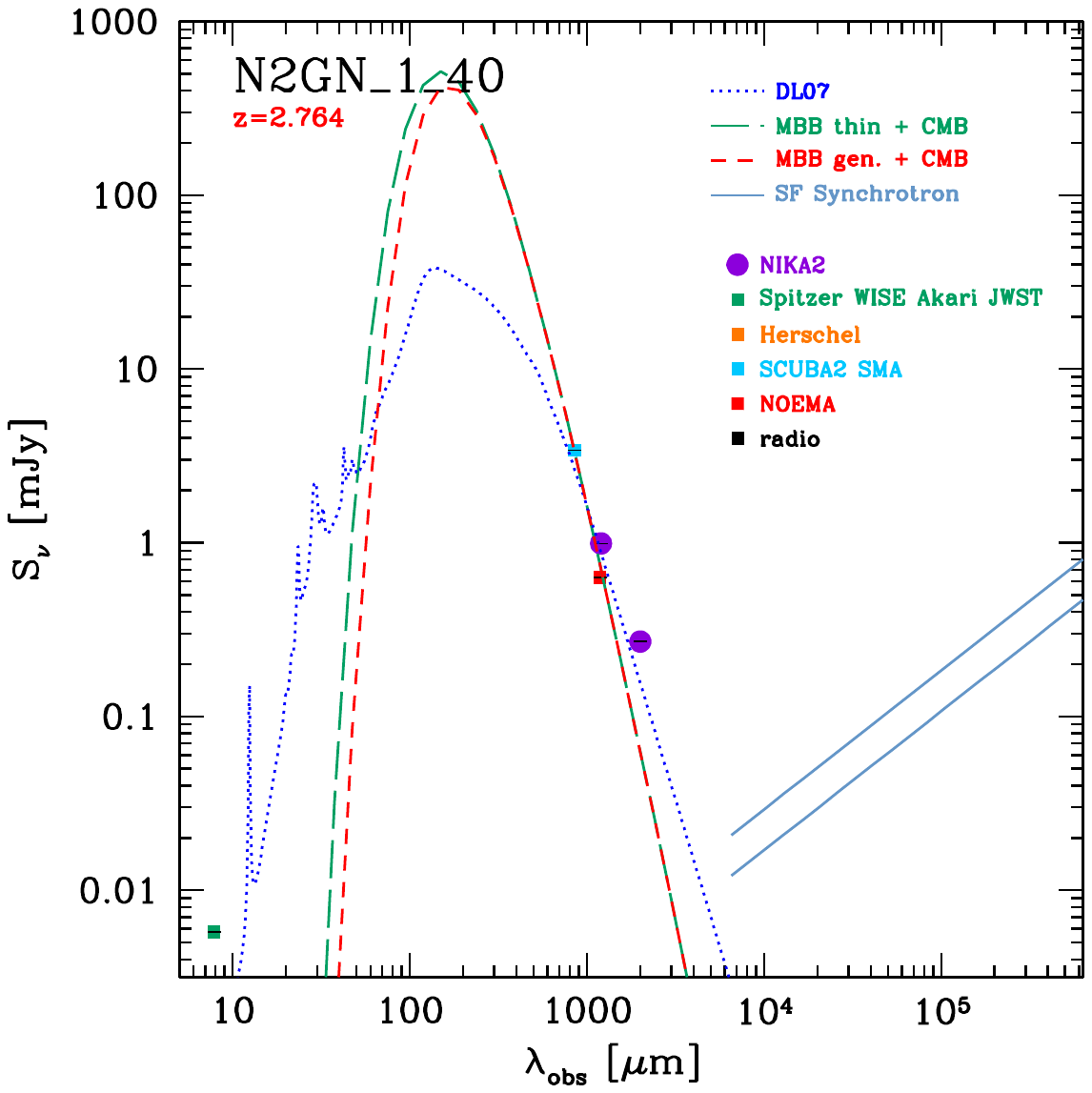}
\includegraphics[align=c,trim=0 0.18\imageheight{} 0 0.075\imageheight{}, clip, width=0.25\textwidth]{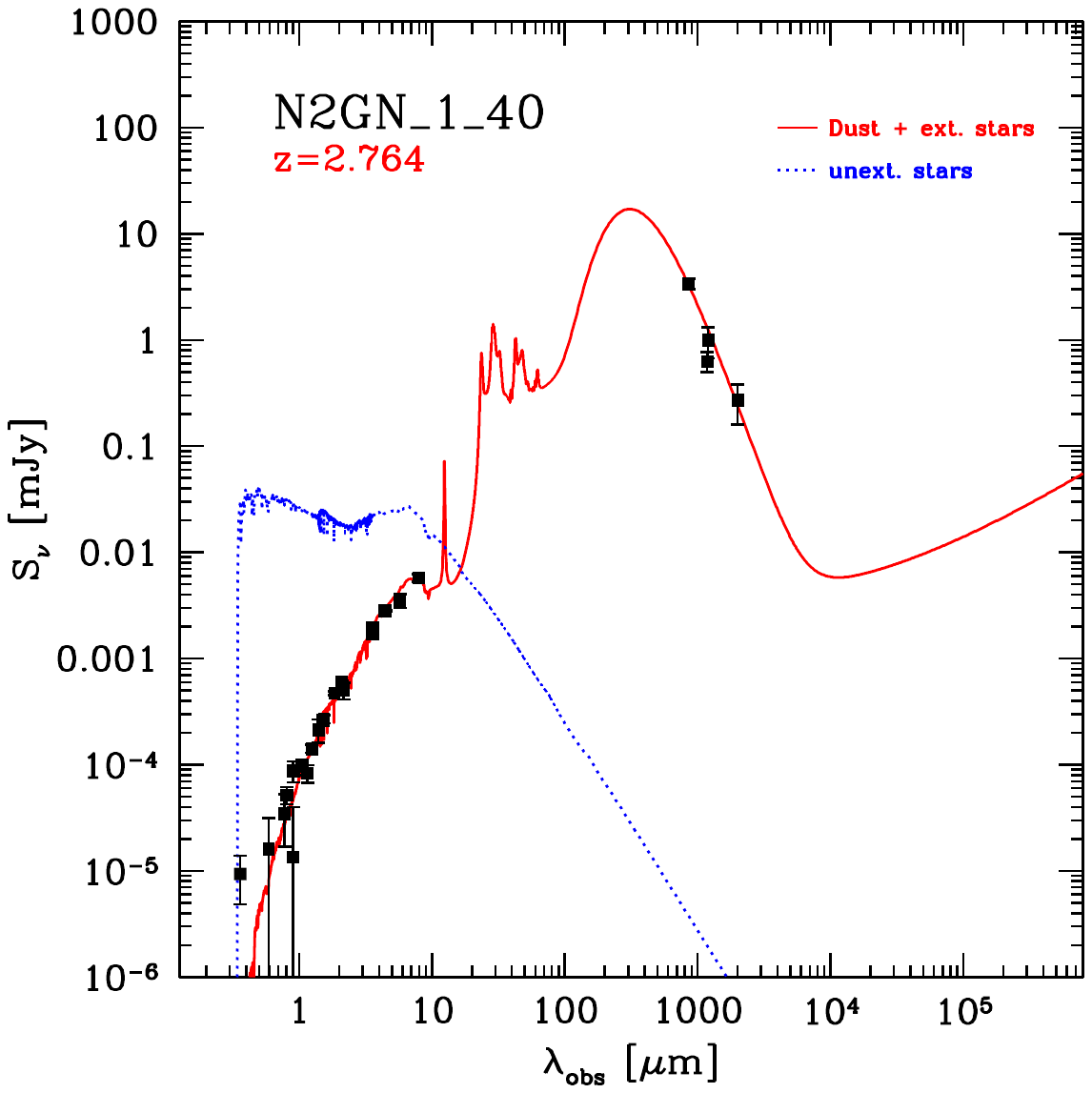}
\includegraphics[align=c,width=0.4\textwidth]{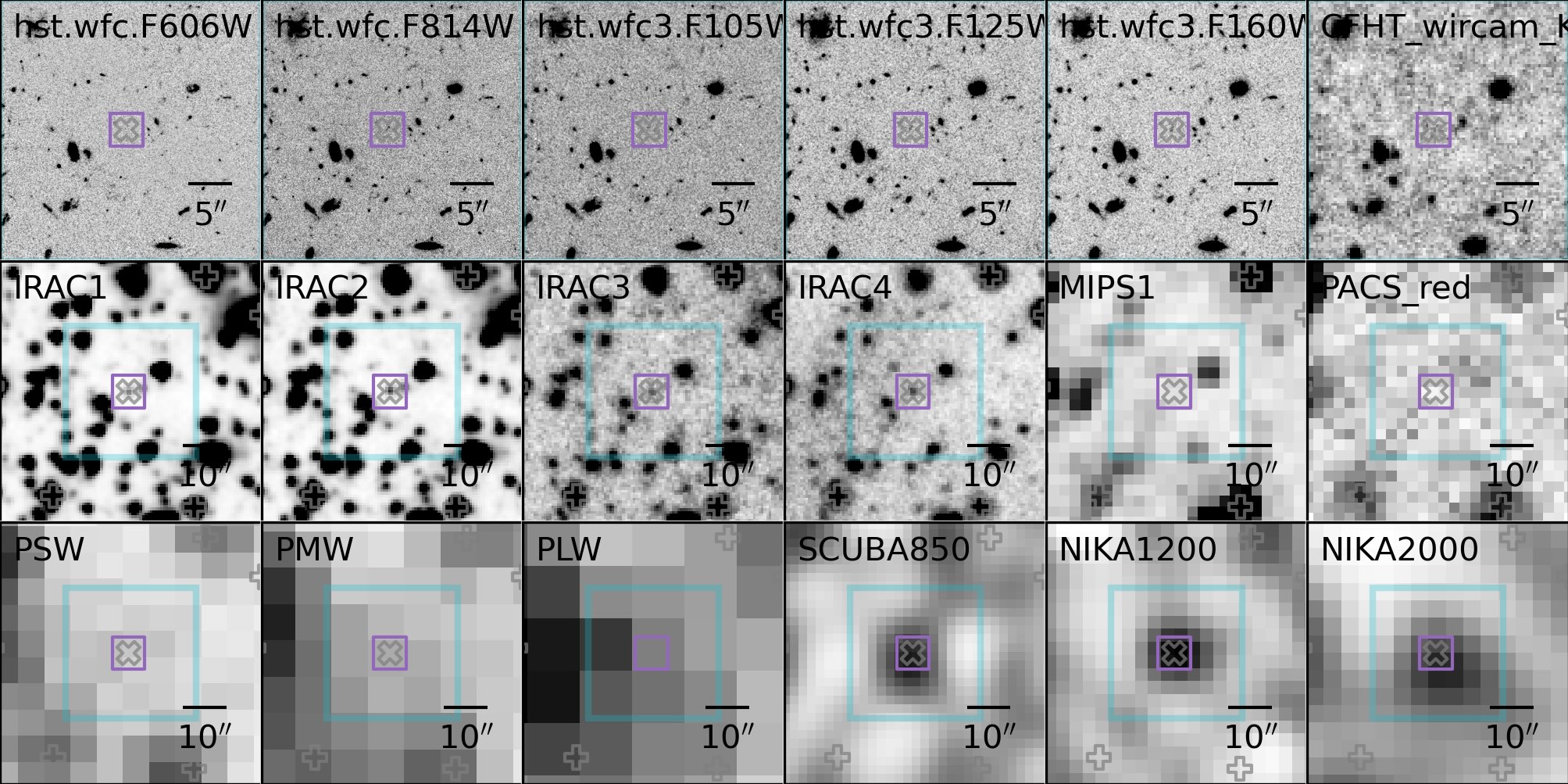}
\includegraphics[align=c,trim=0 0.18\imageheight{} 0 0.075\imageheight{}, clip, width=0.25\textwidth]{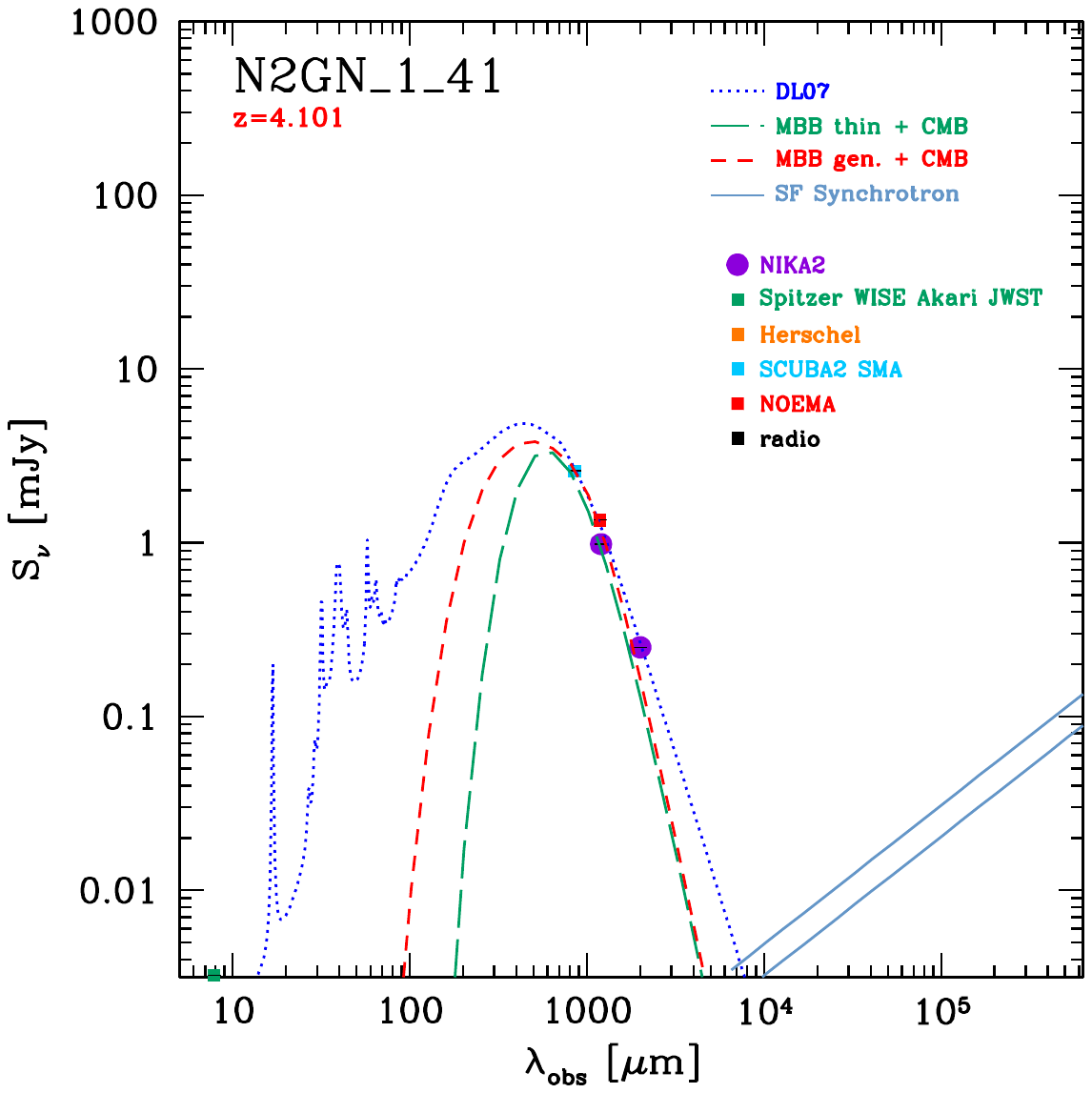}
\includegraphics[align=c,trim=0 0.18\imageheight{} 0 0.075\imageheight{}, clip, width=0.25\textwidth]{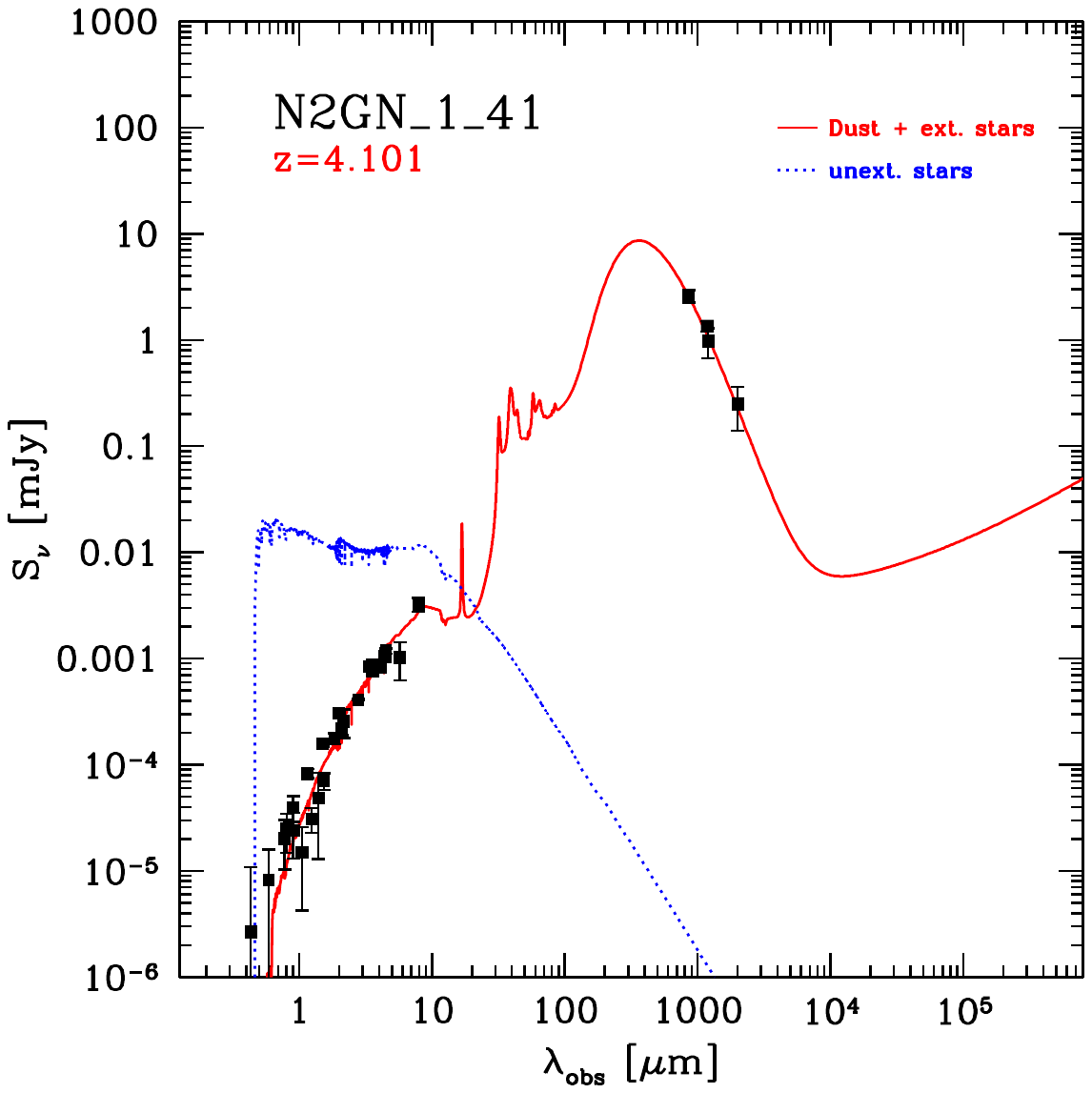}
\includegraphics[align=c,width=0.4\textwidth]{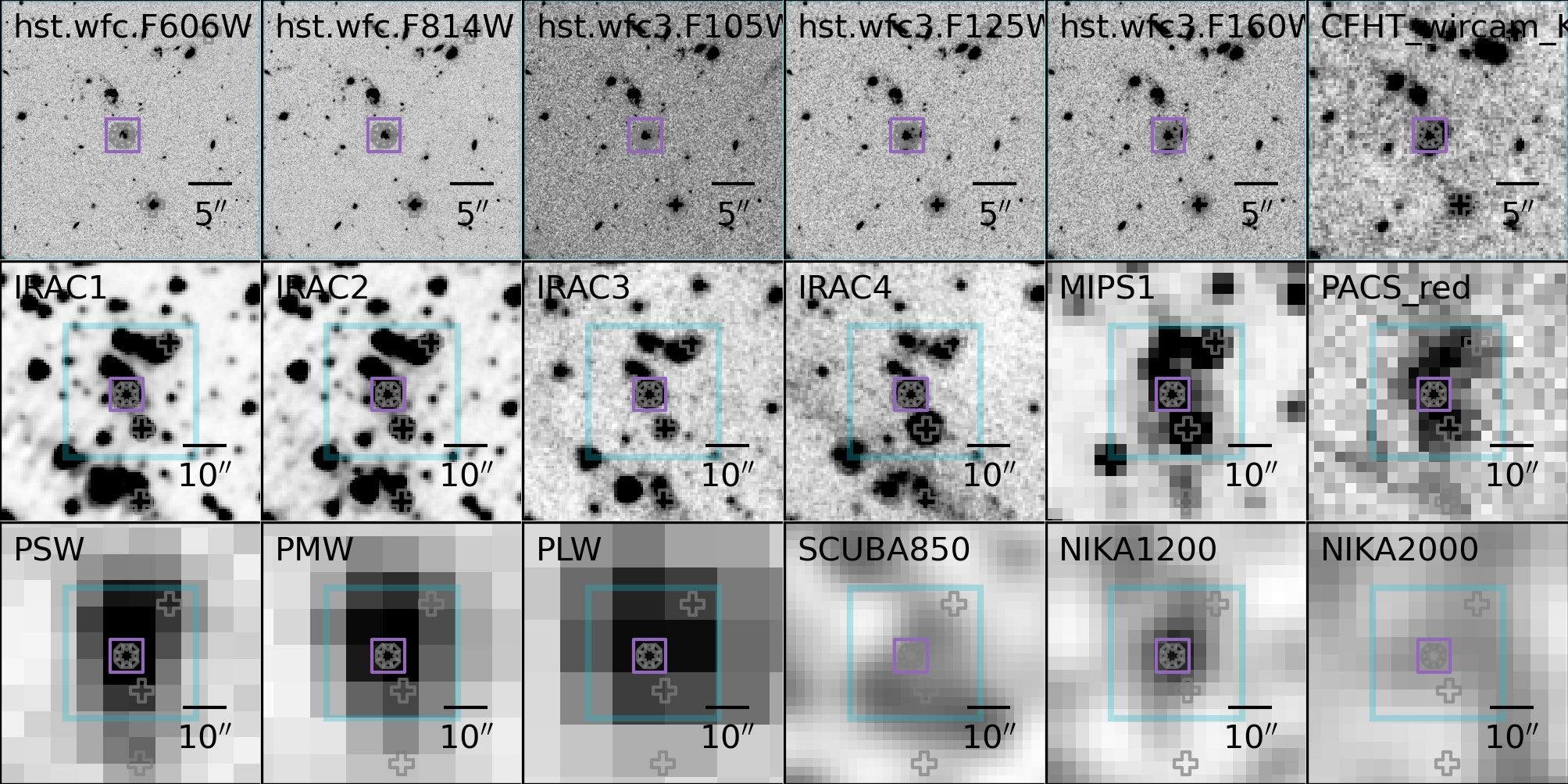}
\includegraphics[align=c,trim=0 0.18\imageheight{} 0 0.075\imageheight{}, clip, width=0.25\textwidth]{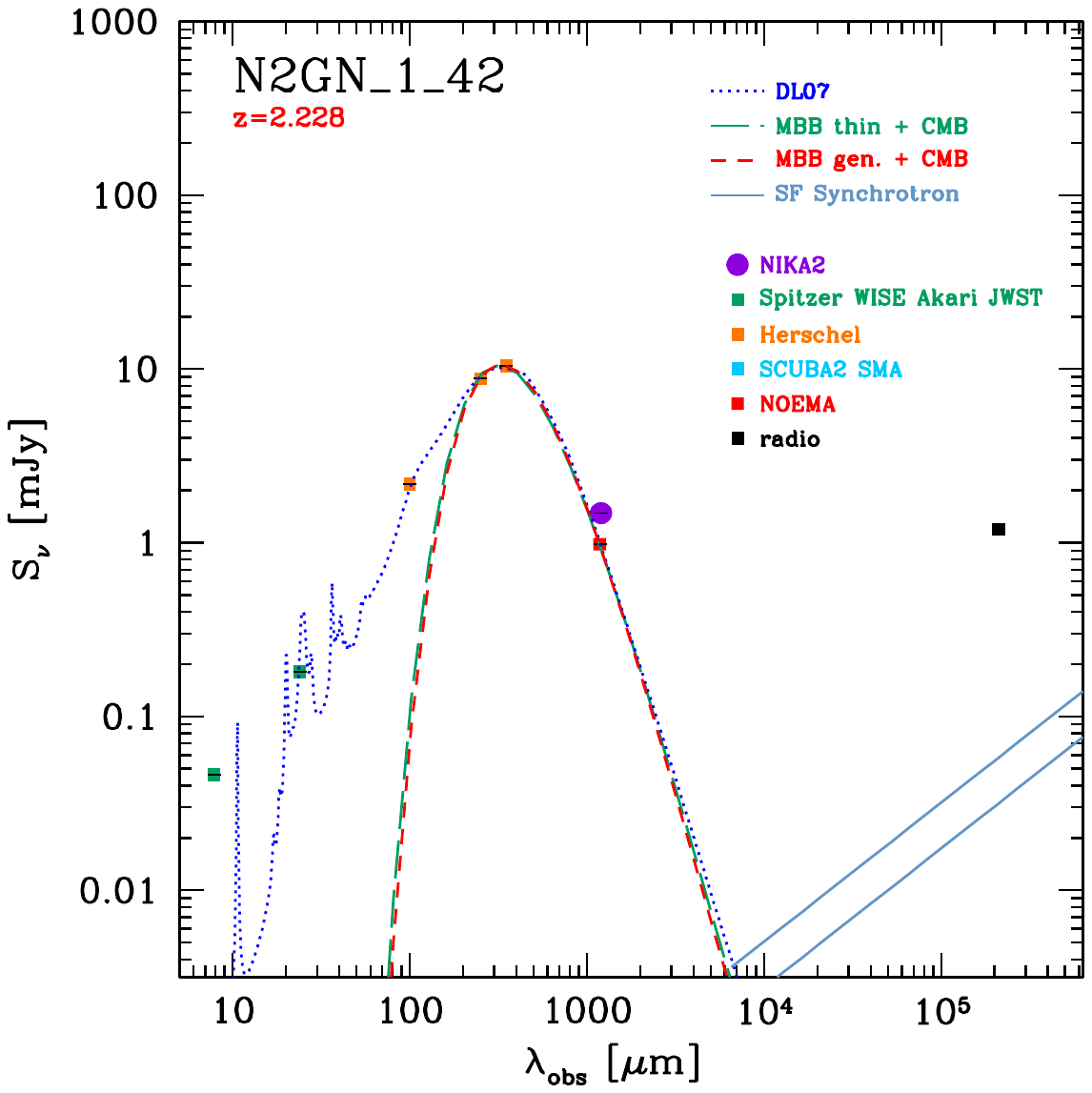}
\includegraphics[align=c,trim=0 0.18\imageheight{} 0 0.075\imageheight{}, clip, width=0.25\textwidth]{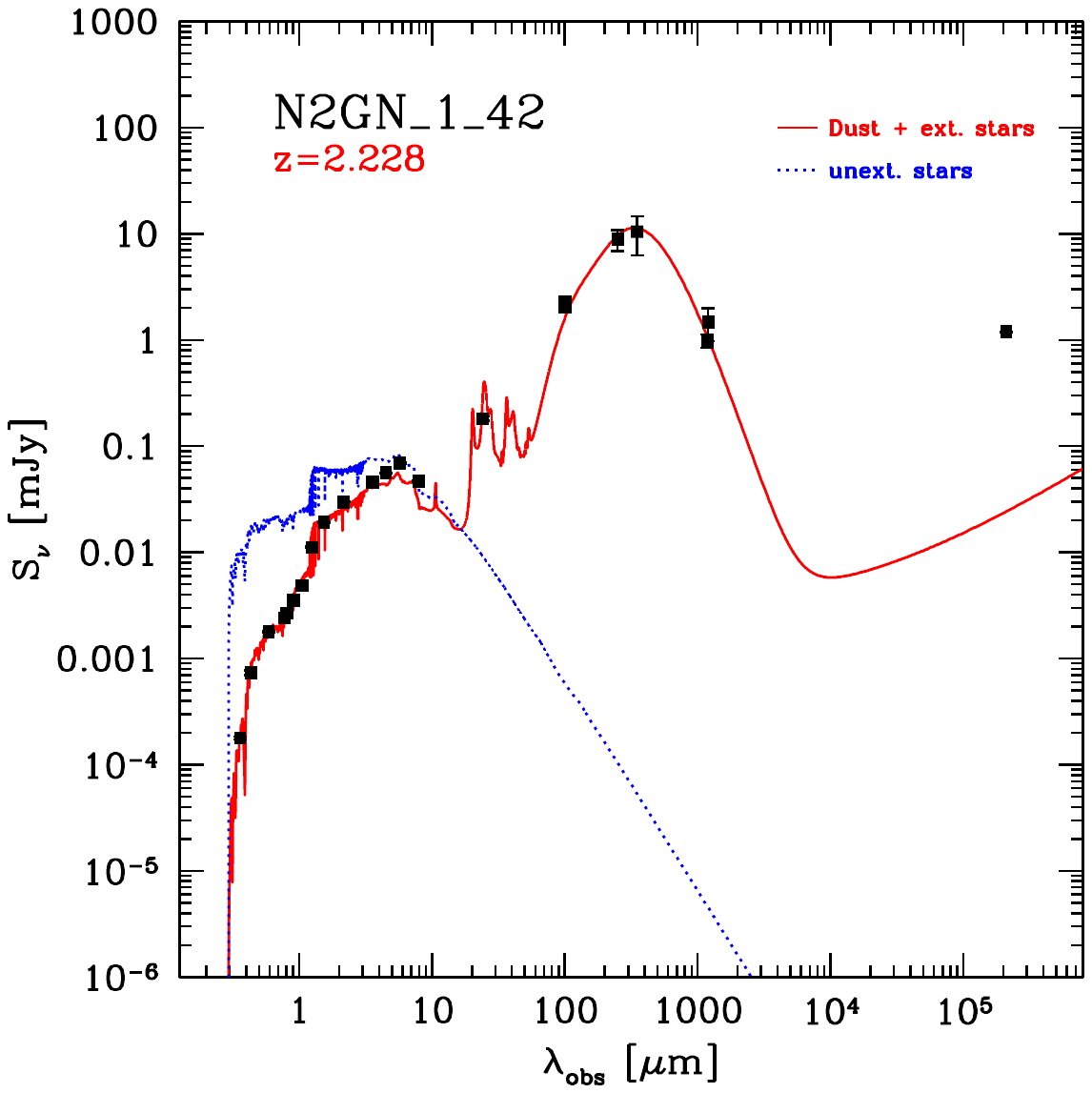}
\includegraphics[align=c,width=0.4\textwidth]{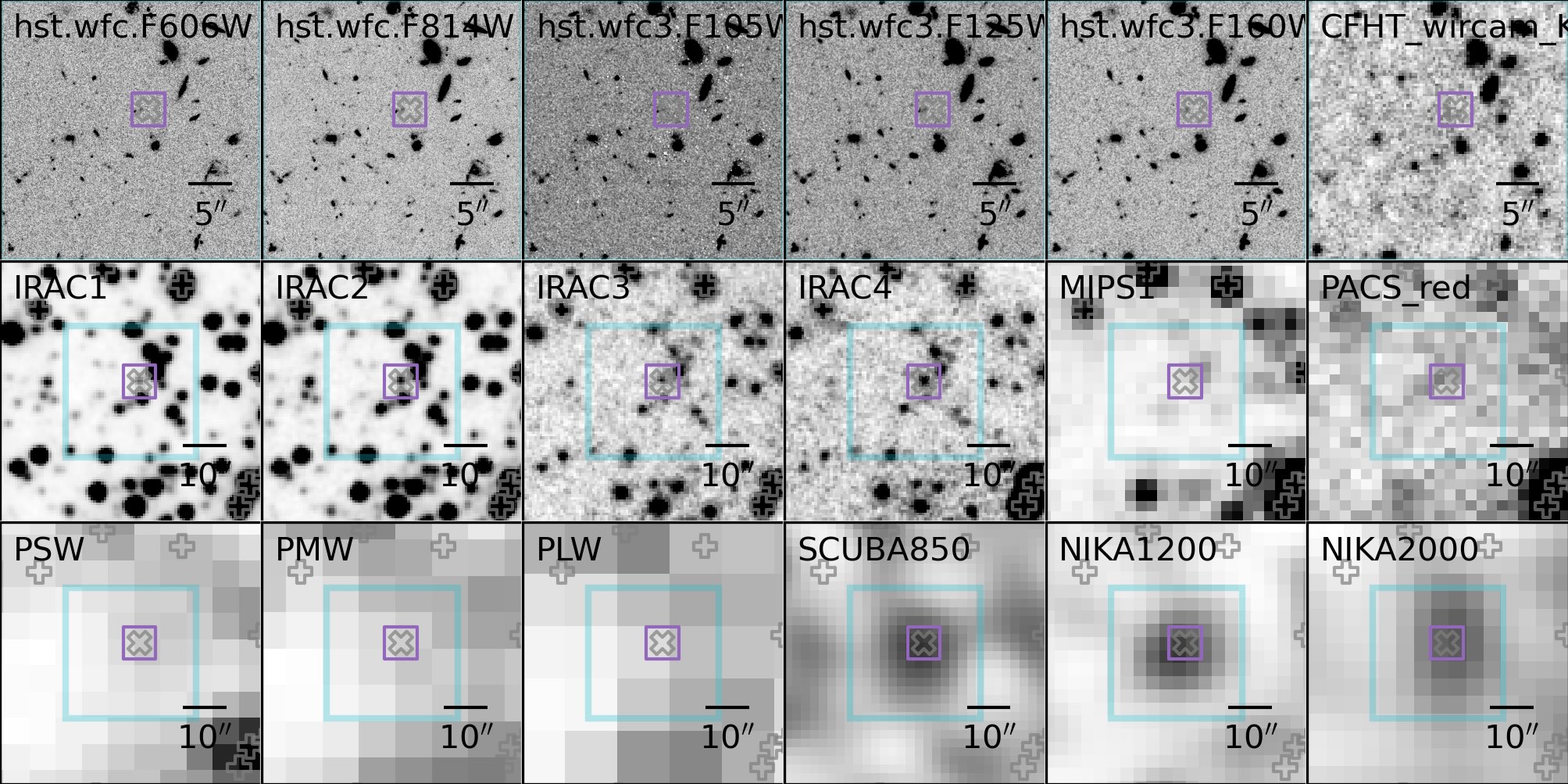}
\includegraphics[align=c,trim=0 0.18\imageheight{} 0 0.075\imageheight{}, clip, width=0.25\textwidth]{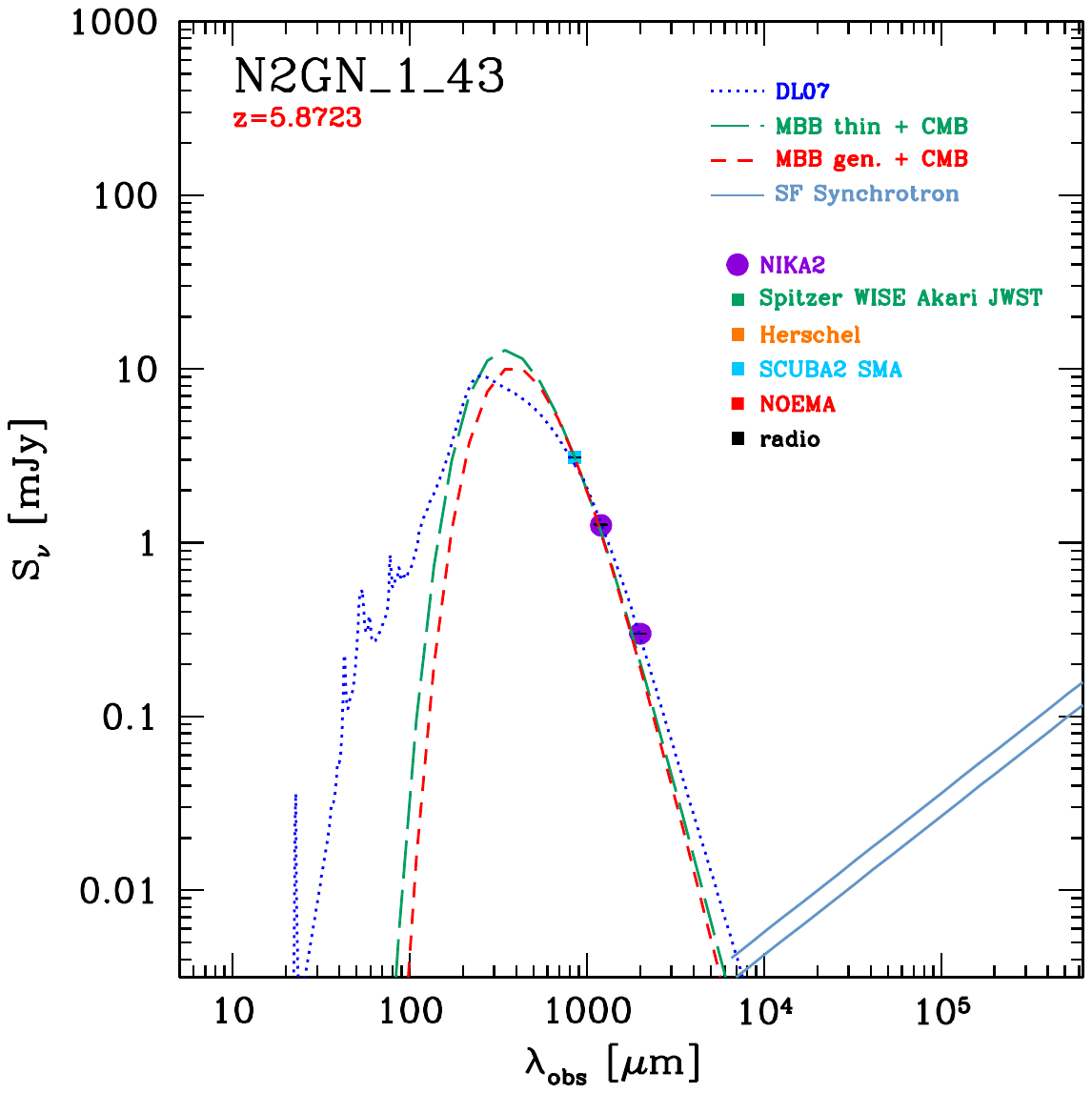}
\includegraphics[align=c,trim=0 0.18\imageheight{} 0 0.075\imageheight{}, clip, width=0.25\textwidth]{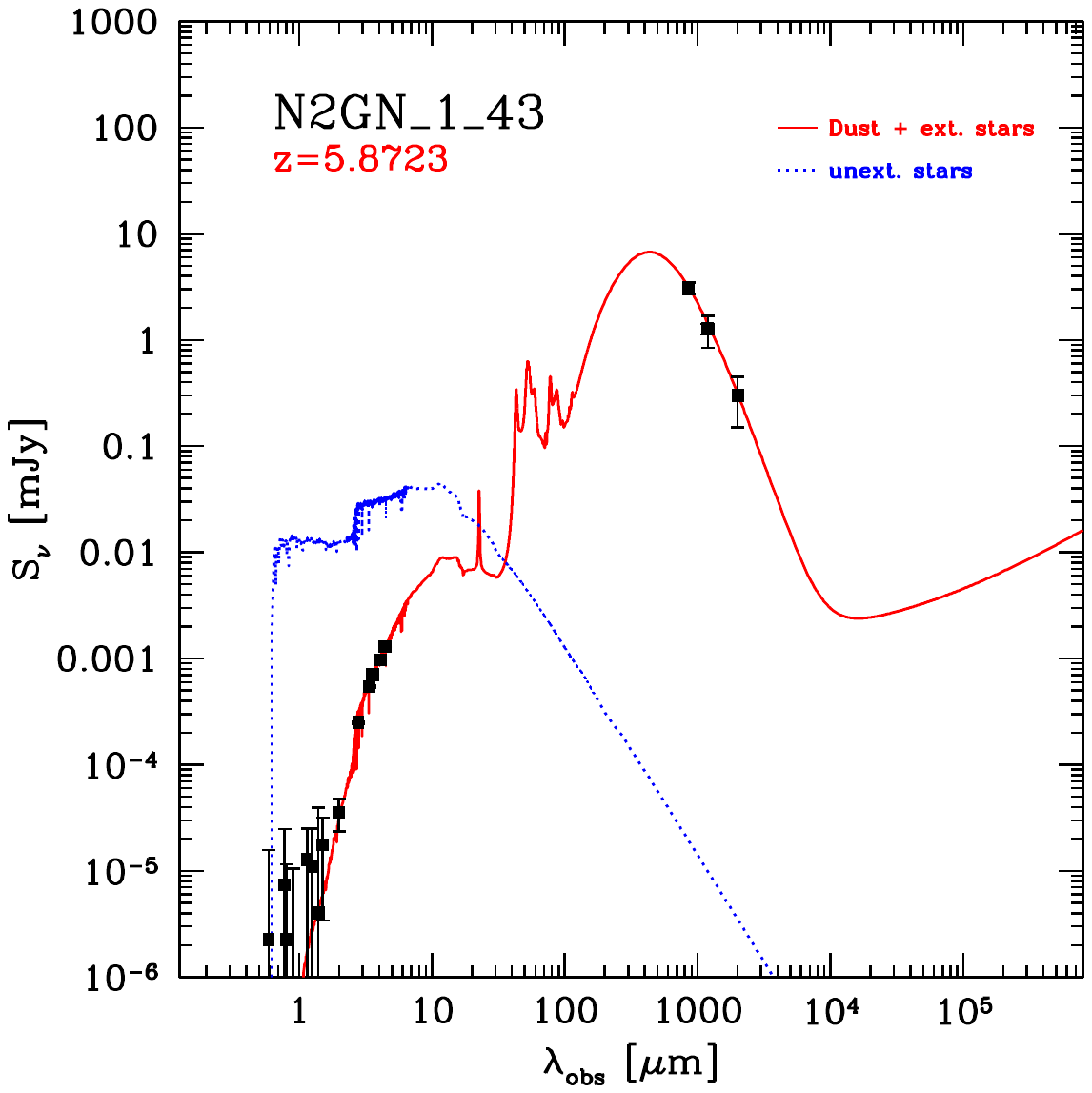}
\includegraphics[align=c,width=0.4\textwidth]{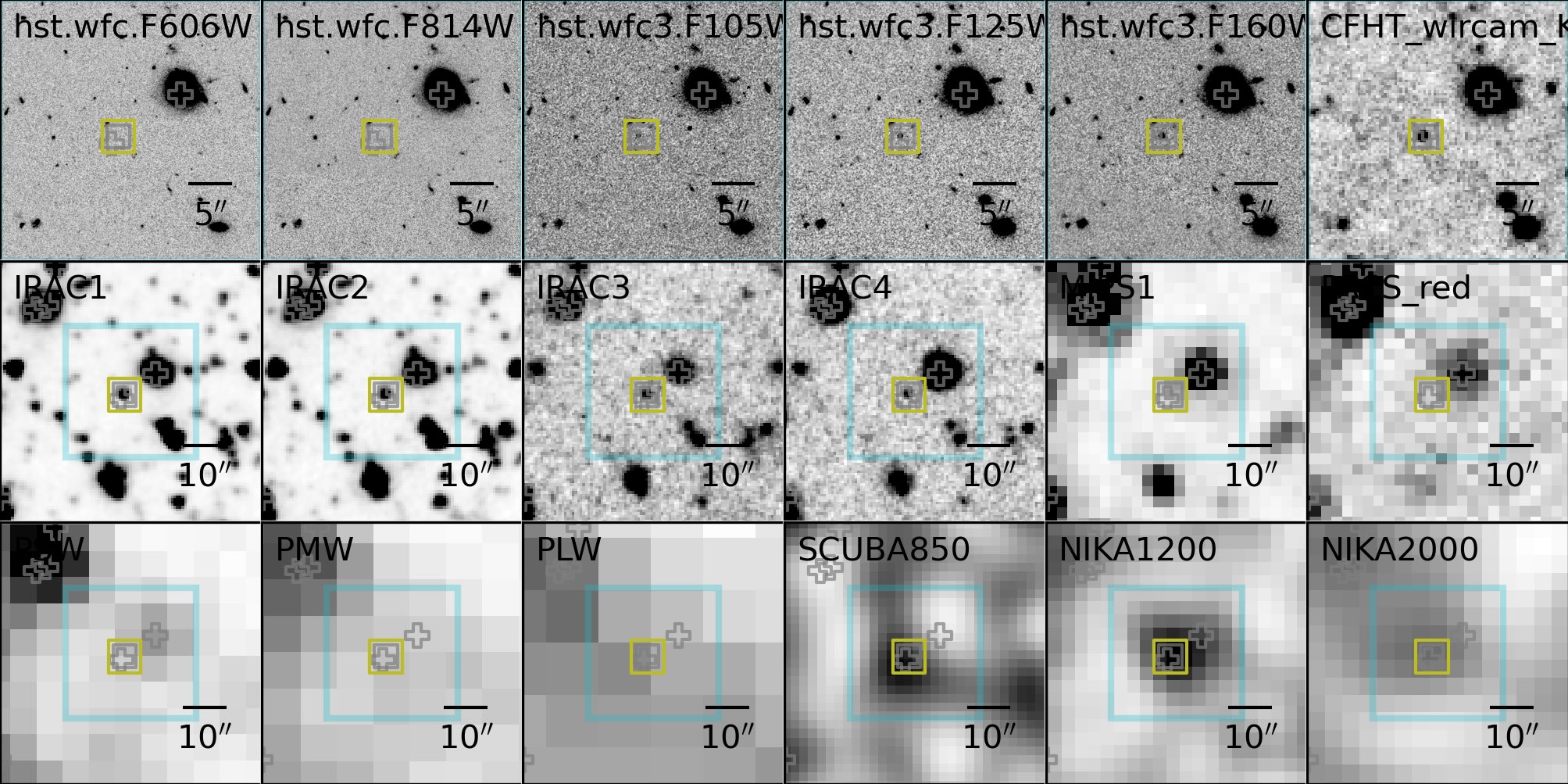}
\includegraphics[align=c,trim=0 0.18\imageheight{} 0 0.075\imageheight{}, clip, width=0.25\textwidth]{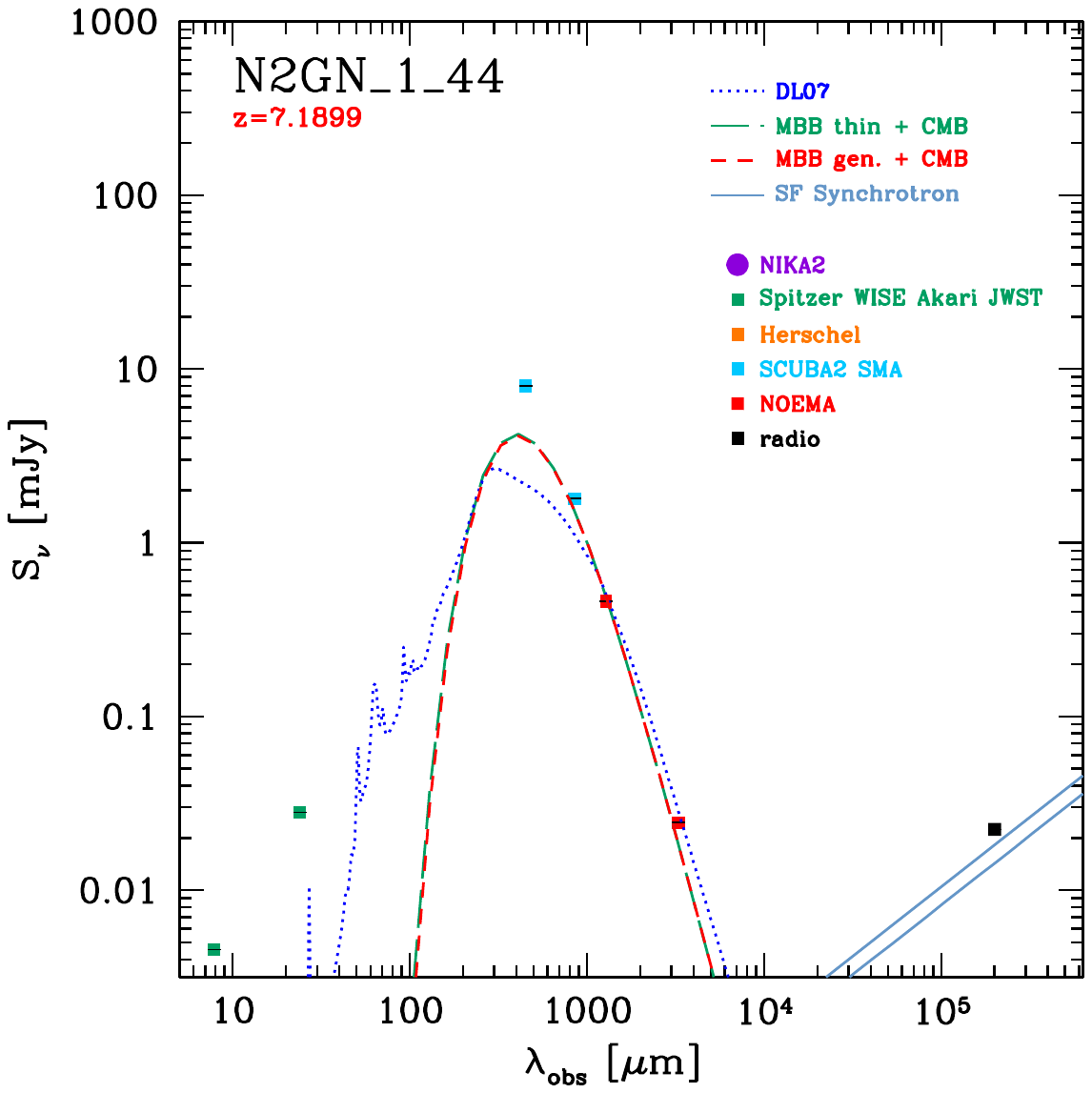}
\includegraphics[align=c,trim=0 0.18\imageheight{} 0 0.075\imageheight{}, clip, width=0.25\textwidth]{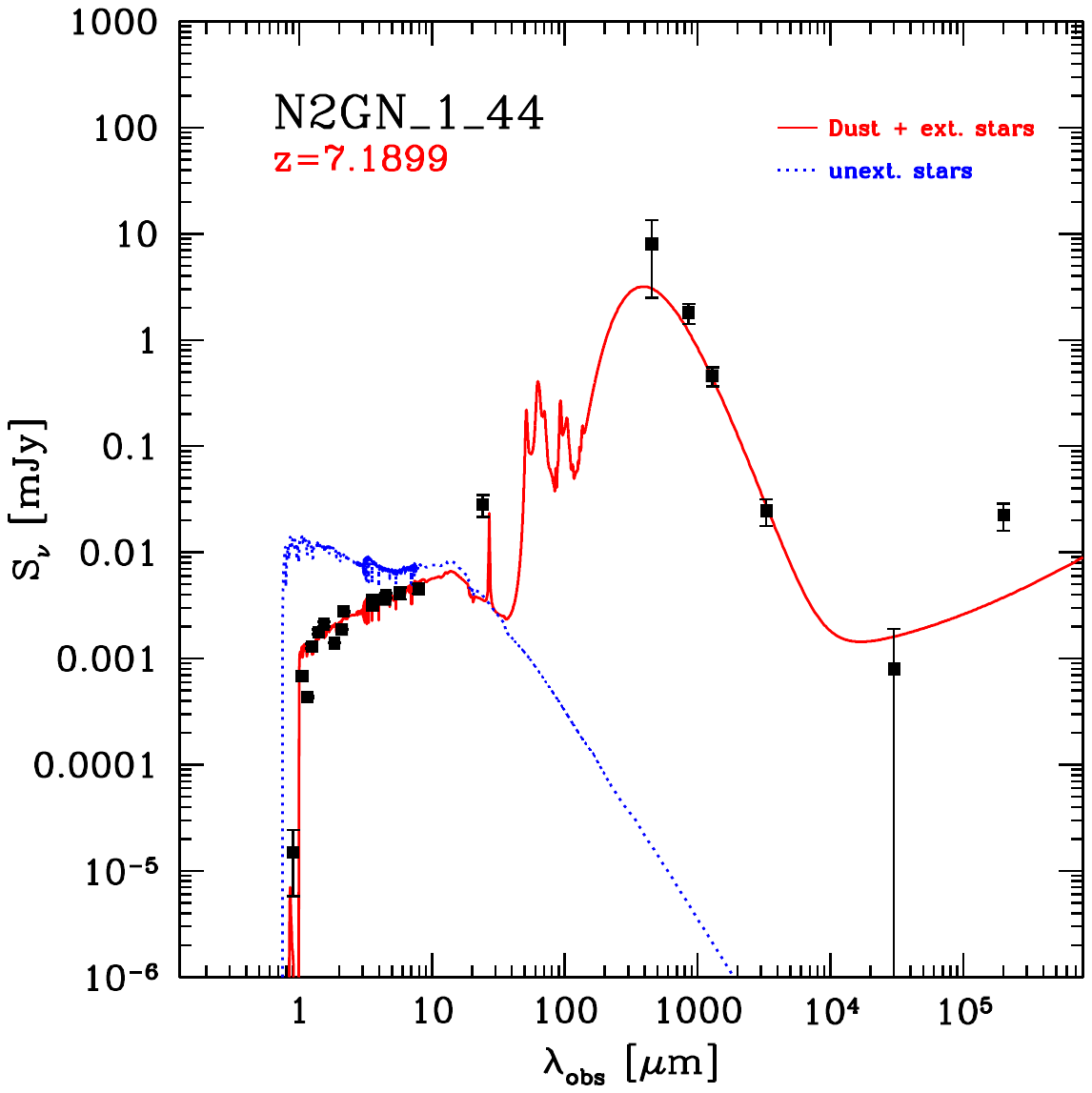}
\caption{continued.}
\end{figure*}

\addtocounter{figure}{-1}
\newpage

\begin{figure*}[t]
\centering
\includegraphics[align=c,width=0.4\textwidth]{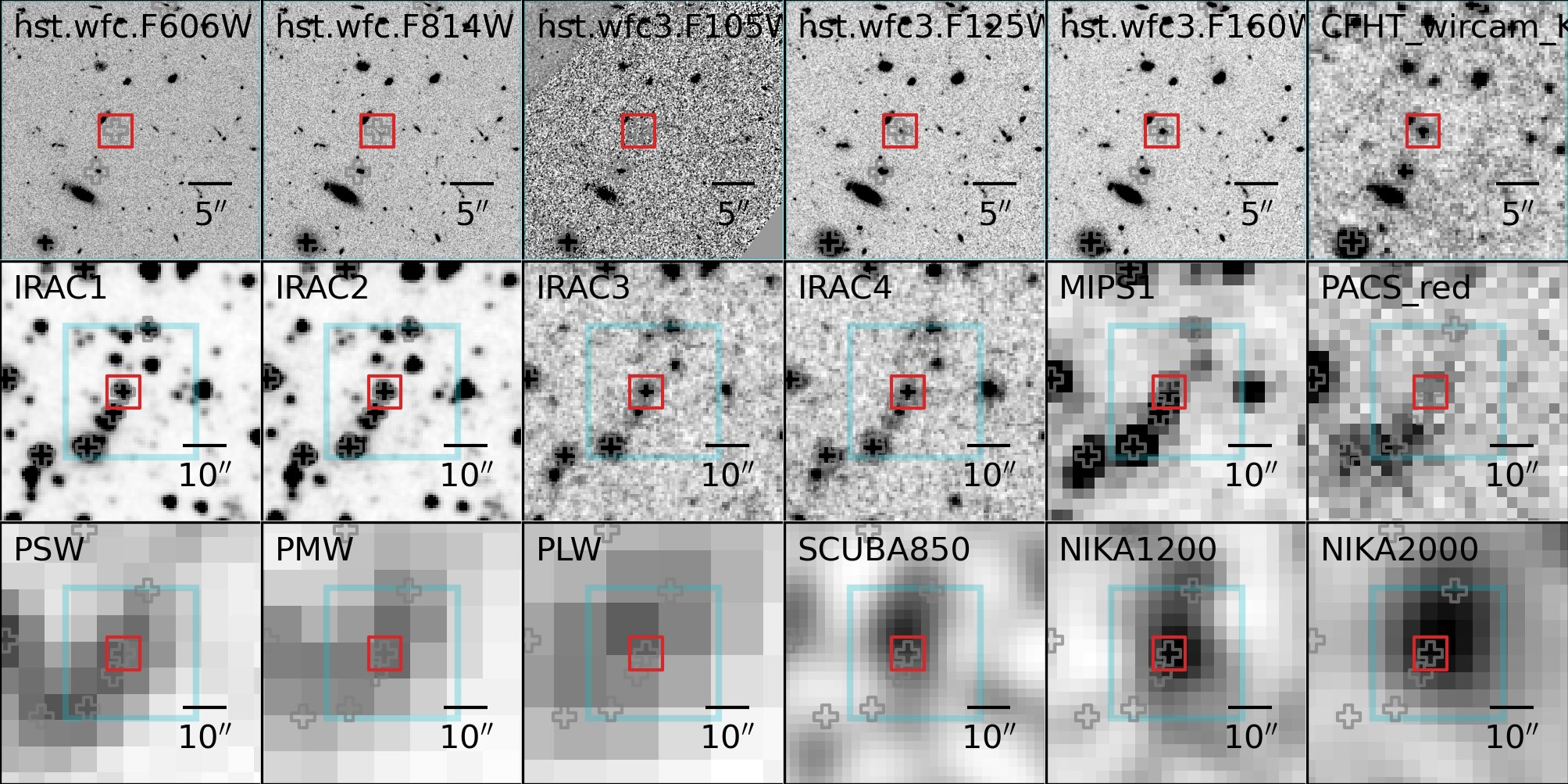}
\includegraphics[align=c,trim=0 0.18\imageheight{} 0 0.075\imageheight{}, clip, width=0.25\textwidth]{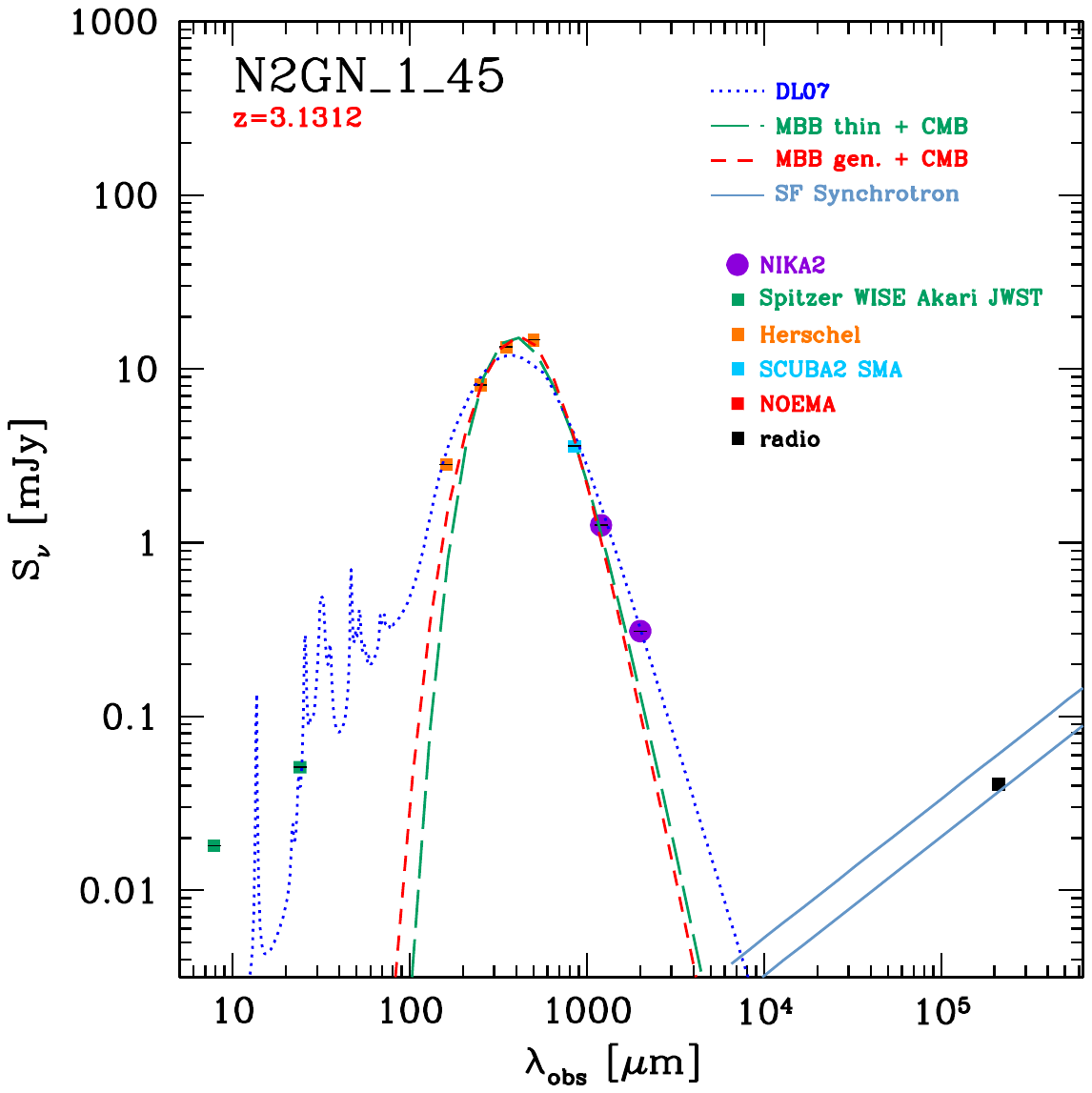}
\includegraphics[align=c,trim=0 0.18\imageheight{} 0 0.075\imageheight{}, clip, width=0.25\textwidth]{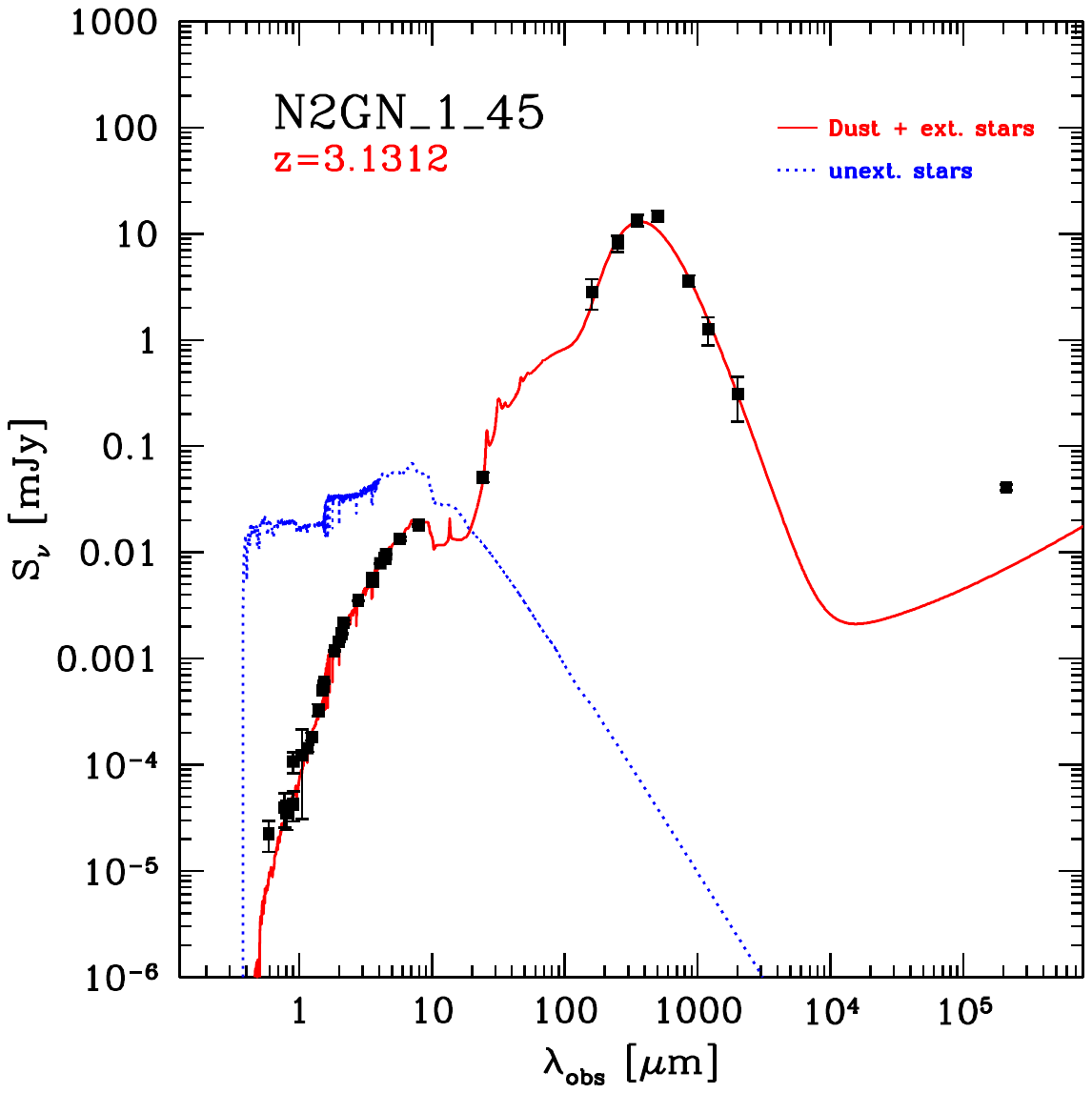}
\includegraphics[align=c,width=0.4\textwidth]{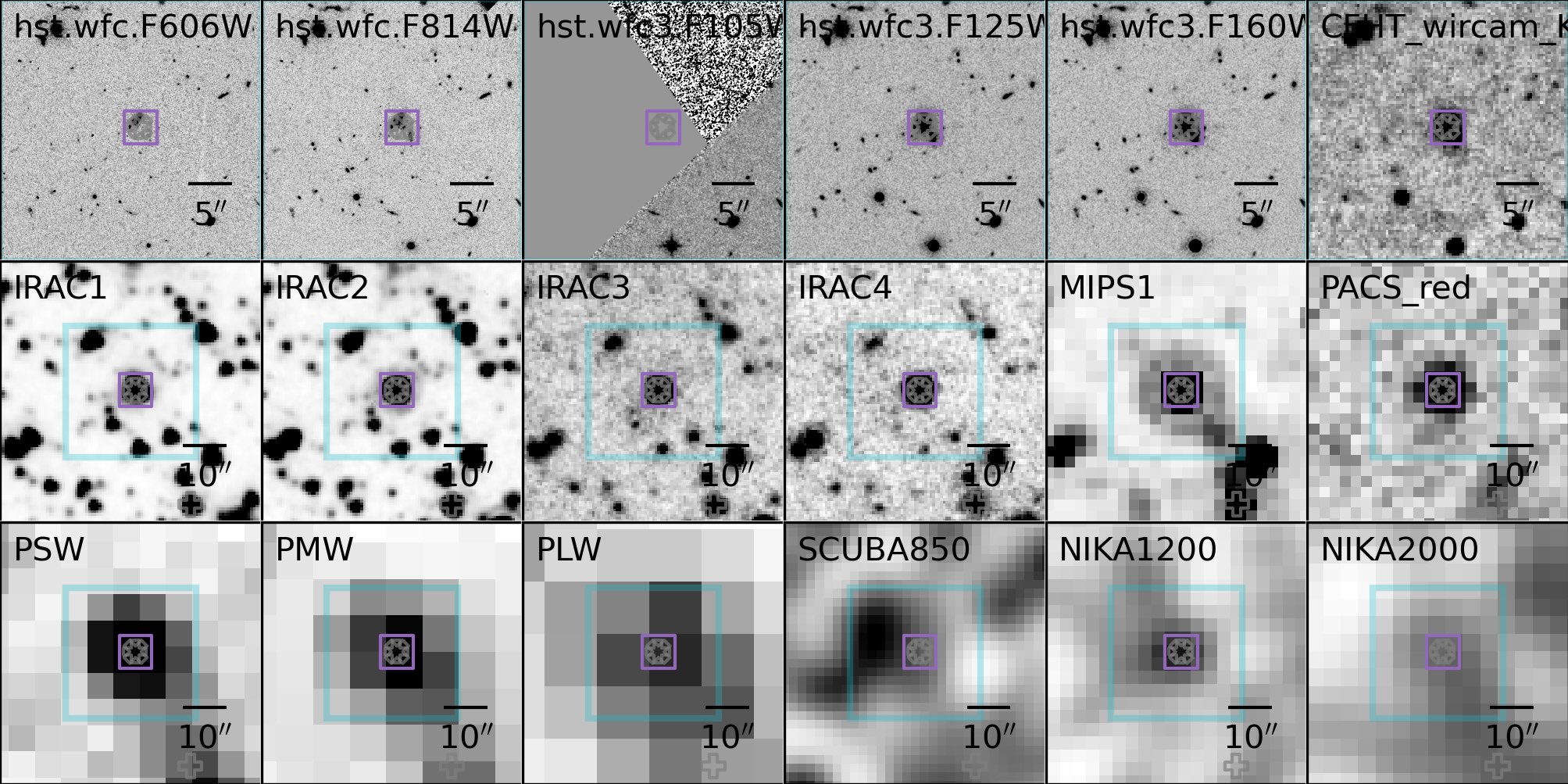}
\includegraphics[align=c,trim=0 0.18\imageheight{} 0 0.075\imageheight{}, clip, width=0.25\textwidth]{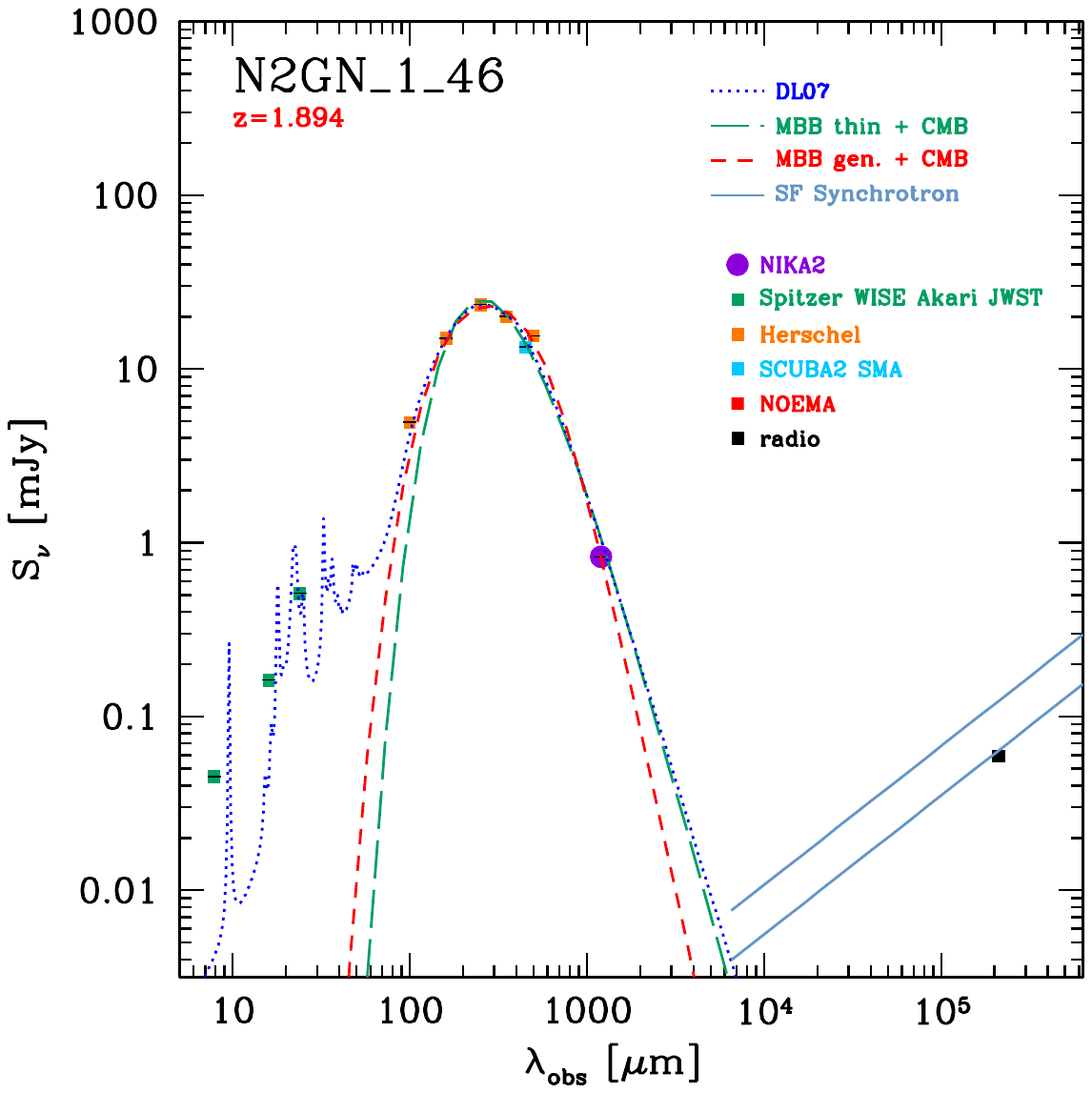}
\includegraphics[align=c,trim=0 0.18\imageheight{} 0 0.075\imageheight{}, clip, width=0.25\textwidth]{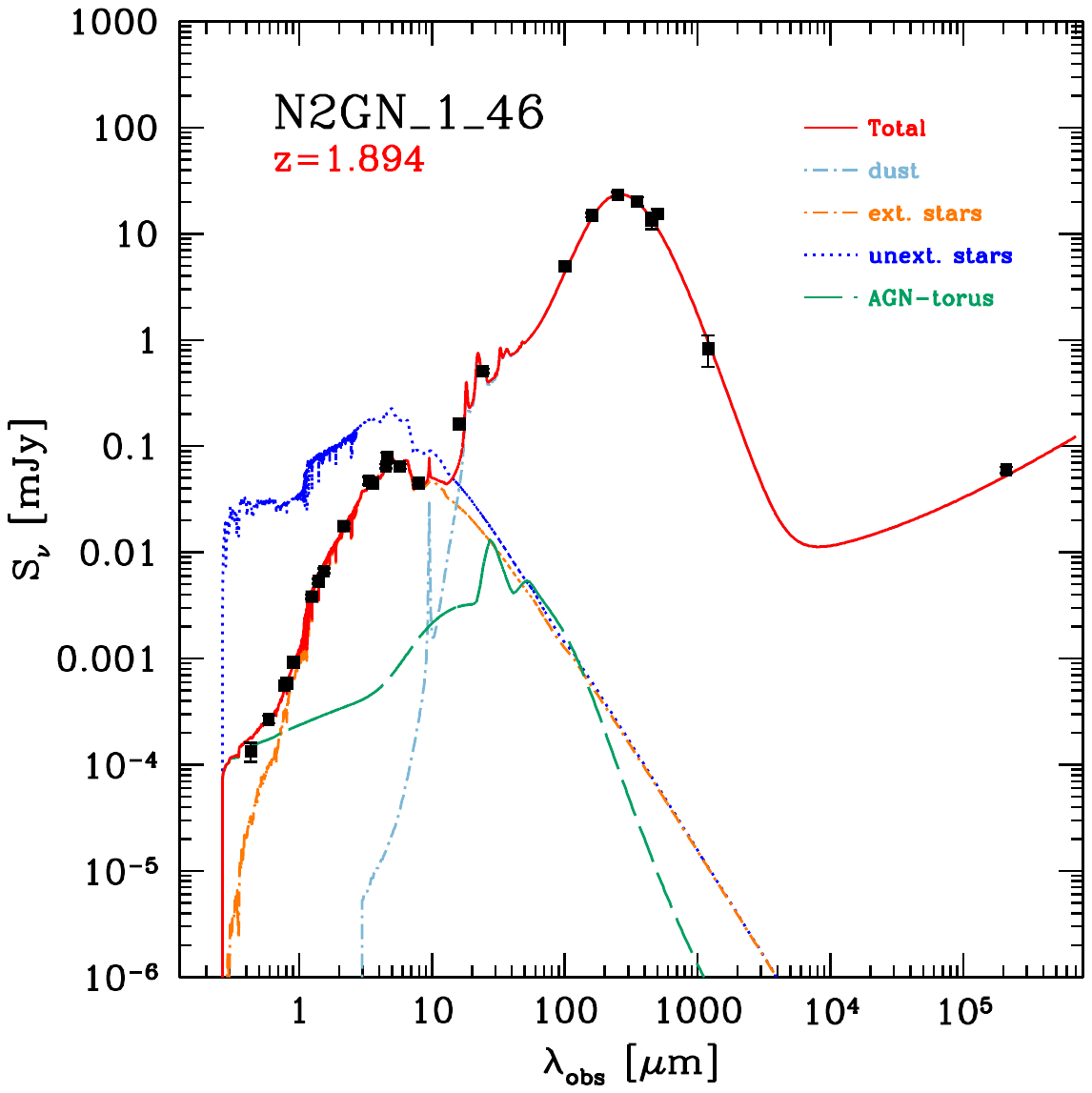}
\includegraphics[align=c,width=0.4\textwidth]{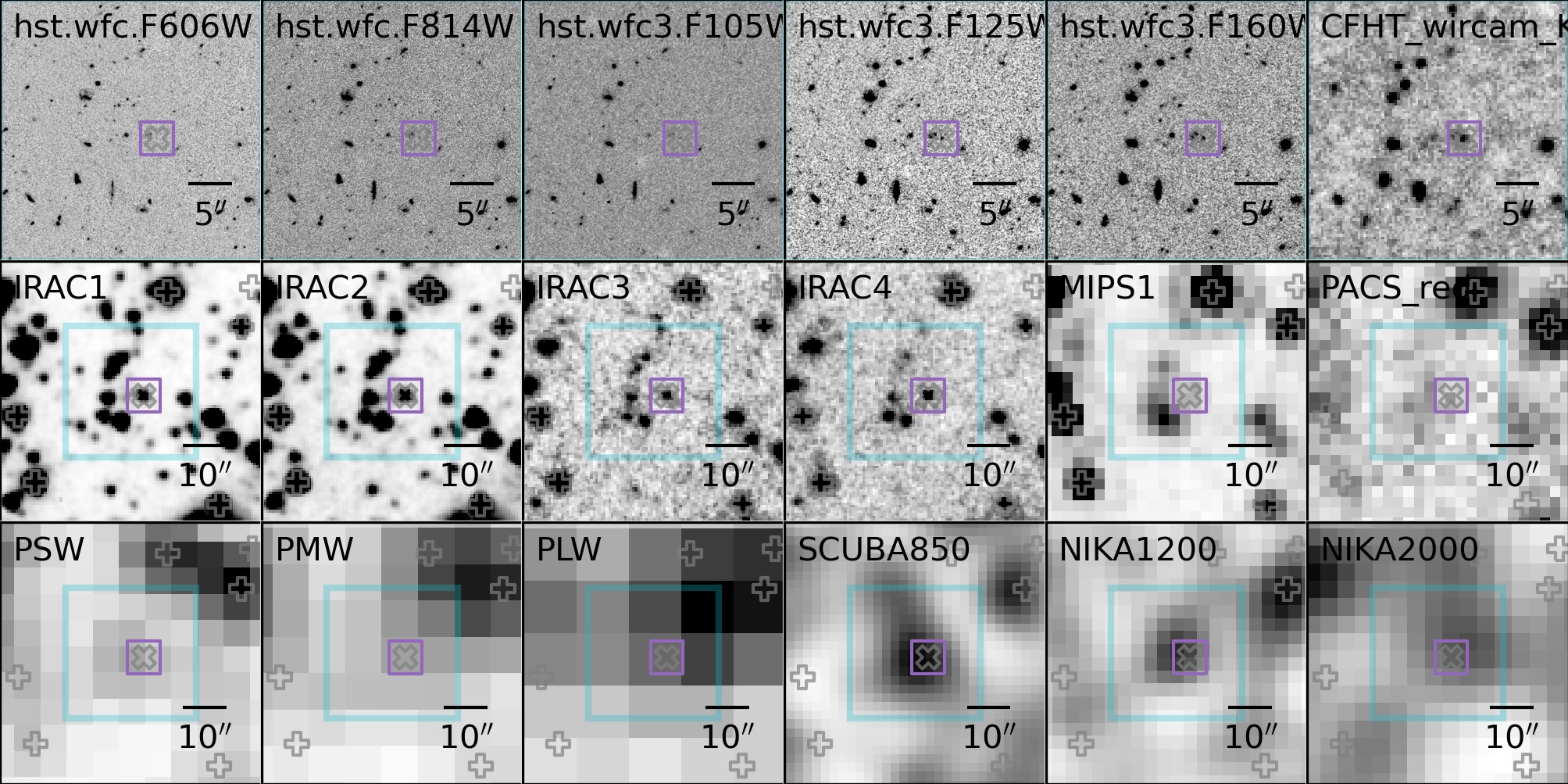}
\includegraphics[align=c,trim=0 0.18\imageheight{} 0 0.075\imageheight{}, clip, width=0.25\textwidth]{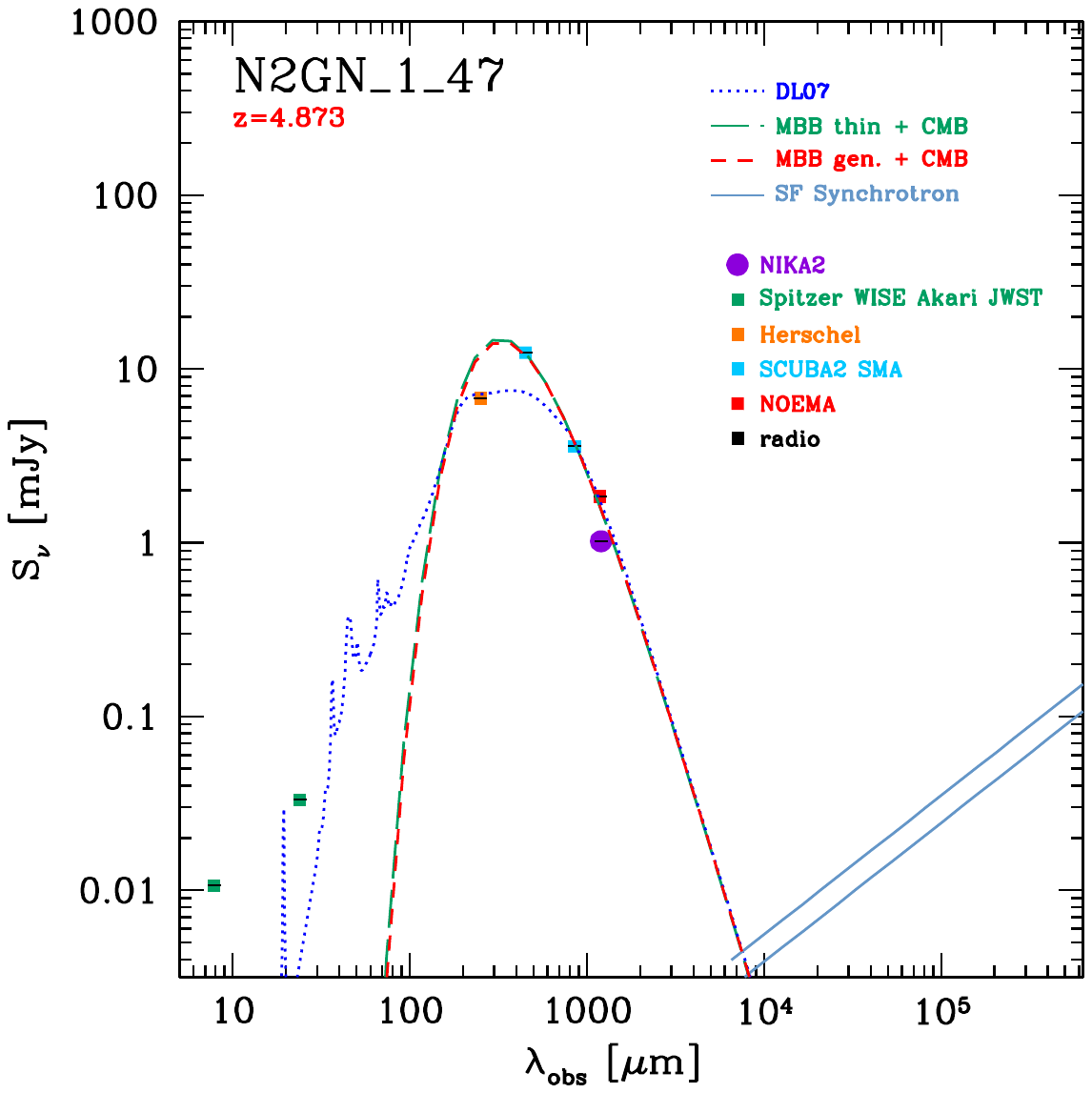}
\includegraphics[align=c,trim=0 0.18\imageheight{} 0 0.075\imageheight{}, clip, width=0.25\textwidth]{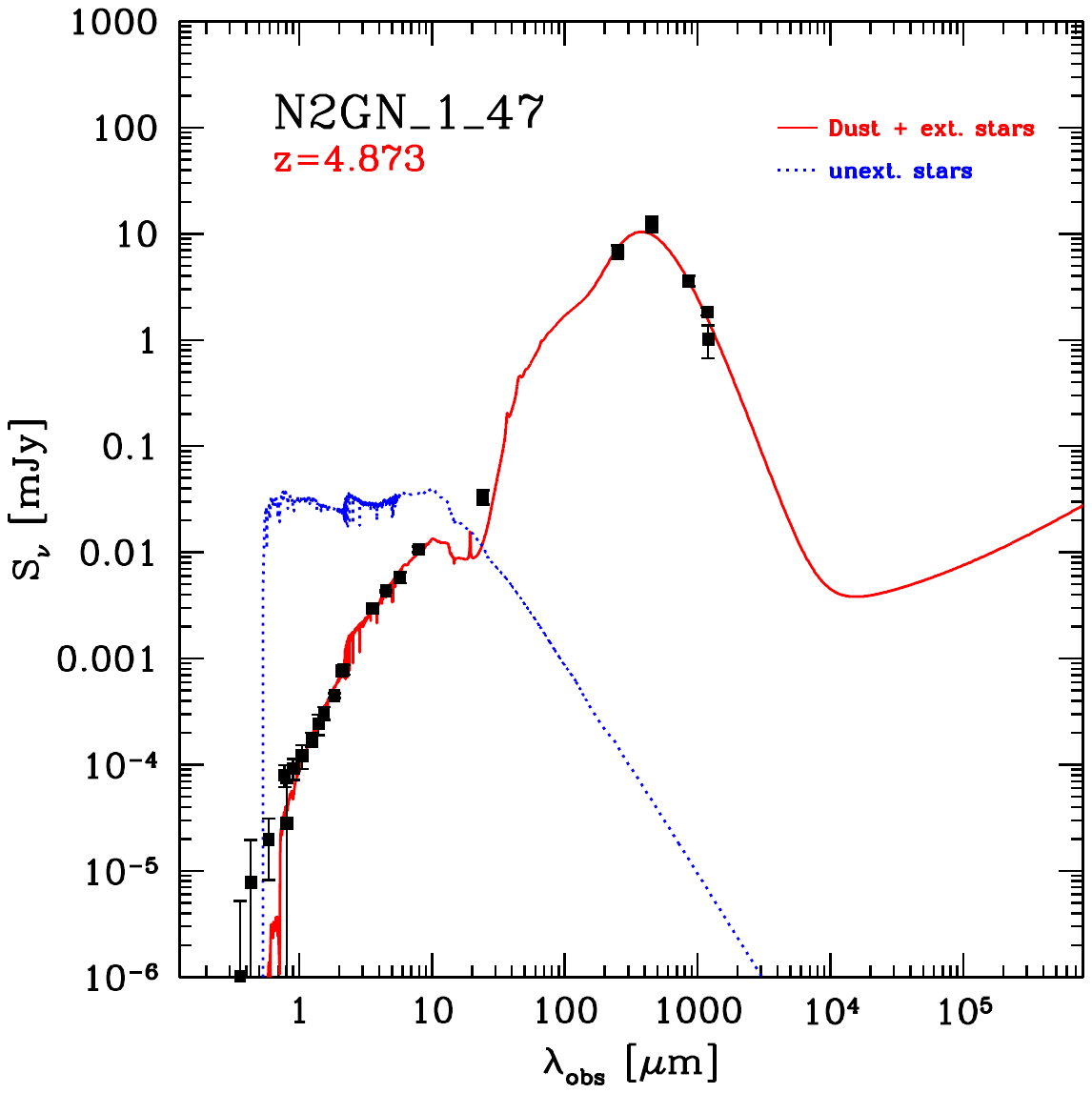}
\includegraphics[align=c,width=0.4\textwidth]{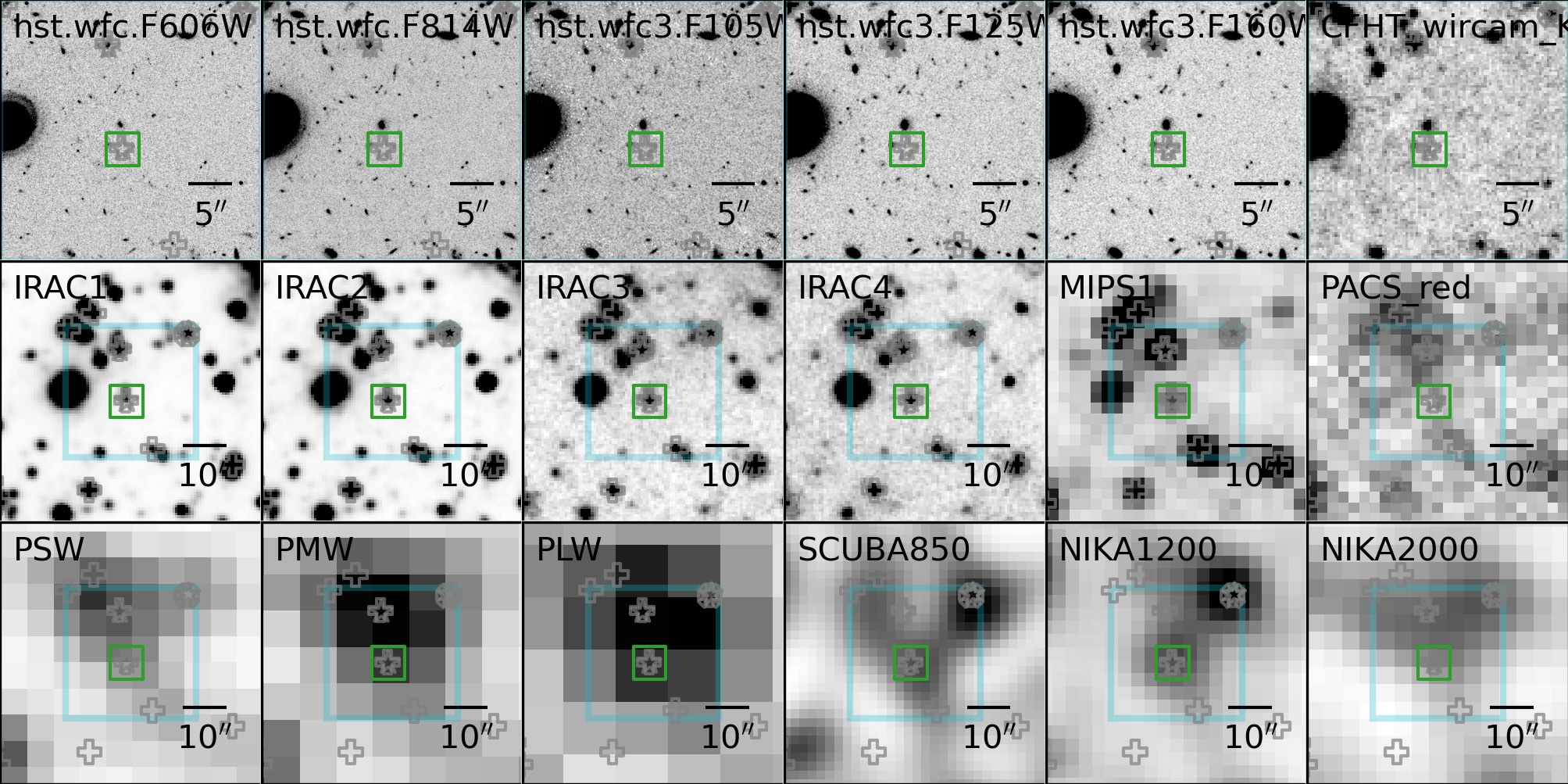}
\includegraphics[align=c,trim=0 0.18\imageheight{} 0 0.075\imageheight{}, clip, width=0.25\textwidth]{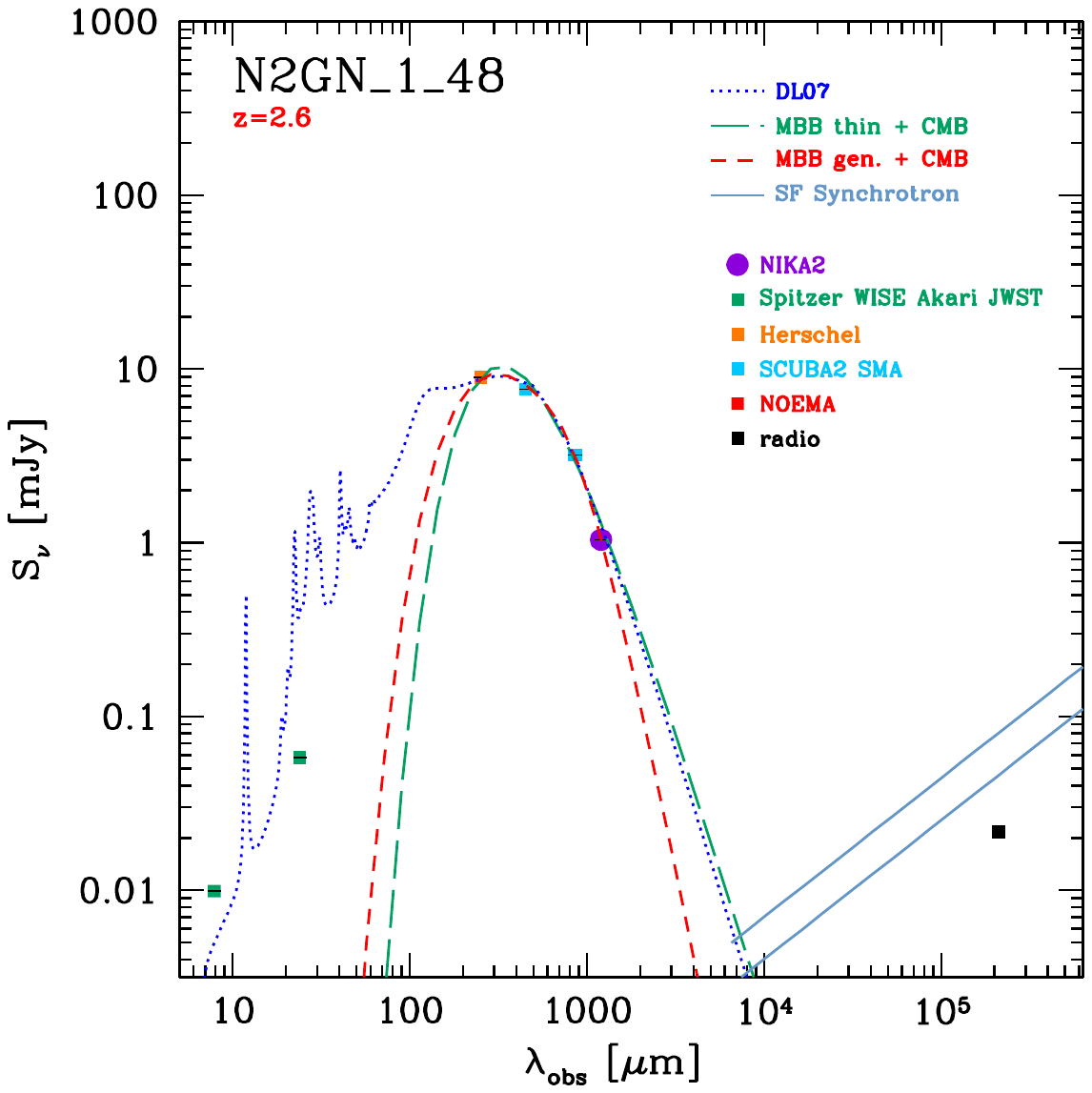}
\includegraphics[align=c,trim=0 0.18\imageheight{} 0 0.075\imageheight{}, clip, width=0.25\textwidth]{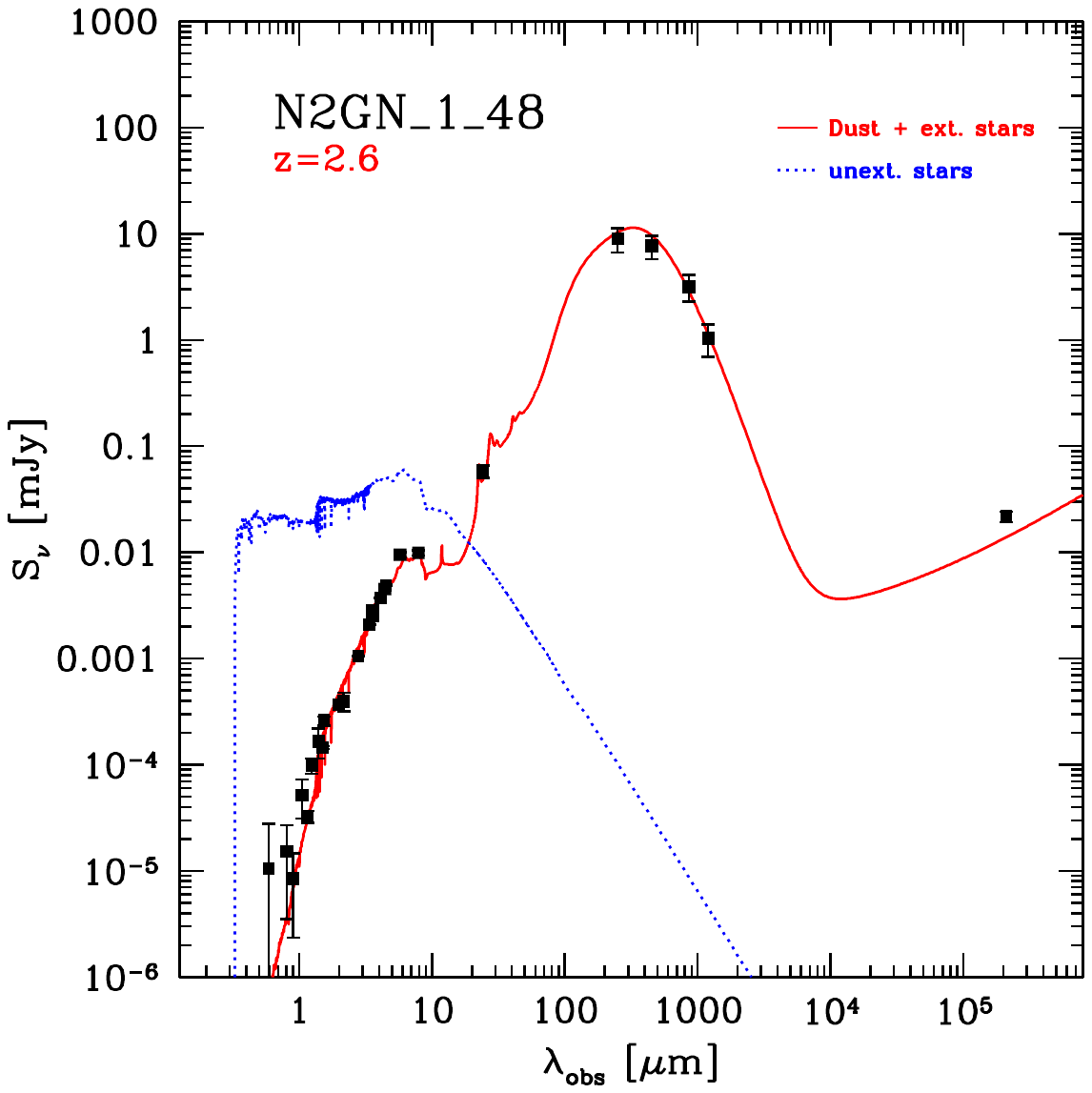}
\includegraphics[align=c,width=0.4\textwidth]{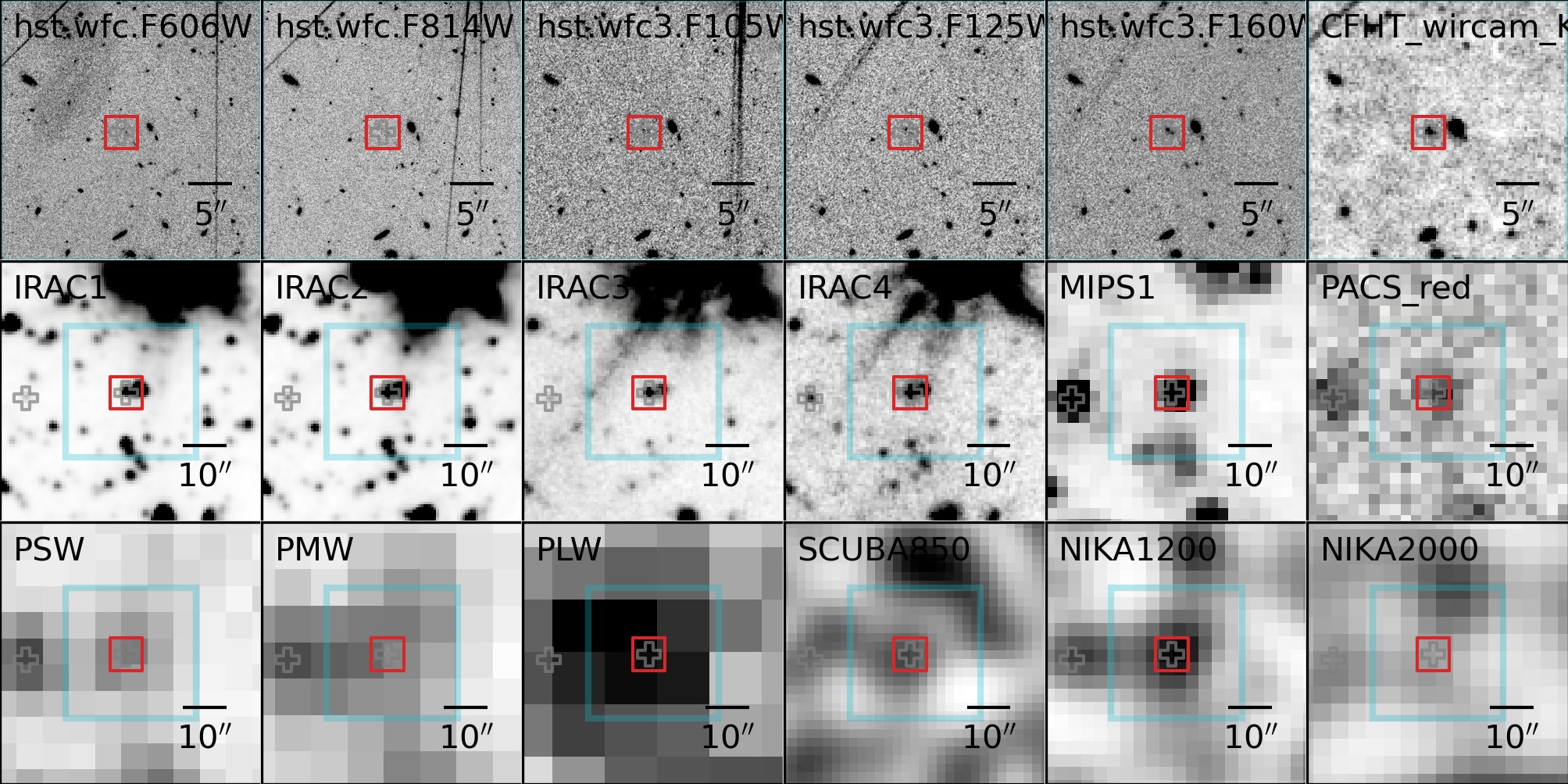}
\includegraphics[align=c,trim=0 0.18\imageheight{} 0 0.075\imageheight{}, clip, width=0.25\textwidth]{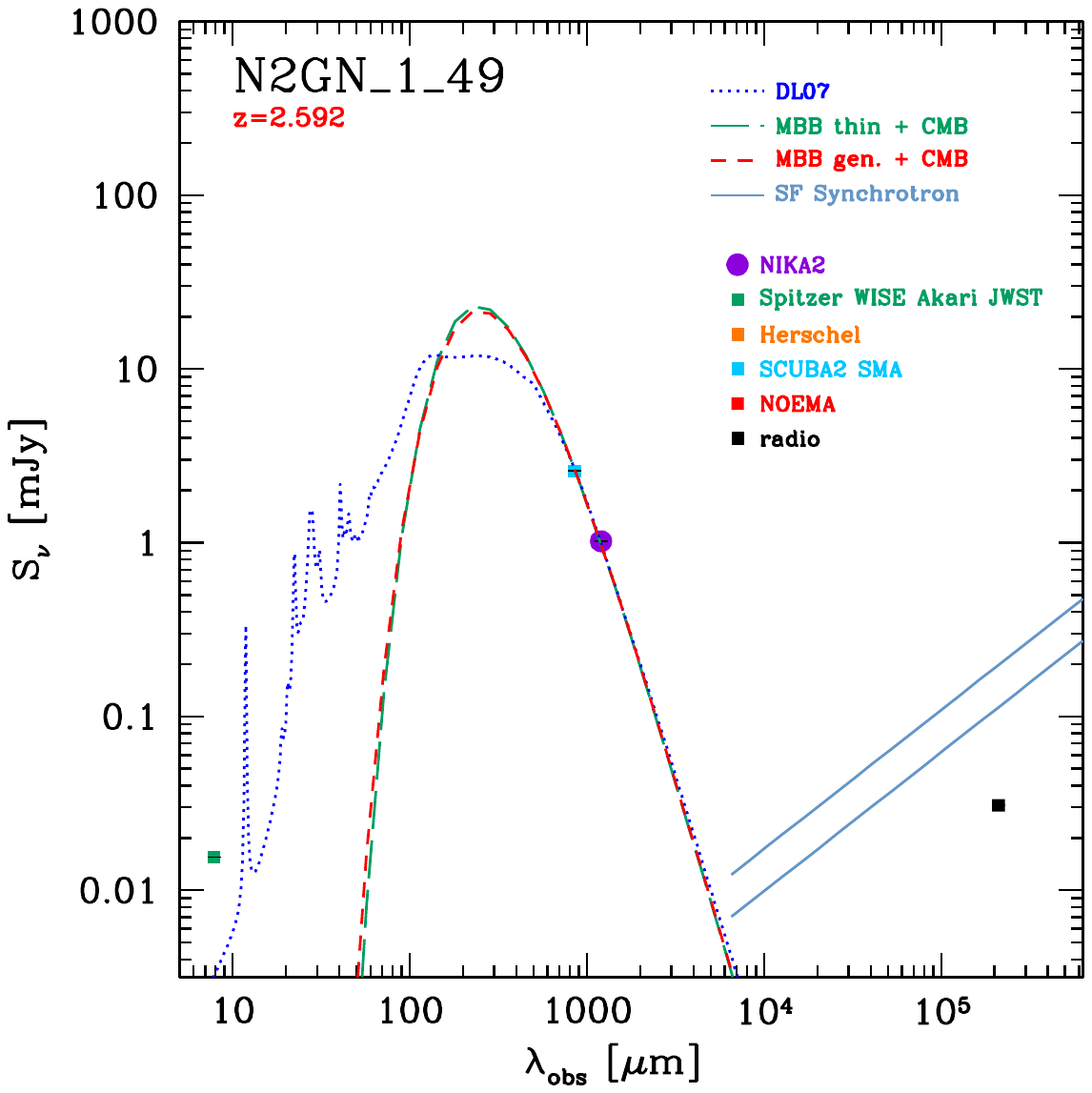}
\includegraphics[align=c,trim=0 0.18\imageheight{} 0 0.075\imageheight{}, clip, width=0.25\textwidth]{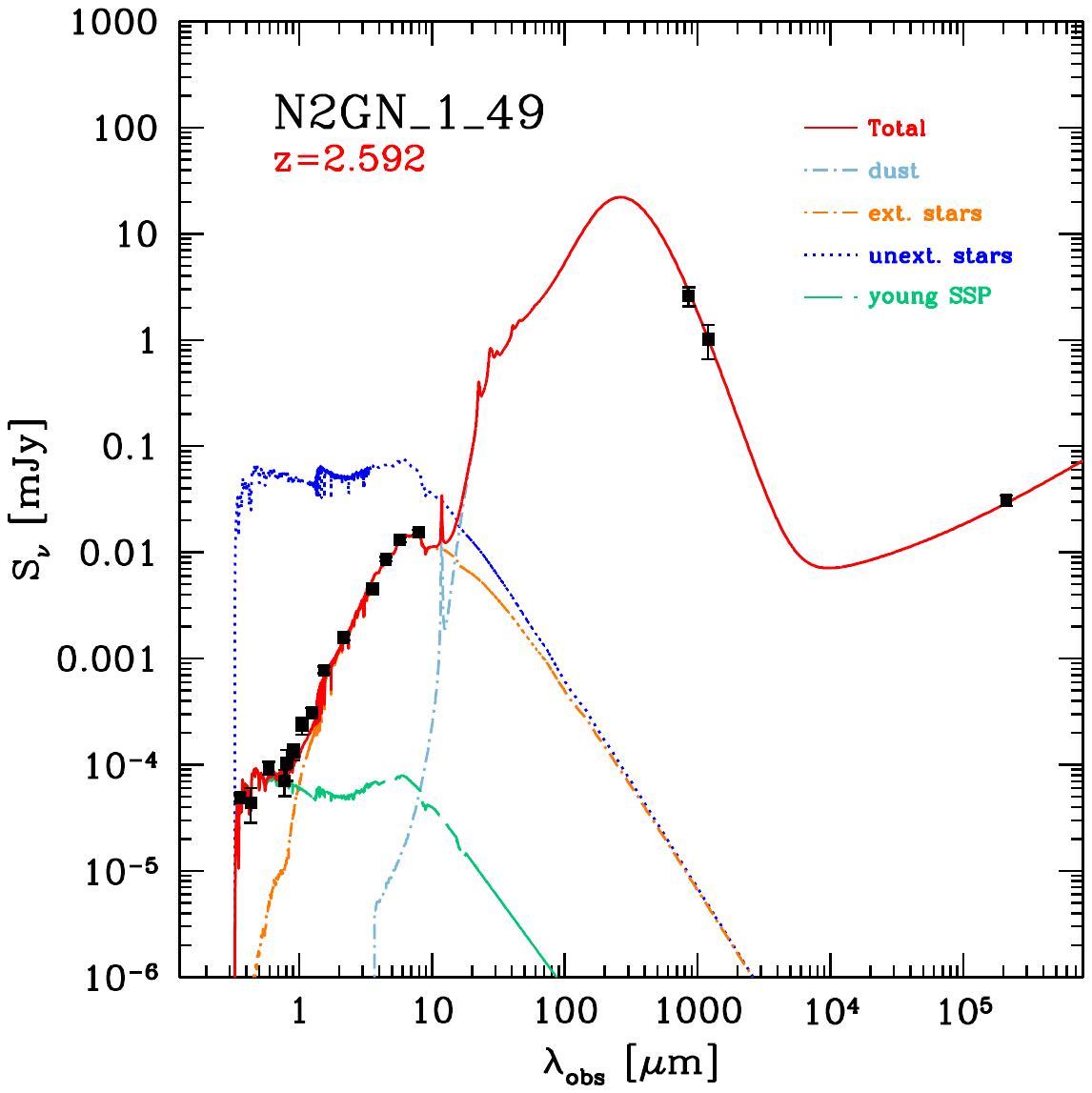}
\caption{continued.}
\end{figure*}

\addtocounter{figure}{-1}
\newpage

\begin{figure*}[t]
\centering
\includegraphics[align=c,width=0.4\textwidth]{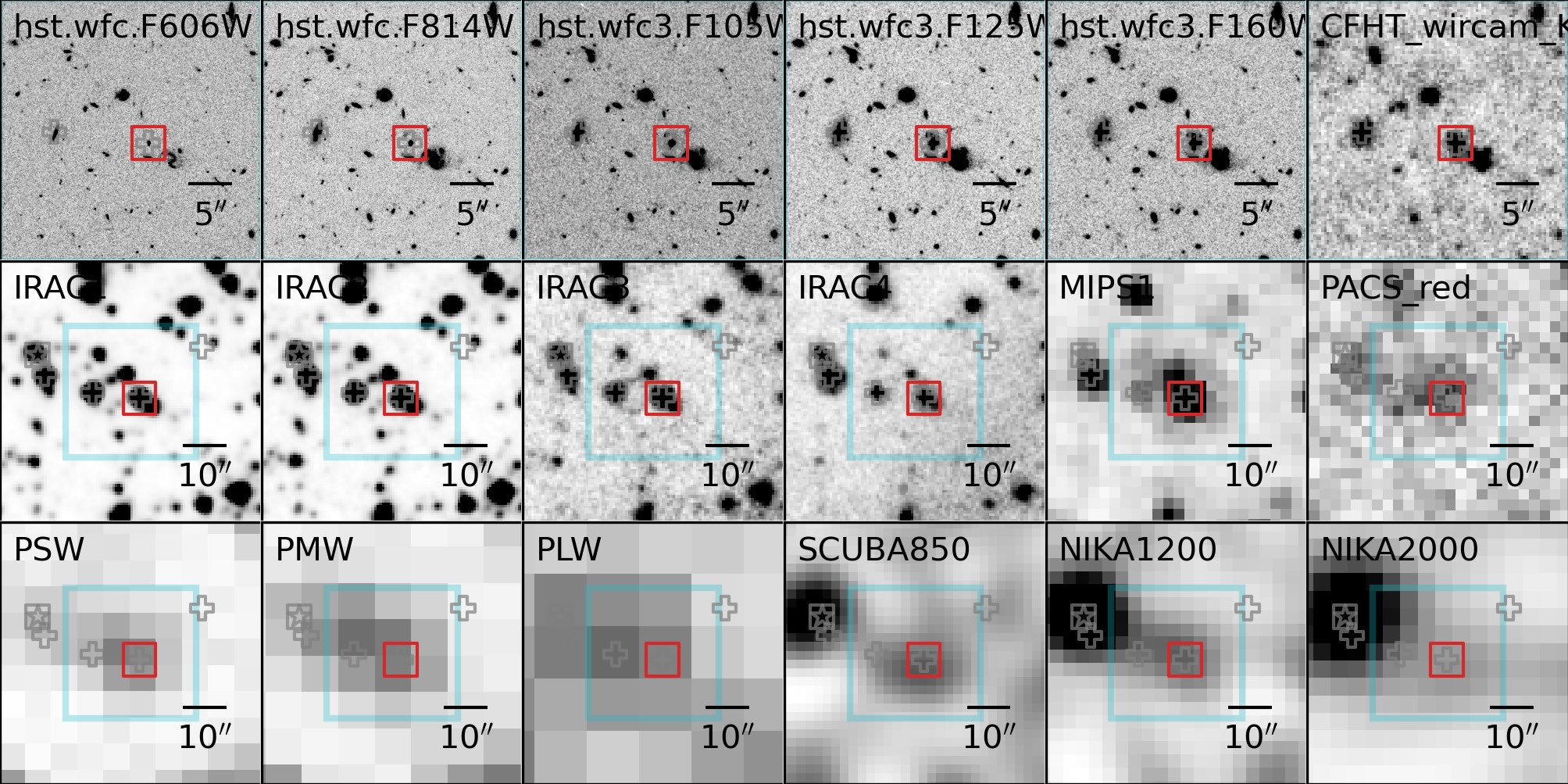}
\includegraphics[align=c,trim=0 0.18\imageheight{} 0 0.075\imageheight{}, clip, width=0.25\textwidth]{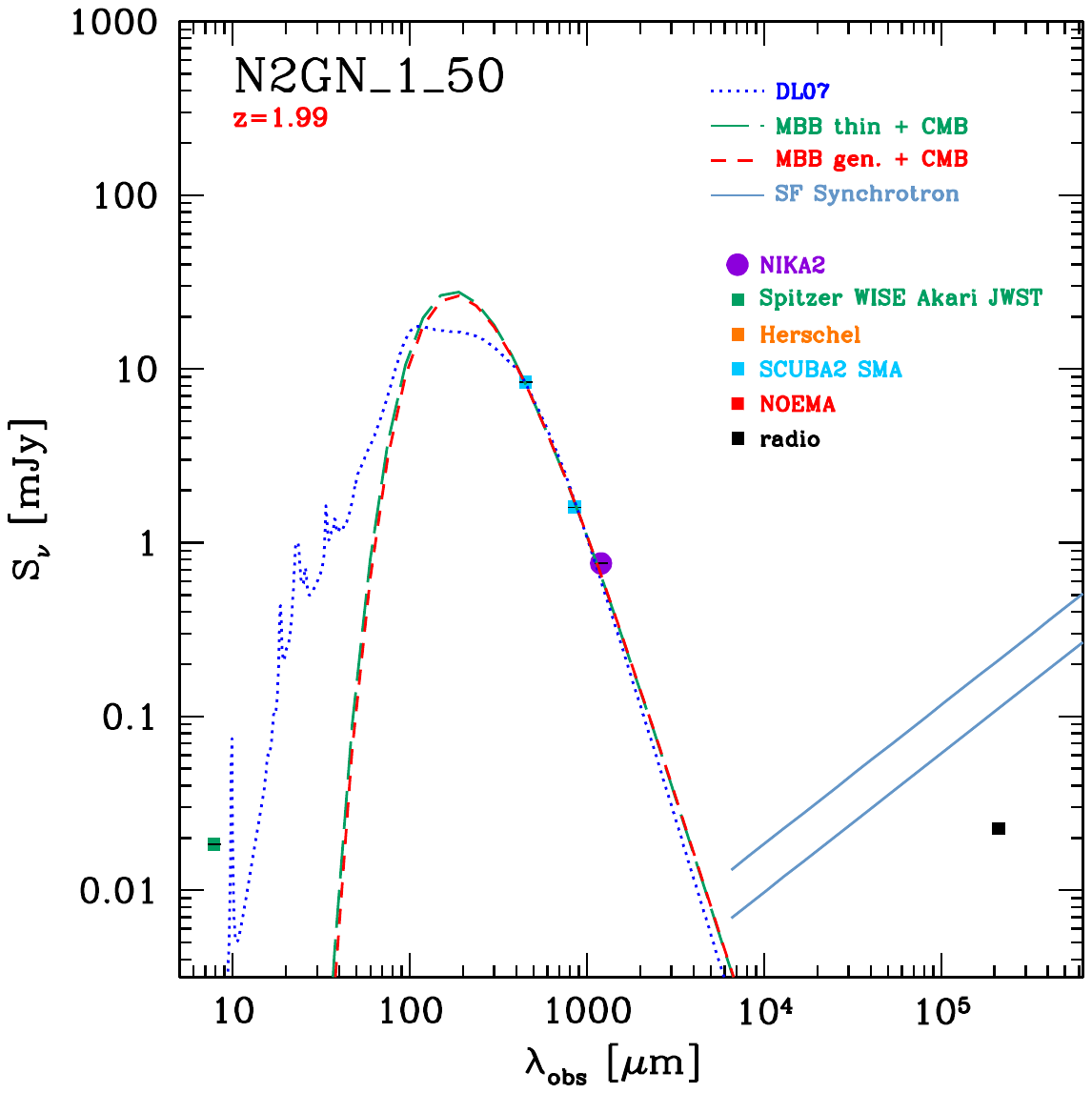}
\includegraphics[align=c,trim=0 0.18\imageheight{} 0 0.075\imageheight{}, clip, width=0.25\textwidth]{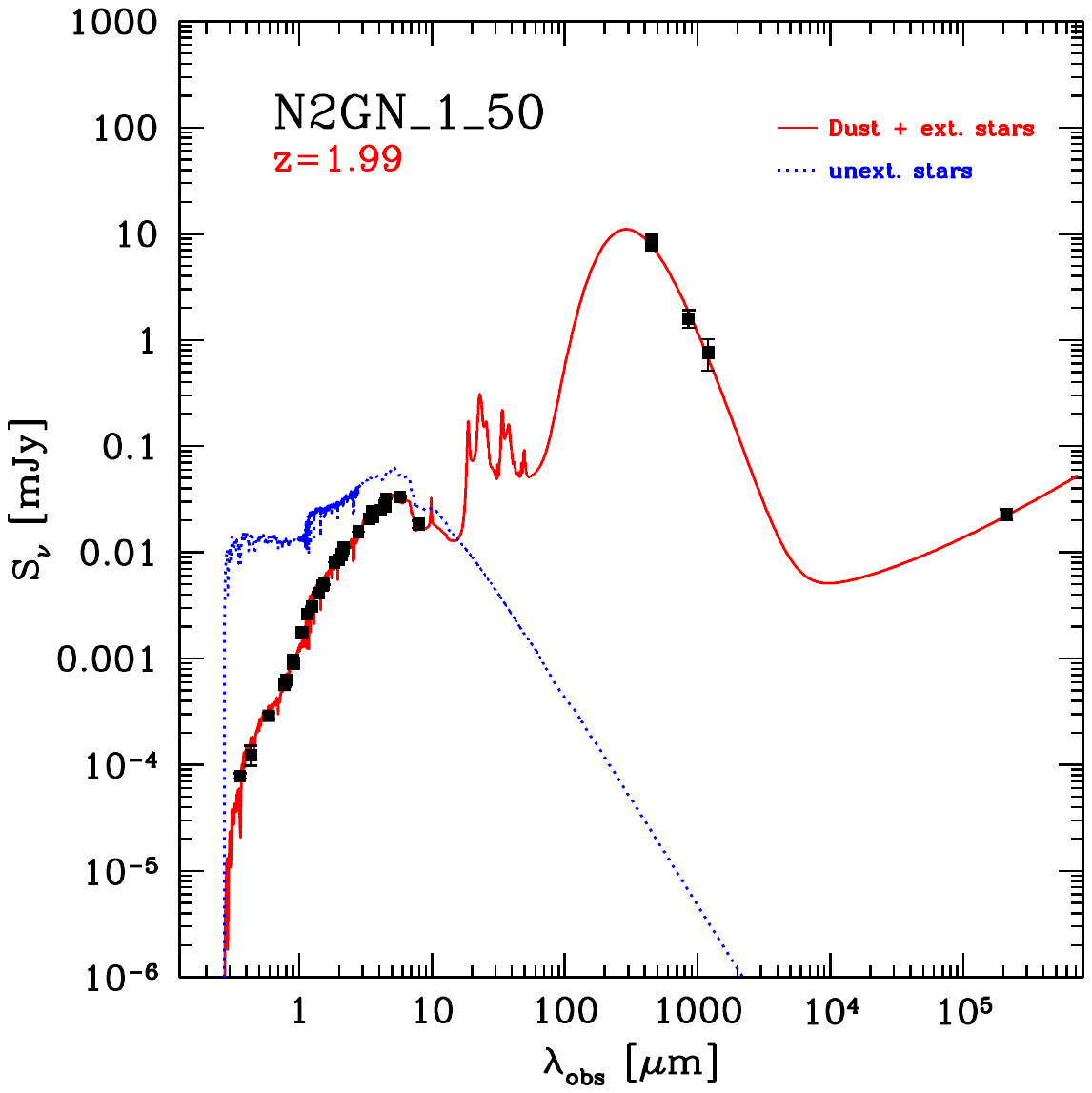}
\includegraphics[align=c,width=0.4\textwidth]{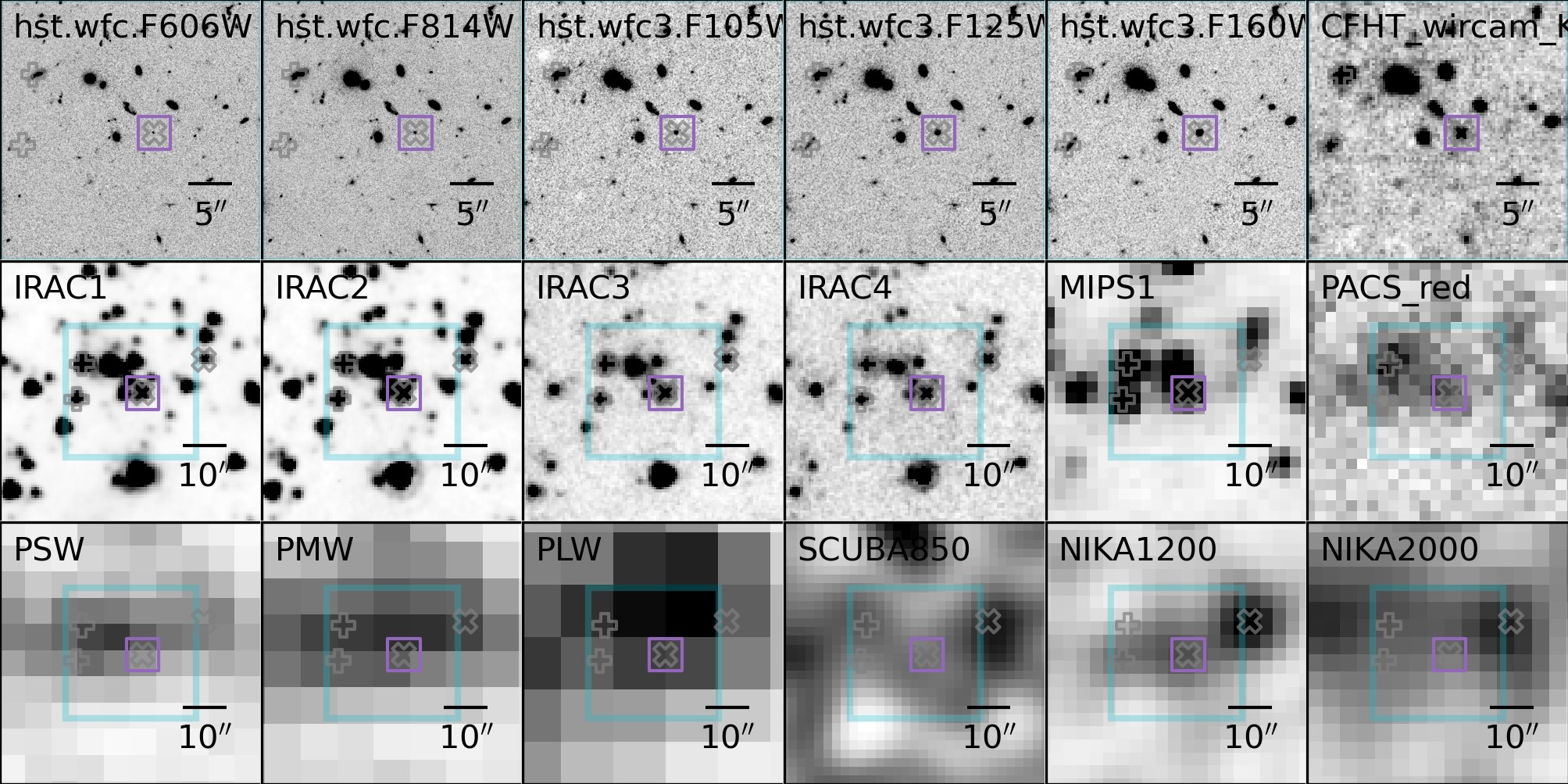}
\includegraphics[align=c,trim=0 0.18\imageheight{} 0 0.075\imageheight{}, clip, width=0.25\textwidth]{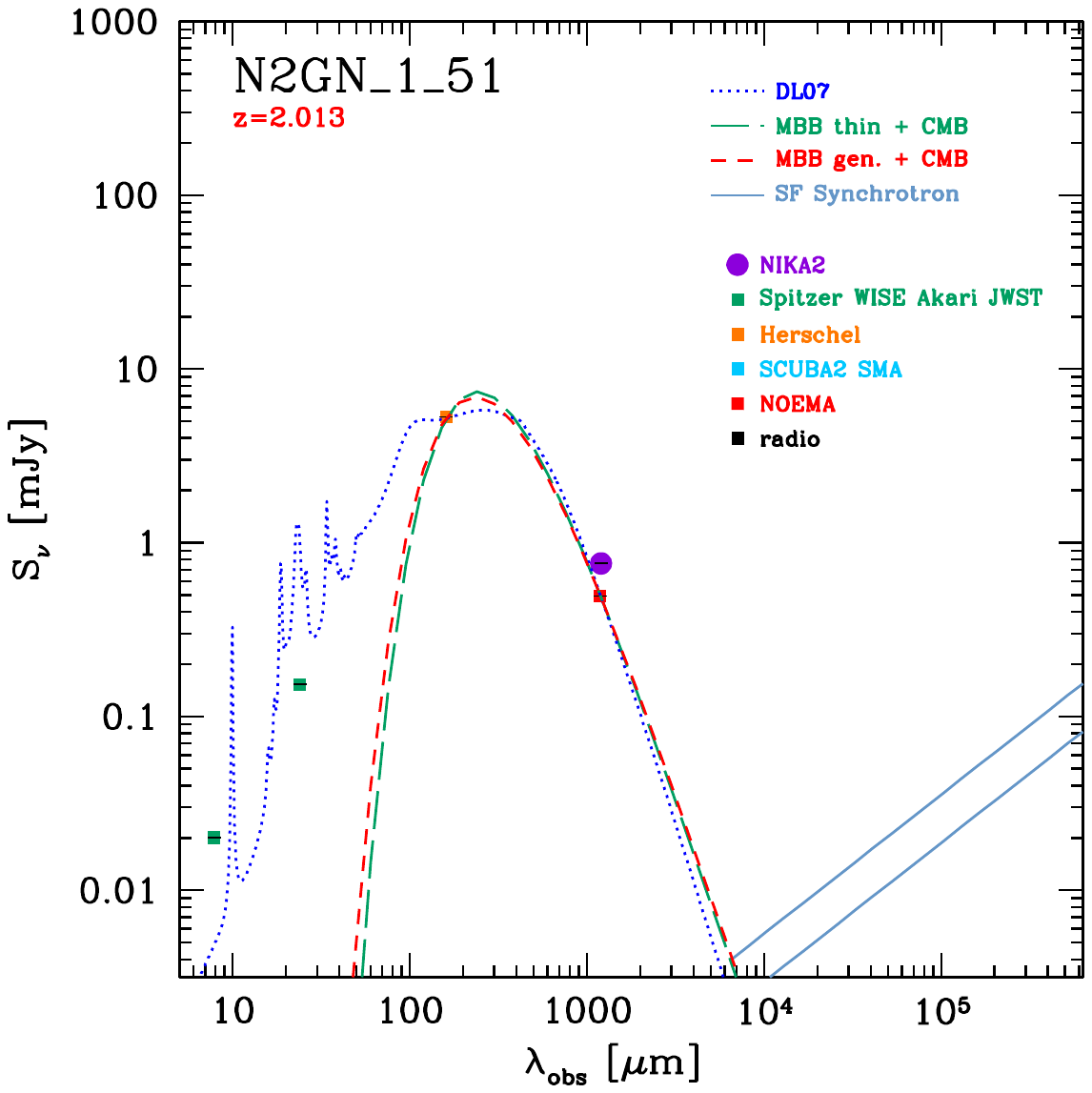}
\includegraphics[align=c,trim=0 0.18\imageheight{} 0 0.075\imageheight{}, clip, width=0.25\textwidth]{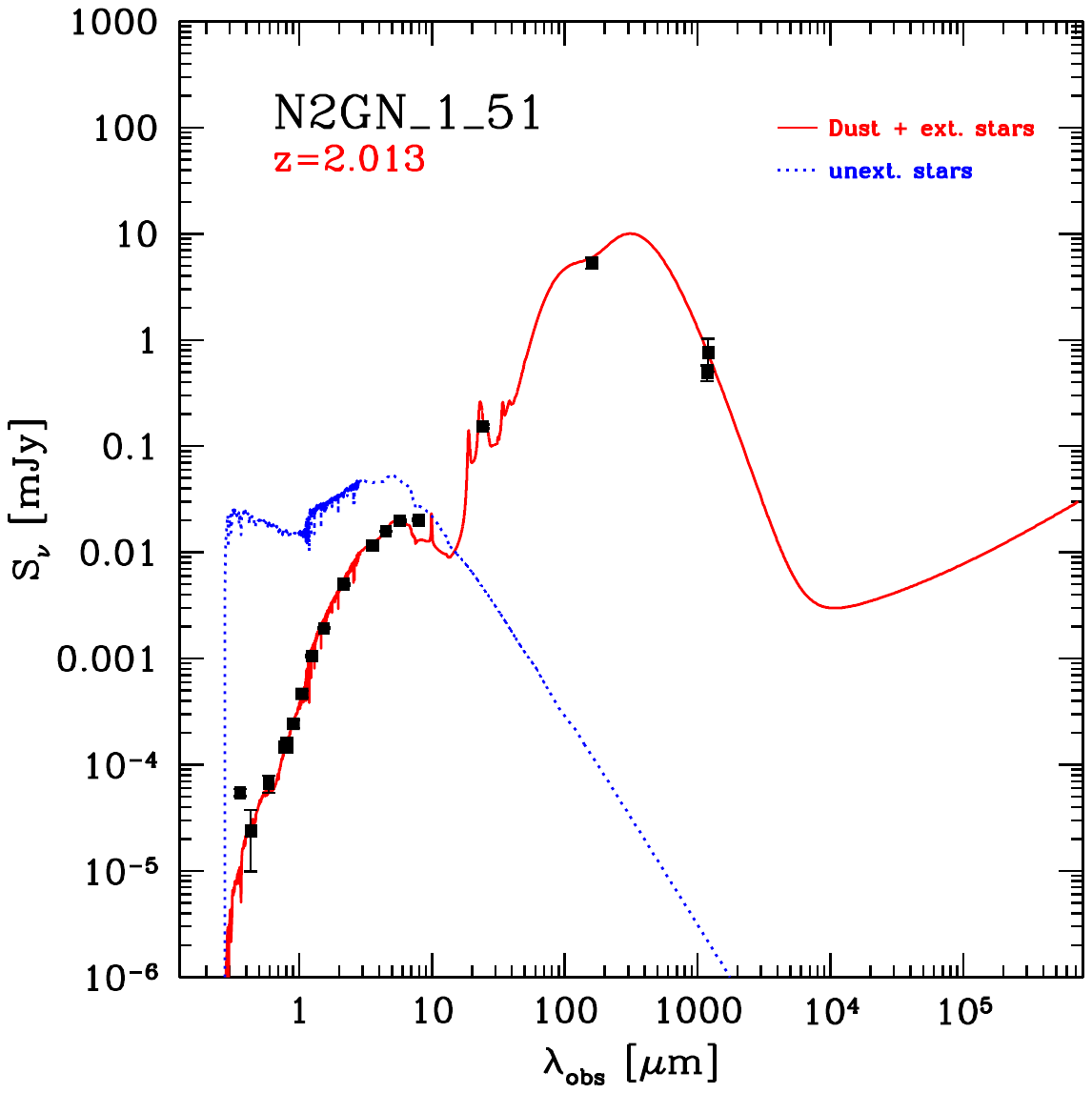}
\includegraphics[align=c,width=0.4\textwidth]{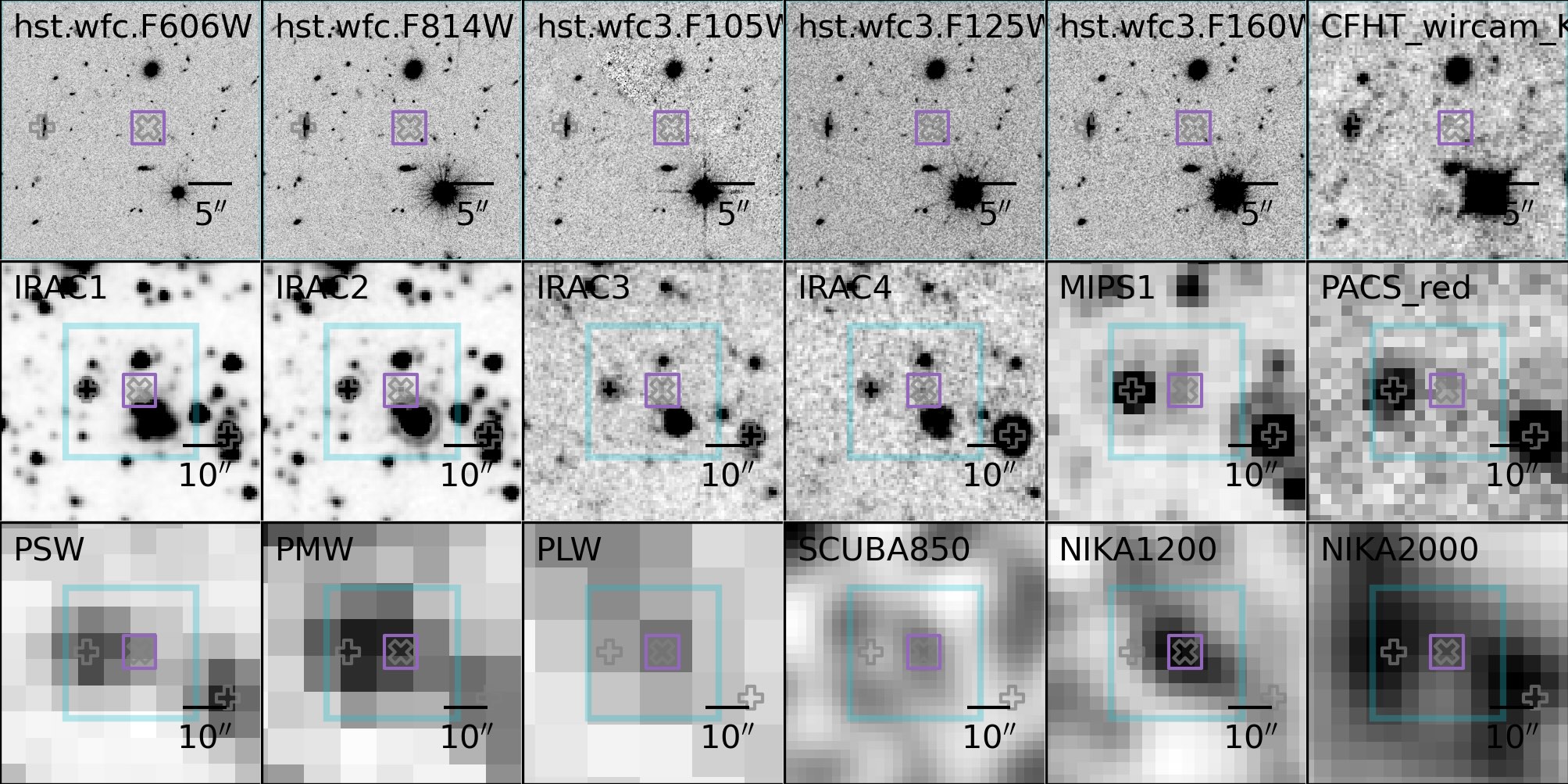}
\includegraphics[align=c,trim=0 0.18\imageheight{} 0 0.075\imageheight{}, clip, width=0.25\textwidth]{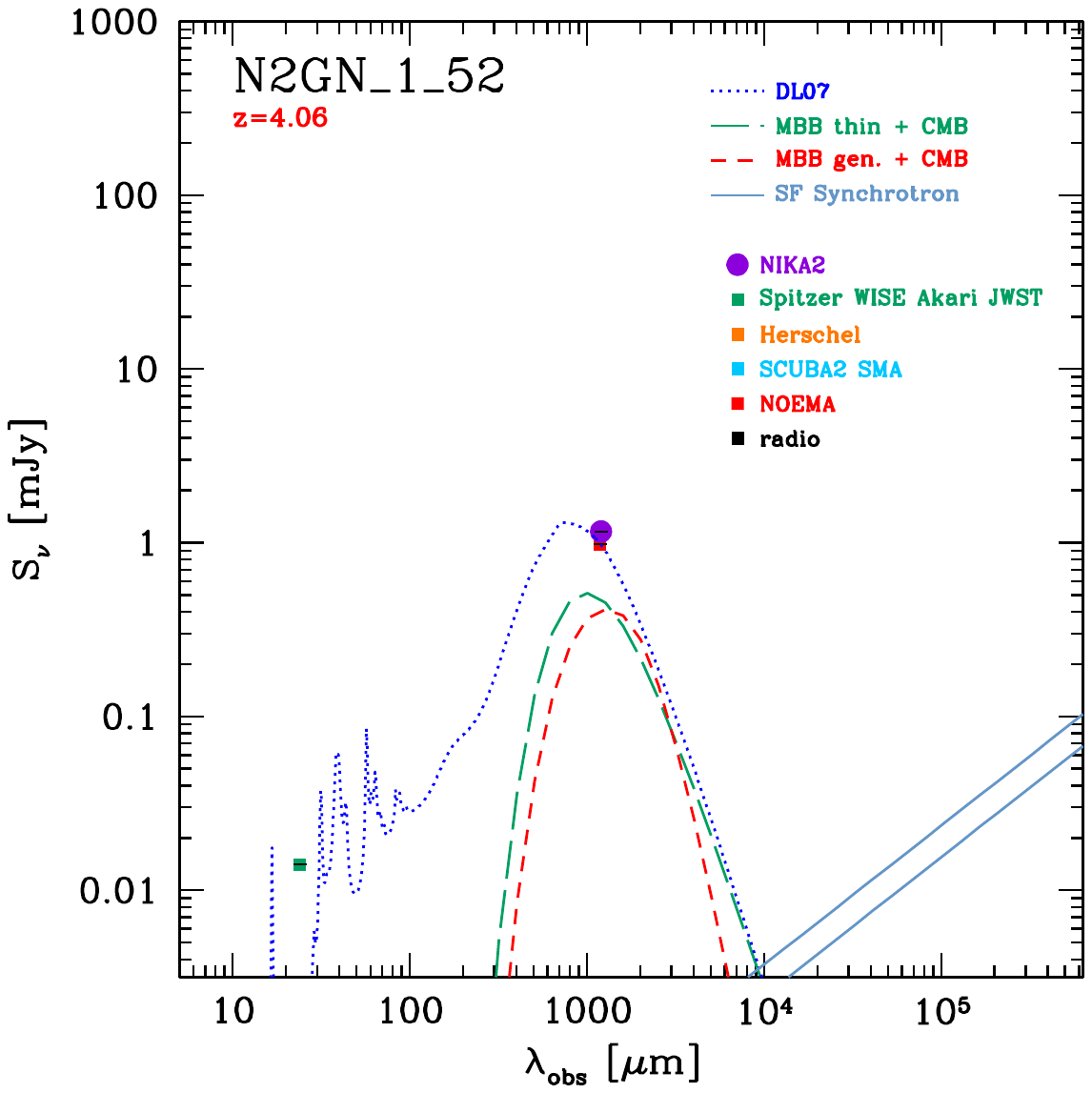}
\includegraphics[align=c,trim=0 0.18\imageheight{} 0 0.075\imageheight{}, clip, width=0.25\textwidth]{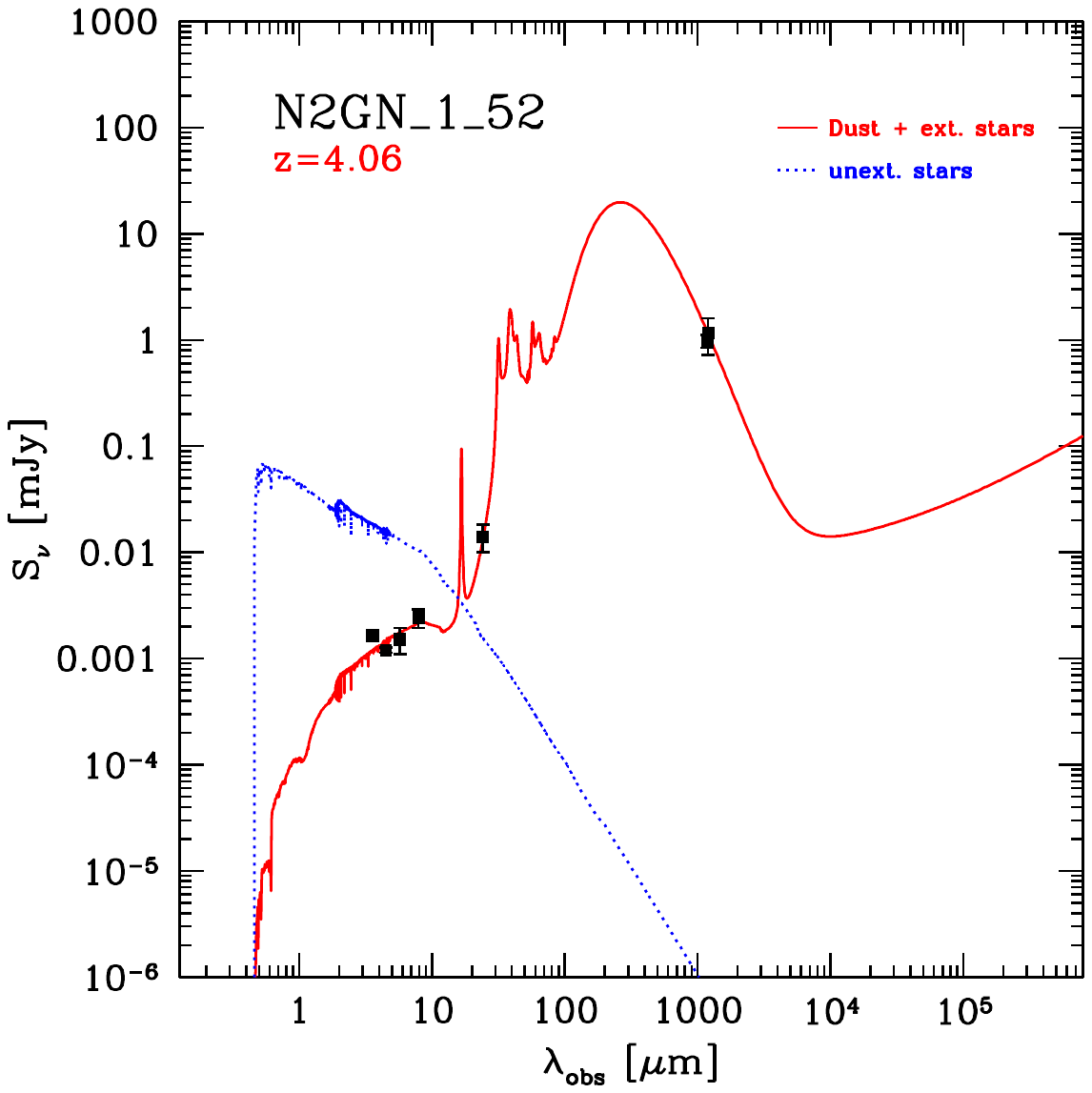}
\includegraphics[align=c,width=0.4\textwidth]{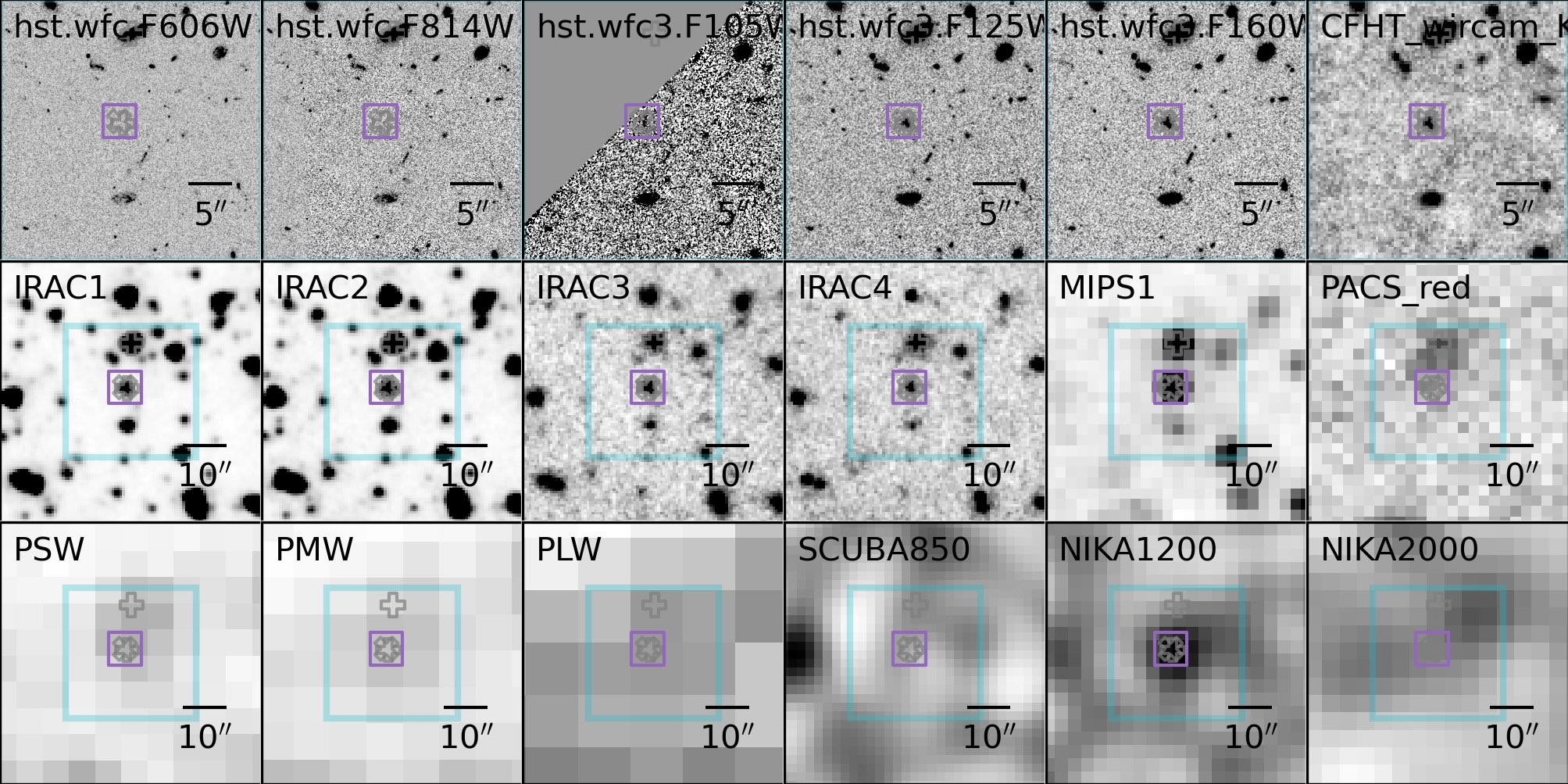}
\includegraphics[align=c,trim=0 0.18\imageheight{} 0 0.075\imageheight{}, clip, width=0.25\textwidth]{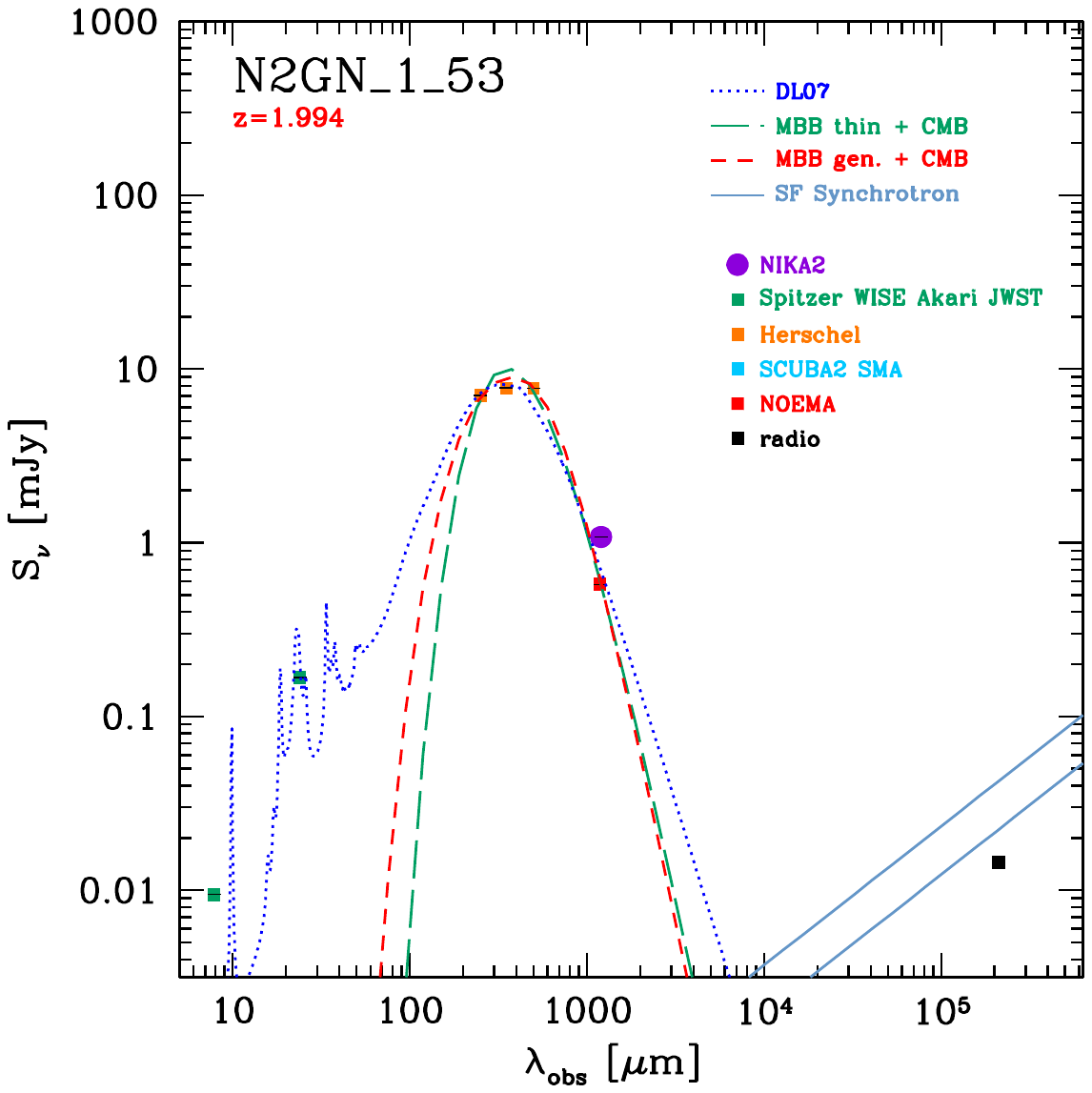}
\includegraphics[align=c,trim=0 0.18\imageheight{} 0 0.075\imageheight{}, clip, width=0.25\textwidth]{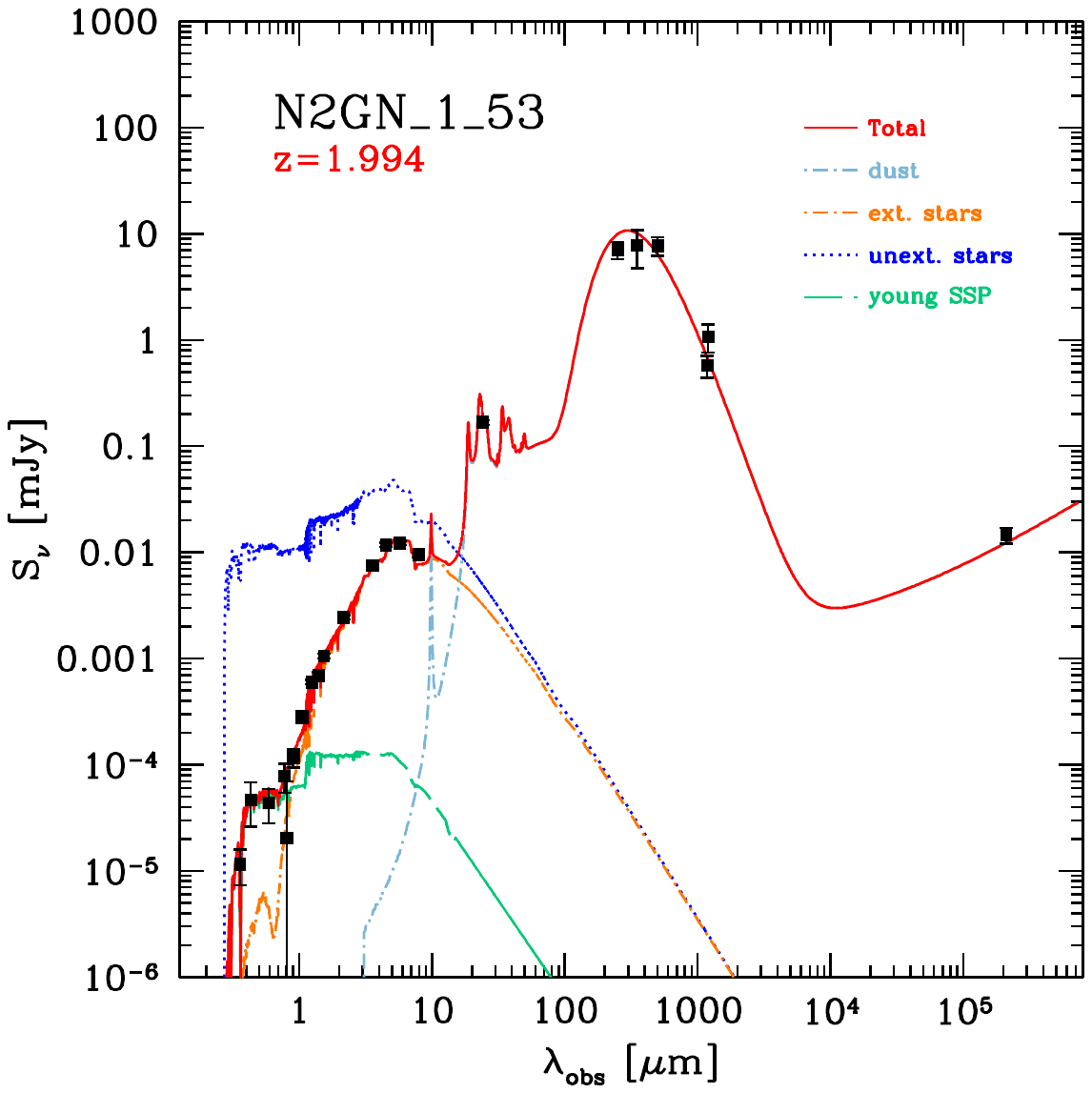}
\includegraphics[align=c,width=0.4\textwidth]{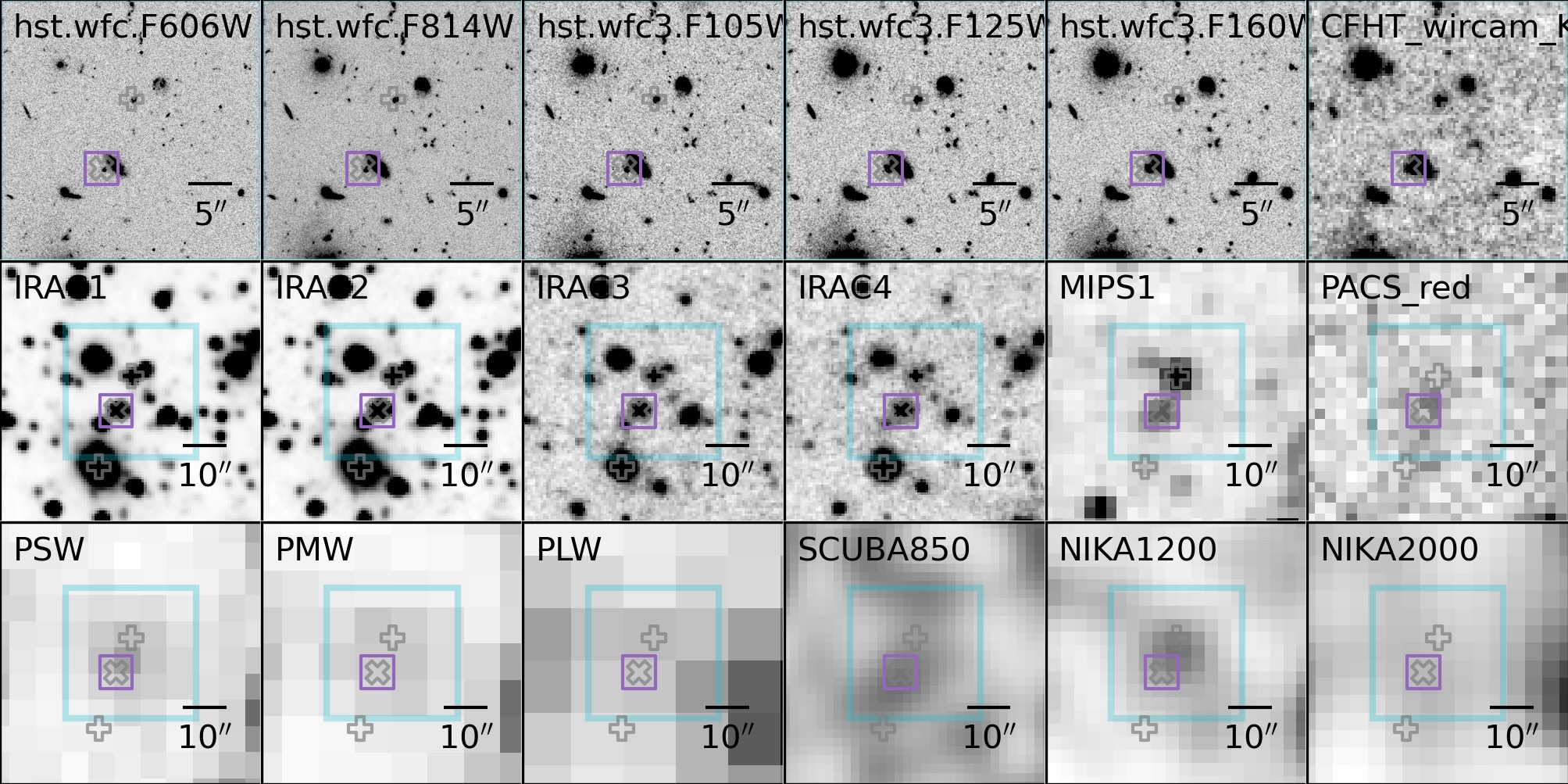}
\includegraphics[align=c,trim=0 0.18\imageheight{} 0 0.075\imageheight{}, clip, width=0.25\textwidth]{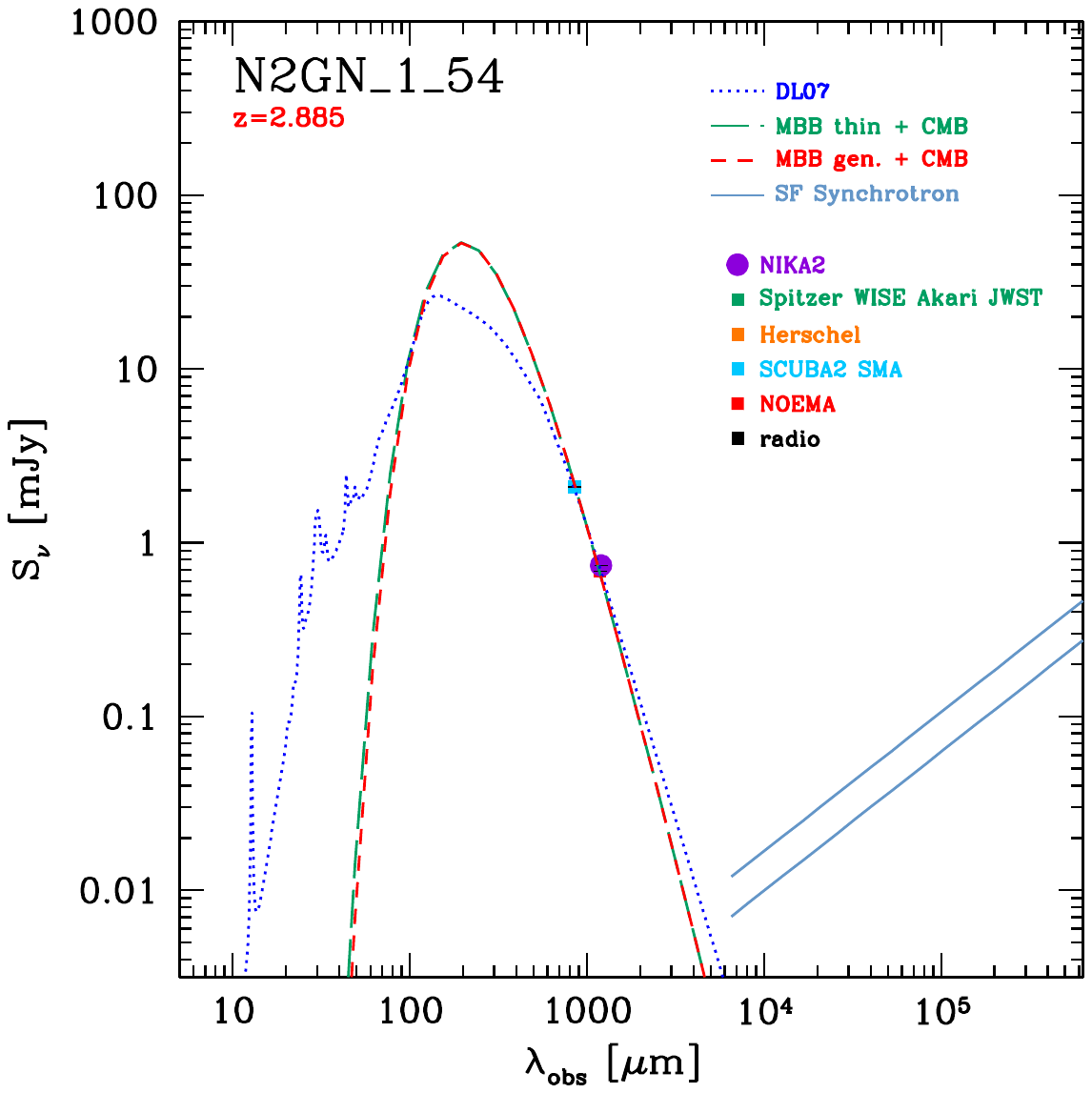}
\includegraphics[align=c,trim=0 0.18\imageheight{} 0 0.075\imageheight{}, clip, width=0.25\textwidth]{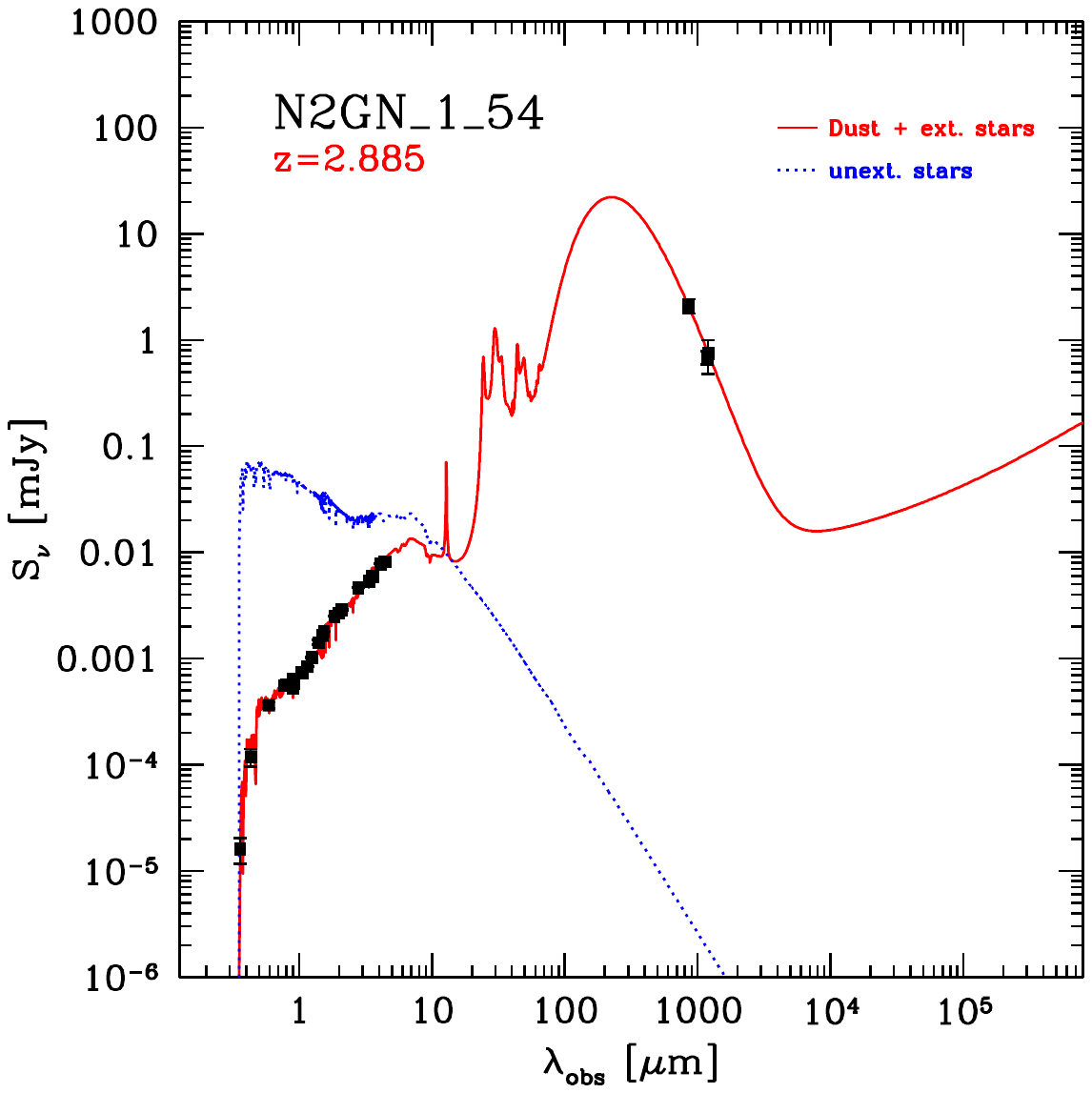}
\caption{continued.}
\end{figure*}

\addtocounter{figure}{-1}
\newpage

\begin{figure*}[t]
\centering
\includegraphics[align=c,width=0.4\textwidth]{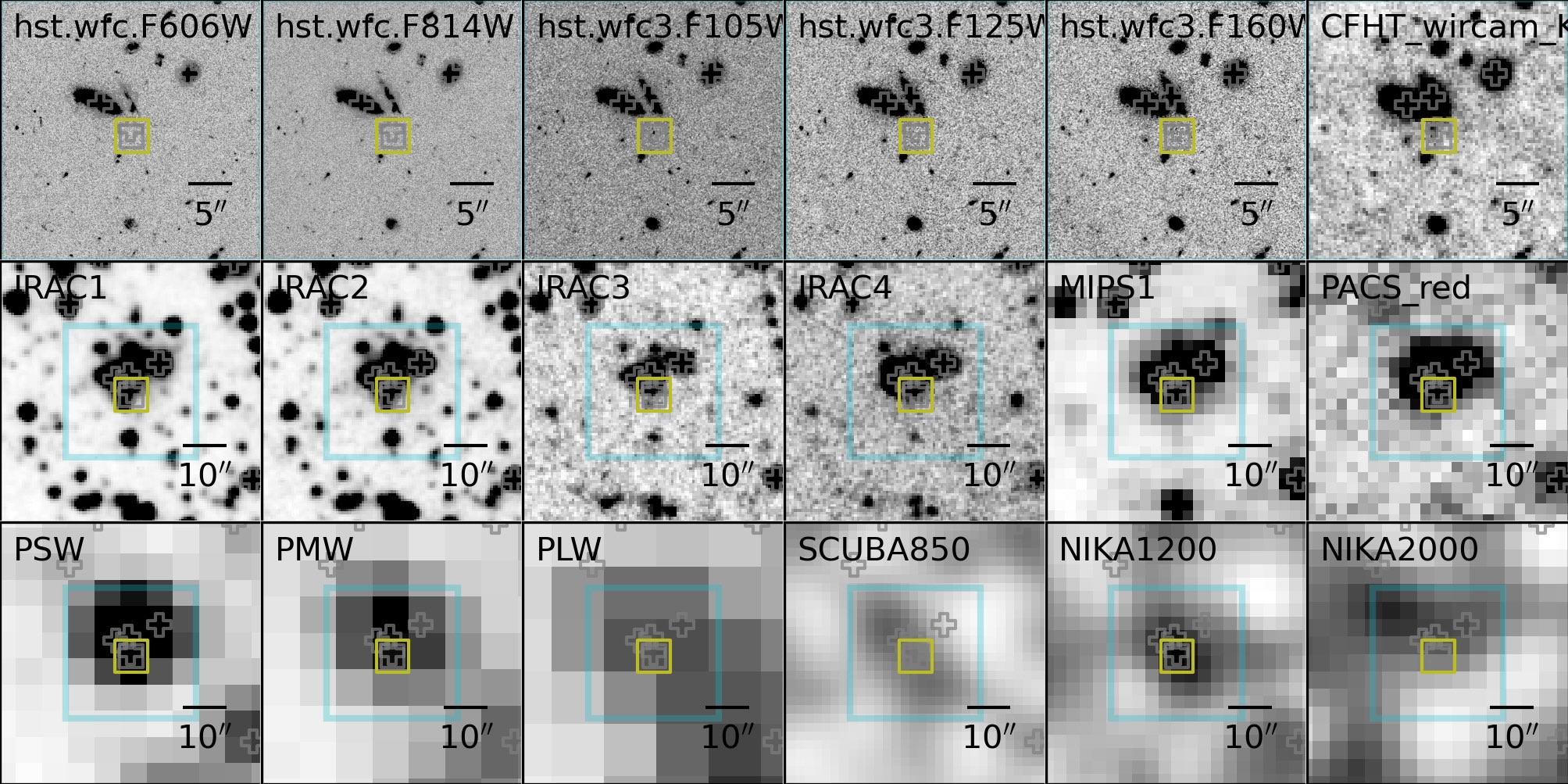}
\includegraphics[align=c,trim=0 0.18\imageheight{} 0 0.075\imageheight{}, clip, width=0.25\textwidth]{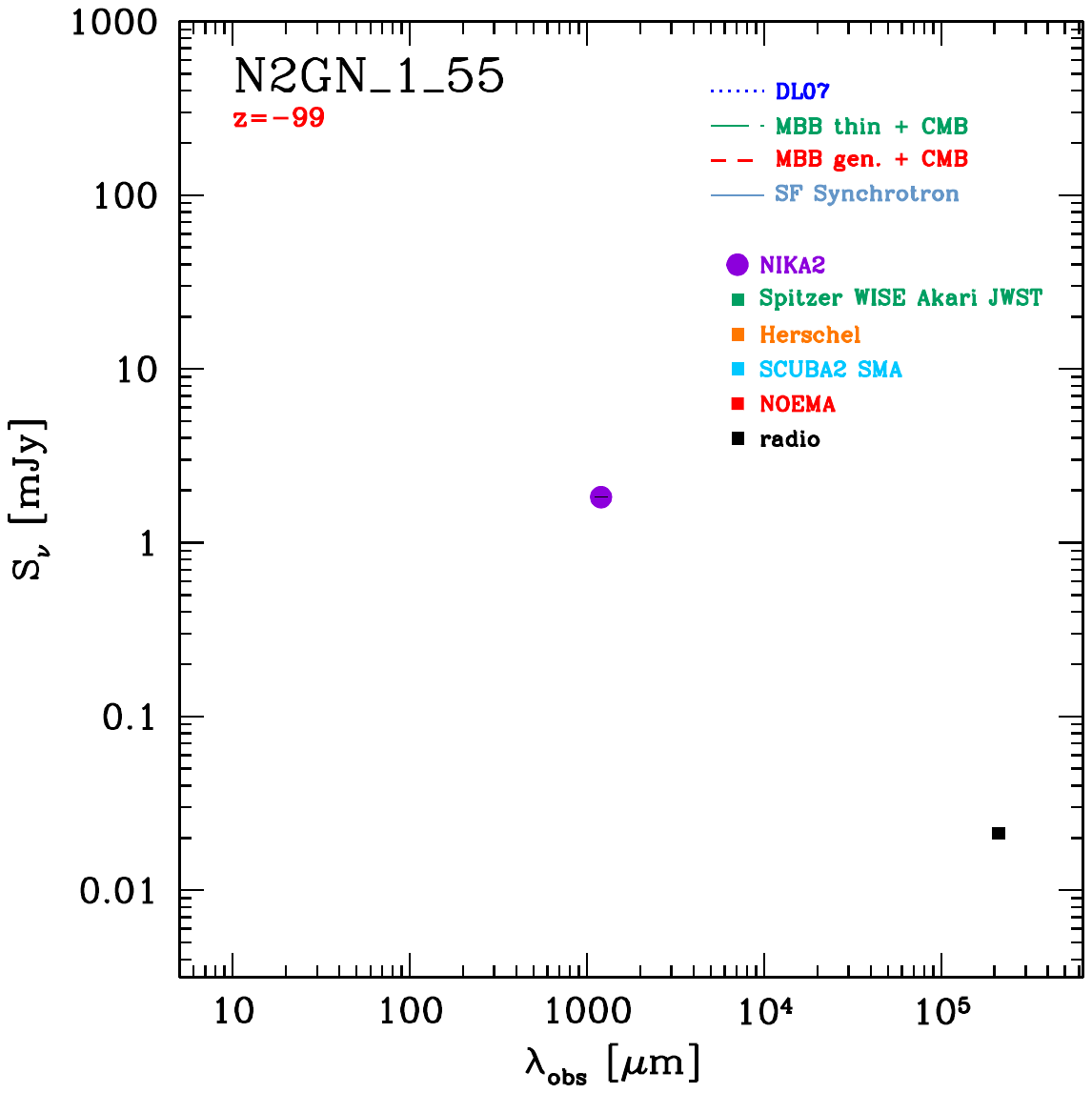}
\includegraphics[align=c,trim=0 0.18\imageheight{} 0 0.075\imageheight{}, clip, width=0.25\textwidth]{figs2_indiv_objs/white.pdf}
\includegraphics[align=c,width=0.4\textwidth]{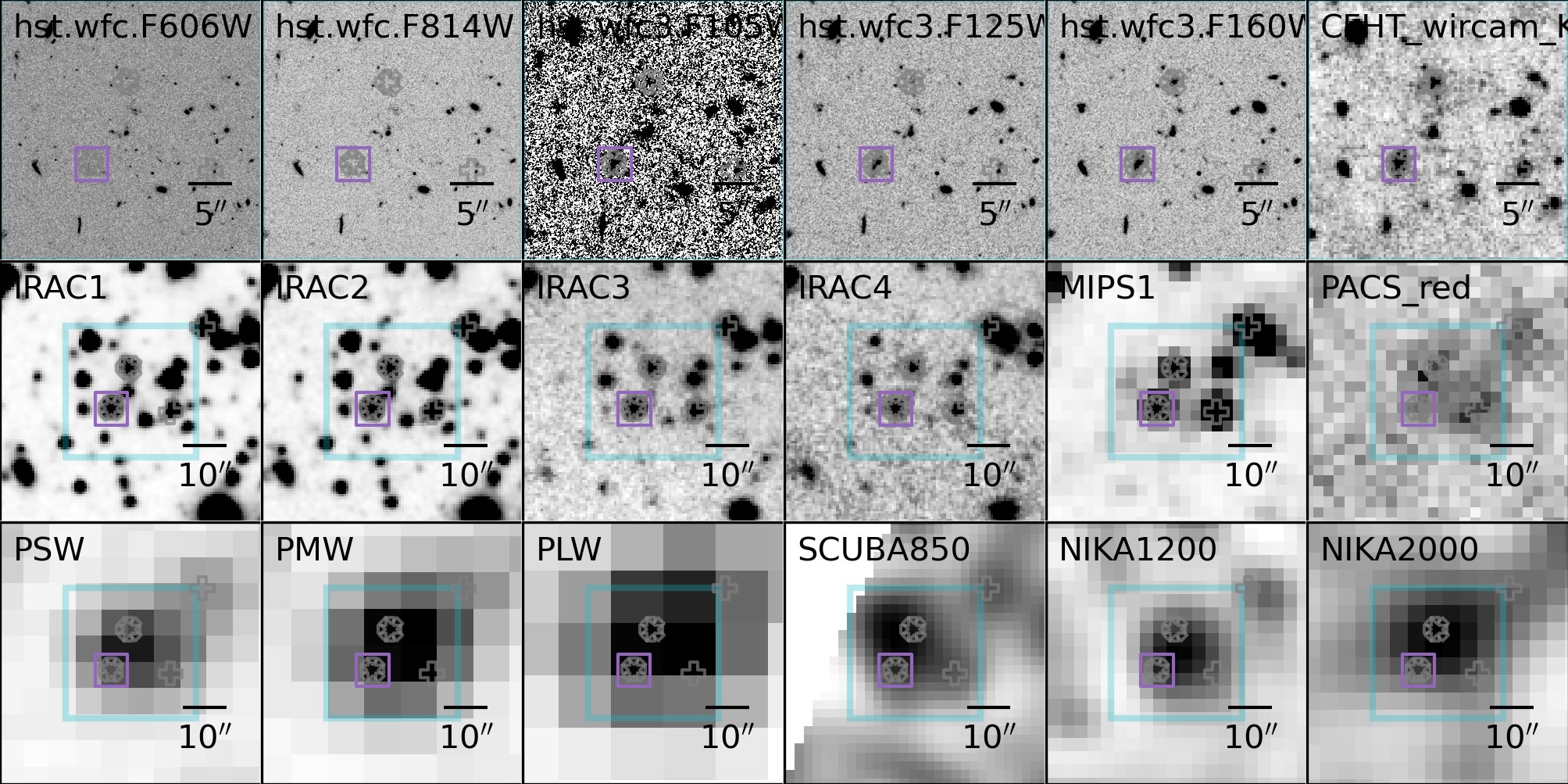}
\includegraphics[align=c,trim=0 0.18\imageheight{} 0 0.075\imageheight{}, clip, width=0.25\textwidth]{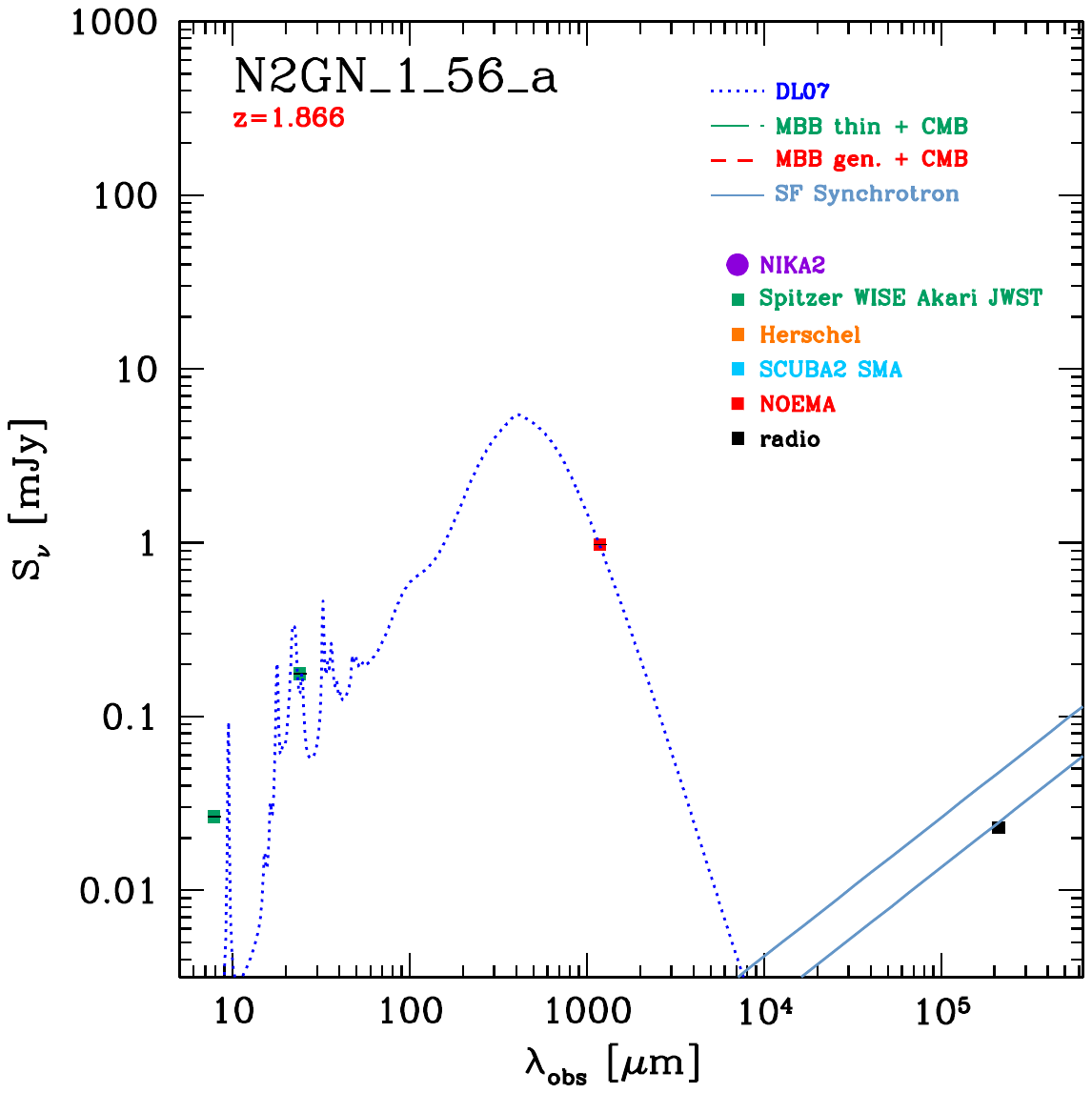}
\includegraphics[align=c,trim=0 0.18\imageheight{} 0 0.075\imageheight{}, clip, width=0.25\textwidth]{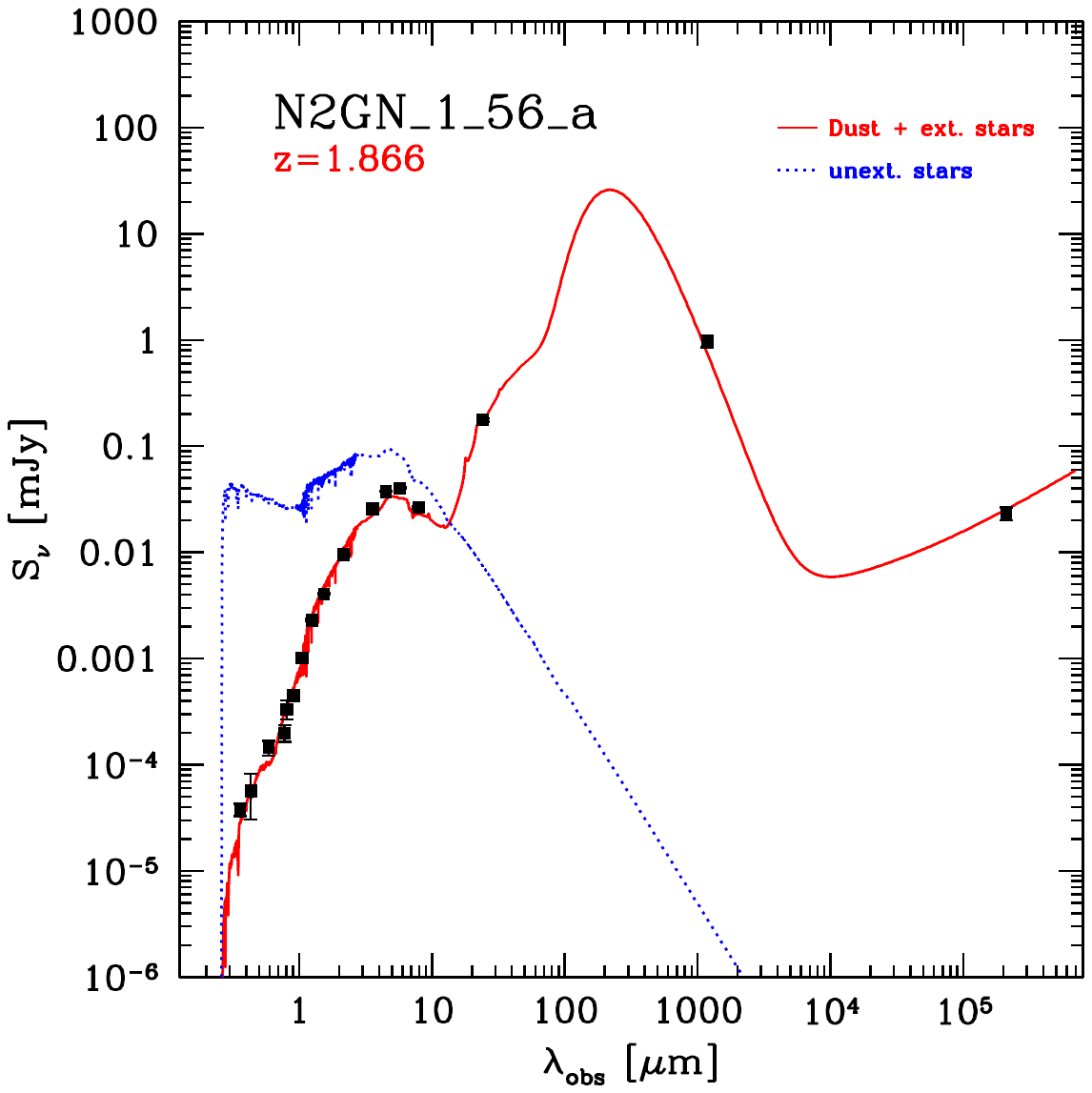}
\includegraphics[align=c,width=0.4\textwidth]{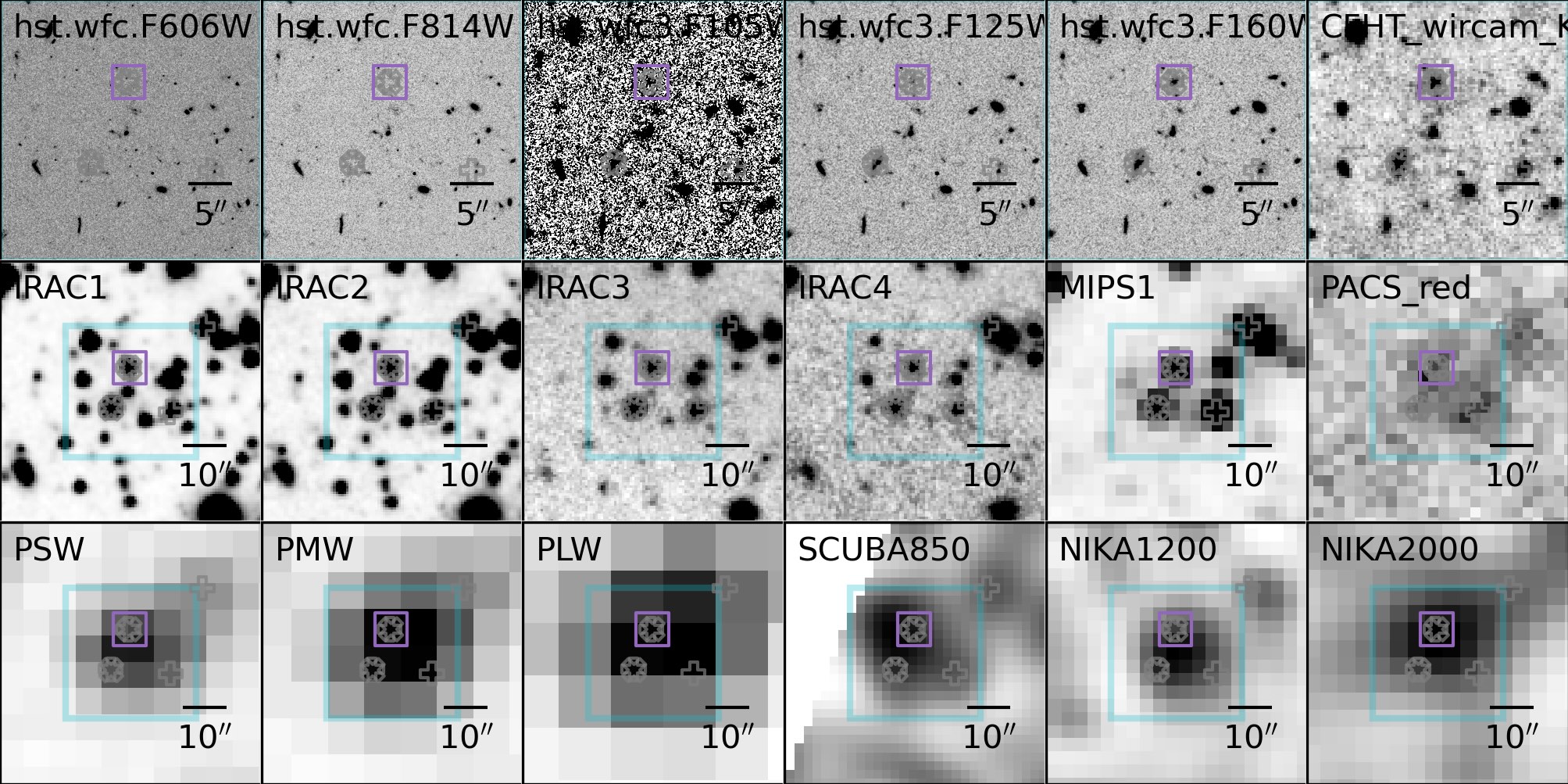}
\includegraphics[align=c,trim=0 0.18\imageheight{} 0 0.075\imageheight{}, clip, width=0.25\textwidth]{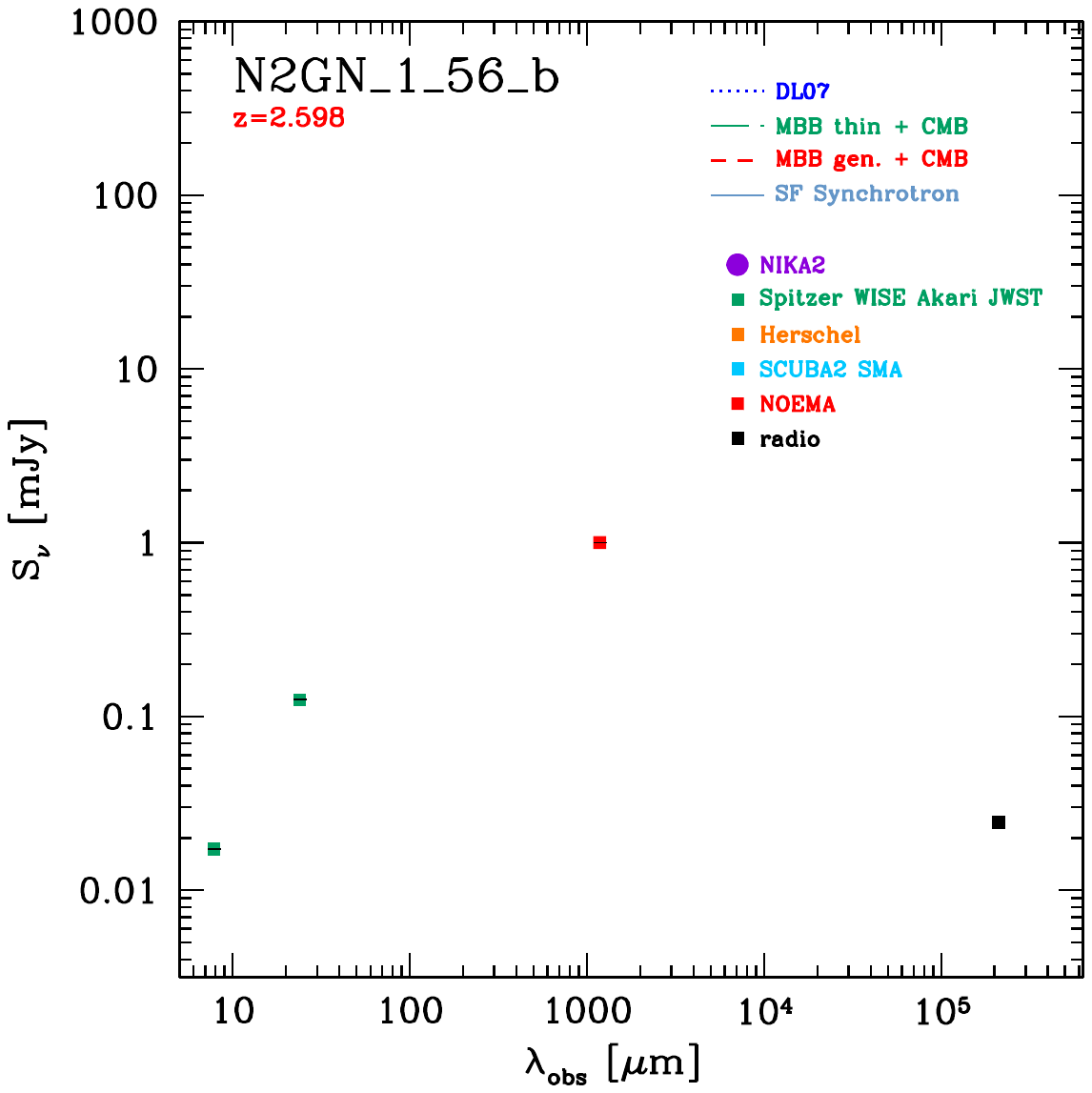}
\includegraphics[align=c,trim=0 0.18\imageheight{} 0 0.075\imageheight{}, clip, width=0.25\textwidth]{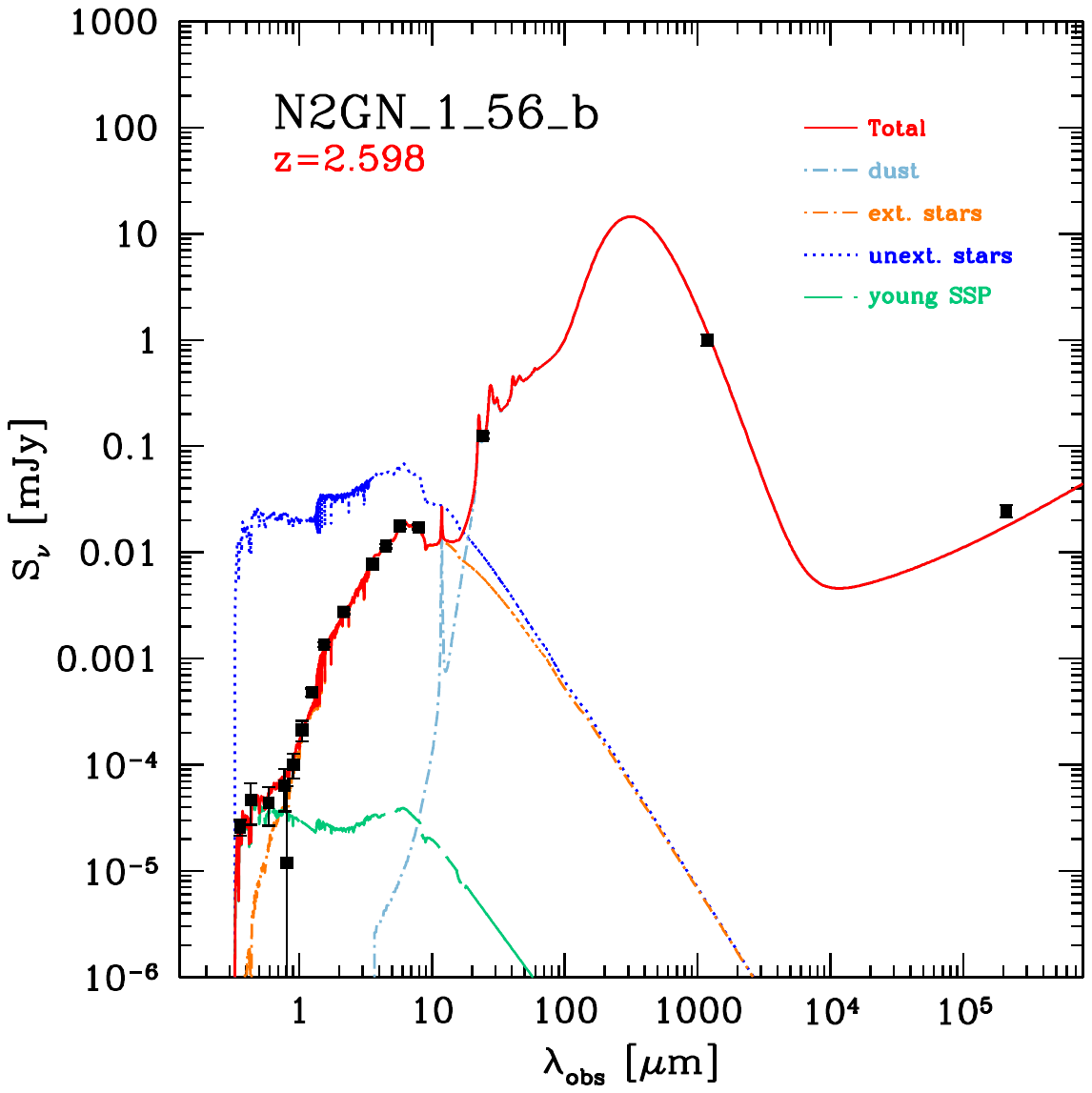}
\includegraphics[align=c,width=0.4\textwidth]{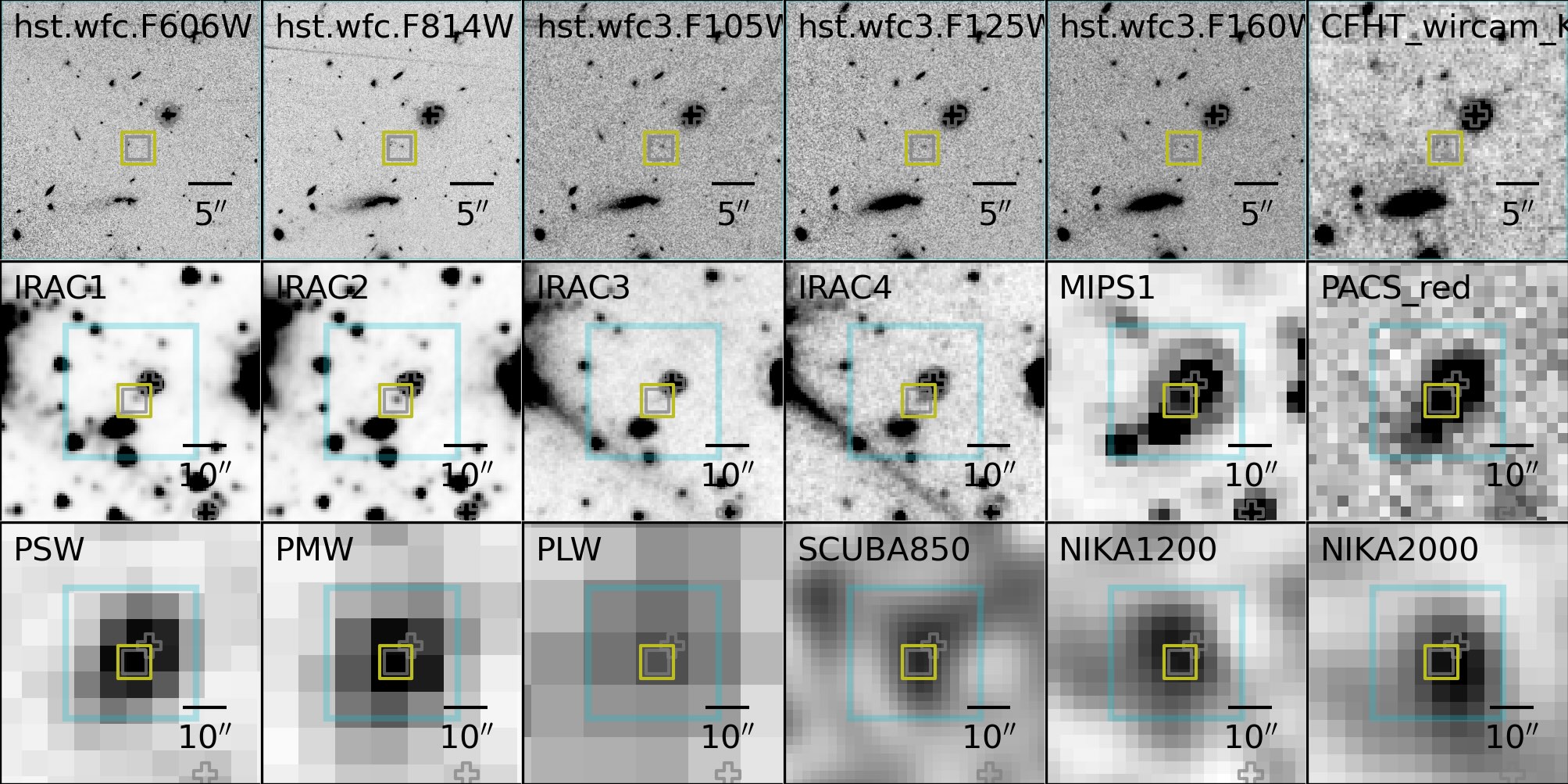}
\includegraphics[align=c,trim=0 0.18\imageheight{} 0 0.075\imageheight{}, clip, width=0.25\textwidth]{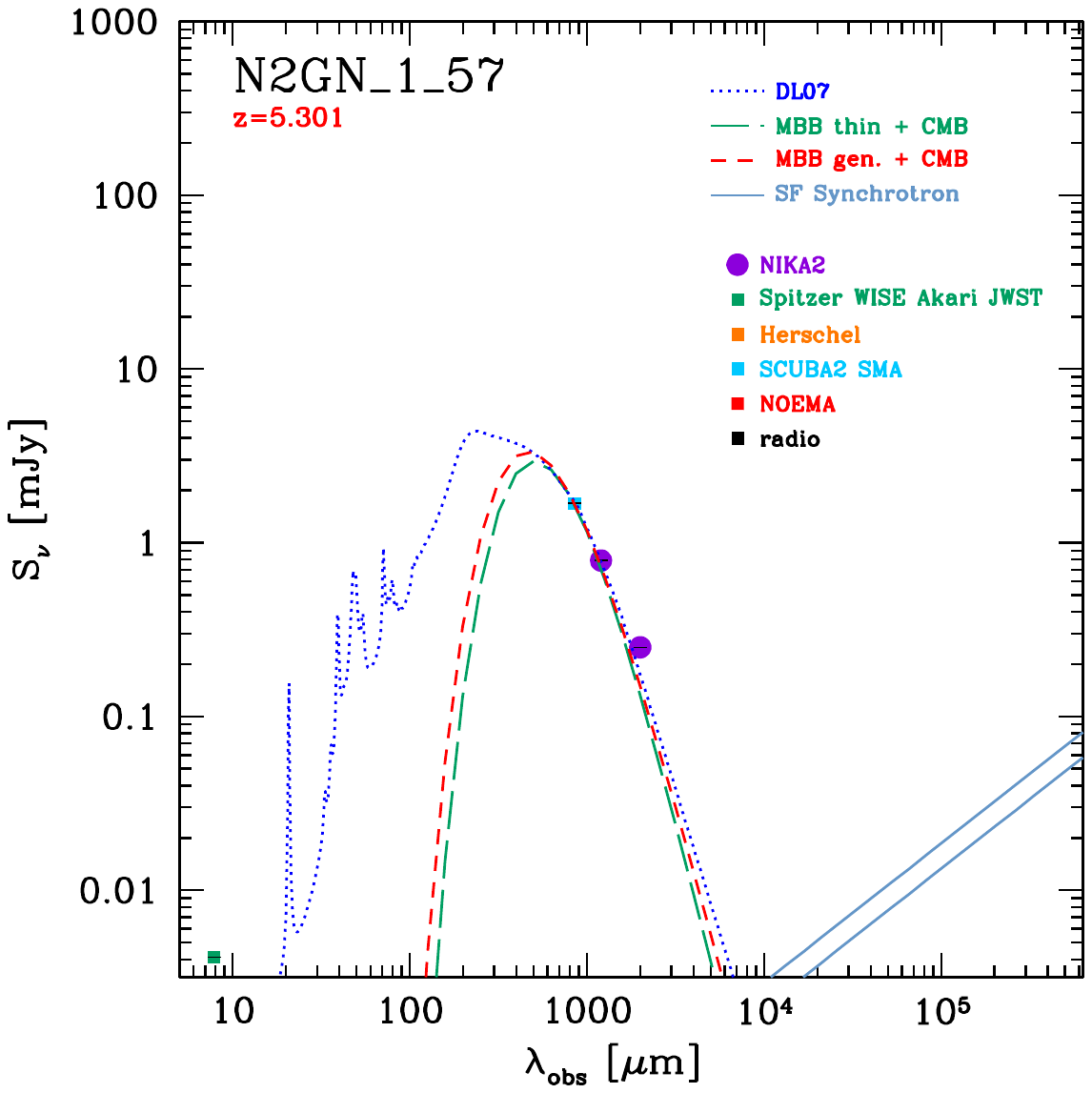}
\includegraphics[align=c,trim=0 0.18\imageheight{} 0 0.075\imageheight{}, clip, width=0.25\textwidth]{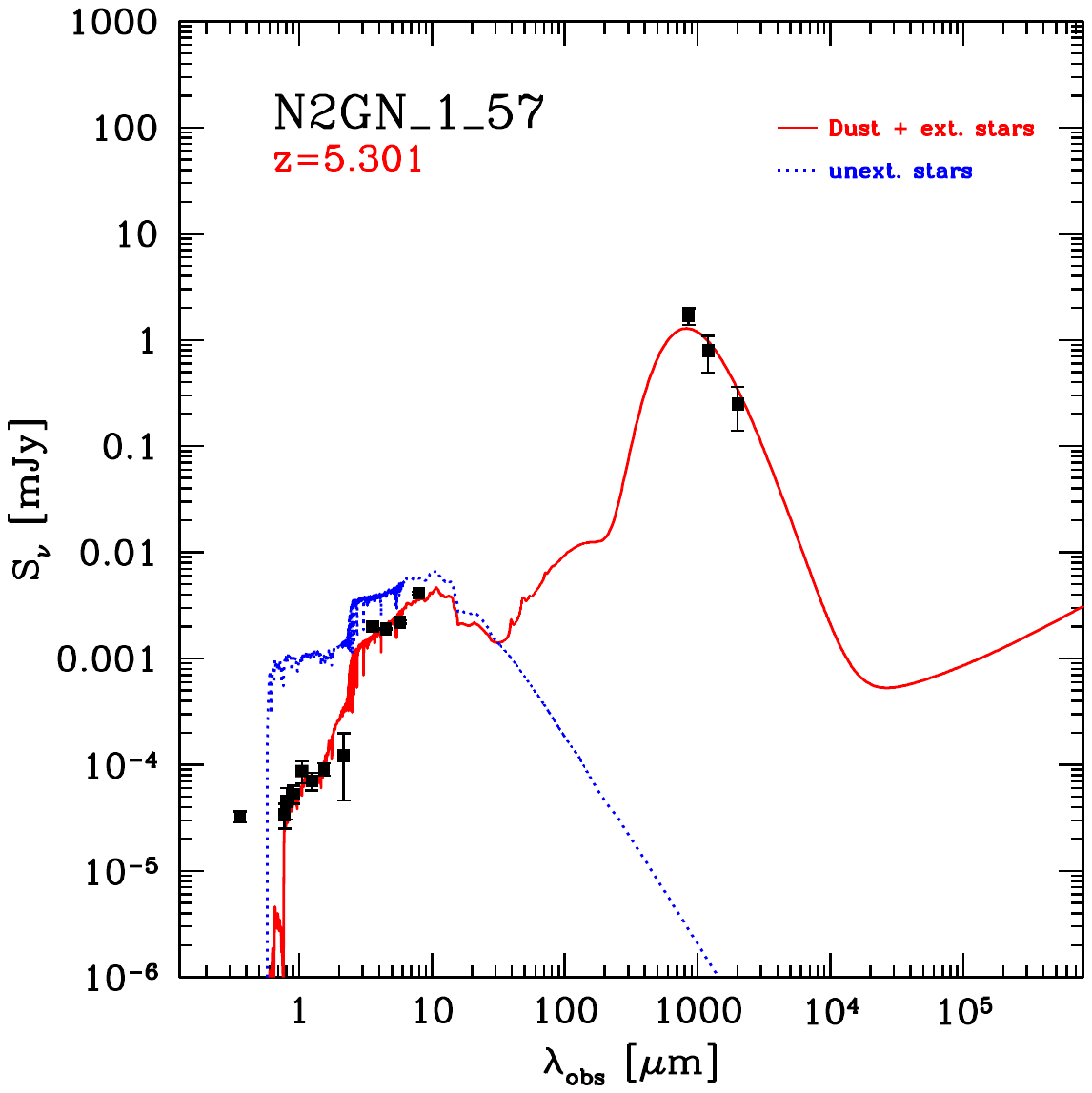}
\includegraphics[align=c,width=0.4\textwidth]{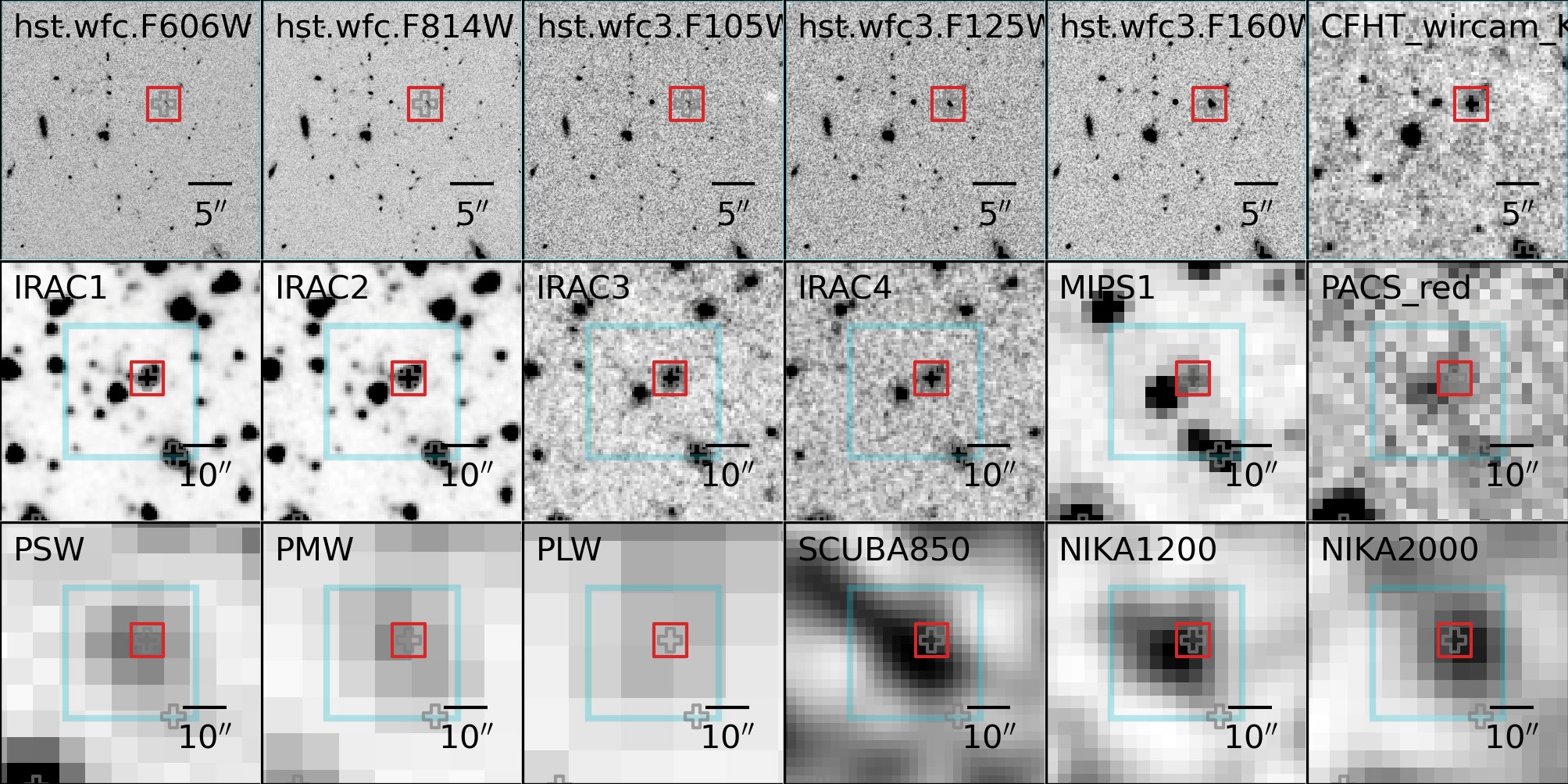}
\includegraphics[align=c,trim=0 0.18\imageheight{} 0 0.075\imageheight{}, clip, width=0.25\textwidth]{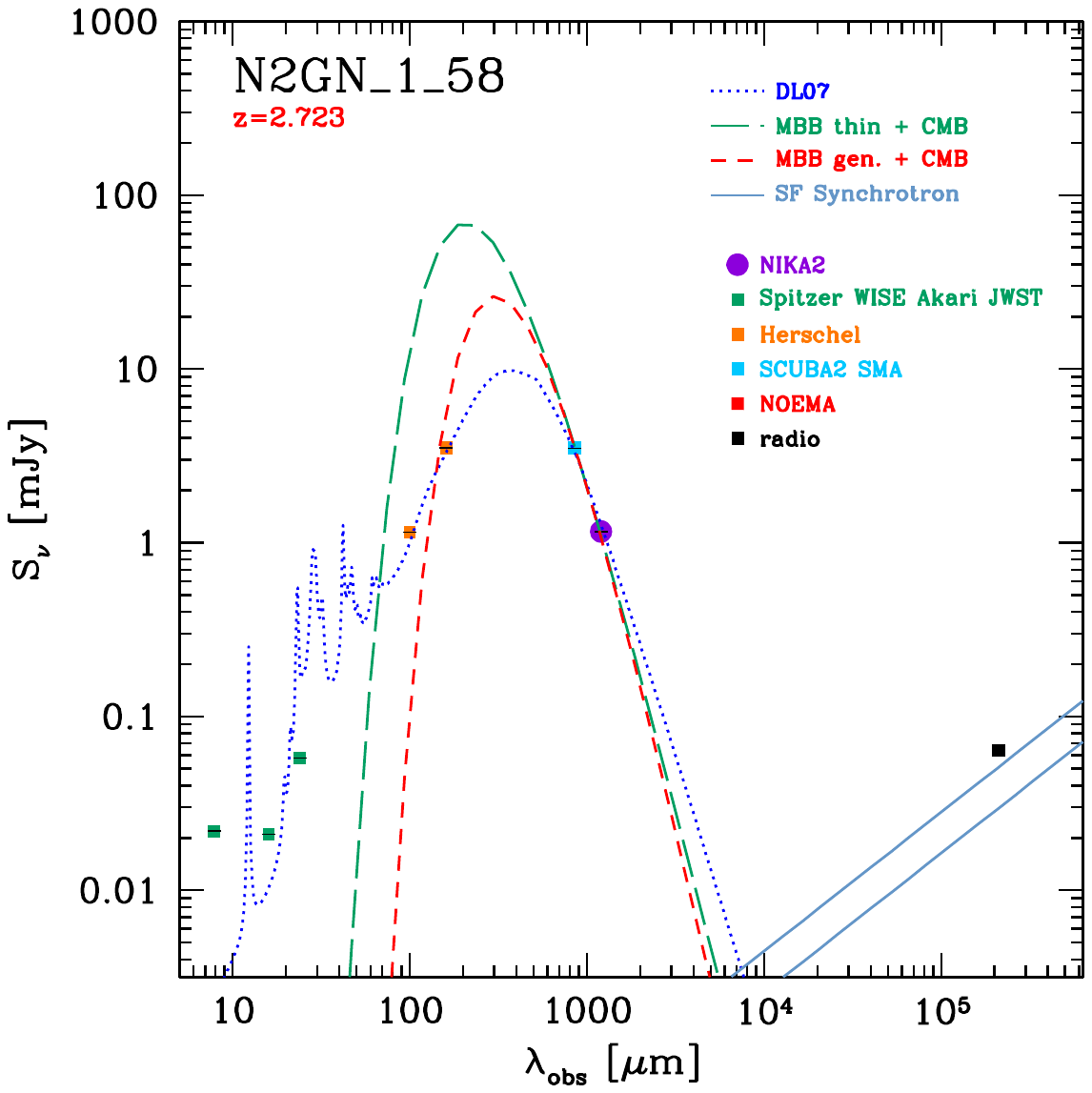}
\includegraphics[align=c,trim=0 0.18\imageheight{} 0 0.075\imageheight{}, clip, width=0.25\textwidth]{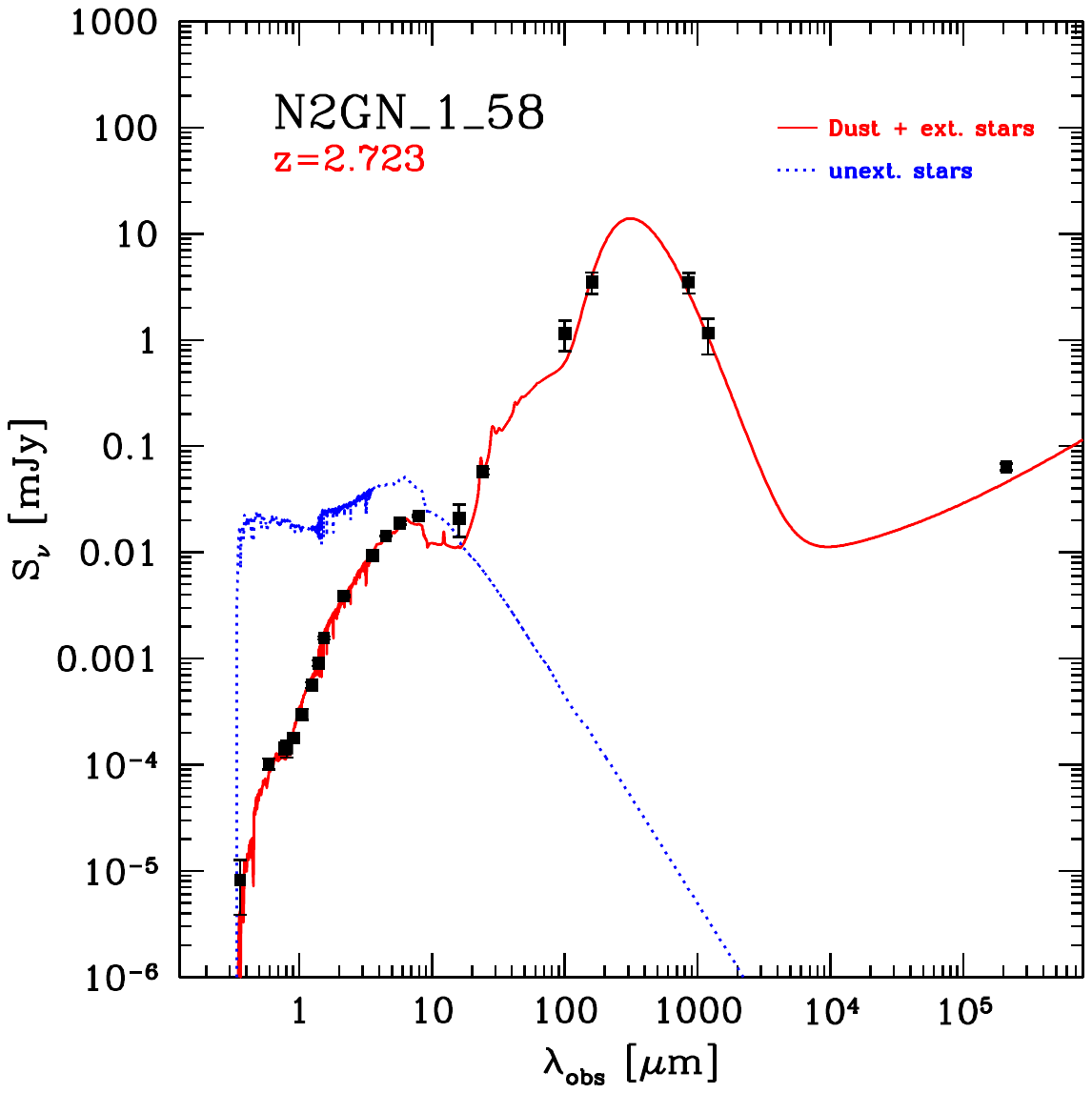}
\caption{continued.}
\end{figure*}

\addtocounter{figure}{-1}
\newpage

\begin{figure*}[t]
\centering
\includegraphics[align=c,width=0.4\textwidth]{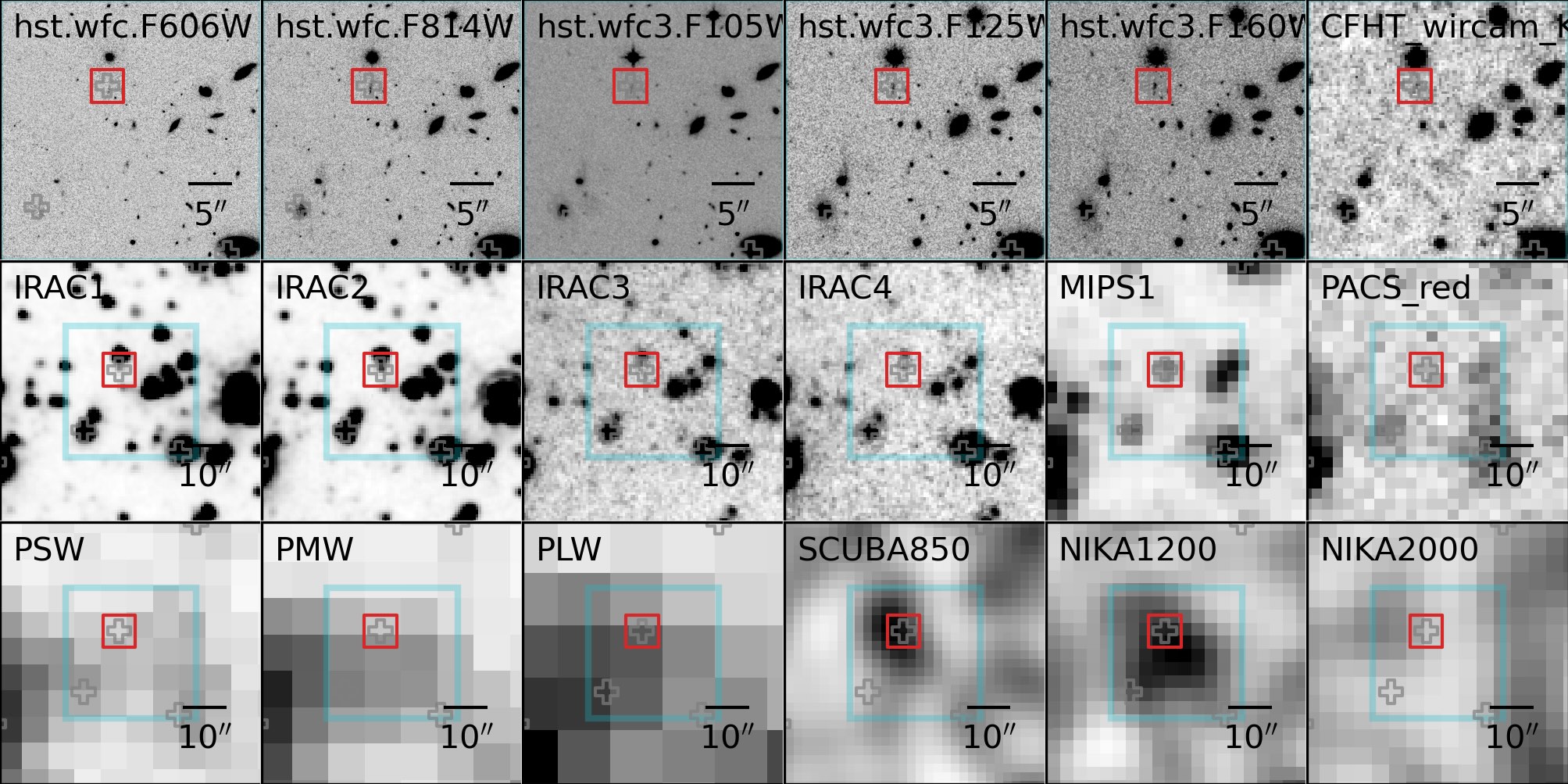}
\includegraphics[align=c,trim=0 0.18\imageheight{} 0 0.075\imageheight{}, clip, width=0.25\textwidth]{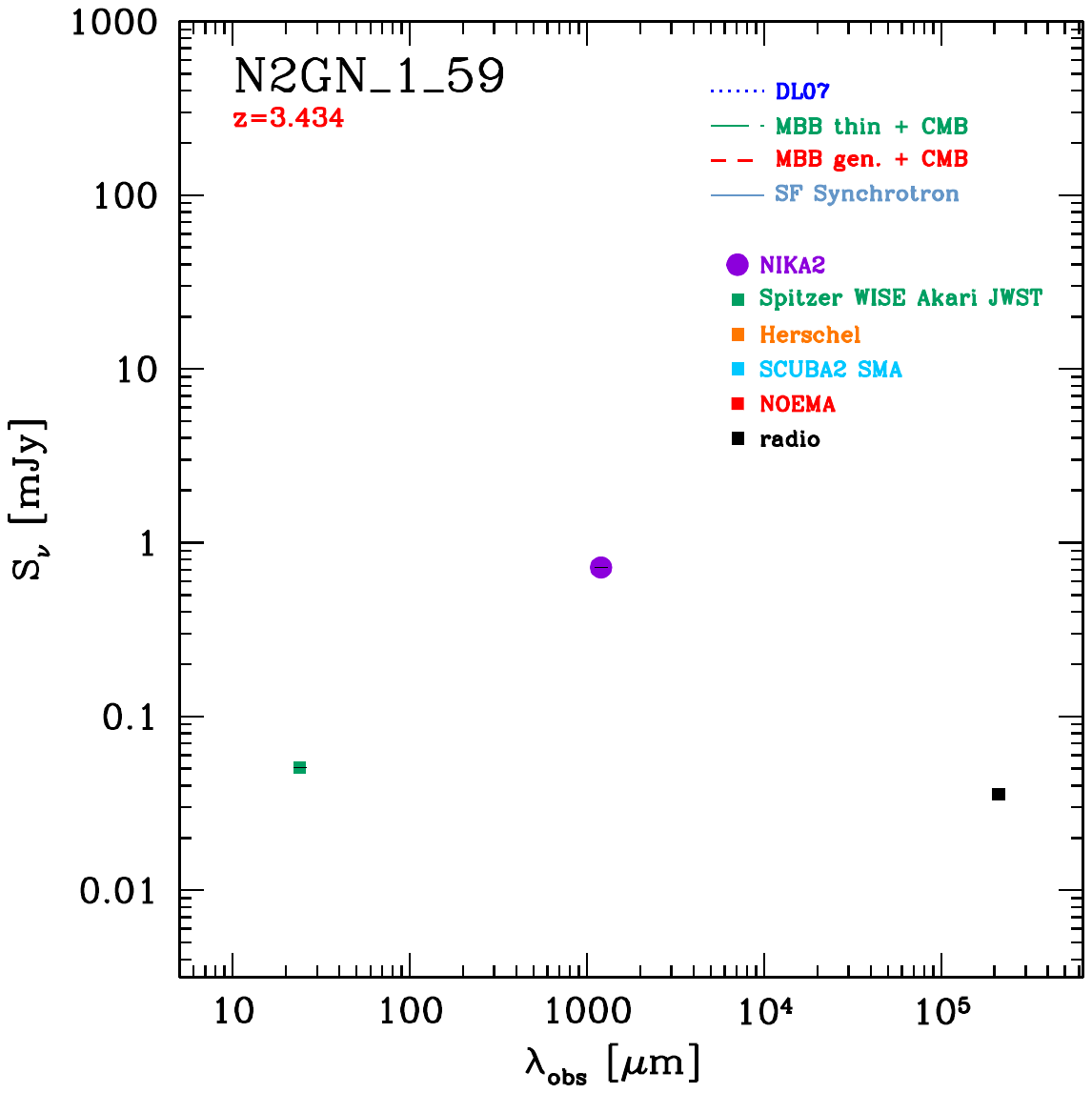}
\includegraphics[align=c,trim=0 0.18\imageheight{} 0 0.075\imageheight{}, clip, width=0.25\textwidth]{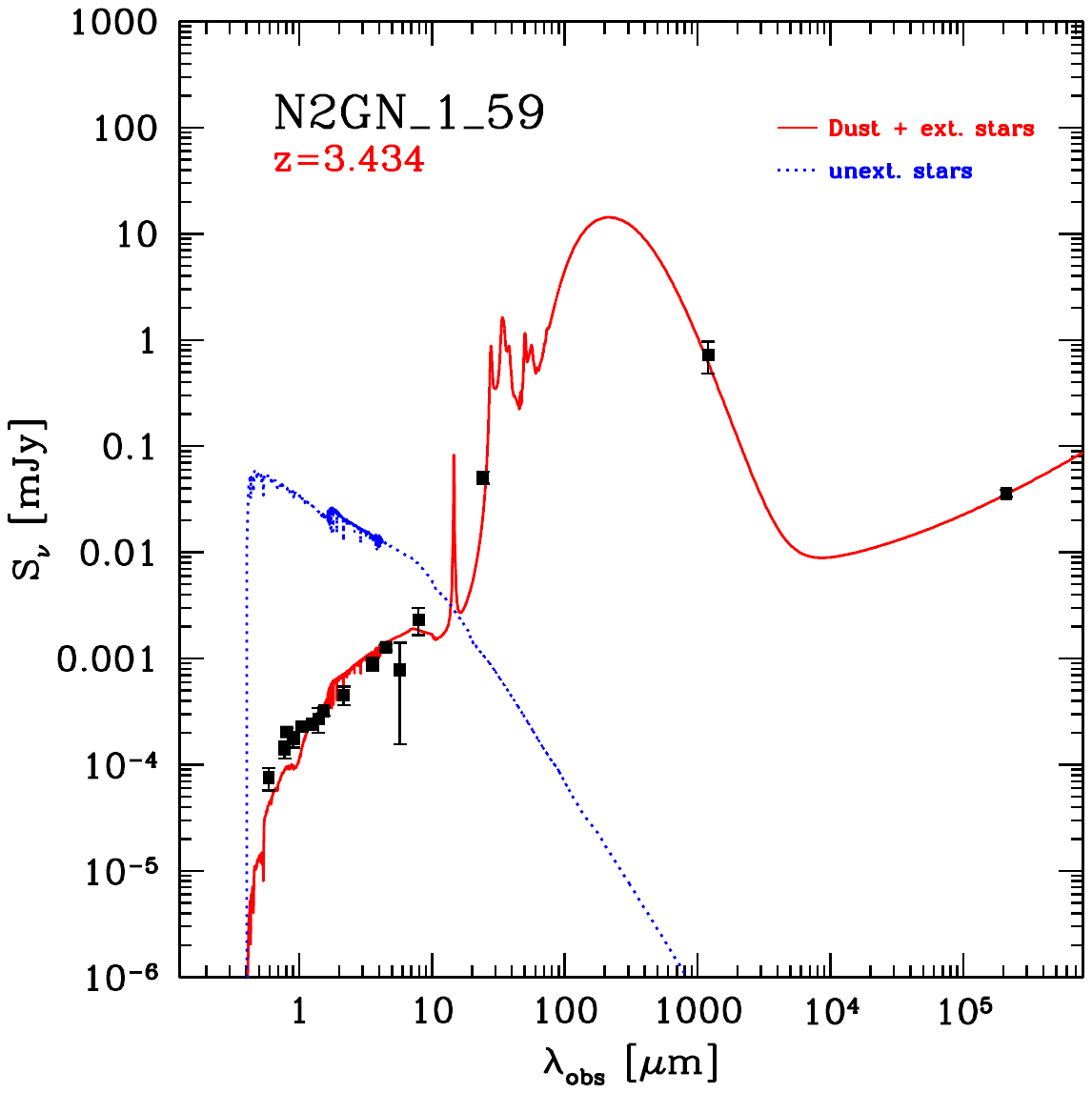}
\includegraphics[align=c,width=0.4\textwidth]{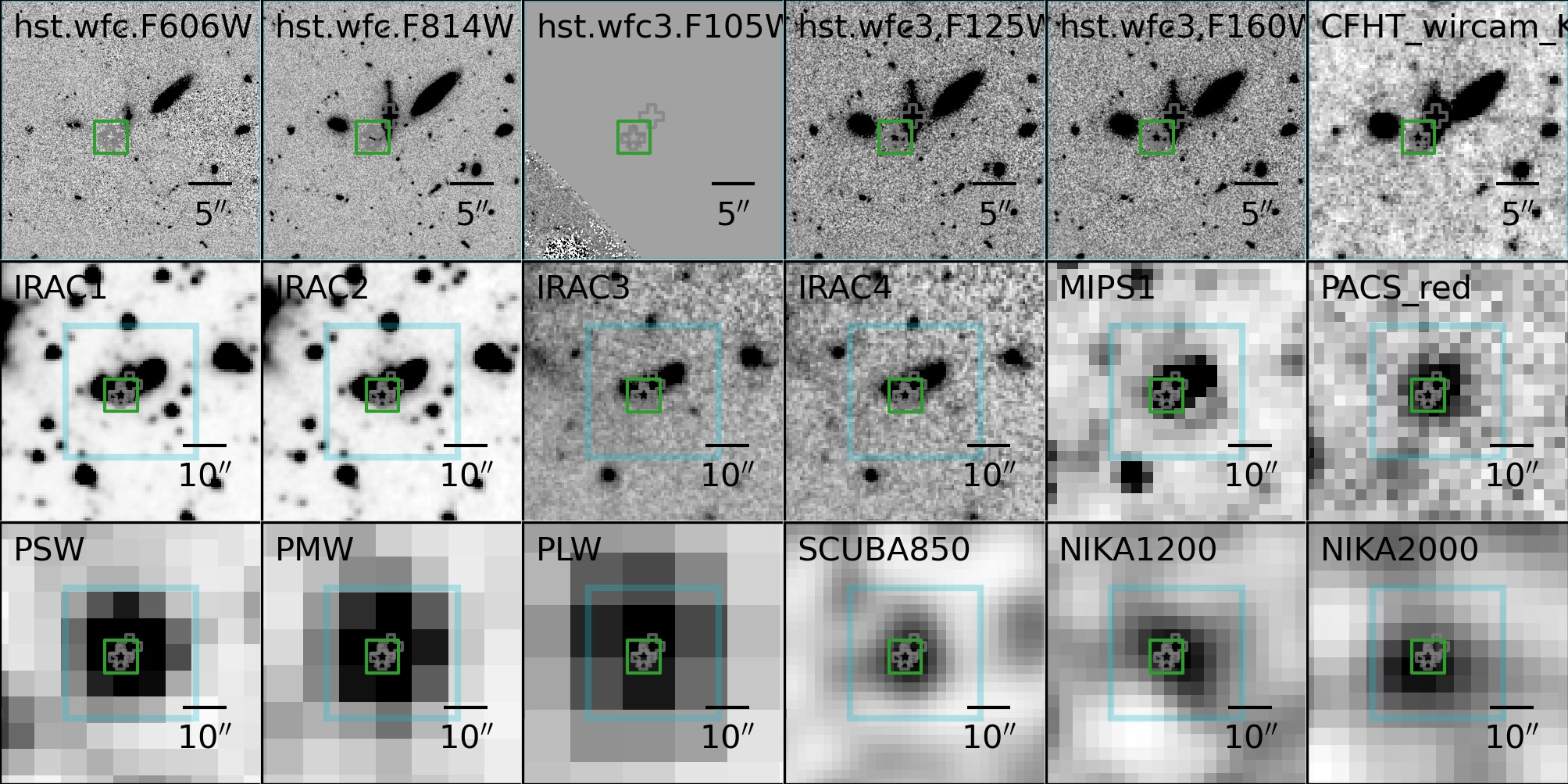}
\includegraphics[align=c,trim=0 0.18\imageheight{} 0 0.075\imageheight{}, clip, width=0.25\textwidth]{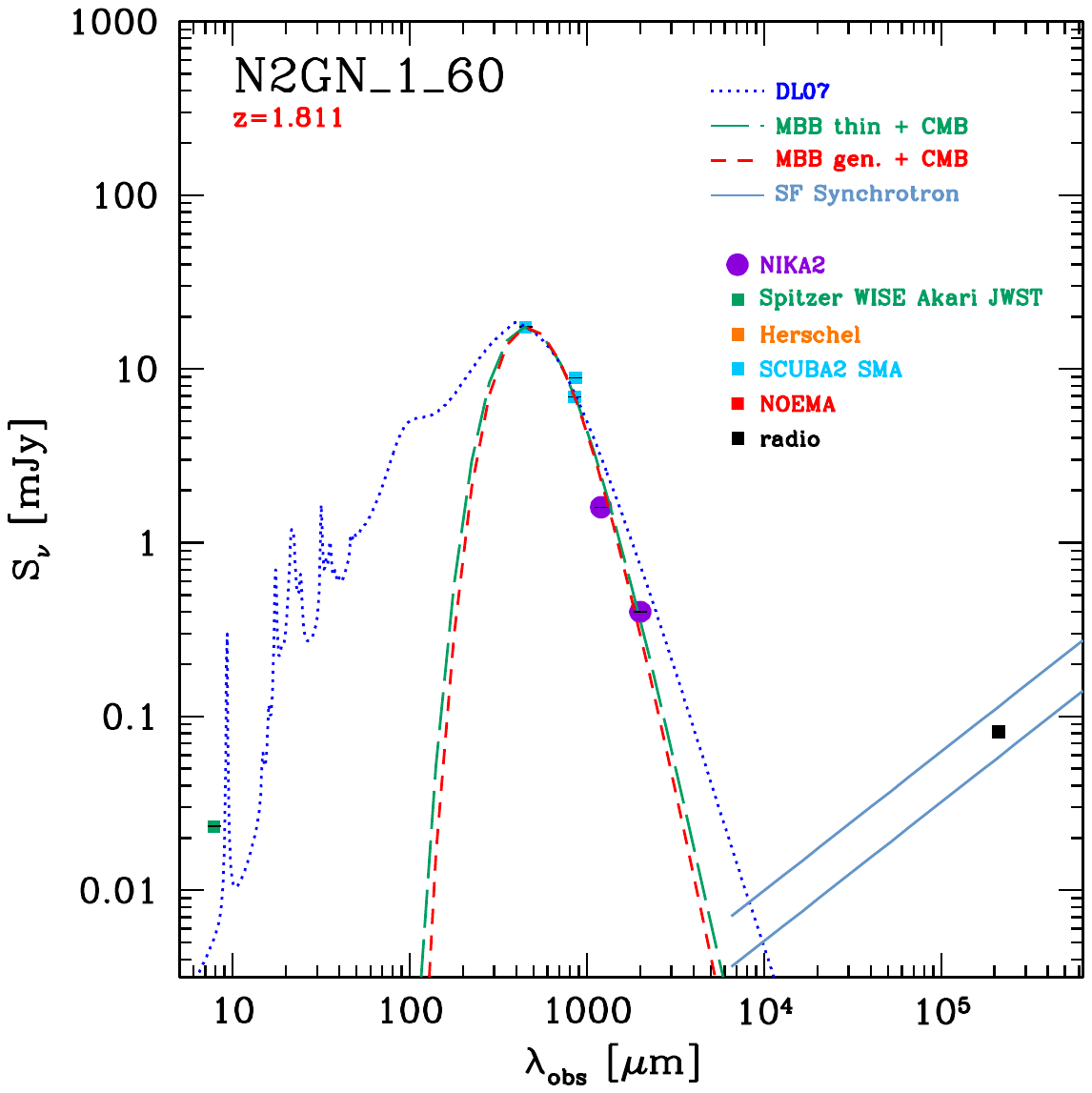}
\includegraphics[align=c,trim=0 0.18\imageheight{} 0 0.075\imageheight{}, clip, width=0.25\textwidth]{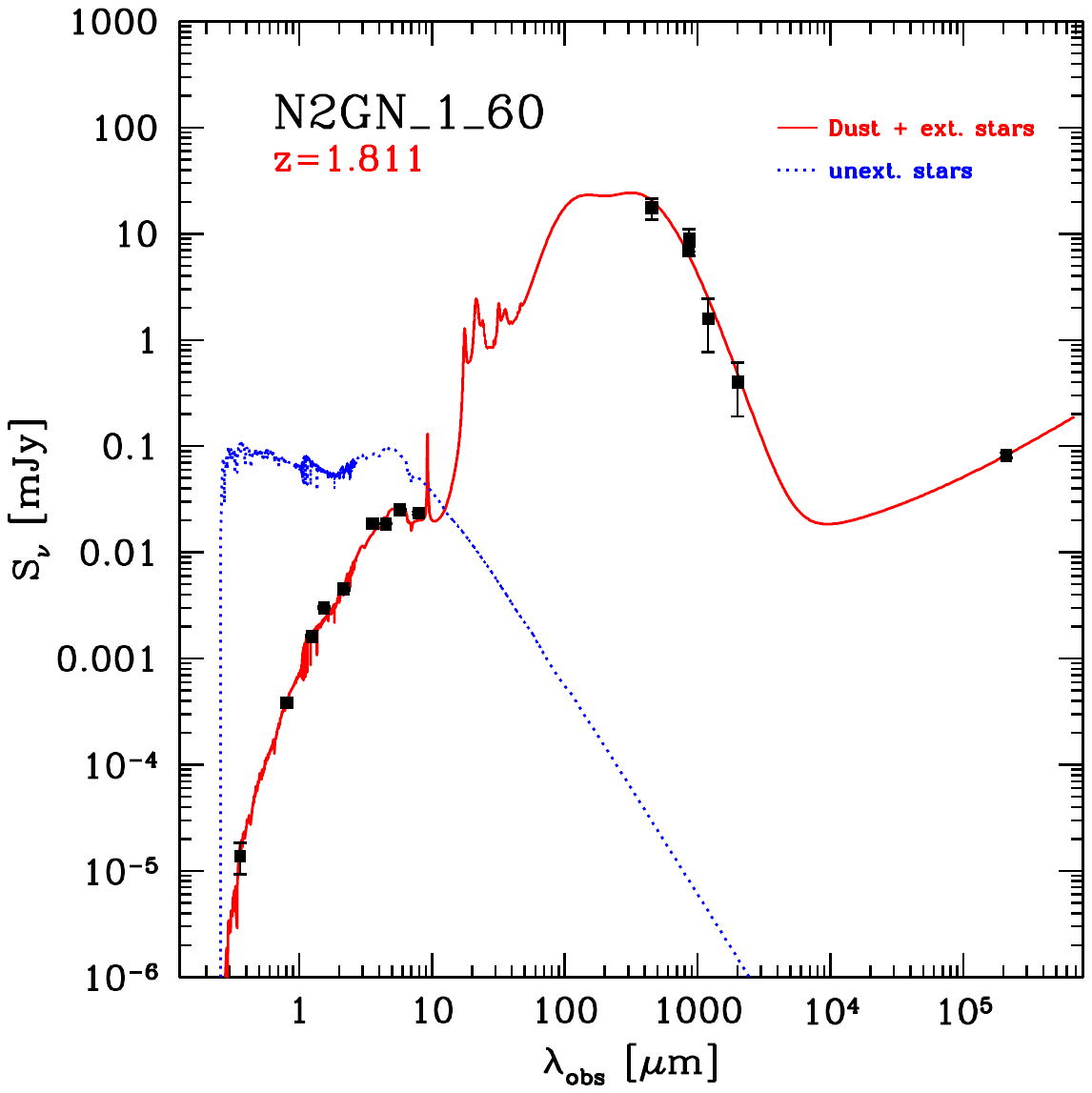}
\includegraphics[align=c,width=0.4\textwidth]{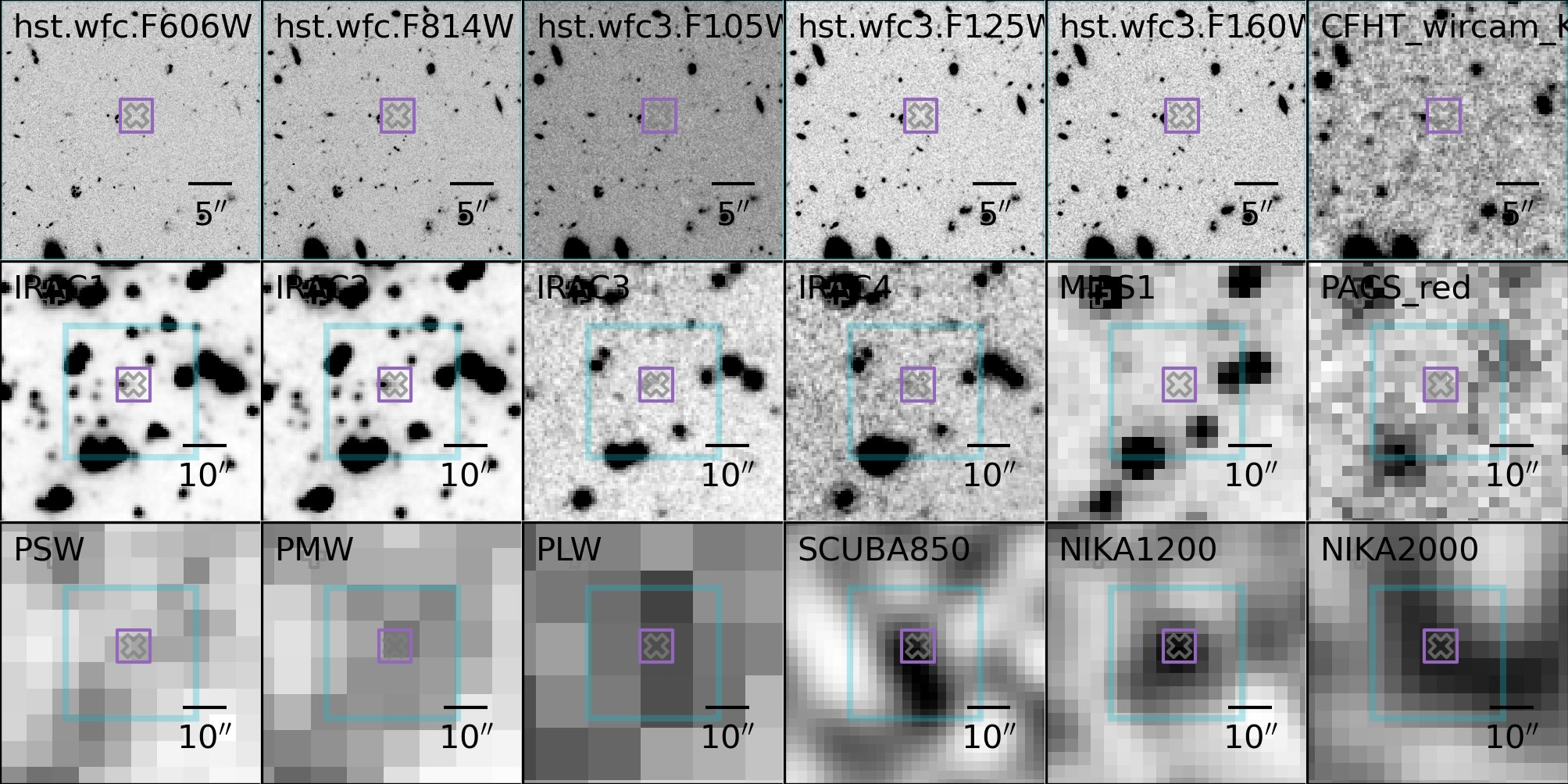}
\includegraphics[align=c,trim=0 0.18\imageheight{} 0 0.075\imageheight{}, clip, width=0.25\textwidth]{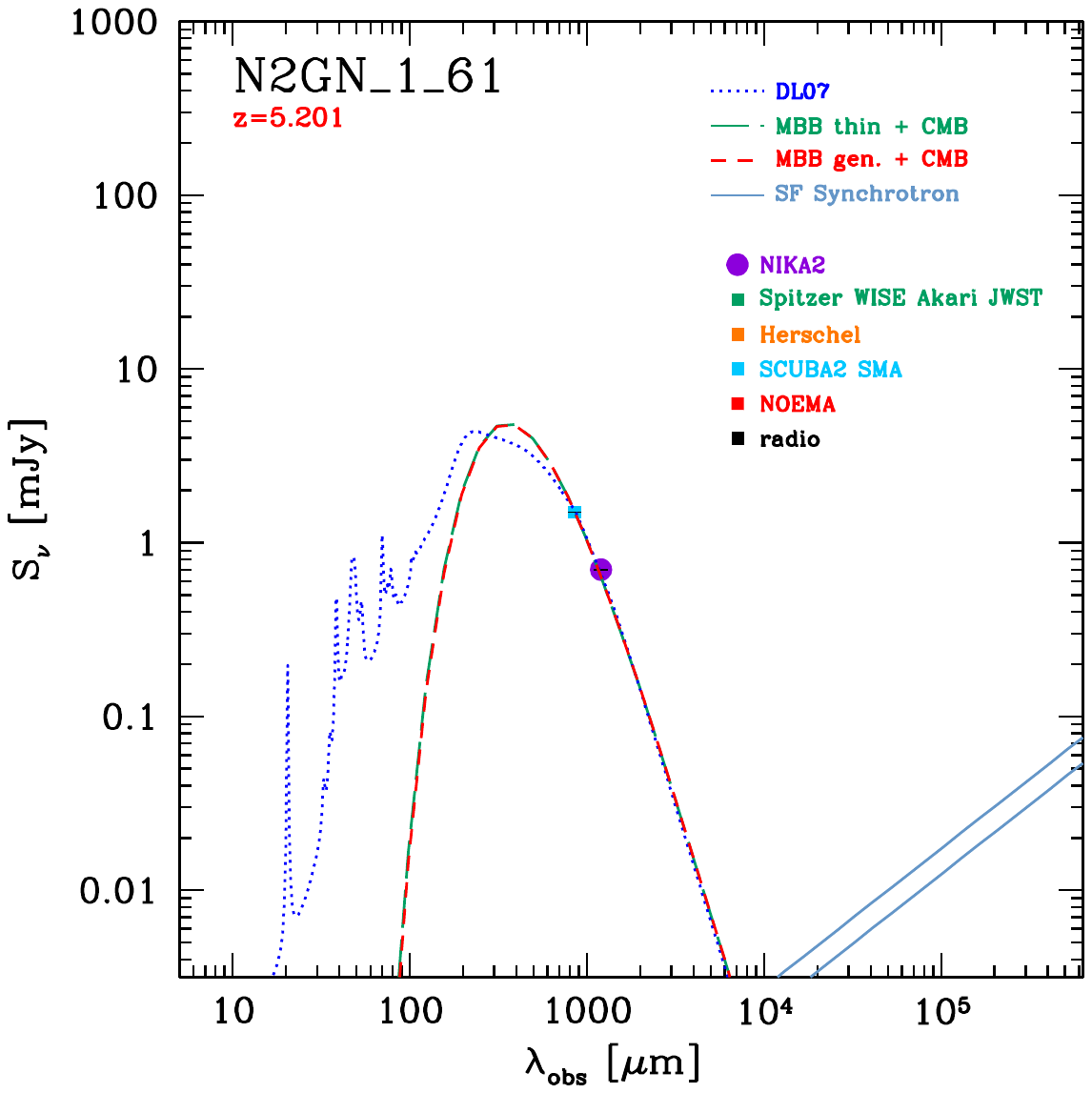}
\includegraphics[align=c,trim=0 0.18\imageheight{} 0 0.075\imageheight{}, clip, width=0.25\textwidth]{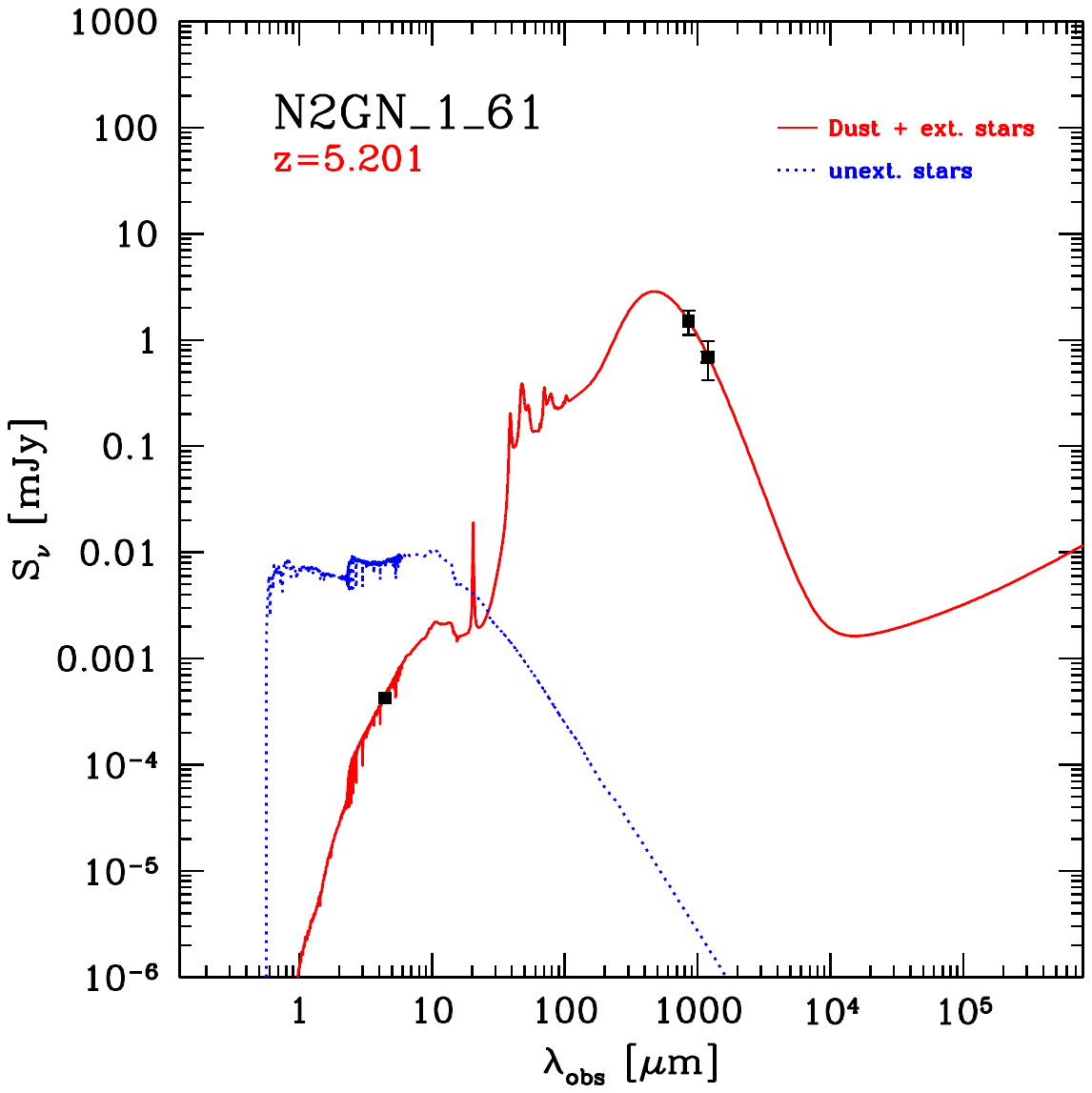}
\includegraphics[align=c,width=0.4\textwidth]{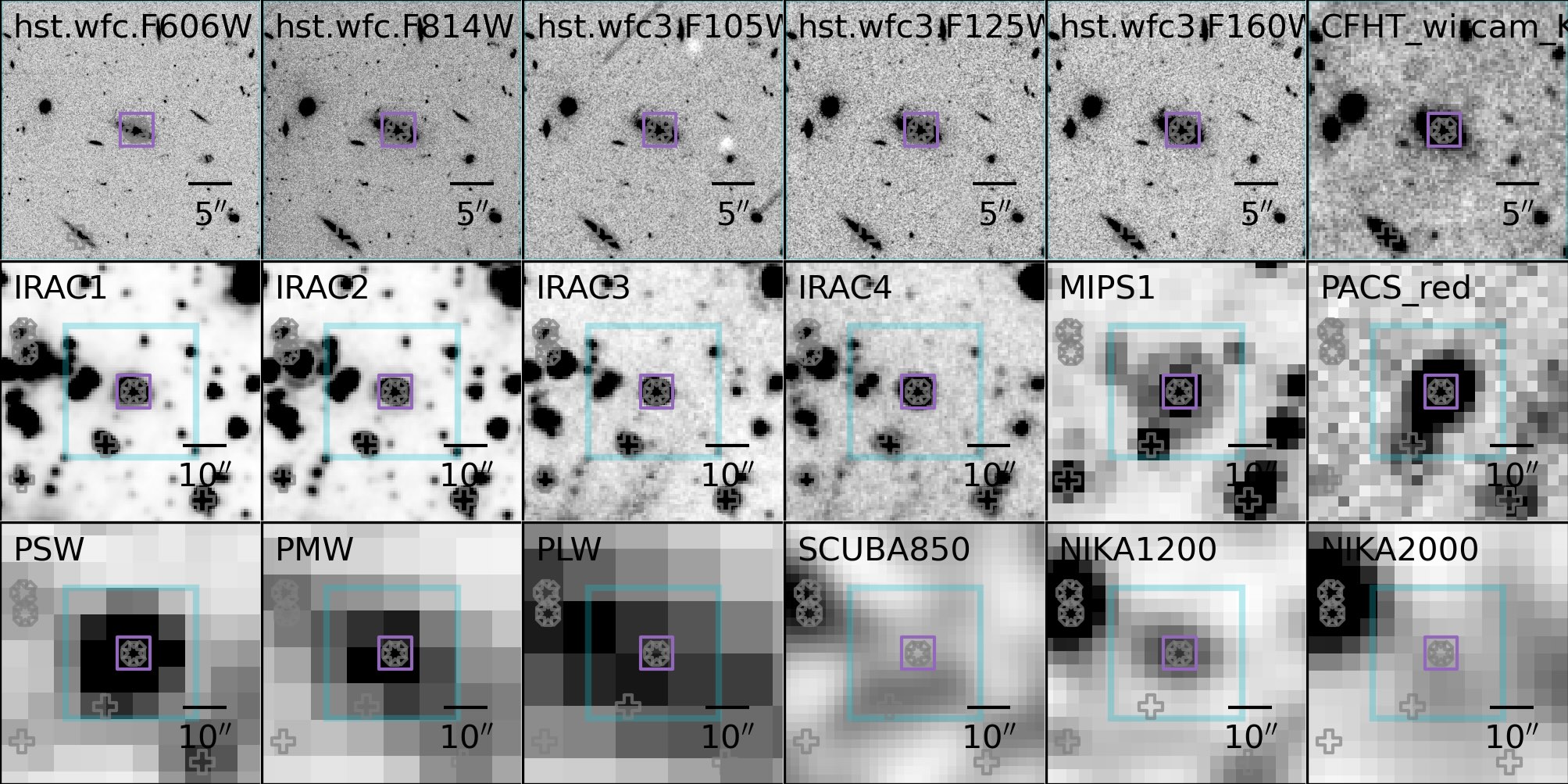}
\includegraphics[align=c,trim=0 0.18\imageheight{} 0 0.075\imageheight{}, clip, width=0.25\textwidth]{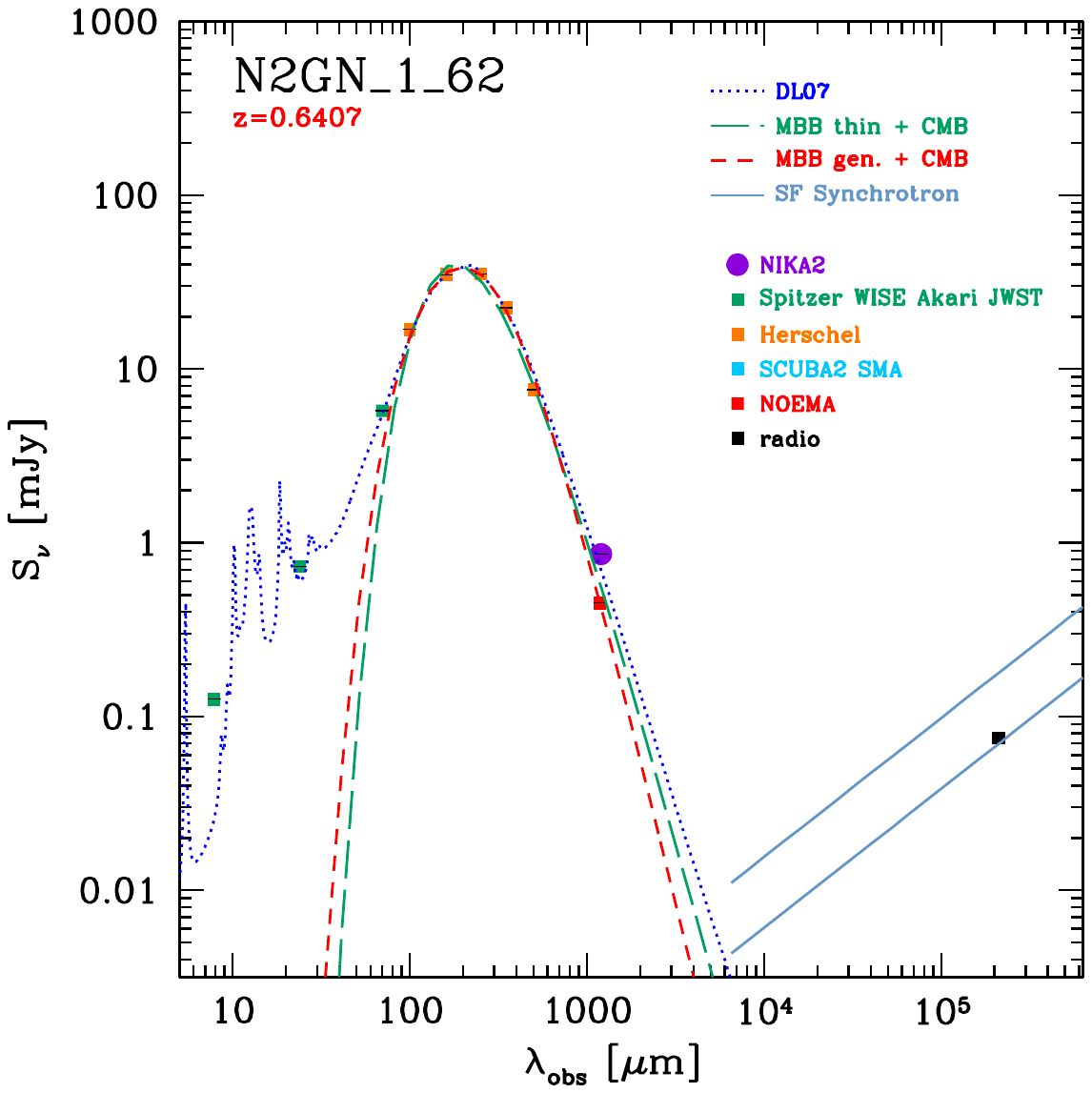}
\includegraphics[align=c,trim=0 0.18\imageheight{} 0 0.075\imageheight{}, clip, width=0.25\textwidth]{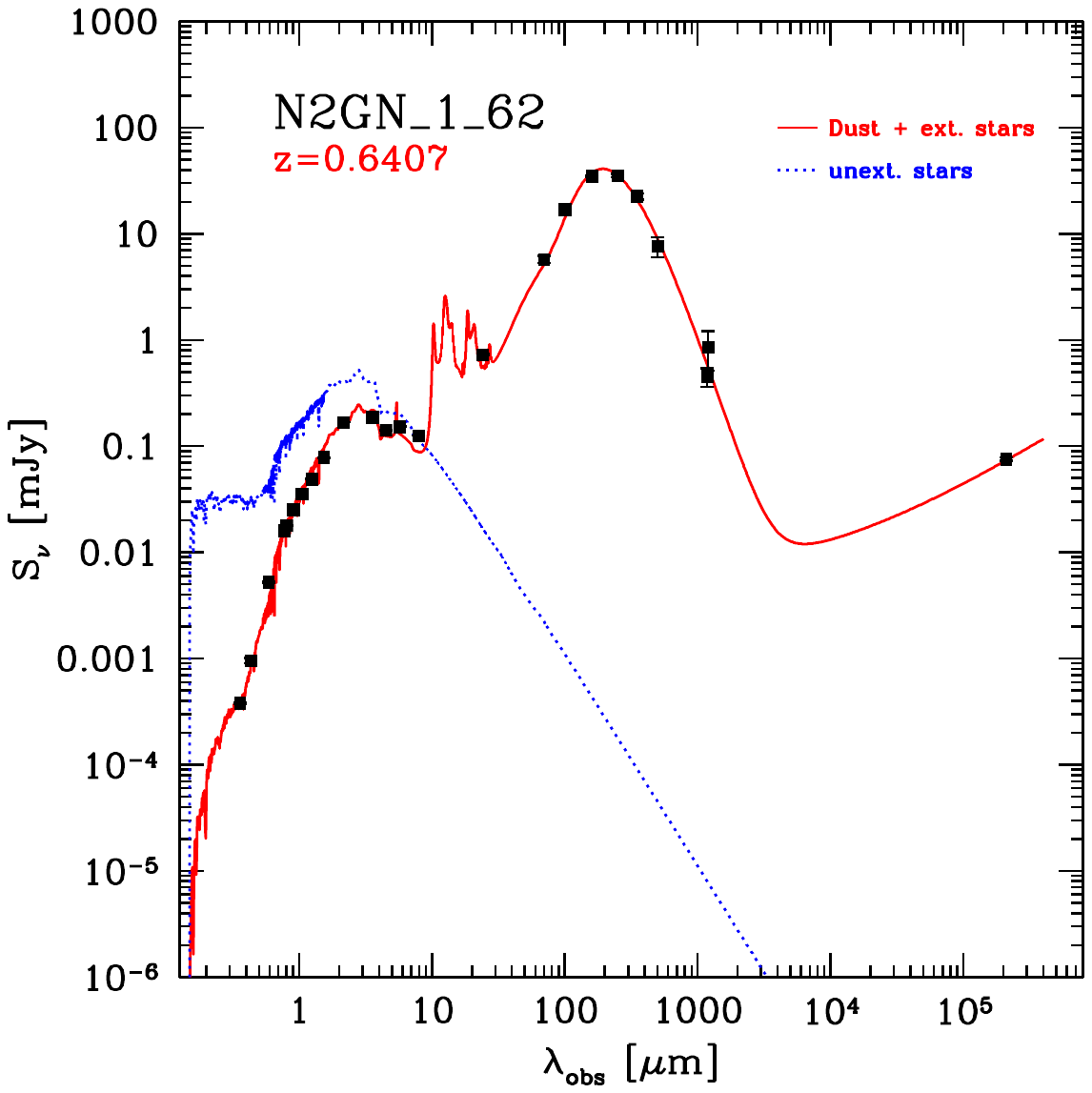}
\includegraphics[align=c,width=0.4\textwidth]{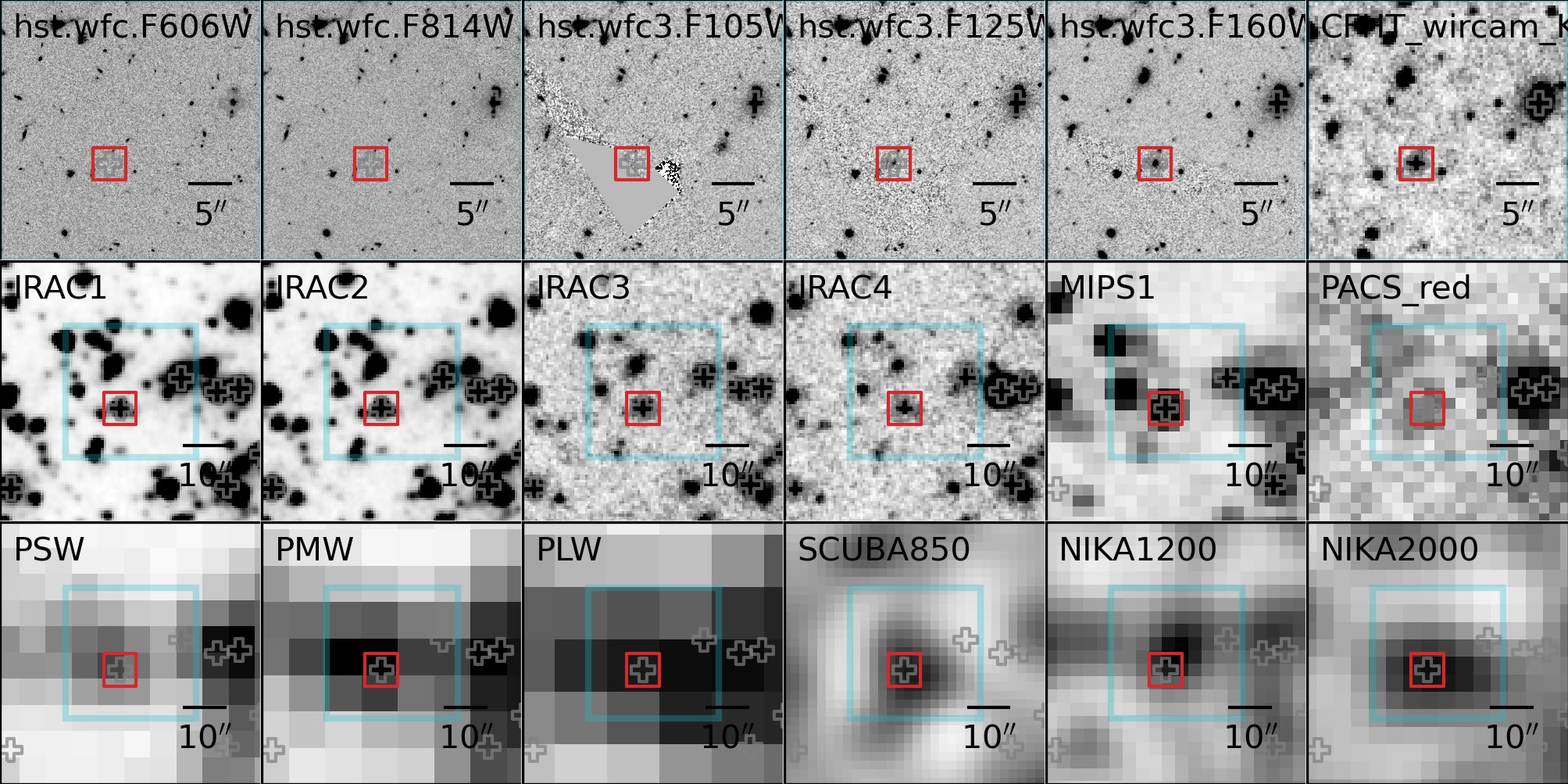}
\includegraphics[align=c,trim=0 0.18\imageheight{} 0 0.075\imageheight{}, clip, width=0.25\textwidth]{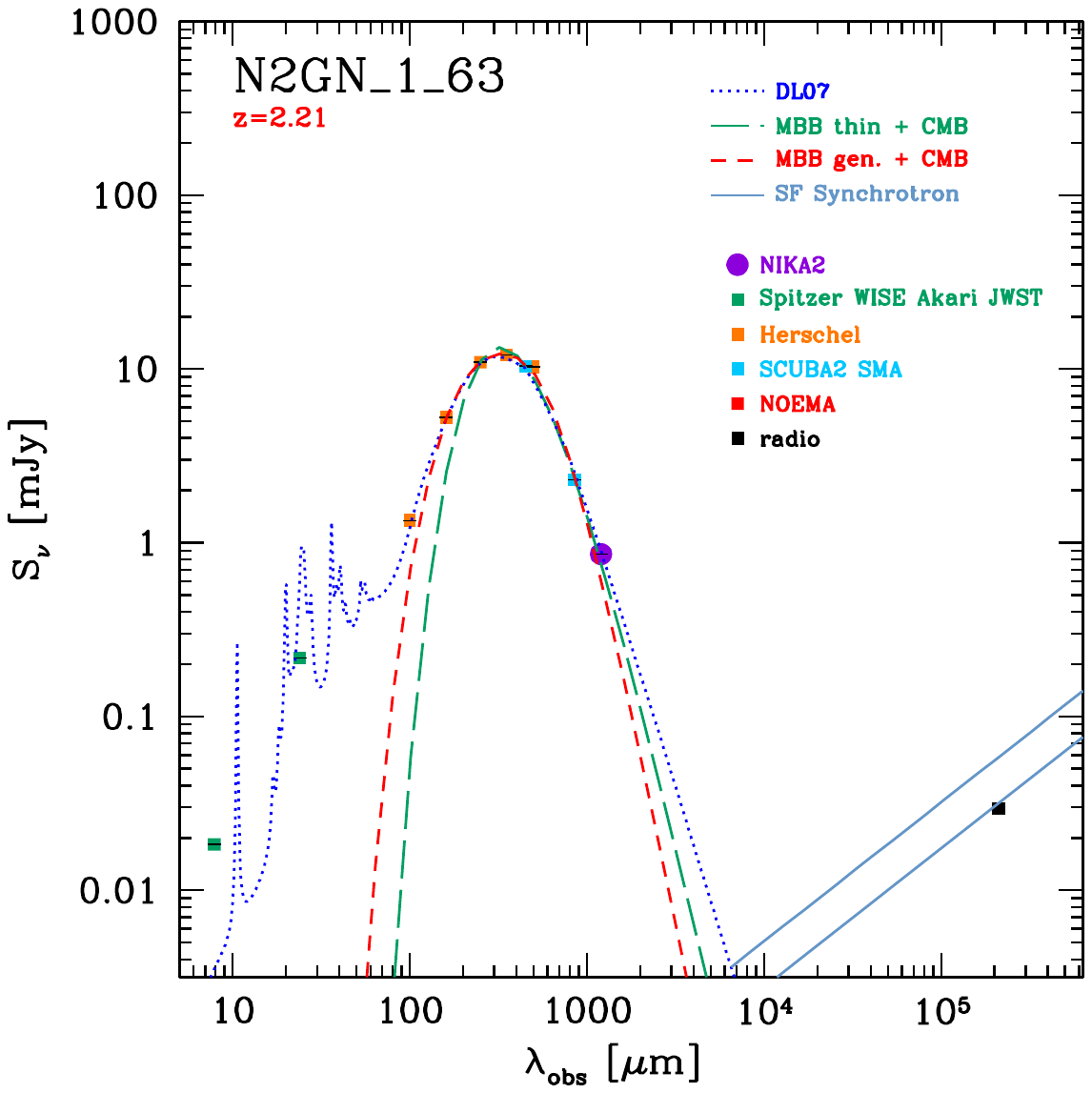}
\includegraphics[align=c,trim=0 0.18\imageheight{} 0 0.075\imageheight{}, clip, width=0.25\textwidth]{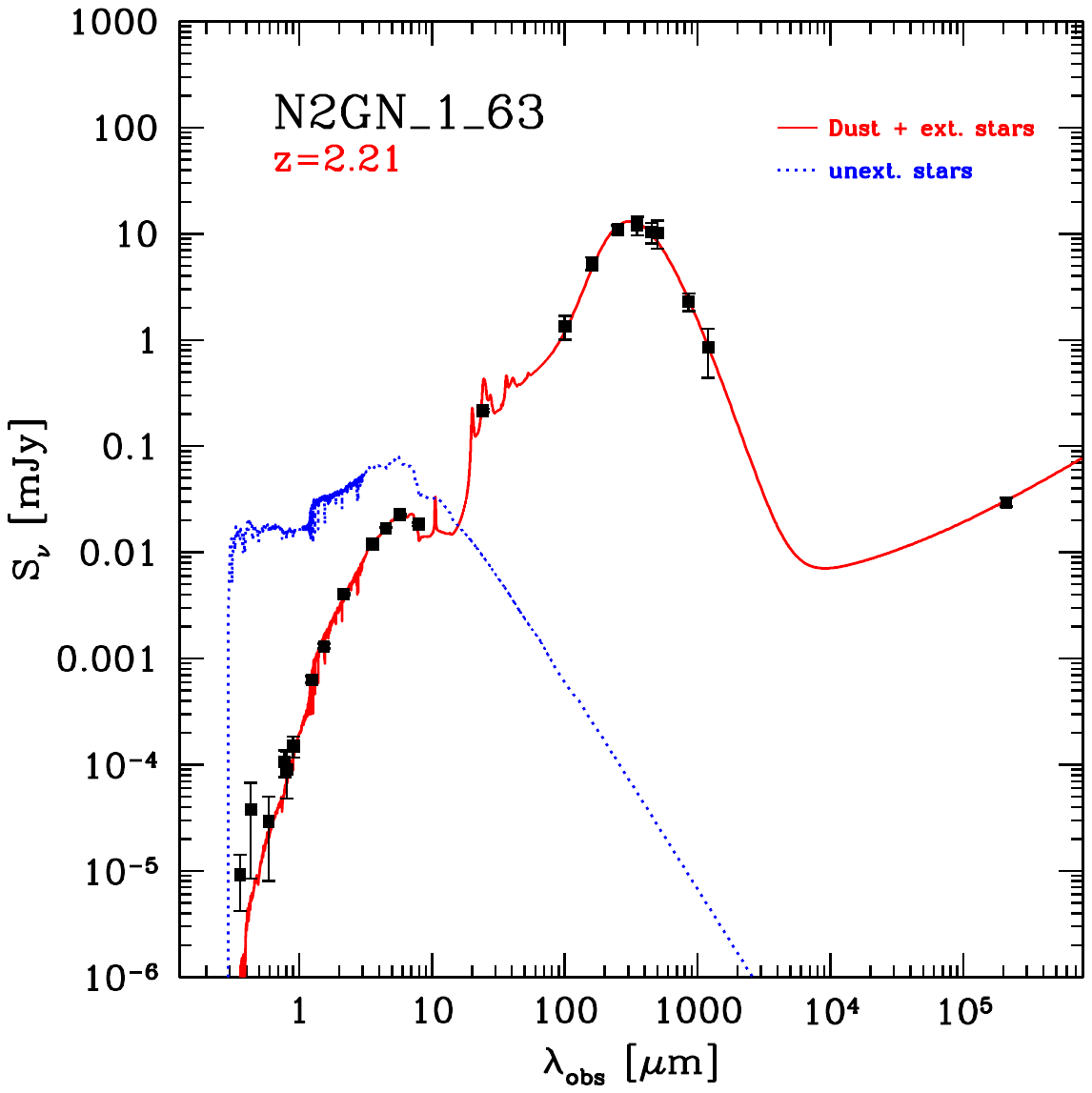}
\caption{continued.}
\end{figure*}

\addtocounter{figure}{-1}
\newpage

\begin{figure*}[t]
\centering
\includegraphics[align=c,width=0.4\textwidth]{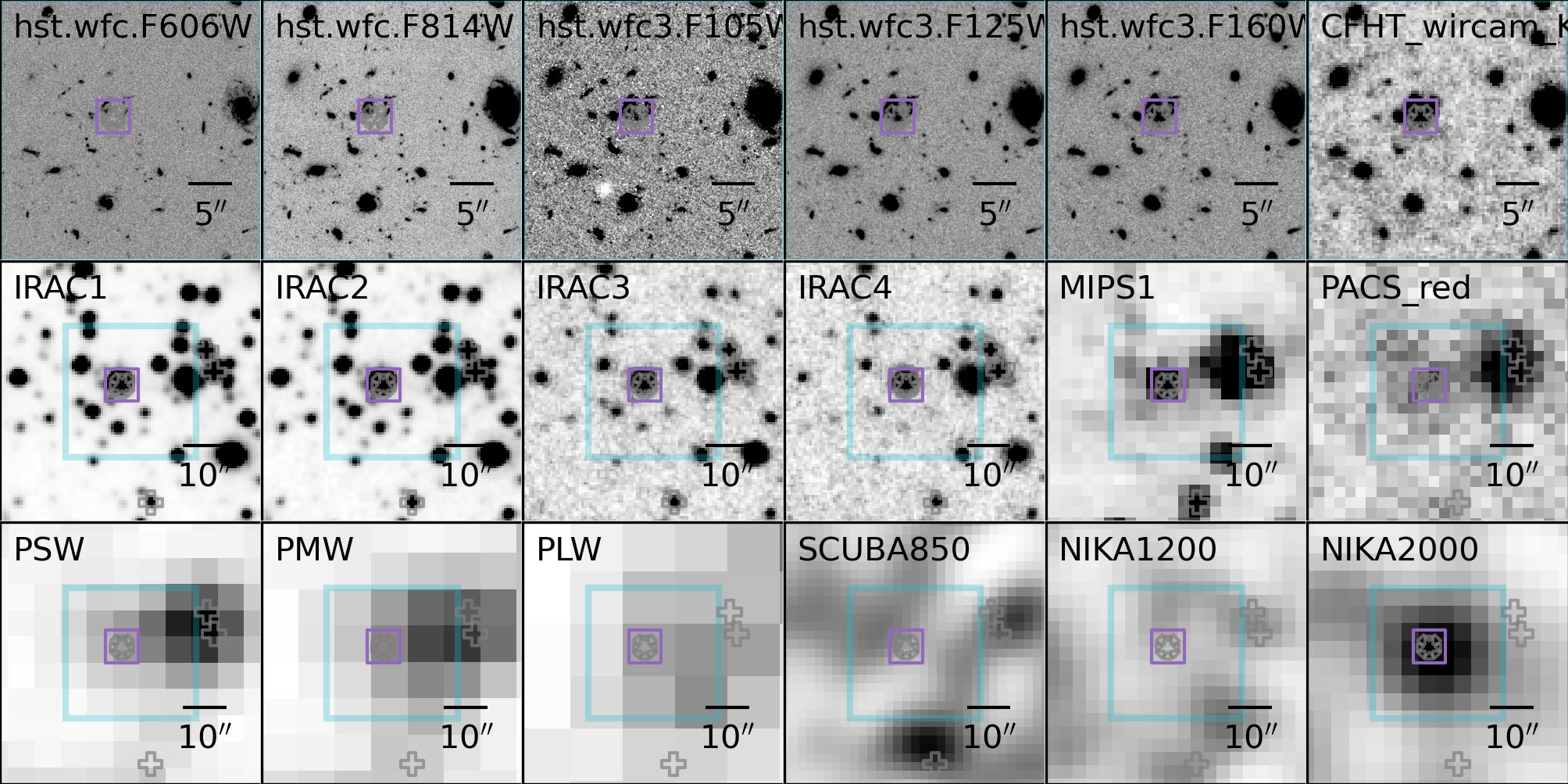}
\includegraphics[align=c,trim=0 0.18\imageheight{} 0 0.075\imageheight{}, clip, width=0.25\textwidth]{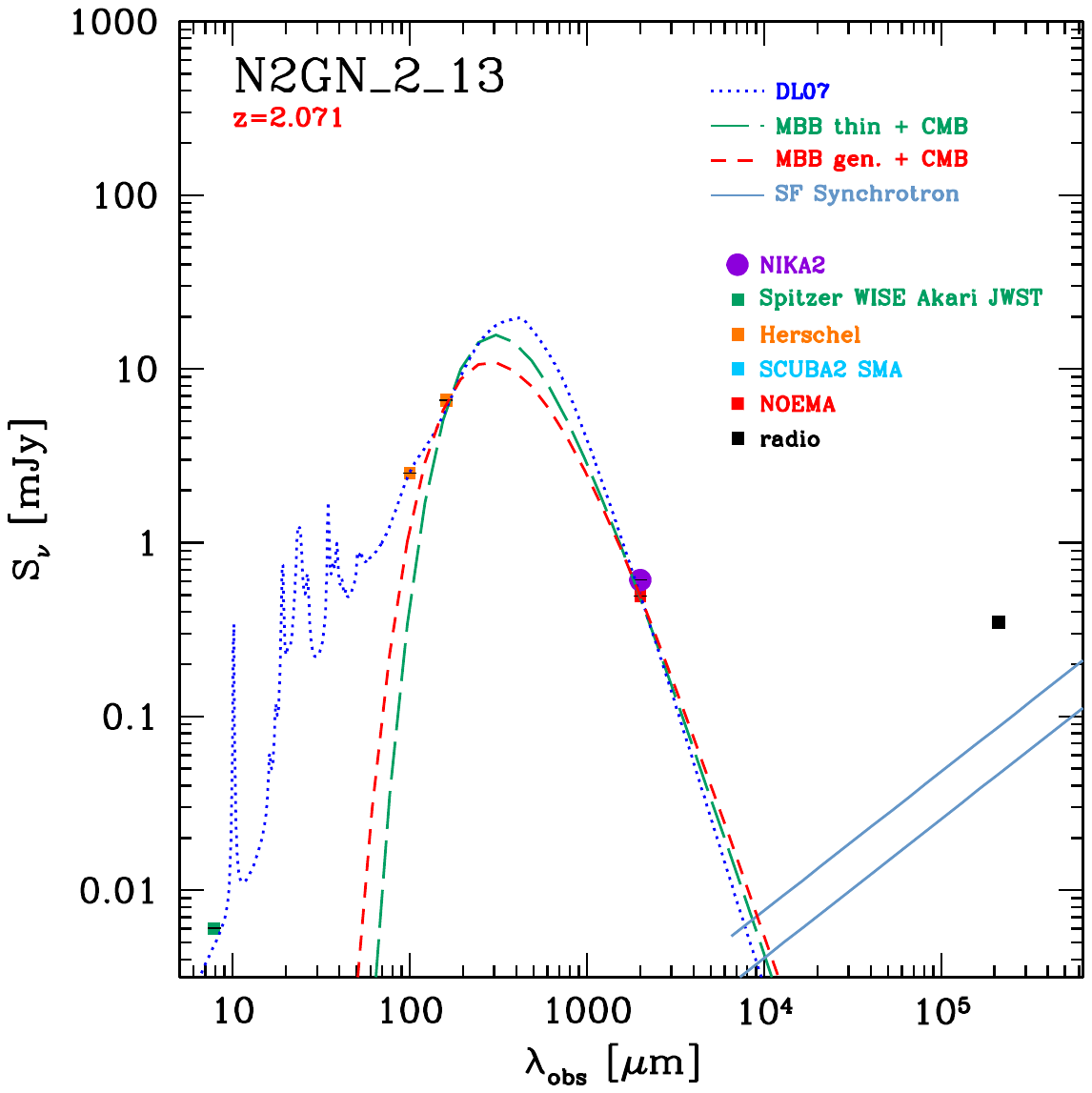}
\includegraphics[align=c,trim=0 0.18\imageheight{} 0 0.075\imageheight{}, clip, width=0.25\textwidth]{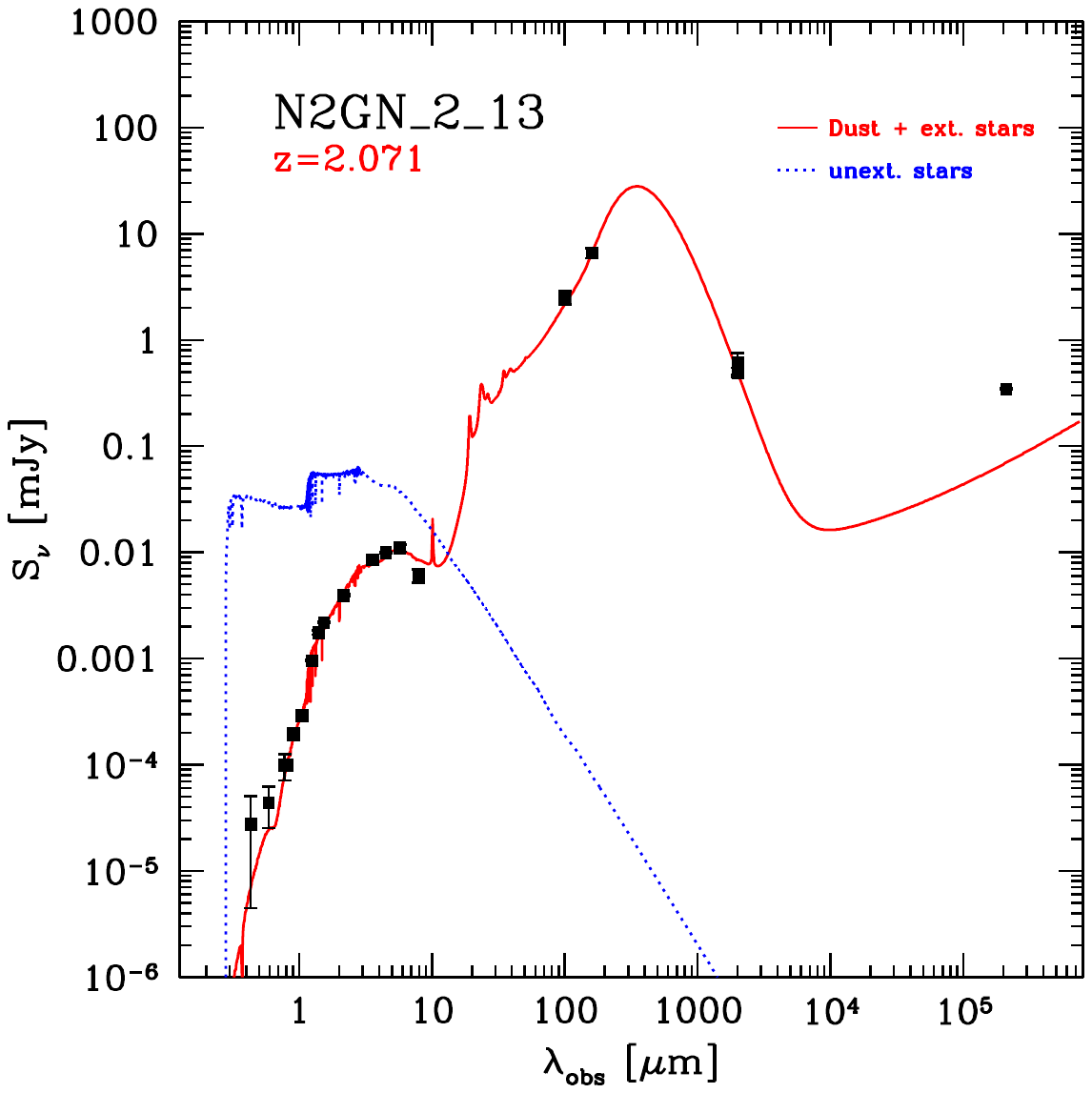}
\includegraphics[align=c,width=0.4\textwidth]{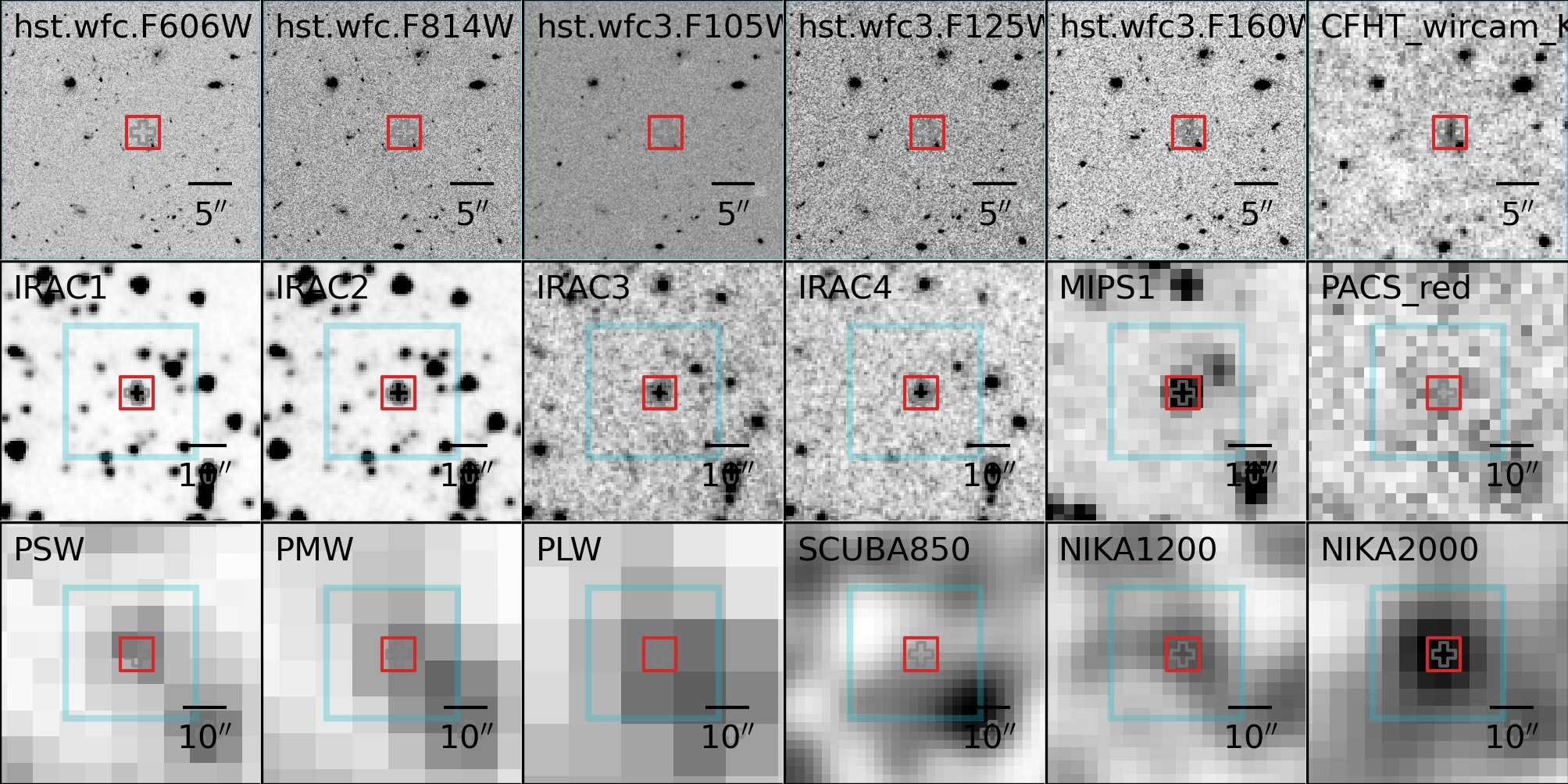}
\includegraphics[align=c,trim=0 0.18\imageheight{} 0 0.075\imageheight{}, clip, width=0.25\textwidth]{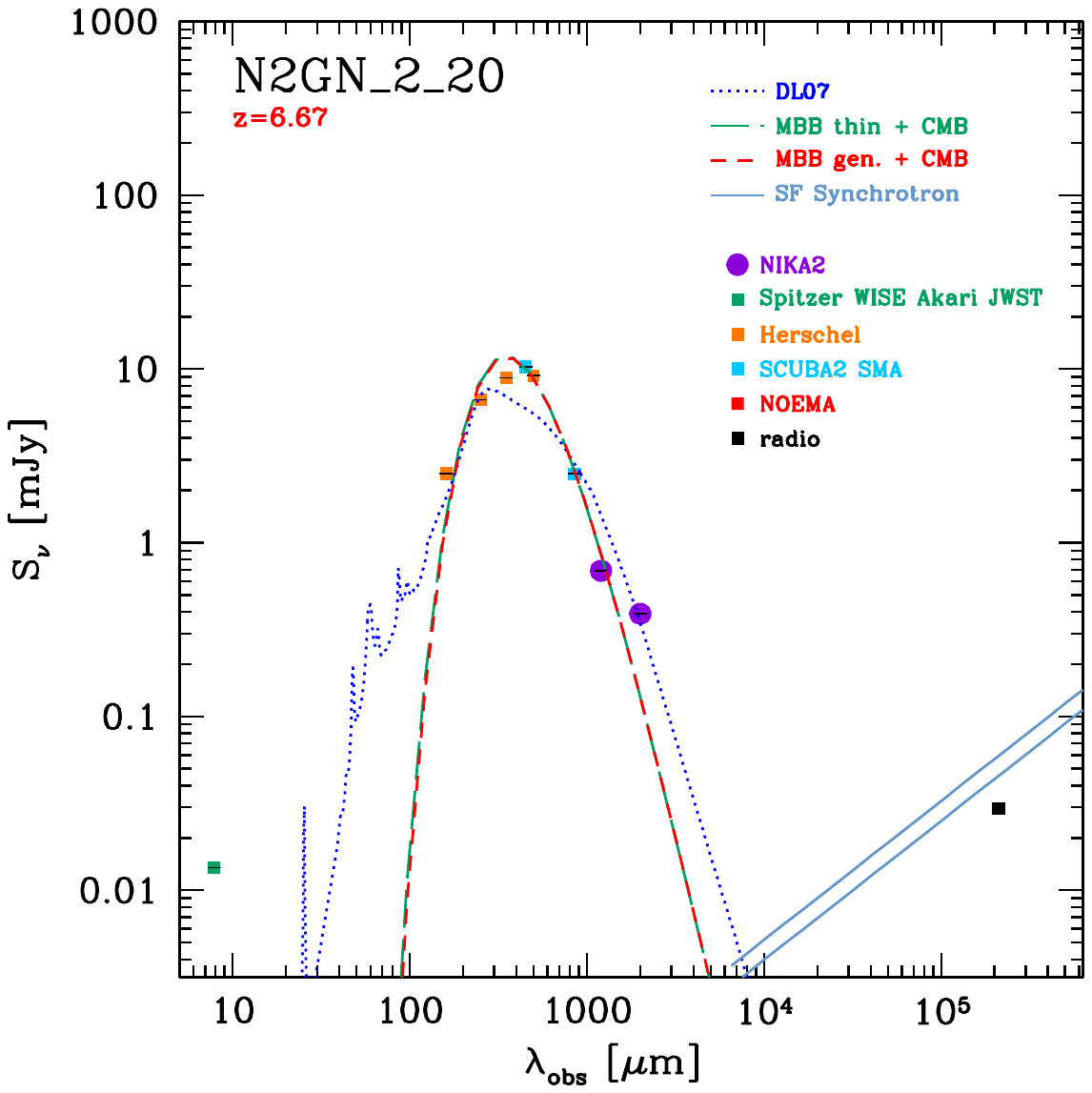}
\includegraphics[align=c,trim=0 0.18\imageheight{} 0 0.075\imageheight{}, clip, width=0.25\textwidth]{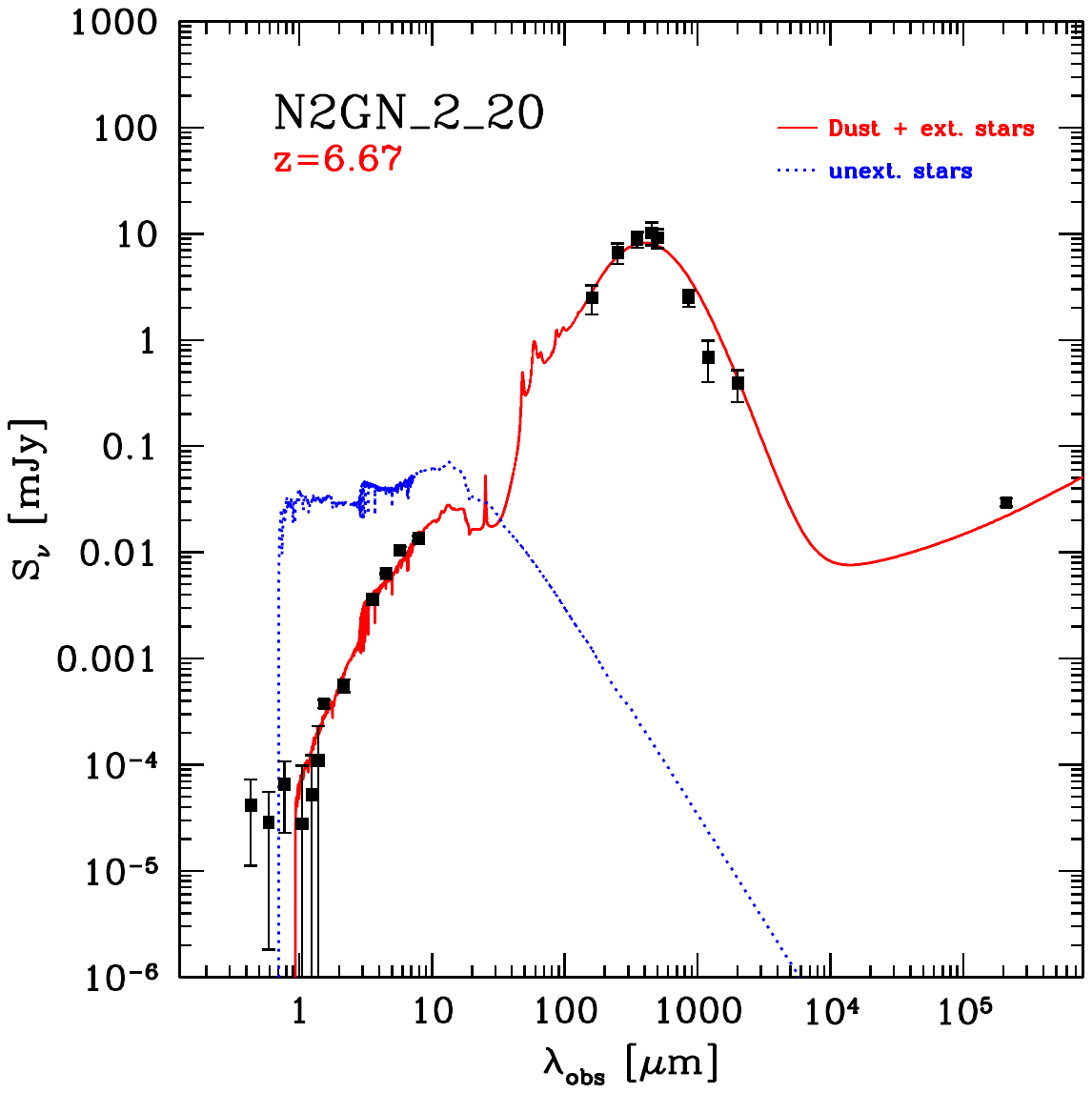}
\caption{continued.}
\end{figure*}

\onecolumn
%\longtab{      %%% A&A wants us to use this (see aa.dem), but Overleaf won't show the table (it works fine on SB's laptop)
\tiny{
\begin{landscape}
\begin{longtable}{lcl|cccc|ccl|cccccc}
\caption{\label{tab:ID_z_Md_LIR} Details of the N2GN sources: identification through high spatial resolution millimeter or radio data; redshift (spectroscopic or photometric); results of SED fitting.}\\
\hline
\hline
\multicolumn{3}{c|}{Name}	& \multicolumn{4}{c|}{Identification}	& \multicolumn{3}{c|}{Redshift}& \multicolumn{6}{c}{SED fitting results}\\
N2CLS ID 	& IAU (NIKA2)	& \multicolumn{1}{c|}{Alt.$^\dagger$} 	& Method & Ref.	& R.A. 		& Dec.		& $z$ 	& Nature 	& Ref. & $M_\textrm{dust}$ & 2.5\% & 97.5\% & $L_\textrm{IR}$ & 2.5\% & 97.5\% \\ 
 	& 	 	&  	&  & & [hh:mm:ss]		& [dd:mm:ss]		& &  	& & $[10^8 M_\odot]$ & $\log[M_\odot]$& $\log[M_\odot]$ &  $[10^{12} L_\odot]$ & $\log[L_\odot]$& $\log[L_\odot]$\\ 
\hline
\endfirsthead
\caption{continued.}\\
\hline\hline
\multicolumn{3}{c|}{Name}	& \multicolumn{4}{c|}{Identification}	& \multicolumn{3}{c|}{Redshift}& \multicolumn{6}{c}{SED fitting results}\\
N2CLS ID 	& IAU (NIKA2)	& \multicolumn{1}{c|}{Alt.$^\dagger$} 	& Method & Ref.	& R.A. 		& Dec.		& $z$ 	& Nature 	& Ref. & $M_\textrm{dust}$ & 2.5\% & 97.5\% & $L_\textrm{IR}$ & 2.5\% & 97.5\% \\ 
 	& 	 	&  	&  & & [hh:mm:ss]		& [dd:mm:ss]		& &  	& & $[10^8 M_\odot]$ & $\log[M_\odot]$& $\log[M_\odot]$ &  $[10^{12} L_\odot]$ & $\log[L_\odot]$& $\log[L_\odot]$\\ 
\hline
\endhead
\hline
\multicolumn{16}{l}{$^\dagger$: The HDF850, (GOODS) 850,  GN, AzGN, GN1200 and P- nomenclatures were defined by \citet{hughes1998}, \citet{wang2004}, \citet{pope2005}, \citet{perera2008}, \cite{greve2008}}\\
\multicolumn{16}{l}{\hspace{0.25truecm} and \cite{Penner2011}, respectively.}\\
\multicolumn{16}{l}{$^\ddagger$: no magnification correction has been applied to the results of N2GN\_1\_06.}\\
\multicolumn{16}{l}{Id. refs.: {\em 1} \citet{cowie2017}; {\em 2} \citet{daddi2009}; {\em 3} this work (Sect\,\ref{sect:data_NOEMA}); {\em 4} \citet{walter2012}; {\em 5} \citet{jin2022}; {\em 6} \citet{bingthesis}; {\em 7} \citet{owen2018}; {\em 8} \citet{xiao2024};}\\
\multicolumn{16}{l}{\hspace{1.0truecm} {\em 9} \citet{fujimoto2022}; {\em 10} \citet{arrabal_haro2018}.}\\
\multicolumn{16}{l}{$z$ refs.: {\em a} \citet{riechers2020}; {\em b} \citet{kodra2023}; {\em c} \citet{barger2012}; {\em d} \citet{daddi2009}; {\em e} \citet{bothwell2010}; {\em f} \citet{neri2014}; {\em g} \citet{maseda2024};  {\em h} this work (Sect.~\ref{sect:redshift});}\\
\multicolumn{16}{l}{\hspace{0.8truecm} {\em i} \citet{jin2022}; {\em j} \citet{cowie2017}; {\em k} \citet{wang2016b}; {\em l} NOEMA Lagache et al. (in prep.); {\em m} \citet{cowie2004}; {\em n} \citet{liu2018};}  \\
\multicolumn{16}{l}{\hspace{0.8truecm} {\em o} \citet{xiao2024}; {\em p} \citet{fujimoto2022}; {\em q} JADES DJA; {\em r} M.~Xiao, priv. comm.; {\em s} \citet{arrabal_haro2018}.}\\ %}
\endfoot
N2GN\_1\_01 		& N2GN J123633+621408	&  GN10 	& SMA 	& 1	& 12:36:33.44 	& 62:14:08.7 	& 5.303	& spec 	& a 		     & 20.92   & 9.18   & 9.36  & 23.90  & 13.22  & 13.46\\ 
N2GN\_1\_02 		& N2GN J123730+621259	&  850-6 	& SMA 	& 1	& 12:37:30.79 	& 62:12:58.9 	& 1.963	& phot 	& b 		     & 60.46   & 9.78   & 9.78  &  4.73  & 12.68  & 12.68\\ 
N2GN\_1\_03 		& N2GN J123707+621408	&  GN19 	& SMA 	& 1	& 12:37:07.21 	& 62:14:08.3 	& 2.484	& spec 	& c 		     & 38.26   & 9.58   & 9.59  &  4.71  & 12.58  & 12.68\\ 
N2GN\_1\_04 		& N2GN J123711+622211	&  GN20 	& PdbI 	& 2	& 12:37:11.90 	& 62:22:12.1 	& 4.055	& spec 	& d 		     & 66.28   & 9.71   & 9.91  & 19.85  & 13.29  & 13.30\\ 
N2GN\_1\_05 		& N2GN J123711+621330	&  GN1200.3	& NOEMA	& 3	& 12:37:11.33 	& 62:13:30.9 	& 1.996	& spec 	& e 		     & 21.48   & 9.31   & 9.59  &  6.20  & 12.76  & 12.82\\ 
N2GN\_1\_06$^\ddagger$ 	& N2GN J123652+621226	&  HDF850.1 	& PdBI 	& 4	& 12:36:51.99 	& 62:12:25.7 	& 5.185	& spec 	& f 		     & 11.52   & 9.03   & 9.06  &  8.27  & 12.92  & 12.96\\ %%% a.k.a. GN14?
N2GN\_1\_07 		& N2GN J123645+621448	& GN12 		& SMA 	& 1	& 12:36:46.09 	& 62:14:48.5 	& 2.960 & spec 	& g 		     & 21.07   & 8.99   & 9.41  &  7.23  & 12.85  & 13.00\\ 
N2GN\_1\_08 		& N2GN J123631+621714	& AzGN2		& SMA 	& 1	& 12:36:31.92 	& 62:17:14.7 	& 4.340	& phot 	& h 		     & 24.05   & 9.10   & 9.38  & 12.45  & 12.79  & 13.24\\ 
N2GN\_1\_09 		& N2GN J123627+621217	& AzGN10	& PdbI 	& 5,6	& 12:36:27.50 	& 62:12:17.8 	& 4.147	& spec 	& i 		     & 12.49   & 8.93   & 9.13  &  3.86  & 12.45  & 12.76\\ 
N2GN\_1\_10 		& N2GN J123635+621423	&  		& VLA 	& 7	& 12:36:35.59 	& 62:14:24.0 	& 2.005	& spec 	& j 		     & 11.21   & 9.01   & 9.11  &  6.25  & 12.73  & 12.80\\ 
N2GN\_1\_11 		& N2GN J123713+621826	& GN40 		& VLA 	& 7	& 12:37:13.89 	& 62:18:26.3 	& 3.990	& phot 	& k 		     &  8.49   & 8.90   & 9.11  & 12.90  & 13.03  & 13.16\\ 
N2GN\_1\_12\_a 		& N2GN J123636+621155	& GN11 		& PdBI 	& 6	& 12:36:37.50 	& 62:11:56.5 	& 2.590	& phot 	& h 		     & 33.69   & 8.39   & 9.11  &  5.05  & 12.27  & 13.89\\ 
N2GN\_1\_12\_b 		& N2GN J123636+621155	& GN11 		& PdBI 	& 6	& 12:36:36.10 	& 62:11:54.4 	& 4.163	& phot 	& b 		     &  1.98   & 8.15   & 9.99  & 12.36  & 13.04  & 13.14\\ 
N2GN\_1\_13 		& N2GN J123658+621451	& GN32		& NOEMA	& 3	& 12:36:58.61 	& 62:14:51.1 	& 5.181	& spec 	& l 		     &  6.34   & 8.80   & 8.88  & 18.13  & 12.71  & 13.26\\ 
N2GN\_1\_14 		& N2GN J123701+621146	& GN17		& SMA 	& 1	& 12:37:01.60 	& 62:11:45.9 	& 1.760	& spec 	& b 		     & 19.16   & 9.26   & 9.44  &  3.69  & 12.53  & 12.57\\ 
N2GN\_1\_15 		& N2GN J123618+621550	& GN6 		& SMA 	& 1	& 12:36:18.32 	& 62:15:50.7 	& 2.000	& spec 	& b 		     & 17.89   & 9.25   & 9.25  &  5.77  & 12.77  & 12.77\\ 
N2GN\_1\_16  		& N2GN J123622+621615	& GN9 		& SMA 	& 1	& 12:36:22.08 	& 62:16:16.2 	& 3.470	& phot 	& h 		     & 12.11   & 8.95   & 9.31  & 17.65  & 12.67  & 13.42\\ 
N2GN\_1\_17\_a 		& N2GN J123738+621734	& GN37 		& NOEMA	& 3	& 12:37:38.11 	& 62:17:37.1 	& 3.190	& spec 	& m 		     & 13.49   & 8.90   & 9.26  &  2.39  & 12.36  & 12.49\\ 
N2GN\_1\_17\_b 		& N2GN J123738+621734	& GN37	 	& NOEMA	& 3	& 12:37:38.09 	& 62:17:32.3 	& -- 	& --	& -- 		     &   --    & 8.20   & 9.26  &  --    & --     & --  \\ 
N2GN\_1\_18 		& N2GN J123702+621425	&  		& NOEMA	& 3	& 12:37:02.59 	& 62:14:26.7 	& 4.390	& phot 	& h 		     &  6.07   & 8.62   & 9.03  &  7.78  & 12.53  & 13.14\\ 
N2GN\_1\_19 		& N2GN J123616+621515	&  GN4		& SMA 	& 1	& 12:36:16.10 	& 62:15:13.7 	& 2.578	& spec 	& b 		     & 11.69   & 9.03   & 9.06  &  9.50  & 12.95  & 12.98\\ 
N2GN\_1\_20 		& N2GN J123550+621041	& AzGN4 	& SMA 	& 1	& 12:35:50.22 	& 62:10:42.4 	& 5.333	& phot 	& n 		     & 10.66   & 8.92   & 9.26  & 30.19  & 12.89  & 13.48\\ 
N2GN\_1\_21 		& N2GN J123634+621241	& GN26 		& VLA 	& 7	& 12:36:34.52 	& 62:12:41.0 	& 1.219	& spec 	& b 		     &  8.84   & 8.94   & 8.94  &  3.61  & 12.56  & 12.56\\ 
N2GN\_1\_22 		& N2GN J123553+621340	& AzGN13	& SMA 	& 1	& 12:35:53.23 	& 62:13:37.9 	& 2.098	& spec 	& b 		     & 33.07   & 9.30   & 9.56  &  2.59  & 12.31  & 12.44\\ 
N2GN\_1\_23 		& N2GN J123656+621207	& GN15		& NOEMA	& 3	& 12:36:56.54 	& 62:12:07.5 	& 5.179	& spec 	& o 		     &  3.84   & 8.52   & 8.76  & 10.42  & 12.87  & 13.31\\ 
N2GN\_1\_24\_a 		& N2GN J123719+621219	& AzGN27 	& NOEMA	& 3	& 12:37:18.94 	& 62:12:17.5 	& 2.697	& phot 	& b 		     &  2.38   & 7.67   & 9.30  &  1.54  & 11.95  & 12.31\\ 
N2GN\_1\_24\_b 		& N2GN J123719+621219	& AzGN27 	& NOEMA	& 3	& 12:37:19.61 	& 62:12:21.0 	& 2.817	& phot 	& b 		     &  8.18   & 7.73   & 9.68  &  1.23  & 11.68  & 12.66\\ 
N2GN\_1\_25 		& N2GN J123712+621212	& GN1200.29	& NOEMA	& 3	& 12:37:12.05 	& 62:12:11.8 	& 2.914	& spec 	& b 		     & 22.95   & 8.96   & 9.36  &  2.44  & 12.37  & 12.55\\ 
N2GN\_1\_26 		& N2GN J123740+621221	& GN38 		& SMA 	& 1	& 12:37:41.14 	& 62:12:20.4 	& 5.310	& phot 	& b 		     & 13.20   & 8.68   & 9.33  & 17.26  & 13.23  & 13.25\\ 
N2GN\_1\_27\_a 		& N2GN J123621+621709	& GN7		& NOEMA	& 3	& 12:36:21.26 	& 62:17:08.5 	& 1.988	& spec 	& b 		     &  6.85   & 8.76   & 9.03  &  3.73  & 12.49  & 12.58\\ 
N2GN\_1\_27\_b 		& N2GN J123621+621709	& GN7 		& NOEMA	& 3	& 12:36:20.98 	& 62:17:09.7 	& 1.989	& spec 	& b 		     & 16.70   & 8.77   & 9.25  &  6.30  & 12.79  & 12.80\\ 
N2GN\_1\_28 		& N2GN J123728+621920	&  		& VLA 	& 7	& 12:37:28.12 	& 62:19:20.2 	& 3.222	& spec 	& g 		     &  9.80   & 8.86   & 9.13  & 5.30  & 12.67  & 12.79\\ 
N2GN\_1\_29 		& N2GN J123702+621302	& GN18 		& VLA	& 7	& 12:37:02.57 	& 62:13:02.4 	& 2.819	& phot 	& b 		     &  6.97   & 8.60   & 8.86  &  3.77  & 12.51  & 12.62\\ 
N2GN\_1\_30 		& N2GN J123723+621714	&  		& NOEMA	& 3	& 12:37:23.52 	& 62:17:15.0 	& 2.613	& phot 	& b 		     &  3.80   & 8.57   & 8.94  &  4.89  & 12.53  & 12.71\\ 
N2GN\_1\_31 		& N2GN J123712+621035	&  AzGN20	& VLA	& 7	& 12:37:12.48 	& 62:10:35.6 	& 2.638	& phot 	& b 		     & 25.38   & 9.09   & 9.45  &  3.89  & 12.33  & 12.71\\ 
N2GN\_1\_32 		& N2GN J123629+621513	&  		& VLA	& 7	& 12:36:29.45 	& 62:15:13.2 	& 3.652	& spec 	& b 		     &  9.84   & 8.49   & 9.09  &  4.34  & 12.62  & 12.66\\ 
N2GN\_1\_33 		& N2GN J123628+621044	& GN25 		& JWST	& 8	& 12:36:29.03	& 62:10:45.5 	& 5.388	& phot 	& b 		     &  4.46   & 8.60   & 8.77  & 19.66  & 13.26  & 13.47\\ 
N2GN\_1\_34\_a 		& N2GN J123644+621938	& AzGN28	& NOEMA	& 3	& 12:36:45.67 	& 62:19:39.9 	& -- 	& --	& -- 		     &   --    & 8.57   & 9.47  &  --    & --     & --  \\ 
N2GN\_1\_34\_b 		& N2GN J123644+621938	& AzGN28 	& NOEMA	& 3	& 12:36:44.04 	& 62:19:38.5 	& 4.120	& phot 	& h 		     & 15.08   & 8.79   & 9.57  & 19.67  & 12.91  & 13.37\\ 
N2GN\_1\_35 		& N2GN J123637+620851	&  		& VLA	& 7	& 12:36:37.04 	& 62:08:52.4 	& 1.950	& phot 	& b 		     & 15.62   & 9.19   & 9.19  &  2.13  & 12.32  & 12.32\\ 
N2GN\_1\_36 		& N2GN J123658+620930	&  		& SMA 	& 1	& 12:36:58.53 	& 62:09:31.6 	& 3.930	& phot 	& h 		     &  5.85   & 8.66   & 9.03  &  8.25  & 12.90  & 12.94\\ 
N2GN\_1\_37 		& N2GN J123631+620956	&  		& VLA 	& 7	& 12:36:31.26 	& 62:09:57.7 	& 2.302	& spec 	& b 		     &  6.57   & 8.81   & 8.89  &  5.11  & 12.70  & 12.76\\ 
N2GN\_1\_38 		& N2GN J123639+621540	&  		& NOEMA	& 3	& 12:36:39.46 	& 62:15:42.6 	& 3.126	& phot 	& b 		     &  8.38   & 8.49   & 9.19  &  3.09  & 11.93  & 12.83\\ 
N2GN\_1\_39 		& N2GN J123713+621546	&  		& NOEMA	& 3	& 12:37:13.66 	& 62:15:45.3 	& 2.301	& spec 	& b 		     & 11.02   & 8.79   & 9.20  &  1.55  & 12.14  & 12.25\\ 
N2GN\_1\_40 		& N2GN J123728+621422	&  		& NOEMA	& 3	& 12:37:28.10 	& 62:14:21.7 	& 2.764	& phot 	& b 		     &  7.66   & 8.76   & 9.29  &  4.21  & 12.29  & 12.78\\ 
N2GN\_1\_41 		& N2GN J123648+621115	&  		& NOEMA	& 3	& 12:36:48.79 	& 62:11:15.5 	& 4.101	& phot 	& b 		     &  3.89   & 8.47   & 9.12  &  4.39  & 12.22  & 12.89\\ 
N2GN\_1\_42 		& N2GN J123746+621739	&  		& NOEMA	& 3	& 12:37:46.68 	& 62:17:38.7 	& 2.228	& phot 	& b 		     & 14.80   & 8.16   & 9.50  &  1.98  & 12.29  & 12.30\\ 
N2GN\_1\_43 		& N2GN J123712+621117	&  		& NOEMA	& 3	& 12:37:11.95 	& 62:11:19.7 	& 5.872	& phot 	& h 		     &  3.26   & 8.44   & 8.84  &  7.84  & 12.48  & 13.11\\  
N2GN\_1\_44 		& N2GN J123616+621232	& GNz7q		& NOEMA	& 9	& 12:36:16.92 	& 62:12:32.1 	& 7.190	& spec 	& p 		     &  0.74   & 7.84   & 8.24  &  5.42  & 12.43  & 12.73\\ 
N2GN\_1\_45 		& N2GN J123608+621435	& GN23 		& VLA 	& 7	& 12:36:08.60 	& 62:14:35.4 	& 3.131	& spec 	& q 		     &  9.41   & 8.82   & 9.10  &  3.90  & 12.49  & 12.64\\ 
N2GN\_1\_46 		& N2GN J123716+621643	&  		& NOEMA	& 3	& 12:37:16.61 	& 62:16:43.4 	& 1.894	& phot 	& b 		     &  7.59   & 8.87   & 8.87  &  3.17  & 12.50  & 12.51\\ 
N2GN\_1\_47 		& N2GN J123634+620942	&  		& NOEMA	& 3	& 12:36:34.46 	& 62:09:42.0 	& 4.873	& phot 	& b 		     &  4.67   & 8.53   & 8.79  & 10.94  & 12.82  & 13.13\\ 
N2GN\_1\_48 		& N2GN J123713+621158	& GN21 		& SMA 	& 1	& 12:37:14.03 	& 62:11:56.4 	& 2.600	& spec 	& q 		     &  9.16   & 8.76   & 9.25  & 3.17  & 12.30  & 12.56\\ 
N2GN\_1\_49 		& N2GN J123739+621558	&  		& VLA 	& 7	& 12:37:39.52 	& 62:15:58.5 	& 2.592	& phot 	& b 		     &  5.18   & 8.22   & 8.71  &  7.21  & 12.85  & 13.12\\ 
N2GN\_1\_50 		& N2GN J123648+621217	& AzGN22 	& VLA	& 7	& 12:36:48.65 	& 62:12:15.8 	& 1.990	& phot 	& r 		     &  5.93   & 8.56   & 9.02  &  1.22  & 11.98  & 12.25\\ 
N2GN\_1\_51 		& N2GN J123725+621707	&  		& NOEMA	& 3	& 12:37:25.54 	& 62:17:07.3 	& 2.013	& phot 	& b 		     &  6.93   & 8.69   & 9.33  &  2.08  & 12.30  & 12.32\\ 
N2GN\_1\_52 		& N2GN J123716+622003	&  		& NOEMA	& 3	& 12:37:16.22 	& 62:20:04.2 	& 4.060	& phot 	& n 		     &  3.00   & 7.47   & 9.24  & 13.98  & 11.39  & 13.13\\ 
N2GN\_1\_53 		& N2GN J123648+620921	&  		& NOEMA	& 3	& 12:36:48.86 	& 62:09:22.1 	& 1.994	& phot 	& b 		     &  5.52   & 8.55   & 8.84  &  1.10  & 11.98  & 12.12\\ 
N2GN\_1\_54 		& N2GN J123651+621501	&  		& NOEMA	& 3	& 12:36:52.34 	& 62:14:56.8 	& 2.885	& phot 	& b 		     &  3.02   & 8.35   & 9.09  &  8.05  & 11.63  & 12.91\\ 
N2GN\_1\_55 		& N2GN J123635+620703	& AzGN11	& VLA	& 7	& 12:36:35.80 	& 62:07:03.0 	& -- 	& --	& -- 		     &   --    & 8.61   & 9.49  &  --    & --     & --  \\ 
N2GN\_1\_56\_a 		& N2GN J123800+621615	& AzGN21 	& NOEMA	& 3	& 12:38:00.86 	& 62:16:11.7 	& 1.866	& phot 	& b 		     &  4.77   & 7.81   & 9.43  &  3.22  & 11.83  & 12.49\\ 
N2GN\_1\_56\_b 		& N2GN J123800+621615	& AzGN21	& NOEMA	& 3	& 12:38:00.26 	& 62:16:21.1 	& 2.598	& phot 	& b 		     &  7.89   & 8.89   & 8.89  &  3.24  & 12.52  & 12.52\\ 
N2GN\_1\_57 		& N2GN J123731+621618	&  		& SHARDS& 10	& 12:37:31.69 	& 62:16:16.5 	& 5.301	& phot 	& b 		     & 12.37   & 8.26   & 9.10  &  0.51  & 11.71  & 12.66\\ 
N2GN\_1\_58 		& N2GN J123702+622021	&  		& VLA	& 7	& 12:37:01.56 	& 62:20:24.8 	& 2.723	& phot 	& b 		     &  6.95   & 8.84   & 9.04  &  3.10  & 12.46  & 12.54\\ 
N2GN\_1\_59 		& N2GN J123701+621729	&  		& VLA 	& 7	& 12:37:02.12 	& 62:17:34.2 	& 3.434	& phot 	& s 		     &  2.13   & 8.15   & 8.87  &  9.22  & 12.37  & 12.97\\ 
N2GN\_1\_60 		& N2GN J123634+621924	& P-34		& SMA 	& 1	& 12:36:34.92 	& 62:19:23.7 	& 1.811	& phot 	& b 		     & 35.51   & 9.25   & 9.67  &  5.62  & 12.55  & 12.90\\ 
N2GN\_1\_61 		& N2GN J123657+621652	& P-40 		& NOEMA	& 4	& 12:36:57.48 	& 62:16:54.4 	& 5.201	& spec 	& r 		     &  2.18   & 8.09   & 9.10  &  2.72  & 11.72  & 12.67\\ 
N2GN\_1\_62 		& N2GN J123734+621723	&  		& NOEMA	& 3	& 12:37:34.51 	& 62:17:23.3 	& 0.641	& spec 	& b 		     &  5.30   & 8.62   & 8.92  &  0.50  & 11.67  & 11.72\\ 
N2GN\_1\_63 		& N2GN J123618+621008	& GN5 		& VLA	& 7	& 12:36:19.12 	& 62:10:04.3 	& 2.210	& spec 	& b 		     &  7.57   & 8.65   & 9.01  &  2.07  & 12.24  & 12.35\\ 
N2GN\_2\_20 		& N2GN J123608+621251	& GN3 		& VLA	& 7	& 12:36:08.67 	& 62:12:51.0 	& 6.670	& phot 	& b 		     &  3.70   & 8.50   & 8.58  & 18.00  & 13.19  & 13.26\\ 
N2GN\_2\_13 		& N2GN J123720+621128	&  		& NOEMA	& 3	& 12:37:21.26 	& 62:11:30.2 	& 2.071	& phot 	& b 		     & 29.18   & 9.41   & 9.59  &  3.28  & 12.50  & 12.52\\ 
\hline
\end{longtable}
\end{landscape}
}
%}
\twocolumn

% --------------------------------------------------------------------

\end{appendix}

% --------------------------------------------------------------------

\end{document}